%% file: main.tex
\pdfoutput=1 

\documentclass[11pt, a4paper, headsepline]{scrbook}

\usepackage{lmodern} 		
\usepackage[T1]{fontenc}	

\usepackage{amssymb, amsmath, color, dsfont, graphicx, float, setspace, tipa}
\usepackage[utf8]{inputenc}
\usepackage[german, english]{babel}
\usepackage[pdfpagelabels,
			pdfstartview = FitH,
			bookmarksopen = true,
			bookmarksnumbered = true,
			linkcolor = black,
			plainpages = false,
			hypertexnames = false,
			citecolor = black,
			breaklinks]{hyperref}
\usepackage{url}
\allowdisplaybreaks 		

\usepackage{rotating}
\usepackage{longtable} 		
\usepackage{xspace}
\usepackage{newclude} 		
\usepackage{pdfpages} 

\usepackage[titletoc, title]{appendix}

\usepackage{caption}
\usepackage{enumitem} 

\usepackage{setspace} 
\usepackage[sc]{mathpazo}
\usepackage{dutchcal} 


\usepackage[]{natbib}
\usepackage{listings}

\usepackage{slantsc}

\setkomafont{title}{\rmfamily\bfseries\boldmath}
\addtokomafont{section}{\rmfamily\bfseries\boldmath}
\addtokomafont{subsection}{\rmfamily\bfseries\boldmath}
\addtokomafont{subsubsection}{\rmfamily\bfseries\boldmath}
\addtokomafont{disposition}{\rmfamily} 
\setkomafont{descriptionlabel}{\rmfamily\bfseries\boldmath}

\newcommand{\del}{\partial}                            
\newcommand{\de}{\mathrm{d}}                           

\newcommand{\CONST}{\mathrm{const.\ }}                 

\newcommand{\msol}{M_\odot}                            
\newcommand{\Lsol}{L_\odot}                            
\newcommand{\order}{\mathcal{O}}                       

\newcommand{\x}{\mathbf{x}}                            
\newcommand{\V}{\mathbf{v}}                            

\newcommand{\deldx}{\frac{\del}{\del x}}				
\newcommand{\ddx}{\frac{\de}{\de x}}					
\newcommand{\DELDX}[1]{\frac{\del  #1}{\del x}}			

\newcommand{\deldr}{\frac{\del}{\del r}}				

\newcommand{\deldt}{\frac{\del}{\del t}}				
\newcommand{\ddt}{\frac{\de}{\de t}}					
\newcommand{\DELDT}[1]{\frac{\del  #1}{\del t}}			
\newcommand{\DDT}[1]{\frac{\de  #1}{\de t}}				

\newcommand{\deldxalpha}{\frac{\del}{\del x^\alpha}}
\newcommand{\DELDXALPHA}[1]{\frac{\del #1}{\del x^\alpha}}
\newcommand{\ddxalpha}{\frac{\de}{\de x^\alpha}}
\newcommand{\DDXALPHA}[1]{\frac{\de #1}{\de x^\alpha}}

\newcommand{\Aij}{$\mathcal{A}_{ij}$\xspace}	
\newcommand{\Aijm}{\ensuremath{\mathcal{A}_{ij}}}	
\newcommand{\U}{\ensuremath{\mathcal{U}}}			
\newcommand{\W}{\ensuremath{\mathcal{W}}}			
\newcommand{\F}{\ensuremath{\mathcal{F}}}			
\newcommand{\uc}{\ensuremath{\mathcal{u}}}			
\newcommand{\wc}{\ensuremath{\mathcal{w}}}			
\newcommand{\fc}{\ensuremath{\mathcal{f}}}			
\newcommand{\Fbf}{\ensuremath{\mathbf{F}}}			
\newcommand{\psitilde}{\ensuremath{\tilde{\psi}}}	
\newcommand{\half}{1/2}                 

\newcommand{\absorbers}{\text{HI, HeI, HeII}}

\newcommand{\lingo}[1]{\texttt{#1}}

\newcommand{\codename}[1]{\textsc{#1}}
\newcommand{\swift}{\codename{Swift}\xspace}
\newcommand{\swiftsimio}{\codename{Swiftsimio}\xspace}
\newcommand{\ramses}{\codename{Ramses}\xspace}
\newcommand{\acacia}{\codename{Acacia}\xspace}
\newcommand{\GEARRT}{\codename{Gear-RT}\xspace}
\newcommand{\grackle}{\codename{Grackle}\xspace}
\newcommand{\meshhydro}{\codename{Mesh-Hydro}\xspace}

\newcommand{\phew}{\codename{Phew}\xspace}

\newcommand{\exc}{\texttt{exclusive}}
\newcommand{\inc}{\texttt{inclusive}}
\newcommand{\sad}{\texttt{strictly bound}}
\newcommand{\nosad}{\texttt{loosely bound}}

\let\oldsum\sum
\renewcommand{\sum}{\oldsum\limits}

\makeatletter

\let\jnl@style=\rm
\def\ref@jnl#1{{\jnl@style#1}}

\def\aj{\ref@jnl{AJ}}                   
\def\actaa{\ref@jnl{Acta Astron.}}      
\def\araa{\ref@jnl{ARA\&A}}             
\def\apj{\ref@jnl{ApJ}}                 
\def\apjl{\ref@jnl{ApJ}}                
\def\apjs{\ref@jnl{ApJS}}               
\def\ao{\ref@jnl{Appl.~Opt.}}           
\def\apss{\ref@jnl{Ap\&SS}}             
\def\aap{\ref@jnl{A\&A}}                
\def\aapr{\ref@jnl{A\&A~Rev.}}          
\def\aaps{\ref@jnl{A\&AS}}              
\def\azh{\ref@jnl{AZh}}                 
\def\baas{\ref@jnl{BAAS}}               
\def\bac{\ref@jnl{Bull. astr. Inst. Czechosl.}}
\def\caa{\ref@jnl{Chinese Astron. Astrophys.}}
\def\cjaa{\ref@jnl{Chinese J. Astron. Astrophys.}}
\def\icarus{\ref@jnl{Icarus}}           
\def\jcap{\ref@jnl{J. Cosmology Astropart. Phys.}}
\def\jrasc{\ref@jnl{JRASC}}             
\def\memras{\ref@jnl{MmRAS}}            
\def\mnras{\ref@jnl{MNRAS}}             
\def\na{\ref@jnl{New A}}                
\def\nar{\ref@jnl{New A Rev.}}          
\def\pra{\ref@jnl{Phys.~Rev.~A}}        
\def\prb{\ref@jnl{Phys.~Rev.~B}}        
\def\prc{\ref@jnl{Phys.~Rev.~C}}        
\def\prd{\ref@jnl{Phys.~Rev.~D}}        
\def\pre{\ref@jnl{Phys.~Rev.~E}}        
\def\prl{\ref@jnl{Phys.~Rev.~Lett.}}    
\def\pasa{\ref@jnl{PASA}}               
\def\pasp{\ref@jnl{PASP}}               
\def\pasj{\ref@jnl{PASJ}}               
\def\rmxaa{\ref@jnl{Rev. Mexicana Astron. Astrofis.}}%
\def\qjras{\ref@jnl{QJRAS}}             
\def\skytel{\ref@jnl{S\&T}}             
\def\solphys{\ref@jnl{Sol.~Phys.}}      
\def\sovast{\ref@jnl{Soviet~Ast.}}      
\def\ssr{\ref@jnl{Space~Sci.~Rev.}}     
\def\zap{\ref@jnl{ZAp}}                 
\def\nat{\ref@jnl{Nature}}              
\def\iaucirc{\ref@jnl{IAU~Circ.}}       
\def\aplett{\ref@jnl{Astrophys.~Lett.}} 
\def\apspr{\ref@jnl{Astrophys.~Space~Phys.~Res.}}
\def\bain{\ref@jnl{Bull.~Astron.~Inst.~Netherlands}}
\def\fcp{\ref@jnl{Fund.~Cosmic~Phys.}}  
\def\gca{\ref@jnl{Geochim.~Cosmochim.~Acta}}   
\def\grl{\ref@jnl{Geophys.~Res.~Lett.}} 
\def\jcp{\ref@jnl{J.~Chem.~Phys.}}      
\def\jgr{\ref@jnl{J.~Geophys.~Res.}}    
\def\jqsrt{\ref@jnl{J.~Quant.~Spec.~Radiat.~Transf.}}
\def\memsai{\ref@jnl{Mem.~Soc.~Astron.~Italiana}}
\def\nphysa{\ref@jnl{Nucl.~Phys.~A}}   
\def\physrep{\ref@jnl{Phys.~Rep.}}   
\def\physscr{\ref@jnl{Phys.~Scr}}   
\def\planss{\ref@jnl{Planet.~Space~Sci.}}   
\def\procspie{\ref@jnl{Proc.~SPIE}}   

\makeatother

\usepackage{lipsum}

\newcommand{\algo}[2]{\begin{mdframed}[style=algo,frametitle={#1}] #2 \end{mdframed}}

\usepackage[framemethod=TikZ]{mdframed}

\definecolor{babyblueeyes}{rgb}{0.63, 0.79, 0.95}
\definecolor{ashgrey}{rgb}{0.7, 0.75, 0.71}
\definecolor{caribbeangreen}{rgb}{0.0, 0.8, 0.6}
\definecolor{bittersweet}{rgb}{1.0, 0.44, 0.37}

\mdfdefinestyle{todo}{%
       innerleftmargin=10,
       innerrightmargin=10,
       frametitlebackgroundcolor=bittersweet,
       frametitlerulewidth=2
}

\mdfdefinestyle{eyecatcher}{%
       innerleftmargin=10,
       innerrightmargin=10,
       frametitlebackgroundcolor=bittersweet,
       frametitlerulewidth=2
}

\mdfdefinestyle{algo}{%
       innerleftmargin=10,
       innerrightmargin=10,
       frametitlebackgroundcolor=ashgrey,
       frametitlerulewidth=2
}

\begin{document}


\setlength{\parindent}{0pt}
\setlength{\parskip}{0pt} 

\frontmatter
\include{head/titlepage}
\include{head/dedication}

\include{head/preprint_note}
\setcounter{page}{0}
\include{head/acknowledgements}

\include{head/abstracts}

\cleardoublepage
\pdfbookmark{\contentsname}{toc}
\tableofcontents

%
%

\setlength{\parskip}{1em}

\mainmatter
\include*{main/introduction}

\part{A Primer On Finite Volume Methods for Hyperbolic Conservation Laws}\label{part:finite-volume}

\include*{main/FV/FV-0-introduction}
\include*{main/FV/FV-1-hyperbolic-conservation-laws}
\include*{main/FV/FV-2-riemann-problem}

\include*{main/FV/FV-3-godunov}

\include*{main/FV/FV-4-higher-order}

\part{Finite Volume Particle Methods}\label{part:meshless}
\include*{main/Meshless/ML-0-introduction}

\include*{main/Meshless/ML-1-partition-of-unity}
\include*{main/Meshless/ML-2-methods}

\include*{main/Meshless/ML-3-implementation}

\part{Radiation Hydrodynamics}\label{part:rt}
\include*{main/RHD/RHD-1-introduction}

\include*{main/RHD/RHD-2-rt-equations}
\include*{main/RHD/RHD-3-rt-numerical-method}

\include*{main/RHD/RHD-4-validation}

\include*{main/RHD/RHD-5-conclusion}

\part{ACACIA} \label{part:ACACIA}
\include*{main/ACACIA/ACA0-introduction}
\include*{main/ACACIA/ACA1-from-halofinding-to-mocks}

\include*{main/ACACIA/ACA2-mergertree_algorithm}

\include*{main/ACACIA/ACA3-mergertree_testing}
\include*{main/ACACIA/ACA4-mergertree_testing_other_codes}
\include*{main/ACACIA/ACA5-SHAM_testing}
\include*{main/ACACIA/ACA6-conclusion}

\part*{Appendix}
\addcontentsline{toc}{part}{Appendix} 
\cleardoublepage

\begin{appendices}
\include*{appendix/app1-primitive-variable-riemann}

\include*{appendix/app2-roes-theorem}
\include*{appendix/app3-psi_gradients_full_expressions}

\include*{appendix/app4-photon-number-to-luminosity}

\include*{appendix/app5-zero-flux-nonzero-energy}

\end{appendices}

\backmatter

\cleardoublepage
\phantomsection
\addcontentsline{toc}{chapter}{Bibliography}
\bibliography{tail/thesis_references,tail/thesis_references_manually_tweaked}
\cleardoublepage

\end{document}

%% file: head/titlepage.tex
\title{GEAR-RT}
\subtitle{Towards Exa-Scale Moment Based Radiative Transfer For Cosmological Simulations Using
Task-Based Parallelism And Dynamic Sub-Cycling with SWIFT }
\author{Mladen Ivkovic}
\date{\today}


\begin{titlepage}

\null\vspace{2cm}
{\huge
    GEAR-RT: Towards Exa-Scale Moment Based

    Radiative Transfer For Cosmological

    Simulations Using Task-Based Parallelism

    And Dynamic Sub-Cycling with SWIFT

}

\vspace{1cm}

\vspace{8cm}

Mladen Ivkovic

18.02.2023

\end{titlepage}

\cleardoubleemptypage

%% file: head/dedication.tex
\cleardoublepage

\thispagestyle{empty}
\vspace*{7cm}

\begin{center}
    Dedicated to my teachers and mentors\\
    Whose blood, sweat, and tears opened doors\\
    So I may walk down this path.
\end{center}

\cleardoublepage

\thispagestyle{empty}

\textit{ 
In the beginning was the source code, and the source code was on gitlab, and the source code was 
good.}

\textit{
And I wrote: ``Let there be light!''
} \\

\textit{And the compiler said:}

\begin{lstlisting}
        In file included from rt.h:38,
                        from space.c:55:
        ./rt/GEAR/rt.h:475:40: error: 'p' undeclared here (not in a function)
        475 |   struct rt_part_data* restrict rtd = &p->rt_data;
             |                                        ^
        ./rt/GEAR/rt.h:478:3: error: expected identifier or '(' before 'for'
        478 |   for (int g = 0; g < RT_NGROUPS; g++) {
             |   ^~~
        ./rt/GEAR/rt.h:478:21: error: expected '=', ',', ';', 'asm' or '__attribute__' before '<' 
        token
        478 |   for (int g = 0; g < RT_NGROUPS; g++) {
             |                     ^
        ./rt/GEAR/rt.h: In function 'rt_kick_extra':
        ./rt/GEAR/rt.h:538:18: error: declaration of 'p' shadows previous non-variable 
        [-Werror=shadow]
        538 |     struct part* p, float dt_therm, float dt_grav, float dt_hydro,
             |     ~~~~~~~~~~~~~^
        In file included from rt.h:39:
        ./rt/GEAR/rt_iact.h: In function 'runner_iact_rt_inject':
        ./rt/GEAR/rt_iact.h:162:15: error: declaration of 'Vinv' shadows a global declaration 
        [-Werror=shadow]
        162 |   const float Vinv = 1.f / pj->geometry.volume;
             |               ^~~~
            ./rt/GEAR/rt.h:475:40: error: 'p' undeclared here (not in a function)
        475 |   struct rt_part_data* restrict rtd = &p->rt_data;
             |                                        ^
\end{lstlisting}

\textit{And the linker said:}

\begin{lstlisting}
        /usr/bin/ld: swift-swift.o: in function `main':
        swift.c:(.text.startup+0x2de2): undefined reference to `rt_cross_sections_init'
        collect2: error: ld returned 1 exit status
\end{lstlisting}

\textit{And the debugger said:}

\begin{lstlisting}
        Program received signal SIGFPE, Arithmetic exception.
        compute_interaction (pi=pi@entry=0x7fffffff7440, pj=pj@entry=0x7fffffff75a0, 
            a=a@entry=0.874422371, H=H@entry=1) at testDistance.c:45
        45	    runner_iact_gradient(r2, dx, pi->h, pj->h, pi, pj, a, H);
\end{lstlisting}

\textit{And the scheduler said:}

\begin{lstlisting}
        [00000.0] scheduler.c:scheduler_ranktasks():1871: Unsatisfiable task dependencies detected.
\end{lstlisting}

\vspace{3mm}

\textit{And I went to make another cup of coffee, for it was going to be a long night.}

%% file: head/preprint_note.tex
\cleardoublepage

\begin{small}

\chapter*{Note on this Preprint}

This is the preprint version of my PhD Thesis submitted to the \'Ecole Polytechnique F\'ed\'erale 
de Lausanne in February 2023, with some very minor changes to the printed version, mostly in order 
to remove personal data floating around the wilderness of the internet.

For quick access, here is a list of links to codes and software libraries used throughout this work.

\subsection*{This Work}

This work is available on \url{https://github.com/mladenivkovic/thesis_public}.

\subsection*{Finite Volume Code}

\meshhydro (mentioned and used in Part~\ref{part:finite-volume}) is publicly available under
\url{https://github.com/mladenivkovic/mesh-hydro}

\subsection*{Finite Volume Particle Methods}

The python module to visualise ``effective surfaces'' discussed in Part~\ref{part:meshless} is 
available on \href{https://github.com/mladenivkovic/astro-meshless-surfaces}{PyPI.org} and on 
\url{https://github.com/mladenivkovic/astro-meshless-surfaces}.

\subsection*{\GEARRT and \swift}

\GEARRT is part of \swift. \swift is available under \url{https://github.com/swiftsim/swiftsim} and 
\url{https://gitlab.cosma.dur.ac.uk/swift/swiftsim}.

Online documentation for \swift can be found on \url{swiftsim.com/docs}, and is also shipped along 
with the git repository \swift comes in. An onboarding guide for a quick start with \swift is 
provided under \url{swiftsim.com/onboarding.pdf} (and is also part of the \swift git 
repository).\\[1em]

Additional radiative transfer related tools and tests for \GEARRT and \swift, including all the 
tests following the \citet{ilievCosmologicalRadiativeTransfer2006} and 
\citet{ilievCosmologicalRadiativeTransfer2009} comparison project presented in 
Sections~\ref{chap:IL6} and \ref{chap:IL9} can be found in 
\url{https://github.com/SWIFTSIM/swiftsim-rt-tools}.\\[1em]

This work made heavy use of the \codename{swiftsimio} \citep{borrowSwiftsimioPythonLibrary2020} 
visualisation and analysis python library. It is available on 
\href{https://pypi.org/project/swiftsimio/}{PyPI.org} and on 
\url{https://github.com/SWIFTSIM/swiftsimio}. Documentation is available under 
\url{https://swiftsimio.readthedocs.io}.

\subsection*{\acacia}

\acacia is part of \ramses, and available under \url{https://bitbucket.org/rteyssie/ramses/}.

\end{small}

%% file: head/acknowledgements.tex
\chapter*{Acknowledgements}
\markboth{Acknowledgements}{Acknowledgements}
\addcontentsline{toc}{chapter}{Acknowledgements}

While my name may be displayed on the title page as the author of this work, none of this would have
been possible without the combined efforts of many others. First and foremost, I want to thank my
supervisors, Yves Revaz and Anne Verhamme, for offering me the opportunity to work on a doctorate
with them. They were nothing but kind and supportive throughout these past four years, and generous
with advice and encouragement. I also want to thank Romain Teyssier, who (aside from innumerable
teachings) referred me to this position, and who took plenty of time to show me how to write a good
paper. Many thanks also to Jean-Paul Kneib, whose support and efforts allowed the position I filled
to exist, and to my thesis defense jury, who in alphabetical order by last name were Michaela
Hirschmann, Joki Rosdahl, and Joachim Stadel, for their many helpful suggestions on how to improve
this manuscript. A very special thank you goes out to Yves Revaz and Joki Rosdahl for their
exceptionally detailed reviews of the first submission of this thesis. \\

The past four years would have been really tough without good coworkers, and I was very lucky to
have excellent ones, which I am happy to call friends today. Lo\"ic Hausammann, who was my senior
by one year and some change, was blessed by the gods with infinite knowledge and had an answer to
life, the Universe, and everything. Whether the answers were correct though was no concern of his;
as long as he had them, things were good. Admittedly, they were quite funny and entertaining though.
He also showed me around the bowels of \swift as I made my first steps into the unknown, and was
eager to help wherever he could. Florian Cabot, who is nothing short of a computational science and
visualization wizard, had the misfortune of sharing an office with me for years. (I'm still not sure
how they managed to stick the only two metalhead vegetarians at the Obs in the same corner office,
but I'm glad they did!). It was great having somebody around with many common interests, and hunting
down ridiculous bugs in our codes was much more fun while overdramatically complaining about them to
each other. And Mahsa Sanati, who has the incredible superpower to make friends in every city she
visits. Unfortunately our individual projects diverged from each other, but she never failed to
infect the room with cheerfulness. Especially when ice cream was involved. Or even mentioned. Thank
you all for all the good times we shared and all the support you offered. It was a pleasure and a
privilege to work with you. \\

I would also like to thank the fantastic people around \swift, in particular to Matthieu Schaller
and Bert Vandenbroucke, for the warm welcome they extended to me when I first started; for the
unending eagerness to offer advice, support, ideas, and help; for the review of what is probably
hundreds of merge requests. Many thanks also to Pedro Gonnet, Peter Draper, Richard Bower, Josh
Borrow, and Jacob Kegerreis. I hope we can continue working together on \swift for many years to
come. \\

My undying gratitude and love is extended to my mother and brother. I do not have the words to
express how grateful I am for all the sacrifices you took upon yourself so I could walk down this
path, so forgive me for keeping it short. The same goes to my dear dear friends, Tony Inchiparambil, Kay B\"ohringer, and Hermann Nienhaus. Without fail, they were there for me with cheer and love, song and drink, an open ear and a shoulder to lean on, in both the best of times and darkest of times. Without you, I wouldn't be who I am, nor where I am today. We partook. \\

Finally, many thanks to the HPC-EUROPA3 project for awarding me a grant to spend time taking
\GEARRT to the next level in Leiden with Matthieu. This work was carried out under the HPC-EUROPA3
project (project number: 730897) with the support of the European Commission Capacities Area -
Research Infrastructures Initiative.

\bigskip
\bigskip

So long, and thanks for all the fish! \\

\noindent\textit{18. February 2023}
\hfill M.~I.

%% file: head/abstracts.tex
\cleardoublepage
\chapter*{Abstract}
\markboth{Abstract}{Abstract}
\addcontentsline{toc}{chapter}{Abstract}

Numerical simulations have become an indispensable tool in astrophysics and cosmology. The constant
need for higher accuracy, higher resolutions, and models of ever-increasing sophistication and
complexity drives the development of modern tools which target largest computing systems and employ
state-of-the-art numerical methods and algorithms. Hence modern tools need to be developed while
keeping optimization and parallelization strategies in mind from the start.

In this work, the development and implementation of \GEARRT, a radiative transfer solver using the
M1 closure in the open source code \swift, is presented, and validated using standard tests for
radiative transfer. \GEARRT is modeled after \codename{Ramses-RT} (\cite{ramses-rt13}) with some key
differences. Firstly, while \codename{Ramses-RT} uses finite volume methods and an adaptive mesh
refinement (AMR) strategy, \GEARRT employs particles as discretization elements and solves the
equations using a Finite Volume Particle Method (FVPM).
Secondly, \GEARRT makes use of the task-based parallelization strategy of \swift, which allows for
optimized load balancing, increased cache efficiency, asynchronous communications, and a domain
decomposition based on work rather than on data.

\GEARRT is able to perform sub-cycles of radiative transfer steps w.r.t. a single hydrodynamics
step. Radiation requires much smaller time step sizes than hydrodynamics, and sub-cycling permits
calculations which are not strictly necessary to be skipped. Indeed, in a test case with gravity,
hydrodynamics, and radiative transfer, the sub-cycling is able to reduce the runtime of a simulation
by over 90\%. Allowing only a part of the involved physics to be sub-cycled is a contrived matter
when task-based parallelism is involved, and it required the development of a secondary time
stepping scheme parallel to the one employed for other physics. It is an entirely novel feature
in \swift.

Since \GEARRT uses a FVPM, a detailed introduction into finite volume methods and finite volume
particle methods is presented. In astrophysical literature, two FVPM methods are written about:
\cite{hopkinsGIZMONewClass2015} have implemented one in their \codename{Gizmo} code, while the one
mentioned in \cite{ivanovaCommonEnvelopeEvolution2013} isn't used to date. In this work, I test an
implementation of the \cite{ivanovaCommonEnvelopeEvolution2013} version, and conclude that in its
current form, it is not suitable for use with particles which are co-moving with the fluid, which
in turn is an essential feature for cosmological simulations.

Finally, the implementation of \acacia, a new algorithm to generate dark matter halo merger trees
with the AMR code \ramses, is presented.
As opposed to most available merger tree tools, it works on the fly during the course of the
$N$-body simulation. It can track dark matter substructures individually using the index of the most
bound particles in the clump.
Once a halo (or a sub-halo) merges into another one, the algorithm still tracks it through the last
identified most bound particle in the clump, allowing to check at later snapshots whether the
merging event was definitive. 
another one.
The performance of the method is compared using standard validation diagnostics, demonstrating that
it reaches a quality similar to the best available and commonly used merger tree tools. As proof of
concept, \acacia is used together with a parameterized stellar-mass-to-halo-mass relation to generate
a mock galaxy catalog that shows good agreement with observational data.

\hspace{1cm}

\textbf{Keywords}: Numerical Methods -- Numerical Simulations -- Finite Volume Methods -- Finite
Volume Particle Methods -- Meshless Methods -- Radiative Transfer -- Parallel Computing -- Galaxies
Formation -- Galaxies Evolution -- Merger Trees -- Dark Matter

\cleardoublepage

\begin{otherlanguage}{german}
\chapter*{Zusammenfassung}
\markboth{Zusammenfassung}{Zusammenfassung}

Numerische Simulationen sind zu einem unabdiglichen Werkzeug in Astrophysik und Kosmologie
geworden. Der best\"andige Bedarf nach h\"oherer Genauigkeit, nach h\"oherer Aufl\"osung, und nach
Modellen mit stetig steigender Rafinesse und Komplexit\"at treibt die Entwicklung von modernen
Werkzeugen an, welche die gr\"ossten Rechnersysteme anzielen und von numerischen Methoden und
Algorithmen des neusten Standes Gebrauch machen. Statt sich nur auf die Physik und die Modelle zu
konzentrieren, m\"ussen moderne Werkzeuge von Beginn an unter Beachtung von Optimisation und
Parallelisationsstrategien entwickelt werden.

In dieser Arbeit wird die Entwicklung und Implementation von \GEARRT pr\"asentiert, das
Strahlungs\"ubertragung mittels der M1 Schliessung l\"ost, und durch Standardtests f\"ur
Strahlungs\"ubertragung validiert. \GEARRT ist nach dem Vorbild von \codename{Ramses-RT}
(\cite{ramses-rt13}) geformt, jedoch mit einigen entscheidenden Unterschieden. Erstents benutzt
\codename{Ramses-RT} Finite-Volumen-Verfahren und eine adaptive Gitterverfeinerungsstrategie
(Adaptive Mesh Refinement, AMR), w\"ahrend \GEARRT Teilchen als Diskretisationselemente einsetzt
und die Gleichungen mittels eines Finite-Volumen-Teilchen-Verfahrens (FVTV) l\"ost. Zweitens
benutzt \GEARRT die aufgabenbasierte Parallelisationsstrategie von \swift, welche eine optimisierte
Lastbalancierung, verbesserte Cache-Effizienz, asynchrone Kommunikationen, und eine arbeitsbasierte
Dom\"anendekomposition statt einer datenbasierten erlaubt.

\GEARRT ist f\"ahig Unterzyklen der Strahlungs\"ubertragung im Bezug auf einzelne hydrodynamische
Schritte zu t\"atigen. Strahlung ben\"otigt viel kleinere Zeitschritte als Hydrodynamik, und der
Einsatz von Unterzyklen erlaubt es Schritte, welche nicht streng n\"otig sind, zu \"uberspringen.
In einem Versuch mit Gravitation, Hydrodynamik, und Strahlungs\"ubertragung war der Einsatz von
Unterzyklen imstande, die Laufzeit der Simulation \"uber 90\% zu reduzieren.
Nur einen Teil der beteiligten Physik in Unterzyklen zu l\"osen ist eine vertrackte Angelegenheit
unter Anbetracht der aufgabenbasierten Parallelisationsstrategie, und erforderte die Entwicklung
einer sekund\"aren Zeitschrittmethode, die parallel zu der anderen (welche die restliche Physik
l\"ost) l\"auft. Der Einsatz von Unterzyklen ist eine vollst\"andig neue Funkion in \swift.

Da \GEARRT ein FVTV einsetzt, ist eine detaillierte Einf\"uhrung in Finite-Volumen- und
Finite-Volumen-Teilchen-Verfahren gegeben. In der Literatur der Astrophysik wird \"uber zwei FVTV
geschrieben: \cite{hopkinsGIZMONewClass2015} haben eine in ihrem \codename{Gizmo} Code implementiert,
w\"ahrend diejenige, welche in \cite{ivanovaCommonEnvelopeEvolution2013} erw\"ahnt wird, bis heute
nicht benutzt wird. In dieser Arbeit teste ich eine Implementation der
\cite{ivanovaCommonEnvelopeEvolution2013}-Version, und komme zum Schluss, dass es in der heutigen
Form unbrauchbar f\"ur den Einsatz mit Teilchen ist, welche mit dem Fluid mitbewegt werden, was
von essentieller Wichtigkeit in kosmologischen Simulationen ist.

Letztendlich wird die Implementation von \acacia, einem neuen Algorithmus um Verschmelzungsb\"aume
(Merger Trees) von Halos bestehend aus dunkler Materie, im AMR Code \ramses pr\"asentiert. Im
Unterschied zu den meisten verf\"ugbaren Merger Tree Werkzeugen arbeitet \acacia noch w\"ahrend die
$N$-Teilchen-Simulation verl\"auft. Es kann Unterstrukturen von dunkler Materie durch die Indizes
von den am st\"arksten gebundenen Teilchen der Klumpen einzeln verfolgen. Wenn ein Halo (oder
Unter-Halo) mit einem anderen verschmilzt, wird es vom Algorithmus durch den zuletzt als am
st\"arksten gebunden identifizierten Teilchen weiterhin verfolgt. Dies erm\"oglicht es in sp\"ateren
Speicherabz\"ugen zu pr\"ufen, ob die Verschmelzung tats\"achlich definitiv war. Die Leistung der
Methode wird mittels Standardvalidationsdiagnostiken verglichen, und zeigt, dass es vergleichbare
Qualit\"at mit den besten verf\"ugbaren und h\"aufig benutzten Merger Tree Werkzeugen liefert. Als
Konzeptnachweis wird \acacia zusammen mit einer parametrisierten Stern-Masse-Zu-Halo-Masse-Beziehung
eingesetzt, um k\"unstliche Galaxienkataloge zu erstellen, welche gute \"Ubereinstimmung mit
observierten Daten aufweisen.

\end{otherlanguage}

%% file: main/introduction.tex
\chapter*{Introduction}
\addcontentsline{toc}{chapter}{Introduction}
\markboth{Introduction}{Introduction}

A particularly intriguing period of the young Universe is the Epoch of Reionization, where the first
light emitting sources begin to form and gradually ionize the neutral gas surrounding them,
eventually fully ionizing the intergalactic medium itself.
Minutes after the Big Bang \citep[e.g.][]{moGalaxyFormationEvolution2010} the Universe cools off and
is rarefied sufficiently for primordial nucleosynthesis to begin, where protons, neutrons, helium,
and traces of other elements begin to form for the first time. As the rapid early expansion of the
Universe progresses, the energy and matter densities everywhere become increasingly diluted, and the
temperature of the Universe proceeds to decrease. A couple of hundreds of thousands of years later,
it has cooled off enough for the ions to combine with the free electrons during the so-called
``recombination'' epoch, making the Universe largely neutral. After recombination, since the number
of free electrons decreased steeply, the present radiation decouples from the thermodynamic
equilibrium with matter held into place through Compton scattering, and is today observable as the
Cosmic Microwave Background \citep[CMB,][]{peeblesPrinciplesPhysicalCosmology1993b}, albeit
redshifted due to the ongoing expansion to a blackbody spectrum with temperature
$\sim2.7$K. The Universe then finds itself in the ``Cosmic Dark Ages'', as no sources of light are present. In the meantime, first cosmological structures begin to form rooted in tiny perturbations and overdensities, which are observable in the CMB, in the initially homogeneous and isotropic matter distribution. The perturbations grow under the attractive influence of gravity, where overdensities gradually accrete matter into what will later become massive halos of dark matter (DM), populated with clusters of galaxies.

The Epoch of Reionization begins with the Cosmic Dawn, the appearance of first luminous sources which emit ionizing UV radiation. The first stars, so-called Pop III stars, are metal-free and predicted to be very massive and very luminous, but short lived with lifetimes in the order of a few Myr \citep[e.g.][]{schaererPropertiesMassivePopulation2002}. First gravitationally bound gas clouds are expected to form around redshift $z \sim 30 - 40$, while first stars are expected a bit later, around $z \sim 30$, due to the inefficient cooling channels of the metal-free primordial gas \citep[e.g.][]{gloverFormationFirstStars2005}. The early phase of the reionization proceeds slowly,
as more and more radiative sources appear and the density of the IGM decreases through the
continuous expansion of the Universe. The sources are clustered in space, and hence reionization
proceeds in a patchy fashion, with ionized HII regions forming first around what are essentially
point sources on cosmological scales. Once the separate ionized bubbles overlap, the reionization
proceeds much faster, as the ionizing radiation can propagate unperturbed over large regions of
space through the optically thin ionized IGM, eventually fully ionizing the entire IGM by redshift
$z \sim 6$ \citep{robertsonGalaxyFormationReionization2022}, around one billion years after the Big
Bang.

The nature of the main drivers of Cosmic Reionization is still debated. Less massive dwarf galaxies are more numerous than bigger and more massive galaxies, but also less luminous. Furthermore, massive galaxies contain denser gas, which may reduce the amount of escaped ionizing radiation because the recombination rate of the ions is directly proportional to their number density, and thus a bigger fraction of radiation is used to combat the recombination of already ionized gas. On the other hand, more massive galaxies are more likely to contain an Active Galactic Nucleus, which is an additional strong source of ionizing radiation. In any case, one key question with regards to Cosmic Reionization is how much ionizing radiation escapes the galaxies and is able to ionize and heat the IGM. Usually the amount of escaped radiation is described as the ``escape fraction''
$f_{\mathrm{esc}}$. Unfortunately, this quantity is notoriously difficult to measure
observationally, calculate theoretically, and to simulate.
Accurate and detailed analytical predictions of the Epoch of Reionization are unthinkable due to
the sheer complexity of the problems involved. Detailed modeling of Cosmic Reionization requires the treatment of non-linear physical processes of galaxy formation and evolution. The most notable of these processes are gravity, fluid dynamics, star formation and stellar feedback, and, in particular with regards to the Epoch of Reionization, radiative transfer and thermochemistry. To this end, we need to resort to numerical simulations to solve the equations that govern the processes of structure formation and evolution. Any direct detection of the escape of ionizing Lyman Continuum (LyC) photons from galaxies at $z > 5$ is impossible due to the increasing opacity of the IGM \citep[e.g.][]{inoueUpdatedAnalyticModel2014}. Hence galaxies at the Epoch of Reionization are not observable in LyC, and we have to rely on indirect measurements and on simulations for theoretical predictions. Furthermore, simulations and the generation of mock galaxy catalogues in the Epoch of Reionization will furthermore be necessary to test and validate proposed indirect tracers of the ionizing photons in observations, as direct observation is not an option.
Recent simulations however face difficulties to find agreement on the escape fraction with respect
to the galaxy mass  \citep[e.g.][]{wiseBirthGalaxyIII2014a,paardekooperFirstBillionYears2015,xuGalaxyPropertiesUV2016,
rosdahlSPHINXCosmologicalSimulations2018,yehTHESANProjectIonizing2023} because the propagation of
the ionization front in the interstellar medium (ISM) of galaxies depends strongly on the
distribution of the ionizing sources and on the structure of the ISM itself. Both these factors are realized in simulations through sub-grid models of star formation and stellar feedback, which can vary broadly between models \citep[see
e.g.][]{rosdahlSnapCracklePop2017,kimAGORAHighresolutionGalaxy2016,
roca-fabregaAGORAHighresolutionGalaxy2021b}. Furthermore, stellar feedback was shown to
have a pivotal role in regulating the escape fraction  \citep{trebitschFluctuatingFeedbackregulatedEscape2017}. Clearly much effort also needs to be provided from the simulation side to tackle the mysteries of the Epoch of Reionization as well in the upcoming years.


In addition to the complexity of the underlying physics that needs to be solved, state-of-the-art
simulation codes need to be fast, efficient, and able to solve huge problems on supercomputing
infrastructure. The codes need to be able to incorporate sophisticated models while maintaining a
sufficiently high mass and spatial resolution of the results. Simultaneously, they need to cover
large enough volumes to be able to generate statistically representative samples. As such, the
underlying optimization strategies are a vital concern for state-of-the-art simulation software,
and need to be taken into account from the very beginning of the development process.

Motivated by the expected new observational insights into the Epoch of Reionization thanks to the
recently launched James Webb Space Telescope and the upcoming Square Kilometer Array (SKA) facility,
as well as the anticipated exa-scale computing facilities in the upcoming years, the goal of my
thesis was to design a radiative transfer solver for the new simulation code \swift. Notably \swift
employs a task-based parallelization strategy, uses asynchronous communications, and is geared
towards making optimal use of current state-of-the-art and future supercomputing infrastructure.
\swift is an open source code and can be downloaded from
\url{https://gitlab.cosma.dur.ac.uk/swift/swiftsim}.

The radiative transfer solver I implemented, named \GEARRT, solves the moments of the equation of
radiative transfer and the closure condition called ``M1''. It is modeled after the method used in
the \ramses adaptive mesh refinement simulation code (\cite{ramses-rt13}). The major advantages of
using this method are that it is a well known and studied method, and that it scales well
independently of the number of sources of radiation in the simulation. However, contrary to the
\ramses implementation, my implementation in \swift uses (Lagrangian) particles. Indeed in its
default modes, \swift utilizes particles as discrete elements. To be able to solve the equations of
radiative transfer in a similar fashion, I make use of the Finite Volume Particle Methods (FVPM).
FVPM allow to solve partial differential equations that take the form of hyperbolic conservation
laws using particles as discretization elements, while not needing to construct a mesh to exchange
fluxes between particles as is done in e.g. moving mesh methods.

The notion of hyperbolic conservation laws and strategies for their solution using numerical
techniques is a core topic throughout this work. Indeed both the equations of fluid dynamics, i.e.
the Euler equations, as well as the equations of radiative transfer and the moments thereof can be
formulated as hyperbolic conservation laws. However, in general exact analytical solutions are not
available, and we need to resort to numerical methods like the finite volume and finite volume
particle methods. They are very well suited for hyperbolic conservation laws. Many discretization
methods and numerical solution strategies have been developed to date, each with its own advantages
and caveats. As such, finite volume methods come with a wealth of complexities and subtleties. Since
they take such a prominent role throughout this work, the first part of my thesis aims to establish
a rudimentary overview and understanding of how finite volume methods in the context of hyperbolic
conservation laws work, and what complexities and limitations they contain. The outline follows
selected topics from \citet{toroRiemannSolversNumerical2009} and
\citet{levequeFiniteVolumeMethods2002}. The theoretical aspects in Part~\ref{part:finite-volume}
are presented along with results of selected experiments using an extensive stand-alone didactic
finite volume solver I have written, called \meshhydro. \meshhydro is intended as a didactic
complement to the theoretical background of finite volume methods, and to provide future students
with a working solver that can be tested, played, and experimented with. To this end, the
implementation is kept simple, a detailed documentation of the code and the implemented equations is
provided, and plenty of examples with reference solutions are available. \meshhydro is open source
software and available on \url{https://github.com/mladenivkovic/mesh-hydro}.

Using finite volumes as a fundamental building block, a detailed introduction to Finite Volume
Particle Methods, which are the ones used by \GEARRT, are given in Part~\ref{part:meshless}.
However, Part~\ref{part:meshless} focuses only on the application of FVPM on hydrodynamics. In
astrophysical literature, two FVPM methods are mentioned: \cite{hopkinsGIZMONewClass2015} have
implemented one in their \codename{Gizmo} code, while the one mentioned in
\cite{ivanovaCommonEnvelopeEvolution2013} isn't used to date. In this work, I test an
implementation of the \cite{ivanovaCommonEnvelopeEvolution2013} version in \swift and compare the
results with the implementation of the method following \cite{hopkinsGIZMONewClass2015}.
Furthermore, a detailed discussion of the implementation of the FVPM applied to fluid dynamics is
provided, as the implementation of the radiative transfer with \GEARRT is intimately coupled to the
hydrodynamics.

\GEARRT and the radiative transfer is then the topic of Part~\ref{part:rt} of this thesis, where a
detailed description of the method and its implementation is given.

As mentioned above, \swift makes use of a task-based parallelism strategy. In short, that means
that it splits the entire problem into any number of subsets (e.g. any number of arbitrary groups
of particles). These subsets define the units of work (tasks) that need to be completed. Alongside
the tasks, conflicts and dependencies between them need also to be defined, which ensures the
correct order of execution of the tasks and establishes which tasks may be run concurrently in any
order. This strategy allows processing units (e.g. cores of the central processing unit) to grab and
work on any tasks that are currently available. An advantage of this strategy is that it minimizes
the idle waiting time of processors because there are no global synchronization points between all
processors during a time step. Furthermore, MPI communications can be run using asynchronous calls
while processors can keep busy with other work until the MPI messages arrive. A caveat of the method
is however that all the underlying tasks, dependencies, and conflicts need to be defined and
implemented manually by the developers first. This implementation is prone to logical errors, as
many special and corner cases may occur during a simulation that require additional attention and
are near impossible to predict. A further complication is that the exact order of execution of the
tasks is in general not reproducible between any two runs using multiple processors. The complexity
to define dependencies and conflicts correctly in a way that covers all possible corner cases
combined with the non-reproducible nature of the execution make the development of the additional
tasking system very tricky and time consuming. The tasks and dependencies required for a
moment-based radiative transfer solver are now in place in \swift, and have been fully tested
extensively and rigorously. Any future radiative transfer solvers for \swift will make use of the
infrastructure I implemented, allowing developers to focus on the physical aspects of the problem
rather than computational ones. While the essentials of task-based parallelism and its application
for hydrodynamics in \swift is discussed in Part~\ref{part:meshless}, the additional tasks and
dependencies required for radiative transfer are described in Part~\ref{part:rt}.

A significant effort during my thesis was dedicated to the additional implementation to allow \swift
to ``sub-cycle'' the radiative transfer integration step with respect to a hydrodynamics integration
step. The maximally allowable time step size is determined by the so-called ``CFL'' stability
condition, which depends on the propagation velocity of the conserved quantity. When comparing the
highest occurring gas velocities to the propagation velocity of radiation, the speed of light, they
will be several orders of magnitude lower, thus making the maximal radiation time steps several
orders of magnitude lower than the hydrodynamic ones. The core idea of sub-cycling is to then
integrate a (high) number of radiation time steps during a single hydrodynamics time step, and in
this way omit all the needlessly performed hydrodynamics integrations we would have to do otherwise.
This required to completely decouple the radiative transfer module in SWIFT from the hydrodynamics,
and add a second independent internal time integration engine to be able to execute the radiative
transfer sub-cycles correctly. The sub-cycling scheme still allows for particles to have individual
independent time step sizes, which are determined by the local state and environment of the
particles as opposed to some global criterion. The sub-cycling is a completely novel feature in
\swift, and has been rigorously tested. While currently only the radiative transfer can be
sub-cycled, the algorithm can also be extended for any other physics that might require it. As
for performance, in idealized tests with hydrodynamics, self-gravity, and radiative transfer, the
sub-cycling is able to reduce the time-to-solution by over 90$\%$. A publication of the sub-cycling
scheme is currently in preparation. The sub-cycling scheme is also discussed in Part~\ref{part:rt}.

Beside the main part of my thesis, some effort was directed towards completing \acacia, a
novel method to create on-the-fly merger trees in the \ramses code, which constitutes the final
Part~\ref{part:ACACIA} of this thesis. It can track dark matter substructures individually using
the index of the most bound particle in the clump. Once a halo (or a sub-halo) merges into another
one, the algorithm still tracks it through the last identified most bound particle in the clump,
allowing to check at later snapshots whether the merging event was definitive, or whether it was
only temporary, with the clump only traversing another one. The same technique can be used to track
orphan galaxies that are not assigned to a parent clump any more because the clump dissolved due to
numerical overmerging. At the time of first submission, \ramses was the only publicly available
simulation code which was able to create merger trees on-the-fly. A paper describing and testing
\acacia is now published. The \ramses code is also publicly available and can be downloaded from
\url{https://bitbucket.org/rteyssie/ramses/}.

%% file: main/FV/FV-0-introduction.tex
\chapter{Introduction}

Conservation laws take a prominent role in physics. They pop up everywhere where some conserved
quantity evolves dynamically, i.e. has fluxes, and allow us to establish a formal description of
physical systems. Conserved quantities themselves are beloved for the same reason, and it helps that
there is no scarcity of them. In fact, the Noether theorem
\citep{noetherInvarianteVariationsprobleme1918} states that for each symmetry of action of a system,
a conserved quantity (and a corresponding conservation law) can be found. For example, time
invariant systems conserve energy, translationally invariant systems conserve momentum, and
rotationally invariant systems conserve angular momentum. Some concrete examples of conservation
laws are the charge conservation in electrical currents, the Jeans equations, the heat flow in an
uniform rod, the traffic flow equation, and the shallow water equations. In this part of the
thesis, I will focus mainly on two sets of conservation laws: The first are the equations of linear advection, whose simple form allows to find exact solutions for many problems arising in numerical
methods which solve conservation laws, such as the finite volume methods. The second set that will
be a major focus in this Part of this work are the Euler equations, which describe the mass,
momentum, and energy conservation of an ideal fluid. While the linear advection equation in this
Part will mostly be used as a model and example for a conservation law, the solution of Euler
equations constitutes an essential building block for simulations of astrophysical systems. Details
of the linear advection equation and the Euler equations are discussed in the subsequent sections.
Finally, in Part \ref{part:rt} the moments of the equations of radiative transfer, which also take
form of a hyperbolic conservation law, are discussed and solved. This part of the work aims to
establish a rudimentary overview and understanding of how finite volume methods in the context of
hyperbolic conservation laws work, and what complexities and limitations they contain. The outline
follows selected topics from \citet{toroRiemannSolversNumerical2009} and
\citet{levequeFiniteVolumeMethods2002}. The content won't be reserved to only theory and equations
though - where applicable, actual results from simulations using the \meshhydro code are
shown and discussed. \meshhydro is a code that I have developed in scope of this work to
serve as a simple didactical tool. It was created to solve linear advection as well as the Euler
equations using various Riemann solvers, slope and flux limiters, integration schemes, and even
external source terms. Furthermore, it is extensively documented, it contains a range of
ready-to-run examples, and comes with a suite of visualization and analysis tools written in Python
3. It is however limited to discretizing space in a regular grid only, and only in one or two
dimensions. It is also not parallelized at all. \meshhydro is open source software and
available on \url{https://github.com/mladenivkovic/mesh-hydro}.

%% file: main/FV/FV-1-hyperbolic-conservation-laws.tex
\chapter{Hyperbolic Conservation Laws}

First, let's get a little more concrete.
Formally, conservation laws are systems of partial differential equations (PDEs) that can be
written in the form
\begin{align}
  \deldt \U + \deldx \F(\U) &= 0 \\[.5em] \label{eq:conservation-law-1D-introduction}
  \text{with } \U = \left( \begin{matrix}
                    \uc_1 \\ \uc_2 \\ \vdots \\ \uc_m \\
                    \end{matrix} \right), & \quad \quad
                \F =\left( \begin{matrix}
                    \fc_1 \\ \fc_2 \\ \vdots \\ \fc_m \\
                    \end{matrix} \right)
\end{align}
where $\U$ is called the vector of conserved states, and $\F$ is the vector of fluxes, and in
general is a function of the state vector. This definition of conservation laws is valid for one
spatial dimension. However, the extension to multiple dimensions is straightforward: Instead of
being a vector, the flux becomes a tensor, and the partial spatial derivative is a divergence
operator instead.

A PDE of the form
\begin{align}
    \deldt &\U + \mathcal{A} \deldx \U + \mathcal{B} = 0 \\[.5em]
  \text{with }  &\U = \left( \begin{matrix}
                    u_1 \\ u_2 \\ \vdots \\ u_m \\
                    \end{matrix} \right),  \quad \quad
                \mathcal{B} = \left( \begin{matrix}
                    b_1 \\ b_2 \\ \vdots \\ b_m \\
                    \end{matrix} \right), & \quad \quad
                \mathcal{A} =\left( \begin{matrix}
                    a_{11} & \ldots & a_{1m} \\
                    \vdots & \ddots & \vdots \\
                    a_{m1} & \ldots & a_{mm} \\
                    \end{matrix} \right)
\end{align}
is called

\begin{itemize}
 \item linear with constant coefficients if $a_{ij} = \CONST$ and $b_i = \CONST$
 \item linear with variable coefficients if $a_{ij} = a_{ij}(x, t)$ and $b_i = b_i(x, t)$ or
$\mathcal{B} = \alpha + \beta \U$
 \item  quasi-linear if $\mathcal{A} = \mathcal{A}(\U)$
 \item homogeneous if $\mathcal{B} = 0$
\end{itemize}

A conservation laws like \ref{eq:conservation-law-1D-introduction} can be written in homogeneous
quasi-linear form:

\begin{align}
 \DELDT{\U} + \DELDX{F} = 0 = \DELDT{\U} + \frac{\del \F}{\del \U} \DELDX{U} = \DELDT{U} +
\mathcal{A}(\U) \DELDX{U}
\end{align}

where $\mathcal{A}$ is the Jacobian matrix of the flux function $\F(\U)$ and has the elements
$a_{ij} = \frac{\del f_i}{\del u_j}$.

Furthermore, a conservation law \ref{eq:conservation-law-1D-introduction} is said to be hyperbolic
at a point $(x, t)$ if $\mathcal{A}$ has $m$ real eigenvalues $\lambda_1$, ..., $\lambda_m$ and a
corresponding set of $m$ linearly independent right eigenvectors $\mathbf{K}^{(1)}$, ...,
$\mathbf{K}^{(m)}$. If the $\lambda_i$ are all distinct, the system is called strictly hyperbolic.
The fact that the conservation laws under consideration are hyperbolic, i.e. have $m$ eigenvalues
and linearly independent eigenvectors, is a crucial element for the construction of the numerical
solution strategy using finite volume methods.

As mentioned before, this part of the work will focus on the linear advection equation and the
Euler equation which govern ideal gases. The linear advection equation in its general,
time-dependent form in three dimensions is given by:

\begin{align}
  \deldt{\uc} +
    a(\x, t) \deldx{\uc} +
    b(\x, t) \frac{\partial}{\partial y} \uc +
    c(\x, t) \frac{\partial}{\partial z} \uc
    = 0 \ .
    \label{eq:linear-advection-general}
\end{align}

If the coefficiens are sufficiently smooth (i.e. differentiable), it can be expressed as a
conservation law with source terms:

\begin{align}
  \deldt{\uc} +
    \deldx{(a \uc)} +
    \frac{\partial}{\partial y} (b\uc) +
    \frac{\partial}{\partial z} (c \uc )
    = \uc \left(
    \deldx{a} +
    \frac{\partial}{\partial y} b +
    \frac{\partial}{\partial z} c
    \right) \ .
\end{align}

In its simplest form, which is also the form that will be made heavy use of in this work, we only
consider one dimension and a constant coefficient $a$:

\begin{align}
    \deldt{\uc} + a \deldx{\uc} = 0 \label{eq:linear-advection-1D-const-coeff}
\end{align}

The Euler equations are given by:

\begin{align}
	\deldt
	\begin{pmatrix}
		\rho \\
		\rho \V \\
		E
	\end{pmatrix}
	+
	\nabla \cdot
	\begin{pmatrix}
		\rho \V\\
		\rho \V \otimes \V + p\\
		(E + p) \V
	\end{pmatrix}
	=
	\begin{pmatrix}
		0\\
		\rho \mathbf{a}\\
		\rho \mathbf{a} \V
	\end{pmatrix}
	\label{eq:euler-equations}
\end{align}

where
\begin{itemize}
	\item $\rho$ is the fluid density
	\item $\V = (v_1, v_2, v_3)^T$ is the fluid velocity at a given point. This is the mean
or bulk velocity of the fluid, not the velocity of individual particles that compose the fluid.
	\item $p$ is the pressure
	\item $E$ is the specific energy. $E = \frac{1}{2} \rho \V^2 + \rho u$, with $u$ being the
specific internal thermal energy.
	\item $\mathbf{a}$ is an acceleration due to some external force.
\end{itemize}

The outer product $\cdot \otimes \cdot$ gives the following tensor:
\begin{align}
	(\V \otimes \V)_{ij} = v_i v_j
\end{align}

Furthermore, for ideal gases, we have the equation of state:
\begin{align}
	p &= n k_B T && \text{ideal gas law} \\
\end{align}
which can also be written as:
\begin{align}
	u &= \frac{1}{\gamma - 1}\frac{p}{\rho} \label{eq:equation-of-state}
\end{align}
and the following relations:
\begin{align}
	c_s &= \sqrt{\frac{\del p}{\del \rho} \bigg{|}_s } = \sqrt{\frac{\gamma p}{\rho}} &&
\text{sound speed of the gas} \\
	s &= c_V \ln \left( \frac{p}{\rho^\gamma} \right) + s_0 && \text{entropy} \label{eq:entropy}\\
	p &= C \rho ^ \gamma && \text{entropy relation for smooth flow, i.e. no shocks}
\end{align}

where
\begin{itemize}
	\item $n$ is the number density of the gas
	\item $k_B$ is the Boltzmann constant
	\item $T$ is the temperature
	\item $s$ is the entropy
	\item $\gamma$ is the adiabatic index
    \item $c_V$ is the specific heat capacity at constant volume
\end{itemize}

In 1D, we can write the Euler equations without source terms ($\mathbf{a} = 0$) and dropping the
index $1$ from the velocity term $v_1$ as:

\begin{align}
	\deldt{
		\begin{pmatrix}
			\rho \\ \rho v \\ E
		\end{pmatrix}
		}
	+ \deldx {
		\begin{pmatrix}
			\rho v\\
			\rho v^2  + p\\
			(E + p) v
		\end{pmatrix}
	} = 0 \ .
	\label{eq:euler-equations-1D}
\end{align}

%% file: main/FV/FV-2-riemann-problem.tex
\chapter{The Riemann Problem and Riemann Solvers} \label{chap:riemann}

\section{The Riemann Problem and Solution Strategy}

\begin{figure}[H]
\centering
\includegraphics[width=.6\linewidth]{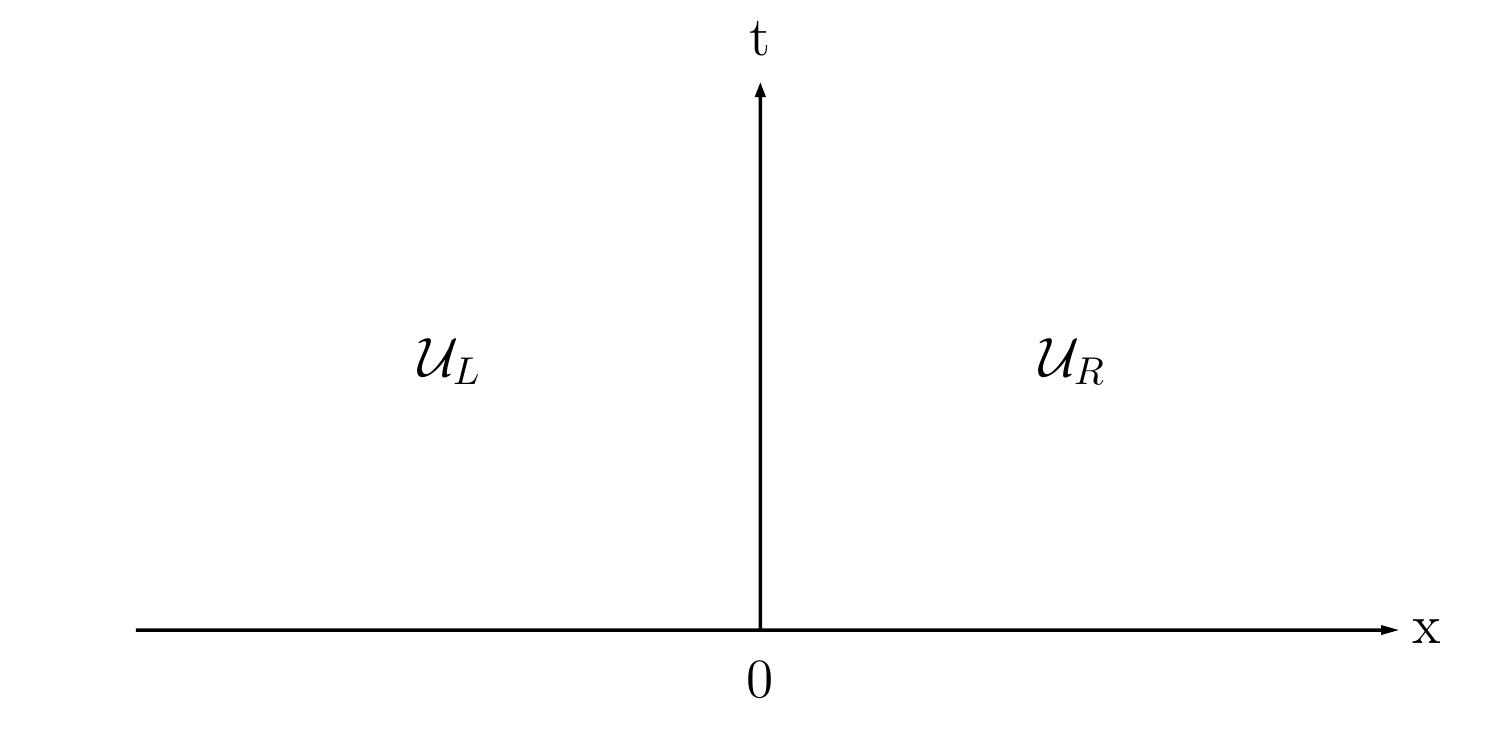}%
\caption{
    The Riemann Initial Value Problem (IVP) in 1D. Two constant initial states, $\U_L$ and $\U_R$,
are initially separated at $x = 0$.
}
\label{fig:riemann-problem}
\end{figure}

At the heart of finite volume fluid dynamics is the solution to a specific initial value problem
(IVP) called the ``Riemann problem'', as we will see shortly. First, let us define the Riemann
problem: For a hyperbolic system of conservation laws of the form

\begin{align}
	\DELDT{\U} + \DELDX{\F(\U)} = 0
\end{align}

the Riemann problem is defined as

\begin{align}
	\U(x, t=0) =
		\begin{cases}
			\U_L & \text{ if } x < 0\\
			\U_R & \text{ if } x > 0\\
		\end{cases}
\end{align}

which is shown in Figure~\ref{fig:riemann-problem}.

In finite volume methods, the fluid (or rather, the conserved quantities) will be discretized
in space by splitting the entire volume into a number of cells (sub-volumes of finite size), each
covering a small fraction of the total volume. Each cell will also keep track of the states $\U$
within their own volume individually. The Riemann problem and the solution thereof enters the game
once we look at what the situation between two adjacent cells is: If we call one of the two adjacent
cells ``the left one'' and the other ``the right one'', then we have \emph{exactly} the Riemann
problem again: Setting $x = 0$ at the interface between two cells, then the state of the left cell
constitutes $\U_L$, while the state of the right cell provides $\U_R$ for the Riemann problem. So in
order to evolve the simulation in time, we need the solution of this Riemann problem in time. In
this chapter, the solution strategies for the Riemann problem are discussed. It begins with the
simple case of the linear advection equation with constant coefficients (eq.
\ref{eq:linear-advection-1D-const-coeff}), and builds up over linear hyperbolic systems all the way
to hyperbolic conservation laws. Later, in Chapter~\ref{chap:godunov}, we will make use of the
Riemann solvers introduced in this Chapter to obtain a numerical method to evolve the Euler
equations for arbitrary initial conditions over any desired interval of time.

\subsection{The Riemann Problem For The Linear Advection Equation With Constant Coefficients}

The solution to the Riemann problem for the linear advection equation with constant coefficients
(eq. \ref{eq:linear-advection-1D-const-coeff}) is somewhat trivial because the equation has an
analytical solution valid for all $t \geq 0$ which is also quite simple:

\begin{align}
    \uc(x, t) = \uc(x - at, 0) \ .
    \label{eq:linear-advection-solution}
\end{align}

\begin{figure}[H]
    \centering
	\includegraphics[width=.8\linewidth]{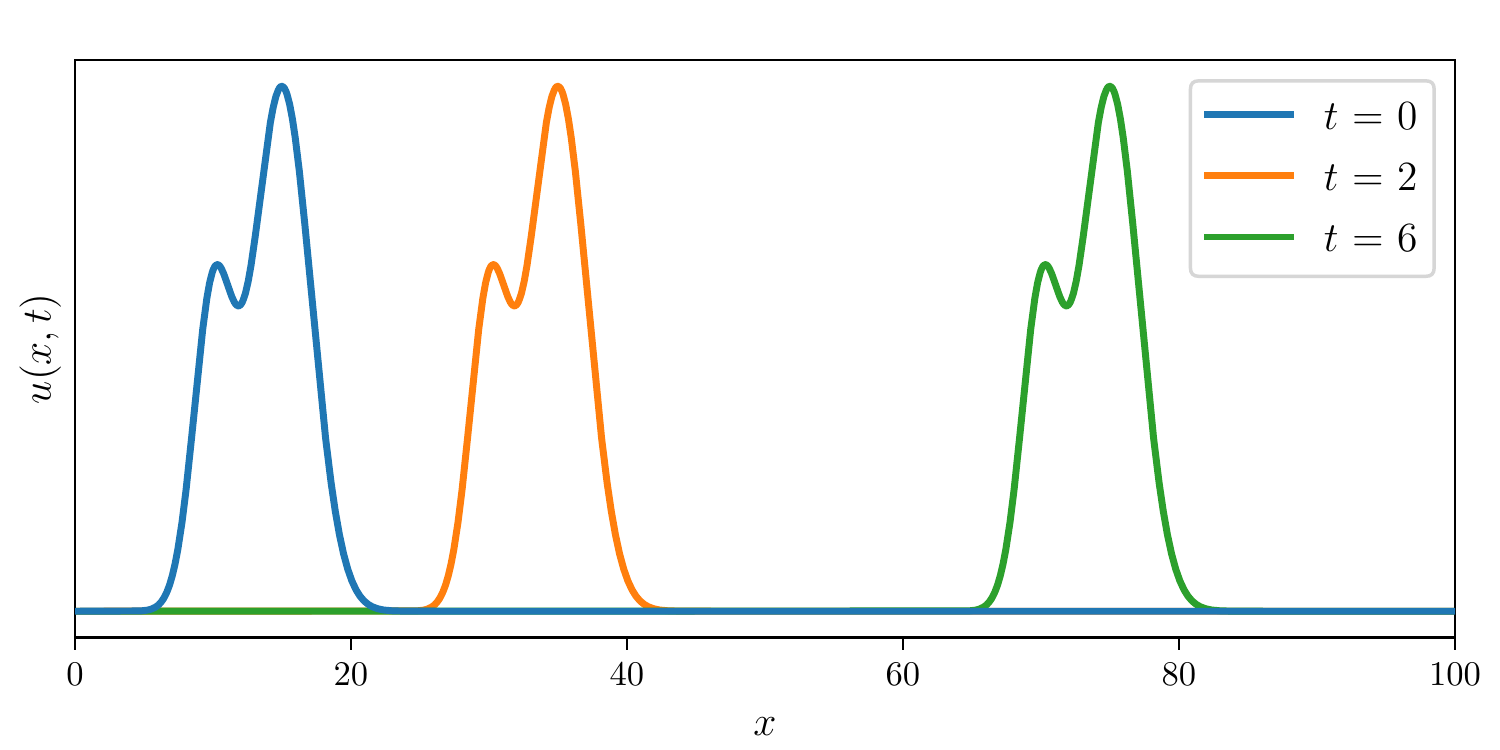}%
	\caption[Analytical solution of the linear advection equation with constant coefficients]{
        The analytical solution to the linear advection equation with constant coefficients (eq.
        \ref{eq:linear-advection-1D-const-coeff}) with $a = 10$ in arbitrary units. The initial
        function $\uc_0(x) = \uc(x, t=0)$ is transported (or rather: \emph{advected}) to the
        position $x = x_0 + a t$.
	}
    \label{fig:linear-advection-theory}
\end{figure}

This is easily verified by applying the chain rule:

\begin{align}
    \deldt \uc(t) = \DELDX{\uc} \DELDT{(x - at)} = - a \deldx{\uc}
\end{align}

which satisfies eq. \ref{eq:linear-advection-1D-const-coeff}. Effectively, this means that the
solution at a time $t > 0$ is just the initial function $\uc_0(x) = \uc(x, t=0)$ transported (or
rather: \emph{advected}) to the position $x = x_0 + a t$. The solution is illustrated in Fig.
\ref{fig:linear-advection-theory}.

This simple case offers a good opportunity to introduce the notion of \emph{characteristics}, which
will be used in more complex cases as well.  Characteristics may be defined as curves $x = x(t)$
in the $x$-$t$ plane along which the PDE becomes an ordinary differential equation (ODE). Consider
the parametrizaton $x = x(t)$, which makes $\uc = \uc(x(t), t)$. Then the total derivative of $\uc$
w.r.t. time is given by:

\begin{align}
    \DDT{\uc} = \DELDT{\uc} + \DDT{x} \DELDX{\uc} \ .
\end{align}

If now $\DDT{x} = a$, it follows that

\begin{align}
    \DDT{\uc} = \DELDT{\uc} + a \DELDX{\uc} = 0
\end{align}

where the second equality follows from the definition of the linear advection equation. The more
interesting part of this equation is that it gives $\DDT{\uc} = 0$ along the curve $x = x(t)$: it
means that along the characteristic curve which satisfies $\DDT{x} = a$, $\uc$ is constant. On the
$x - t$ plane, $a$ determines the slope of the characteristic. For constant coefficients $a$, all
characteristics are parallel. To determine the solution at some $(x, t > 0)$, all we need to do is
follow the characteristic that goes through that point back to $t = 0$.

Coming back to the Riemann problem for linear advection equations with constant coefficients, it
can explicitly be written as:

\begin{align}
    & \deldt \uc + a \deldx \uc = 0 &&\\
    & \uc(x, 0) = \uc_0 (x) = \begin{cases}
                             \uc_L & \text{ if } x < 0 \\
                             \uc_R & \text{ if } x > 0
                            \end{cases} &&
\end{align}

\begin{figure}[H]
    \centering
	\includegraphics[width=.4\linewidth]{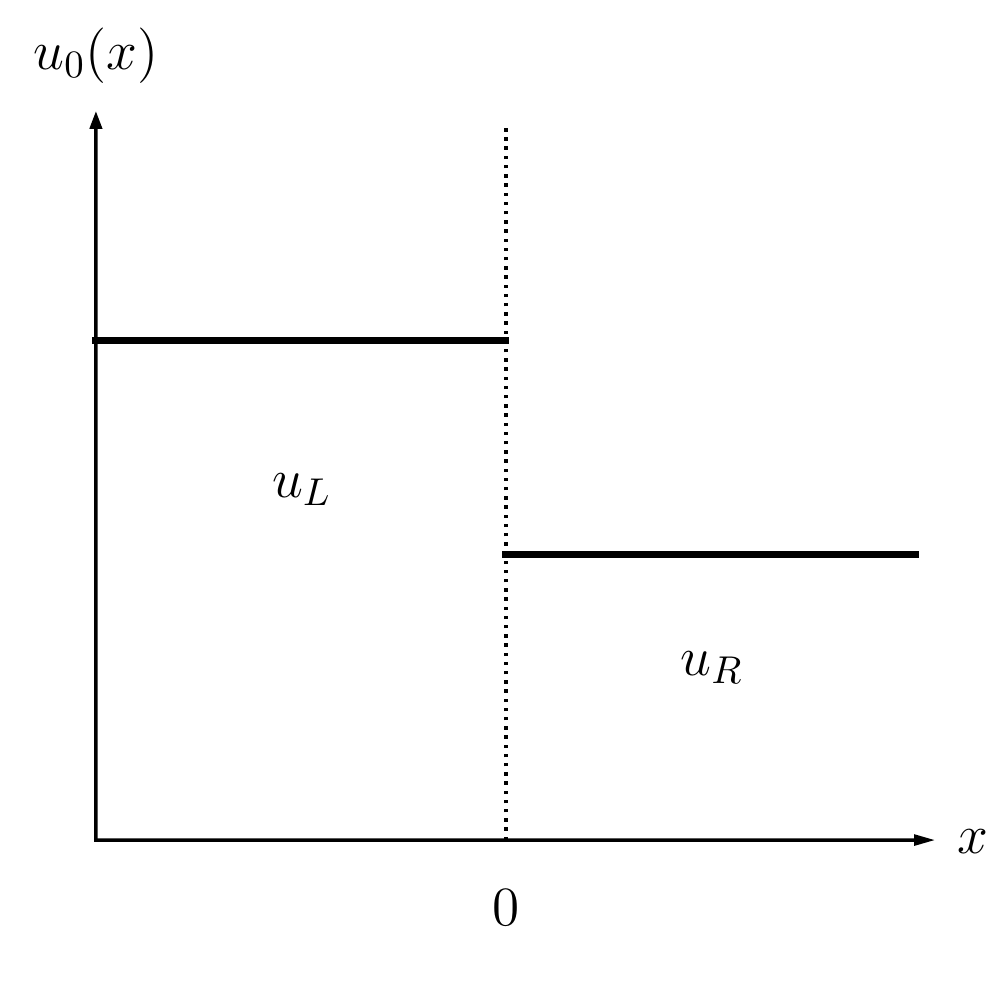}%
	\includegraphics[width=.4\linewidth]{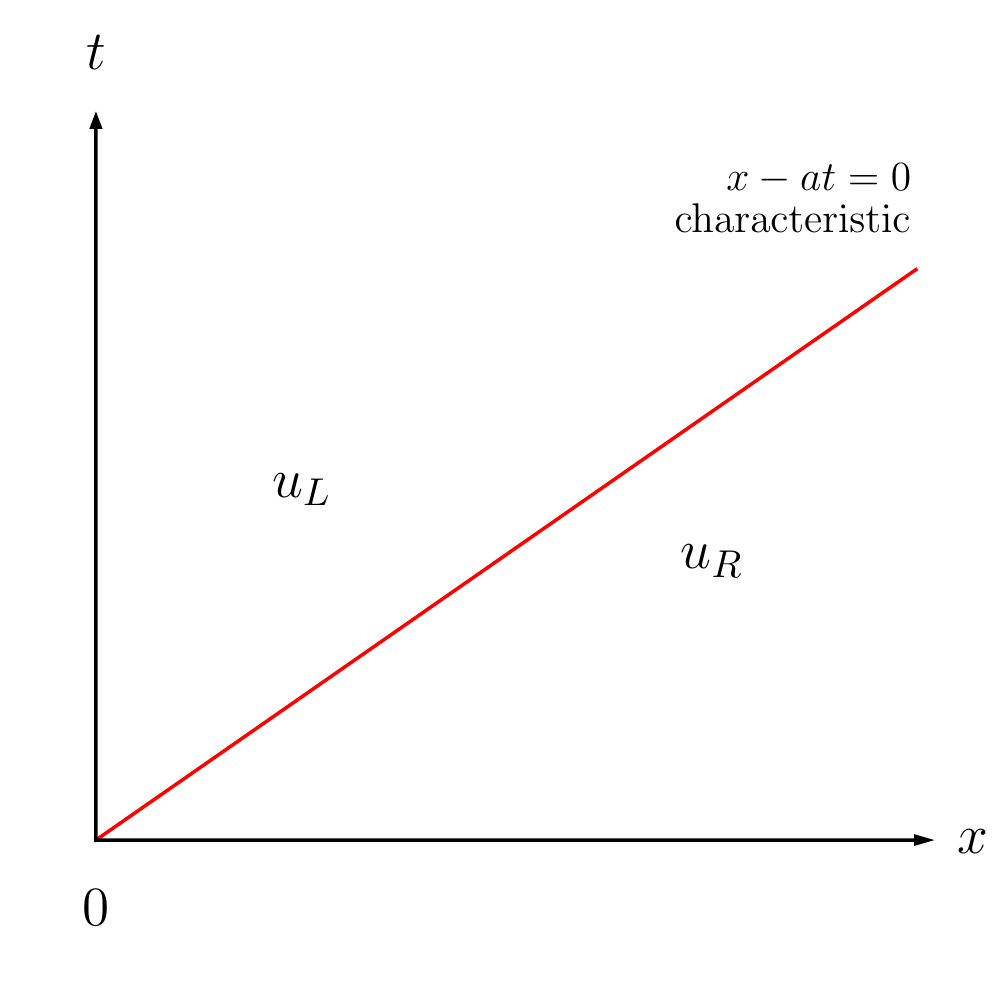}%
	\caption[Riemann problem and solution for the linear advection equation with constant
coefficients]{
    Left: The initial Riemann problem for the linear advection equation with constant
coefficients. Right: The solution to the Riemann problem. For $t > 0$, the two initial states
$\uc_L$ and $\uc_R$ will be separated by the $x - at = 0$ characteristic.
	}
    \label{fig:riemann-linear-advection}
\end{figure}

This IVP is illustrated in Fig.~\ref{fig:riemann-linear-advection}. Using the general solution
\ref{eq:linear-advection-solution}, it follows that the solution for the Riemann problem is given
by:

\begin{align}
    \uc(x, t) = \uc_0(x - at) = \begin{cases}
                                    \uc_L & \text{ if } x - at < 0 \\
                                    \uc_R & \text{ if } x - at > 0
                                \end{cases}
\end{align}

which is illustrated in Fig. \ref{fig:riemann-linear-advection}. In other words: The two initial
states $\uc_L$ and $\uc_R$ will be separated on the $x - t$ plane by the characteristic that
satisfies $\DDT{x} = a$.

\subsection{The Riemann Problem For Linear Hyperbolic Systems}

Let's now extend the analysis to linear hyperbolic systems, i.e. sets of $m$ hyperbolic PDEs of the
form

\begin{align}
    \deldt \U + \mathcal{A} \deldx \U = 0 \label{eq:linear-hyperbolic-system}
\end{align}

with $\mathcal{A} = \CONST$
By definition of a hyperbolic system, $\mathcal{A}$ has $m$ real
eigenvalues $\lambda_i$ and $m$ linearly independent eigenvectors $\mathbf{K}_i$. This allows
$\mathcal{A}$ to be expressed in diagonalized form, i.e. in terms of a diagonal matrix $\Lambda$
and the matrix $\mathcal{K}$:

\begin{align}
 \mathcal{A} = \mathcal{K} \ \Lambda \ \mathcal{K}^{-1} \quad \text{or} \quad
 \Lambda = \mathcal{K}^{-1} \mathcal{A} \mathcal{K}
\end{align}

where $\Lambda$ is a diagonal matrix whose diagonal elements are the eigenvalues $\lambda_i$, and
$\mathcal{K}$ is the matrix whose columns $\mathcal{K}^{(i)}$ are the right eigenvectors
$\mathbf{K}_i$ of $\mathcal{A}$ corresponding to the eigenvalues $\lambda_i$. To make use of
$\mathcal{K}$, we first note that since $\mathcal{A}$ is constant, $\mathcal{K}$ must be too, and
therefore $\deldt \mathcal{K} = \deldx \mathcal{K} = 0$.

By defining

\begin{align}
 \W = \mathcal{K}^{-1} \U \quad \text{ or } \quad \U = \mathcal{K} \W
\end{align}

we can write eq.~\ref{eq:linear-hyperbolic-system} as

\begin{align}
  \mathcal{K} \deldt \W + \mathcal{A} \mathcal{K} \deldx \W = 0
\end{align}

Applying $\mathcal{K}^{-1}$ from the left on the entire equation gives us a neat result:

\begin{align}
  \mathcal{K}^{-1} \mathcal{K} \deldt \W + \mathcal{K}^{-1} \mathcal{A} \mathcal{K} \deldx \W  =
  \deldt \W + \Lambda \deldx \W = 0 \label{eq:canonical-form}
\end{align}

Eq.~\ref{eq:canonical-form} is called the canonical or characteristic form of the system. In this
form, the system is decoupled: Recall that $\Lambda$ is a diagonal matrix, and hence the $i$-th PDE
of the system is given by

\begin{align}
    \deldt \wc_i + \lambda_i \deldx \wc_i = 0 \quad \quad i = 1, \dots, m
\end{align}

and in this form is identical to the linear advection equation with constant coefficients, for
which we have a solution. The characteristic speed is now $\lambda_i$, and the characteristic curves
need to satisfy $\DDT{x} = \lambda_i$. So we have effectively transformed a system of $m$ coupled
PDEs into a system of $m$ independent linear advection equation with constant coefficients.

In order to obtain the solution in terms of the original variables $\U$, the transformation

\begin{align}
    \U(x,t) = \mathcal{K} \W(x,t)
\end{align}

or component-wise

\begin{align}
    \U_j(x,t) = \sum_i^m \wc_i(x,t) \mathcal{K}_{ij}
\end{align}

must be calculated. The solution for $t > 0$ is then given by applying the characteristic solution:

\begin{align}
    \U_j(x,t) = \sum_i^m \wc_i(x,t) \mathcal{K}_{ij} = \sum_i^m \wc_i(x - \lambda_i t,t=0)
\mathcal{K}_{ij}
\end{align}

So given a point $(x, t)$ on the $x-t$ plane, the solution $\U(x,t)$ at this point depends again
only on the initial data. However, contrary to the linear advection equation with constant
coefficients, this time we were dealing with a system of $m$ equations, and so it is natural that
the solution at a point $(x, t)$ depends on $m$ points of the initial data, or more precisely the
points $x_0^{(i)} = x - \lambda_i t$. These are the intersections of the characteristics with
velocities $\lambda_i$ with the $x$-axis. The solution for $\U(x,t)$ can be seen as the
superposition of $m$ waves, each of which is advected independently and without change in shape due
to the constraint that $\mathcal{A} = \CONST$
The $i$-th wave has the shape $\wc_i(t=0)
\mathbf{K}^{(i)}$ and speed $\lambda_i$.

Let us now formulate the solution for the Riemann problem for linear hyperbolic systems, which is
given by:

\begin{align}
    & \deldt \U + \mathcal{A} \deldx \U = 0 && \\
    & \U(x, 0) = \U^{(0)} (x) = \begin{cases}
                               \U_L & \text{ if } x < 0 \\
                               \U_R & \text{ if } x > 0
                              \end{cases} &&
\end{align}

\begin{figure}[H]
    \centering

\includegraphics[width=.7\linewidth]{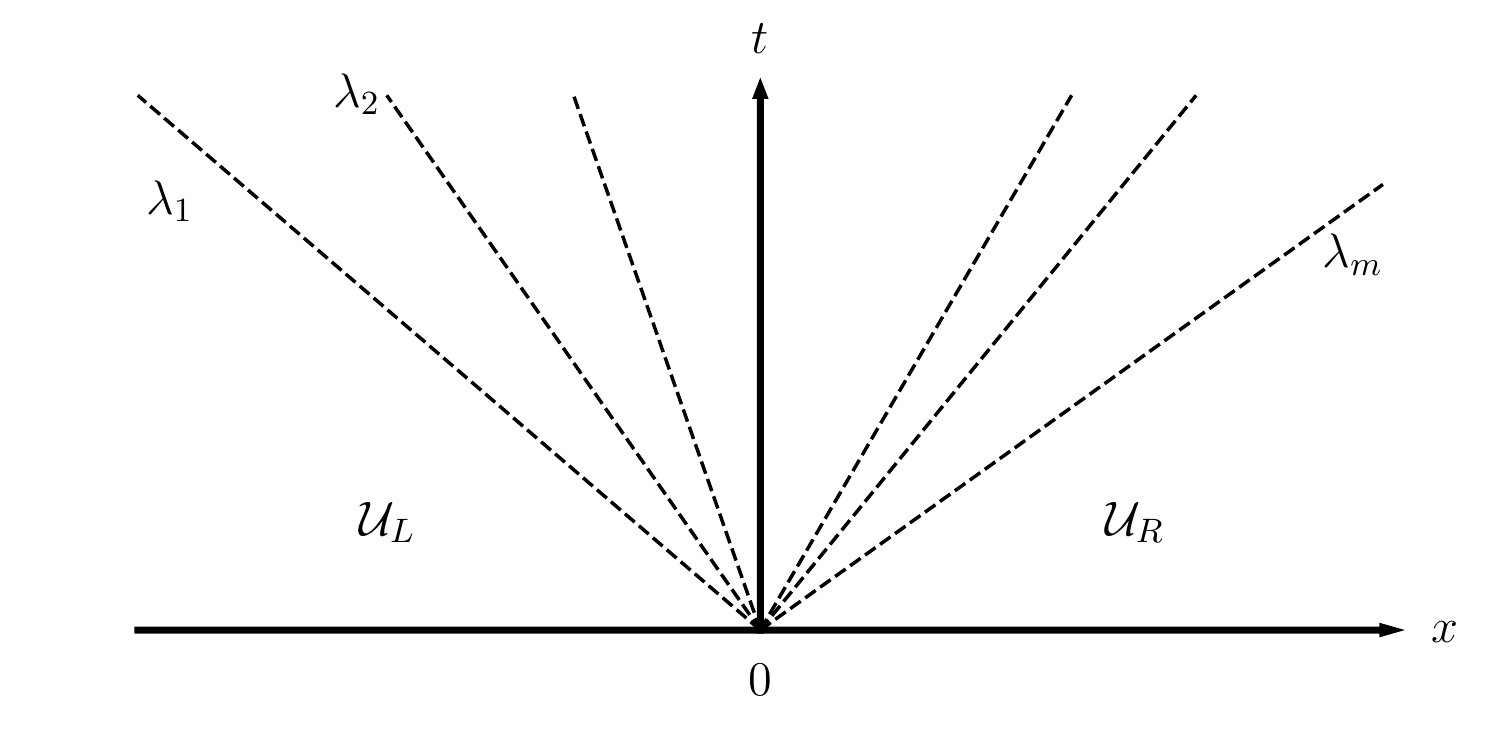}%
	\caption[Structure of the solution of the Riemann problem for linear hyperbolic systems]{
	The structure of the solution of the Riemann problem for linear hyperbolic systems. For each of
the $m$ eigenvalues $\lambda_i$ of the problem a characteristic with speed $\lambda_i$ will form a
jump discontinuity between two states.}
    \label{fig:riemann-linear-hyperbolic-system}
\end{figure}

If we assume that the system is strictly hyperbolic, we can index the eigenvalues
$\lambda_i$ such that $\lambda_1 < \lambda_2 < \ldots < \lambda_m$. The structure of the solution
consists of $m$ waves emanating from the origin, one for each eigenvalue $\lambda_i$ (see
Fig.~\ref{fig:riemann-linear-hyperbolic-system}). Waves for which $\lambda_i > 0$ will advect the
component $\wc_{i,L} = \oldsum_j \mathcal{K}_{ij}^{-1}\U_{L,j}$ into regions where $x > 0$ and
where the initial state was $\U_R$, while waves for which $\lambda_i < 0$ will advect the component
$\wc_{i,R} = \oldsum_j \mathcal{K}_{ij}^{-1}\U_{R,j}$ into regions where $x < 0$ and where the
initial state was $\U_L$.
To find a precise expression for the general solution, we can make use of the fact that the
eigenvectors $\mathbf{K}^{(i)}$ are linearly independent, meaning that we can express any state $\U
= \oldsum_i^m \gamma_i \mathbf{K}^{(i)}$ as linear combinations of the eigenvectors. That allows us
to define

\begin{align}
  \U_L = \sum_i^m \alpha_i \mathbf{K}^{(i)} \quad \text{ and } \quad
  \U_R = \sum_i^m \beta_i \mathbf{K}^{(i)}
\end{align}

From the general solution, we know that $\U(x, t) = \oldsum_i \wc_i \mathbf{K}^{(i)}$ with

\begin{align}
    \wc_i(x, t) = \wc_i(x - \lambda_i t, 0) = \begin{cases}
                                               \alpha_i & \text{ if } x - \lambda_i t < 0 \\
                                               \beta_i & \text{ if } x - \lambda_i t > 0 \\
                                              \end{cases}
\end{align}

Alternatively, it can be rewritten by defining the index $I$ for any given point $(x, t)$ such that
the eigenvalues $\lambda_I \leq \frac{x}{t} < \lambda_{I+1}$, i.e. $x - \lambda_i t > 0\ \forall i
\leq I$. Then the solution is given by

\begin{align}
    \U(x, t) = \sum_{i = I+1}^m \alpha_i \mathbf{K}^{(i)} + \sum_{i=1}^I \beta_i \mathbf{K}^{(i)}
\end{align}

\subsection{The Riemann Problem For Hyperbolic Conservation Laws}

Finally, let's look into the Riemann problem for hyperbolic conservation laws, whose solution we
will require to construct numerical methods to solve fluid dynamics and radiative transfer. For
simplicity, let's consider only a single equation instead of an entire system of equations to start
with, i.e.

\begin{align}
    \deldt \uc + \deldx \fc(\uc) = 0
\end{align}

which can also be written in the form

\begin{align}
    \deldt \uc + \alpha(\uc) \deldx \uc = 0
\end{align}

where $\alpha(\uc) = \frac{\partial \fc}{\partial \uc}$ is not constant any longer, but a function
of $\uc$. We can once again apply the method of characteristics and set $\DDT{x} = \alpha(\uc)$,
giving us

\begin{align}
    \DDT{\uc} = \deldt \uc + \alpha(\uc) \deldx \uc = 0
\end{align}

\begin{figure}[H]
    \centering
    \includegraphics[width=\linewidth]{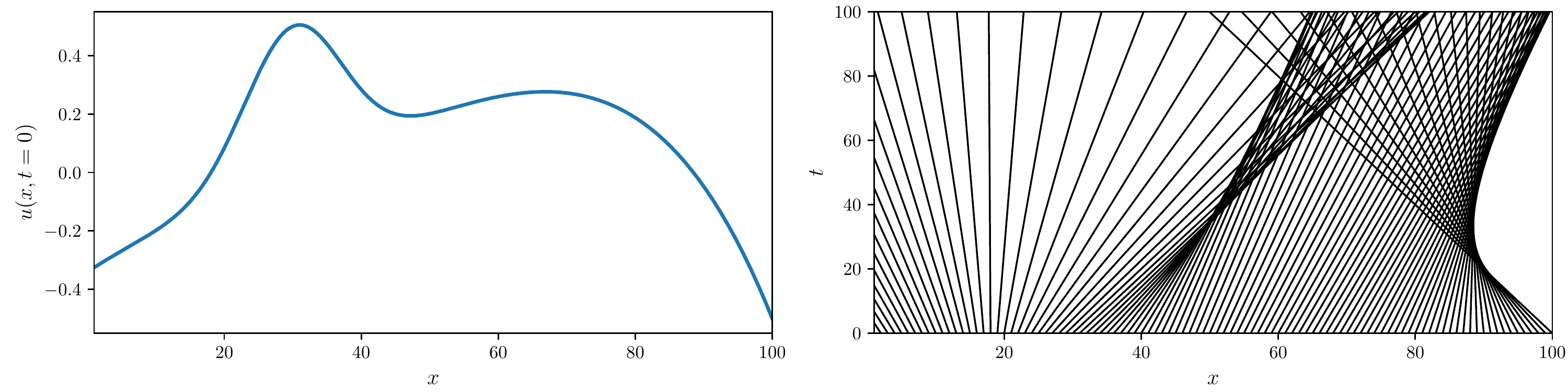}%
    \caption[Characteristics for the Burgers equation]{
    Left: Some initial conditions $\uc_0(x, t=0)$ for the Burgers'
equation~(\ref{eq:inviscid-burgers}). Right: The corresponding characteristics that satisfy
$\DDT{x} = \uc$. All units are arbitrary.
}
    \label{fig:burgers-characteristics}
\end{figure}

Just like in the previous cases, $\uc$ is constant along these characteristics. Even though the
characteristic speeds $\alpha(\uc)$ depend on the state $\uc$, given that $\uc$ is constant along
the characteristic, characteristic speeds themselves are constant as well. Therefore, once again
the characteristic curves are straight lines. However, they aren't identical over all $x$ any more,
and the translation distorts the the initial function as time evolves. This is a distinguishing
feature of non-linear problems. To illustrate why and how the complications arise, let's consider a
concrete example of a conservation law: The (inviscid) Burgers equation, given by:

\begin{align}
    \deldx \uc + \uc \deldx \uc = 0  \label{eq:inviscid-burgers}
\end{align}

which has characteristic speeds $\alpha(\uc) = \uc$. The Burgers equation can be obtained from the
momentum equation of the Euler equations and the additional assumption that the fluid density (and
therefore also pressure) variations are negligible. Figure~\ref{fig:burgers-characteristics} shows
an example of some initial state $\uc(x, t=0)$ on the left side, and the corresponding
characteristic lines on the right. The complications are evident: The characteristics can now fan
out and diverge in so-called ``expansive regions'' as is the case around $x \sim 20$ in
Figure~\ref{fig:burgers-characteristics}. Or they can get narrower and steeper as time evolves in
so-called ``compressive regions'', as is happens around $x \sim 50$ and $x \sim 90$ in
Figure~\ref{fig:burgers-characteristics}. This steepening, called ``wave steepening'', results in
the characteristics eventually intersecting, at which point our solution breaks down: there is no
single-valued solution any longer. The initial $\uc$ should be constant along their
characteristics, but that cannot be the case at the point of an intersection, where they are
supposed two have more than one value simultaneously. And we know  for a fact that the two
intersecting characteristics must stem from two different initial values $\uc$, because the slope of
the characteristics is $\alpha(\uc)$. In order for characteristics to intersect, the slopes can't be
equal, otherwise the characteristics would be parallel and never intersect.

In general, expansive regions can be found in regions where $\deldx \alpha(\uc)
> 0$, while compressive regions are regions where initially $\deldx \alpha(\uc) > 0$. In regions of
constant characteristic speeds $\deldx \alpha(\uc) = 0$, the solution is the same as for linear
hyperbolic systems.

There are two approaches to deal with this occurrence. One option is to modify the equation (and
the physical system) that is being solved to not allow wave steepening, e.g. by introducing a
``viscous'' or diffusive term that is proportional to the second derivative of $\uc$ w.r.t. $x$.
This term will then have the exact opposite, wave-easing, effect: as the wave steepens, $\deldx
\uc$ increases, and so does $\frac{\del^2}{\del x^2} \uc$, having a proportionally stronger effect
in the opposite direction of the steepening, and counteracting it in this manner.

\begin{figure}[H]
    \centering
    \includegraphics[width=\linewidth]{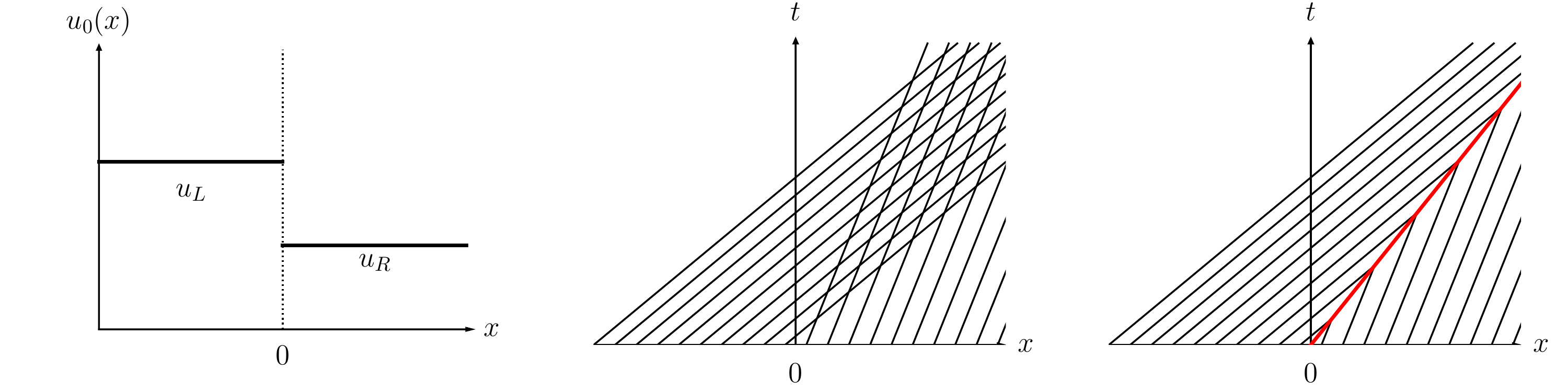}%
    \caption[Characteristics and shock wave for the Riemann problem of Burgers equation]{
    Left: A shock generating Riemann problem initial conditions $\uc_0(x)$ for the inviscid Burgers
equation~(\ref{eq:inviscid-burgers}). Middle: The corresponding characteristics. Right: The
resulting shock wave (in red), which is a jump discontinuity.
}
    \label{fig:burgers-riemann-shock}
\end{figure}

The other alternative is to stick to the inviscid equation, but to allow for discontinuous
solutions, i.e. shocks, to be formed as a process of increasing compression. The states left and
right of the shock wave are then determined ``as usual'' by the method of characteristics, while
the shock wave itself is modeled as a jump discontinuity at the points where the characteristics
intersect. To illustrate this approach, the shape of the solution for a Riemann problem for the
Burgers equation is depicted in Fig.~\ref{fig:burgers-riemann-shock}.

Modeling a shock wave as a jump discontinuity is all good and well, but an important question
remains: What is the velocity of the shock wave going to be? To answer this question we need to
resort to using the integral form of conservation laws, so we can deal with the discontinuity in an
appropriate manner. The integral form is given by integrating the entire conservation law equation
(eq.~\ref{eq:conservation-law-1D-introduction}) over both a space interval $x \in [x_0, x_1]$ and
time interval $t \in [t_0, t_1]$:

\begin{align}
    \int\limits_{x_0}^{x_1} \int\limits_{t_0}^{t_1} \left[\deldt \U + \deldx \F \right] \de x \de t
= 0 \ .
\end{align}

The right hand side equality is trivially satisfied by virtue of the integrand,
eq.~\ref{eq:conservation-law-1D-introduction}, always satisfying it. We can immediately deal
away with the partial derivatives:

\begin{align}
    \int\limits_{x_0}^{x_1} \int\limits_{t_0}^{t_1} \left[\deldt \U + \deldx \F \right] \de x \de t
=
    \int\limits_{x_0}^{x_1} \left[ \U(x, t_1) - \U(x, t_0) \right] \de x +
    \int\limits_{t_0}^{t_1} \left[ \F(x_1, t) - \F(x_0, t) \right] \de t \ .
    \label{eq:rankine-hugeniot-integral}
\end{align}

\begin{figure}[H]
    \centering
    \includegraphics[width=.5\linewidth]{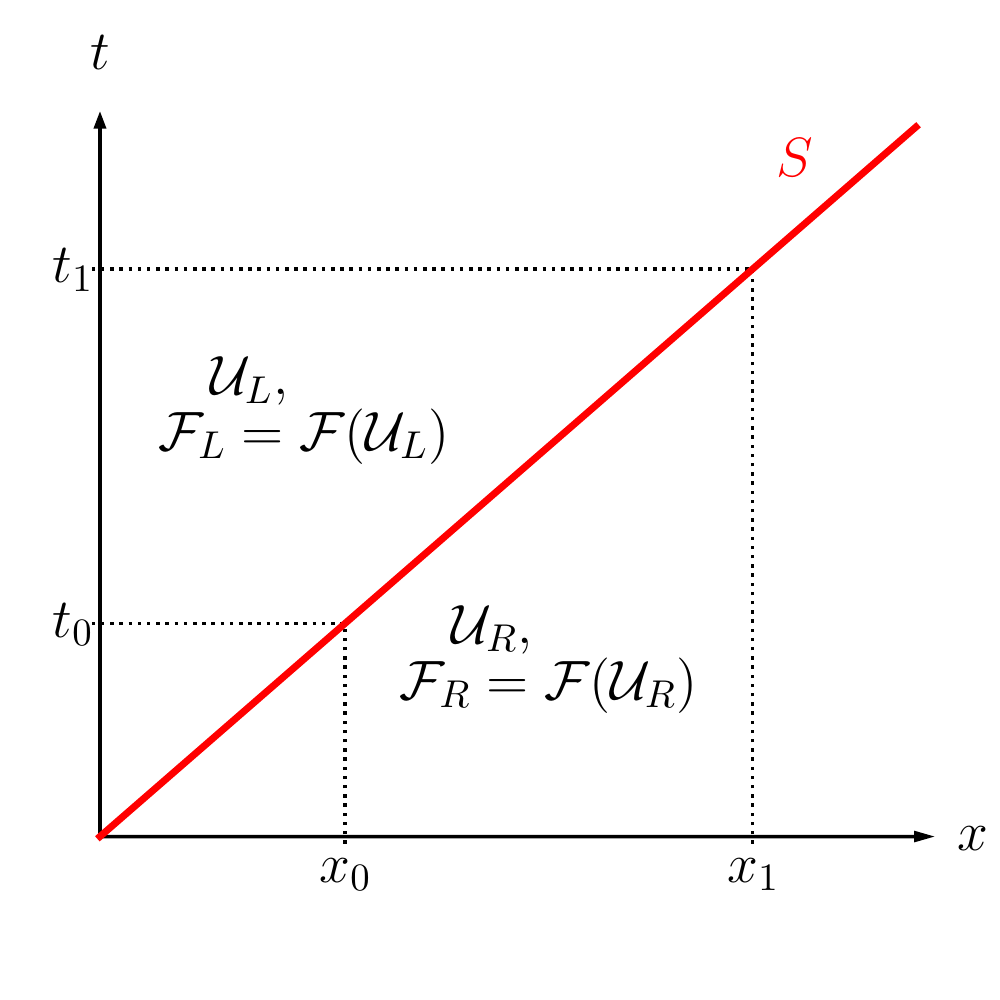}%
    \caption[Setup for the Rankine-Hugeniot conditions]{
Setup to derive the Rankine-Hugeniot conditions with a constant left state $\U_L$ and flux
$\F_L$ and a right state $\U_R$ and corresponding flux $\F_R$, which are separated by a jump
discontinuity which propagates at constant velocity $S$.
}
    \label{fig:rankine-hugeniot}
\end{figure}

To determine the shock speed $S$, consider a solution $\U(x,t)$ such that $\U(x,t)$ and $\F(x,t)$
are continuous everywhere except on a line $S = S(t)$ on the $x-t$ plane. Across this line,
$\U(x,t)$ has a jump discontinuity. Let us refer to states and fluxes ``left'' of $S(t)$, i.e.
where $x < S(t)$, as $\U_L$ and $\F_L$. Similarly, let the states and fluxes to the ``right'', i.e.
where $x > S(t)$, be $\U_R$ and $\F_R$. Furthermore, select two points in space, $x_0$ and $x_1$
with $x_0 < x_1$, and two points in time, $t_0$ and $t_1$ with $t_0 < t_1$, such that both points
$(x_0, t_0)$ and $(x_1, t_1)$ coincide with the shock wave $S(t)$ on the $x-t$ plane, and
consequently the position of the shock can be inferred by $x_1 = x_0 + S (t_1 - t_0)$. This
situation is illustrated in Figure~\ref{fig:rankine-hugeniot}.

If the shock is indeed infinitely thin, the states $\U_{L,R}$ are homogeneous over space, and the
fluxes $\F_{L,R}$ are homogeneous over time, then we can see that

\begin{align}
    \U(t_1, x) = \U_L &&  \U(t_0, x) = \U_R && \forall x \in [x_0, x_1] \\
    \F(t, x_1) = \F_R &&  \F(t, x_0) = \F_L && \forall t \in [t_0, t_1]
\end{align}

using these relations, the integral \ref{eq:rankine-hugeniot-integral} is easily solved:

\begin{align}
    &
    \int\limits_{x_0}^{x_1} \left[ \U(x, t_1) - \U(x, t_0) \right] \de x +
    \int\limits_{t_0}^{t_1} \left[ \F(x_1, t) - \F(x_0, t) \right] \de t = &&\\
    &
    \int\limits_{x_0}^{x_1} \U(x, t_1) \de x -
    \int\limits_{x_0}^{x_1} \U(x, t_0) \de x +
    \int\limits_{t_0}^{t_1} \F(x_1, t) \de t -
    \int\limits_{t_0}^{t_1} \F(x_0, t) \de t = &&\\
    &
    \int\limits_{x_0}^{x_1} \U_L \de x -
    \int\limits_{x_0}^{x_1} \U_R \de x +
    \int\limits_{t_0}^{t_1} \F_R \de t -
    \int\limits_{t_0}^{t_1} \F_L \de t = &&\\
    &
    - ( \U_R - \U_L)(x_1 - x_0) + (\F_R - \F_L) (t_1 - t_0) = 0 &&
\end{align}

which finally leads to

\begin{align}
    \F_R - \F_L &= \frac{x_1 - x_0}{t_1 - t_0} \left( \U_R - \U_L \right) \\
                &= S \left( \U_R - \U_L \right) \label{eq:rankine-hugeniot}
\end{align}

Expression~\ref{eq:rankine-hugeniot} is called the Rankine-Hugeniot conditions, which gives us the
shock propagation speed $S$ as well as a relation between states and fluxes across discontinuities.

\subsubsection{Requirements For Physically Meaningful Shocks and Rarefaction Waves}

An additional point needs to be made for shocks in order for them to be physical solutions: They
need to arise from compressive regions. Following the sketch in Figure~\ref{fig:rankine-hugeniot},
that translates to the condition that the characteristic speeds $\alpha_L$ and $\alpha_R$ must obey
$\alpha_L > \alpha_R$, i.e. they must converge into the discontinuity. This condition is called
the ``entropy condition''. The inverse case, where $\alpha_L < \alpha_R$ for two states $\U_L$ and
$\U_R$ separated by a jump discontinuity, is mathematically permissible, but physically
nonsensical: In case of these ``rarefaction shocks'' characteristics diverge from the discontinuity.
Given that the characteristics carry information about the state they emanate from, this would mean
that new information must be generated out of nowhere along the discontinuity to be carried. This is
an entropy violating condition, as in the case of gas dynamics, it contradicts the second law of
thermodynamics. The second law of thermodynamics states that the entropy of a system must be
non-decreasing. This is the case for compressive shocks, across which the entropy of a system
increases. Since the entropy at each point can be computed as a function of pressure and density
(see eq.~\ref{eq:entropy}), and it increases for compressive shocks, it follows that for the case
of rarefaction shocks, where the inverse behavior to compressive shocks occurs, the entropy must
decrease. This is an unphysical solution.

Furthermore, the rarefaction shock solution is unstable under small perturbations, in the sense
that small perturbations of the initial data lead to large changes in the solution. To demonstrate
that point, let's modify the initial conditions from a rarefaction generating Riemann problem,
i.e. from two constant states $\uc_L$ and $\uc_R$ with diverging characteristics. Take two points
$x_L$ and $x_R$, which are separated by some distance $\Delta x$. Then let the states with $x < x_L$
be constant and called $\uc_L$, the states with $x > x_R$ be constant and called $\uc_R$, while the
state in the intermediate region $\Delta x$ between them increase linearly between $\uc_L$ and
$\uc_R$, i.e.

\begin{align}
    \uc_0(x) = \begin{cases}
                 \uc_L                                      & \text{ if } x \leq x_L \\
                 \uc_L + (\uc_R - \uc_L)\frac{x - x_L}{x_R - x_L} & \text{ if } x_L < x < x_R \\
                 \uc_R                                      & \text{ if } x \geq x_R   \\
               \end{cases} \label{eq:rarefaction-setup}
\end{align}

\begin{figure}[H]
    \centering
    \includegraphics[width=\linewidth]{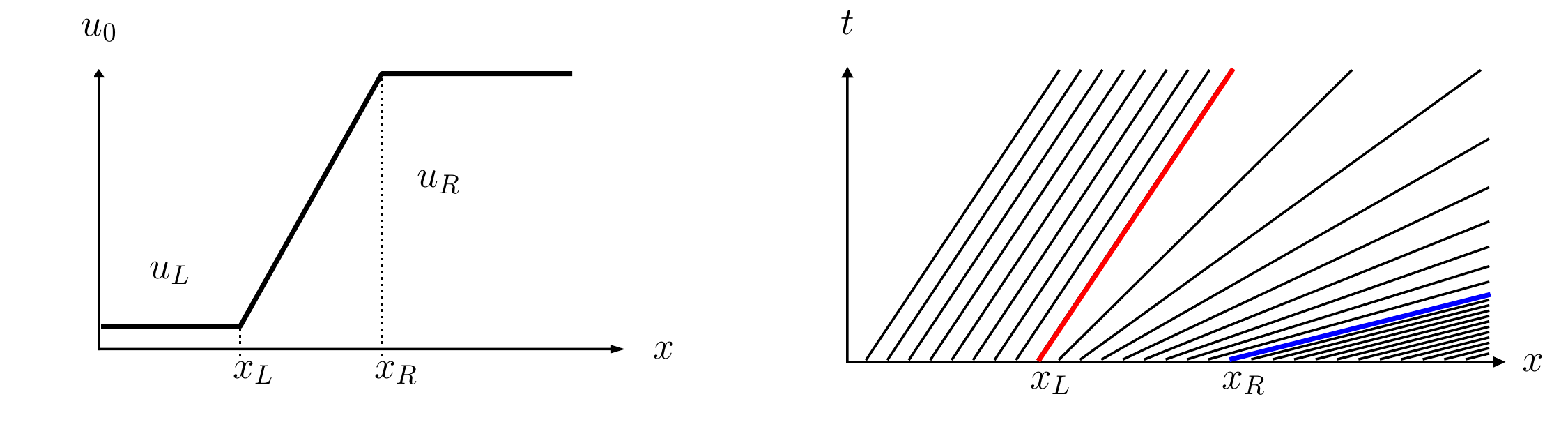}%
    \caption[Rarefaction wave characteristics]{
Left: The initial conditions described by eq.~\ref{eq:rarefaction-setup}, in arbitrary units.
Right: The corresponding characteristics. The two outermost characteristics of the intermediate
region $x_L < x < x_R$ are highlighted: The tail of the rarefaction wave is in red, while the head
is in blue.
}
    \label{fig:rarefaction-setup}
\end{figure}

These initial states and the corresponding characteristics (assuming once again the inviscid
Burgers' equation for simplicity) are shown in Fig.~\ref{fig:rarefaction-setup}. Two special
characteristics at the outer edges of the intermediate region can be identified, namely the ones
that overlap with the characteristics of the outer constant regions. Assuming a convex flux, i.e.
$\deldx \alpha(\uc) > 0$  (as is done in Fig.~\ref{fig:rarefaction-setup}), larger values of $\uc$
propagate faster that lower values, hence the characteristics of $\uc_R$ propagate faster than
those of $\uc_L$. The rightmost characteristic of the intermediate region will have the same
propagation speed as the right constant region, i.e. satisfy $x = x_R + \alpha(\uc_R)t$. This wave
is called the \emph{head} of the rarefaction wave, which is marked blue in
Figure~\ref{fig:rarefaction-setup}. Similarly, the leftmost characteristic satisfies $x = x_L +
\alpha(\uc_L) t$, and is called the \emph{tail} of the rarefaction wave, marked in red in
Figure~\ref{fig:rarefaction-setup}. As for all the other characteristics between the head and the
tail, a region called the ``rarefaction fan'', mathematically they pose no new problem. There are
no intersections nor undefined regions there. It's true that they diverge, but every point $(x, t)$
enclosed by the head and the tail characteristics can still be solved by tracing back the
characteristics to the initial time, just as for linear hyperbolic systems. A major difference to
linear systems however remains the fact that the propagation velocities aren't identical across the
intermediate section: The closer we are to the right state $\uc_R$ as seen from the inside of the
fan, the higher $\alpha(\uc_R)$ will be, and the initial state will transform over time. The result
is a smooth transition between the initial left and right state inside the fan enclosed by the head
and the tail of the rarefaction wave. The faster characteristics will cover a smaller ``volume'' of
the fan, and as time evolves, an increasing fraction of the rarefaction fan will be closer in value
to $\uc_L$ than $\uc_R$. The full solution is given by

\begin{align}
\uc(x, t) = \begin{cases}
            \uc_L                   & \text{ if } \frac{x - x_L}{t} \leq \alpha_L \\
            \uc_0(x - \alpha_I t)   & \text{ if } \alpha_L < \frac{x}{t} < \alpha_R
                \text{ with } \alpha_I = \frac{x - x_L}{t} \\
            \uc_R                   & \text{ if } \frac{x - x_L}{t} \geq \alpha_R
            \end{cases} \label{eq:rarefaction-solution-theory}
\end{align}

This full solution shows how the ``rarefaction shock'' solution is unstable: The initial separation
$\Delta x = x_R - x_L$ has no influence in the solution~\ref{eq:rarefaction-solution-theory}. It
can as well be set to zero, which is the identical setup where a rarefaction shock is a
mathematically permissible solution. In fact, taking the limit $\Delta x \rightarrow 0$ and setting
the initial discontinuity at $x_L = x_R = 0$ leads to what is called a ``centered rarefaction wave''
with the solution

\begin{align}
\uc(x, t) = \begin{cases}
            \uc_L                           & \text{ if } \frac{x}{t} \leq \alpha_L \\
            \alpha_I = \frac{x - x_L}{t}    & \text{ if } \alpha_L < \frac{x}{t} <  \alpha_R \\
            \uc_R                           & \text{ if } \frac{x}{t} \geq \alpha_R
            \end{cases}
            \label{eq:centered-rarefaction-solution-theory}
\end{align}

For any small perturbation however, e.g. any small nonzero $\Delta x$, the rarefaction shock isn't a
permissible solution any longer, while solution~\ref{eq:rarefaction-solution-theory} is still valid.
But solution~\ref{eq:rarefaction-solution-theory} is fundamentally different from the rarefaction
shock solution, and therefore a small perturbation leads to a very different results for
rarefaction shocks, making them an unstable solution.

The attentive reader will have noticed that the exact solution inside the rarefaction fan, where
$\alpha_L < \frac{x}{t} <  \alpha_R$, has not been provided in
eq.~\ref{eq:centered-rarefaction-solution-theory} for the centered rarefaction wave. The reason is
that a general solution valid for any conservation law is not readily available, and it will depend
on the conservation law at hand which is being solved. However, given the previous derivation with
an intermediate region $\Delta x$, an observation can be made: At $x = t= 0$, the point of the
initial discontinuity for the centered rarefaction wave, the initial state $\uc_0(0)$ must take on
all values between $\uc_{L}$ and $\uc_R$, with the corresponding interval of propagation velocities
$\alpha_L$, $\alpha_R$.

\subsubsection{Contact Waves}

Aside from shocks and rarefaction waves, there is a third class of waves that can arise. These
waves are somewhat similar to shock waves in that they are also a jump discontinuity separating two
states. The difference lies in the properties of the eigenvalues and eigenvectors.
\footnote{In contrast to contact discontinuities, one of the defining properties of shock waves is
that they only occur in so-called genuinely nonlinear fields, which are defined by
\begin{align*}
    \nabla \lambda_i (\U) \cdot \mathbf{K}^{(i)}(\U) \neq 0 \quad \forall \U
\end{align*}
}
A $\lambda_i(\U)$-characteristic field is said to be linearly degenerate if

\begin{align}
    \nabla \lambda_i (\U) \cdot \mathbf{K}^{(i)}(\U) = 0 \label{eq:linear-degenerate-field}
\end{align}

for all real-valued vectors $\U$. For a linearly degenerate $\lambda_i$-characteristic field,
should two initially constant left and right states $\uc_L$ and $\uc_R$ have the same propagation
velocities of characteristics, then so-called ``contact discontinuities'' occur. Their
characteristics are parallel on the $x-t$ plane, and the solution is simply that the contact
discontinuity propagates with the same characteristic speed as $\uc_L$ and $\uc_R$. Left and
right of the discontinuity are the original states $\uc_L$ and $\uc_R$.
A simple example of such a contact discontinuity wave is given with the linear advection with
constant coefficients (eq.~\ref{eq:linear-advection-1D-const-coeff}): The eigenvalues of the
Jacobian matrix are $\lambda_i(\U) = \CONST$, and so $\nabla \lambda_i = 0$ is trivially satisfied.
The solution (shown in Figure~\ref{fig:linear-advection-theory}) is a translation of the initial
state with a constant velocity and without changing the original shape. The same is the case for
the contact discontinuities.

\subsubsection{Solution Strategy For The Riemann Problem For Hyperbolic Conservation Laws}

With the three possible wave types discussed, we can now look at a solution strategy for a Riemann
problem for non-linear hyperbolic conservation laws. The Riemann problem is given by the $m \times
m$ system with initial data $\U_L$ and $\U_R$:

\begin{align}
    & \deldt \U + \deldx \F(\U) = 0 && \\
    & \U(x, t=0) = \begin{cases}
                    \U_L & \text{ if } x < 0 \\
                    \U_R & \text{ if } x > 0
                   \end{cases}
\end{align}

From the method of characteristics we know that the solution will consist of $m$ waves whose
propagation velocities will be the determined by the eigenvalues $\lambda_i$ of the Jacobian matrix
$\frac{\del \F}{\del \U}$.
The emanating waves can be classified in one of three families:

\begin{itemize}
    \item Shocks, where the states left and right of the wave, $\U_L^{w}$ and $\U_R^{w}$, are
connected through a single jump discontinuity. The wave propagates at a speed $S_i$, and the
characteristics must converge: $\lambda_i(\U_L^w) > S_i > \lambda_i(\U_R^{w})$ (entropy condition).
    \item Contact waves, where the states left and right of the wave,  $\U_L^{w}$ and $\U_R^{w}$,
are also connected through a jump discontinuity, the characteristics are parallel, i.e.
$\lambda_i(\U_L^w) = S_i = \lambda_i(\U_R^{w})$, and the characteristics form a linearly degenerate
field (Eq. \ref{eq:linear-degenerate-field}).
    \item Rarefaction waves, where the left and right states of the wave are connected through a
smooth transition. The characteristics are diverging, i.e.  $\lambda_i(\U_L^w) <
\lambda_i(\U_R^{w})$.
\end{itemize}

The combination of knowing how many waves there will be and what wave types each wave may be is
called the ``elementary-wave solution'' of the Riemann problem.
The exact solution of a problem depends on both the exact form of the conservation law to be
solved, as well as the initial state. In general, it needs to be solved for every conservation law
individually. However, the elementary-wave solution gives us a great concrete approach in how to do
so. In the following section, the solution for the Riemann problem for the Euler equations is
discussed.

\section{Riemann Solvers for the Euler Equations}

As mentioned before, the solution of the Riemann problem is at the heart of finite volume methods
to solve hyperbolic conservation laws. By discretizing the simulated space into a multitude of
cells, at any given time step of a simulation the problem we are solving is a collection of Riemann
problems centered at the interface of any two adjacent cells. So if we want to solve the Euler
equations using a finite volume method, we need a solver for Riemann problems for the Euler
equations. In this Section, such solvers are discussed. How \emph{exactly} the Riemann solvers are
used in finite volume methods to solve the Euler equations will be the topic of
Chapters~\ref{chap:godunov}~and~\ref{chap:higher-order-schemes}.

There is an exact solution for the Riemann problem for the Euler equations, which will be
schematically derived in the following Section. However, this is not necessarily also the
solver which is actually used in simulations. As we will see, there is no closed form exact
analytical solution, and the solution needs to be found iteratively. This may come with a
significant additional expense, which is why approximate Riemann solvers have also been developed
and are being used. For example, \swift lets users choose between the exact solver, the HLLC solver,
and the TRRS solver. \meshhydro additionally offers the TSRS solver. All these Riemann solvers
are described in this Chapter. We start with the exact solver, while approximate solvers like the
HLLC, TRRS, and TSRS will be discussed later in Section~\ref{chap:riemann-approximate}.

\subsection{The Exact Riemann Solver for the Euler Equations}\label{chap:exact-riemann-solver}

To find an exact solution for the one dimensional Euler equations, we're going to make use of the
previously discussed elementary wave solution for hyperbolic conservation laws. The one dimensional
Euler equations are given by

\begin{align}
    & \deldt \U + \deldx\F(\U) = 0 && \\[.5em]
    & \U = \begin{pmatrix}
             \uc_1 \\ \uc_2 \\ \uc_3
           \end{pmatrix} =
         \begin{pmatrix}
           \rho \\ \rho v \\ E
         \end{pmatrix}, \quad \quad
    \F = \begin{pmatrix}
          \fc_1 \\ \fc_2 \\ \fc_3
         \end{pmatrix} =
         \begin{pmatrix}
           \rho v \\ \rho v^2 + p \\ (E + p) v
         \end{pmatrix} &&
\end{align}

\begin{figure}[H]
    \centering
    \includegraphics[width=.8\linewidth]{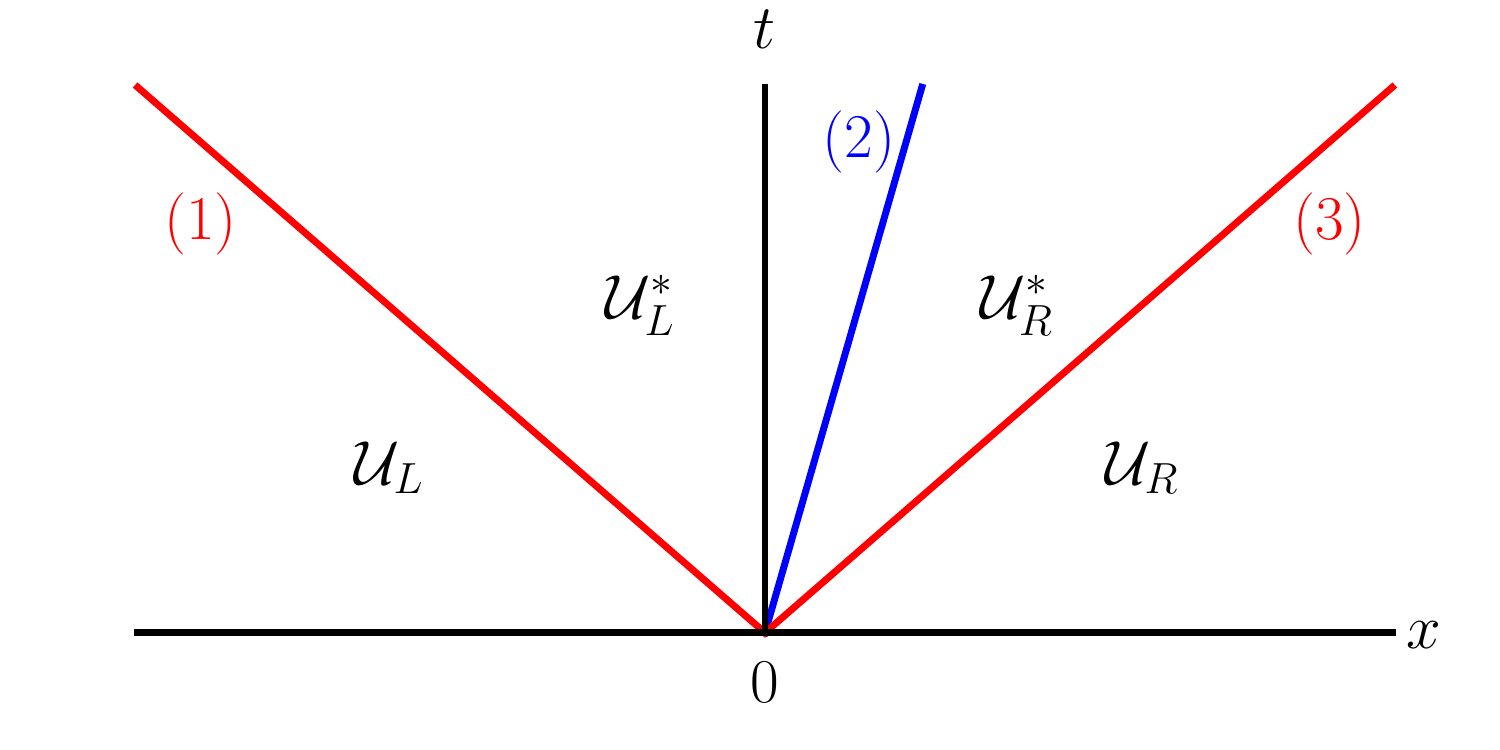}%
    \caption[Structure of the solution to the Riemann problem for Euler equations]{
The structure of the solution to the Riemann problem for Euler equations:
Three waves, (1), (2), and (3), arise from the origin as time evolves.
(2) is always a contact wave, (1) and (3) can be either rarefaction or shock waves in each
case, depending on the initial conditions.\\
The initial states $\U_L$ and $\U_R$ are separated through the waves (1) and (3) from the
two new arising ``star states'' $\U_L^*$ and $\U_R^*$, which themselves are separated by the contact
wave (2).
    }
    \label{fig:riemann-solution}
\end{figure}

Since we have a system of 3 equations, we know that the solution will also contain 3 waves, which
separate the $x-t$ plane into 4 states, as is shown in Figure~\ref{fig:riemann-solution}.
Defining the initial state for the Riemann problem as

\begin{align}
  \U(x, t=0) = \U_0(x) = \begin{cases}
                          \U_L & \text{ if } x < 0 \\
                          \U_R & \text{ if } x > 0
                         \end{cases}
\end{align}

then we adopt the following naming convention:  The leftmost state is the same as the initial
state $\U_L$, while the rightmost state remains $\U_R$. The two new states between them, separated
by the middle wave, are referred to as the ``star states'' $\U_L^*$ and $\U_R^*$, respectively.
The fact that $\U_L$ and $\U_R$ remain a part of the solution as it evolves over time can be easily
understood by the fact that the emanating waves travel at a finite speed. Until the wave doesn't
reach a certain position $x$, the initially constant states $\U_L$ and $\U_R$ at that position have
no reason to evolve.

It now remains to determine which type of wave those 3 emanating waves are. In order to do so, we
need to look at the eigenvalues $\lambda_i$  and the corresponding eigenvectors
$\mathbf{K}^{(i)}$ of the Jacobian matrix  $\frac{\del \F}{\del \U}$. They are given by

\begin{align}
    &\lambda_1 = v - c_s, &&
    \lambda_2 = v,  &&
    \lambda_3 = v + c_s && \\
    &\mathbf{K}^{(1)} = \begin{pmatrix}
                         1 \\ v - c_s \\ \frac{E + p}{\rho} - v c_s
                        \end{pmatrix} &&
    \mathbf{K}^{(2)} = \begin{pmatrix}
                         1 \\ v \\ \frac{1}{2} v^2
                        \end{pmatrix} &&
    \mathbf{K}^{(3)} = \begin{pmatrix}
                         1 \\ v - c_s \\ \frac{E + p}{\rho} - v c_s
                        \end{pmatrix} &&
\end{align}

We can now see that the $\lambda_2$ characteristic field is always linearly degenerate, i.e.
satisfy eq.~\ref{eq:linear-degenerate-field}:
\begin{align}
  \lambda_2 &= v = \frac{\uc_2}{\uc_1} \\
  \nabla \lambda_2 &= \begin{pmatrix}
                       \frac{\del \lambda_2}{\del \uc_1} \\
                       \frac{\del \lambda_2}{\del \uc_2} \\
                       \frac{\del \lambda_2}{\del \uc_3}
                      \end{pmatrix}
                   =  \begin{pmatrix}
                        -\frac{\uc_2}{\uc_1^2} \\
                        \frac{1}{\uc_1} \\
                        0
                      \end{pmatrix}
                   =  \begin{pmatrix}
                        -\frac{v}{\rho} \\
                        \frac{1}{\rho} \\
                        0
                      \end{pmatrix}
                      \\
 \nabla \lambda_2 \cdot \mathbf{K}^{(2)} &= -\frac{v}{\rho} + \frac{v}{\rho} + 0 = 0
\end{align}

while the other two, $\lambda_1$ and $\lambda_3$, are genuinely non-linear:
\begin{align}
  \lambda_{1,3} &= v \pm c_s = \frac{\uc_2}{\uc_1} \pm c_s \\
  \nabla \lambda_{1,3} &= \begin{pmatrix}
                       \frac{\del \lambda_{1,3}}{\del \uc_1} \\
                       \frac{\del \lambda_{1,3}}{\del \uc_2} \\
                       \frac{\del \lambda_{1,3}}{\del \uc_3}
                      \end{pmatrix}
                   =  \begin{pmatrix}
                        -\frac{\uc_2}{\uc_1^2} \\
                        \frac{1}{\uc_1} \\
                        0
                      \end{pmatrix}
                   =  \begin{pmatrix}
                        -\frac{v}{\rho} \\
                        \frac{1}{\rho} \\
                        0
                      \end{pmatrix}
                      \\
 \nabla \lambda_{1,3} \cdot \mathbf{K}^{(1,3)} &= -\frac{v}{\rho} + \frac{v \pm c_s}{\rho} + 0 \neq
0
\end{align}

They are genuinely non-linear for all energies $E$ and fluid velocities $v$. The vacuum case
$\rho = 0$ will require some special treatment though, which will be discussed later in
Section~\ref{chap:vacuum}.

\begin{figure}
\centering
\includegraphics[width=\textwidth]{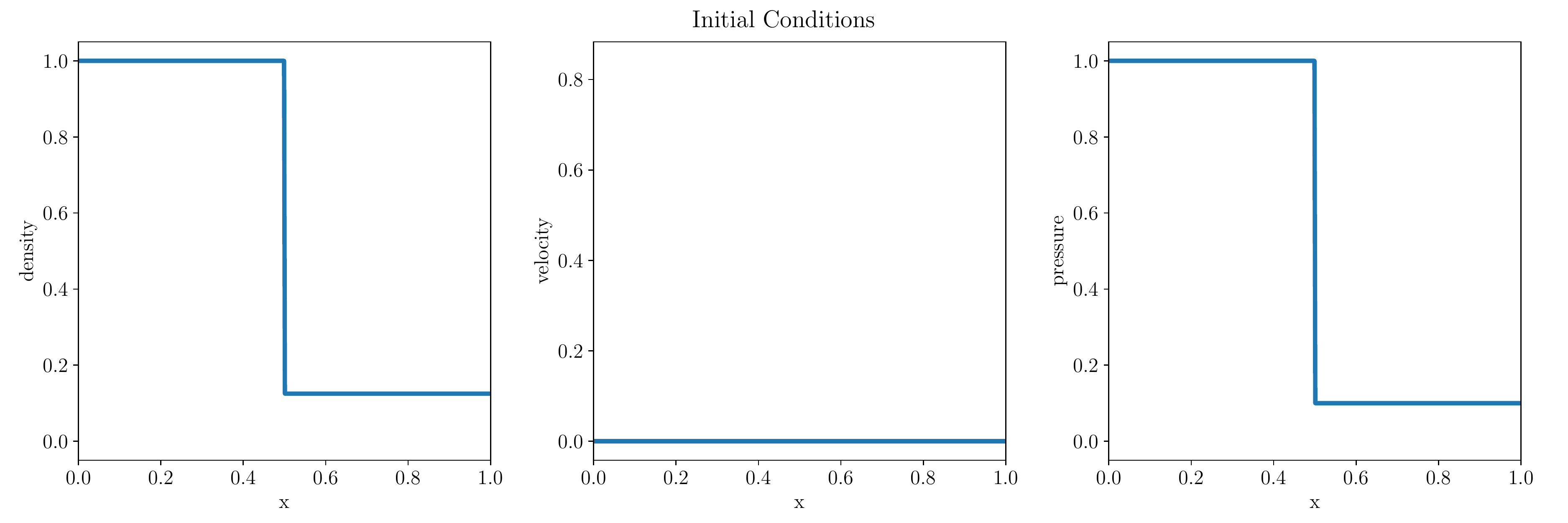}%
\\
\includegraphics[width=\textwidth]{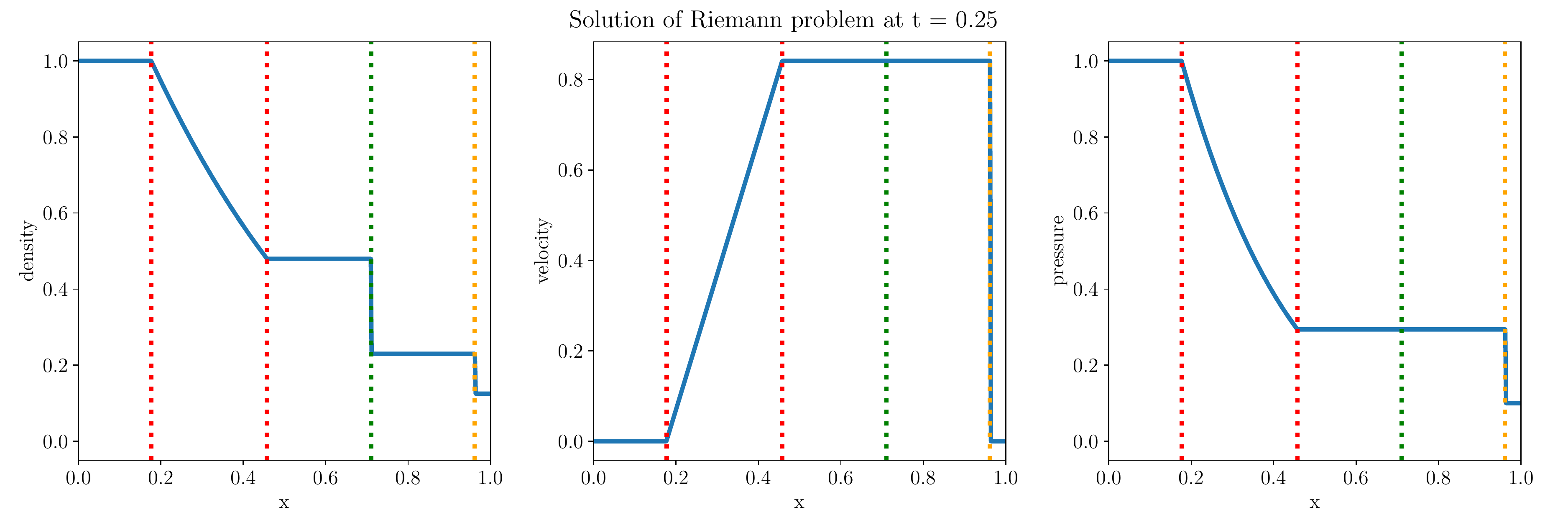}%
\caption[Sod shock initial conditions and solution]{
    Top row: The initial conditions to a classical Riemann problem, called the Sod shock, in
    arbitrary units.
    Bottom row: The exact solution of the problem at $t = 0.25$. The solution consists of a left
    facing rarefaction wave (between the two red dotted lines), easily recognizable due to being a
    smooth transition, not a jump discontinuity.
    To the right (orange dotted line) is a shock wave, across which all three
    primitive quantities (density, pressure, velocity) change as a jump discontinuity.
    The two waves enclose the third middle wave (green dotted line), which is a contact wave.
    The contact wave is a jump discontinuity just like a shock wave, but only the density changes;
    Velocity and pressure remain constant.
    }%
    \label{fig:riemann-solved}
\end{figure}

These considerations tell us the following: The middle wave, associated with $\lambda_2$, will
always be a contact wave, while the other two will either be shocks or rarefactions. Whether the
outer waves are shocks or rarefactions is determined by the new star states $\U_L^*$ and $\U_R^*$.
A model problem containing all three waves is shown in figure \ref{fig:riemann-solved}.
Unfortunately, in order to find the resulting star states $\U_L^*$ and $\U_R^*$ we need to know
which wave type separates them from the outer states $\U_L$ and $\U_R$, respectively, so relations
across the waves, such as the Rankine-Hugeniot relations
\footnote{There are also other very helpful relations, like the Generalized Riemann Invariants,
which have been omitted from this introduction for brevity. In a nutshell, they establish how some
derived quantities change between adjacent states separated by a wave. Under some specific
circumstances, the Generalized Riemann Invariants are constants, and allow us to find relations
between the states.}
(eq.~\ref{eq:rankine-hugeniot}) can be applied correctly. This leaves us with a circular relation:
The specific wave type determines the star states, while the star states determine the wave type.
For this reason, an exact closed form solution for the Euler equations is not available. It is
however possible to make use of an iterative scheme to compute the exact solution numerically,
which is what is at the core of the exact Riemann solver.

A full derivation of the exact Riemann solver is way out of scope for this introduction into the
topic. I refer an interested reader to the works of \citet{toroRiemannSolversNumerical2009}
and \citet{levequeFiniteVolumeMethods2002} for more details. Instead, only an outline of the
derivation and the final result will be discussed.

The outline is as follows: We first establish relations between the four different states $\U_L$,
$\U_L^*$, $\U_R^*$, and $\U_R$ for each permissible wave type between them, which is a contact wave
between $\U_L^*$ and $\U_R^*$, and both shocks and rarefactions between $\U_L$ and $\U_L^*$, and
$\U_R$ and $\U_R^*$, respectively. Experience\footnote{It turns out that using the primitive variables, in particular the pressure $p$ and velocity $v$, is very practical because these two quantities are constant across the middle wave.} shows that a more concise solution can be written in terms of ``primitive variables'' density $\rho$, velocity $v$, and pressure $p$ instead of the
conserved variables density $\rho$, momentum $\rho v$, and energy $E$. Using $\W$ to denote the
state vector of primitive variables, the change of variables is straightforward:


\begin{align}
    \U &=
        \begin{pmatrix}
        (\rho) \\
        (\rho \V) \\
        (E)
        \end{pmatrix} =
        \begin{pmatrix}
        (\rho) \\
        (\rho) \cdot (\V) \\
        \frac{1}{2} (\rho) (\V)^2 + \frac{(p)}{\gamma - 1}
        \end{pmatrix}
    \\
    \W &=
        \begin{pmatrix}
        (\rho) \\
        (\V) \\
        (p)
        \end{pmatrix} =
        \begin{pmatrix}
        (\rho) \\
        \frac{(\rho \V )}{(\rho)} \\
        (\gamma - 1)  \left((E) - \frac{1}{2} \frac{(\rho \V)^2}{(\rho)} \right)
        \end{pmatrix}
\end{align}

Now let's look at the relations between the states across waves:

\subsubsection{Contact Wave}

It can be shown\footnote{It can be shown using the aforementioned Generalized Riemann Invariants,
which have been omitted for brevity.} that across a contact wave, the pressure and fluid velocity
are constant, i.e.

\begin{align}
	p^*_L &= p^*_R = p^*\\
	v^*_L &= v^*_R = v^*
\end{align}

This means that for the Euler equations, the contact wave is a jump discontinuity in the density
$\rho$ only. For this reason, the star state pressure and velocity will have no index indicating
whether they are the left or right star state, and will be referred to as $p^*$ and $v^*$,
respectively. The contact wave also propagates with velocity $\lambda_2(\U_L^*) = \lambda_2(\U_R^*)
= v^*$.

\subsubsection{Shock Wave}

A shock wave is, just like the contact wave, a jump discontinuity. In contrast to a contact wave
however, all three primitive variables $\rho$, $p$, and $u$ change across a shock wave. The
relations between two states separated by a shock wave can be found using the Rankine-Hugeniot
relations (eq. \ref{eq:rankine-hugeniot}).
Explicitly, if the \emph{leftmost} wave (wave (1) in Fig.~\ref{fig:riemann-solution}) is a shock
wave, for a given star state pressure $p^*$ we have:

\begin{align}
	\rho^*_L &=
		\frac{\frac{p^*}{p_L} + \frac{\gamma - 1}{\gamma+1}}{\frac{\gamma - 1}{\gamma+1}
\frac{p^*}{p_L} + 1} \rho_L \label{eq:rho-shock-first}\\
	v^* &=
		v_L - \frac{p^* - p_L}{\sqrt{\frac{p^* + B_L}{A_L}}}
		= v_L - f_L(p^*) \label{eq:velocity-shock-left} \\
    f_L(p^*) &=
        \frac{p^* - p_L}{\sqrt{\frac{p^* + B_L}{A_L}}} \\
	A_L &=
		\frac{2}{(\gamma + 1) \rho_L}\\
	B_L &=
		\frac{\gamma - 1}{\gamma + 1} p_L
\end{align}

The shock speed is
\begin{align}
	S_L = u_L - c_{s,L} \left[\frac{\gamma + 1}{2 \gamma} \frac{p^*}{p_L} + \frac{\gamma -
1}{2\gamma} \right]^{\half} \label{eq:shock-left-speed}
\end{align}

where $c_{s,L}$ is the sound speed in the left state $U_L$.

For a \emph{right shock wave}, i.e. when wave (3) is a shock wave, for a given star state
pressure $p^*$ we have the relations

\begin{align}
	\rho^*_R &=
		\frac{\frac{p^*}{p_R} + \frac{\gamma - 1}{\gamma+1}}{\frac{\gamma - 1}{\gamma+1}
\frac{p^*}{p_R} + 1} \rho_R \\
	v^* &=
		v_R + \frac{p^* - p_R}{\sqrt{\frac{p^* + B_R}{A_R}}}
		= v_R + f_R(p^*) \label{eq:velocity-shock-right} \\
    f_R(p^*) &=
        \frac{p^* - p_R}{\sqrt{\frac{p^* + B_R}{A_R}}} \\
	A_R &=
		\frac{2}{(\gamma + 1) \rho_R}\\
	B_R &=
		\frac{\gamma - 1}{\gamma + 1} p_R
\end{align}

and the shock speed is
\begin{align}
	S_R = v_R + c_{s,R} \left[\frac{\gamma + 1}{2 \gamma} \frac{p^*}{p_R} + \frac{\gamma -
1}{2\gamma} \right]^{\half} \label{eq:shock-right-speed}
\end{align}

where $c_{s,R}$ is the sound speed in the right state $U_R$.

\subsubsection{Rarefaction Wave}

Rarefaction waves are smooth transitions, not infinitesimally thin jump discontinuities.
This makes them really easy to spot in the solutions of Riemann problems (compare with Fig.~
\ref{fig:riemann-solved}).

The rarefaction waves are enclosed by the head and the tail of the wave, between which we have a
smooth transition which is called the ``fan''.
The head is the ``front'' of the wave, i.e. the part of the wave that gets furthest away from the
origin as time progresses.
The tail is the ``back'' of the wave, i.e. the part of the wave that stays closest to the origin as
time progresses.

If we have a \emph{left-facing} rarefaction, i.e. if wave (1) is a rarefaction wave,
the wave speeds of the head, $S_{HL}$, and the tail, $S_{TL}$, are given by

\begin{align}
	S_{HL} &= u_L - c_{s,L}\\
	S_{TL} &= u^* - c_{s,L}^*\\
	c_{s,L}^*  &= c_{s,L} \left( \frac{p^*}{p_L} \right) ^ \frac{\gamma - 1}{2 \gamma}
\end{align}

The star state $\W_L^*$ for a given pressure $p^*$ is determined by

\begin{align}
	\rho^*_L &=
		\rho_L \left( \frac{p^*}{p_L} \right) ^ \frac{1}{\gamma}\\
	v^* &=
		v_L - \frac{2 c_{s,L}}{\gamma - 1} \left[ \left( \frac{p^*}{p_L} \right) ^ \frac{\gamma -
1}{2 \gamma} -1  \right]
		= v_L - f_L(p^*) \label{eq:velocity-rarefaction-left} \\
    f_L(p^*) &=
        \frac{2 c_{s,L}}{\gamma - 1} \left[ \left( \frac{p^*}{p_L} \right) ^ \frac{\gamma - 1}{2
\gamma} -1  \right]\\
\end{align}

where and $c_{s,L}$ is the sound speed in the left state $U_L$.

The solution \emph{inside} the rarefaction fan, i.e. in regions where $S_{HL} \leq \frac{x}{t}
\leq S_{TL}$, is

\begin{align}
	\rho_{\text{fan}, L} &=
		\rho_L \left[ \frac{2}{\gamma + 1} + \frac{\gamma - 1}{\gamma + 1} \frac{1}{c_{s,L}}
\left(v_L - \frac{x}{t}\right) \right] ^ \frac{2}{\gamma -1 } \label{eq:rho-rarefaction-fan-left}\\
	v_{\text{fan}, L} &=
		\frac{2}{\gamma + 1} \left[ \frac{\gamma - 1}{2} v_L + c_{s,L} + \frac{x}{t}  \right] \\
	p_{\text{fan}, L} &=
		p_L \left[ \frac{2}{\gamma + 1} + \frac{\gamma - 1}{\gamma + 1} \frac{1}{c_{s,L}} \left(v_L
- \frac{x}{t}\right) \right] ^ \frac{2 \gamma}{\gamma -1} \label{eq:pressure-rarefaction-fan-left}
\end{align}

If we have a \emph{right-facing rarefaction}, i.e. if wave (3) is a rarefaction wave, we have

\begin{align}
	\rho^*_R &=
		\rho_R \left( \frac{p^*}{p_R} \right) ^ \frac{1}{\gamma}\\
	v^* &=
		v_R - \frac{2 c_{s,R}}{\gamma - 1} \left[ 1 - \left( \frac{p^*}{p_R} \right) ^ \frac{\gamma
- 1}{2 \gamma}  \right]
		= v_R + f_R(p^*)
    \label{eq:velocity-rarefaction-right}
\\
    f_R(p^*) &= \frac{2 c_{s,R}}{\gamma - 1} \left[ 1 - \left( \frac{p^*}{p_R} \right) ^
\frac{\gamma -
1}{2 \gamma}  \right]
\end{align}

where $c_{s,R}$ is the sound speed in the left state $U_R$.

The wave speeds of the head, $S_H$, and the tail, $S_T$, for the left facing wave are
\begin{align}
	S_{HR} &= v_R + c_{s,R}\\
	S_{TR} &= v^* + c^*_{s,R}\\
	c^*_{s,R}  &= c_{s,R} \left( \frac{p^*}{p_R} \right) ^ \frac{\gamma - 1}{2 \gamma}
\end{align}

Finally, the solution inside the rarefaction fan, i.e. in regions where $S_{HL} \leq \frac{x}{t}
\leq S_{TL}$, is

\begin{align}
	\rho_{\text{fan}, R} &=
		\rho_R \left[ \frac{2}{\gamma + 1} - \frac{\gamma - 1}{\gamma + 1} \frac{1}{c_{s,R}}
\left(v_R - \frac{x}{t}\right) \right] ^ \frac{2}{\gamma -1 } \label{eq:rho-rarefaction-fan-right}
\\
	v_{\text{fan}, R} &=
		\frac{2}{\gamma + 1} \left[ \frac{\gamma - 1}{2} v_R - c_{s,R} + \frac{x}{t}  \right]
    \label{eq:velocity-rarefaction-fan-right}
\\
	p_{\text{fan}, R} &=
		p_R \left[ \frac{2}{\gamma + 1} - \frac{\gamma - 1}{\gamma + 1} \frac{1}{c_{s,R}} \left(v_R
- \frac{x}{t}\right) \right] ^ \frac{2 \gamma}{\gamma -1} \label{eq:pressure-rarefaction-fan-right}
\end{align}

\subsubsection{Which wave type do we have?}

As noted before, the middle wave (wave (2) in Fig.~\ref{fig:riemann-solution}) is always a
contact wave, while the other two waves are any combination of rarefaction and/or shock wave.
It turns out that the condition for a rarefaction or shock wave is remarkably simple:

For the left wave (wave (1)):

\begin{align}
	p^* > p_L: && &\quad \text{ (1) is a shock wave}\\
	p^* \leq p_L: && &\quad \text{ (1) is a rarefaction wave}
\end{align}

and for the right wave (wave (3)):

\begin{align}
	p^* > p_R: && & \quad \text{ (3) is a shock wave} \\
	p^* \leq p_R: && & \quad \text{ (3) is a rarefaction wave}
\end{align}

A way of understanding these conditions is to again consider the behavior of the characteristics,
and to take into account that the pressure of an fluid has no preferential direction. So if the
pressure in the star region is greater than the pressure across an outer wave, it'll ``push'' the
fluid from the star region stronger than in the outer regions. Interpreting this ``push'' as the
behavior of the characteristics of the fluid, the characteristics of the star regions will have a
steeper slope, and eventually cross with the characteristics from the outer regions on the $x-t$
plane. This is precisely the condition of converging characteristics required for shock waves in a
compressive region.

\subsubsection{Solution for $p^*$}

The final missing element to have a complete exact solution to the Riemann problem for the Euler
equations is an expression how to obtain $p^*$, the pressure in the star region, depending on the
initial conditions $\W_L$ and $\W_R$. We make use of the fact that $p^*$ and $v^*$ are constant
across the star region states $\W_L^*$ and $\W_R^*$. For both shock and rarefaction waves on either
side, we have equations for $v^*$ depending on the outer states  $\W_L$ and $\W_R$ and $p^*$ (eqns.
\ref{eq:velocity-shock-left}, \ref{eq:velocity-shock-right}, \ref{eq:velocity-rarefaction-left},
and \ref{eq:velocity-rarefaction-right}). By setting $v^*_L - v^*_R = 0$, which must hold across
the contact wave, we obtain the equation

\begin{align}
	f(p^*, \W_L, \W_R) \equiv f_L(p^*, \W_L) + f_R(p^*, \W_R) + (v_R - v_L) = 0
\label{eq:riemann-pressure-equation}
\end{align}

with

\begin{align}
	f_{L,R} &=
		\begin{cases}
			(p^* - p_{L,R}) \left[ \frac{A_{L,R}}{p^* + B_{L,R}} \right]^{\frac{1}{2}}
				& ~\text{ if } ~ p^* > p_{L,R} ~ \quad \text{(shock)} \\
			\frac{2 c_{s,L,R}}{\gamma - 1} \left[ \left( \frac{p^*}{p_{L,R}} \right)^ \frac{\gamma
-1}{2 \gamma} - 1 \right]
				& ~\text{ if } ~ p^* \leq p^*_{L,R} ~ \quad \text{(rarefaction)}
\label{eq:riemann-pstar}\\
		\end{cases} \\
	A_{L,R} &=
		\frac{2}{(\gamma + 1) \rho_{L,R}}\\
	B_{L,R} &=
		\frac{\gamma - 1}{\gamma + 1} p_{L,R}
\end{align}

Since the expressions for $f_{L,R}$ are analytical, we can also compute their derivatives w.r.t.
$p^*$:

\begin{align}
	\frac{\del f_{L,R}}{\del p^*} &=
		\begin{cases}
			\left[
                \frac{A_{L,R}}{p^* + B_{L,R}}
            \right]^{\frac{1}{2}} \left( 1 - \frac{1}{2}
            \frac{p^* - p_{L,R}}{p + B_{L,R}} \right)
				& ~\text{ if } ~ p^* > p_{L,R} ~ \quad \text{(shock)} \\
			\frac{a_{L,R}}{\gamma p_{L,R}}
			\left( \frac{p^*}{p_{L,R}} \right)^ \frac{-(\gamma+1)}{2 \gamma}
				& ~\text{ if } ~ p^* \leq p_{L,R} ~ \quad \text{(rarefaction)}
\label{eq:riemann-pstar-dp}\\
		\end{cases}
\end{align}

giving the complete derivative of eq.~\ref{eq:riemann-pstar}:

\begin{align}
  \frac{\del}{\del p^*} f(p^*, \W_L, \W_R)= \frac{\del f_L(p^*, \W_L)}{\del p^*} + \frac{\del
f_{R}(p^*, \W_R)}{\del p^*} \ ,
\end{align}

and make use of it to find the $p^*$ that solves eq.~\ref{eq:riemann-pstar} using the iterative
Newton-Raphson method. The method prescribes that the $n + 1$-th iteration is determined by

\begin{align}
	p^*_{n+1} = p^*_n - \frac{f(p^*_n, \W_L, \W_R)}{\frac{\del}{\del p^*}f(p*_n, \W_L, \W_R)}
\end{align}

and is re-iterated until it converges, i.e. when the relative pressure change

\begin{align}
	\frac{|p_k - p_{k+1}|}{\frac{1}{2} | p_k + p_{k+1} | } < \epsilon
\end{align}

where $\epsilon$ is some tolerance, e.g. $10^{-6}$. To begin the iteration, a first guess $p_0^*$
is necessary. Taking the average pressure:

\begin{align}
	p_0^* = \frac{1}{2} (p_L + p_R)
\end{align}

gives acceptable results. A better first guess, i.e. a guess that typically leads to fewer required
iteration steps for the method to converge, is by taking the solution for the star state pressure
of the linearized primitive variable Riemann solver (see
Appendix~\ref{app:riemann-primitive-variables}):

\begin{align}
	p_{PV} &=
        \frac{1}{2} (p_L + p_R) - \frac{1}{8} (v_R - v_L)(\rho_L + \rho_R)(c_{s,L} + c_{s,R})\\
	p_0^* &= \max(\epsilon, p_{PV})
\end{align}

With this, the exact Riemann solver is complete. A method to obtain the star state pressure $p^*$
is given, and with the pressure known, the outer wave types are determined. The exact star states,
$\W_L^*$ and $\W_R^*$ can then be determined by using the appropriate relations for the
corresponding wave type, which are given by
eqns.~\ref{eq:rho-shock-first}~-~\ref{eq:pressure-rarefaction-fan-right}.

\subsubsection{Sampling the Solution}

\begin{sidewaysfigure}
	\includegraphics[width=\textheight]{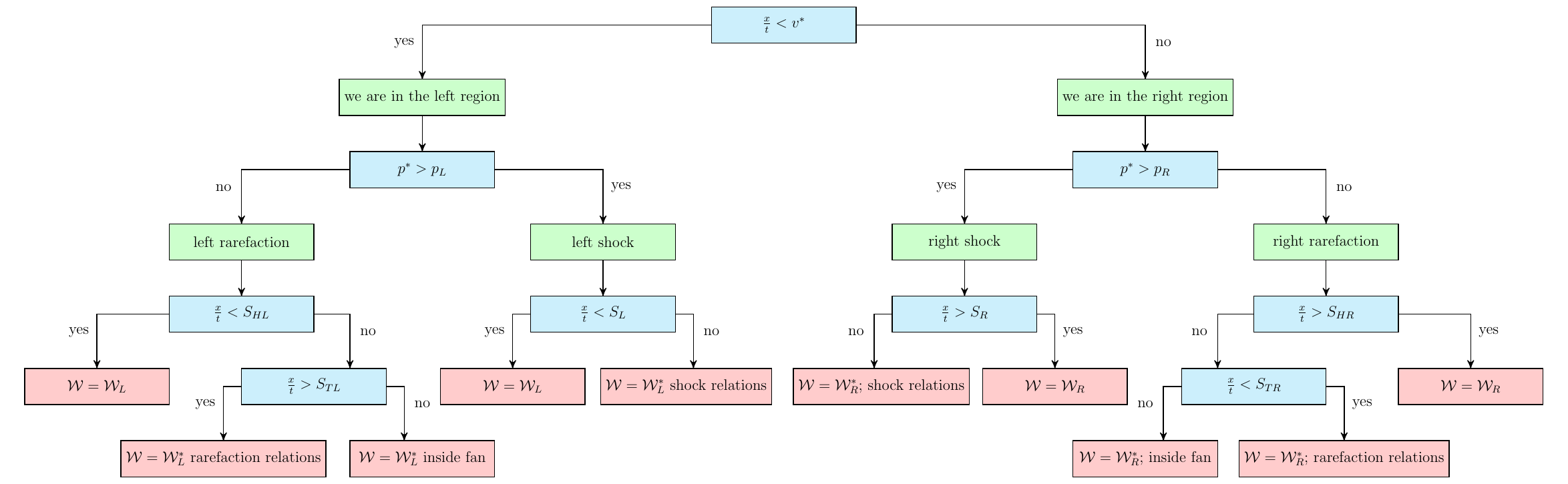}%
	\caption[Flowchart on sampling the solution]{Flow chart to sample the solution of the Riemann
problem for the Euler equations at a given point $(x, t)$.
		\label{fig:sampling-solution}
	}
\end{sidewaysfigure}

With the exact solver readily available, the final task is to find the solution at some desired
point $(x, t)$. This can be achieved by sampling the solution. Assuming all the star region state
variables are computed, then all four states $\U_L$, $\U_L^*$, $\U_R^*$, and $\U_R$ are known. What
is left to do is to determine in which region the point $(x, t)$ is located. This is done by
comparing $x/t$ with the wave velocities: If $x / t < v^*$, then the point $(x, t)$ must be located
in the left region, i.e. in either $\U_L$ or $\U_L^*$. Further comparison of $x / t$ with the wave
speed of the left wave, which is either a shock or a rarefaction, determines whether $(x, t)$ is in
$\U_L$ or $\U_L^*$. If the left wave is a rarefaction, then another possible solution is for $(x,t)$
to be located inside the rarefaction fan instead. The analogous distinction process is performed if
$(x, t)$ is located in the right region. A full flow chart of decision making and finally which
relations to use is shown in figure \ref{fig:sampling-solution}.

\subsection{Approximate Solvers}\label{chap:riemann-approximate}

As mentioned before, the solution of the Riemann problem is at the heart of finite volume fluid
dynamics methods. It needs to be solved over and over again at every time step and between every
pair of interacting cells. For example, if a three dimensional simulation volume is divided up in
only 128 cells of equal size in each dimension, it would require $6.29 \times 10^6$ Riemann
problems to be solved \emph{each time step}, assuming that each interacting pair of cells requires
the solution of only one Riemann problem.\footnote{Some methods, like the MUSCL-Hancock scheme,
require more than one Riemann problem to be solved per interaction per cell in order to increase
the accuracy of the numerical method. More details are given in Section \ref{chap:MUSCL-Hancock}}

Given that the exact solver requires iterations in order to find the pressure of the star states
$p^*$, it can become a considerable computational expense. In order to cut some costs, approximate
non-iterative Riemann solvers have been developed. While they only provide approximate solutions,
tests show that they are indeed sufficiently accurate to be made use of in actual simulations. How
exactly the Riemann solvers are used in finite volume methods to solve the Euler equations will be
the topic of Chapters~\ref{chap:godunov}~and~\ref{chap:higher-order-schemes}, while the influence
of approximate Riemann solvers on simulations will be discussed in
Section~\ref{chap:godunov-application}. In what follows, some popular approximate Riemann solvers
are introduced.

\subsubsection{Two-Rarefaction Riemann Solver (TRRS)}\label{chap:riemann-trrs}

The Two-Rarefaction Riemann Solver (TRRS) is a modification of the exact Riemann solver. The
approach used to skip over the iteration for the star state pressure $p^*$ is to assume that both
outer waves are rarefaction waves, and to use that assumption to get an expression for $p^*$ and
$v^*$, the pressure and velocity in the star region, respectively. The expressions are given by:

\begin{align}
	\beta &\equiv
		\frac{\gamma - 1}{2 \gamma} \\
	v^* &=
		\frac{
			\frac{2}{\gamma - 1} \left[\left(\frac{p_L}{p_R} \right) ^ \beta - 1\right]+
\frac{v_L}{c_{s,L}} \left(\frac{p_L}{p_R} \right) ^ \beta  + \frac{v_R}{c_{s,R}}
		}{
			\frac{1}{c_{s,R}} + \frac{1}{c_{s,L}}\left(\frac{p_L}{p_R} \right) ^ \beta
		} \\
	p^* &=
		\frac{1}{2} \left[
			p_R \left[ \frac{\gamma - 1}{2 c_{s,R}} (v^* - v_R) + 1 \right] ^ \frac{1}{\beta} +
			p_L \left[ \frac{\gamma - 1}{2 c_{s,L}} (v_L - v^*) + 1 \right] ^ \frac{1}{\beta}
		\right] \label{eq:pstar-trrs} \\
		&=
		\left[
			\frac{ c_{s,L} + c_{s,R} - \frac{\gamma - 1}{2} (v_R - v_L)}{\frac{c_{s,L}}{p_L^\beta}
+         \frac{c_{s,R}}{p_R^\beta}}
		\right] ^ \frac{1}{\beta}
\end{align}

To remain consistent with the two-rarefaction assumption, the star state densities can be obtained
using
\begin{align}
    \rho^*_{L,R} = \rho_{L,R} \left(\frac{p^*}{p_{L,R}} \right)^{\frac{1}{\gamma}}
\end{align}

However, an improved solution can be obtained by using the approximation only to estimate the star
state pressure, $p^*$, and then proceeding in the same manner as the exact solver does: By
determining what type of waves the two outer waves are based on $p^*$, and using the appropriate
relations for the wave types to determine the other star state variables and wave velocities.

\subsubsection{Two-Shock Riemann Solver (TSRS)}\label{chap:riemann-tsrs}

Similar to the TRRS solver, the Two-Shock Riemann Solver (TSRS) is also a modification of the exact
Riemann solver. As the name suggests, in order to determine the star region pressure $p^*$, it is
assumed a priori that both the left and right waves are going to be shock waves.

The equation for the pressure in the star region (eq. \ref{eq:riemann-pressure-equation}) then is given by

\begin{align}
	f(p) &= (p - p_L) g_L(p) + (p - p_R) g_R(p) + v_R - v_L = 0 \label{eq:riemann-pressure-tsrs} \\
	g_{L,R}(p) &= \left[ \frac{A_{L,R}}{p + B_{L,R}} \right] ^{\half} \\
	A_{L,R} &=
		\frac{2}{(\gamma + 1) \rho_{L,R}}\\
	B_{L,R} &=
		\frac{\gamma - 1}{\gamma + 1} p_{L,R}
\end{align}

Unfortunately, this approximation does not lead to a closed form solution, and a further
approximation is needed. A first estimate for the pressure, $p_0$, is used as the argument for
$g_{L,R}(p)$ in eq.~\ref{eq:riemann-pressure-tsrs} to get a better approximation for $p^*$:

\begin{align}
	p^* = \frac{g_L(p_0) p_L + g_R(p_0)p_R - (v_R - v_L)}{g_L(p_0) + g_R(p_0)} \label{eq:pstar-tsrs}
\end{align}

The star region velocity which is consistent with the two-shock assumption is given by

\begin{align}
	v^*  = \frac{1}{2} (v_L + v_R) + \frac{1}{2} \left[ (p^* - p_R) g_R(p_0) - (p^* - p_L) g_L(p_0)
\right]
\end{align}

A good choice for $p_0$ is coming from the solution for the star state pressure of the linearized
primitive variable Riemann solver (Appendix~\ref{app:riemann-primitive-variables}):

\begin{align}
	p_{PV} &= \frac{1}{2} (p_L + p_R) - \frac{1}{8} (v_R - v_L)(\rho_L + \rho_R)(c_{s,L} +
c_{s,R})\\
	p_0 &= \max(0, p_{PV})
\end{align}

Just as was the case for the TRRS solver, an improved solution can be found by only using the TSRS
approximation to estimate $p^*$ and determining the other star region quantities using the
relations of the exact Riemann solver.

\subsubsection{HLLC Solver}\label{chap:riemann-hllc}

The HLLC (Harten, Lax, van Leer, contact) Riemann solver is derived by assuming that the solution
consists of three waves that are jump discontinuities. The three waves are traveling with speeds
denoted by $S_L$, $S^*$, and $S_R$, respectively. In the context of the Euler equations, it is an
improved version of the HLL (Harten, Lax, van Leer) solver, which doesn't take into account a middle
contact wave. While the missing contact wave of the HLL approach is sub-optimal for the Euler
equations, it can be a good approximation for other systems of hyperbolic conservation laws that
only consist of two equations. Indeed, the approximate HLL solver will be used later in the context
of the moments of the equation of radiative transfer in Section~\ref{chap:riemann-rt}.


To find relations across the three contact discontinuities which the HLLC approach assumes, the
Rankine-Hugeniot relations (eq.~\ref{eq:rankine-hugeniot}) can be applied. Further relations can be
found by integrating the conservation law equations over both a space and time interval, and using
that integral to obtain the integrated average value in the star region. The solution takes the form

\begin{align}
	\U(x, t) = \begin{cases}
		\U_L ~~~~ &\text{ if }		~~	\frac{x}{t} \leq S_L \\
		\U^*_L ~~~~ &\text{ if }	~~  S_L \leq \frac{x}{t} \leq S^* \\
		\U^*_R ~~~~ &\text{ if }	~~	S^* \leq \frac{x}{t} \leq S_R \\
		\U_R ~~~~ &\text{ if }		~~	S_R \leq \frac{x}{t} \\
	\end{cases}
\end{align}

By additionally using properties of the contact wave from the exact solver, i.e. the fact that $p^*
= p^*_L = p^*_R$ and $v^* = v^*_L = v^*_R = S^*$, an expression for $S^*$ can be found that only
depends on the initial left and right states:

\begin{align}
	S^* &=
		\frac{p_R - p_L  + \rho_L v_L (S_L - v_L) - \rho_R v_R (S_R - v_R)}
			{\rho_L (S_L - v_L) - \rho_R (S_R - v_R) } \label{eq:hllc-sstar}
\end{align}

and ultimately

\begin{align}
	\U^*_{L,R} &=
		\rho_{L,R} \frac{ S_{L,R} - v_{L,R}}{S_{L,R} - S^*}
		\begin{pmatrix}
			1\\
			S^*\\
			\frac{E_{L,R}}{\rho_{L,R}} + (S^* - v_{L,R}) \left( S^* +
\frac{p_{L,R}}{\rho_{L,R}(S_{L,R} - v_{L,R})} \right)
		\end{pmatrix} \\
	\F^*_{L,R} &=
		\F_{L, R} + S_{L,R} ( \U^*_{L,R} - \U_{L,R} )
\end{align}

It remains to find estimates for the left and right wave speeds, $S_L$ and $S_R$.
There are multiple ways to get good and robust estimates. A very simple one is:

\begin{align}
	S_L  &= v_L - c_{s,L} q_L \\
	S_R  &= v_R + c_{s,R} q_R \\
	q_{L,R} &=
		\begin{cases}
			1	~~~~ & \text{ if } p^* \leq p_{L,R} ~~~~ \text{(rarefaction)}\\
			\sqrt{1 + \frac{\gamma + 1}{2 \gamma} \left(\frac{p^*}{p_{L,R}} - 1 \right)}
			~~~~ & \text{ if } p^* > p_{L,R} ~~~~ \text{(shock)}\\
		\end{cases} \\
	p^* &= \max(0, p_{PV})\\
	p_{PV} &= \frac{1}{2} (p_L + p_R) - \frac{1}{8} (v_R - v_L)(\rho_L + \rho_R)(c_{s,L} + c_{s,R})
\label{eq:pstar-pv}.
\end{align}

A better estimate using an adaptive wave speed estimate method can be obtained.
In order to do so, first  the primitive variable solution for the star state pressure $p^*$
following eq. \ref{eq:pstar-pv} is computed (see Appendix~\ref{app:riemann-primitive-variables}).
That first guess is kept as long as the ratio $\frac{p_{max}}{p_{min}}$ is small enough (typically
$\sim 2$), where $p_{max} = \max \{ p_L, p_R \}$ and $p_{min} = \min \{ p_L, p_R \}$. Furthermore,
we also demand that $p_{min} \leq p_{PV} \leq p_{max}$. Should one of these two conditions not be
satisfied, then we switch to the star state solutions of other approximate Riemann solvers: If
$p_{PV} \leq p_{min}$, then it's reasonable to expect two rarefactions to form, and applying the
star state estimates of the Two Rarefaction Riemann Solver (TRRS, eq. \ref{eq:pstar-trrs}) is a good
choice. Otherwise, we expect at least one shock should be present, so the star state estimates of
the Two Shock Riemann Solver (TSRS, eq. \ref{eq:pstar-tsrs}) will likely provide a better estimate.

\subsubsection{On the Accuracy of Approximate Riemann Solvers}

\begin{figure}
\centering
\includegraphics[width=\textwidth]{
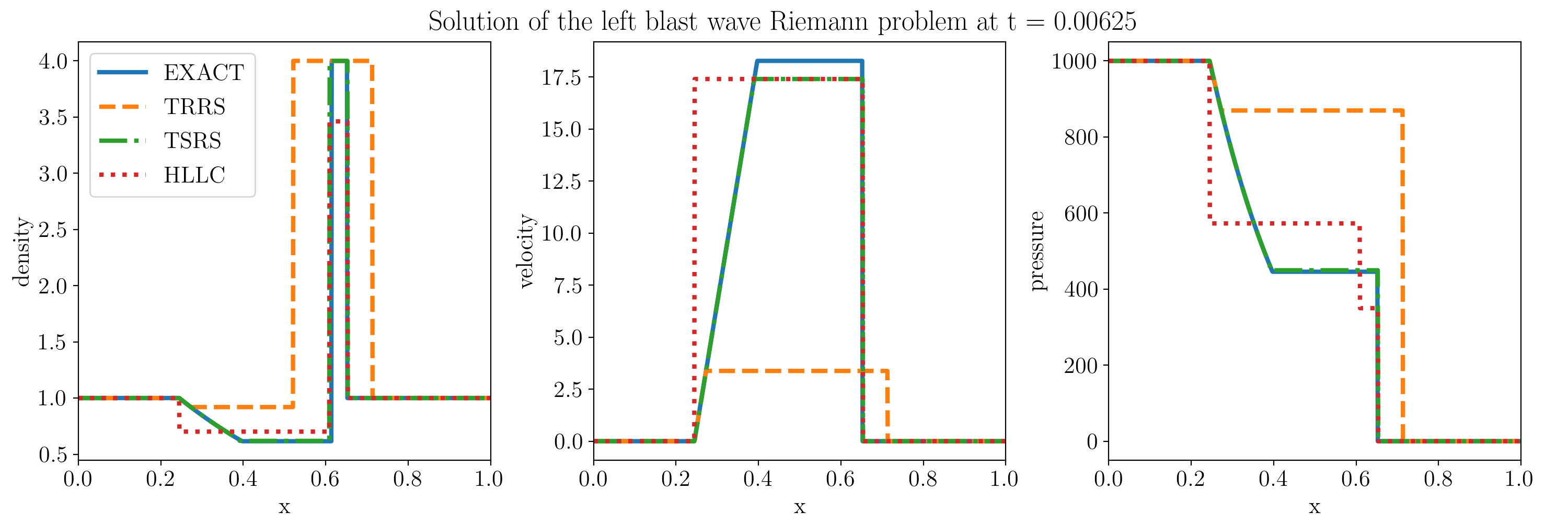} %
\\
\includegraphics[width=\textwidth]{
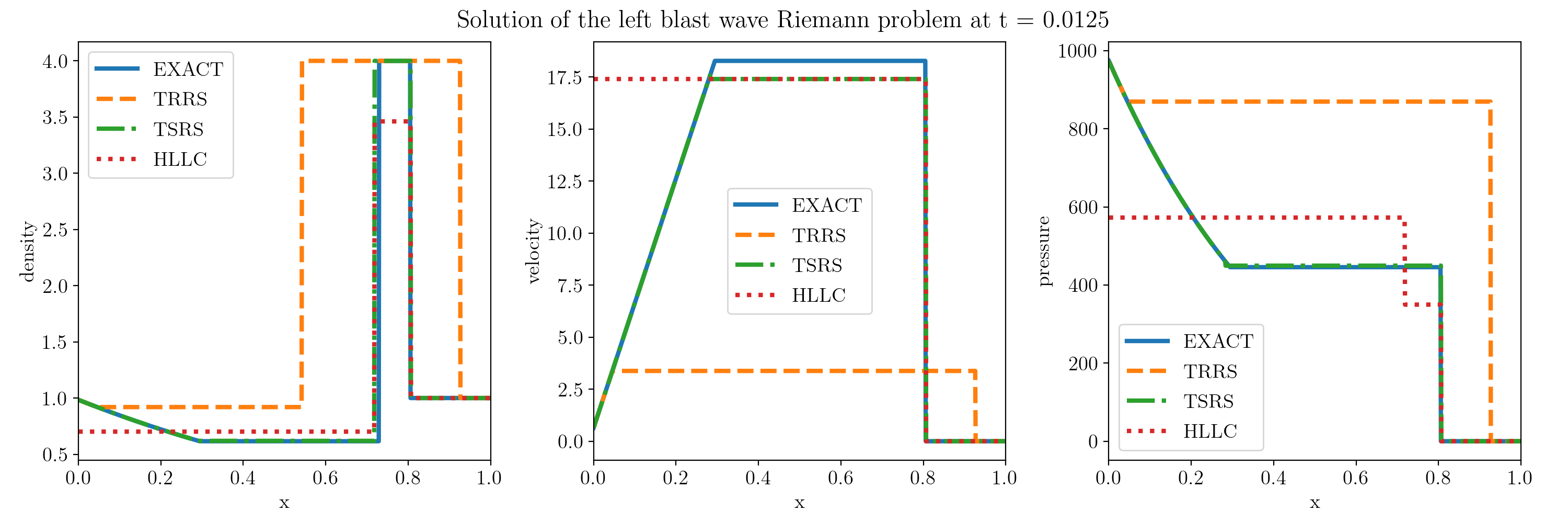} %
\caption[Left blast wave solution with approximate Riemann solvers]{
    The solution to the left blast wave test (eq. \ref{eq:left-blast-wave-ICs}) using different
    approximate and the exact Riemann solver. The exact solution consists of a left-facing
    rarefaction and a right facing shock. To showcase the wave originating on the left side of the
    interface (located at $x=0.5$) and traveling to the right, the solution is shown at two
    different times $t = 0.00625$ and $t = 0.0125$.
    }%
    \label{fig:approximate-riemann-left-blast-wave}
\end{figure}

To conclude the chapter on approximate Riemann solvers, let's have a look at how accurate the
solutions of the approximate solvers are. In what follows, Riemann problems are solved using the
exact solver as well as the approximate TRRS, TSRS, and HLLC solvers. All quantities are in
arbitrary units, and the tests are performed in one dimension using the \meshhydro code's option
to run it as a standalone Riemann solver. For each test, an initial left and right state are
specified, which are separated at the position $x = 0.5$. The solution is obtained using the various
Riemann solvers, and then sampled over the region $x \in [0, 1]$ over $10^3$ equally spaced points.
It is important to note that the solution the Riemann solvers provide for the star
region states do not evolve over time. The only thing that changes over time is the front of the
waves, but the states remain the same for all $t > 0$. This can be seen well in
Figure~\ref{fig:approximate-riemann-left-blast-wave}, where the solution at two different times is
shown. Hence it makes little sense to compare how well the approximate solvers perform as time
evolves, and the results for each test are shown for only a single time.

\begin{figure}
\centering
\includegraphics[width=\textwidth]{
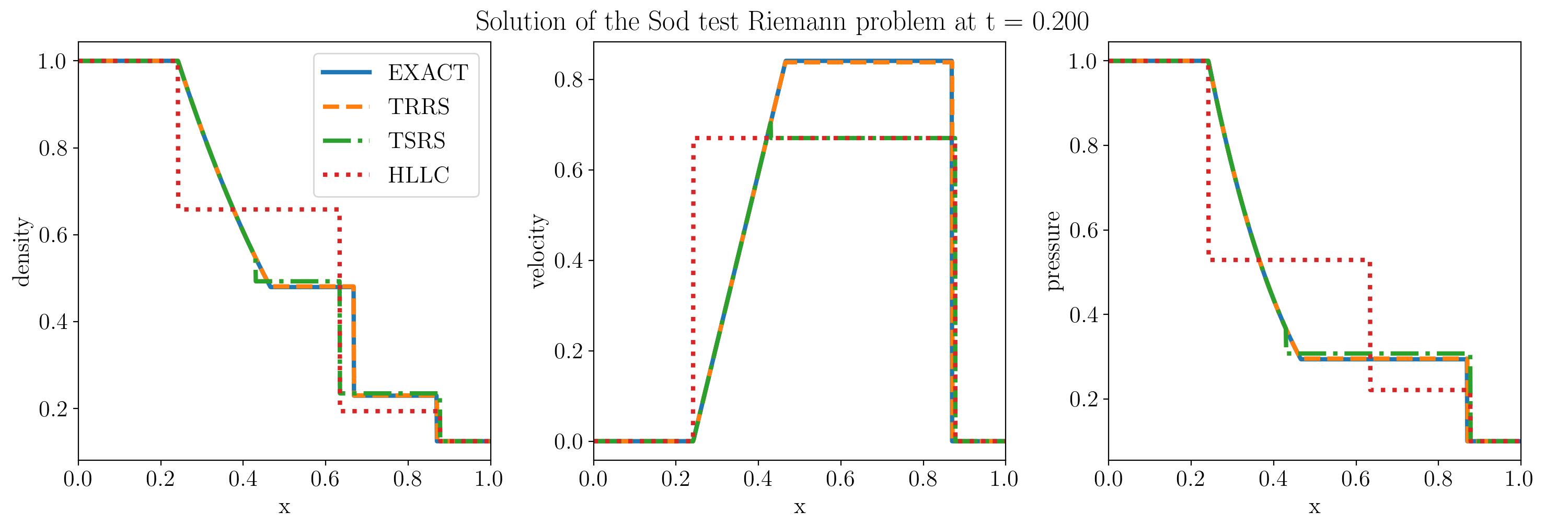} %
\caption[Sod test solution with approximate Riemann solvers]{
    The solution to the Sod test (eq. \ref{eq:sod-test-ICs}) using different approximate and the
    exact Riemann solver. The exact solution consists of a left-facing rarefaction and a right
    facing shock.
    }%
\label{fig:approximate-riemann-sod-test}
\end{figure}

Figure~\ref{fig:approximate-riemann-sod-test} shows the results of the so-called ``Sod test''
Riemann problem with initial conditions

\begin{align}
    \rho_L = 1 && v_L = 0 && p_L = 1 \label{eq:sod-test-ICs} \\
    \rho_R = 0.125 && v_R = 0 && p_R = 0.1      \nonumber
\end{align}

The exact solution consists of a left-facing rarefaction and a right facing shock. The TRRS
solver gives a result which is nearly identical to the exact solution, which makes sense given the
prominent rarefaction wave in the solution. All solvers except the HLLC find approximately the
same pressure in the star region, which indicates that all of them will ``identify'' the correct
wave type when sampling the solution.
Both the HLLC and the TSRS solver find nearly identical velocities in the star region compared to
each other, but offset compared to the exact solution, which is due to the similarity they share in
how the star region velocity (and pressure) is initially estimated.
The HLLC solver, which by construction can't handle the
smooth transition of the rarefaction since it assumes waves are only jump discontinuities, has some
trouble getting the correct solution on the left side of the problem, and the errors propagate
throughout the star region. A further noticeable feature of the HLLC solver is that it's the only
one that has a pressure change over the star region, i.e. across the contact wave. This is not in
agreement with what the properties of the contact wave dictate, and is a consequence of the way the
outer wave speeds $S_L$ and $S_R$ are estimated. To avoid an iterative scheme (which is the
entire point of approximate solvers), either the identical pressure over the star region can be
enforced, or the correct relations across the waves with velocities $S_L$ and $S_R$ can be
enforced. But to have both the pressure and the relations across the wave be consistent can't be
done without an iterative scheme.

Figure~\ref{fig:approximate-riemann-left-blast-wave} shows the results of the so-called ``Left
blast wave'' Riemann problem with initial conditions

\begin{align}
    \rho_L = 1 && v_L = 0 && p_L = 1000 \label{eq:left-blast-wave-ICs} \\
    \rho_R = 1 && v_R = 0 && p_R = 0.01      \nonumber
\end{align}

Similarly to the Sod test, the solution consists of a left-facing rarefaction and a right facing
shock. However in this test the shock is much stronger and much more prominent. The compression
wave travels from the left to the right. Figure~\ref{fig:approximate-riemann-left-blast-wave}
depicts the solution at two times to showcase this behavior.

In this test, the TSRS solver gives a result very close to the exact solution. The HLLC solver
handles the jump discontinuities on the right hand side of the solution quite well, but struggles
again with the rarefaction on the left. The TRRS solver struggles this time around to get close to
the exact solution, which makes sense due to the prominent shock present, which is contrary to its
underlying two-rarefaction assumption.

\begin{figure}
\centering
\includegraphics[width=\textwidth]{
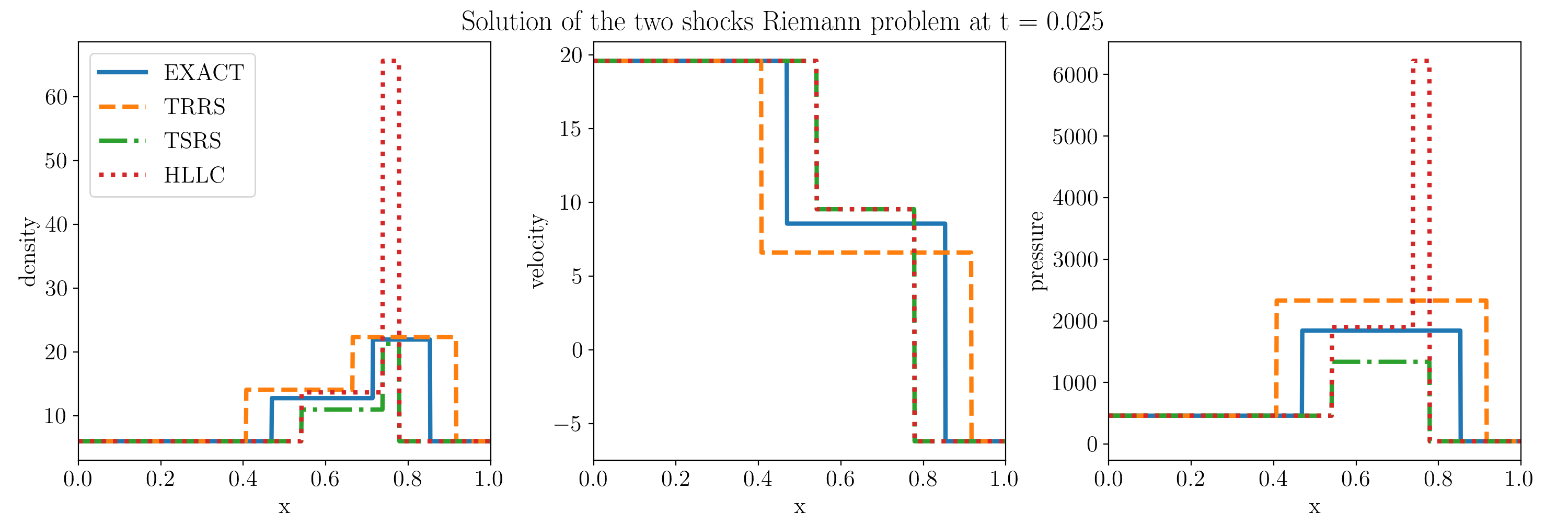} %
\caption[Two shocks solution with approximate Riemann solvers]{
    The solution to the two shock test (eq. \ref{eq:left-blast-wave-ICs}) using different
    approximate and the exact Riemann solver. The exact solution consists of a two shocks on either
    side of the contact wave.
    }%
    \label{fig:approximate-riemann-two-shocks}
\end{figure}

A further test, shown in Figure~\ref{fig:approximate-riemann-two-shocks} where all the approximate
solvers struggle to reproduce the exact solution is the two-shock test, given by the initial
conditions

\begin{align}
    \rho_L = 5.99924 && v_L = 19.5975 && p_L = 460.894 \label{eq:two-shock-ICs} \\
    \rho_R = 5.99242 && v_R = -6.19633 && p_R = 46.095      \nonumber
\end{align}

As the name suggests, the solution consists of two shock waves, one on either side of the central
contact wave. While it can be expected from the TRRS solver to have issues reproducing the exact
result, both the HLLC and the TRRS solvers struggle as well. They fail to obtain a close enough
initial guess for the central region, and the offsets from the exact solution consequently
propagate into the resulting states and wave speeds. A silver lining is that once again all solvers
were at least able to ``identify'' that both outer waves are shock waves.

The errors of the results that approximate solvers yield may seem troubling at first, but it turns
out that when employed as solvers in simulations, those errors aren't as severe as the results
shown may suggest. Indeed the results aren't off by orders of magnitudes. The most extreme
deviations in these examples are off by a factor of $\sim 2 - 3$. Note however that these are errors
on the solution of the Riemann problems specified by the initial conditions only. In simulations,
the the state of the fluid is evolved over  many small time steps, typically much smaller than the
end times shown in the previous plots. This means that initially large errors can get smoothed out
over subsequent time steps. Additionally, numerical effects like diffusion, and the application of
flux and slope limiters for higher order schemes lead to the errors resulting from approximate
solvers being in many cases negligible. The influence of the approximate Riemann solvers is shown
in e.g. Section~\ref{chap:godunov-application}. Which Riemann solver to choose depends on factors
like what physical case is going to be simulated, as well as specifics of the underlying method.
For example, very smooth flows will likely have little to no shocks, and a TRRS solver could be
beneficial. Conversely, in turbulent flows or in cases where a lot of energy is being injected into
the gas, the TRRS solver can be practical. For a general purpose application, the HLLC solver is
recommended, or even the exact one, if one can afford it.

\subsection{Dealing with Vacuum}\label{chap:vacuum}

A final topic that requires some discussion is the special case of vacuum. Vacuum is characterized
by the condition $\rho = 0$. Given the equation of state for ideal gas (eq.
\ref{eq:equation-of-state}), it follows that $E = 0$ as well.
The structure of the solution to the Riemann problem is different when vacuum is present. The
general solution structure of three waves separating four distinct states doesn't hold any longer:
There is no more star region. Instead, a non-vacuum state will be separated from the vacuum state
through a rarefaction fan, which in turn is separated from the vacuum state through a contact wave
which overlaps with the tail of the rarefaction. A shock wave cannot be adjacent to a vacuum state,
which can be shown using the Rankine-Hugeniot relations: Let a left non-vacuum state $\U_L =
(\rho_L, v_L, E_L)^T$ with the corresponding flux $\F_L = (\rho_L v_L, \rho_L v_L^2 + p_L, (E_L +
p_L) v_L)^T$ be adjacent to a vacuum state $\U_0 = (\rho_0, v_0, E_0)^T$ with the corresponding flux
$\F_0 = (\rho_0, \rho v_0^2 + p_0, (E_0 + p_0) v_0)^T$. Assuming the left and the vacuum state are
separated by a jump discontinuity with velocity $S$, then the Rankine-Hugeniot conditions give us

\begin{align}
    \F_L - \F_0 &= S (\U_L - \U_0) \\
    \rho_L v_L - \rho_0 v_0  &= S( \rho_L - \rho_0) \label{eq:vacuum1} \\
    \rho_L v_L^2 + p_L - \rho_0 v_0^2 - p_0  &= S( \rho_L v_L - \rho_0 v_0) \label{eq:vacuum2} \\
    v_L (E_L + p_L) - v_0 (E_0 + p_0) &= S( E_L - E_0) \label{eq:vacuum3}
\end{align}

By assuming $\rho_0 = E_0 = 0$ and that $v_0$ is finite, we obtain

\begin{align}
    \rho_L v_L  &= S \rho_L \label{eq:vacuum4} \\
    \rho_L v_L^2 + p_L - p_0  &= S \rho_L v_L \label{eq:vacuum5} \\
    v_L (E_L + p_L) - v_0 p_0 &= S E_L \label{eq:vacuum6}
\end{align}

These three equations give us the following relations:

\begin{align}
    v_L &= S = v_0 \label{eq:vacuum-v0}\\
    p_L &= p_0 \label{eq:vacuum-p0}
\end{align}

The equal pressures across the wave in eq.~\ref{eq:vacuum-p0} indicate that a shock wave is not a
possible solution, since a shock wave requires unequal pressures between the waves. However,
this solution allows for a contact wave. Eq.~\ref{eq:vacuum-v0} states that the wave will propagate
at the velocity dictated by the non-vacuum state, which physically makes sense if one interprets the
wave as the boundary between the vacuum and the non-vacuum states whose position evolves over time.
Given eq.~\ref{eq:vacuum-p0}, the pressure at the contact wave must be the same as the pressure in
the vacuum state, which is zero. This means that for any left state $\U_L$, the only applicable
solution for $t > 0$ between the left state itself and the contact wave must be a rarefaction fan,
since $p_L \geq p_0$. This leads to the conclusion that the solution structure must be that the
initial left state $\U_L$ is separated from the vacuum state through a rarefaction fan, which ends
with a contact wave overlapping with its tail at the vacuum state.

With the structure of the solution known, the previously found relations can be applied again, and
the solution to the Riemann problem with a left non-vacuum state $\U_L$ and right vacuum state
$\U_0$ is given by

\begin{align}
    S_{vac, L} &= v_L + \frac{2 c_{s,L}}{\gamma - 1} \\
    \U_{L, \text{ with vacuum }}(x,t) &=
        \begin{cases}
            \U_L & \quad \text{ if } \frac{x}{t} \leq v_L - c_{s,L} \\
            \U_{L, \text{inside fan}} & \quad \text{ if } v_L - c_{s,L} < \frac{x}{t} < S_{vac, L}
\\
            \U_{vac} & \quad \text{ if } \frac{x}{t} \geq S_{vac, L}\\
        \end{cases}
\end{align}

The solution $\U_{L, \text{inside fan}}$ inside the rarefaction fan is given by
eqns.~\ref{eq:rho-rarefaction-fan-left}~-~\ref{eq:pressure-rarefaction-fan-left}.
The inverse case, with a right non-vacuum state $\U_R$ and a left vacuum state $\U_{vac}$, the
solution is given by

\begin{align}
    S_{vac, R} &= v_R - \frac{2 c_{s,R}}{\gamma - 1} \\
    \U_{R, \text{ with vacuum }} &=
        \begin{cases}
            \U_{vac} & \quad \text{ if } \frac{x}{t} \leq S_{vac, R}\\
            \U_{R, \text{inside fan}} & \quad \text{ if } S_{vac, R} < \frac{x}{t} < v_R + c_{s,R}\\
            \U_R & \quad \text{ if } \frac{x}{t} \geq v_R + c_{s,R} \\
        \end{cases}
\end{align}

The solution $\U_{R, \text{inside fan}}$ inside the rarefaction fan is given by
eqns.~\ref{eq:rho-rarefaction-fan-right}~-~\ref{eq:pressure-rarefaction-fan-right}.

\begin{figure}
\centering
\includegraphics[width=\textwidth]{
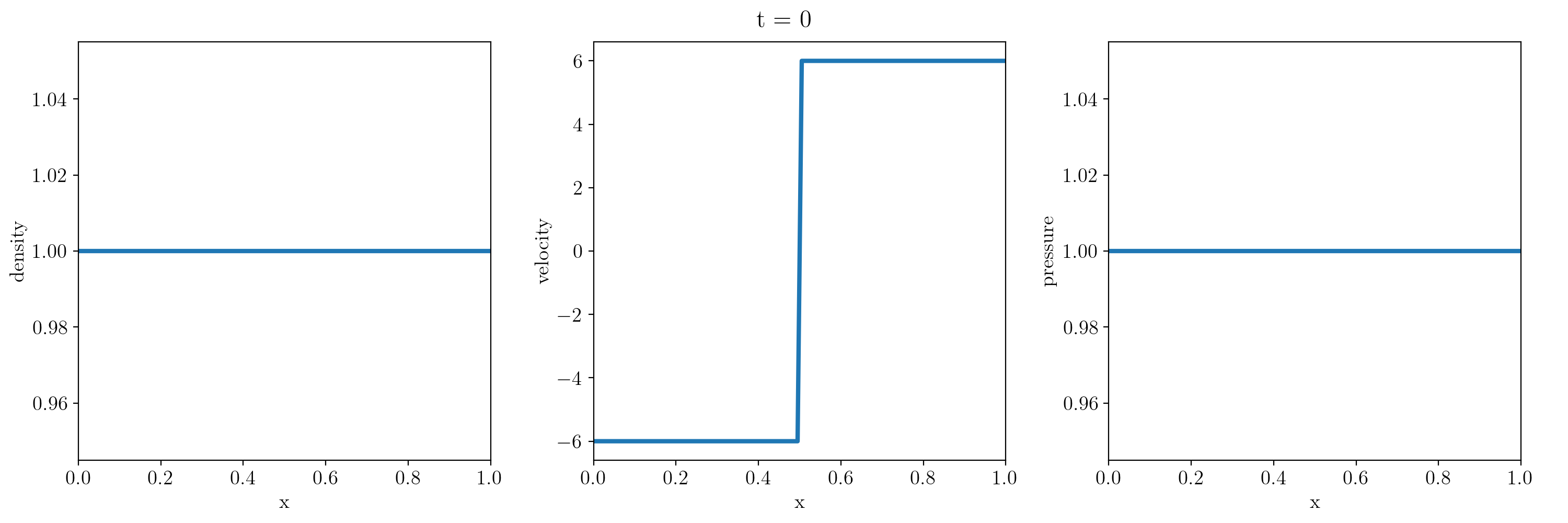} %
\includegraphics[width=\textwidth]{
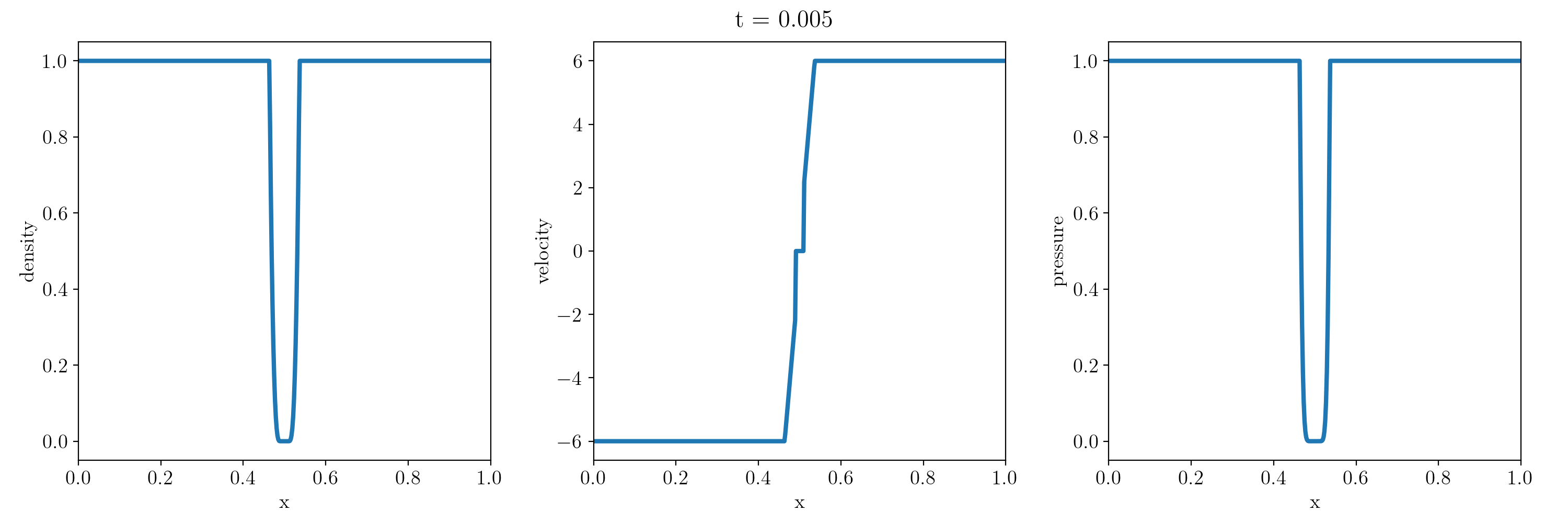} %
\includegraphics[width=\textwidth]{
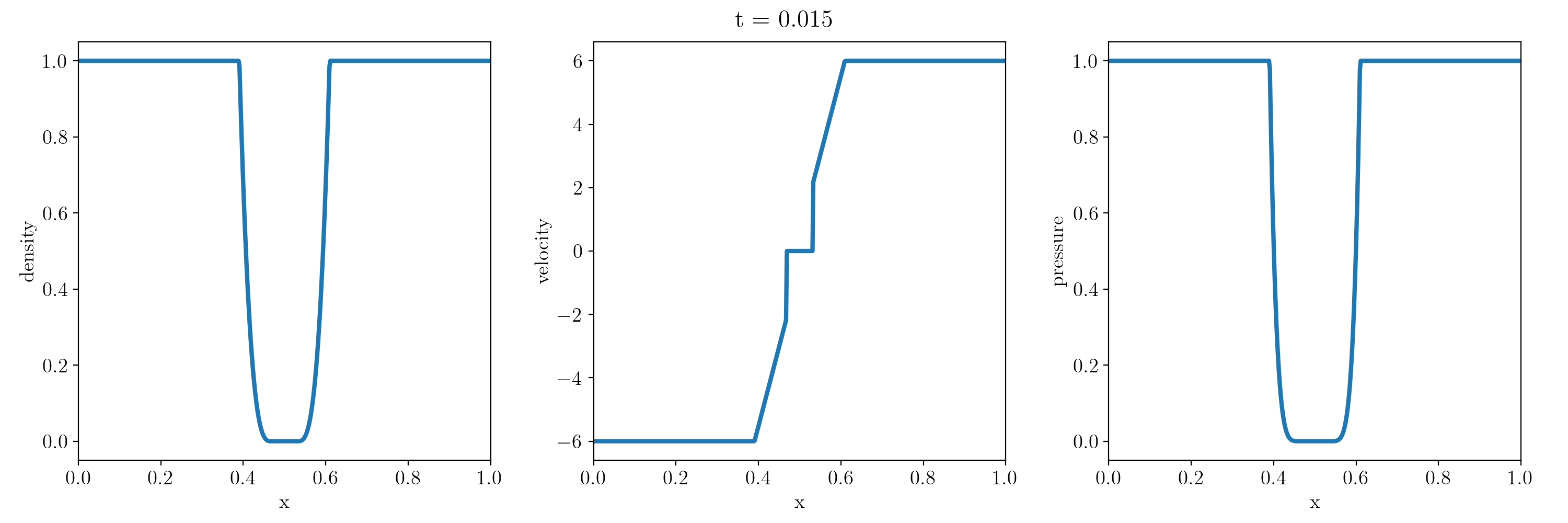} %
\includegraphics[width=\textwidth]{
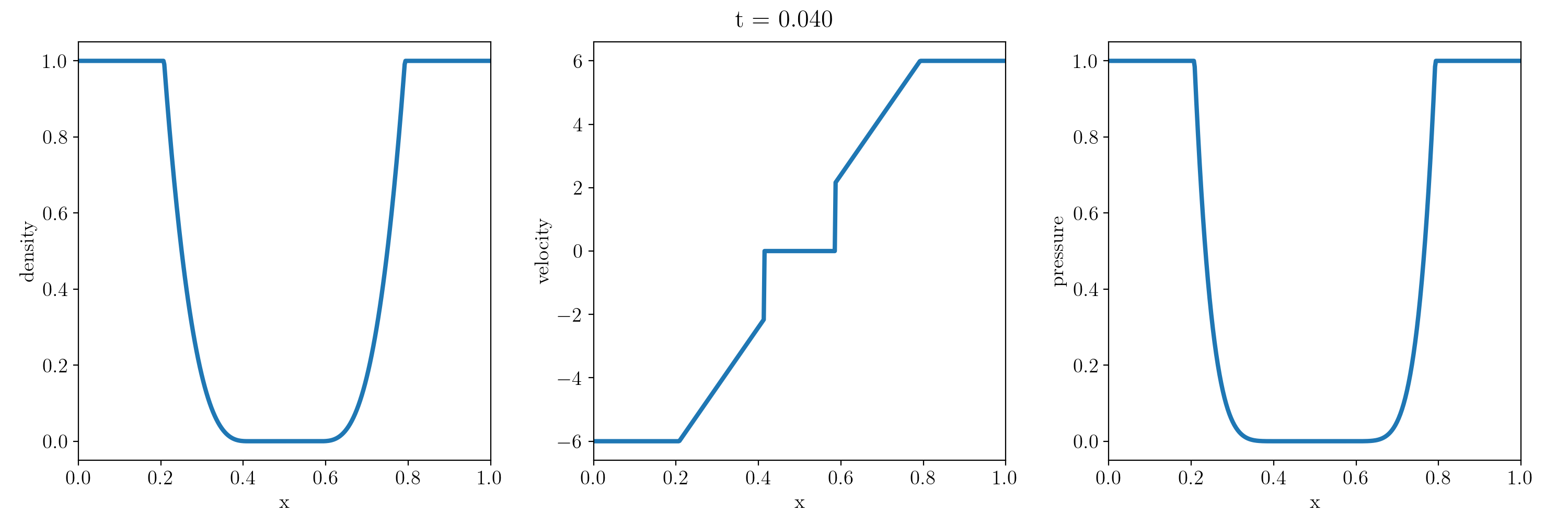} %
\caption[Riemann problem that generates a central vacuum]{
    Initial conditions and time evolution of a Riemann problem that generates vacuum in the central
    region.
    }%
    \label{fig:vacuum-generating}
\end{figure}

In certain cases, with both the left and the right state being non-vacuum states, vacuum
can be generated in central regions of the solution. This can occur when the difference between the
left and right fluid velocities have the opposite direction and are too high for the material to
keep up, generating a vacuum between the two initial states as time evolves. The structure of the
solution contains a left-facing wave connected to a central vacuum region over a rarefaction fan,
and a second left-facing rarefaction fan that connects the central vacuum region with the
right-facing wave. This scenario is shown in Figure~\ref{fig:vacuum-generating}.  To find an
explicit solution, we can make use us the fact that
there must be two rarefaction tails, with speeds $S_{vac,L}$ and $S_{vac,R}$, respectively, just
like in the vacuum-adjacent cases before. The only difference is that now there are two of them.
For two rarefactions fans to develop, $S_{vac, L} \leq S_{vac, R}$ must hold, and hence a condition
for a vacuum generating case can be written as

\begin{align}
    \Delta v_{crit} \equiv \frac{2 c_{s,L}}{\gamma - 1 } + \frac{2 c_{s,R}}{\gamma - 1 } \leq v_R -
v_L \label{eq:vacuum-generating-condition}
\end{align}

If a Riemann problem, like in Figure~\ref{fig:vacuum-generating}, satisfies
condition~\ref{eq:vacuum-generating-condition}, then the full solution is given by

\begin{align}
    \U_{\text{vacuum generating}} =
        \begin{cases}
            \U_{L, \text{ with vacuum }} & \quad \text{ if } \frac{x}{t} \leq S_{vac, L}\\
            \U_{vac} & \quad \text{ if } S_{vac, L} < \frac{x}{t} <  S_{vac, R}\\
            \U_{R, \text{ with vacuum }} & \quad \text{ if } \frac{x}{t} \geq S_{vac, R} \\
        \end{cases}
\end{align}

%% file: main/FV/FV-3-godunov.tex
\chapter{Godunov's Method For Non-Linear Schemes}\label{chap:godunov}

\section{Notation}

In the following chapters, numerical methods which discretize both space and time are discussed.
Space will be discretized in cells of equal size (or more precisely, of equal length, as only one
dimensional cases will be discussed). Integer indices describe their position. The following
convention is used: The index $0$ also represents the leftmost cell. Spatial indices are denoted as
subscripts. The time variable will be discretized in integer-valued time steps. The time steps may
have variable lengths from one time step to the next, but the time nevertheless progresses step by
step, which is denoted by integer superscripts.

Explicitly, the following notation is used:

\begin{itemize}
\item integer subscript: Value of a quantity at the cell, i.e. the center of the cell. Example:
$\U_i$, $\U_{i-2}$
\item non-integer subscript: Value at the cell faces (or interfaces), e.g. $\F_{i-\half}$ is
the flux at the interface between cell $i$ and $i-1$, i.e. the left cell as seen from cell $i$.
\item integer superscript: Indication of the time step. E.g. $\U ^ n$: State at time step $n$
\item non-integer superscript: (Estimated) value of a quantity in between time steps. E.g.
$\F^{n + \half}$: The flux at the middle of the time between steps $n$ and $n + 1$.
\end{itemize}

\section{The Method}

Armed with Riemann solvers for the Euler equations, we can now turn to solving other initial
value problems besides Riemann problems. Godunov's method is an excellent starting point to look
into some essential underlying concepts of finite volume methods. For simplicity, let's stick to
one dimensional problems to start with.

Let $\tilde{\U}(x, t^n)$ be some continuous state with $0 \leq x \leq L$ which we'd
like to evolve over time. The spatial domain is discretized into $N$ computing cells of regular
size $\Delta x = \tfrac{L}{N}$. Cell $i$'s center is located at $x_i = (i + \tfrac{1}{2}) \Delta
x$, while the cells' left and right boundaries are located at $x_{i - \half} = (i - 1) \Delta x$
and $x_{i+\half} = i\Delta x$, respectively. The state inside each cell is assumed to be piece-wise
constant, i.e. constant within the cell boundaries, which can be realized by taking integral
averages of the continuous state:

\begin{align}
    \U_i^n = \frac{1}{\Delta x} \int\limits_{x_{i-\half}}^{x_{i+\half}} \tilde{\U}(x, t^n) \de x \ .
\label{eq:godunov-averaging-state}
\end{align}

\begin{figure}
    \includegraphics[width=\textwidth]{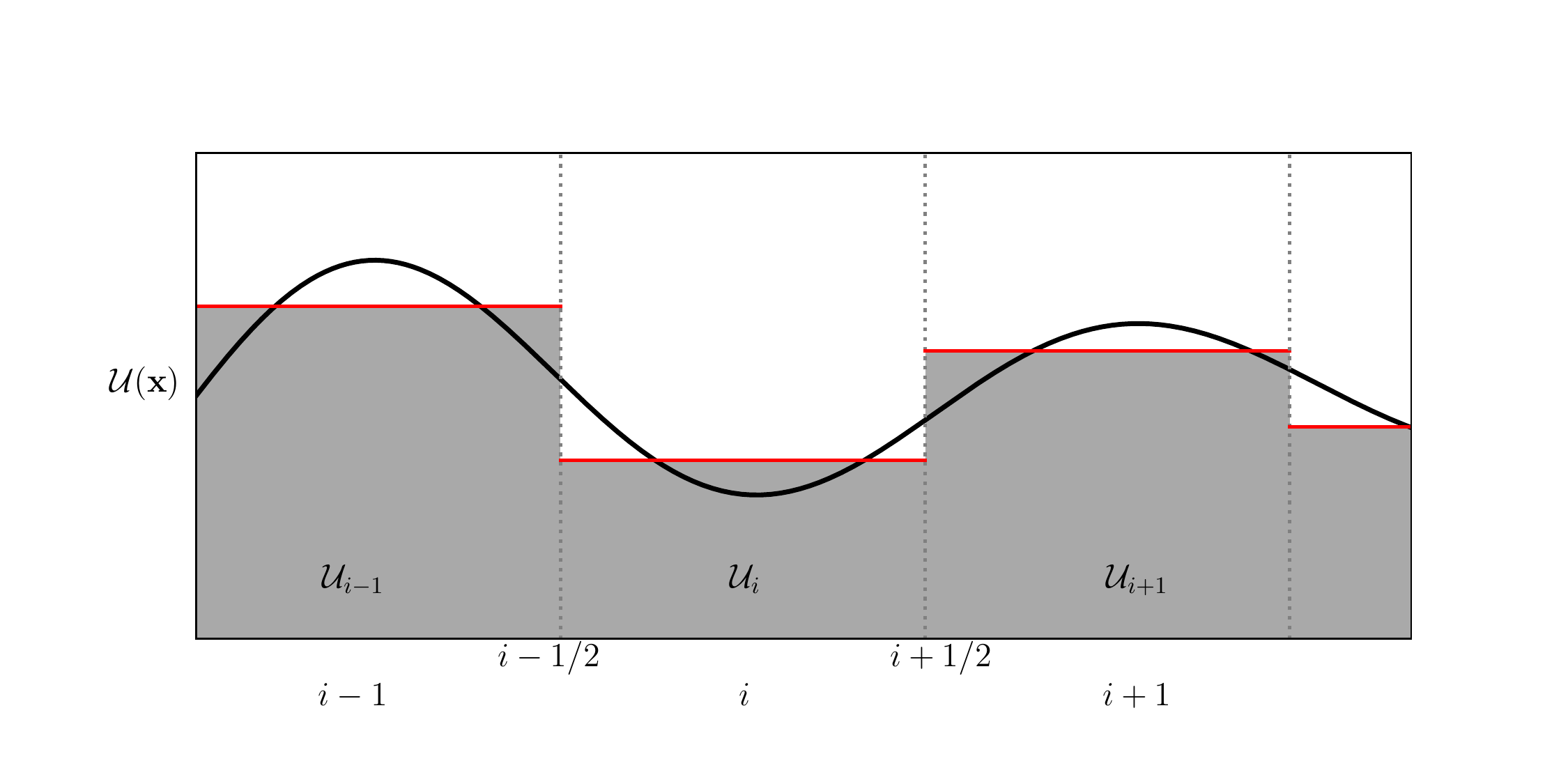}%
    \caption[A piece-wise constant representation of continuous data among cells.]
    {
A piece-wise constant representation (in red) of initially continuous data (black line)
among cells.
    }
\label{fig:piecewise-constant}
\end{figure}

Obviously the states are allowed to vary from cell to cell. The piece-wise constant representation of
continuous data through integral averages as prescribed by eq.~\ref{eq:godunov-averaging-state} is
shown in  Fig.~\ref{fig:piecewise-constant}.

Godunov's method is based on the integral form of conservation laws, i.e.

\begin{align}
 \int\limits_{x_1}^{x_2} \int\limits_{t_0}^{t_1} \de x \de t \left[ \deldt \tilde{\U} + \deldx \F
\right] = 0
\end{align}

and as such allows for discontinuous solutions. Applying this integral form to the volume of a
single cell, i.e. $x_1 = x_{i-\half}$ and $x_2 = x_{i+\half}$ and between two times $t_0 = t^n$ and
$t_1 = t^{n+1}$ gives

\begin{align}
 \int\limits_{x_1}^{x_2} \tilde{\U}(x, t^{n+1}) \de x  =
 \int\limits_{x_1}^{x_2} \tilde{\U}(x, t^{n}) \de x  +
 \int\limits_{t^n}^{t^{n+1}} \F(\tilde{\U}(x_{i-\half}, t)) \de t -
 \int\limits_{t^n}^{t^{n+1}} \F(\tilde{\U}(x_{i+\half}, t)) \de t
\label{eq:conservation-law-integral-form}
\end{align}

At this point, we use our knowledge of the general structure of the solution of the Riemann problem
at the cell boundaries $x_{i \pm \half}$. At the left boundary, $x = x_{i-\half}$, the state
$\U(x_{i-\half}, t)$ for $t > 0$ is given by the solution of the Riemann problem with the left
state $\U_L = \U_{i-1}$ and the right state $\U_R = \U_{i}$ centered at the boundary $x =
x_{i-\half}$. The elementary wave solution of the Riemann problem tells us that as time evolves,
there will be three waves emanating from that point, separating the two initial states into four
constant states. The structure of the solution is shown in Figure~\ref{fig:riemann-solution}. What
Figure~\ref{fig:riemann-solution} also shows very well is that for all $t > 0$, the state at $x' =
0$
($x'$ being the $x$ coordinate in the frame of reference in Figure~\ref{fig:riemann-solution}, i.e.
of the centered Riemann problem) will remain constant: The only thing that varies over time is
the position of the wave fronts, which travel at constant speeds and are therefore straight lines on
the $x'-t$ plane. So once the state for $x' = 0$ is determined for $t > 0$, it will remain
constant. This is also true when $x' = 0$ is inside a rarefaction fan, as the state is determined
by the characteristics that satisfy $x'/t = 0$ (compare also the explicit solutions given in
eqns.~\ref{eq:rho-rarefaction-fan-left} ~-~\ref{eq:pressure-rarefaction-fan-right}). Therefore the
state $\U(x_{i-\half}, t^n)$ at the left cell boundary $x = x_{i-\half}$, which corresponds to the
center $x' = 0$ of the centered Riemann problem, will also be constant for all $t > 0$. The same
holds true for the right boundary $x = x_{i+\half}$ for the Riemann problem with left state $\U_L =
\U_i$ and the right state $\U_R = \U_{i+1}$.

Making use of the fact that the states at the cell boundaries $\U(x_{i \pm \half})$ are constant,
the corresponding fluxes $\F(\U(x_{i \pm \half}))$ are constant as well, and the integrals over
time in eq.~\ref{eq:conservation-law-integral-form} are trivially solved:

\begin{align}
 \int\limits_{t^n}^{t^{n+1}} \underbrace{\F(\tilde{\U}(x_{i-\half}, t))}_{\equiv \F_{i-\half}
= \CONST} \de t-
 \int\limits_{t^n}^{t^{n+1}} \underbrace{\F(\tilde{\U}(x_{i+\half}, t))}_{\equiv \F_{i+\half}
= \CONST} \de t &=
 \underbrace{(t^{n+1} - t^n)}_{\equiv \Delta t} \left[ \F_{i-\half} - \F_{i+\half}  \right] \\
 &= \Delta t  \left[ \F_{i-\half} - \F_{i+\half}  \right]
\end{align}

leaving us with

\begin{align}
 \int\limits_{x_1}^{x_2} \tilde{\U}(x, t^{n+1}) \de x  =
 \int\limits_{x_1}^{x_2} \tilde{\U}(x, t^{n}) \de x  +
 \Delta t  \left[ \F_{i-\half} - \F_{i+\half}  \right] \label{eq:godunov-intermediate-step}
\end{align}

Dividing eq.~\ref{eq:godunov-intermediate-step} by $\Delta x$, we obtain

\begin{align}
 \frac{1}{\Delta x} \int\limits_{x_1}^{x_2} \tilde{\U}(x, t^{n+1}) \de x  =
 \frac{1}{\Delta x}  \int\limits_{x_1}^{x_2} \tilde{\U}(x, t^{n}) \de x  +
 \frac{\Delta t}{\Delta x}  \left[ \F_{i-\half} - \F_{i+\half}  \right]
\end{align}

and we recognize the integral averages (eq.~\ref{eq:godunov-averaging-state}), allowing us to write
the final form of Godunov's scheme:

\begin{align}
\boxed{
 \U^{n+1}_i = \U^{n}_i +
 \frac{\Delta t}{\Delta x}  \left[ \F_{i-\half} - \F_{i+\half}  \right]  \label{eq:godunov-scheme}
}
\end{align}

It is noteworthy that this is an exact solution to a piece-wise constant initial state. More
sophisticated cases, where assumptions of the states to be non-constant over the duration of
a time step, or for the states to be non-constant over the volume of a cell (e.g. a piece-wise
linear reconstruction instead of a piece-wise constant one) are relaxed, will be the topic of
Chapter~\ref{chap:higher-order-schemes}.

Something that was glanced over so far is that we assumed that the states and therefore the fluxes
at the cell boundaries remain constant indefinitely. While in theory that is true for isolated
Riemann problems that don't experience any outside perturbations nor influences, in this scenario
we have a collection of Riemann problems that are all a cell width $\Delta x$ apart from each
other. While the states at the boundaries, which represent the center of the Riemann problem,
remain constant in an isolated case, the waves emanating from that origin propagate over time, and
eventually cross the distance $\Delta x$ both towards the left and the right. Once they do, the
states at the distance $\Delta x$ from the origin will change. However, those are also the
positions of other cell boundaries, which we used as the origins for other Riemann problems and
assumed to remain constant. Clearly in this case the assumption of constancy of the states at cell
boundaries is violated. In order to prevent this violation from occurring, we must impose a
condition: The time step size $\Delta t$ must not be large enough for a wave to travel the distance
$\Delta x$, i.e.

\begin{align}
 \Delta t \leq \frac{\Delta x}{S_{max}^n}
\end{align}

where $S_{max}^n$ is the maximal wave speed among all waves emanating from the solutions of the
Riemann problems at each cell boundary. This condition is called the Courant-Friedrichs-Levy (CFL)
condition, and is usually expressed as

\begin{align}
 \Delta t = C_{CFL} \frac{\Delta x}{S_{max}^n}, \quad\quad C_{CFL} \in [0, 1) \label{eq:godunov-cfl}
\end{align}

$C_{CFL}$ is often referred to as the ``Courant number''.

Some more points need to be raised regarding the limitation of the time step size $\Delta t$.
Firstly, while the straightforward argumentation to prohibit waves from reaching adjacent cell
boundaries works for the comparatively simple Godunov scheme, matters can get more complicated for
higher order schemes and when more than one dimension is treated. In addition to physical
arguments, further restrictions must be included in order to ensure the stability of the scheme.
``Stability'' in this context refers to ensuring that spurious oscillations around discontinuities
such as shocks don't develop, as these oscillations can grow exponentially and quickly lead to
positive and negative infinite values in the solution. The development of oscillations around
discontinuities is a consequence of extending the method to higher orders of accuracy, and will be
discussed in Chapter~\ref{chap:higher-order-schemes}.

Doing proper stability analyses of the methods presented in this work would go way beyond the
scope of this thesis. To make matters worse, in more contrived cases like for higher order methods
and non-linear conservation laws, a rigorous proof of stability is not available to date. In
practice, the findings from simpler cases like linear conservation laws are applied to non-linear
ones as well, and are known to yield adequate results. Nevertheless, to provide an impression of how
stability analysis can be done, the outlines of two well-known methods of stability analyses are
given below:

\begin{itemize}
 \item The \emph{Von Neumann Stability Analysis} approach looks into the Fourier transform of the
underlying conservation law that is to be solved, and looks how the solution of the numerical method
evolves in Fourier space. Stability conditions are found by requiring that the solution in Fourier
space mustn't grow exponentially, but remain bounded for all times, and by ensuring that the phase
doesn't allow for too high propagation velocities.

\item The \emph{Lax-Richtmeyer Stability Analysis} relies on comparing the error resulting from the
numerical method being applied to a conservation law to the exact solution of the conservation law
at each time step. Since the method is applied repeatedly as the simulation progresses in order to
advance the initial state in time, the errors accumulate and propagate each time step. The stability
criteria arise from the requirement for the (cumulative) error to remain bounded.

\end{itemize}

Secondly, depending on the method, the exact wave speeds may not always be known. In Godunov's
method, where only a single Riemann problem is solved, the correct solution can be obtained, but
this is not always possible in higher order methods. Instead, an approximate estimate can be used. A
reliable choice is

\begin{align}
	S^n_{max} = \max_i\{ |v_i^n| + c_{s,i}^n \} \label{eq:wavespeed-estimate-godunov}
\end{align}

This estimate may underestimate the shock speeds (compare to eqns.~\ref{eq:shock-left-speed} and
\ref{eq:shock-right-speed}), but a reasonable choice of $C_{CFL}$ can ensure that no instabilities
occur. A value of $C_{CFL} \lesssim 0.8 - 0.9$ is recommended.

To conclude the chapter on Godunov's method, let's explicitly write down the algorithm to evolve a
system from some $t_{start}$ to some $t_{end}$:

\algo{Evolving A System From $t_{start}$ To $t_{end}$ Using Godunov's Method}
{
To start, set $t_{current} = t^0 = t_{start}$ and set up the initial states $\U_i^0$ for
each cell $i$.\\[.5em]
While $t_{current} < t_{end}$, solve the $n$-th time step:\\[.5em]
\indent~~~~Compute inter-cell fluxes: For every cell pair $(i, i+1)$, solve the Riemann \\
\indent~~~~~~~~problem to find the flux
$\F^n_{i+\half} = \F(\U_i^n, \U_{i+1}^n)$. \\[.5em]
\indent~~~~Find the maximal permissible time step $\Delta t$ using eq.~\ref{eq:godunov-cfl} \\[.5em]
\indent~~~~Find the updated states $\U_i^{n+1}$ using eq.~\ref{eq:godunov-scheme} \\[.5em]
\indent~~~~Update the current time: $t_{current} = t^{n+1} = t^n + \Delta t$
}

\section{Applications of Godunov's Method}\label{chap:godunov-application}

\begin{figure}
    \centering
    \includegraphics[width=\textwidth]{
    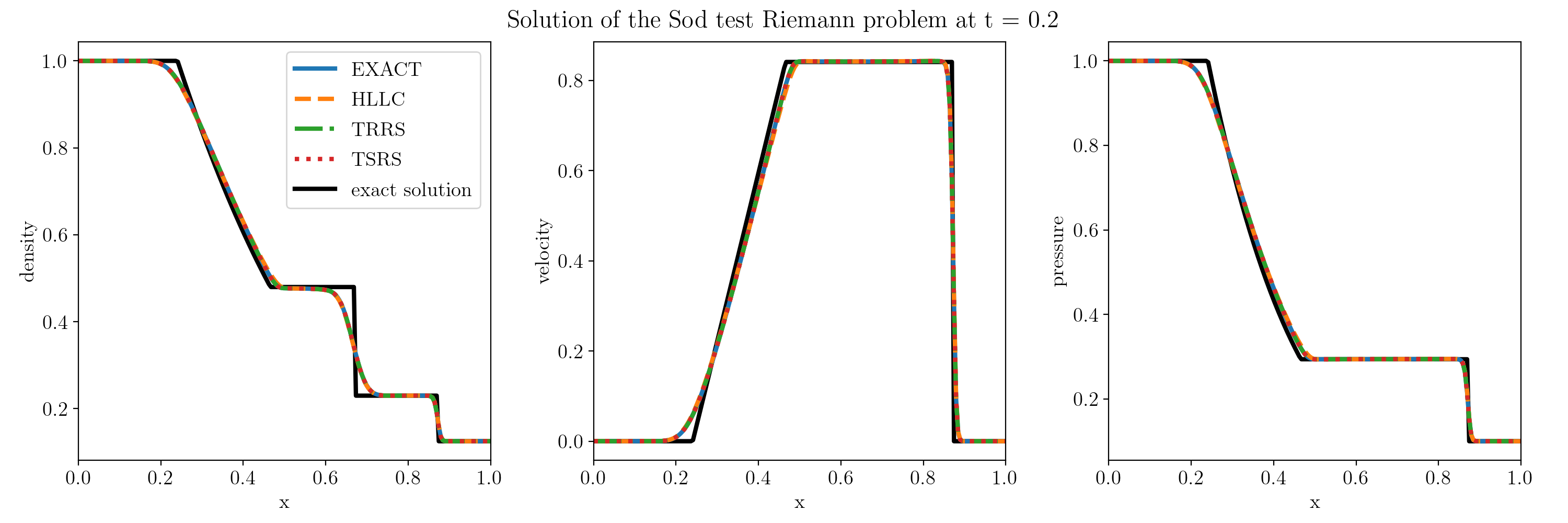} %
    \includegraphics[width=\textwidth]{
    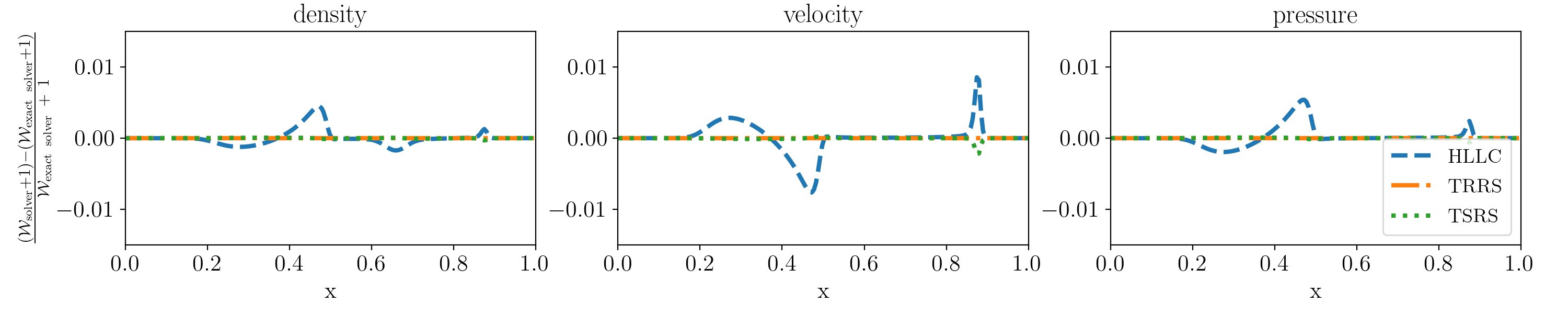} %
    \caption[Sod test using Godunov's method]{
Top: The solution to the Sod test (eq. \ref{eq:sod-test-ICs}) using Godunov's method with the exact,
TRRS, TSRS, and HLLC Riemann solver, respectively, and the exact solution of the problem.
The solution consists of a left-facing rarefaction and a right facing shock. \\
Bottom: The relative differences of the solution of approximate Riemann solvers compared to the
solution using the exact Riemann solver with Godunov's method. The relative differences are
computed with all primitive quantities $\W = (\rho, v, p)$ being increased by 1 to avoid
divisions by zero.
    }%
    \label{fig:godunov-sod-test}
\end{figure}
\begin{figure}
    \centering
    \includegraphics[width=\textwidth]{
    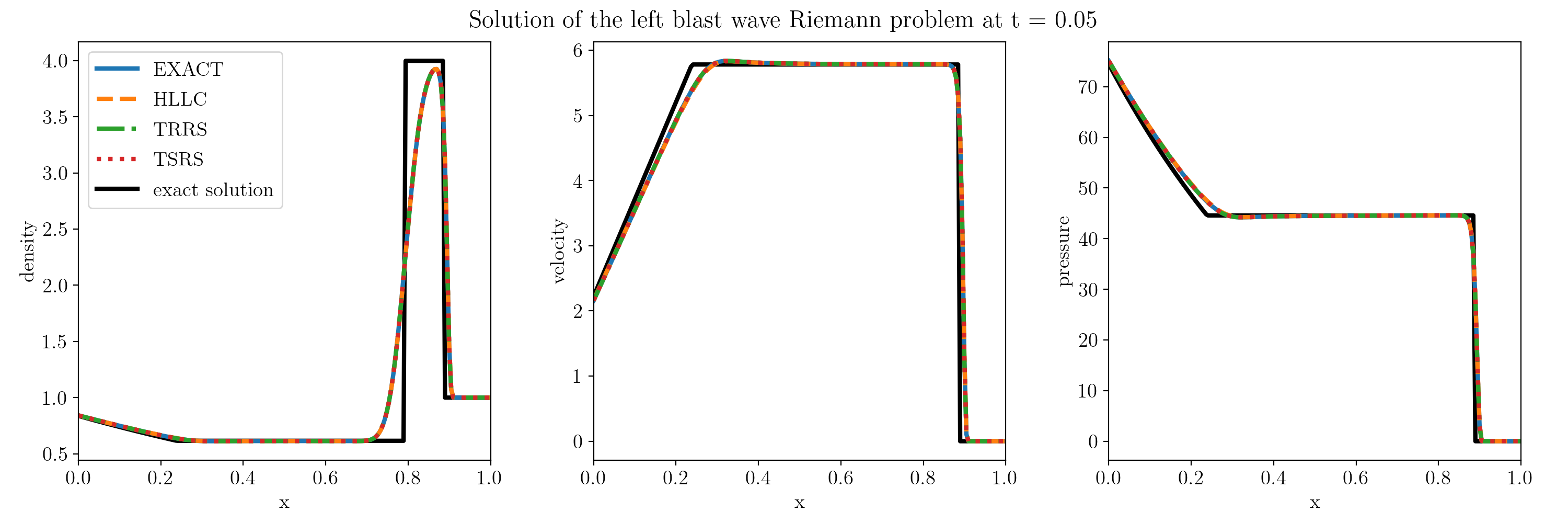} %
    \includegraphics[width=\textwidth]{
    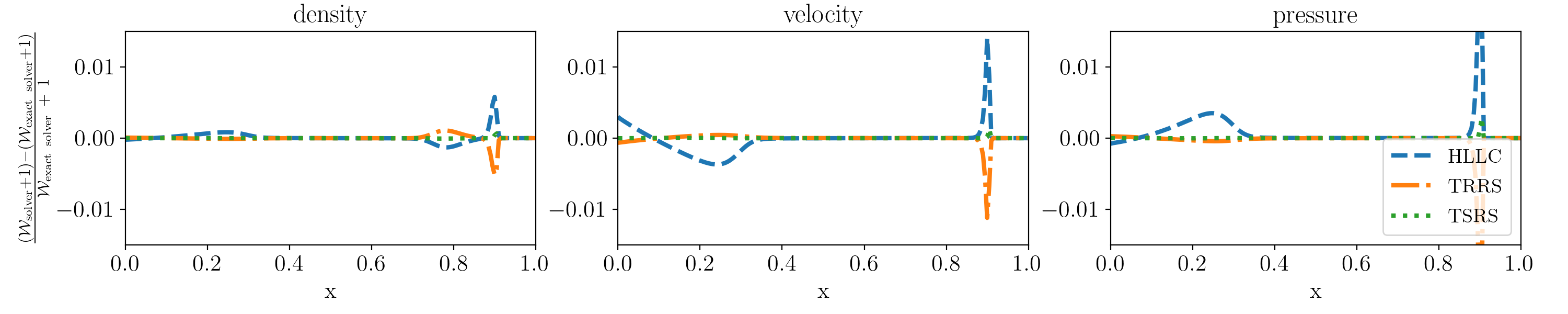} %
    \caption[Left blast wave solution using Godunov's method]{
Top: The solution to the left blast wave test (eq. \ref{eq:left-blast-wave-ICs}) using Godunov's
method with the exact, TRRS, TSRS, and HLLC Riemann solver, respectively, and the exact solution
of the problem.  The solution consists of a left-facing rarefaction and a right facing shock. \\
Bottom: The relative differences of the solution of approximate Riemann solvers compared to the
solution using the exact Riemann solver with Godunov's method. The relative differences are
computed with all primitive quantities $\W = (\rho, v, p)$ being increased by 1 to avoid
divisions by zero.
    }%
    \label{fig:godunov-left-blast-wave}
\end{figure}
\begin{figure}
    \centering
    \includegraphics[width=\textwidth]{
    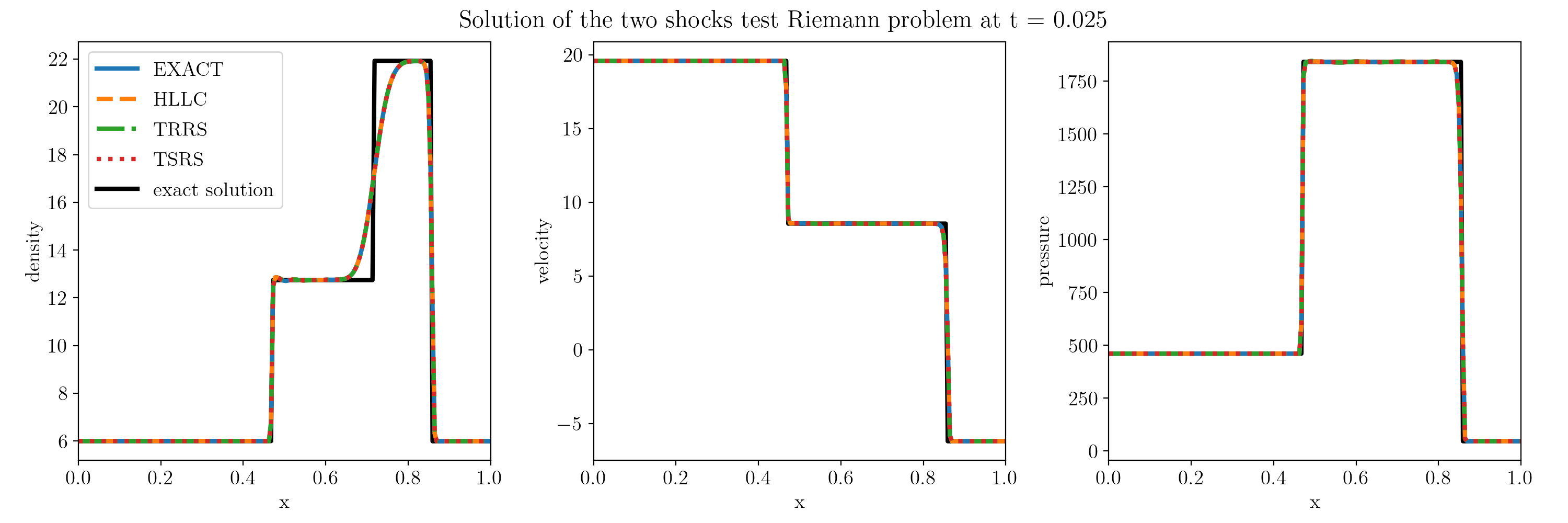} %
    \includegraphics[width=\textwidth]{
    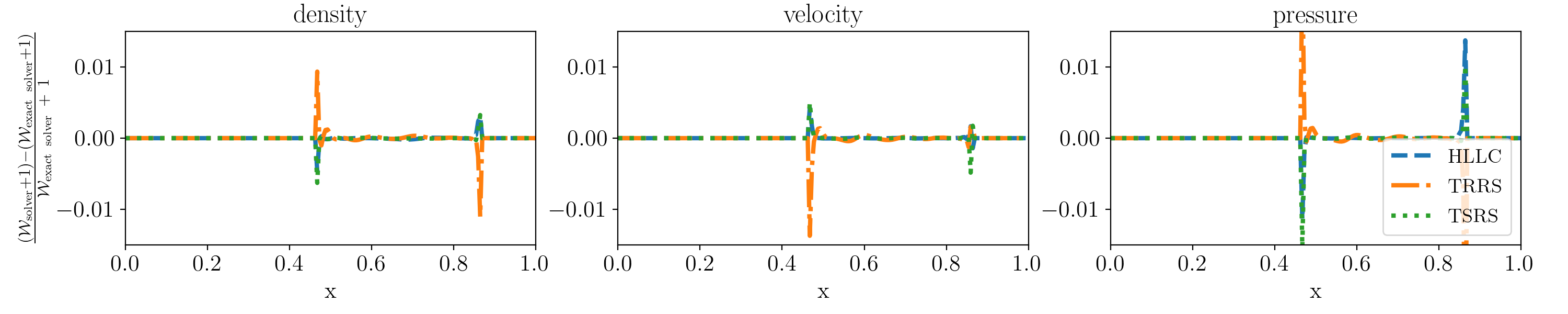} %
    \caption[Two shocks solution using Godunov's method]{
Top: The solution to the two shock test (eq. \ref{eq:two-shock-ICs}) using Godunov's method with
the exact, TRRS, TSRS, and HLLC Riemann solver, respectively, and the exact solution of the problem.
The exact solution consists of a two shocks on either side of the central contact wave. \\
Bottom: The relative differences of the solution of approximate Riemann solvers compared to the
solution using the exact Riemann solver with Godunov's method. The relative differences are
computed with all primitive quantities $\W = (\rho, v, p)$ being increased by 1 to avoid
divisions by zero.
    }%
    \label{fig:godunov-two-shocks}
\end{figure}

At this point, we're ready to put Godunov's method to use for the Euler equations. We can
make use of the fact that exact solutions to Riemann problems exist, and use them as a reference to
gauge the performance of the scheme. The results presented in what follows have been obtained using
the \meshhydro code.

Figure~\ref{fig:godunov-sod-test} shows the results of the ``Sod test'' Riemann problem with initial
conditions given in eq.~\ref{eq:sod-test-ICs}, where the exact solution consists of a left-facing
rarefaction and a right facing shock. Figure~\ref{fig:godunov-left-blast-wave} shows the results of
the so-called ``Left blast wave'' Riemann problem with initial conditions given in
eq.~\ref{eq:left-blast-wave-ICs}. The solution consists of a left-facing rarefaction and a right
facing shock. However in this test the shock is much stronger and much more prominent.
Figure~\ref{fig:godunov-two-shocks} shows the result of the ``two shock'' test with initial
conditions~\ref{eq:two-shock-ICs}. The solution consists of two shock waves, one on either side of
the central contact wave. In all three examples, the results using the exact Riemann solver as well
as the approximate TSRS, TRRS, and HLLC solvers are shown, along with relative differences of the
solutions obtained with approximate Riemann solvers compared to the solution obtained using the
exact Riemann solver in Godunov's method.

The choice of Riemann solver has negligible influence on the solution using Godunov's method,
affirming that the use of approximate solvers is an adequate approach. Aside from a few exceptions
around sharp discontinuities, the solutions with approximate Riemann solvers agree with the
solution using the exact Riemann solver to a level of below 1\%. In the presented tests, the
TRRS and the TSRS solver appear to perform better than the HLLC solver. This is however not
generally valid, and is due to the simple nature of the Riemann problem initial conditions that
have been used as the example test cases. The full solutions of the test problem consist of only
elementary waves, which the TRRS and TSRS solvers are able to handle rather well. In more complex
cases, the HLLC solver is may provide a better solution.

The method is able to successfully solve for the general structure of the exact solution, and all
emanating waves can be found. However, a noticeable feature is the lack of sharp discontinuities in
the solution using Godunov's method, where the exact solution should have some. Instead, Godunov's
method predicts smooth transitions. This is particularly noticeable for the density across contact
waves. The shock fronts aren't affected as much, but are nevertheless ``rounded off''.
On first sight, this may appear puzzling given that Godunov's method makes use of the integral form
of conservation laws, which explicitly allow for discontinuous solutions. So why doesn't it predict
solutions with sharp discontinuities? The answer is \emph{numerical diffusion}, which will be the
subject of the succeeding section.

\section{On Numerical Diffusion and Order Of Accuracy}\label{chap:numerical_diffusion}

\subsection{Numerical Diffusion}

To demonstrate and quantify the effect of numerical diffusion, let's apply Godunov's method to the
linear advection equation with constant coefficients, which greatly simplifies the required
computations.

The one dimensional linear advection equation with constant coefficients is given by

\begin{align}
    \deldt \uc + \deldx \fc = \deldt \uc + a \deldx \uc = 0, && a = \CONST
    \label{eq:numerical-diffusion-advection-equation}
\end{align}

The analytical solution is given by

\begin{align}
    \uc(x, t) = \uc(x - at, t=0)
\end{align}

which can be used to determine the solution of the Riemann problem with the initial left state
$\uc_L$ and right state $\uc_R$:

\begin{align}
    \uc(x, t>0) = \begin{cases}
                    \uc_L & \text{ if } x - at < 0 \\
                    \uc_R & \text{ if } x - at > 0
                  \end{cases}
\end{align}

Applied to the cell boundary $\uc_{i-\half}$ with initial left state $\uc_L = \uc_{i-1}$ and
right state $\uc_R = \uc_i$, and centering the problem at the boundary position $x_{i-\half}$,
the solution at the boundary is given by

\begin{align}
    \uc_{i-\half}(t>0) =
                \begin{cases}
                    \uc_{i-1} & \text{ if } - at < 0  \quad \Leftrightarrow \text{ if } a > 0  \\
                    \uc_{i} & \text{ if } - at > 0 \quad \Leftrightarrow \text{ if } a < 0
                \end{cases}
\end{align}

The solution for the right boundary is similarly given by

\begin{align}
    \uc_{i+\half}(t>0) =
                \begin{cases}
                    \uc_{i} & \text{ if } - at < 0  \quad \Leftrightarrow \text{ if } a > 0  \\
                    \uc_{i+1} & \text{ if } - at > 0 \quad \Leftrightarrow \text{ if } a < 0
                \end{cases}
\end{align}

In the interest of clarity, let's assume $a > 0$ in what follows.
Plugging the solution of the Riemann problem at the boundaries into definition into Godunov's
method, eq.~\ref{eq:godunov-scheme}, we get the expression

\begin{align}
    \uc_i^{n+1} = \uc_i^n + \frac{\Delta t}{\Delta x} a (\uc^{n}_{i-1} - \uc^{n}_{i})
\label{eq:godunov-advection}
\end{align}

\begin{figure}
    \centering
    \includegraphics[width=.5\textwidth]{
    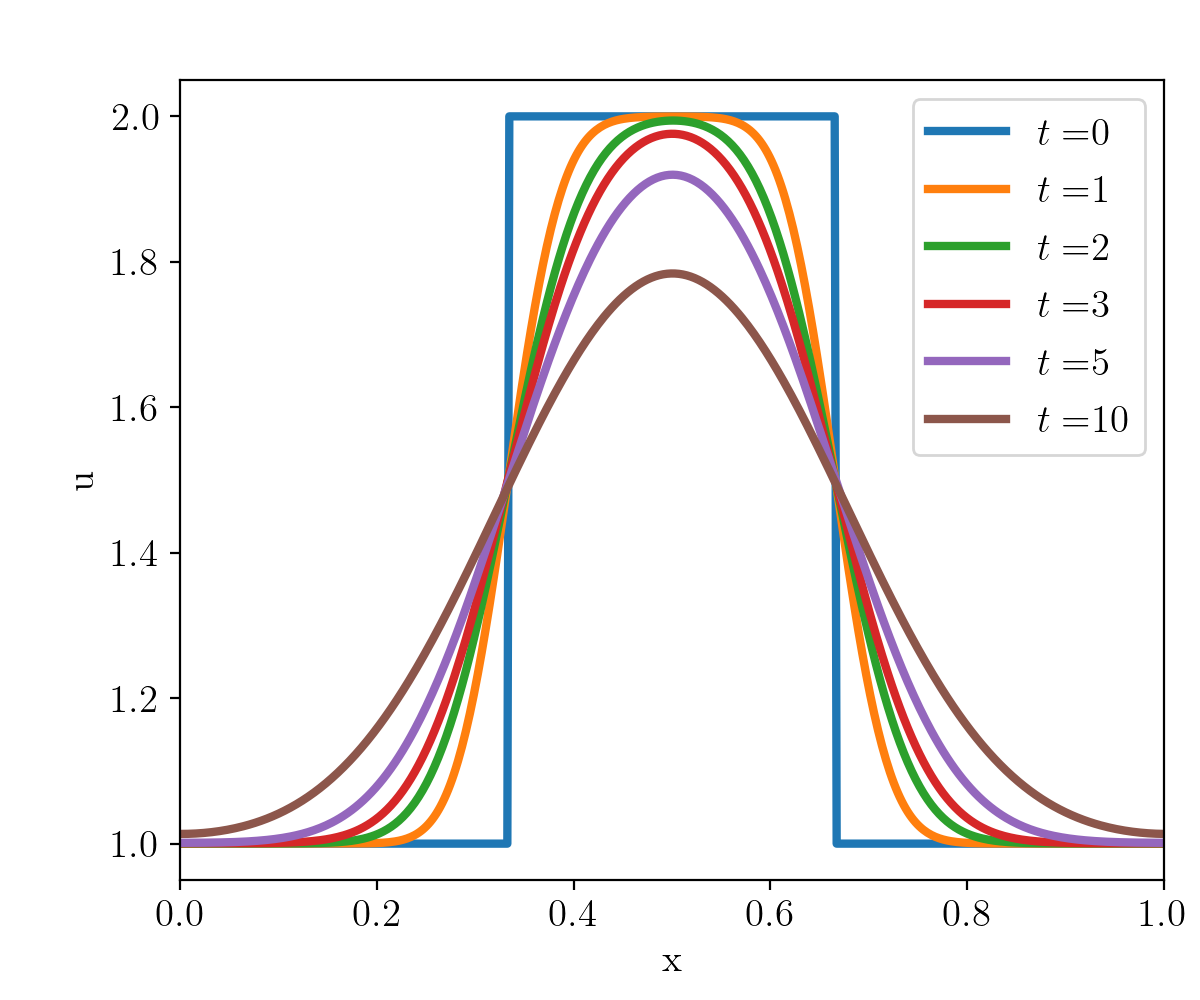}%
    \includegraphics[width=.5\textwidth]{
    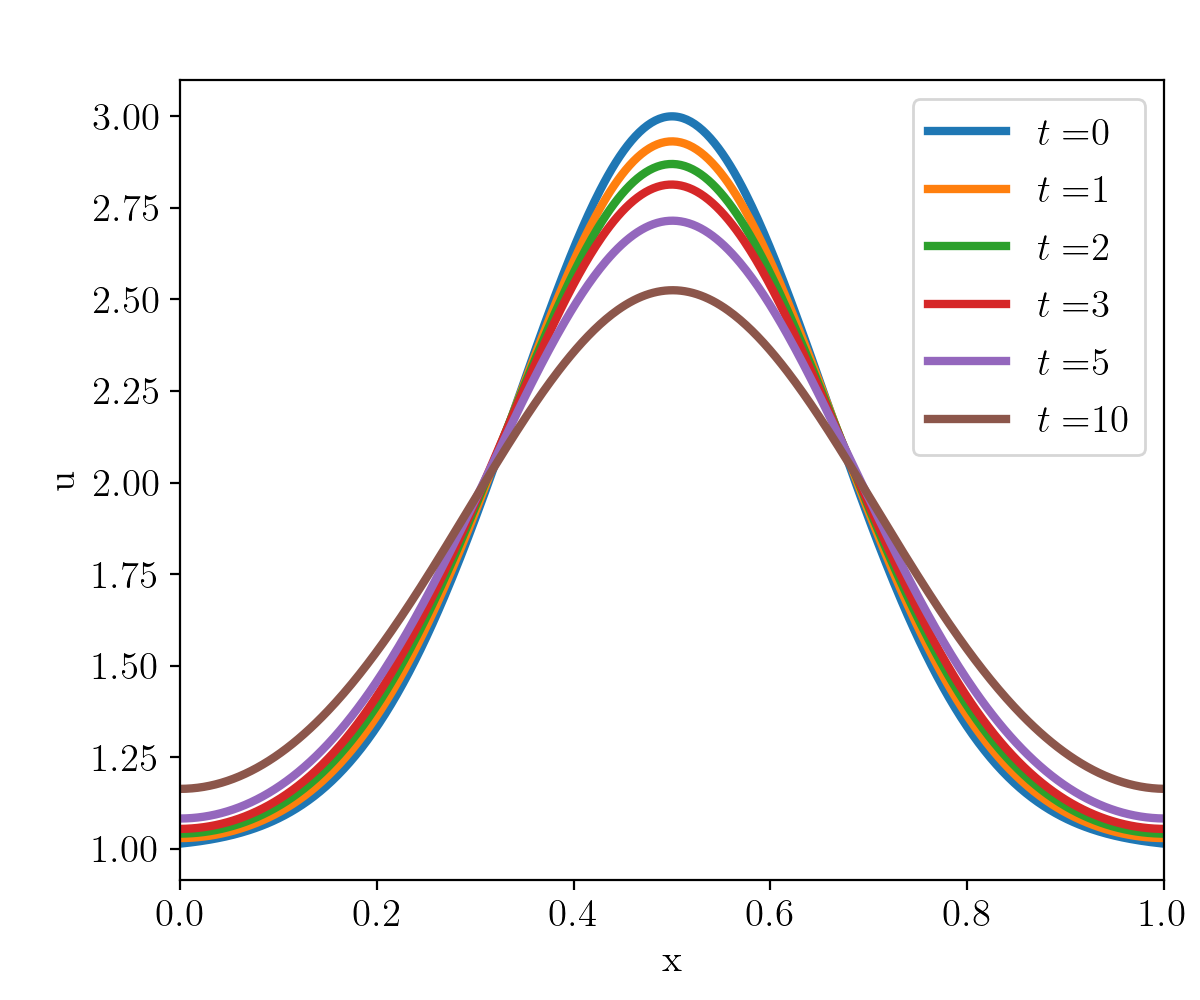}%
    \caption[Linear advection with Godunov's method]{
The solution of the linear advection equation with constant coefficient $a = 1$ (in arbitrary
units). On the left, the initial conditions are a step function, on the right, they are a
Gaussian. The results at $t > 0$ have been repositioned to the original position to demonstrate
how the initial shape changes over time, which according to the analytical solution shouldn't
happen.
    }%
    \label{fig:linear-advection-godunov}
\end{figure}

The analytical solution for the linear advection equation with constant coefficients is
readily available, and dictates that the solution at any $t > 0$ should be just the initial state
$\uc (t = 0)$ translated to a new position $x' = x(t=0) + at$. This makes it easy to compare
the results obtained using Godunov's method to the exact solution of any initial conditions.
Figure~\ref{fig:linear-advection-godunov} shows the solution using Godunov's method for a step
function and a Gaussian at different times, moved back to the original position to demonstrate how
the initial shape changes over time. Evidently the same diffusive effects around discontinuities as
was the case for the Euler equations in
Figures~\ref{fig:godunov-sod-test}~-~\ref{fig:godunov-two-shocks} appear.

To understand what is happening, we first take note of the fact that Godunov's scheme for linear
advection with constant coefficients is equivalent to a first order finite difference
discretization. Approximating

\begin{align}
    \deldt \uc &\approx \frac{\uc_i^{n+1} - \uc_i^n}{\Delta t} \\
    \deldx \fc &\approx \frac{\fc^{n}_{i} - \fc_{i-1}^n}{\Delta x}
        = a\frac{\uc^{n}_{i} - \uc_{i-1}^n}{\Delta x}
\end{align}

and plugging these approximations into the conservation
law~\ref{eq:numerical-diffusion-advection-equation} gives

\begin{align}
    \frac{\uc_i^{n+1} - \uc_i^n}{\Delta t} + a\frac{\uc^{n}_{i} - \uc_{i-1}^n}{\Delta x} = 0
    \label{eq:numerical-diffusion-finite-difference}
\end{align}

which is identical to Godunov's scheme~\ref{eq:godunov-advection}.

The terms $\uc_i^{n+1}$ and $\uc_{i-1}^n$ can be Taylor-expanded to second order in $\Delta t$ and
$\Delta x$:

\begin{align}
    \uc_i^{n+1} &=
        \uc_i^n + \Delta t \DELDT{\uc} +
        \frac{\Delta t^2}{2} \frac{\del^2 \uc}{\del t^2} + \mathcal{O}(\Delta t^3)
    \label{eq:taylor-un+1}\\
    \uc_{i-1}^{n} &=
        \uc_i^n - \Delta x \DELDX{\uc} +
        \frac{\Delta x^2}{2} \frac{\del^2 \uc}{\del x^2} + \mathcal{O}(\Delta x^3)
    \label{eq:taylor-ui-1}
\end{align}

Inserting these expansions into the discretized
equation~\ref{eq:numerical-diffusion-finite-difference} gives

\begin{align}
   & \frac{1}{\Delta t} \left[
        \uc_i^n + \Delta t \DELDT{\uc} +
        \frac{\Delta t^2}{2} \frac{\del^2 \uc}{\del t^2} + \mathcal{O}(\Delta t^3) - \uc_i^n
   \right] + \nonumber \\
   & \frac{a}{\Delta x} \left[
        \uc_i^n - \uc_i^n + \Delta x \DELDX{\uc} -
        \frac{\Delta x^2}{2} \frac{\del^2 \uc}{\del x^2} + \mathcal{O}(\Delta x^3)
   \right] = 0 \\
   = & \DELDT{\uc} + \frac{\Delta t}{2} \frac{\del^2 \uc}{\del t^2}
   + a \DELDX{\uc} - a \frac{\Delta x}{2} \frac{\del^2 \uc}{\del x^2}
    + \mathcal{O}(\Delta t^2) + \mathcal{O}(\Delta x^2)
\end{align}

keeping only the first order terms in $\Delta t$ and $\Delta x$, this can be rearranged to

\begin{align}
   \DELDT{\uc} + a \DELDX{\uc} =
   a \frac{\Delta x}{2} \frac{\del^2 \uc}{\del x^2}
   - \frac{\Delta t}{2} \frac{\del^2 \uc}{\del t^2} \label{eq:num-diff-intermediate}
\end{align}

The left hand side of eq.~\ref{eq:num-diff-intermediate} is identical to the conservation law we
are solving, and should be zero. Hence we can interpret everything on the right hand side as the
highest order error term that the numerical scheme introduces. Let's name it $Err$:

\begin{align}
    Err =
    a \frac{\Delta x}{2} \frac{\del^2 \uc}{\del x^2}
    - \frac{\Delta t}{2} \frac{\del^2 \uc}{\del t^2} \label{eq:num-diff-error}
\end{align}

To proceed, we express $\frac{\del^2 \uc}{\del t^2}$ as a function of $\frac{\del^2 \uc}{\del x^2}$
by differentiating the analytical advection equation once w.r.t. $t$:

\begin{align}
    \deldt \left( \deldt \uc + a \deldx \uc \right) =
    \frac{\del^2 \uc}{\del t^2} + a \frac{\del^2 \uc}{\del x \del t} = 0
\end{align}

and once w.r.t. $x$:
\begin{align}
    \deldx \left( \deldt \uc + a \deldx \uc \right) =
    \frac{\del^2 \uc}{\del x \del t} + a \frac{\del^2 \uc}{\del x^2} = 0
\end{align}

Relating these two derivatives over their common term $\frac{\del^2 \uc}{\del x \del t}$ gives us
the required relation:

\begin{align}
  -a \frac{\del^2 \uc}{\del x \del t}  =
      \frac{\del^2 \uc}{\del t^2} = a^2 \frac{\del^2 \uc}{\del x^2}
\end{align}

Which allows us to express the error $Err$ as

\begin{align}
    Err &=
    a \frac{\Delta x}{2} \frac{\del^2 \uc}{\del x^2}
    - \frac{\Delta t}{2} \frac{\del^2 \uc}{\del t^2}
    =
    a \frac{\Delta x}{2} \frac{\del^2 \uc}{\del x^2}
    + a^2 \frac{\Delta t}{2} \frac{\del^2 \uc}{\del x^2} \\
    &= \frac{a \Delta x}{2} \left( 1 - \frac{a \Delta t}{\Delta x} \right)
        \frac{\del^2 \uc}{\del x^2} \\
    &= \frac{a \Delta x}{2} \left( 1 - C_{CFL} \right)
        \frac{\del^2 \uc}{\del x^2}
\end{align}

Comparing this result and eqns.~\ref{eq:num-diff-intermediate} and \ref{eq:num-diff-error} with the
advection-diffusion equation\footnote{also called the ``convection-diffusion equation''}

\begin{align}
    \deldt \uc + a \deldx \uc = D \frac{\del^2 \uc}{\del x^2}  \label{eq:advection-diffusion}
\end{align}

it is clear that the error term that is introduced by the discretization of the equations is in
fact a diffusion term with the diffusion coefficient $D = \frac{a \Delta x}{2} \left( 1 - C_{CFL}
\right)$. This expression for $D$ also tells us how the diffusivity of the method will behave:

\begin{itemize}
 \item $D \propto \Delta x$: The diffusivity decreases with smaller grid spacing $\Delta x$
 \item $D \propto (1 - C_{CFL})$: The diffusivity decreases with bigger (maximal) time step sizes
    $C_{CFL}$.
\end{itemize}

\begin{figure}
    \centering
    \includegraphics[width=.5\textwidth]{
    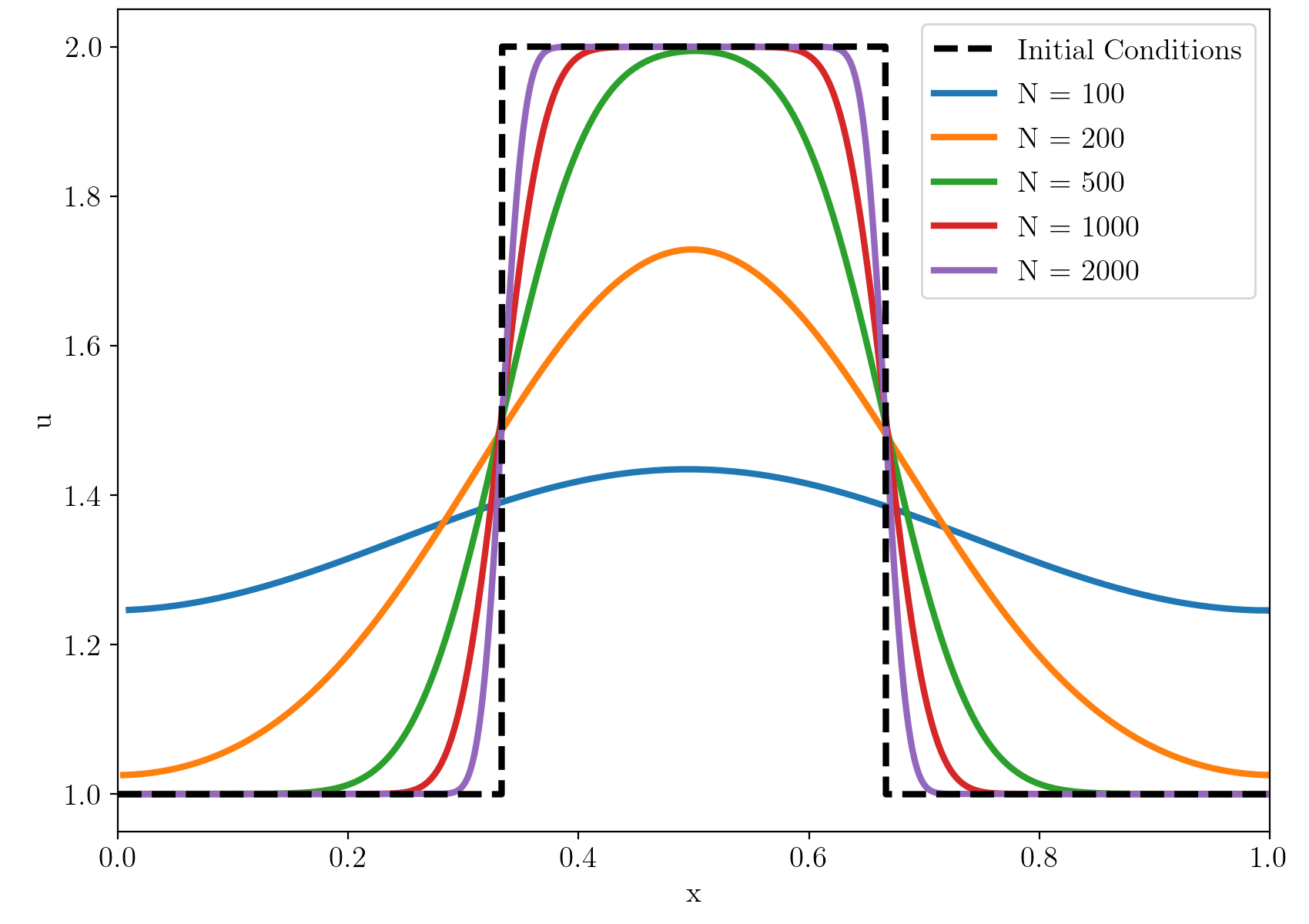}%
    \includegraphics[width=.5\textwidth]{
    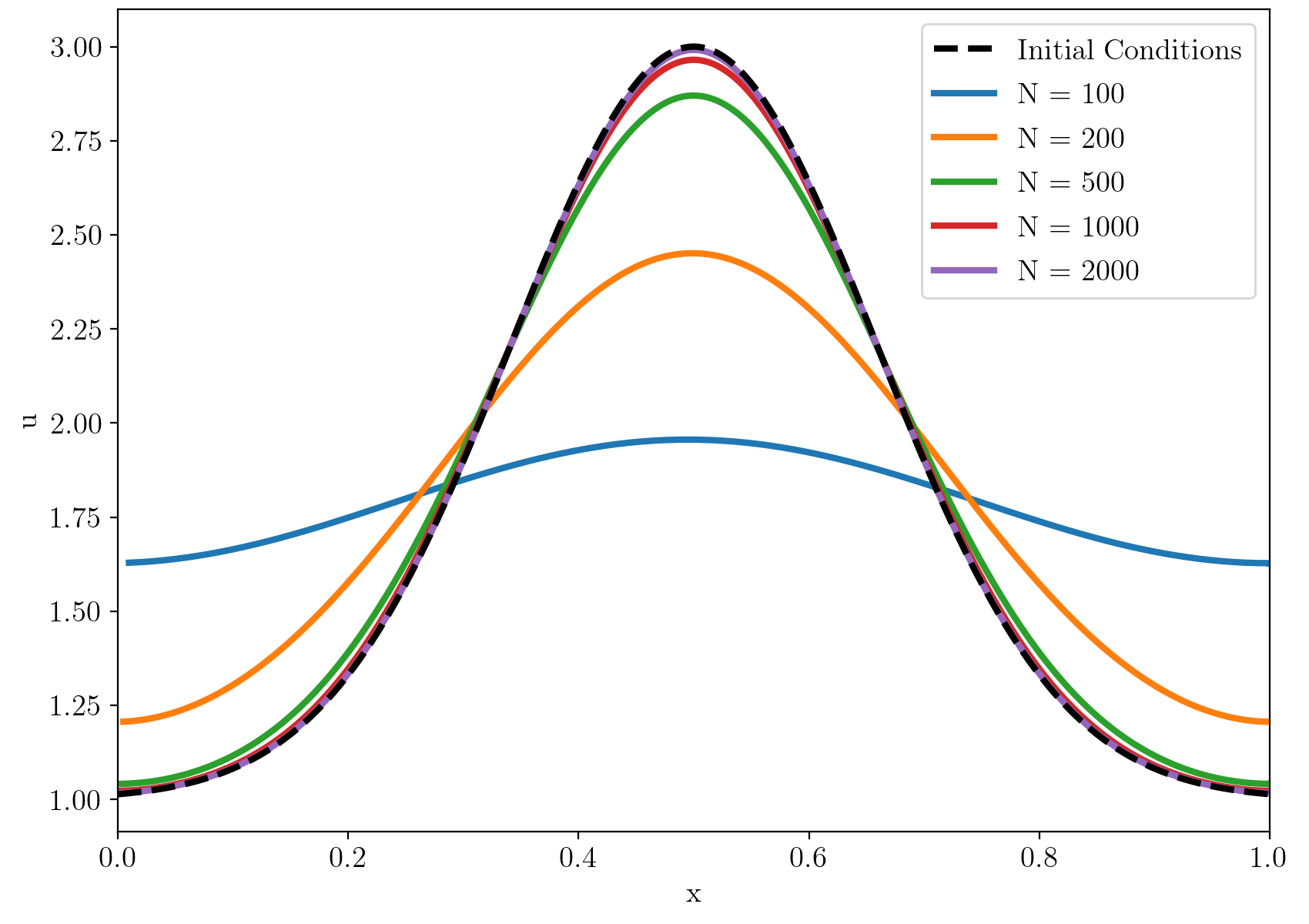}%
    \caption[Dependence of the numerical diffusion on $\Delta x$.]{
The solution of the linear advection equation with constant coefficient $a = 1$ (in arbitrary
units) for a fixed $C_{CFL} = 0.1$ and varying cell sizes $\Delta x = 1/N$ after $10^4$ time
steps. On the left, the initial conditions are a step function, on the right, they are a
Gaussian. The results have been repositioned to the original position at $t=0$ to demonstrate
how the initial shape changes over time.
}%
    \label{fig:linear-advection-diffusivity-dx}
\end{figure}

\begin{figure}
    \centering
    \includegraphics[width=.5\textwidth]{
    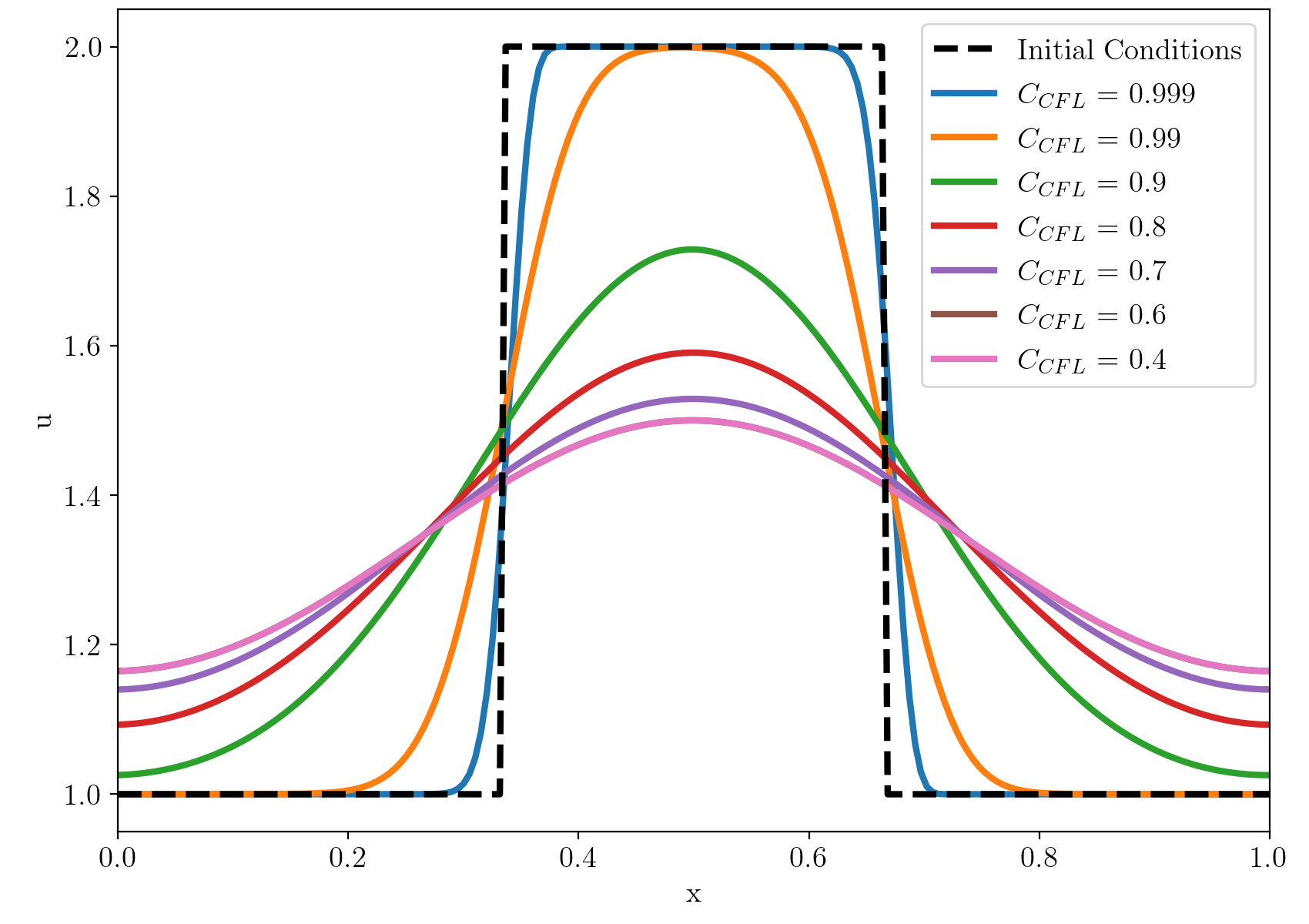}%
    \includegraphics[width=.5\textwidth]{
    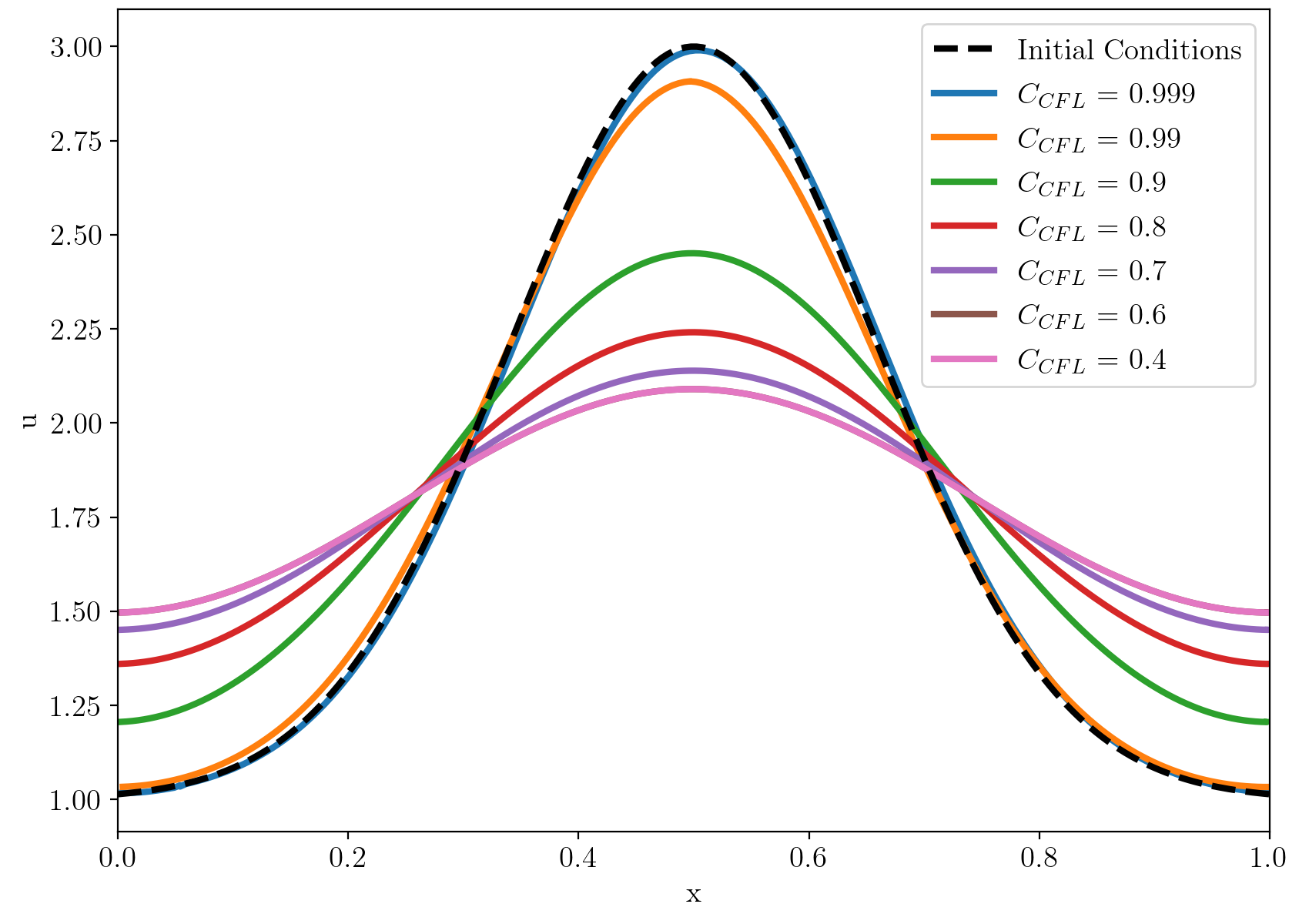}%
    \caption[Dependence of the numerical diffusion on $C_{CFL}$.]{
The solution of the linear advection equation with constant coefficient $a = 1$ (in arbitrary units)
for fixed cell sized $\Delta x = 1/200$ and varying $C_{CFL}$ after $10^4$ time steps. On the left,
the initial conditions are a step function, on the right, they are a Gaussian. The results have been
repositioned to the original position at $t=0$ to demonstrate how the initial shape changes over
time.
    }%
    \label{fig:linear-advection-diffusivity-cfl}
\end{figure}

\begin{figure}
    \centering
    \includegraphics[width=.5\textwidth]{
    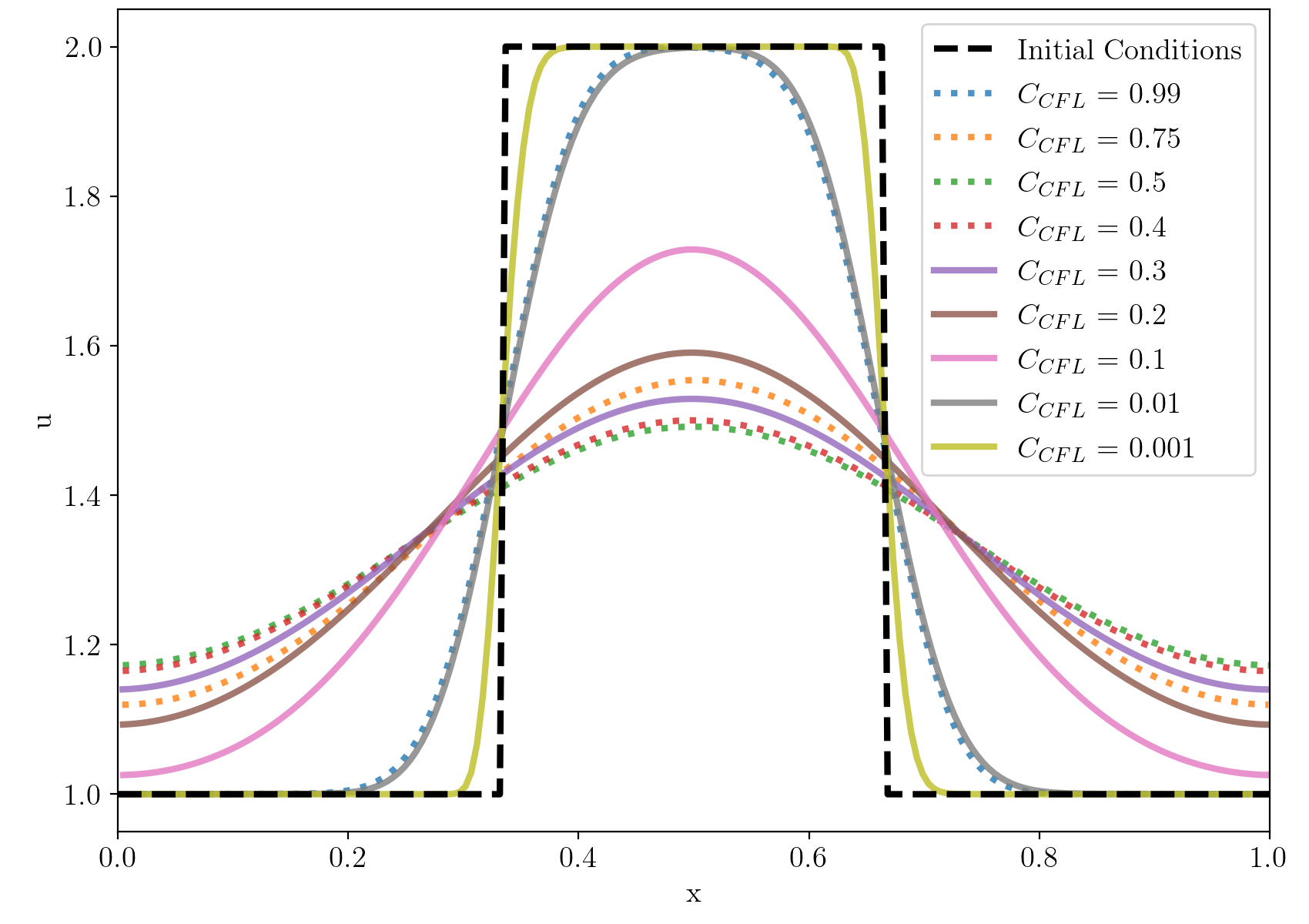}%
    \includegraphics[width=.5\textwidth]{
    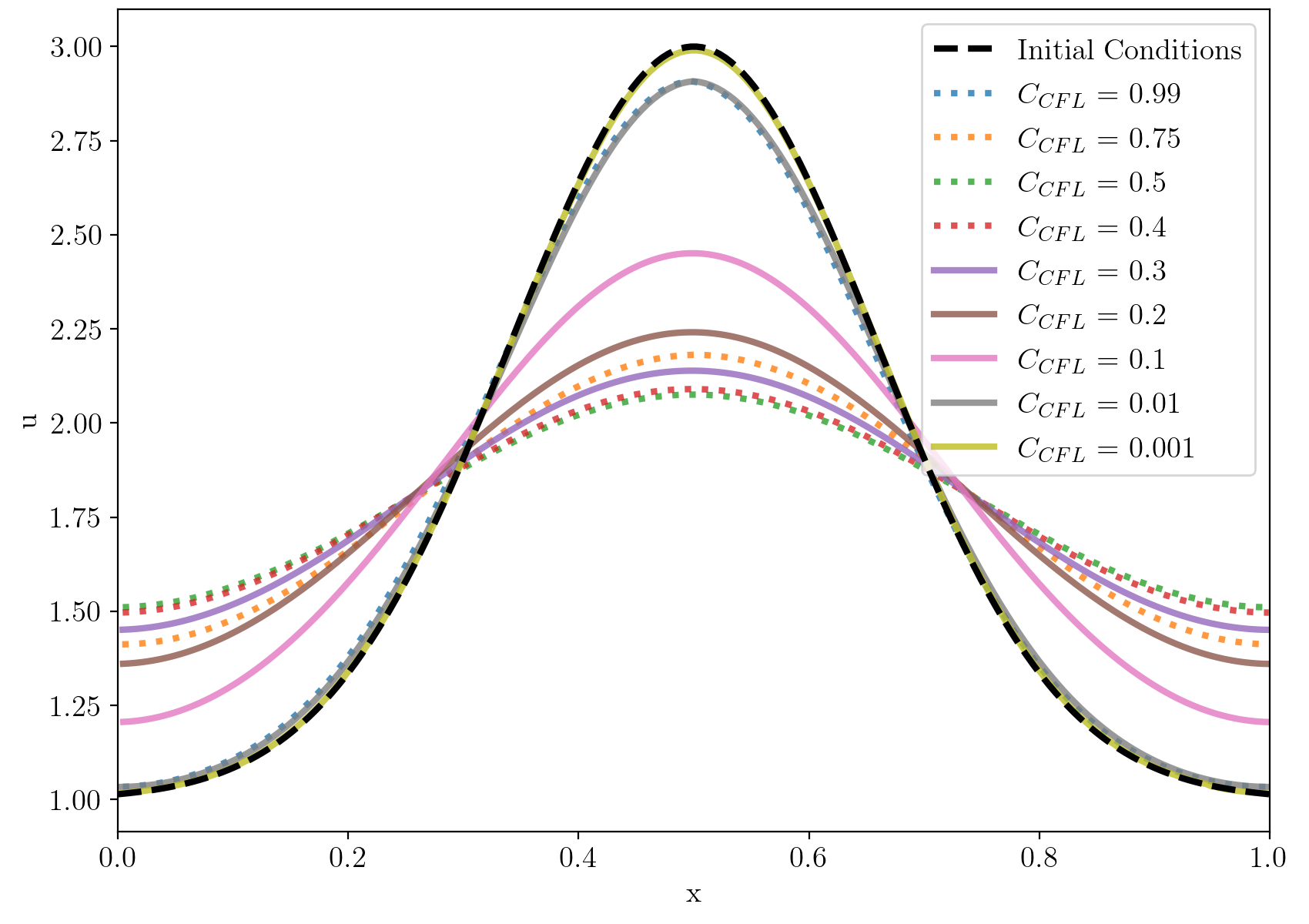}%
    \caption[Dependence of the numerical diffusion on $C_{CFL}$ with very low $C_{CFL}$.]{
Same as Figure~\ref{fig:linear-advection-diffusivity-cfl}, but additionally going down to very low
$C_{CFL}$. At some point around $C_{CFL} \sim 0.5$, the solution begins to improve again.
    }%
    \label{fig:linear-advection-diffusivity-low-cfl}
\end{figure}

\begin{figure}
    \includegraphics[width=\textwidth]{
    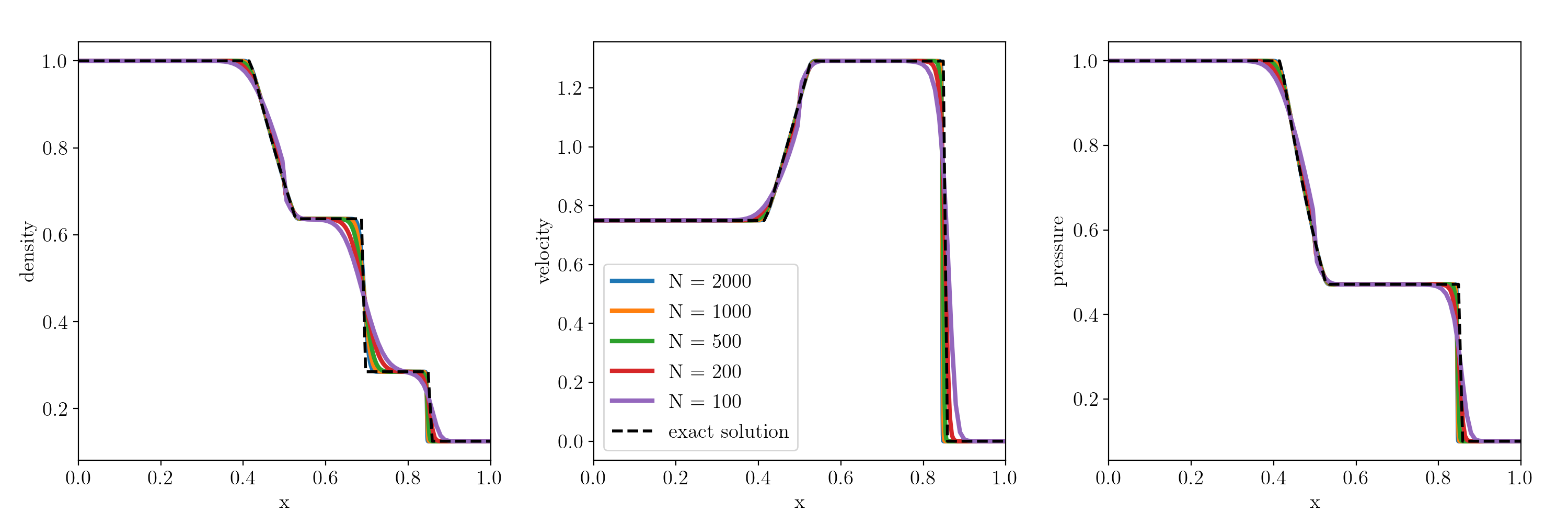}%
    \caption[Dependence of the numerical diffusion on $\Delta x$ for the Sod test with
    Godunov's method]{
The Sod test problem (eq.~\ref{eq:sod-test-ICs}) for the Euler equations solved using varying
cell sizes $\Delta x = 1/N$ at time $t = 0.15$ (in arbitrary units).
    }%
    \label{fig:godunov-sod-test-diffusivity-dx}
\end{figure}

These findings can readily be tested, albeit with a caveat: The choices of $\Delta x$, $\Delta t$,
and $C_{CFL}$ are not independent. They are related through eq.~\ref{eq:godunov-cfl}, where in the
case of linear advection with constant coefficients $S_{max}^n = a$.
This means that if $C_{CFL}$ is kept fixed while $\Delta x$ is varied, the time step $\Delta t$ will
change as well. Naturally the same happens when $C_{CFL}$ is varied while $\Delta x$ is kept fixed.
If $\Delta x$ or $C_{CFL}$ decrease, the time step size $\Delta t$ decreases as well, meaning that
more time steps will be necessary to reach a given end time. The caveat here is that the total
error increases with the number of time steps, as deviations introduced each individual time step
accumulate. This can clearly be seen in Fig.~\ref{fig:linear-advection-godunov}, where $\Delta x$
and $C_{CFL}$ are constant, and the further the simulation progresses in time, the worse the
results become. So in order to obtain a meaningful and accurate comparison with respect to time
step size, we need to compare the results after a fixed number of time steps, and not at fixed
end times.

Figure~\ref{fig:linear-advection-diffusivity-dx} shows how the diffusivity decreases with decreasing
$\Delta x$, while Figure~\ref{fig:linear-advection-diffusivity-cfl} shows how the diffusivity
decreases as $C_{CFL}$ increases, confirming our previous findings. To confirm that the findings
also hold for Godunov's method applied to the Euler equations, the solution of the Sod test with
varying $\Delta x$ is shown in Figure~\ref{fig:godunov-sod-test-diffusivity-dx}.
However, for the sake of clarity the solutions here are shown for a fixed end time $t_{end} =
0.15$. While that is not ideal to showcase how the diffusivity behaves depending on $\Delta x$ and
$C_{CFL}$\footnote{
In fact, the accumulated error with the increased number of required time steps to get to a set
$t_{end}$ nearly exactly cancels out the variations when varying $C_{CFL}$, and all resulting
curves are nearly identical. For this reason, a figure containing how the diffusivity behaves  with
varying $C_{CFL}$ for Godunov's method for the Euler equations is omitted. Compare also to
Figure~\ref{fig:godunov-accuracy}.
}, the alternative where the total number of time steps is kept constant would
require to visualize solutions at different end times. Contrary to the solution of the linear
advection, the shape of the exact solution changes over time, so each result would take a different
shape, which leads to convoluted and unwieldy plots.

\subsection{Order of Accuracy}

An interesting thing occurs if we repeat the same experiment as in
Figure~\ref{fig:linear-advection-diffusivity-cfl}, but go down to even lower values of $C_{CFL}$.
The solution \emph{improves} again around $C_{CFL} \sim 0.5$, as is shown in
Figure~\ref{fig:linear-advection-diffusivity-low-cfl}. To understand this behavior, we need to look
into the order of accuracy of the scheme. To estimate the error introduced each step by the method,
which we call the one step error $Err_{OS}$, we compute the difference between the exact solution
$\uc_{i,e}^{n+1}$ at time $t^{n+1}$ and the predicted solution $\uc_i^{n+1}$ which is evaluated
using Godunov's method and exact initial conditions $\uc_i^n$:

\begin{align}
    Err_{OS}
    &= \uc_i^{n+1} - \uc_{i,e}^{n+1}  \\
    &= \uc_{e,i}^n - \frac{a \Delta t}{\Delta x} (\uc_{i,e}^n - \uc_{i-1,e}^n) - \uc_{i,e}^{n+1}
\end{align}

Using the Taylor expansions \ref{eq:taylor-un+1} and \ref{eq:taylor-ui-1} to express
$\uc_{i,e}^{n+1}$ and $\uc_{i-1,e}^n$, and keeping only second order terms in $\Delta x$ and
$\Delta t$, we can write

\begin{align}
    Err_{OS}
    &= \uc_{i,e}^n
        - a \frac{\Delta t}{\Delta x}
        \left( \uc_{i,e}^n -
            \uc_{i,e}^n +
            \DELDX{\uc_{i,e}^n } \Delta x -
            \frac{\Delta x^2}{2} \frac{\del ^2 \uc_{i,e}^n }{\del x^2}
            \order(\Delta x^3) \right)  - &&
\\
    &\quad  - \left(
        \uc_{i,e}^n +
        \DELDT{\uc_{i,e}^n } \Delta t +
        \frac{\Delta t^2}{2} \frac{\del ^2 \uc_{i,e}^n }{\del t^2}
        + \order(\Delta t^3)  \right) && \\
    &= - \Delta t \left(
        a \left(
            \DELDX{\uc_{i,e}^n } - \frac{\Delta x}{2} \frac{\del ^2 \uc_{i,e}^n }{\del x^2} +
\order(\Delta x^2)         \right)
        + \DELDT{\uc_{i,e}^n } + \frac{\Delta t}{2} \frac{\del ^2 \uc_{i,e}^n }{\del t^2} +
\order(\Delta t^2)
    \right) \\
    &= - \Delta t \left(
        \underbrace{\DELDT{\uc_{i,e}^n } + a\DELDX{\uc_{i,e}^n }}_{= 0} + \order(\Delta x) +
\order(\Delta t)
    \right)
\end{align}

Finally giving us:
\begin{align}
    Err_{OS} \propto \Delta t \left( \order(\Delta x) + \order(\Delta t) \right)
\label{eq:one-step-error}
\end{align}

While the one step error is proportional to $\Delta t^2$, reducing $\Delta t$ also means that more
steps need to be carried out to reach some end time $t_{end}$. If for some initially set $\Delta
t_0$ one needs $N$ steps to reach $t_{end}$, i.e. $t_{end} = N \Delta t_0$, then for some $\Delta
t_1 < \Delta t_0$ we require $t_{end} = N (\Delta t_0 / \Delta t_1)$ steps. It is then convenient to
define the Local Truncation Error $Err_{LT}$

\begin{align}
    Err_{LT} = \frac{1}{\Delta t} Err_{OS}
\end{align}

to describe how a scheme scales with cell spacing $\Delta x$ and time step size $\Delta t$. For
this scheme,

\begin{align}
    Err_{LT} = \order(\Delta t) + \order(\Delta x) \label{eq:local_truncation_error}
\end{align}

so Godunov's scheme is first order accurate in $\Delta x$ and $\Delta t$.

This property explains why in Figure~\ref{fig:linear-advection-diffusivity-low-cfl} the results
begin to improve once $C_{CFL}$ becomes small enough. The diffusion term $D \propto (1 - C_{CFL})$
has an upper boundary at $C_{CFL} = 0$. Once $C_{CFL}$ is small enough, the diffusion term stops
increasing noticeably. For example, going from $C_{CFL} = 0.01$ to $C_{CFL} = 0.001$ changes the
amplitude of the diffusion term form $0.99$ to $0.999$, or less than one per cent. However, since
the Courant number directly determines the time step size $\Delta t$ and the scheme is first order
accurate in time, the expected result when going from $C_{CFL} = 0.01$ to $C_{CFL} = 0.001$ should
improve by a factor of 10. In the case of the solution of the linear advection equation using
Godunov's method, the turnaround point, where the results begin to improve again, occurs at
$C_{CFL} \sim 0.5$ (see Figure~\ref{fig:linear-advection-diffusivity-low-cfl}).

Another interesting point can be made when comparing the results for the step function with the
results of the smooth Gaussian: It appears that the deviations from the expected solution for the
step function are somewhat larger, in particular in cases with lower diffusivity, i.e. for small
$\Delta x$ in Figure~\ref{fig:linear-advection-diffusivity-dx} and large $C_{CFL}$ in
Figure~\ref{fig:linear-advection-diffusivity-cfl}.

To understand that phenomenon, we need to look into a different approach to estimate the local
truncation error using a formalism that allows for discontinuous solutions. The previous expression
assumed a smooth (differentiable) state, which is not the case any longer with discontinuities
being present. Instead, we make use of the analytical solution for the advection-diffusion
equation~\ref{eq:advection-diffusion} as the analytical expression for the solution Godunov's
scheme will provide. This solution is given by

\begin{align}
    \uc^n(x) = \uc_0(x - a t^n) \ \mathrm{erfc}\left(\frac{x - at^n}{\sqrt{4 D t}} \right)
\end{align}

with

\begin{align}
    \mathrm{erfc}(x) = \frac{2}{\sqrt{\pi}} \int_x^\infty \exp (-z^2) \de z
\end{align}

To estimate the error at time $t^n$, we take the 1-norm of the difference between the exact
solution $\uc_e = \uc_0{x - at^n}$ and the analytical expression of the solution of Godunov's
scheme:

\begin{align}
    Err &= || \uc_e(x, t^n) - \uc^n(x) ||_1
        = \int\limits_{-\infty}^{\infty} | \uc_e(x, t^n) - \uc^n(x) | \de x \\
        &= \int\limits_{-\infty}^{\infty}
            | \uc_0(x - at) - \uc_0(x - at) \ \mathrm{erfc}\left(\frac{x - at^n}{\sqrt{4 D t}}
\right)) | \de x \\
        &= \int\limits_{-\infty}^{\infty}
            | \uc_0(x') (1 - \mathrm{erfc}\left(\frac{x'}{\sqrt{4 D t}} \right)) | \de x'
\end{align}

In the last step, the substitution $x' = x - at$ was used. Choosing the discontinuous initial
conditions

\begin{align}
    \uc_0 = \begin{cases}
            1 & \text{ if } x < 0\\
            0 & \text{ if } x > 0\\
            \end{cases}
\end{align}

and using a second substitution $y = \frac{-x'}{\sqrt{4 D t}}$ the integral can be written as

\begin{align}
    Err &= \int\limits_{-\infty}^{0}
        | (1 - \mathrm{erfc}\left(\frac{x'}{\sqrt{4 D t}} \right)) | \de x' \\
        &= \sqrt{4 D t} \int\limits^{\infty}_{0}
        | (1 - \mathrm{erfc}(-y)) | \de y
\end{align}

The integral can be further simplified, but since it will in any case yield a result independent of
$x$ and $t$, we don't actually care for its exact evaluation, and instead write

\begin{align}
    Err = C_1 \sqrt{D t} = C_2 \sqrt{\Delta x t} = C_2 \sqrt{\Delta x N \Delta t}
\label{eq:error-discontinuity}
\end{align}

where $C_1$ and $C_2$ are some constants independent of $\Delta x$ and $\Delta t$. Note that in
order to derive this result, we chose the initial conditions to be discontinuous. So this
result tells us that in the presence of discontinuities, the error scales with $\sqrt{\Delta t}$ and
$\sqrt{\Delta x}$, while for smooth conditions, it scales with $\Delta t$ and $\Delta x$. The exact
factor of $\sqrt{\Delta t}$ and $\sqrt{\Delta x}$ difference to the smooth condition isn't
generally valid for other methods and initial conditions, but it is a fact that the presence of
discontinuities reduces the order of accuracy of a scheme just like it did in this case.

Finally, to conclude the chapter on Godunov's method, we can verify our findings of the order of
accuracy of the method by conducting numerical experiments and measuring how the error scales with
$\Delta x$ and $C_{CFL} \propto \Delta t$. The error is estimated using

\begin{align}
    Err = \frac{1}{N} \sum_i^N |\uc_i - \uc_{i,exact}|
\end{align}

\begin{figure}
    \centering
    \includegraphics[width=.5\textwidth]{
    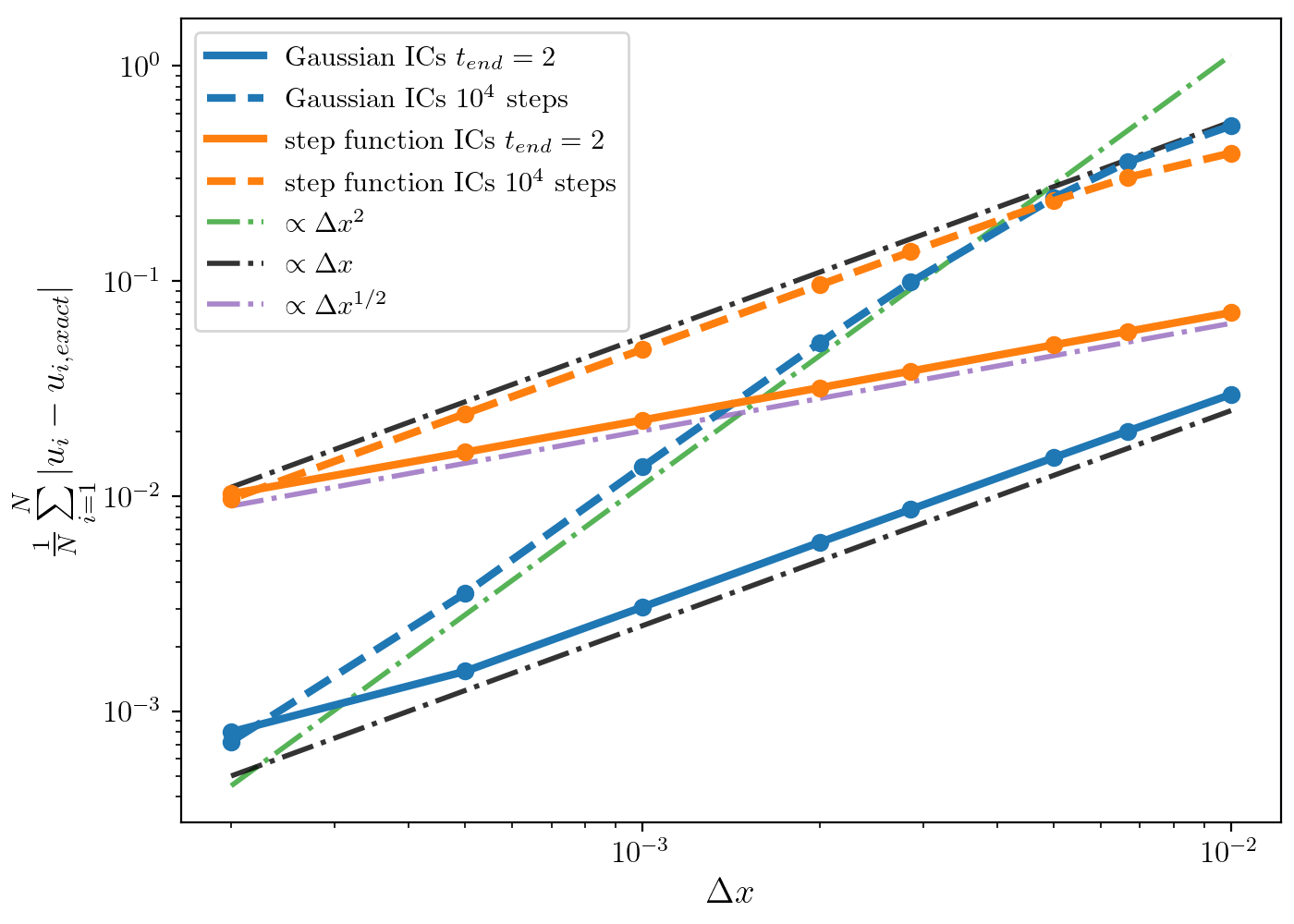}%
    \includegraphics[width=.5\textwidth]{
    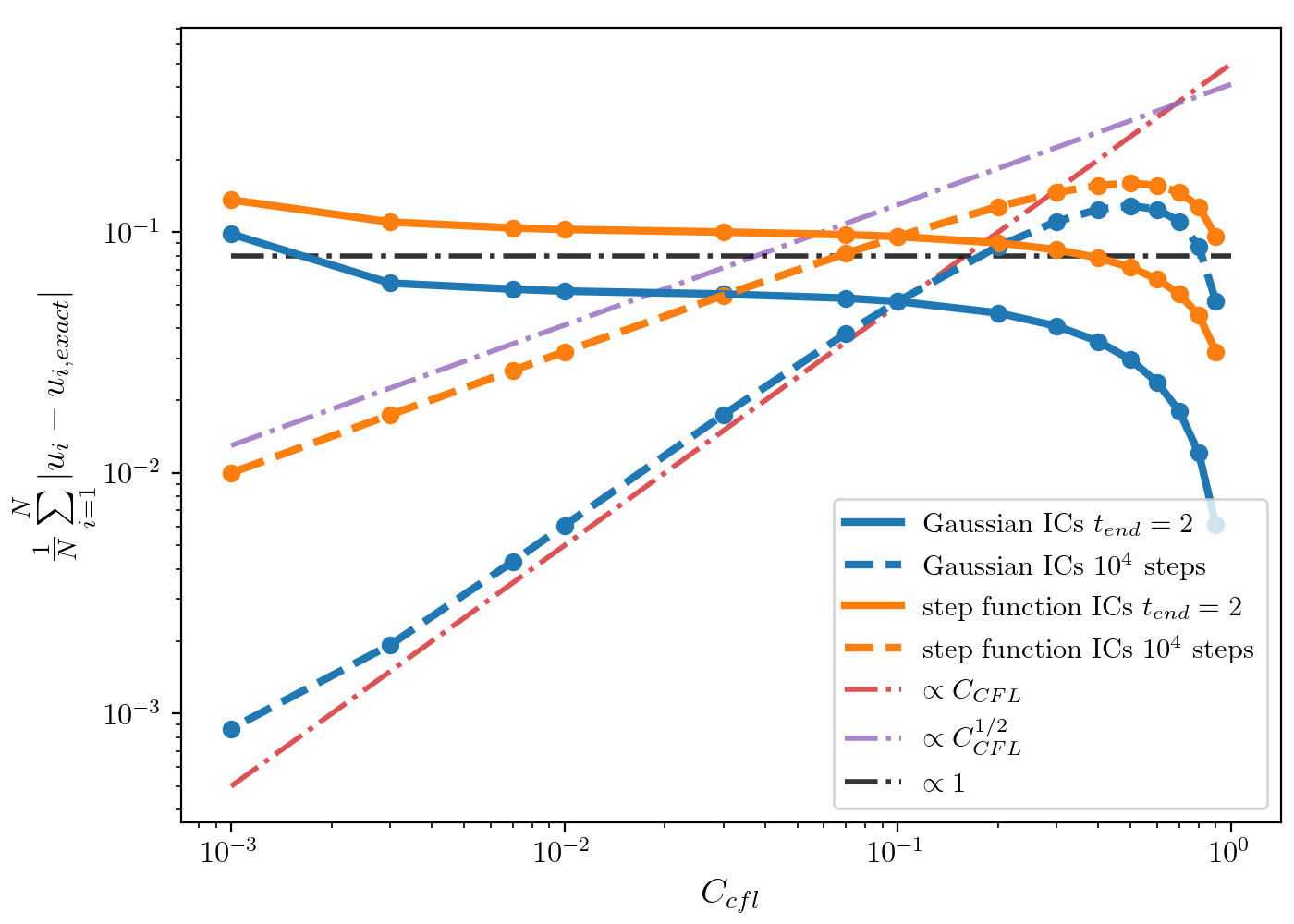}%
    \caption[Order of accuracy w.r.t $\Delta x$ and $C_{CFL}$ for linear advection using Godunov's
method.]{
The order of accuracy w.r.t. $\Delta x$ (left) and the Courant number $C_{CFL}$ (right) of
the linear advection equation with constant coefficients using Godunov's method for both a step
function and a Gaussian initial conditions. For comparison, lines with slopes $1/2$, $1$, and $2$
are overplotted. The experiments are run for both a fixed end time $t_{end} = 2$ as well as for a
fixed total number of time steps individually.
    }%
    \label{fig:linear-advection-accuracy}
\end{figure}

Figure~\ref{fig:linear-advection-accuracy} shows the results for the linear advection equation with
constant coefficients using Godunov's method. Both a step function containing discontinuities and
smooth Gaussian initial conditions are used. Also results for both a fixed number of time steps and
for a fixed end time are shown, along with guiding lines for varying powers of $\Delta x$. Let's
look at the results of each of the four cases:

\begin{itemize}
\item For the smooth Gaussian and a fixed number of total time steps (blue dashed line), the error
scales with $\sim \Delta x^2$. The reason is that for a fixed $C_{CFL}$, reducing $\Delta x$
also reduces the time step $\Delta t$. We have seen that the local truncation
error~\ref{eq:local_truncation_error} depends on both $\Delta t$ and $\Delta x$, and in this
scenario, \emph{both} are reduced, leading to a net scaling with a power greater than $1$. Another
way of understanding this phenomenon is by considering the fact that the one step
error~\ref{eq:one-step-error} is proportional to $\Delta t^2$. By keeping the total number of time
steps fixed for all $\Delta x$, the amount of errors that can be accumulated over the course of the
simulation is kept constant, resulting in a scaling $\propto \Delta t^2 = (C_{CFL} \Delta x / a)^2
\propto \Delta x^2$ for constant $a$ and $C_{CFL}$.

\item For the smooth Gaussian and a fixed end time $t_{end} = 2$ (blue solid line), the error
scales with $\sim \Delta x$, which is exactly what the local truncation
error~\ref{eq:local_truncation_error} predicts. In this scenario, the accumulation of individual
one step errors due to the larger number of total time steps necessary with decreasing $\Delta x$
reduces the total accuracy. However, this line is the one that is of higher interest for practical
applications: After all, typically a simulation is run until a specified end time, not for a
certain number of steps.

\item The discontinuous step function and a fixed end time $t_{end} = 2$ (orange solid line), the
error is as predicted by eq.~\ref{eq:error-discontinuity}: Since the end time $t_{end} = N \Delta
t$ is kept constant, the result should be precisely an error $\propto \sqrt{\Delta x}$.

\item The discontinuous step function and a fixed total number of steps time (orange dashed line),
the error scales with $\sim \Delta x$, which is again higher than for the case with a fixed end
time. The reason is that since the end time is not being kept constant, the factor $\sqrt{N \Delta
t}$ is not constant any more. $N$ is constant, but $\Delta t \propto \Delta x$ for a fixed
$C_{CFL}$, and the error scales in total $\propto \sqrt{\Delta x \Delta x} = \Delta x$, which is
exactly what we observe.
\end{itemize}

\begin{figure}
    \centering
    \includegraphics[width=.5\textwidth]{
    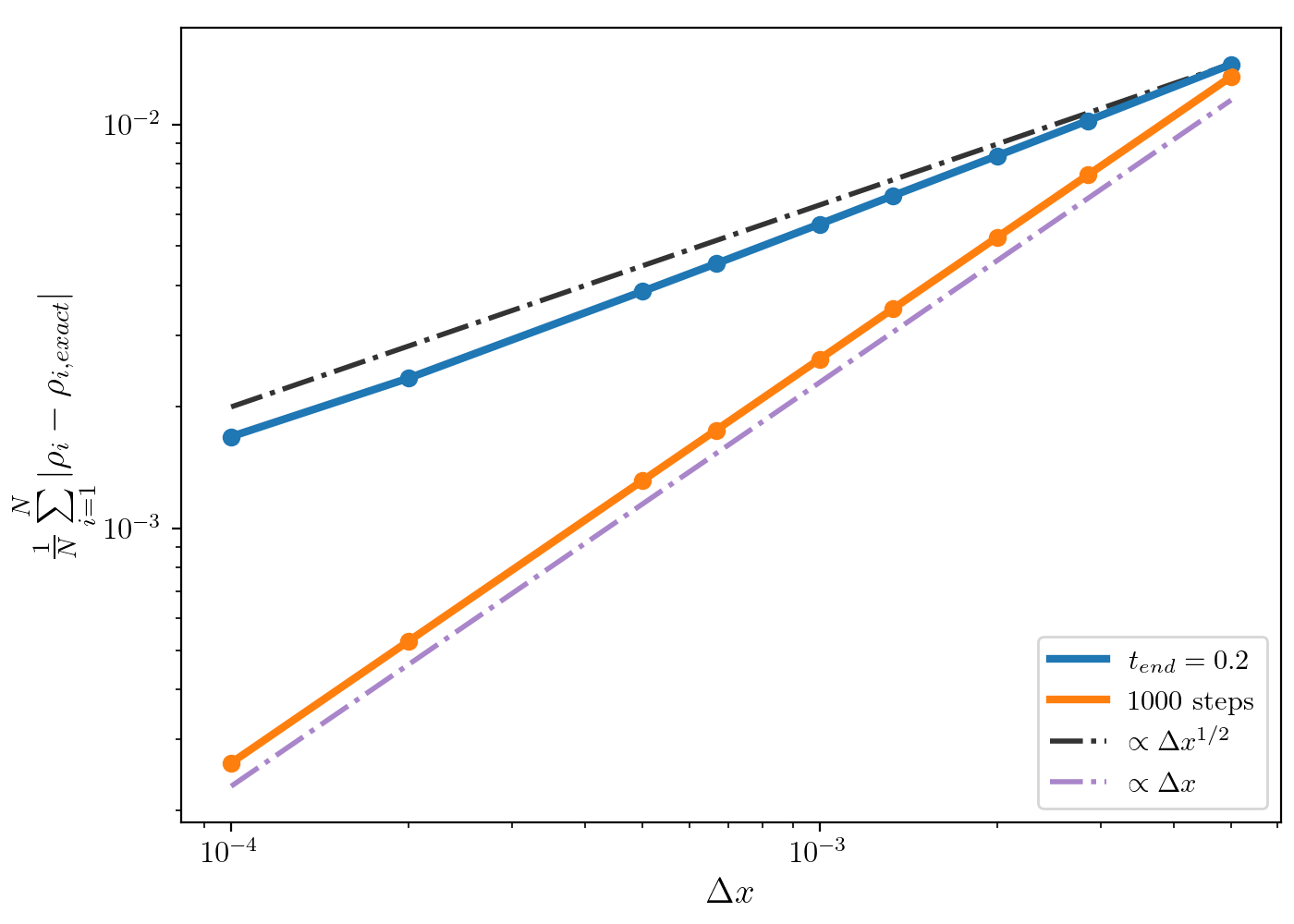}%
    \includegraphics[width=.5\textwidth]{
    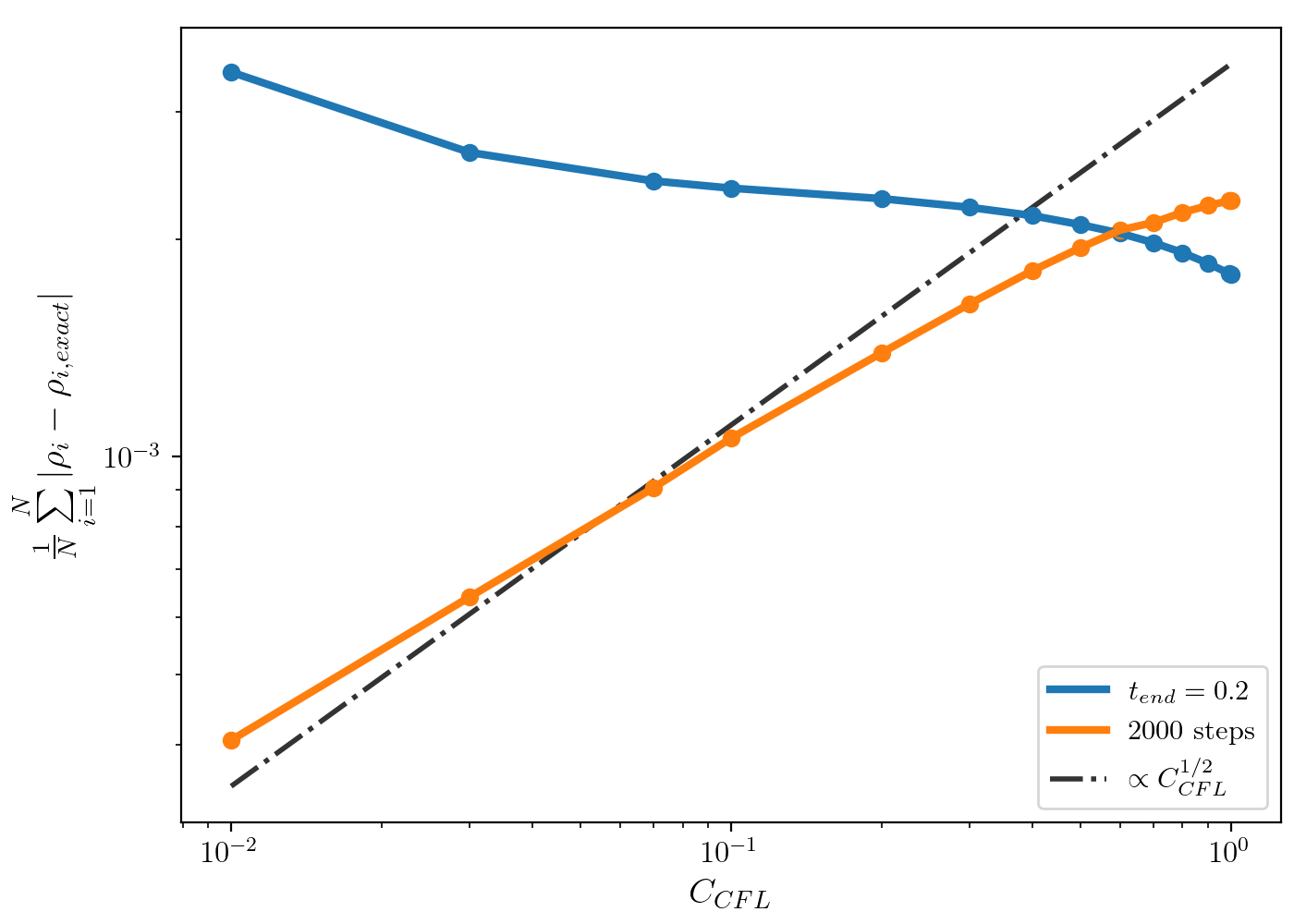}%
    \caption[Order of accuracy w.r.t $\Delta x$ and $C_{CFL}$ for Euler equations using Godunov's
method.]{
The order of accuracy w.r.t. $\Delta x$ (left) and the Courant number $C_{CFL}$ (right) of
the Euler equations using Godunov's method for the Sod test (eq.~\ref{eq:sod-test-ICs}). For
comparison, lines with slopes $1/2$, and $1$ are over-plotted. The experiments are run for both a
fixed end time $t_{end} = 0.2$ as well as for a fixed total number of time steps individually. The
fixed total number of steps was chosen such that all three waves remain within the boundary for all
choices of $\Delta x$ and $C_{CFL}$.
    }%
    \label{fig:godunov-accuracy}
\end{figure}

The case for varying $C_{CFL}$ and fixed $\Delta x$ in Figure~\ref{fig:linear-advection-accuracy}
is somewhat simpler to interpret, since the choice of $C_{CFL}$ doesn't affect the cell spacing
$\Delta x$. For a fixed number of time steps, the expected slopes $\propto C_{CFL}$ and $\propto
C_{CFL}^{1/2}$ for the smooth and the discontinuous, respectively, are reproduced. In the cases for
a fixed end time, the accumulation of one step errors cancels out the improved accuracy per step,
and the net scaling is $\propto 1$. In all cases, the influence of the diffusion term $D \propto (1
- C_{CFL})$ is clearly seen for $C_{CFL} \gtrsim 0.1$. The same behavior can be seen for Godunov's
method applied to the Euler equations in Figure~\ref{fig:godunov-accuracy}, since the solved
problem, the Sod test (eq.~\ref{eq:sod-test-ICs}), contains discontinuities as part of its exact
solution.

%% file: main/FV/FV-4-higher-order.tex
\chapter{Higher Order and TVD Schemes}\label{chap:higher-order-schemes}

Higher order schemes, i.e. schemes whose accuracy scales with cell spacings $\Delta x$ and time
step sized $\Delta t$ with a power greater than $1$, are very desirable for practical applications.
While they require more calculations for a single cell update $\uc^{n+1}_i$, their improved scaling
makes them attractive since increasing the resolution, i.e. reducing $\Delta x$ and $\Delta t$,
leads to improved results. There is always a turning point at which the additional computational
cost per update is overtaken by the better scaling, leading to more accurate solutions at a
comparatively lower computational cost. For example, say a second order accurate solution
$\uc_{2}^{n+1}$ requires 4 times more computations in order to be second order accurate, i.e. scale
with $\Delta x^2$, compared to the first order solution $\uc_{1}^{n+1}$. If $N$ is the number of
computations required to obtain the full solution $\uc_{1}^{n+1}$ for all cells, then decreasing
$\Delta x$ by a factor of $8$ means also increasing the number of required cells to cover the same
volume. The computational cost of the first order method increases linearly to $8N$, and the error
will be reduced by a factor of $1/8$. For the second order scheme, the initial cost would be $4 N$,
and the cost after reducing $\Delta x$ by a factor of $8$ increases linearly to $32 N$. However the
error is reduced by a factor $\propto \Delta x^2$, i.e. by a factor of $1/64$. In order to obtain
this sort of improvement of accuracy using the first order accurate scheme, we'd need to decrease
$\Delta x$ by a factor of $64$, leading to a required computation workload of $64 N$, or twice as
much as the second order scheme. Additionally, the first order scheme would have much higher memory
requirements in order to reproduce the same accuracy, as it would need to store $8$ times more
cells.\footnote{The factor 8 stems from the assumption that the problem is one dimensional, and
that the number of cells is directly proportional to the cell size $\Delta x$. For two and three
dimensions and assuming cells of equal size, an additional factor of $8$ needs to be added
for each additional dimension. This leads to $64$ and $512$ more cells required to be stored,
respectively.}

Extending Godunov's method to higher order accuracy can at first glance seem straightforward. In
this chapter, two distinct approaches to do so are discussed:

\begin{itemize}
\item A method that relaxes the assumption that the states inside a cell are piece-wise constant,
and instead assumes that they are piece-wise linear, called the MUSCL-Hancock method
\item A method that attempts to find an improved expression for the inter-cell fluxes $\F_{i\pm
\half}$ by relaxing the assumption that the states left and right of the boundary remain constant
throughout the entire time step, called the Weighted Average Flux (WAF) method.
\end{itemize}

However, higher order accurate methods come with a plethora of new problems that weren't present
for first order methods. In particular, if left untreated, second order methods will always develop
spurious oscillations in the solutions, which typically are violently unstable. There are ways to
handle this issue, namely by applying slope and flux limiters, but the way they work can sometimes
appear like witchcraft. To showcase the problems of second order methods and how the limiters work,
we first have a look at second order schemes for the ol' reliable linear advection with constant
coefficients.

\section{Higher Order Schemes For Scalar Equations}

\subsection{The WAF Scheme}\label{chap:WAF}

The core idea of the Weighted Average Flux (WAF) scheme is to improve the estimate for the fluxes
at the cell boundaries, $\F_{i \pm \half}$.
When deriving Godunov's method, we started off by using the integral form of the conservation law.
A key part of solving the integral of the fluxes over time was to make use of the fact that
solution to the Riemann problem centered between two adjacent cells, $\uc_i^n$ and $\uc_{i+1}^n$,
predicts that for all $t > 0$ the solution at the boundary will remain constant.
The general solution structure of the Riemann problem consists of waves that will emanate from the
center and travel along characteristics, as is shown in e.g.
Figures~\ref{fig:riemann-linear-hyperbolic-system} and \ref{fig:riemann-solution}. However, given
that the states are represented as integral averages of continuous states, i.e.

\begin{align}
    \uc_i^n = \frac{1}{\Delta x} \int\limits_{x_{i-\half}}^{x_{i+\half}} \uc(x, t^n) \de x
\end{align}

a consequence of waves emanating from cell boundaries and propagating through the cells is that the
integral averaged cell states won't be constant for $t > t^n$ precisely \emph{because} there are
waves propagating through them.
The WAF method tries to get a better estimate of the fluxes that would actually pass between
a boundary if the underlying problem weren't discretized, but continuous. To do so, the constant
fluxes $\fc_{i + \half}$ of Godunov's scheme are replaced by fluxes averaged over space.
Explicitly, the fluxes are estimated at the midpoint in time between two time steps, $t^{n+\half} =
t^n + \frac{1}{2} \Delta t$, and averaged from the center of the left cell, $x_i$, to the center of
the right cell, $x_{i+1}$:

\begin{align}
    \fc_{i+\half}^{WAF} = \frac{1}{\Delta x} \int\limits_{x_i}^{x_{i+1}} \fc(\uc_{i+\half}(
t^{n+\half})) \de x
\end{align}

\begin{figure}
    \centering
    \includegraphics[width=.8\textwidth]{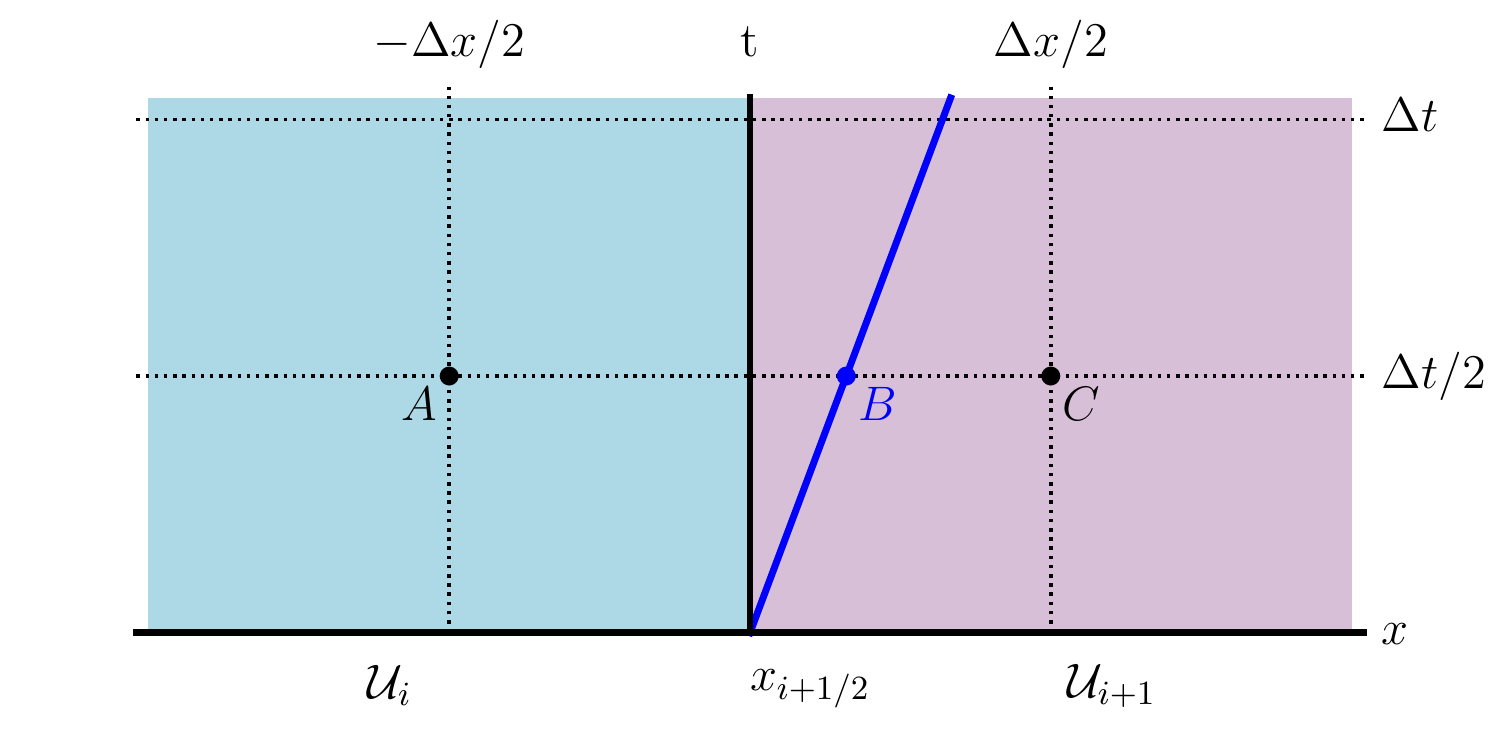}%
    \caption[WAF flux estimate for linear advection]{
The set-up for computing the WAF inter-cell flux at the mid-step in time. For $t > 0$ and a
coefficient $a > 0$, the emanating wave (blue line) will intersect the $t = \Delta t/2$ line at
some point $B > x_{i+\half}$
    }
    \label{fig:advection-WAF}
\end{figure}

$\uc_{i+\half}$ is the solution to the Riemann problem with initial data $\uc_i$ and $\uc_{i+1}$.
Figure~\ref{fig:advection-WAF} illustrates the situation assuming a positive coefficient $a > 0$.
Making use of the analytical solution of the Riemann problem,

\begin{align}
    \uc(x, t) = \begin{cases}
                 \uc_i & \text{ if } \frac{x - x_{i-\half}}{t} < a \\
                 \uc_{i+1} & \text{ if } \frac{x - x_{i-\half}}{t} > a
                \end{cases}
\end{align}

we can subdivide the interval $[x_i, x_{i+1}]$ into two subdomains along the $t = t^{n + \half}$
line in Figure~\ref{fig:advection-WAF}: Let point $A$ be at $(x_i, t^{n+\half})$, point $C$ at
$(x_{i+1}, t^{n+\half})$, and point $B$ at the position of the discontinuity, $(x_{i+\half}+a
\Delta t / 2, t^{n+\half})$. The lengths of the subdomains $\overline{AB}$ and $\overline{BC}$ are
given by
\begin{align}
    \overline{AB} &= \frac{\Delta x}{2} + a \frac{\Delta t}{2}
        = \frac{1}{2}\left(1 + \frac{a \Delta t}{\Delta x}\right) \Delta x
        = \frac{1}{2}\left(1 + C_{CFL} \right) \Delta x  \\
    \overline{BC} &= \frac{\Delta x}{2} - a \frac{\Delta t}{2}
        = \frac{1}{2}\left(1 - C_{CFL} \right) \Delta x
\end{align}

Since the states between
$\overline{AB}$ and $\overline{BC}$ are constant, the integral averaged flux is trivial to compute:

\begin{align}
\fc_{i+\half}^{WAF}
    &= \frac{1}{\Delta x} \int\limits_{x_i}^{x_{i+1}} \fc(\uc_{i+\half}( t^{n+\half})) \de x \\
    &= \frac{1}{\Delta x} \left[
    \int\limits_{A}^{B} \fc(\uc_{i+\half}^{n+\half}) \de x +
    \int\limits_{B}^{C} \fc(\uc_{i+\half}^{n+\half}) \de x
    \right] \\
    &= \frac{1}{\Delta x} \left[
    \overline{AB} \fc(\uc_{i+\half}^{n+\half}) |_{x = A}^B +
    \overline{BC} \fc(\uc_{i+\half}^{n+\half}) |_{x = B}^C
    \right] \\
    &= \frac{1}{\Delta x} \left[
    \overline{AB} (a \uc_i) + \overline{BC} (a \uc_{i+1})
    \right] \\
    &= \frac{1}{\Delta x} \left[
        \frac{1}{2}\left(1 + C_{CFL} \right) \Delta x (a \uc_i) +
        \frac{1}{2}\left(1 - C_{CFL} \right) \Delta x  (a \uc_{i+1})
    \right] \\
    &=  \frac{1}{2}\left(1 + C_{CFL} \right) (a \uc_i) +
        \frac{1}{2}\left(1 - C_{CFL} \right) (a \uc_{i+1}) \label{eq:WAF-flux}
\end{align}

For $a > 0$, the flux is a weighted average of the upwind flux $\fc_i = a \uc_i$ and the downwind
flux $\fc_{i+1} = a \uc_{i+1}$, hence the name of the method.

The final update scheme reads as follows:

\begin{align}
    \uc_{i}^{n+1} =
        \uc_i + \frac{\Delta t}{\Delta x}\left(\fc_{i-\half}^{WAF} - \fc_{i+\half}^{WAF} \right)
\end{align}

Formally, this equation looks identical to Godunov's method, given in eq.~\ref{eq:godunov-scheme}.
The big difference which results in a higher order accuracy is ``hidden'' in the expression for the
fluxes $\fc^{WAF}_{i\pm \half}$. They are now estimated using an integral average of cell states
which are no longer assumed to remain constant during a time step $\Delta t$, but are allowed to
change over the duration thereof.

\subsection{The MUSCL-Hancock Scheme}\label{chap:MUSCL-Hancock}

The approach behind MUSCL (Monotone Upstream-Centered Schemes for Conservation Laws) schemes to
achieve higher order accuracy is to depart from the piece-wise constant reconstruction of the data,
and use a higher order interpolation instead. The simplest way of modifying the piece-wise constant
data is to replace the constant states $\uc_i^n$ with a linear local reconstruction

\begin{align}
    \uc_i^n(x) = \uc_i^n + (x - x_i) s_i, \quad x \in [0, \Delta x]
\end{align}

where $s_i$ is a suitably chosen slope of $\uc_i(x)$. The integral average of $\uc_i^n(x)$ in each
cell is identical to the piece-wise constant state, and hence the reconstruction remains
conservative. A general way of defining a slope is using a free parameter $\omega \in [-1, 1]$ and
write

\begin{align}
    s_i = \frac{1}{\Delta x} \left[
    \frac{1}{2} ( 1 + \omega ) (\uc_{i}^n - \uc_{i-1}^n) +
    \frac{1}{2} (1 - \omega)(\uc_{i+1}^n - \uc_{i}^n)
    \right]
\end{align}

Some special values for $\omega$ are:
\begin{align}
    \omega = 0: && \text{Centered slope (Fromm) } &&
        &s_i^n = \frac{\uc_{i+1}^n - \uc_{i-1}^n}{ 2
\Delta x}\\
    \omega = 1: && \text{Upwind slope (Beam-Warming) } &&
        &s_i^n = \frac{\uc_{i}^n - \uc_{i-1}^n}{\Delta x} \\
    \omega = -1: && \text{Downwind slope (Lax-Wendroff) } &&
        &s_i^n = \frac{\uc_{i+1}^n -
\uc_{i}^n}{\Delta x}
\end{align}

\begin{figure}
    \centering
    \includegraphics[width=\textwidth]{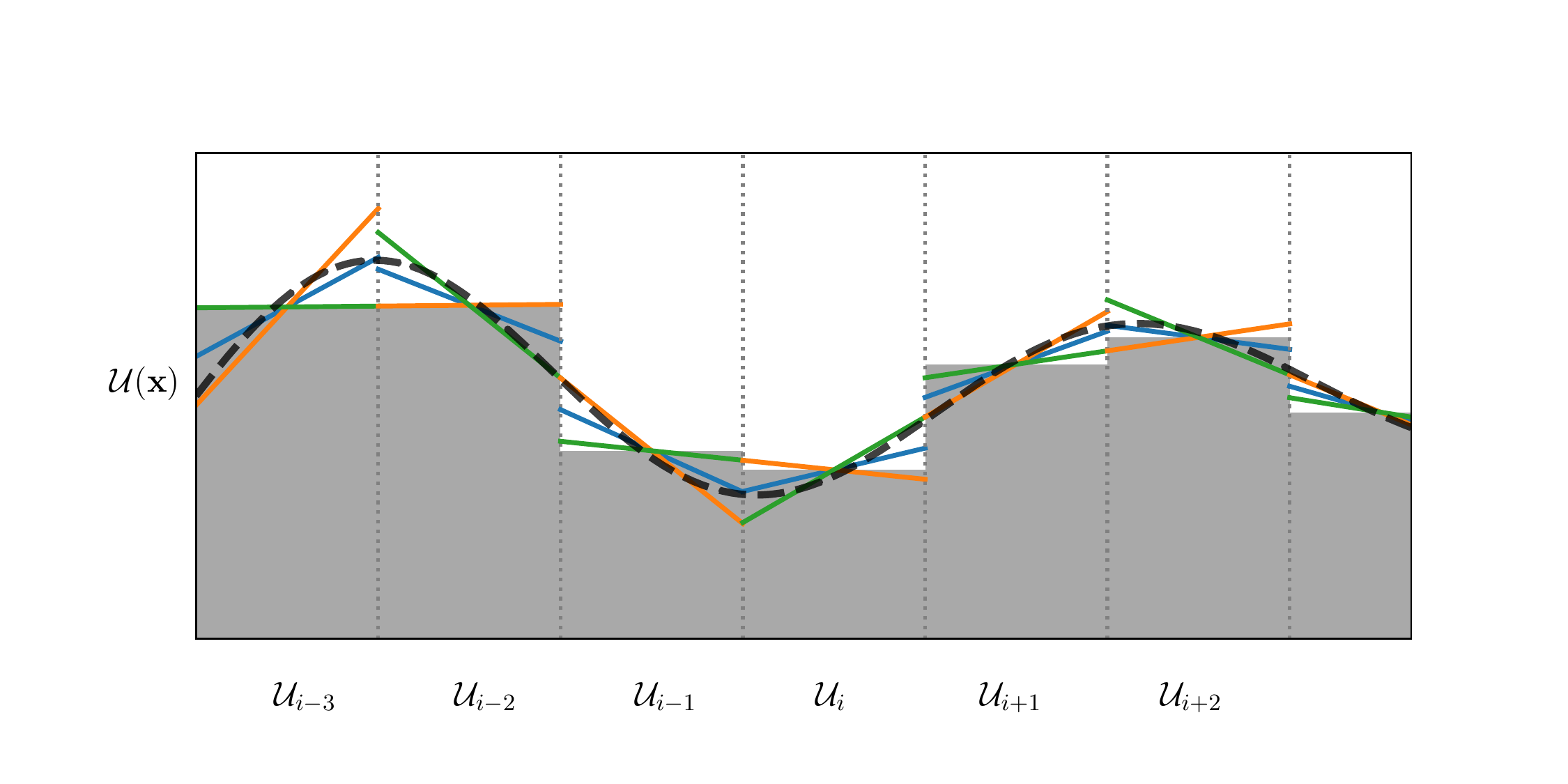}%
    \caption[Piece-wise linear reconstruction of data]{
Piece-wise linear reconstruction of the data (black dashed line), using different choices for the
slope. The blue lines show the reconstruction using a centered slope, the orange lines are using an
upwind slope, while the green lines use a downwind slope. The gray background represents the the
piece-wise constant reconstruction of the data.
    }
    \label{fig:piecewise-linear}
\end{figure}

The piece-wise linear reconstruction of the data using these three slope choices are illustrated in
Figure~\ref{fig:piecewise-linear}.

A complication that arises as a consequence of moving away from the piece-wise constant
reconstruction of the data is that the solution of the conventional Riemann problem centered at cell
boundaries is not applicable any longer. Instead, the problem at hand is a generalized Riemann
problem with non-constant left and right states $\uc_{i-1}(x)$ and $\uc_{i}(x)$, respectively.
The solution no longer contains uniform regions, and the characteristics are not straight lines any
longer. Unfortunately, the solution for generalized Riemann problems is not available for all
conservation laws, and further approximations need to be made in order to proceed. The
MUSCL-Hancock method prescribes the following approximation: First evolve the boundary extrapolated
values $\uc_L = \uc_i - \frac{\Delta x}{2} s_i$ and $\uc_R = \uc_i + \frac{\Delta x}{2} s_i$ over
half a time step by approximating the state \emph{inside} the cell $i$ as a conventional Riemann
problem with constant states $\uc_L$ and $\uc_R$, i.e. find

\begin{align}
    \overline{\uc}_L &= \uc_L + \frac{\Delta t}{2 \Delta x} (\fc(\uc_L) - \fc(\uc_R))
\label{eq:boundary-extrapolated-L}\\
    \overline{\uc}_R &= \uc_R + \frac{\Delta t}{2 \Delta x} (\fc(\uc_L) - \fc(\uc_R))
\label{eq:boundary-extrapolated-R}
\end{align}

These evolved boundary extrapolated values are then used to approximate the generalized Riemann
problem by as a conventional Riemann problem. Their evolved states serve as an approximation of the average state on the cell boundaries throughout the time step. Consequently the left and right evolved states $\overline{\uc}_{L}$ and $\overline{\uc}_{R}$ are used as the constant states between cell boundaries over the entire time step $\Delta t$. The final update formula is hence given by

\begin{align}
 \uc_i^{n+1} &= \uc_i^n + \frac{\Delta t}{\Delta x} (\fc_{i-\half} - \fc_{i+\half}) \\
 \fc_{i-\half} &= RP(\overline{\uc}_{i-1, R},\ \overline{\uc}_{i,L}) \\
 \fc_{i+\half} &= RP(\overline{\uc}_{i, R},\ \overline{\uc}_{i+1,L})
\end{align}

where $RP(l, r)$ represents the solution of the centered Riemann problem with left state $l$ and
right state $r$ at $x = 0$.

\subsection{On Monotonicity, Data Compatibility, and Total Variation}

\begin{figure}
    \centering
    \includegraphics[width=.33\textwidth]{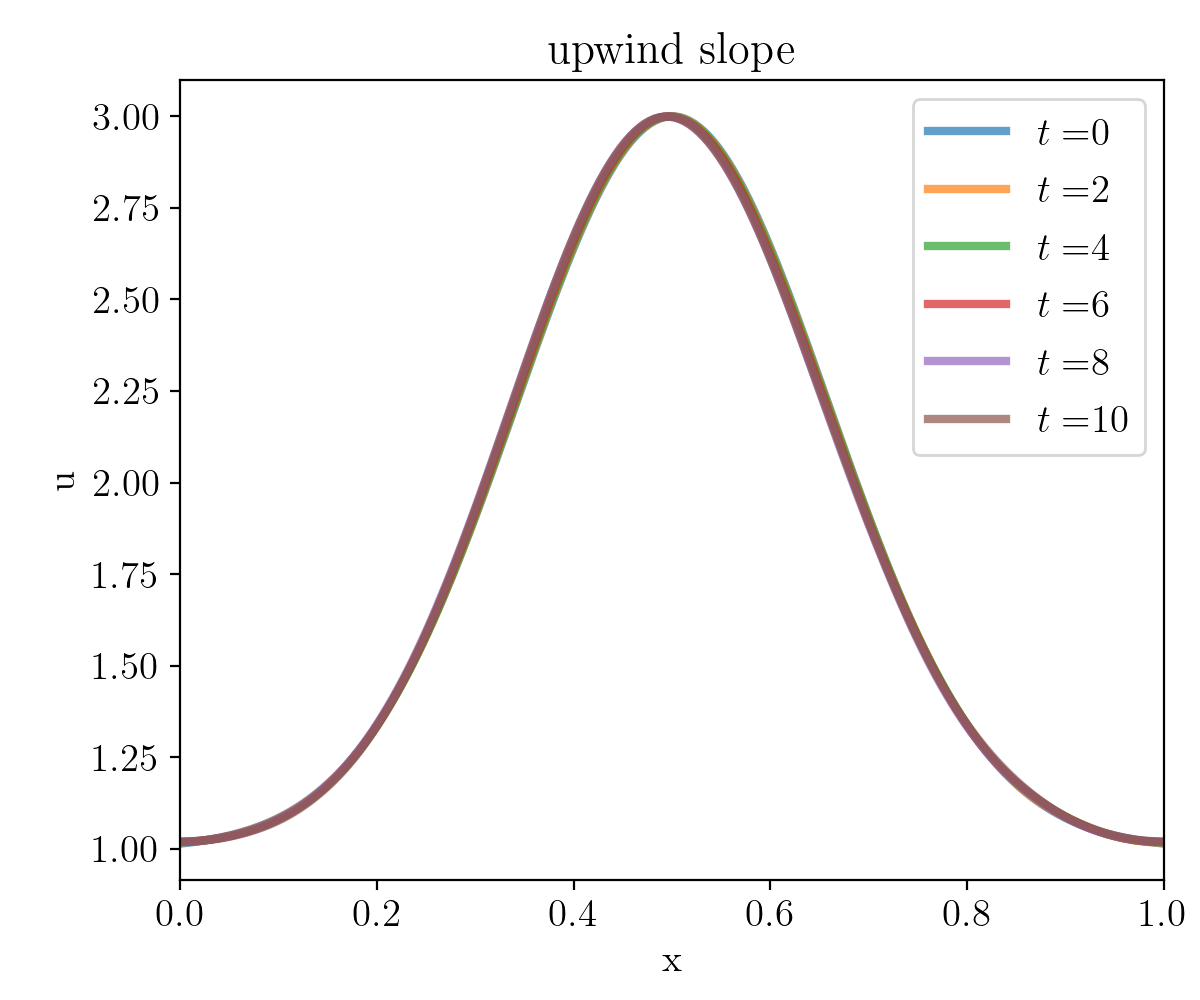}%
    \includegraphics[width=.33\textwidth]{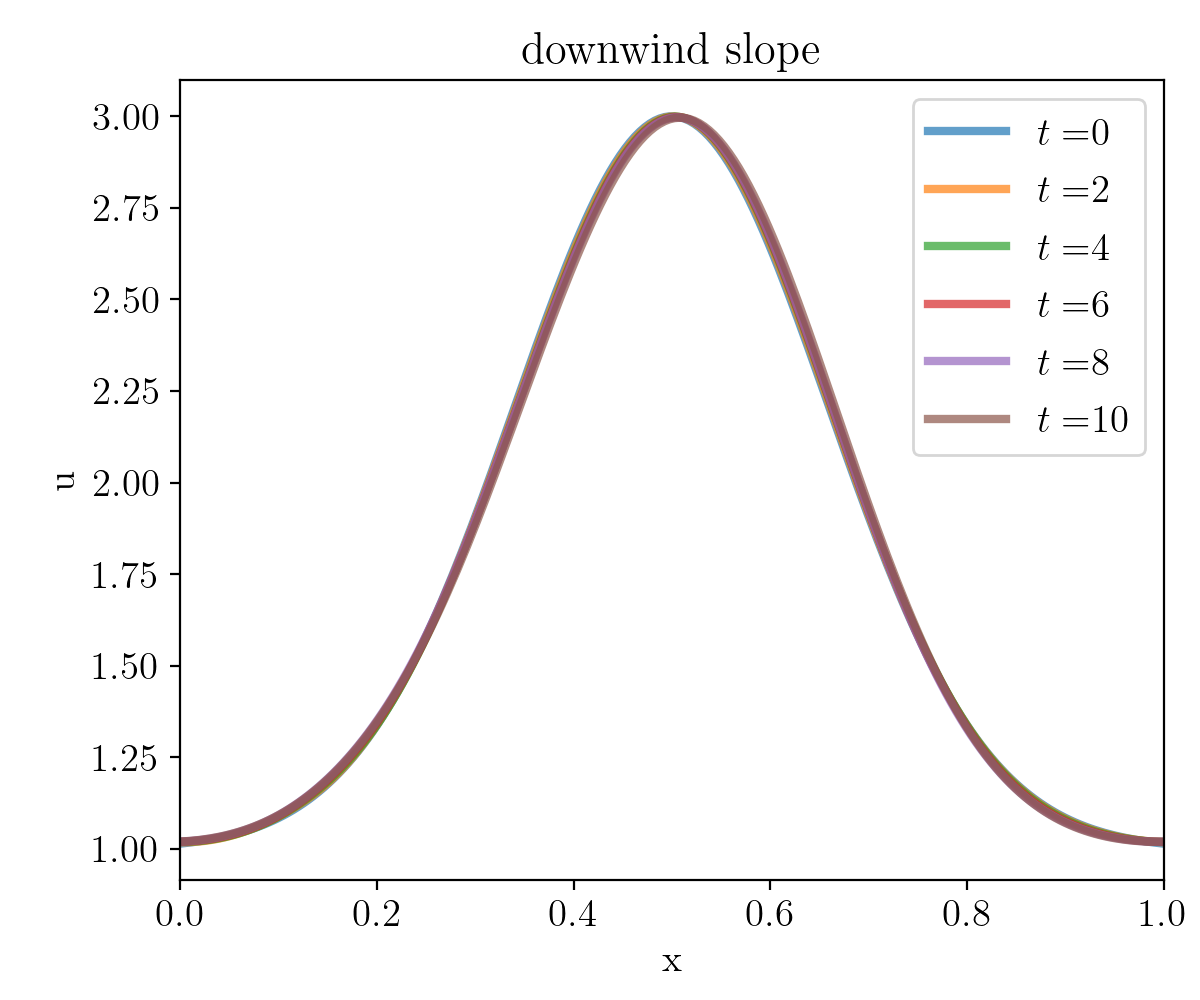}%
    \includegraphics[width=.33\textwidth]{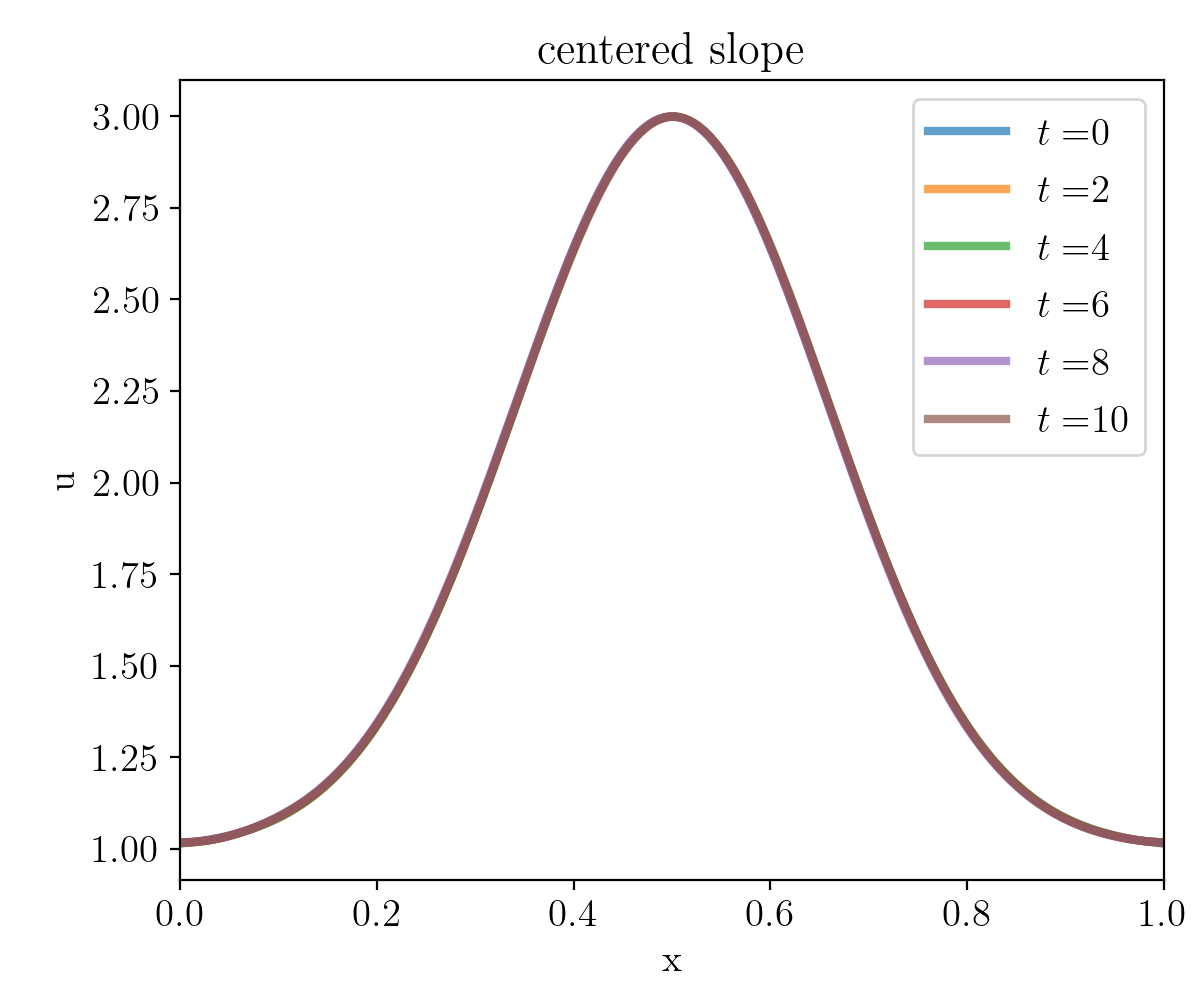}%
    \caption{
Solution of the linear advection equation using the MUSCL-Hancock method and an upwind (left)
slope, a downwind slope (center), and a centered slope (right) for a smooth Gaussian initial
state with advection coefficient $a = 1$.
    }
    \label{fig:pwlin-gaussian-no-limiter}
\end{figure}

\begin{figure}
    \centering
    \includegraphics[width=.33\textwidth]{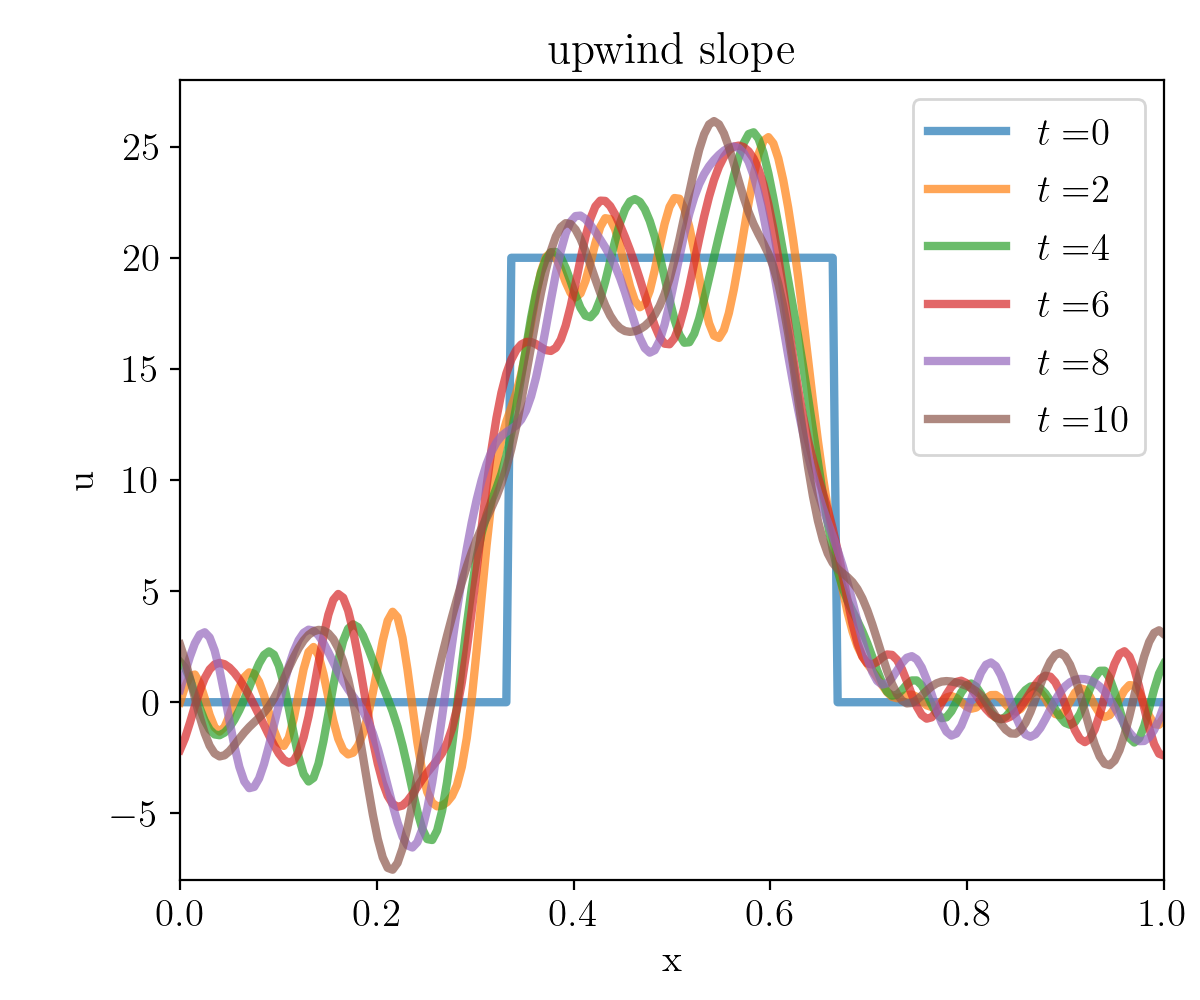}%
    \includegraphics[width=.33\textwidth]{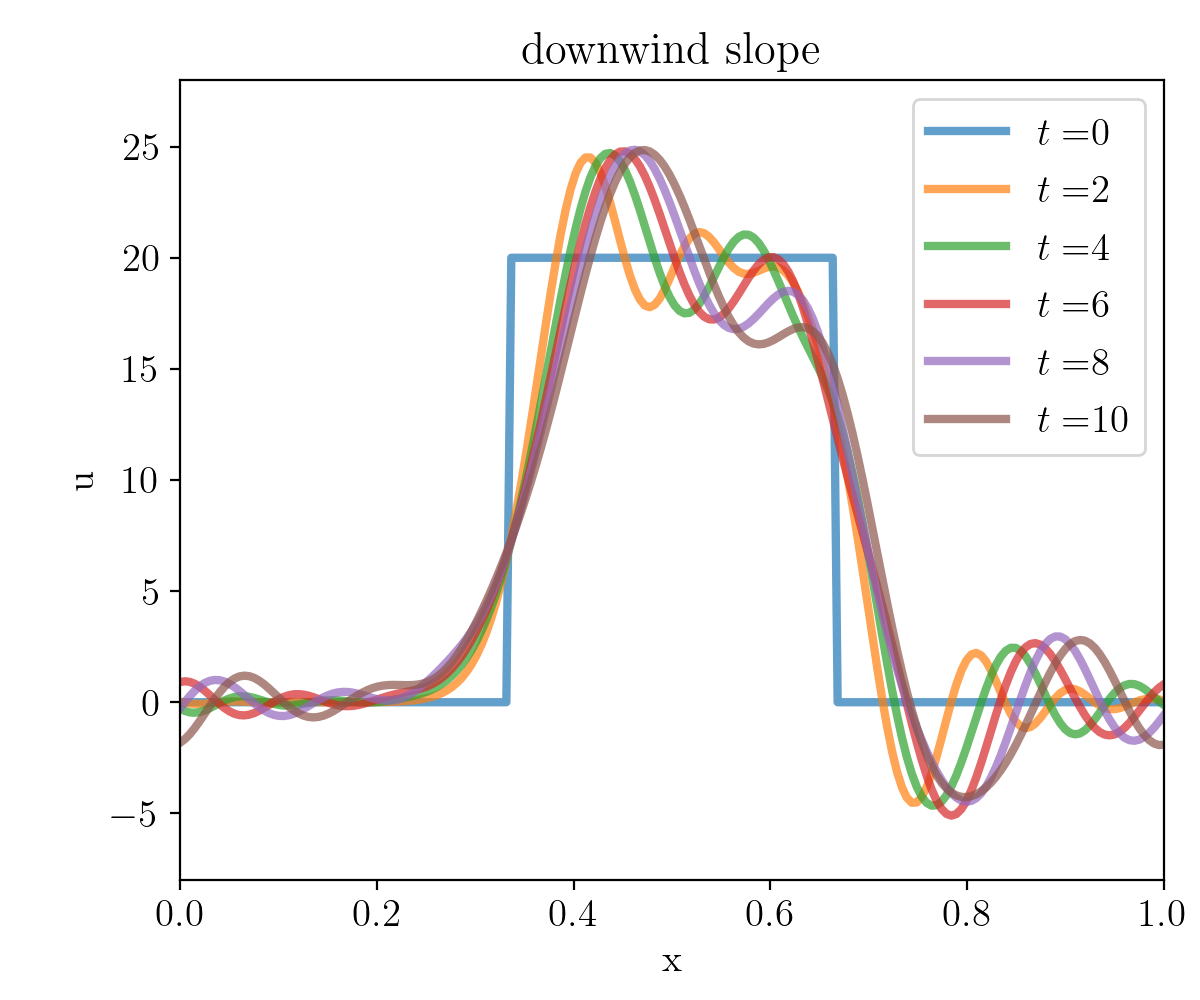}%
    \includegraphics[width=.33\textwidth]{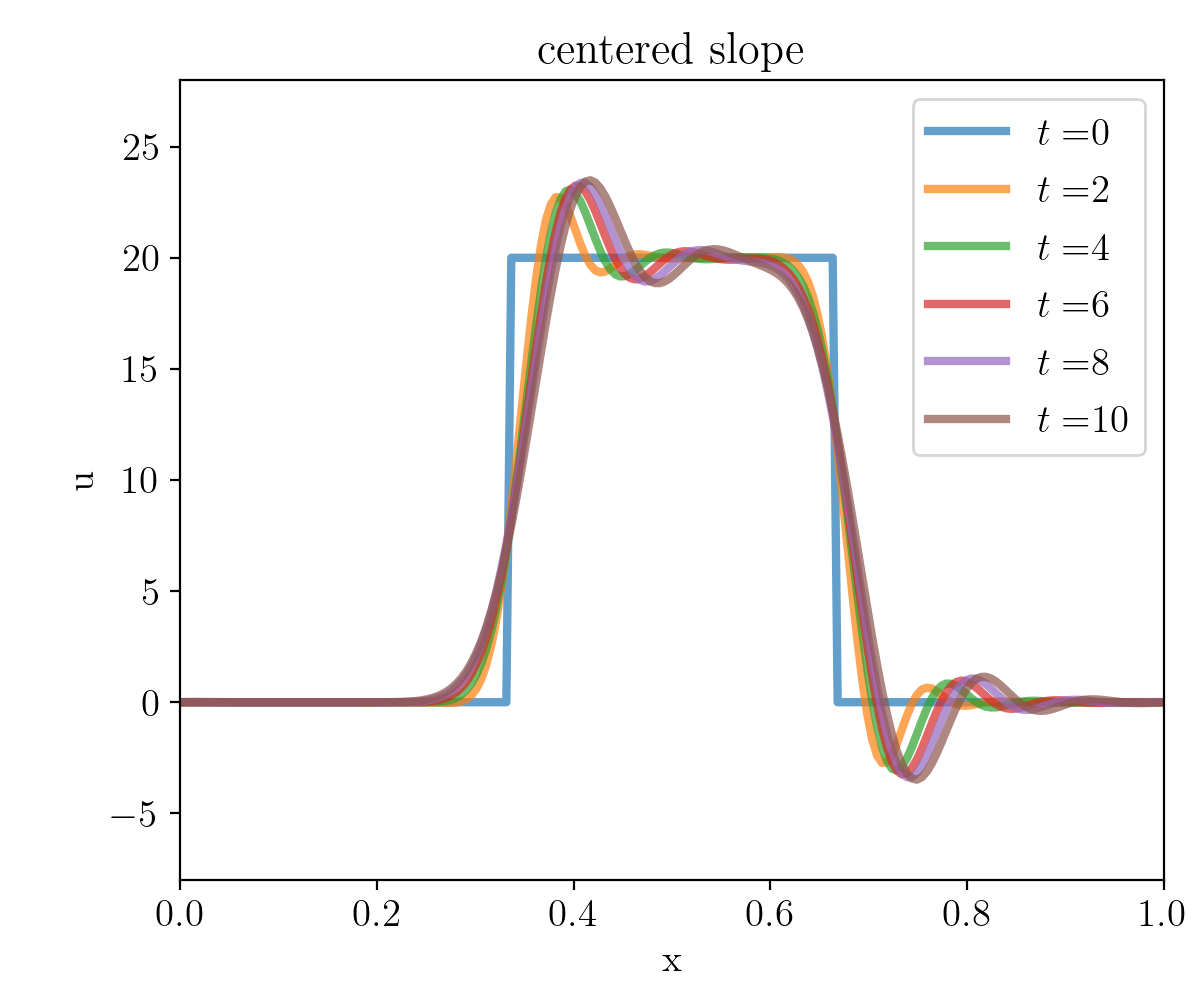}%
    \caption{
Solution of the linear advection equation using the MUSCL-Hancock method and an upwind (left)
slope, a downwind slope (center), and a centered slope (right) for step function initial
state with advection coefficient $a = 1$.
    }
    \label{fig:pwlin-step-no-limiter}
\end{figure}

Figure~\ref{fig:pwlin-gaussian-no-limiter} shows the results of the linear advection using the
MUSCL-Hancock method for the upwind, downwind, and the centered slope on smooth initial conditions.
Compared to the results of the first order method in Figure~\ref{fig:linear-advection-godunov}, it
is obvious that the solution has improved substantially. In this example there is no significant
diffusion any more, which is to be expected since the diffusion term entered as a second order error
in the first order accurate method. This cannot occur in the second order accurate method, as the
second order terms are handled explicitly. However, the situation changes drastically when
discontinuities are involved, as is shown in Figure~\ref{fig:pwlin-step-no-limiter}. The solution
now includes spurious oscillations that evolve and grow over time, which was not the case for the
first order method in Figure~\ref{fig:linear-advection-godunov}. Even worse, because the new
minima and maxima grow with time, they may keep growing to infinity, or other unphysical values.
Clearly they are a serious problem.

These oscillations are a typical phenomenon in higher order schemes. In fact, they are a
consequence of the extension to higher orders. For scalar non-linear conservation laws, this can be
shown by making use of the notion of monotone schemes. A scheme

\begin{align}
    \uc_i^{n+1} = M(\uc_{i-k_l}^n, \hdots, \uc_{i+k_R}^n) = \sum_{k=-k_L+1}^{k_R} b_k \uc_{i+k}^n
\label{eq:monotone-scheme}
\end{align}

with $k_L$, $k_R$ being two non-negative integers is called monotone if

\begin{align}
    \frac{\del M}{\del \uc_j^n} \geq 0 \quad \forall \ j
    \quad \Leftrightarrow \quad
    b_k \geq 0 \quad \forall \ j, k
\end{align}

Since $M$ is a non-decreasing function of all of its arguments, it is equivalent to the following
property:

\begin{align}
    \text{if } v_i^n \geq \uc_i^n \ \forall i, \quad \text{ then } \
    v_i^{n+1} = M(v_i^n) \geq \uc_i^{n+1} = M(\uc_i^n) \ \forall i \label{eq:monotone-2}
\end{align}

If a monotone scheme is applied on some given data set $\{\uc_i^n\}$, then the monotone scheme will
introduce no new minima or maxima, i.e.

\begin{align}
    \max_i \{ \uc_i^{n+1} \} \leq \max_i \{ \uc_i^n \} \label{eq:monotone-max} \\
    \min_i \{ \uc_i^{n+1} \} \geq \min_i \{ \uc_i^n \} \label{eq:monotone-min}
\end{align}

To demonstrate this property, we define $v_i = \max_j \{ u_j^n \} = \CONST \ \forall i$. From the
application of the scalar conservation law

\begin{align}
    \deldt v + \deldx \fc(v) = 0 = \deldt v + \frac{\del \fc}{\del v} \underbrace{\DELDX{v}}_{= 0}
\end{align}

it follows  that $\deldt v = 0$, and $v$ is constant w.r.t. time  as well. Since $v_i = \CONST
\ \forall i$, we can write

\begin{align}
    v_i^{n+1} = M(v_{i - k_L + 1}, \hdots, v_{i+k_R}) = M(v_i) = b_0 v_i
\end{align}

and because it must be constant in time as well, it follows that $b_0 = 1$ and $v_i^{n+1} = v_i^n$.
Using property~\ref{eq:monotone-2} and by definition of $v_i^n = \max_j \{ \uc_j^n \} $,
it follows that

\begin{align}
    v_i^{n+1} =
        v_i^n \geq u_i^{n+1} \quad
        &\Rightarrow \max_j \{ \uc_j^n \} \geq \uc_i^{n+1} \ \forall i,j \\
        &\Rightarrow \max_j \{ \uc_j^n \} \geq \max_j \{ \uc_j^{n+1} \}
\end{align}

The proof for the equivalent relation between minima follows analoguely. For any given time $t^n$
the properties~\ref{eq:monotone-max} and \ref{eq:monotone-min} can be applied recursively back to
$t^0$, which shows that no new minima and maxima will be created by monotone schemes regardless of
the number of time steps taken, and hence no spurious oscillations like in
Figure~\ref{fig:pwlin-step-no-limiter} will be generated. Conversely, it also means that monotone
schemes will clip extrema, as the minima will always increase, while maxima will always decrease.

Unfortunately, there are no monotone, linear schemes for non-linear scalar hyperbolic conservation
laws of second order of accuracy or higher. This fact is known as Godunov's theorem.  It can be
shown using Roe's theorem (see Appendix~\ref{app:roe}), which states that a scheme of the
form~\ref{eq:monotone-scheme} is
$p$-th order accurate if and only if

\begin{align}
    \sum_{k = -k_L}^{k_R} k^q b_k = (-C_{CFL})^q \quad , \quad 0 \leq q \leq p
\end{align}

where $b_k$ are the constant coefficients of the linear scheme, which must be $\geq 0$ for a
monotone scheme. Let

\begin{align}
    S_q \equiv \sum_{k = -k_L}^{k_R} k^q b_k
\end{align}

denote the summation for a single $q$.  Then for a scheme of second order accuracy, we require

\begin{align}
    S_0 = 1, && S_1 = -C_{CFL}, && S_2 = C_{CFL}^2
\end{align}

Explicitly computing $S_2$ reveals

\begin{align}
    S_2
    &= \sum_{k = -k_L}^{k_R} k^2 b_k
    = \sum_{k = -k_L}^{k_R} ([k + C_{CFL}]^2 - 2kC_{CFL} - C_{CFL}^2) b_k \\
    &= \sum_{k = -k_L}^{k_R} [k + C_{CFL}]^2  b_k
    - 2 C_{CFL} \underbrace{\sum_{k = -k_L}^{k_R} k b_k}_{= S_1 = -C_{CFL}}
    - C_{CFL}^2 \underbrace{\sum_{k = -k_L}^{k_R} b_k}_{= S_0 = +1} \\
    &= \sum_{k = -k_L}^{k_R} (k + C_{CFL})^2 b_k + C_{CFL}^2 \label{eq:godunov-theorem-S2}
\end{align}

In order for the scheme to be second order accurate, $S_2$ needs to be equal to $C_{CFL}^2$, which
is only satisfied in two cases. The first case is if $b_k = 0\ \forall k$, which isn't really a
method, since all it does is zero out any initial state. The other case is $C_{CFL} = -k \ \forall
k$. Since there can be only a single Courant number, there can also be only one coefficient index
$k = k_0$, and we need $b_k = 0 \ \forall k \neq k_0$. Furthermore the Courant number must be an
integer, which cannot be the case given the stability requirement $0 < C_{CFL} < 1$. It then
follows that a monotone scheme can't be second order accurate or higher.

If we nevertheless want to make use of schemes of higher order, some adjustments are necessary in
order to avoid spurious oscillations and instabilities around sharp gradients and discontinuities.
However, ideally the adjustments should only take effect close to sharp gradients and
discontinuities, while the second order accurate method should remain unmodified in smooth parts of
the problem. This way we can reap the most benefits from the higher order accuracy. Requiring the
method to behave differently depending on the current state of the problem, i.e. differently in
smooth and discontinuous regions, leads to the core of the approach: The method needs to adapt to
the current state of the problem. This means that a) the coefficients $b_k$ can't be constant any
longer, as they need to adapt according to the situation, and b) we need to find a way to quantify
when and how to modify the method. A way of illustrating the core concept would be to take a closer
look at the results for different choices of the slope in Figure~\ref{fig:pwlin-step-no-limiter}.
In Figure~\ref{fig:pwlin-step-no-limiter}, the advection coefficient is $a = 1$, and the initial
step function travels to the right. The solution using the centered slope for example does quite
well behind the step wave, i.e. at $x \sim 0.3$: there are no spurious oscillations there. But it
introduces a new peak directly in front of the discontinuities at $x \sim 0.4$ and $x \sim 0.7$.
The downwind slope behaves similarly. The upwind slope however deals a bit better with the state in
front of the step wave, i.e. at $x \sim 0.7$, albeit not perfectly. The idea behind an ``adaptive''
method could be to modify the reconstruction of the data by selecting between the different choices
of the slope and pick the ones that promise to give the best results depending on the region. For
example, we could select an upwind slope reconstruction in front of the wave, and a centered slope
behind the step wave. In regions where no linear reconstruction yields acceptable results, for
example around $x \sim 0.4$, we can always fall back on the monotone first order method
(Fig.~\ref{fig:linear-advection-godunov}) which never introduces new extrema.

It remains to establish concrete expressions for the requirements above. One way of ensuring that
a method doesn't develop new extrema is by demanding the algorithm to be ``data compatible''. A
scheme is called compatible with a data set $\{\uc_i^n\}$ if the solution $\{\uc_i^{n+1}\}$ is
bounded by the upwind pair $(\uc_{i-d}^n, \uc_i^n)$, where $d = \text{sign}(C_{CFL}) =
\text{sign}(a)$. This definition can also be written as

\begin{align}
    \min\{\uc_{i-d}^n, \uc_i^n\} \leq \uc_i^{n+1} \leq \max\{\uc_{i-d}^n, \uc_i^n\}
\end{align}

or alternatively

\begin{align}
    0 \leq \frac{\uc_i^{n+1} - \uc_i^n}{\uc_{i-d} - \uc_i^n} \leq 1 \label{eq:data-compatible}
\end{align}

A different way of ensuring that no new extrema form, nor existing extrema grow, is to demand that
the method is total variation diminishing. The total variation $TV$ for a mesh function is defined
as

\begin{align}
    TV(\uc^n) = \sum_i |\uc_{i+1}^n - \uc_i^n |
\end{align}

and a total variation diminishing (TVD) scheme satisfies

\begin{align}
    TV(\uc^{n+1}) \leq TV(\uc^{n}) \label{eq:TVD}
\end{align}

It can be shown that monotone schemes are TVD. An additional neat property of TVD schemes is that
they converge. Equipped with the constraints~\ref{eq:data-compatible} and~\ref{eq:TVD}, we can now
go on to find expressions for modified WAF and MUSCL-Hancock schemes which are both second order
accurate in smooth regions and don't develop spurious oscillations.

\subsection{TVD Version of the WAF Scheme}\label{chap:WAF-tvd-linear-advection}

To obtain a TVD version of the WAF scheme, we start off with a more general form of the WAF flux
(eq.~\ref{eq:WAF-flux}) for $a \geq 0$:

\begin{align}
    \fc_{i+\half} = \frac{1}{2} (1 + \phi_{i+\half})(a \uc_i^n) + \frac{1}{2} (1 - \phi_{i+\half})
(a \uc_{i+1}^n)
\end{align}

The goal is to find valid ranges for $\phi$ such that the method is TVD. A first upper and lower
limit for $\phi$ comes from the physical limitations in the extreme case $\phi = \pm 1$, which
corresponds to the characteristic having the maximally allowable velocity, therefore

\begin{align}
    -1 \leq \phi \leq 1
\end{align}

Applying the data compatibility condition~\ref{eq:data-compatible} and after some algebra we arrive
at the inequalities

\begin{align}
    -1 \leq \frac{1}{r_{i+\half}} (1 - \phi_{i+\half}) + \phi_{i-\half} \leq \frac{2 -
|C_{CFL}|}{|C_{CFL}|}
\end{align}

which is valid for both positive and negative $a$, and with the ``flow parameter'' $r_{i+\half}$

\begin{align}
    r_{i+\half} = \begin{cases}
                   \frac{\uc_i^n - \uc_{i-1}^n}{\uc_{i+1}^n - \uc_i^n} & \text{ if } a > 0 \\[6pt]
                   \frac{\uc_{i+2}^n - \uc_{i+1}^n}{\uc_{i+1}^n - \uc_i^n} & \text{ if } a < 0 \\
                  \end{cases}
\end{align}

\begin{figure}
    \centering
    \includegraphics[width=.6\textwidth]{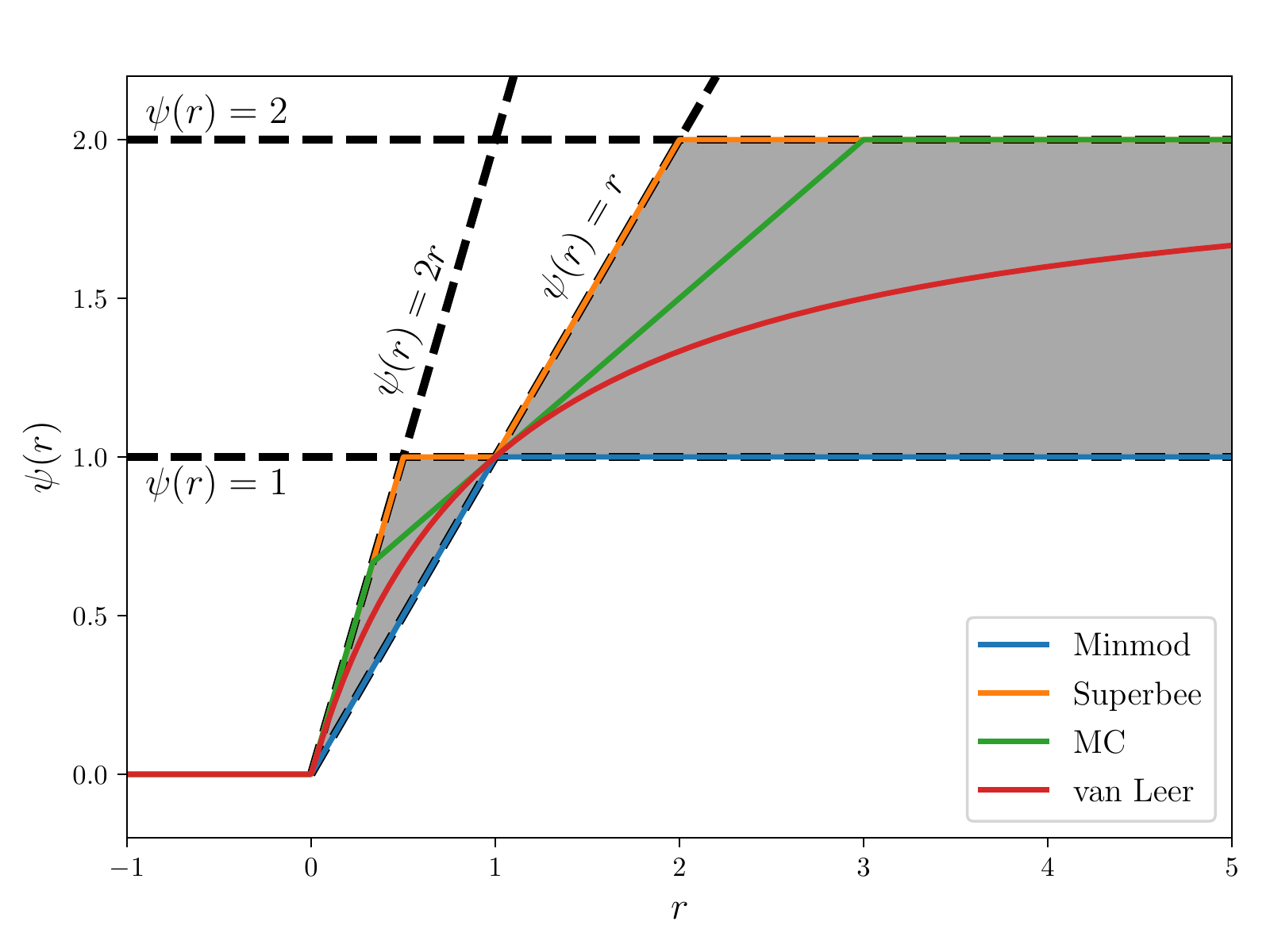}%
    \caption[Flux limiters and TVD region]{
Some popular flux limiter functions $\psi(r)$ for the flow parameter $r$. The grey region is the
entire permissible region for the flux limiters within which the inequalities guaranteeing the
method to be TVD are satisfied.
    }
    \label{fig:flux-limiters-TVD-region}
\end{figure}

Additionally, we impose that

\begin{align}
    \phi_{i\pm\half}(r = 1) = |C_{CFL}|
\end{align}

to ensure that the method remains second order accurate in smooth regions where the flow parameter
$r \sim 1$, i.e. where the upwind and downwind slope are nearly identical. As long as
$\phi_{i\pm\half}$ satisfies the above inequalities, the resulting method will be TVD. This means
however that there is any number of possible choices for the ``flux limiter'' $\phi(r)$ that
ensures that the scheme is TVD. To conform with a more general derivation of flux limited methods
by \citet{swebyHighResolutionSchemes1984}, it is practical to write the flux limiter in the form

\begin{align}
    \phi(r) = 1 - (1 - |C_{CFL}|) \psi(r)
\end{align}

to make use of well known expressions for $\psi(r)$ comparable with other methods.
The permissible values for $\psi(r)$ for which it satisfies the inequalities above can be
interpreted as regions on the $r - \psi(r)$ plane. These regions are are shown in
Figure~\ref{fig:flux-limiters-TVD-region} as the gray background, along with some well known limiter
functions:

\begin{flalign}
	\text{Minmod} 								&&\quad \psi(r) &= \mathrm{minmod}(1, r)\\
	\text{Superbee} 							&&\quad \psi(r) &= \max(0, \min(1, 2r), \min(2, r))
\\
	\text{MC (monotonized central-difference)} 	&&\quad \psi(r) &= \max(0, \min ((1+r)/2, 2, 2r))\\
	\text{van Leer}								&&\quad \psi(r) &= \frac{r + |r|}{1 + |r|}
\end{flalign}

where

\begin{align}
	\mathrm{minmod}(a, b) =
		\begin{cases}
			a	& \quad \text{ if } |a| < |b| \text{ and } ab > 0\\
			b	& \quad \text{ if } |a| > |b| \text{ and } ab > 0\\
			0	& \quad \text{ if } ab \leq 0\\
		\end{cases}
\end{align}

\begin{figure}
    \centering
\includegraphics[width=.7\textwidth]{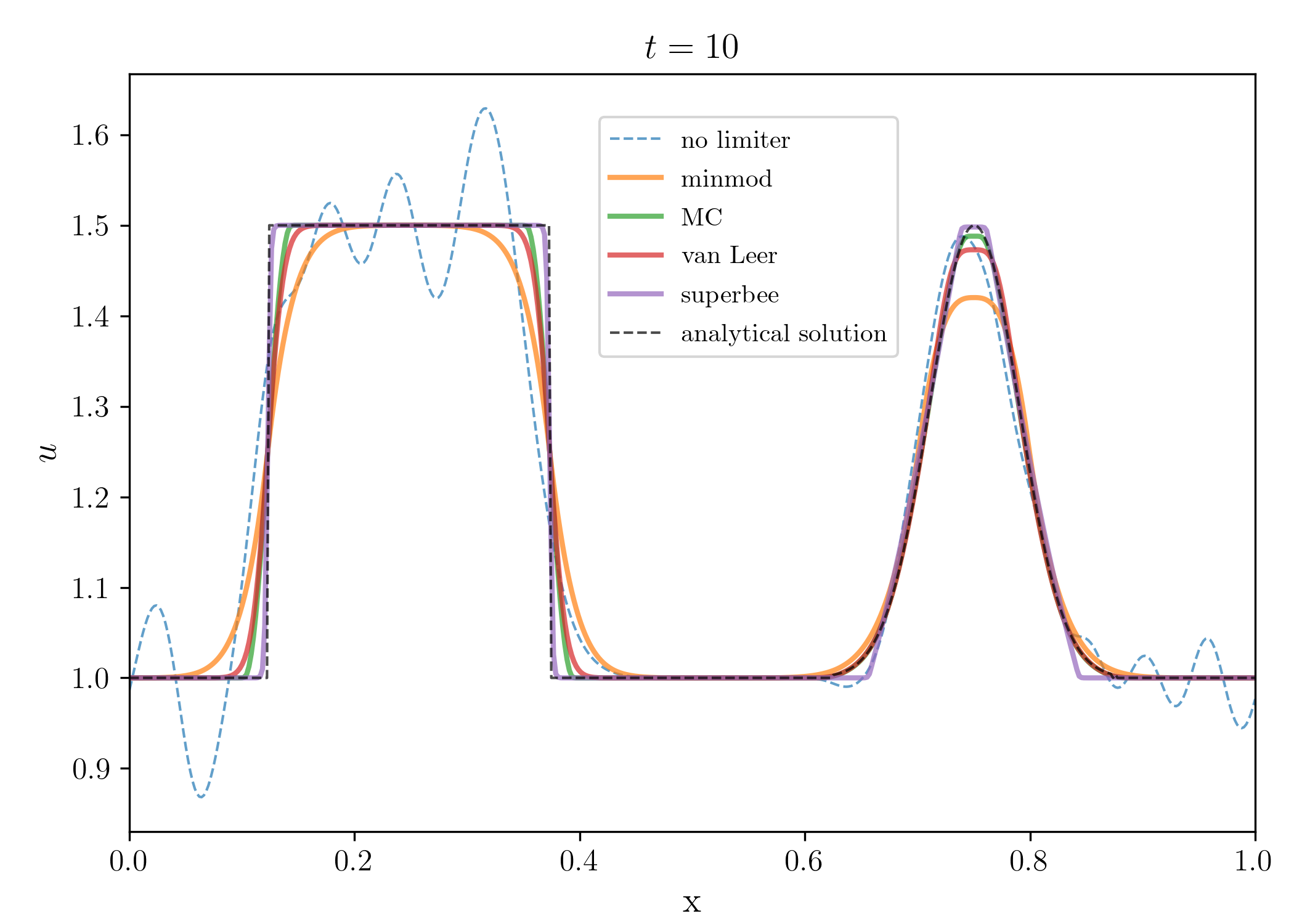}
    \caption[WAF scheme for linear advection with limiters]{
The results for linear advection using the WAF scheme and various flux limiters, compared to the
analytical solution (dashed black line) and the solution without limiters (dashed blue line) for a
step function (left) and a smooth Gaussian (right).
    }
    \label{fig:advection-WAF-solution}
\end{figure}

Figure~\ref{fig:advection-WAF-solution} shows the solution of the linear advection equation using
the WAF method and the various flux limiters described above. None of the solutions that employ a
flux limiter develop spurious oscillations, validating the flux limiting approach. The limiters
display a variety of diffusivity, comparable to the results of the first order method in
Figure~\ref{fig:linear-advection-godunov}. The minmod limiter is the most diffusive one among the
selected limiters, while the superbee limiter displays nearly no diffusivity around discontinuities
at all. However, the superbee limiter tends to clip minima and maxima in smooth regions as well,
and so the top of the Gaussian profile becomes flattened, while the smooth wings get clipped into
discontinuities. This different behavior suggests that there is no general best choice for which
flux limiter to use, and it should be decided based on the underlying problem to be solved.

\begin{figure}
    \centering
\includegraphics[width=.49\textwidth]{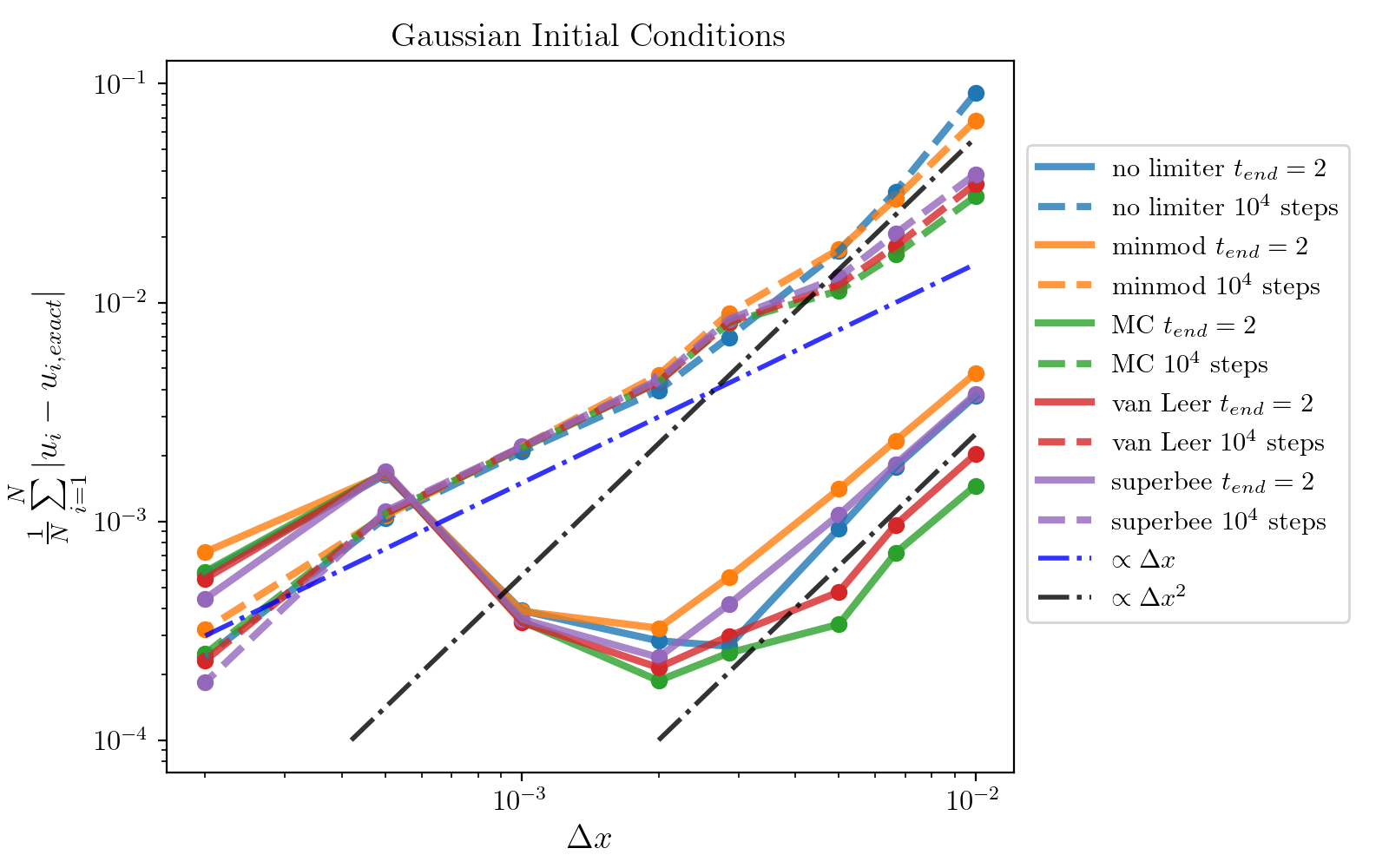}%
\includegraphics[width=.49\textwidth]{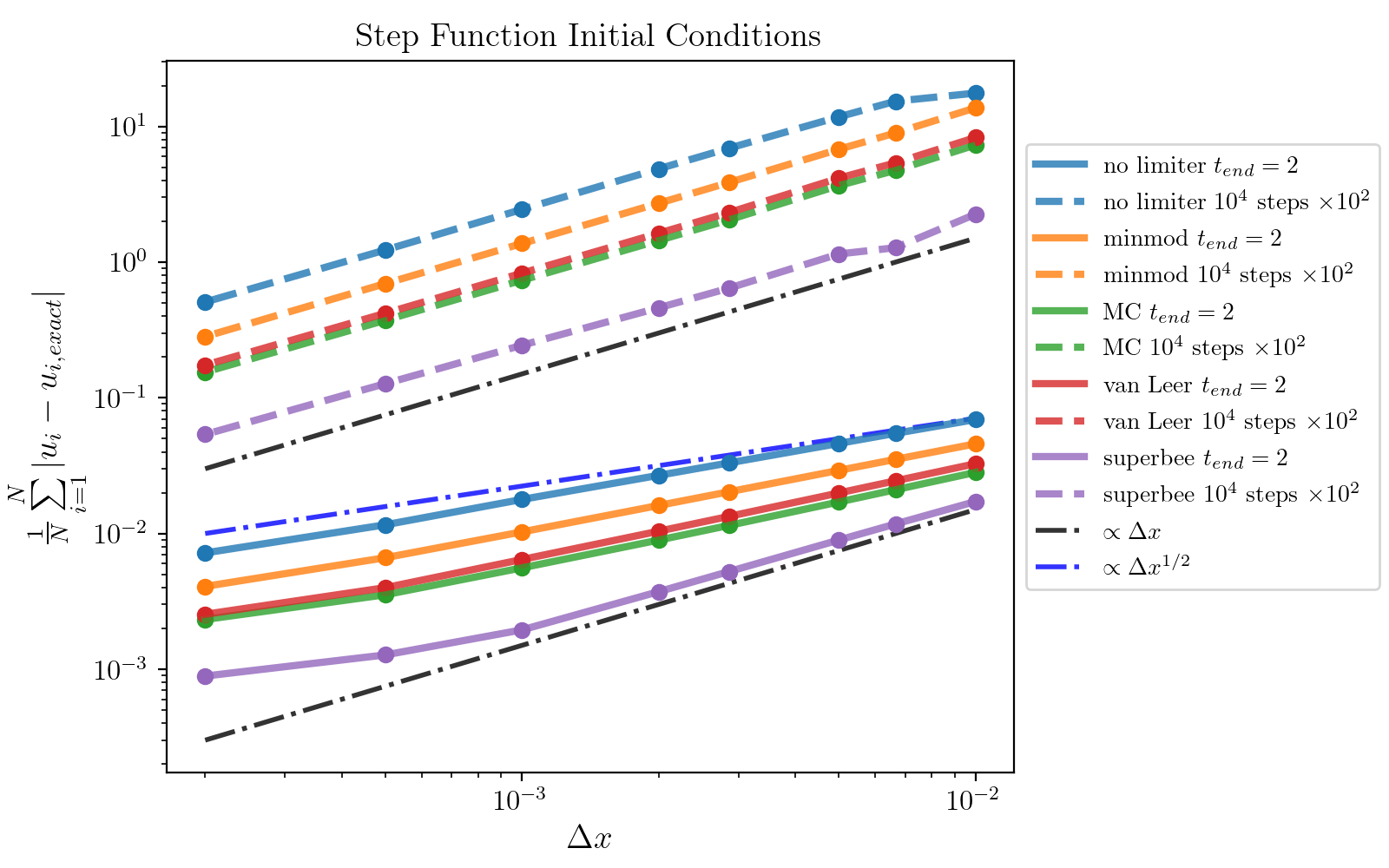}
    \caption[Order of Accuracy w.r.t. $\Delta x$ for the WAF scheme.]{
The order of accuracy w.r.t $\Delta x$ for the WAF method used for the linear advection equation
with a Gaussian initial condition (left) and a step function (right). The experiments are run for
both a fixed end time $t_{end} = 2$ as well as for fixed number of time steps individually, while
using varying flux limiters. Lines with slopes $1/2$, $1$, and $2$ are over-plotted for comparison.
The results for the step function after a fixed number of steps have been scaled by a factor of
$10^2$ for clarity.
    }
    \label{fig:advection-WAF-accuracy-dx}
\end{figure}

Finally, let's look at how the order of accuracy of the WAF scheme.
Figure~\ref{fig:advection-WAF-accuracy-dx} shows the average error using various limiters for
smooth Gaussian initial conditions and for a step function for varying $\Delta x$ while keeping
$C_{CFL}$ fixed. Again the errors are measured for both a fixed end time $t_{end} = 2$ of the
simulation, as well as after a fixed number of steps have been completed. For the fixed end time,
the error for the smooth Gaussian indeed decreases with $\propto \Delta x^2$, as the method
promises. However, it quickly converges at $\Delta x \sim 2 \times 10^3$ around the error $10^{-4}$
where it edges on the limits of single precision floating point numbers. After this point, any
further reduction of $\Delta x$ doesn't improve the solution. Since for a fixed $C_{CFL}$ a
decrease of $\Delta x$ means the identical increase in time step size $\Delta t$, decreasing
$\Delta x$ only leads to more time steps being required to reach the specified end time, and
accumulates more errors over more steps. The same applies to the case where the errors are compared
after a fixed number of time steps have been executed: After the point of convergence is reached,
the accuracy doesn't decrease proportional to $\Delta x^2$ any longer, and the order of accuracy is
reduced. While the order of $\Delta x^2$ for the fixed number of time steps is the same as for the
first order method in Figure~\ref{fig:linear-advection-accuracy} for the Gaussian, the order of
$\Delta x^2$ is a clear improvement over the order $\Delta x$ of the first order method. In
particular, note that the absolute value of the error estimate is lower compared to the first order
method by nearly a full order of magnitude even for large $\Delta x$.

In the case of the step function, the improvement in order of accuracy compared to the first order
Godunov method are evident as well. For the experiments run until a fixed end time, the errors
decrease with a slope between $\Delta x^{1/2}$ and $\Delta x$, while in the first order case the
trend was nearly exactly $\Delta x^{1/2}$. For the experiments run over a fixed number of time
steps, the slope of $\Delta x$ is the same in both cases because the presence of the discontinuity
dominates the error term. The fact that discontinuities reduce the order of accuracy of the solution
however persists, and the orders remain lower than in the case of the smooth Gaussian initial
condition.

\subsection{TVD Version of the MUSCL-Hancock Scheme}\label{chap:muscl-hancock-advection-tvd}

The condition for the MUSCL-Hancock scheme to be TVD can be written as

\begin{align}
    \min_j{\uc_j^n} \leq \uc_i^{n+1} \leq \max_j {\uc_j^n}
\end{align}

with $j \in [i-1, i+1]$.
This condition can be translated to a condition on the boundary extrapolated values
$\overline{\uc}_{i,L}$ and $\overline{\uc}_{i,R}$
(eqns.~\ref{eq:boundary-extrapolated-L}-\ref{eq:boundary-extrapolated-R}).

\begin{align}
    \min \{ \uc_{i-1}^n, \uc_i^n \} &\leq \overline{\uc}_{i,L} \leq \max \{ \uc_{i-1}^n, \uc_i^n \}
\quad \forall i \\
    \min \{ \uc_{i}^n, \uc_{i+1}^n \} &\leq \overline{\uc}_{i,R} \leq
    \max \{ \uc_{i}^n, \uc_{i+1}^n \} \quad \forall i
\end{align}

To derive more concrete expressions for the inequalities, let's assume that $\uc_{i-1}^n \leq
\uc_i^n$ to start with. Then the inequality reads as

\begin{align}
    \uc_{i-1}^n \leq \overline{\uc}_{i,L} = \uc_i^n - \frac{1}{2}(1 + C_{CFL}) \sigma_i \leq \uc_i^n
\end{align}

where the equality in the middle stems from the definition of $\overline{\uc}_{i,L}$ and $\sigma_i
= s_i \Delta x$. From the left hand side inequality, we find

\begin{align}
    \uc_{i-1}^n - \uc_{i}^n &\equiv -\Delta \uc_{i-\half} \leq - \frac{1}{2}(1 + C_{CFL}) \sigma_i \\
    \sigma_i &\leq \frac{2}{1 + C_{CFL}} \Delta \uc_{i-\half}
\end{align}

where we defined $\Delta \uc_{i-\half} = \uc_{i}^n - \uc_{i-1}^n$. From the second inequality, we
obtain

\begin{align}
    \uc_{i}^n - \frac{1}{2}(1 + C_{CFL}) \sigma_i &\leq \uc_i^n\\
    - \frac{1}{2}(1 + C_{CFL}) \sigma_i &\leq \uc_i^n - \uc_i^n = 0\\
    \sigma_i &\geq 0
\end{align}

Finally giving us

\begin{align}
 0 \leq \sigma_i \leq \frac{2}{1 + C_{CFL}} \Delta \uc_{i-\half} &&
 \text{ if } \Delta \uc_{i-\half} \geq 0
\end{align}

The same exercise can be repeated by assuming $\uc_i^n \leq \uc_{i-1}^n$, i.e. $\Delta
\uc_{i-\half} \leq 0$ to obtain

\begin{align}
 0 \geq \sigma_i \geq \frac{2}{1 + C_{CFL}} \Delta \uc_{i-\half} &&
 \text{ if } \Delta \uc_{i-\half} \leq 0
\end{align}

and by using the second inequality for $\overline{\uc}_{i,R}$ and again considering both cases
$\Delta \uc_{i+\half} \geq 0$ and $\Delta \uc_{i+\half} \leq 0$, two more conditions are found:

\begin{align}
 0 \leq \sigma_i &\leq \frac{2}{1 - C_{CFL}} \Delta \uc_{i+\half} &&
 \text{ if } \Delta \uc_{i+\half} \geq 0 \\
 0 \geq \sigma_i &\geq \frac{2}{1 - C_{CFL}} \Delta \uc_{i+\half} &&
 \text{ if } \Delta \uc_{i+\half} \leq 0
\end{align}

An immediate first result is that in case $\Delta \uc_{i-\half}$ and $\Delta \uc_{i+\half}$ have
opposite signs, the conditions reduce to $0 \geq s_i \leq 0$, and the only possible solution is
$s_i = 0$. If they have the same sign however, the above inequalities give us upper limits, which
can be combined into a single upper boundary:

\begin{align}
    \sigma_i \leq \min \left\{\frac{2}{1 - C_{CFL}} | \Delta \uc_{i+\half}|, \frac{2}{1 + C_{CFL}}
|\Delta \uc_{i-\half}| \right\}
\end{align}

To find methods that satisfy these conditions, the estimated slopes $\sigma_i$ of the MUSCL-Hancock
method are modified to slope limited slopes $\overline{\sigma}_i$ that contain a slope limiter
$\xi(r)$.
For a general slope $\sigma_i$ of the form

\begin{align}
    \sigma_i =
    \frac{1}{2} (1 + \omega) \Delta \uc_{i-\half} +
    \frac{1}{2} (1 - \omega) \Delta \uc_{i+\half}
\end{align}

the limited slope is given by

\begin{align}
    \overline{\sigma}_i  = \xi(r) \sigma_i
\end{align}

where $r$ is the flow parameter

\begin{align}
    r = \frac{\Delta \uc_{i-\half}}{\Delta \uc_{i+\half}}
\end{align}

and the TVD conditions can hence be written in terms of $\xi(r)$ as

\begin{align}
\xi(r) &= \begin{cases}
          0 & \text{ if }  r \leq 0 \\
          \leq \min \{ \xi_L, \xi_R \} & \text{ if } r > 0
         \end{cases} \\
\xi_L(r) &= \frac{\frac{4r}{(1 + C_{CFL})}}{1 - \omega + (1 + \omega) r} \\
\xi_R(r) &= \frac{\frac{4}{(1 - C_{CFL})}}{1 - \omega + (1 + \omega) r}
\end{align}

Once again any number of functions $\xi(r)$ that satisfy the above conditions are imaginable, and
hence many possible choices for slope limiters exist. Some popular limiters are:

\begin{flalign}
	\text{Minmod} 								&&\quad \xi(r) &=
					\begin{cases}
						0, & r \leq 0\\
						r, & 0 \leq r \leq 1\\
						\min\{1, \xi_R(r)\}, & r \geq 1\\
					\end{cases}
			\\
	\text{Superbee} 							&&\quad \xi(r) &=
						\begin{cases}
							0, & r \leq 0\\
							2r, & 0 \leq r \leq \frac{1}{2}\\
							1, & \frac{1}{2} \leq r \leq 1 \\
							\min\{r, \xi_R(r), 2\}, & r \geq 1\\
						\end{cases}
				\\
	\text{van Leer}								&&\quad \xi(r) &=
						\begin{cases}
							0, & r \leq 0\\
							\min\{\frac{2r}{1+r},  \xi_R(r)\}, & r \geq 0\\
						\end{cases}
\end{flalign}

Note that these slope limiters are not equivalent to the flux limiters bearing the same name, but
are analogous. For example, the Superbee limiters both follow the upper edge for of their
respective TVD regions, but are not identical expressions.

\section{Higher Order Schemes For The Euler Equations}

\subsection{The WAF Scheme}

\begin{figure}
    \centering
    \includegraphics[width=.8\linewidth]{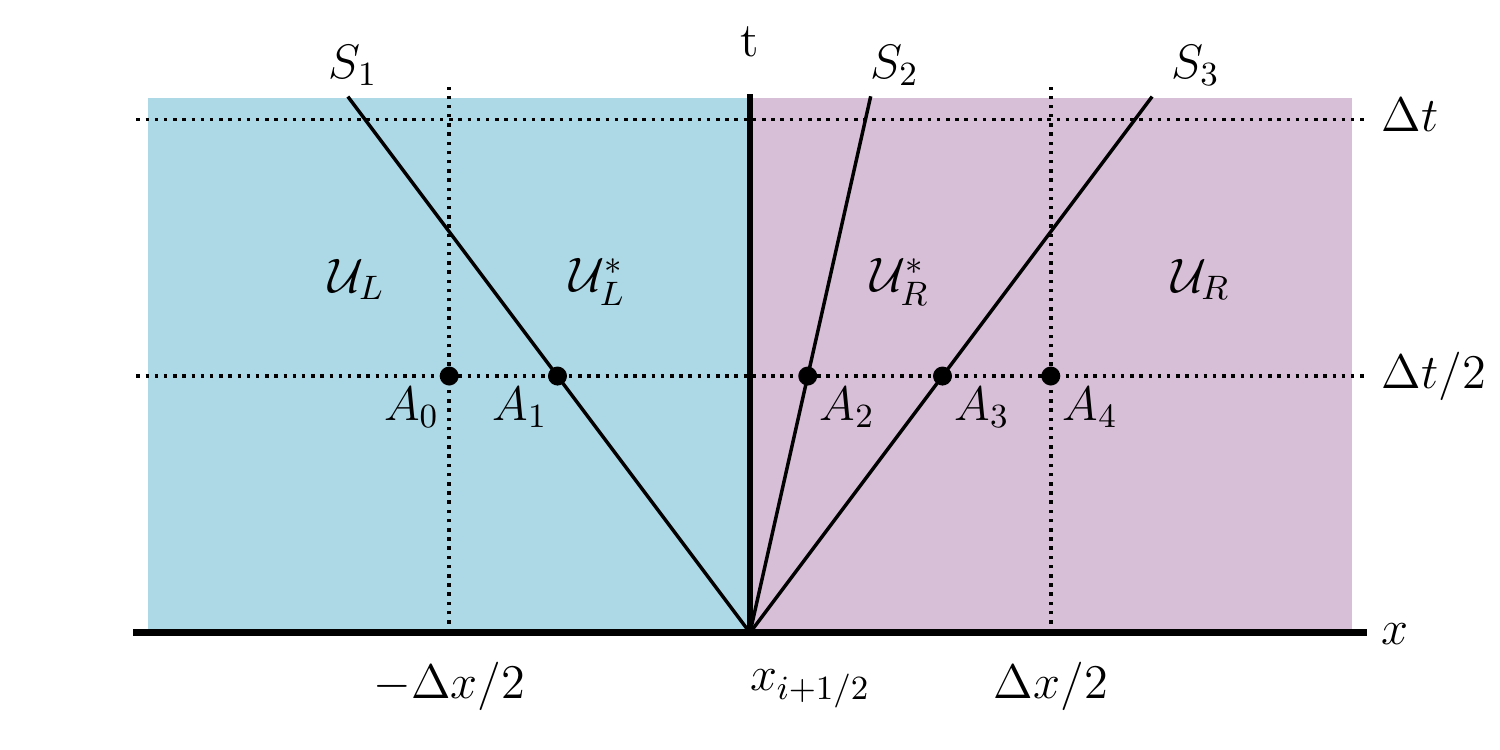}
    \caption[Setup for the WAF method for Euler equations]{
        Sketch for the set-up for the WAF method for Euler equations.
    }%
    \label{fig:hydro-WAF-setup}
\end{figure}

The WAF scheme for the Euler equations also estimates the fluxes at the midpoint in time by using
an integral average

\begin{align}
\F_{i+\half}^{WAF} = \frac{1}{\Delta x} \int\limits_{-\half \Delta x}^{\half \Delta x}
    \F(\U_{i+\half}) \de x \label{eq:WAF-flux-euler}
\end{align}

where $\U_{i+\half}$ is the solution of the Riemann problem with piece-wise constant data $\U_i^n$
and $\U_{i+1}^n$ centered at the position $x = x_{i+\half}$. Instead of a single wave emanating as
the solution of the Riemann problem for the linear advection equation, for the Euler equations we
obtain three waves, which separate the $x-t$ plane into 4 regions $\U_L$, $\U_L^*$, $\U_R^*$, and
$\U_R$. Naming the points $(x_k,\Delta t/2)$ plane where the characteristics of the three waves $k$
are situated at the time $\Delta t / 2$ as $A_k$, (see Figure~\ref{fig:hydro-WAF-setup}), and
choosing the indices $k$ such that the wave velocities $S_1 < S_2 < S_3$, then the distances
between these points are given by

\begin{align}
    \overline{A_0 A_1} &= \frac{\Delta x}{2} (1 + C_1) \\
    \overline{A_1 A_2} &= (C_2 - C_1) \frac{\Delta x}{2} \\
    \overline{A_2 A_3} &= (C_3 - C_2) \frac{\Delta x}{2} \\
    \overline{A_3 A_4} &= (1 - C_3) \frac{\Delta x}{2} \\
    C_k &\equiv \frac{S_k \Delta t}{\Delta x}
\end{align}

The integral~\ref{eq:WAF-flux-euler} can then be performed separately between each two adjacent
points $A_k$, $A_{k+1}$. Rarefaction waves require some special attention, since they contain a
fan between the head and the tail waves, which is composed of a smooth transition, not a constant
state. While technically an analytical solution to the integral over the rarefaction fan can be
found, experience shows that an approximation is sufficiently accurate. The approximation is to
simply ``lump'' the fan together with the closest adjacent constant state, where ``closest'' is
with regards to the distance from the center, i.e. the $t$-axis. This special treatment for
rarefactions is not necessary for HLL type Riemann solvers, whose solution is the flux itself, not
the individual states.

The WAF flux then takes the form

\begin{align}
    \F_{i+\half}^{WAF} &= \sum_{k=1}^4 \frac{1}{2} (C_k - C_{k-1}) \F^{(k)} \\
    &= \frac{1}{2} \left( \F_i + \F_{i+1} \right) -
    \frac{1}{2} \sum_{k=1}^3 C_k (\F^{(k+1)} - \F^{(k)})
\end{align}

A TVD modification of the WAF flux is

\begin{align}
    \F_{i+\half}^{WAF} &=
        \frac{1}{2} \left( \F_i + \F_{i+1} \right) -
        \frac{1}{2} \sum_{k=1}^3 \mathrm{sign}(C_k) \phi_{i+\half}^{(k)} (\F^{(k+1)} - \F^{(k)})
\end{align}

where

\begin{align}
    \phi_{i+\half} ^{(k)} &= \phi_{i+\half}(r^{(k)}) \\
    r^{(k)} &= \begin{cases}
                \frac{\Delta q_{i-\half}^{(k)}}{\Delta q_{i+\half}^{(k)}} & \text{ if } C_k > 0 \\
                \frac{\Delta q_{i+3/2}^{(k)}}{\Delta q_{i+\half}^{(k)}} & \text{ if } C_k < 0
               \end{cases}
\end{align}

$q$ is a single quantity which is known to change across each wave $k$, for example density or
internal energy. For the limiters $\phi_{i+\half}$, the same expressions as in
Section~\ref{chap:WAF-tvd-linear-advection} may be used.
This extension for to a TVD form of the WAF method for Euler equations is somewhat empirical, and
there is no rigorous proof that it will actually remain TVD in every conceivable case, but is found
to work well in practice. The reason is that in order to demonstrate the TVD properties of the
scheme for the linear advection, we made heavy use of the analytical solution of the linear
advection equation. Doing the same for Euler equations while covering all possible outcomes for any
initial conditions for the Riemann problems that need to be solved is a daunting task, and so we
need to rely on the findings in the simpler case of the linear advection.

\subsection{The MUSCL-Hancock Scheme}

The MUSCL-Hancock Scheme for Euler equations follows the same steps as the one for linear
advection, with the difference that instead of a single scalar quantity $\uc$, now a state vector
$\U$ is being represented through piece-wise linear reconstructions of the field, i.e.

\begin{align}
    \U_i = \U_i(x) = \U_i^n + \frac{x - x_i}{\Delta x} S_i \quad, \quad x \in [0, \Delta x]
\end{align}

where $S_i$ is now a vector containing a slope of each component of $\U$. The values of $\U_i$ at
the boundaries $x_{i \pm \half}$ are analoguely given by

\begin{align}
    \U_L = \U_i^n - \frac{1}{2} S_i && \U_R = \U_i^n + \frac{1}{2} S_i
\end{align}

Again the boundary extrapolated values are evolved for half a time step using

\begin{align}
    \overline{U}_L &= \U_{i,L} + \frac{1}{2}\frac{\Delta t}{\Delta x} [\F(\U_{i,L})- \F(\U_{i,R})]\\
    \overline{U}_R &= \U_{i,R} + \frac{1}{2}\frac{\Delta t}{\Delta x} [\F(\U_{i,L})- \F(\U_{i,R})]\\
\end{align}

These evolved boundary extrapolated values are then used to approximate the generalized Riemann
problem by solving a conventional Riemann problem with constant left and right evolved states
$\U_{L}$ and $\U_{R}$ between cell boundaries over the entire time step $\Delta t$. The final
update formula is hence given by

\begin{align}
 \U_i^{n+1} &= \U_i^n + \frac{\Delta t}{\Delta x} (\F_{i-\half} - \F_{i+\half}) \\
 \F_{i-\half} &= RP(\overline{\U}_{i-1, R},\ \overline{\U}_{i,L}) \\
 \F_{i+\half} &= RP(\overline{\U}_{i, R},\ \overline{\U}_{i+1,L})
\end{align}

where $RP(l, r)$ represents the solution of the centered Riemann problem with left state $l$ and
right state $r$ at $x = 0$. The TVD version of the method is obtained by limiting the slopes $S_i$
for each component individually in the same manner it is done for the linear advection equation in
Section~\ref{chap:muscl-hancock-advection-tvd}: The slope $S_i$ is replaced by the slope limited
version $\overline{S}_i$

\begin{align}
    S_i^k = \xi_i^k S_i^k
\end{align}

for each component $k$ of the state vector $\U$. The same slope limiters as listed in
Section~\ref{chap:muscl-hancock-advection-tvd} may be employed.

\begin{figure}
   \centering
   \includegraphics[width=\linewidth]{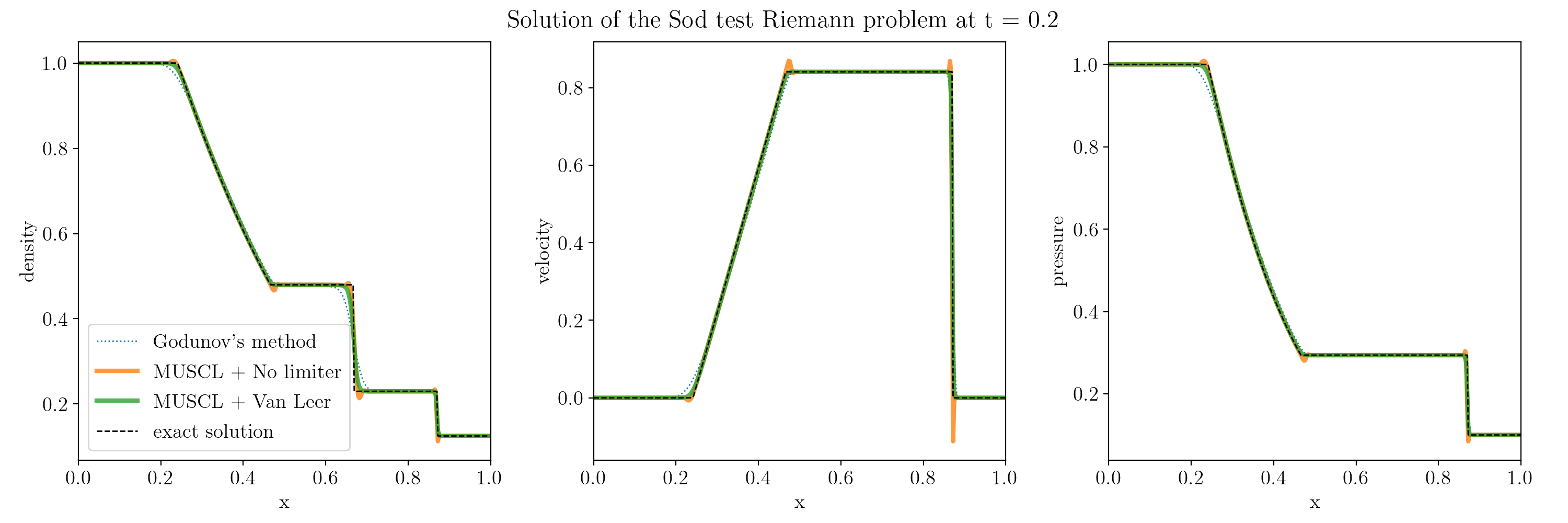}
   \caption[Sod test using MUSCL-Hancock method]{
The solution to the Sod test (eq. \ref{eq:sod-test-ICs}) using Godunov's method, the MUSCL-Hancock
method with a limiter, and the MUSCL-Hancock method with the Van Leer limiter.
}
    \label{fig:MUSCL-sod-test}
\end{figure}

\begin{figure}
   \centering
\includegraphics[width=\linewidth]{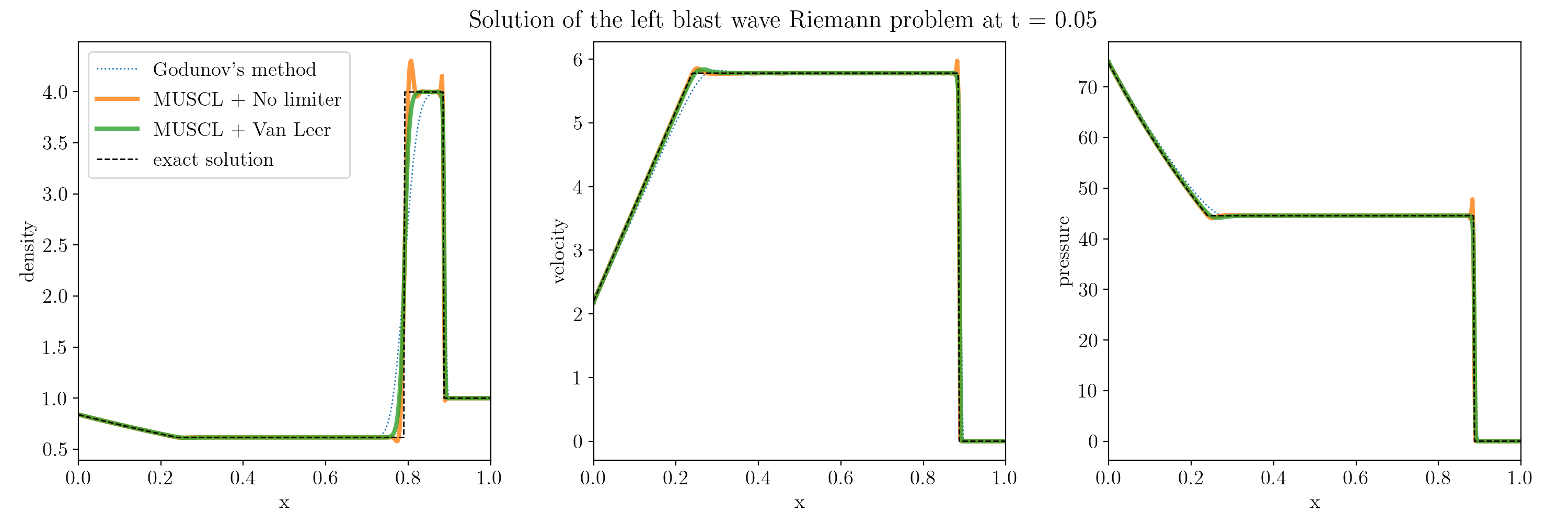}
   \caption[Left Blast Wave test using MUSCL-Hancock method]{
The solution to the Left Blast Wave test (eq. \ref{eq:left-blast-wave-ICs}) using Godunov's method,
the MUSCL-Hancock method with a limiter, and the MUSCL-Hancock method with the Van Leer limiter.
}
   \label{fig:MUSCL-left-blast-wave}
\end{figure}

\begin{figure}
   \centering
   \includegraphics[width=\linewidth]{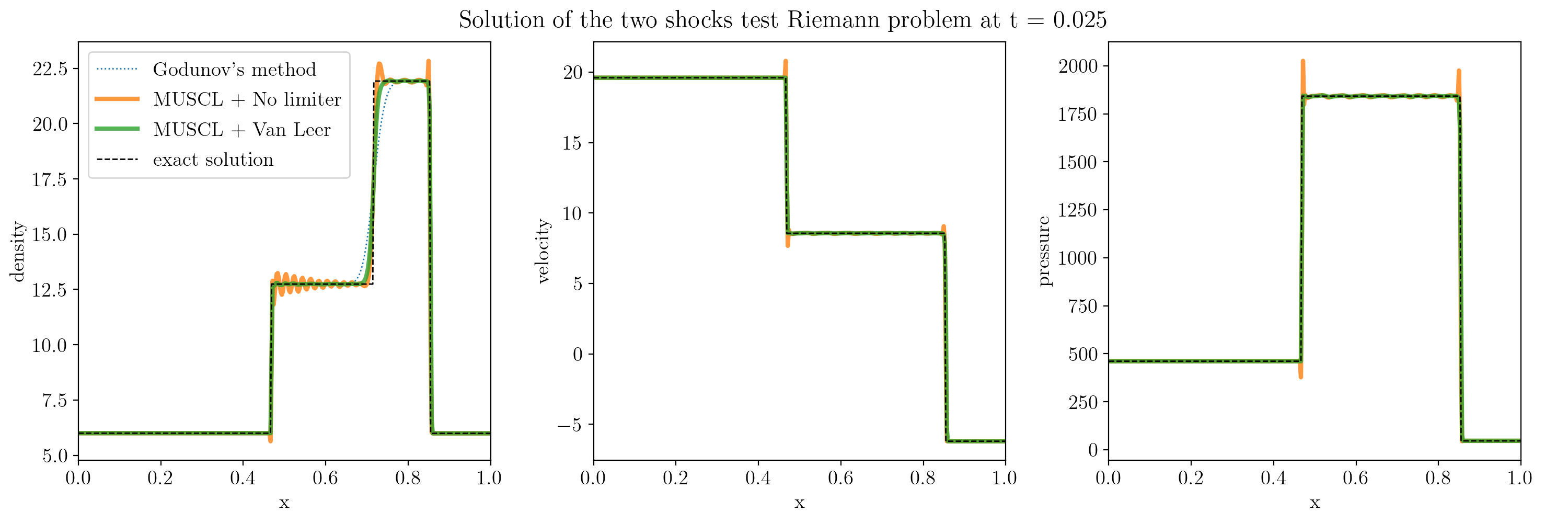}
   \caption[Two Shocks test using MUSCL-Hancock method]{
The solution to the Two Shocks test (eq. \ref{eq:two-shock-ICs}) using Godunov's method, the
MUSCL-Hancock method with a limiter, and the MUSCL-Hancock method with the Van Leer limiter.
}
    \label{fig:MUSCL-two-shocks}
\end{figure}

To showcase the improvements that follow using a second order method with limiters
for the Euler equations, Figures~\ref{fig:MUSCL-sod-test},~\ref{fig:MUSCL-left-blast-wave},
and~\ref{fig:MUSCL-two-shocks} show the solution of various Riemann problems using the MUSCL-Hancock
method and Godunov's method. For clarity, only the results using the Van Leer slope limiter are
shown, along with the solution when using no limiter at all. In all cases, the better treatment of
discontinuities compared to Godunov's first order method are obvious, and the spurious oscillations
visible in the solution without a slope limiter disappear as well.

The difference in results between a first and second order scheme can perhaps be better illustrated
using a two-dimensional example where mixing of several fluid phases occurs. A fitting example is
the Kelvin-Helmholtz instability, which arises when two fluid phases shear past each other on a
slightly perturbed interface. The initial conditions are taken from
\citet{mcnallyWellposedKelvinHelmholtzInstability2012} and prescribe the following density inside a
periodic two dimensional box of size unity in each dimension:

\begin{align}
\rho(x,y) = \begin{cases}
            \rho_1 - \overline{\rho} \exp \left( \frac{y - 1/4}{L} \right) &
                \text{ if } 0 \leq y < 1/4 \\
            \rho_2 + \overline{\rho} \exp \left( \frac{1/4 - y}{L} \right) &
                \text{ if } 1/4  \leq y < 1/2 \\
            \rho_2 + \overline{\rho} \exp \left( \frac{y - 3/4}{L} \right) &
                \text{ if } 1/2  \leq y < 3/4 \\
            \rho_1 - \overline{\rho} \exp \left( \frac{3/4 - y}{L} \right) &
                \text{ if } 3/4 \leq y \leq 1
            \end{cases}
\end{align}

where $\overline{\rho} = (\rho_1 - \rho_2) / 2$, $\rho_1 = 1$, $\rho_2 = 2$, and $L = 0.025$. To
obtain a shearing motion of the phases, the velocity of the fluid in $x$ direction is set up as

\begin{align}
v_x(x,y) = \begin{cases}
            v_1 - \overline{v} \exp \left( \frac{y - 1/4}{L} \right) &
                \text{ if } 0 \leq y < 1/4 \\
            v_2 + \overline{v} \exp \left( \frac{1/4 - y}{L} \right) &
                \text{ if } 1/4  \leq y < 1/2 \\
            v_2 + \overline{v} \exp \left( \frac{y - 3/4}{L} \right) &
                \text{ if } 1/2  \leq y < 3/4 \\
            v_1 - \overline{v} \exp \left( \frac{3/4 - y}{L} \right) &
                \text{ if } 3/4 \leq y \leq 1
            \end{cases}
\end{align}

where again $\overline{v} = (v_1 - v_2) / 2$ and $v_1 = 0.5$, $v_2 = -0.5$. Finally, the small
perturbation in the shear is added by giving a small velocity in $y$ direction:

\begin{align}
    v_y(x,y) = 0.01 \sin(4\pi x)
\end{align}

\begin{figure}
\centering
\includegraphics[width=.5\linewidth]{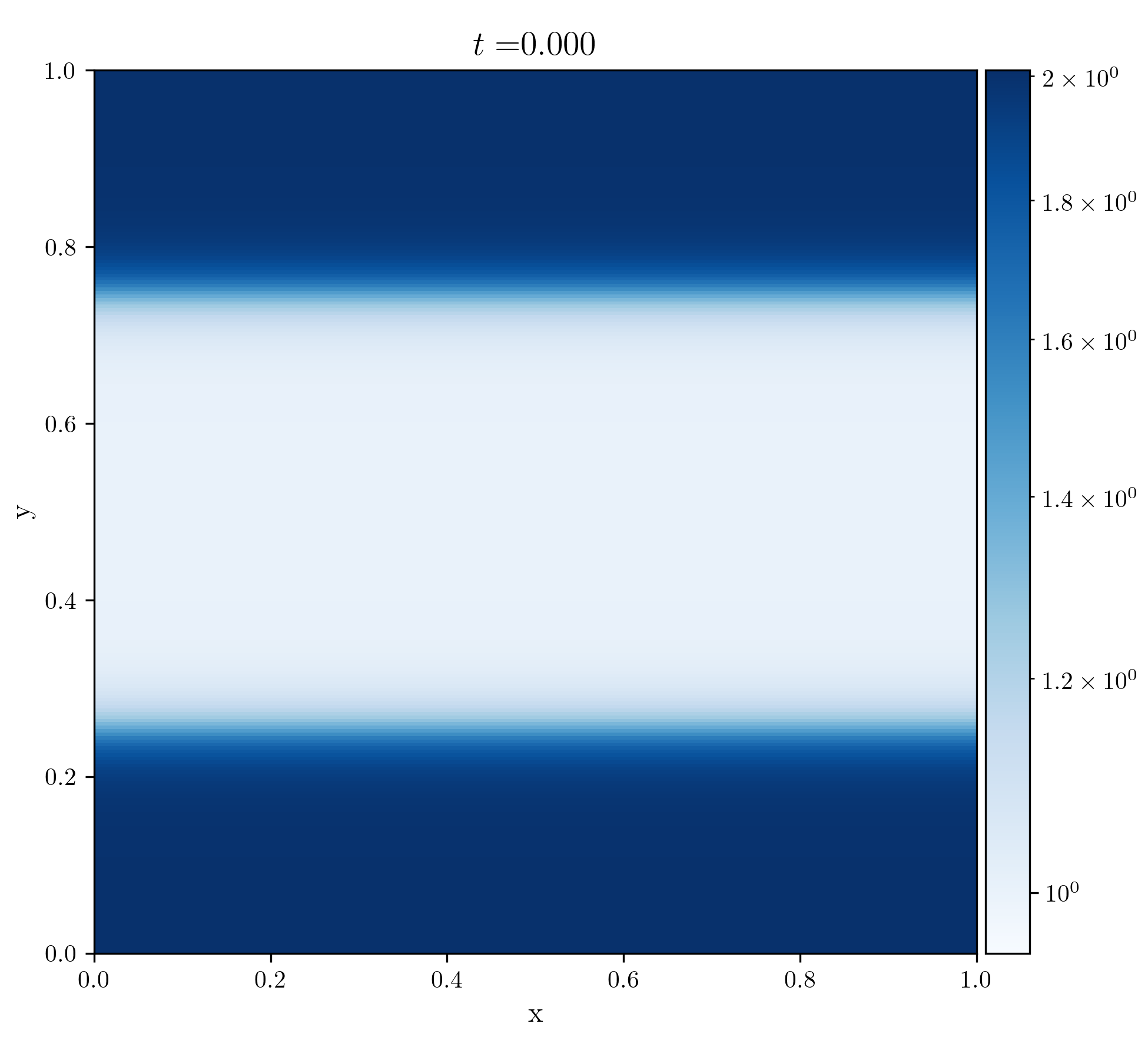}%
\includegraphics[width=.5\linewidth]{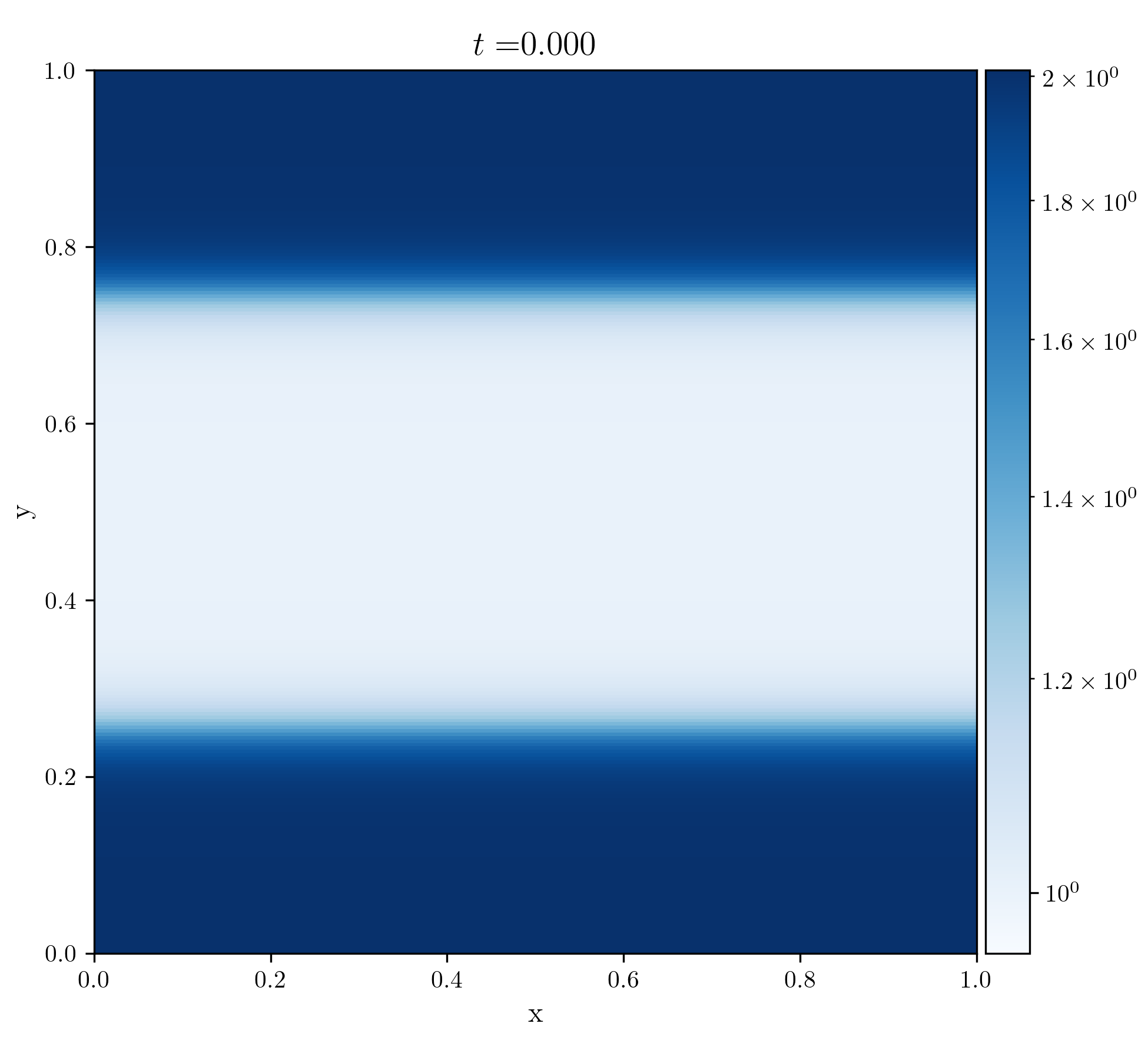}%
\\
\includegraphics[width=.5\linewidth]{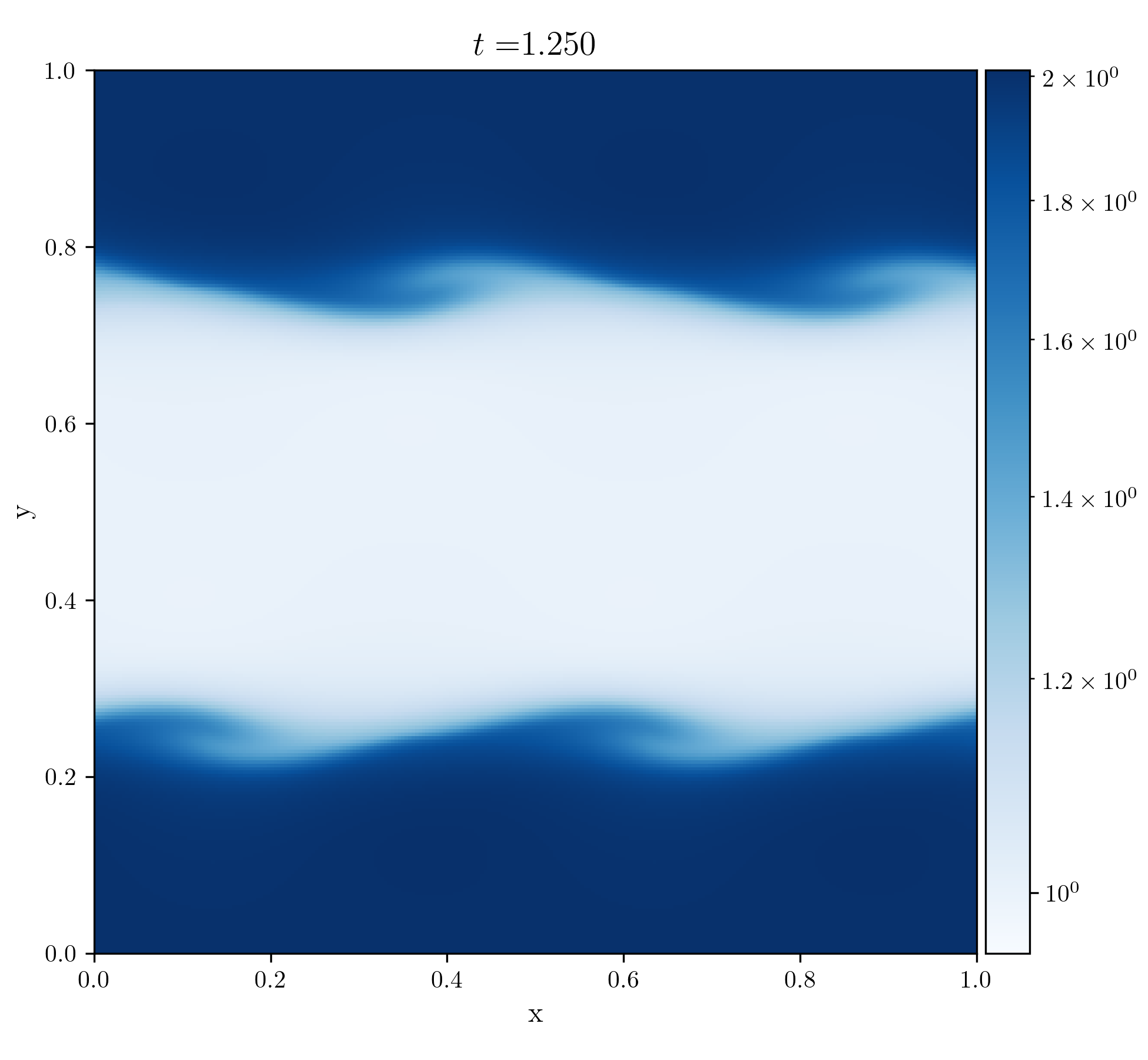}%
\includegraphics[width=.5\linewidth]{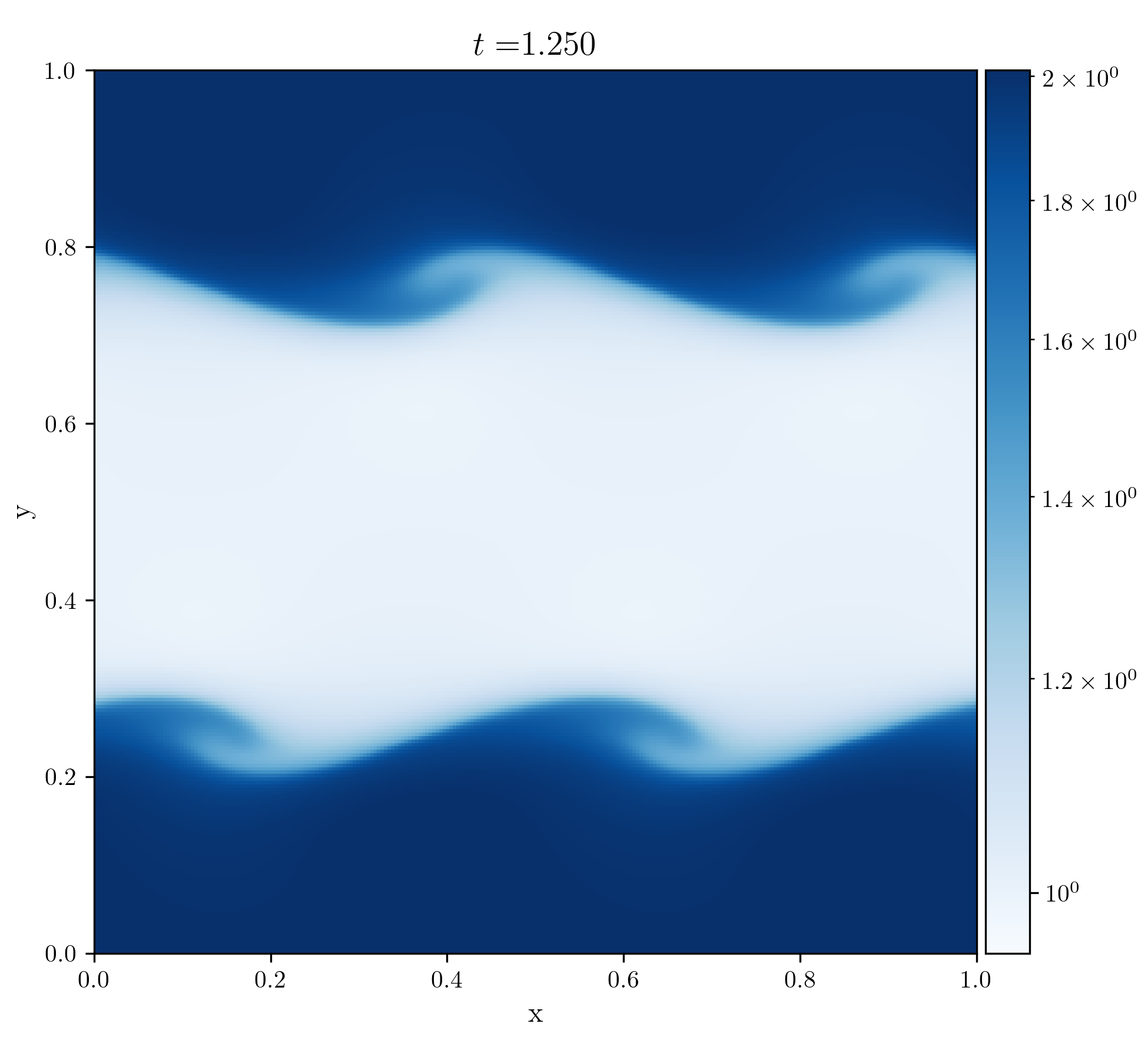}%
\\
\includegraphics[width=.5\linewidth]{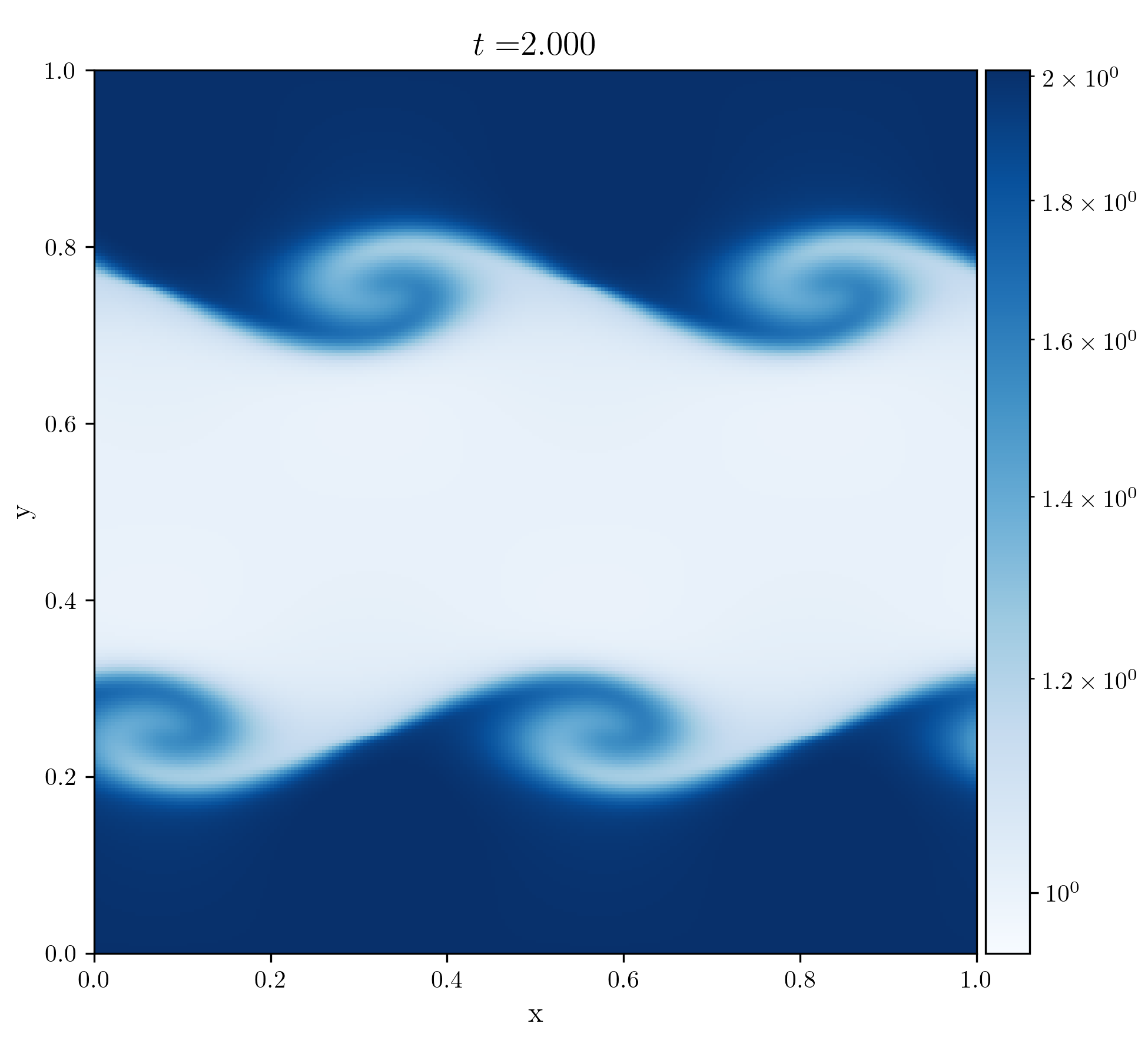}%
\includegraphics[width=.5\linewidth]{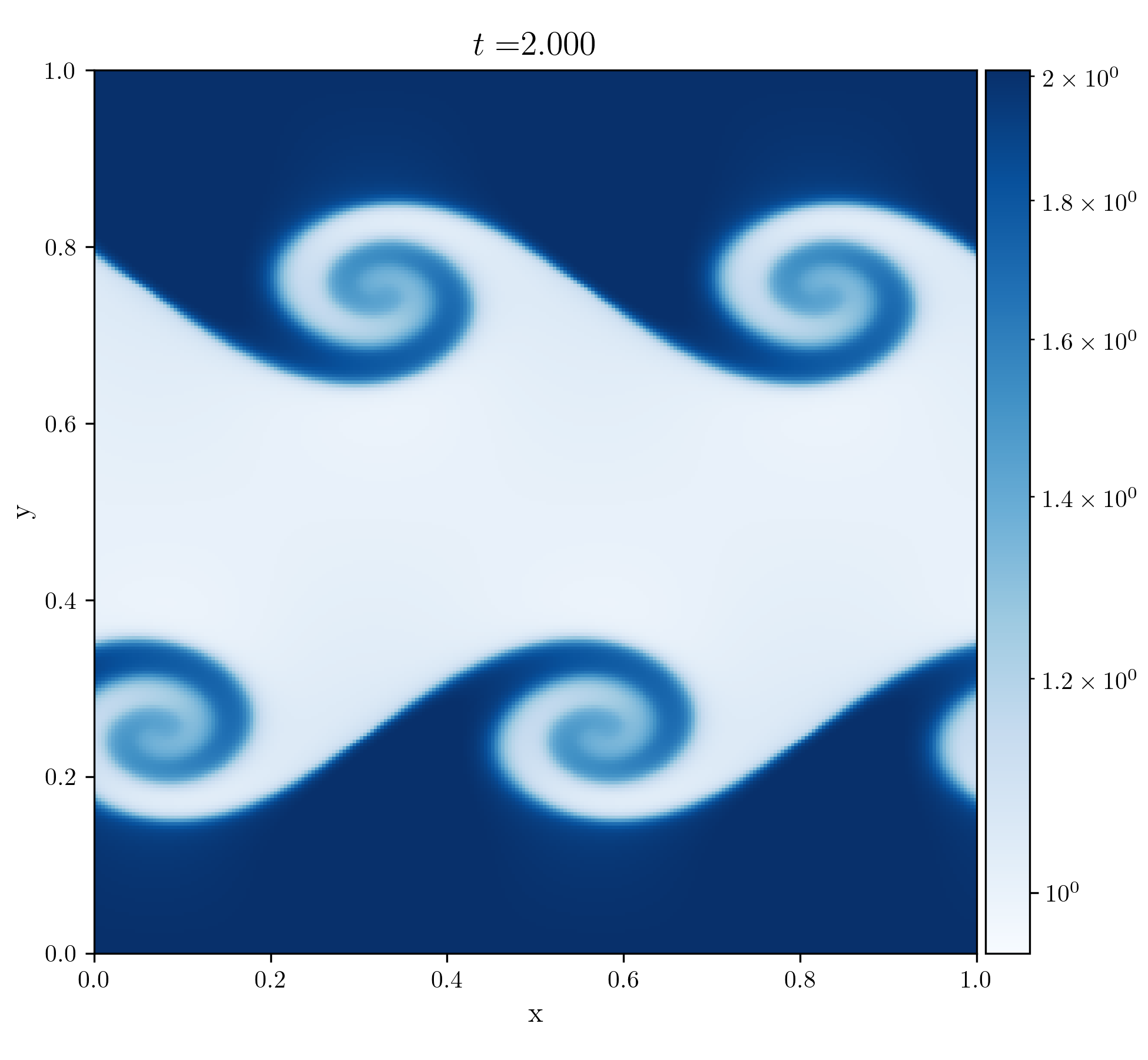}%
\caption[Kelvin-Helmholtz Instability with Godunov's scheme and MUSCL-Hancock 1]{
Density evolution for the Kelvin-Helmholtz instability problem solved with Godunov's method (left)
and the MUSCL-Hancock method (right) at times $t = 0, 1.25, 2$ in arbitrary units.
}
\label{fig:kelvin-helmholtz-1}
\end{figure}

\begin{figure}
\centering
\includegraphics[width=.5\linewidth]{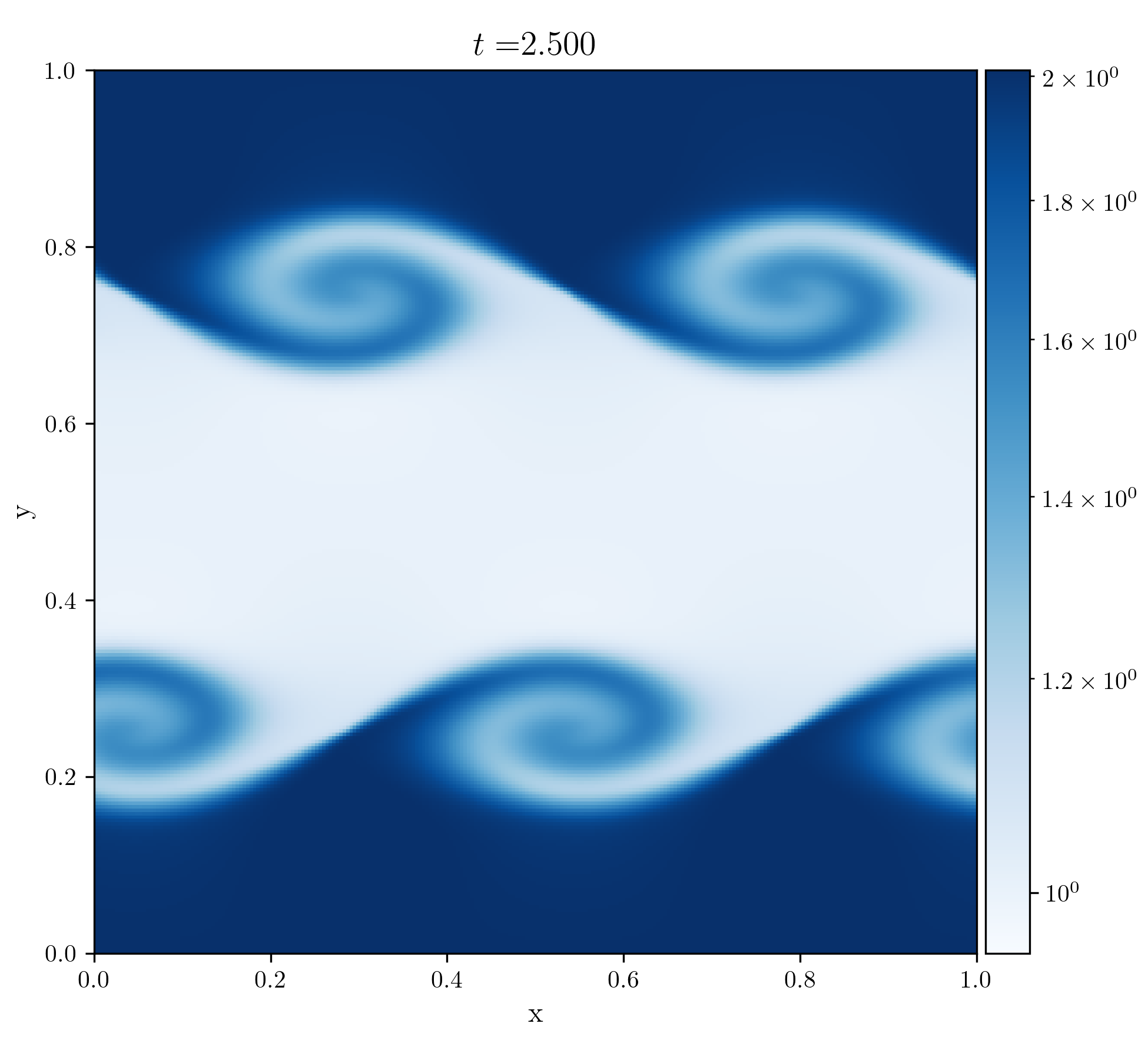}%
\includegraphics[width=.5\linewidth]{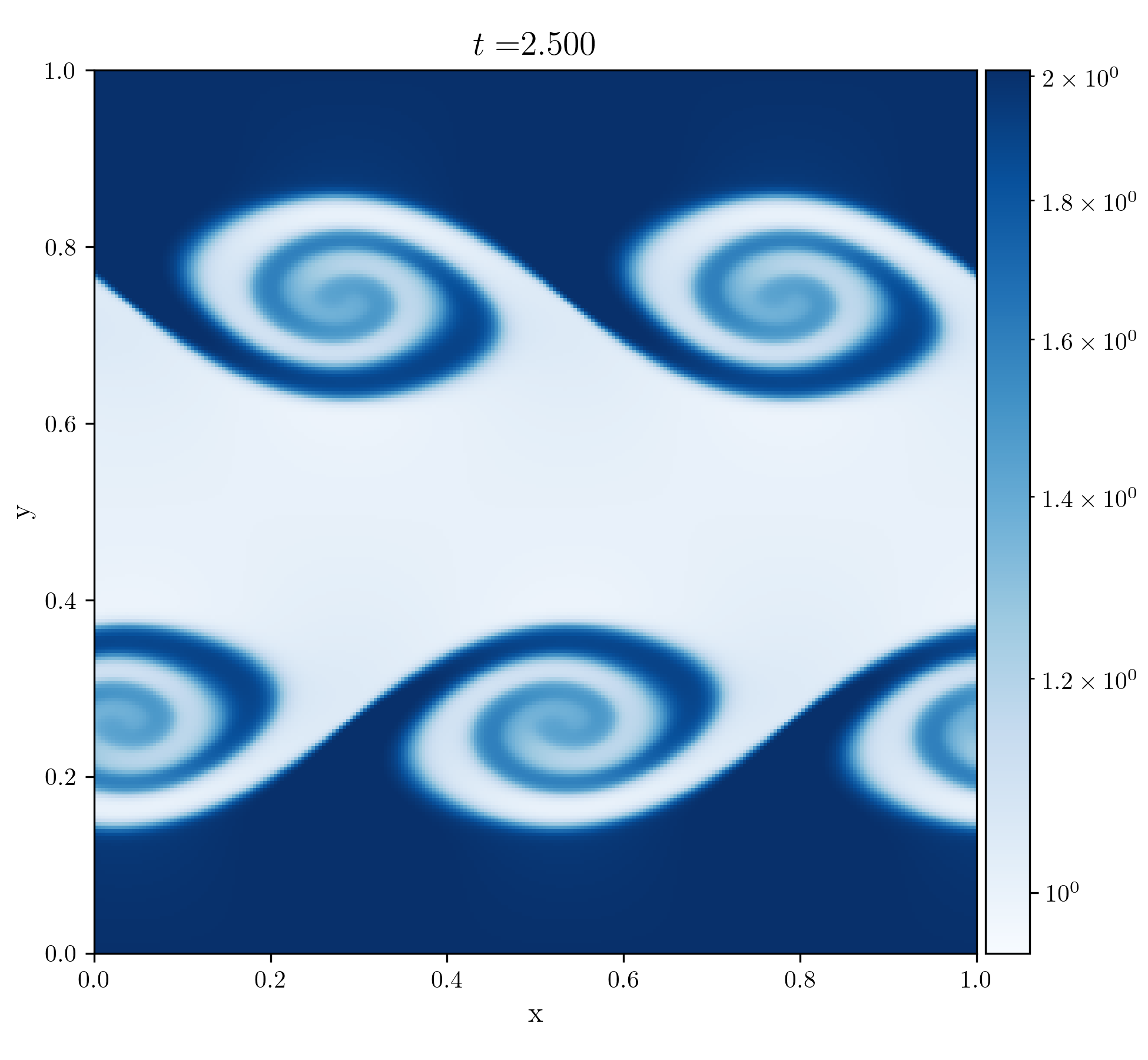}%
\\
\includegraphics[width=.5\linewidth]{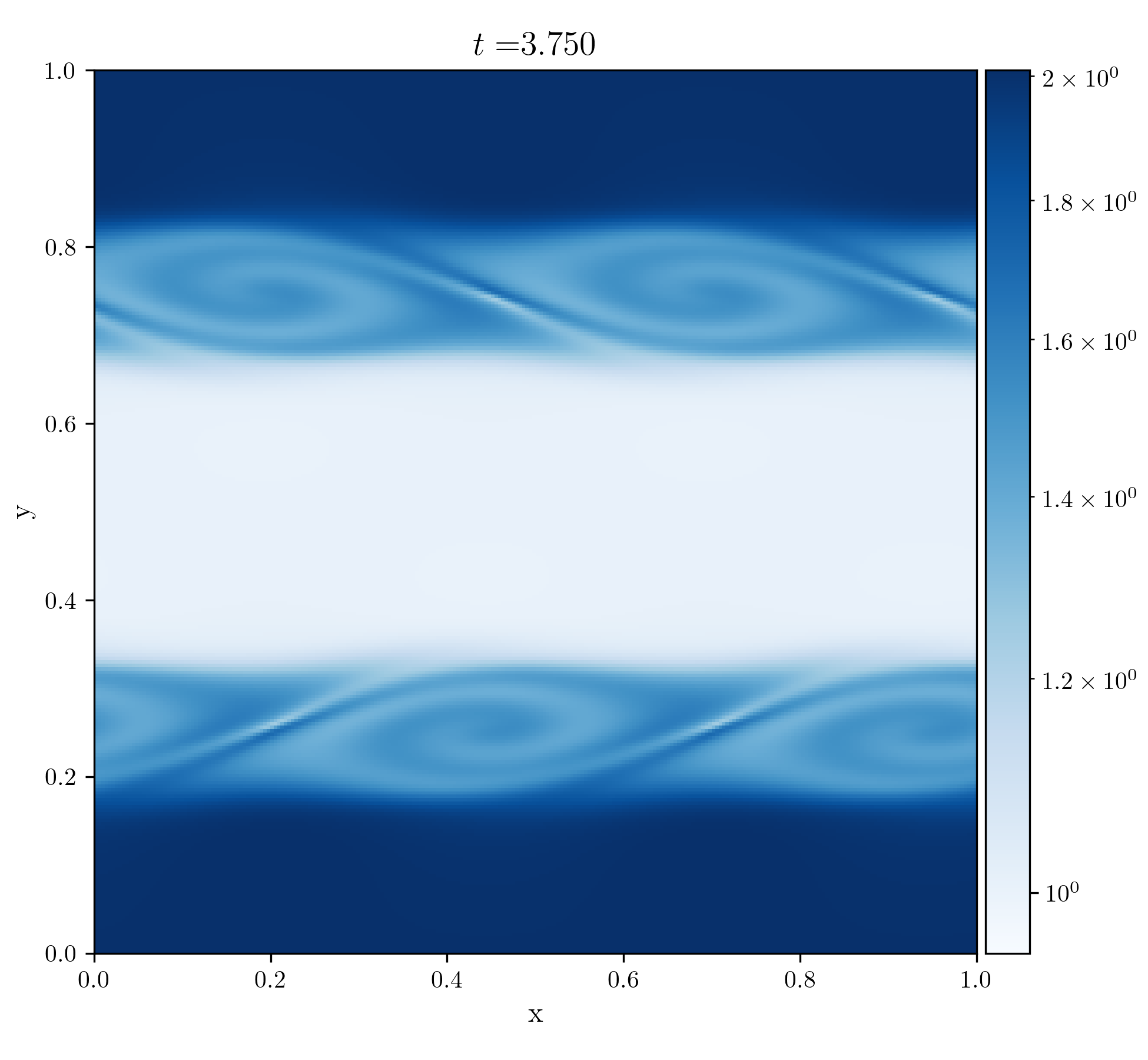}%
\includegraphics[width=.5\linewidth]{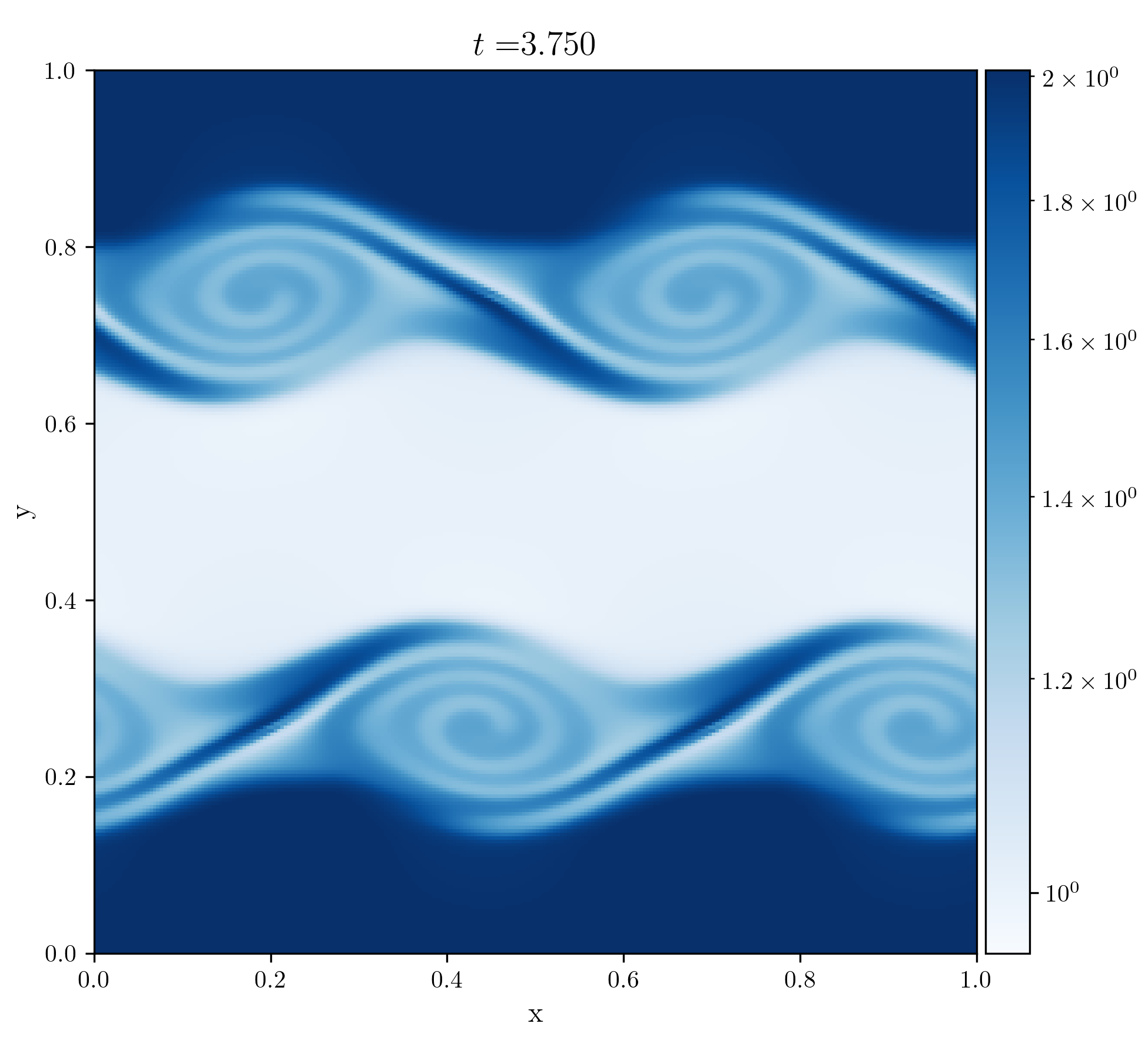}%
\\
\includegraphics[width=.5\linewidth]{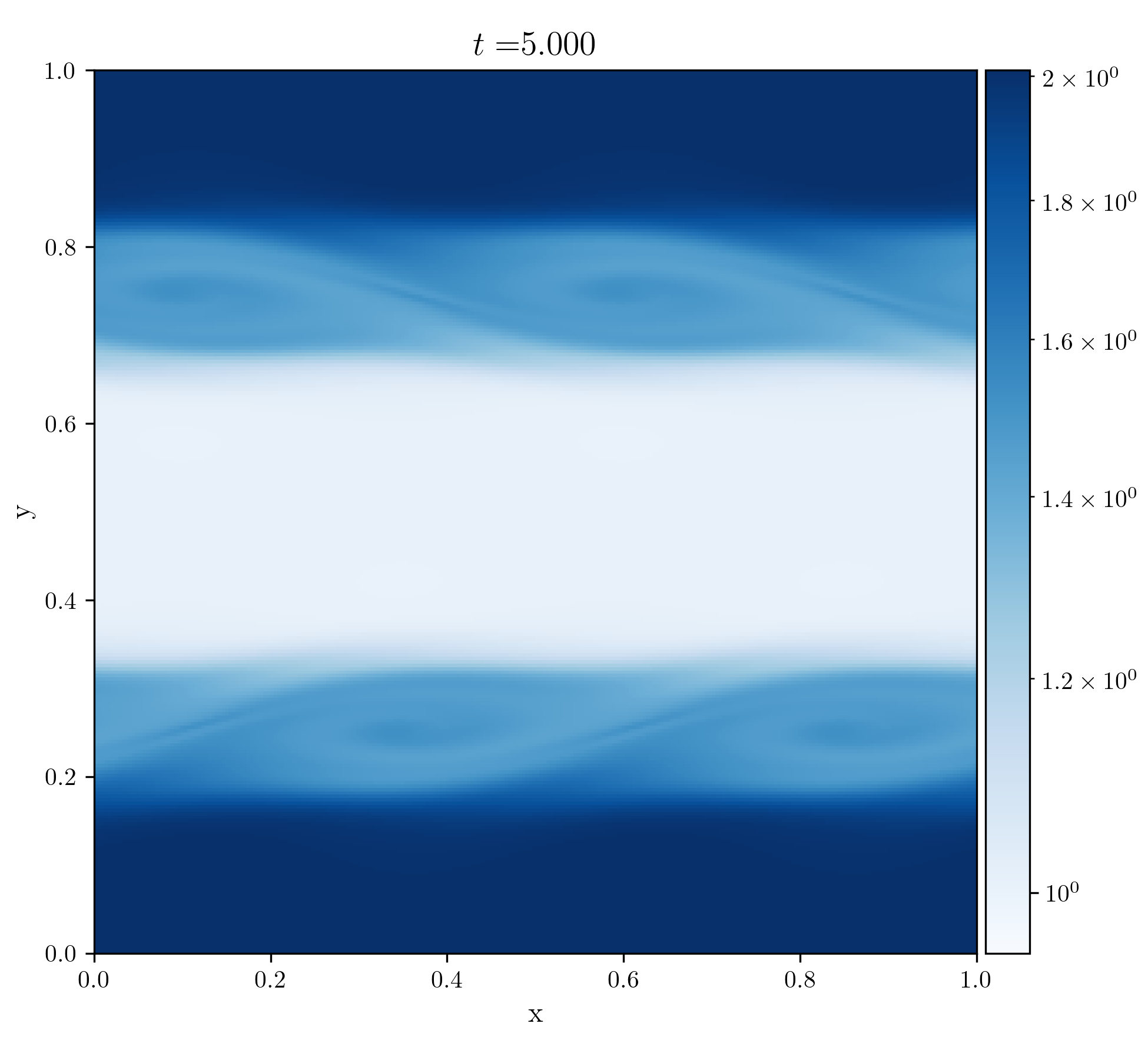}%
\includegraphics[width=.5\linewidth]{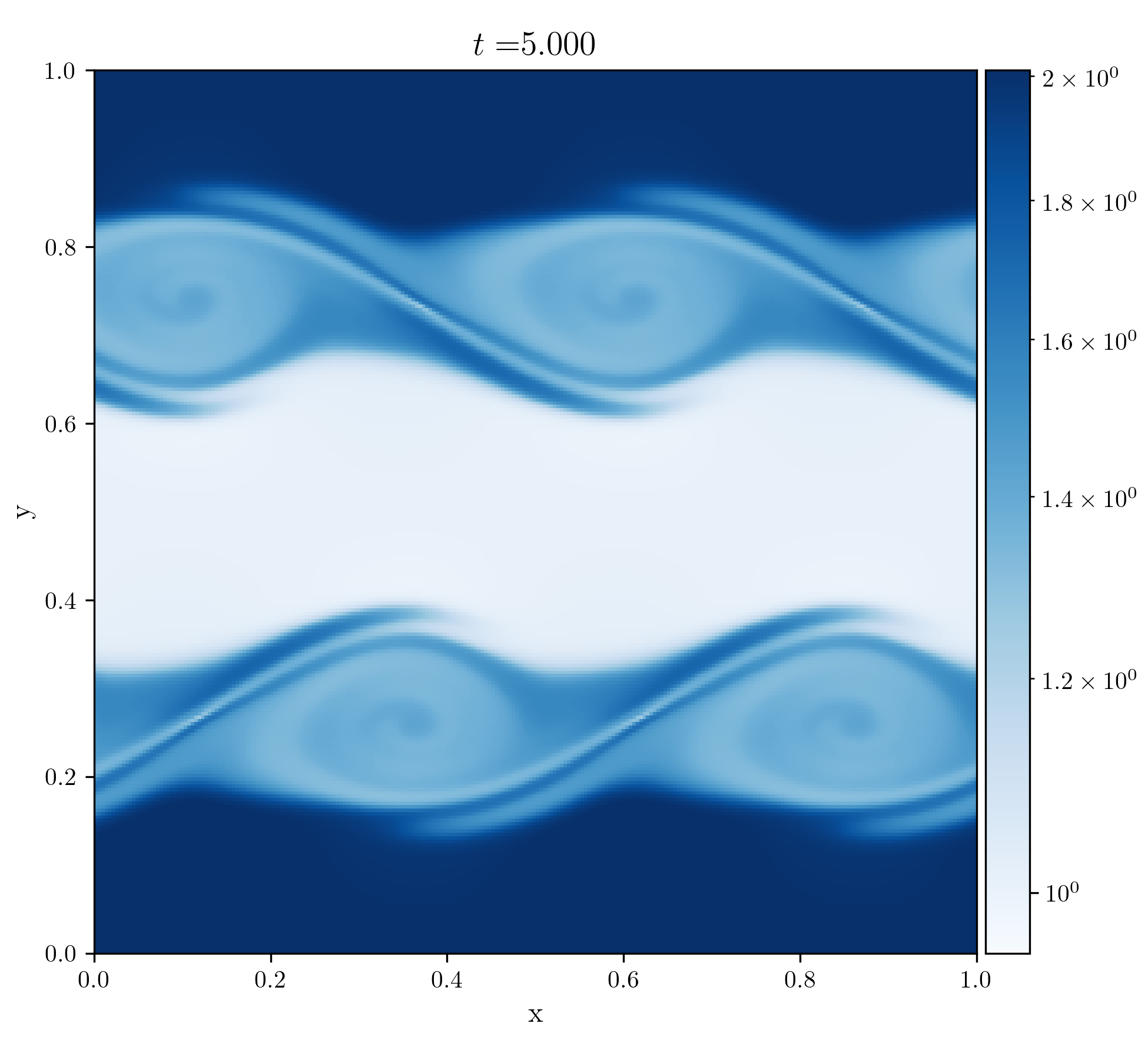}%
\caption[Kelvin-Helmholtz Instability with Godunov's scheme and MUSCL-Hancock 2]{
Density evolution for the Kelvin-Helmholtz instability problem solved with Godunov's method (left)
and the MUSCL-Hancock method (right) at times $t = 2.5, 3.75, 5.$ in arbitrary units.
}
\label{fig:kelvin-helmholtz-2}
\end{figure}

The initial pressure is set everywhere to $p = 2.5$. The results using these initial conditions
with the \meshhydro code on a grid of size $256 \times 256$ for Godunov's method and the
MUSCL-Hancock scheme with the Van Leer limiter are shown in
Figures~\ref{fig:kelvin-helmholtz-1}~and~\ref{fig:kelvin-helmholtz-2}. As time evolves, the initial
perturbations grow, and peaks arise at the interface of the two fluid phases, which penetrate into
the other phase. Eventually these peaks begin to curl up, giving rise to the typical
Kelvin-Helmholtz whirls, as can be seen in the bottom row of Figure~\ref{fig:kelvin-helmholtz-1}.
Both Godunov's scheme and the MUSCL-Hancock scheme are able to reproduce this behavior, although
the contact discontinuities in the results of the MUSCL-Hancock scheme are clearly sharper. However
at later times, shown in Figure~\ref{fig:kelvin-helmholtz-2}, differences become much more obvious.
The MUSCL-Hancock method is able to resolve much more fine structure, while Godunov's method cannot
due to its diffusivity. By the end of the simulation, Godunov's method predicts a nearly fully
mixed interface region, while the MUSCL-Hancock scheme still finds some fine structure along the
interface and inside the whirls.


%% file: main/Meshless/ML-0-introduction.tex
\chapter{Introduction}

This section is dedicated to the discussion of a particular class of finite volume methods and
their application to the Euler equations in the context of cosmological simulations of galaxy
formation - the so-called ``finite volume particle methods''.
Cosmological simulations are challenging in several aspects. One of them is the sheer size of the
problem. For example, suppose we want to simulate a volume of 20 co-moving Mpc, a relatively modest
size for cosmological purposes. Say we wanted to be able to resolve galaxies with typical radii of
$\sim 20$ kpc with at least 100 cells across the radius, giving us a cell size of $\Delta x \sim
200$ pc. Then a simulation software that uses a regular grid as the underlying discretization
technique would need to evolve $10^{15}$ cells each time step until the desired end time is reached.
Assuming each cell only stores three fluid variables as single precision floating point numbers,
e.g. the primitive variables density, fluid velocity, and pressure, that would require roughly 10.7
Petabytes of memory just for the primitive variables alone. And this completely neglects the most
dominant force on cosmological scales, which is gravity, as well as everything else besides the gas.

This problem needs to be addressed on several fronts. On one hand, clearly more sophisticated
methods than discretizing the simulation volume using a regular grid with cells of equal sizes is
necessary. On the other hand, a single personal computer is not sufficient to solve problems on
cosmological scales, and we need to turn towards dedicated high performance computing facilities.
Such supercomputers allow us to solve a single problem using a multitude of processors (and other
hardware specialized for performance like GPUs) with both shared and distributed memory
architectures. This type of computing, where a program makes use of several processors
simultaneously to solve a single problem, is called parallel computing. Parallel computing permits
us to solve a problem with a smaller time-to-solution by using more processing power for a single
problem. The basic idea is similar to children's maths problems like ``If it takes one painter eight hours to paint a house, how long would it take four painters''? Parallel computing also allows to
significantly increase the total problem size. There is an upper limit to how much memory is
available per processor, but using more processors with their respective memories can increase the
maximal solvable problem size. For example, the recent IllustrisTNG-300
(\cite{pillepichFirstResultsIllustrisTNG2018}) simulation was able to complete a simulation with a
volume of 300 co-moving Mpc in each dimension using $2500^3 \approx 1.6 \times 10^{10}$ gas
resolution elements on 24000 processors, which took 34.9 Million CPU hours.

While the usage of supercomputing facilities enables us to complete incredible simulations like
IllustrisTNG-300 and generate amazing results, it comes at the price of having to invest significant
effort into developing simulation software that is capable of efficient parallel computing, and is
able to utilize the capabilities of these facilities and take advantage of the benefits they offer.
This effort is by no means trivial, and state of the art simulation software needs to be designed
with parallel computing in mind from the start.
However, simply throwing more processing power at a problem is not sufficient. Parallel computing
comes with overheads, as additional computational work needs to be done to enable the parallelism.
The overheads increase with the number of computing nodes used, and eventually will dominate the
total workload. Hence there is an upper limit to how much extra performance can be gained. In a
similar manner, not all parts of the simulation program are parallelizable, and there is an upper
theoretical limit to the maximal possible speedup, which is known as Amdahl's law
(\cite{amdahlValiditySingleProcessor1967}). So we need to also look for more sophisticated methods
to solve the hydrodynamics in a cosmological context.

For example, we can make use of the fact that the Universe evolves from an initially remarkably
uniform state to large structures forming over time from initially minute perturbations. As the
structures such as dark matter halos and subhalos, filaments between such halos, galaxies and galaxy
clusters form, increasingly more mass is tied up in gravitationally bound structures. So as time
evolves, only the relatively small regions where the mass is situated become of prime interest,
particularly so for the topic of galaxy formation and evolution. This motivates the approach to not
treat all regions of the Universe equally: We can for example insist on an adequate resolution in
and around galaxies, like for a cell size of $\Delta x \sim 200$pc in the example above, while we
can reduce the spatial resolution in void regions which can span several Mpc. The ansatz is then to
use cells of different sizes whenever and wherever necessary, and adaptively refine cells to
sufficiently small sizes according to the current state of the medium. This approach is called
adaptive mesh refinement (AMR), and is widely used in astrophysical simulation codes
\citep[e.g.][]{teyssierCosmologicalHydrodynamicsAdaptive2002, stoneAthenaAdaptiveMesh2020,
hayesSimulatingRadiatingMagnetized2006, kravtsovAdaptiveRefinementTree1997,
mignonePLUTOCodeAdaptive2012, bryanENZOAdaptiveMesh2014}. An additional advantage of
AMR codes is that the cell refinement strategy can be made use of not only to guarantee a desired
resolution in interesting regions, but can also be applied to ensure a required level of stability,
accuracy, and reduce truncation errors \citep[see
e.g.][]{teyssierGridBasedHydrodynamicsAstrophysical2015a}). For example, take a cell with size
$\Delta x$ and density $\rho$. To achieve the desired mass resolution within a cell, we can impose
a minimal mass, $m_{\min}$, which we want a cell to contain. The refinement criterion is then given
by:

\begin{align}
\text{If } \quad \Delta x > \left(\frac{m_{\min}}{\rho}\right)^{\frac{1}{3}}, \quad \text{ refine.}
\end{align}

However, we can add an additional refinement criterion based on the numerical diffusion, which
depends on the cell size $\Delta x$ (see Chapter~\ref{chap:numerical_diffusion}) and the second
derivative of the conserved states w.r.t. space. The refinement criterion is then motivated by
requiring the numerical diffusion term to be ``small enough''. This effectively puts an upper
threshold on the local truncation error, and is expressed by the diffusion term being smaller than
the actual gradients of the conserved states (which are terms we make use of in our method).
Explicitly, the refinement criterion can be written as:

\begin{align}
\text{If } \quad \Delta x \left| \frac{\del^2 \U}{\del x^2} \right| > C_{thresh}
\left|\frac{\del \U}{\del x} \right|, \quad \text{ refine,}
\end{align}

where $C_{thresh}$ constitutes a dimensionless threshold we choose.

Unfortunately, grid based methods aren't without caveats. For example, it is known that grid
methods are in general not Galilei-invariant \citep[e.g.][]{wadsleyTreatmentEntropyMixing2008}, and
poorly resolved disc galaxies can show alignment of the disc with along the grid
\citep[e.g.][]{hahnLargescaleOrientationsDisc2010}. This motivates to look towards other methods
that don't suffer from the same issues. However, no method is without caveats and limitations, as
they all offer approximate discrete solutions to continuous problems. Even though various methods
are employed on the same underlying conservation laws, the solutions they find can differ,
especially so in strongly nonlinear regimes \citep[e.g.][For an example of difference in results
of two different methods on the same problem in this work, see Figures~\ref{fig:kelvin-helmholtz-1}
and~\ref{fig:kelvin-helmholtz-2} ]{frenkSantaBarbaraCluster1999,
agertzFundamentalDifferencesSPH2007a, braspenningSensitivityNonradiativeCloudwind2022a}. So it is
desirable to have several various well-studied and well-understood methods in order to confirm
the findings beyond doubt.

Other approaches, like the so-called ``Moving Mesh'' methods \citep[e.g][]{springelPurSiMuove2010,
vandenbrouckeMovingMeshCode2016, gaburovMagneticallyLevitatingAccretion2012} depart from the use of
a static (Eulerian) grid and construct irregular cells based on the local fluid flow. Typically
particles, i.e. a collection of points which may change their individual positions  over time, are
used as mesh generating points, which are also advected along with the fluid in a co-moving
(Lagrangian) fashion. By tracing the fluid's motion, the resolution naturally follows the flows and
in this fashion increases the resolution of regions where the fluid accumulates. It should be noted
that while in principle the mesh \emph{can} be constructed in a Lagrangian fashion, it doesn't
\emph{need} to be. The mesh generating points may as well be kept at fixed positions, and the
simulation executed in an Eulerian fashion. For this reason, such methods are called Arbitrary
Langrangian-Eulerian (ALE).

A third approach used in astrophysical simulations, called ``meshless methods'', completely departs
from cells as an underlying discretization technique, and uses particles instead. A famous class of
meshless methods are Smooth Particle Hydrodynamics (SPH) methods, where particles are typically
given some constant mass, and are evolved using the Lagrangian equations of fluid dynamics. SPH is
based on estimating the local fluid density as a weighted sum over neighboring particles, where the
weights are smoothly decreasing functions such that the noise in the density estimate introduced by
distant neighboring particles is reduced. Using this density estimate and the Lagrangian equations
of motion, the system can be evolved in time. While SPH is technically a meshless method, in
astrophysical circles it is usually talked about as a class of methods on its own, and is widely
used \citep[e.g.][]{springelCosmologicalSimulationCode2005, wadsleyGasoline2ModernSmoothed2017,
rosswogLagrangianHydrodynamicsCode2020, menonAdaptiveTechniquesClustered2015,
gonnetSWIFTFastAlgorithms2013}.

Recently, another form of ``meshless method'' has made its entrance for astrophysical applications,
which are the so-called ``finite volume particle methods''. Similarly to moving mesh methods, they
can be used arbitrarily in a Lagrangian or Eulerian manner, but they do not require the
constructions of any cells to solve the hydrodynamics. Instead, similarly to SPH, they make use of
particles as interpolation points, and determine the volumes associated with particles as a local
estimate depending on the neighboring particles. However, in contrast to SPH, they solve the
integral form of the fluid equations, which is more akin to finite volume methods, albeit without
mutually exclusive volume elements such as cells. Another (more formal) difference to SPH is that
the convergence and stability of finite volume particle methods has been demonstrated in
\cite{lansonRenormalizedMeshfreeSchemes2008} and \cite{lansonRenormalizedMeshfreeSchemes2008a},
which is yet to be shown for SPH (see e.g. \cite{vacondioGrandChallengesSmoothed2021}). Some
astrophysical simulation codes \citep[e.g.][]{gaburovAstrophysicalWeightedParticle2011,
hopkinsGIZMONewClass2015, hubberGANDALFGraphicalAstrophysics2018,
grothCosmologicalSimulationCode2023a, alonsoasensioMeshfreeHydrodynamicsPKDGRAV32023} apply the
finite volume particle methods for hydrodynamics. These codes use the scheme described by
\cite{lansonRenormalizedMeshfreeSchemes2008} and \cite{lansonRenormalizedMeshfreeSchemes2008a}.
However, there are other formulations \citep[e.g.][]{hietelFiniteVolumeParticleMethodConservation2001,
hietelMeshlessMethodsConservation2005, ivanovaCommonEnvelopeEvolution2013}, which to date have not been explored in literature in an astrophysical context. This second Part of my thesis is dedicated to first providing a full derivation of the two formulations of finite volume particle methods, and to qualitatively compare their performance. Then the implementation of finite volume particle methods for hydrodynamics in the task-based parallelized code \swift \cite{gonnetSWIFTFastAlgorithms2013} is presented.

%% file: main/Meshless/ML-1-partition-of-unity.tex
\chapter{Partition Of Unity}

\section{Constructing A Partition Of Unity}

\begin{figure}
 \centering
 \includegraphics[width=.7\textwidth]{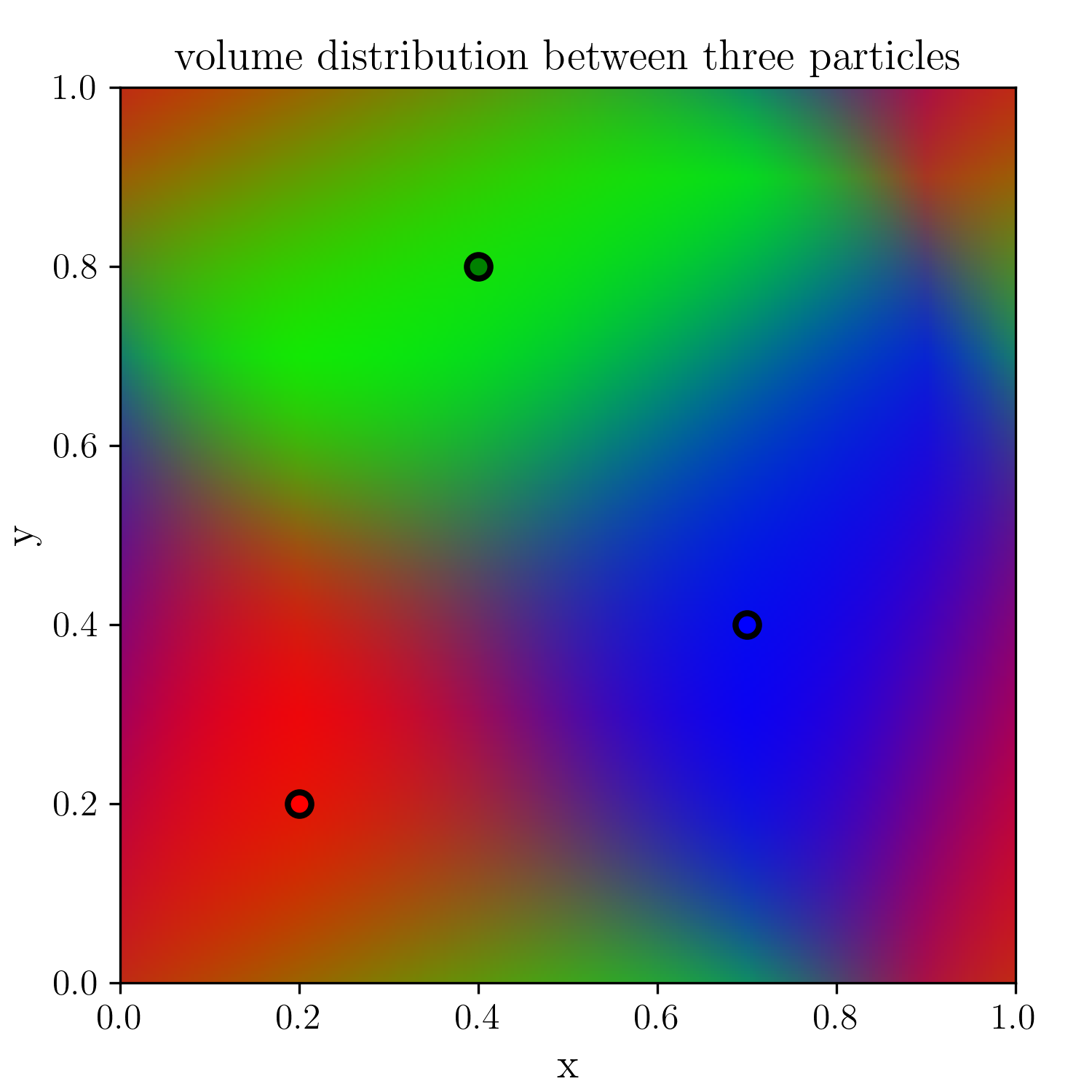}%
 \caption{
The volume distribution among three particles on a periodic two-dimensional domain with side
length of unity in arbitrary units and periodic boundary conditions. The color at each point of the
domain is determined by assigning RGB values of $\psi$ of the red, green, and blue particle at that
point. Some vertical and horizontal features appear at half the box size's distance from each
particle due to the periodic wrapping: In order for every particle to have a nonzero weight at any
point for this illustrative example, their compact support radii have been chosen to be greater
than half the box size. However this leads to a sharp, non-smooth local minimum in the particle's
weight at that distance in each dimension.
}
\label{fig:psi-volume-distribution}
\end{figure}

A key component in deriving the finite volume particle methods is to make use of a ``partition of
unity'' to distribute the spatial domain among particles, which are used as discretization
elements for the scheme. In a sense, each point in space is ``shared'' among all particles in a
weighted fashion, which will be discussed in detail later. Specifically, at any point $\x$ in space
in the domain, we assign a partition $\psi_i(\x)$ to each particle $i$ such that

\begin{equation}
    \sum_i \psi_i(\x) = 1 \label{eq:partition-of-unity}
\end{equation}

where the sum is assumed to include \emph{all} particles $i$. A visualization of a partition of
unity where a volume is divided among three particles is shown in
Figure~\ref{fig:psi-volume-distribution}. The three particles have been assigned a color red, green,
and blue, respectively. Each point in space is given a color, where the RGB value of the color is
determined by the particle of the corresponding color's contribution to the partition of unity at
that point.

This requirement for the partitions $\psi_i(\x)$ to always sum up to exactly one everywhere is what
lead to the name ``partition of unity''. It can be interpreted as an interpolation technique, where
the real values of a field are distributed among particles and the partition of unity allows the
interpolation at any other point in space (see eq.~\ref{eq:psi-interpolation}).
Condition~\ref{eq:partition-of-unity} can be enforced by choosing $\psi$ to take the following form:

\begin{align}
    \psi_i(\x) &\equiv \frac{1}{\omega(\x)} W(\x, \x_i) \label{eq:psi}\\
    \omega(\x) &= \sum_j W(\x, \x_j) \label{eq:omega}
\end{align}

Where $W(\x, \x_i)$ is in principle some arbitrary function depending on the particle position
$\x_i$. In practice, we want apply some constraints to it motivated by both physical arguments,
mathematical requirements, and computational efficiency. In particular, we want:

\begin{itemize}
 \item As we are going to make use of the derivatives, $W(\x, \x_i)$ shall be continuous and
differentiable, and its first derivative shall also be continuous and differentiable.
 \item $W(\x, \x_i)$ shall be spherically symmetric, i.e. $W(\x, \x_i) = W(|\x - \x_i|)$, to avoid
any preferential direction.
 \item $W(\x, \x_i)$ shall have compact support, i.e. there is some distance $H$ for which
$W(|\x - \x_i| > H) = 0$. $H$ is called the compact support radius. We want to be able to treat the
system locally, and only close-by particles should have an influence on a given point $\x$.
\end{itemize}

Functions that satisfy these demands are well known as ``kernels'' in Smooth Particle
Hydrodynamics methods \citep[e.g.][]{monaghanSmoothedParticleHydrodynamics1992,
priceSmoothedParticleHydrodynamics2012}, and are typically computationally inexpensive
polynomials. For example, throughout this work the cubic B-spline kernel
\citep{monaghanRefinedParticleMethod1985} is used, which is given by

\begin{align}
    q &\equiv \frac{r}{H} \\
    W(\x, H) &= \frac{\sigma}{H^\nu}
        \begin{cases}
            (1 - q)^3 - 4\left(\frac{1}{2} - q \right)^3
                & \text{ if } 0 < q \leq \frac{1}{2} \\
            (1 - q)^3
                & \text{ if } \frac{1}{2} \leq q \leq 1 \\
            0 & \text{ otherwise }
        \end{cases}
    \label{eq:cubic-spline-kernel}
\end{align}

where $\nu$ denotes the number of spatial dimensions, and $\sigma$ is a normalization coefficient
given by

\begin{align}
    \sigma = \frac{8}{3} \text{ for } \nu = 1 &&
    \sigma = \frac{80}{7 \pi} \text{ for } \nu = 2 &&
    \sigma = \frac{16}{\pi} \text{ for } \nu = 3
\end{align}

Table 1 in \cite{dehnenImprovingConvergenceSmoothed2012c} lists some other popular choices for
kernels.

The number of particles that have a non-zero $\psi_i(\x)$  at any $\x$, and thus contribute to the
partition of unity at that point, is determined by the compact support radius $H$. Simultaneously,
the number of particles that contribute to the partition of unity at a given point determines the
accuracy and resolution of the method. Hence it is desirable to demand that (roughly) the same
number of particles is enclosed within the compact support radius at any point. In fact, demanding
a number of neighboring particles to be enclosed within the compact support radius of a given
particle is the criterion which will be used to define the particle's compact support radius. This
will be discussed in more detail in Chapter~\ref{chap:meshless-full}. However, if we want to
maintain a (roughly) equal number of particles having some weight at a given $\x$, then the compact
support radius $H$ can't be a fixed constant which is equal for all particles and for all time.
Instead, it will depend on the current particle configuration, and hence will have a dependence on
the position of evaluation, i.e. $H = H(\x)$.  It is hence appropriate to adapt the notation

\begin{align}
    W(\x, \x_i, H) = W(\x, \x_i, H(\x)) = W_i(\x, H)
\end{align}

Furthermore, it is conventional to talk about the ``smoothing length'' $h$ rather
than the compact support radius $H$ in SPH circles. They are related quantities, with $H \approx
2h$, where the exact value of the proportionality coefficient depends on the kernel choice and
dimension. The reason we use $h$ rather than $H$ is that the smoothing length actually has a direct
physical correspondence: \cite{dehnenImprovingConvergenceSmoothed2012c} have shown that the smoothing
length is directly proportional to the minimal wavelength of a sound wave that can be resolved with
SPH. The exact proportionality coefficients to convert between a smoothing length and the compact
support radius for various kernels and dimensions is given in Table 1 of
\cite{dehnenImprovingConvergenceSmoothed2012c}. To keep the same convention, we'll use the notation

\begin{align}
    W(\x, \x_i, h) = W(\x, \x_i, h(\x)) = W_i(\x, h)
\end{align}

\begin{figure}
 \centering
 \includegraphics[width=\textwidth]{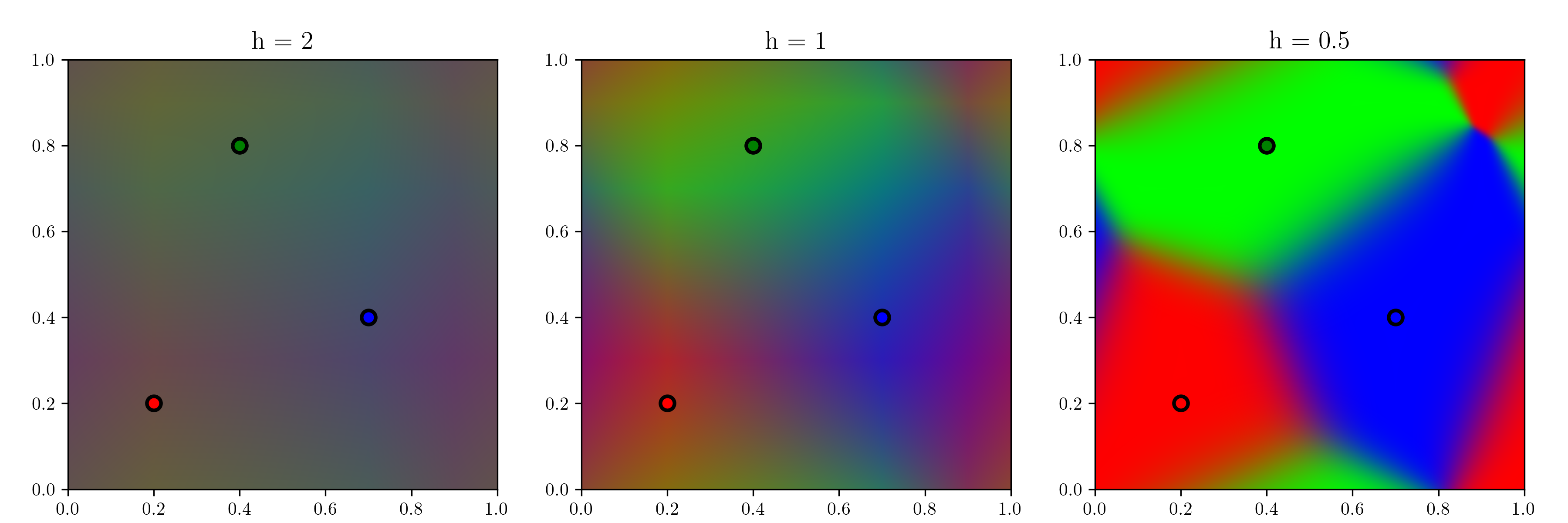}
 \caption{
Same as Figure~\ref{fig:psi-volume-distribution}, but the volume distributions for three different
smoothing lengths $h = 2$, 1, and 0 are shown. As the smoothing length $h$ increases, the value of
the used cubic spline kernel $W(r)$ increases for a fixed distance $r$, and the relative
contribution to the partition of unity at some given point of particles further away increases
accordingly.
}
\label{fig:psi-volume-distribution-vary-h}
\end{figure}

Our demands for the spherically symmetric kernels $W(r)$ to be smooth as well as have compact
support implies that on average they will be a decreasing function of $r$, as they need to go from
some initial value $W(0)$ to zero at $W(r=H)$. Indeed the cubic spline kernel used in this work is
a monotonously decreasing function of distance. This means that each particle's contribution to the
partition of unity is strongest close to the particle position, but it also means that a particle is
likely to be the most dominant contributor to the partition of unity in its immediate vicinity.
This is clearly seen in Figure~\ref{fig:psi-volume-distribution}, where a cubic spline kernel was
used. However, contrary to what the bright red, green, and blue fields close to the particles of
the corresponding color in Figure~\ref{fig:psi-volume-distribution} may suggest, in general no point
in space is assigned to only one particle. The closest particle may be dominant in its
contribution, but not the only contributor to the partition of unity. To underline this point, the
same visualization technique has been applied in Figure~\ref{fig:psi-volume-distribution-vary-h}
for varying smoothing lengths $h$. As $h$ increases, the decline of the kernel value $W_i(r)$
decreases for a specific distance $r$, and the individual particles have more weight at larger
distances. The resulting partition of the volume is much less clearly dominated by any single
particle.

\begin{figure}
 \centering
 \includegraphics[width=\textwidth]{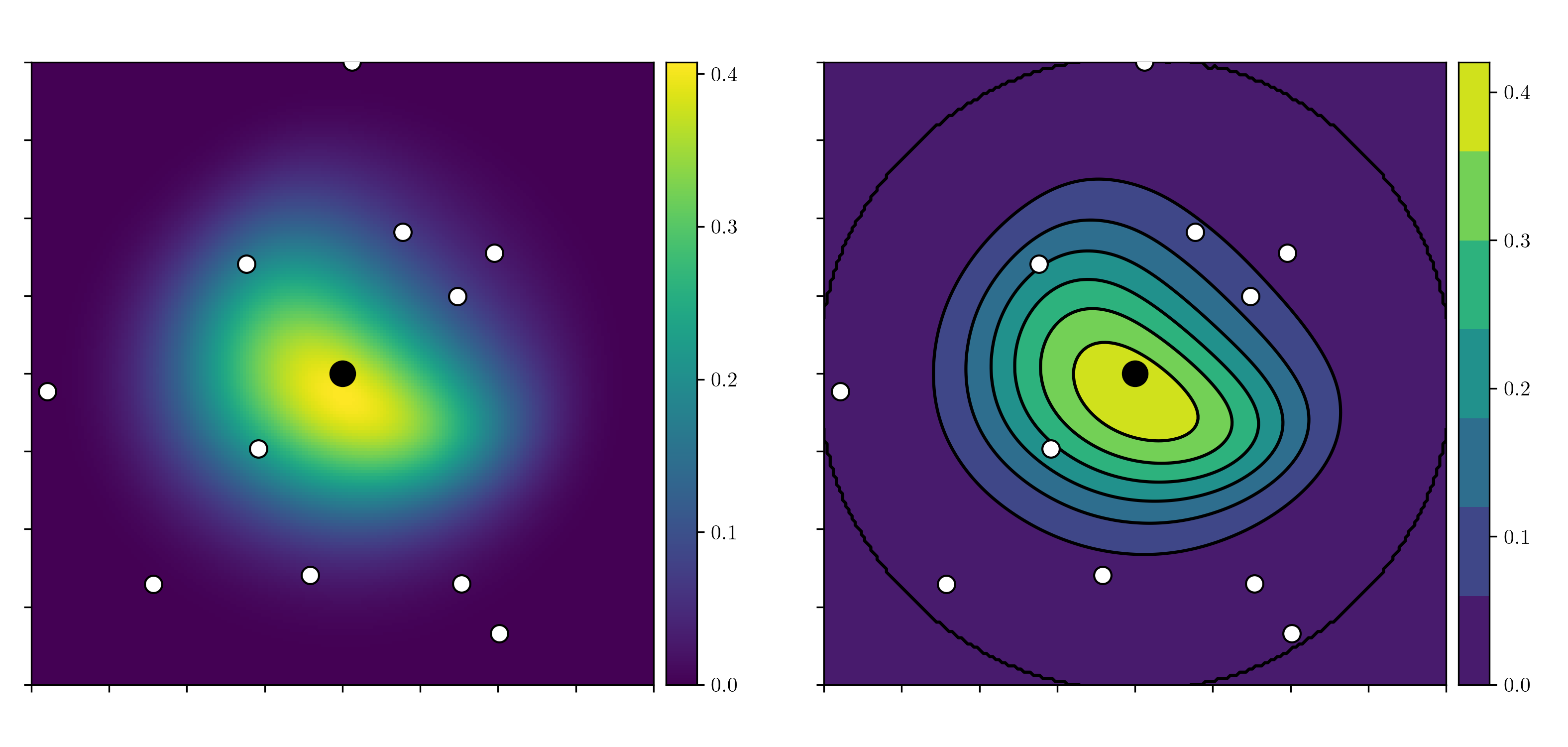}
 \caption{ The partition of unity $\psi_i(\x)$ of a single particle (black dot). The left plot
shows the actual values, the right plot shows a contour plot of the same. Even though the used
kernel functions $W$ are spherically symmetric, the resulting partition of unity of a particle
isn't because of the presence of other particles (white dots).
}
\label{fig:psi-of-x-contour}
\end{figure}

A further notable point is that even though the kernels $W_i$ are assumed to be spherically
symmetric, the partitions of unity of individual particles, $\psi_i$, in general do not possess the
same symmetry. This is due to the normalization $\omega(\x)$ which depends on the particle
configuration. To illustrate that point, an exemplary partition $\psi_i(\x)$ for a single particle
is shown in Figure~\ref{fig:psi-of-x-contour}.

\section{Particle Volume}\label{chap:particle-volume}

Suppose we have data defined on a set of interpolation points $f_i$, which correspond to function
values at particle positions $\x_i$. Then the interpolated value at some arbitrary $\x$ is given by

\begin{align}
    f(\x) = \sum_i f_i \psi_i(\x) \label{eq:psi-interpolation}
\end{align}

and the gradient is given by

\begin{align}
    \nabla f(\x) = \nabla \left( \sum_i f_i \psi_i(\x) \right) = \sum_i f_i \nabla \psi_i(\x)
\end{align}

since the interpolation points $f_i$ are constant. Using this definition and the volume average of
a function given by

\begin{align}
    V \overline{f} = \int_V f(\x) \de V
\end{align}

we can write

\begin{align}
    V \overline{f}
        = \int_V f(\x) \de V
        = \int_V  \sum_i f_i \psi_i(\x)
        = \sum_i f_i \int_V \psi_i(\x) \de V
        = \sum_i f_i V_i
\end{align}

where

\begin{align}
    V_i \equiv \int_V \psi_i(\x) \de V \label{eq:psi-integral-volume}
\end{align}

are the associated particle volumes.

If we set $f(\x) = f_i = 1$, we obtain

\begin{align}
    V \overline{f} &= V = \sum_i f_i V_i = \sum_i V_i \\
    V &= \sum_i V_i
\end{align}

so by construction, the associated particle volumes conserve the total system's volume in a closed
system.

We can find an expression for the $V_i$ in relation to the kernel function $W(\x, \x_i, h(\x))$. In
what follows, we assume that

\begin{itemize}
    \item $W = W(|\x - \x_i|)$ is symmetric, and
    \item $W$ is normed, i.e. $\int_V W(\x, \x_i, h(\x)) \de V = 1$.
\end{itemize}

We also make use of the following:
\begin{enumerate}[label=\roman*.]
\item $f(\x) = \sum_i f(\x_i) \psi_i(\x)$
\item $\psi_i(\x) = \frac{1}{\omega} W(\x - \x_i) = \frac{W(\x - \x_i)}{\sum_j W(\x - \x_j)}$
\item $V_i = \int_V \psi_i(\x) \de V$
\end{enumerate}
Then
\begin{align}
V_i \
    &\overset{(i)}{=} \ \sum_j \psi_j(\x_i) V_j
    \overset{(iii)}{=} \ \sum_j \psi_j (\x_i) \int_V \psi_j(\x) \de V \\
    &\overset{(ii)}{=} \ \sum_j \frac{W(\x_i - \x_j)}{\sum_k W(\x_i - \x_k)} \int_V \psi_j(\x) \de
V \\
    &= \frac{1}{\sum_k W(\x_i - \x_k)}  \sum_j W(\x_i - \x_j) \int_V \psi_j(\x) \de V \\
    &= \frac{1}{\sum_k W(\x_i - \x_k)}  \sum_j \int_V W(\x_i - \x_j) \psi_j(\x) \de V \\
    &= \frac{1}{\omega(\x_i)}  \sum_j \int_V W(\x_i - \x_j) \psi_j(\x) \de V \\
    &\overset{(i)}{=} \
        \frac{1}{\omega(\x_i)}\underbrace{\int_V W(\x_i - \x) \de V}_{=1 \text{, normalized}} \\
    &= \frac{1}{\omega(\x_i)} \label{eq:particle-volume-exact}
\end{align}

Expression~\ref{eq:particle-volume-exact} is only valid assuming that the
interpolation~\ref{eq:psi-interpolation} and the volume integral

\begin{align}
    \int_V f(\x) \de V = \sum_i f_i V_i \label{eq:psi-volume-integral-step}
\end{align}

are exact, which in general they are not. As will be shown below, the volume integral is
$\order(h^2)$ accurate, giving us

\begin{align}
    V_i = \frac{1}{\omega(\x_i)} + \order(h^2) \label{eq:particle-volume}
\end{align}

To demonstrate the accuracy of the integral~\ref{eq:psi-volume-integral-step}, we multiply the
integrand by $1 = \sum_i \psi_i(\x)$ (which is exact by construction) and use a Taylor expansion
around $\x = \x_i$:

\begin{align}
    \int_V f(\x) \de V
        &= \int_V f(\x) \underbrace{\sum_i \psi_i(\x)}_{= 1} \de V
        = \sum_i \int_V f(\x) \psi_i(\x) \de V \\
        &= \sum_i \left[
            f(\x_i) + (\x - \x_i)\nabla f(\x_i) + \order(\x - \x_i)^2
            \right] \psi_i (\x) \\
        &= \sum_i f(\x_i) \int_V \psi_i(\x) \de V +
            \underbrace{\sum_i \int_V (\x - \x_i) \nabla f(\x_i) \psi_i(\x) \de V}_{=0 \text{, will
be shown later}} + \order((\x - \x_i)^2) \label{eq:integral-psi-interemediate}\\
        &= \sum_i f(\x_i) \int_V \psi_i(\x) \de V + \order((\x - \x_i)^2)
\end{align}

So the integral is $\order((\x - \x_i)^2)$ accurate. Considering that the kernels we use have a
compact support radius $H$ which determine the maximal distance $\x - \x_i$ from a particle at
which the kernel value and hence the partition of unity $\psi_i(\x)$ is non-zero, we can express
the error term using $H$ as the upper boundary for the error, i.e. $\max_i (\x - \x_i) = \max_i H_i
= \max_i H(\x_i)$. Equivalently, since the smoothing length $h$ is directly proportional to $H$, we
might just as well express the error in terms of $h$, and write $\order((\x - \x_i)^2) =
\order(h^2)$ like in eq.~\ref{eq:particle-volume}.

It remains to show that the integral of the first order term in
eq.~\ref{eq:integral-psi-interemediate} is indeed zero. Since the gradient term is evaluated at a
fixed $\x_i$, we can write

\begin{align}
    \sum_i \int_V (\x - \x_i) \nabla f(\x_i) \psi_i(\x) \de V =
    \sum_i \nabla f(\x_i)  \int_V (\x - \x_i)\psi_i(\x) \de V
\end{align}

and it suffices to show that

\begin{align}
    \int_V (\x - \x_i)\psi_i(\x) \de V = 0
\end{align}

For simplicity, we demonstrate this only for the one dimensional case. The extension to several
dimensions is straightforward. Furthermore, we set the integration boundaries to be $(-L, L)$,
where $L$ may be infinity.\footnote{In case $L$ isn't infinity, we ignore for now what happens with
particles which are close enough to the borders $\pm L$ such that the border is within its compact
support radius.} The integral in one dimension is then

\begin{align}
    \int_{-L}^{L} (x - x_i) \psi_i(x) \de x
    = \int_{-L}^{L} (x - x_i) \psi(x - x_i) \de x
    = \int_{-L+x_i}^{L+x_i} s \psi(s) \de s
 \end{align}

where first equality makes use of the other notation for $\psi$, and the last equality follows from
using the substitution $s = x - x_i$. We can now integrate by parts and show that the integral
evaluates to zero:

\begin{align}
    \int_{-L+x_i}^{L+x_i} s \psi(s) \de s &=
    \left[
        s \underbrace{\int_{-L+x_i}^{L+x_i} \psi(s) \de s}_{=\int_{-L}^{L} \psi_i(x) \de x = V_i }
    \right]_{s = -L+x_i}^{L+x_i} -
    \int_{-L+x_i}^{L+x_i} \left[
    \underbrace{\frac{\del}{\del s} s}_{= 1}
    \underbrace{\int_{-L+x_i}^{L+x_i} \psi(s) \de s}_{= V_i} \right] \de s \\
    &= \left[ s V_i \right]_{s = -L+x_i}^{L+x_i} - \int_{-L+x_i}^{L+x_i} V_i \de s
    = \left[ s V_i  - s V_i \right]_{s = -L+x_i}^{L+x_i} = 0
\end{align}

Using the partitions of unity, we can now derive concrete expressions for finite volume particle
methods. The partitions of unity will be applied to discretize the conservation laws we want to
solve onto particles.

%% file: main/Meshless/ML-2-methods.tex
\chapter{Finite Volume Particle Methods}\label{chap:FVPM}

Having discussed the partition of unity and the associated particle volumes, we can now move on
towards deriving the actual finite volume particle methods for hyperbolic conservation laws. The
goal is to derive mesh-free methods to solve hyperbolic conservation laws using arbitrary
Lagrangian or Eulerian particles as discretization elements.

To my knowledge, two versions of finite volume particle methods are mentioned in astrophysical
literature to date: One which is based on the works of
\citet{lansonRenormalizedMeshfreeSchemes2008a} and \citet{lansonRenormalizedMeshfreeSchemes2008} ,
and is used in the astrophysical context by \citet{gaburovAstrophysicalWeightedParticle2011},
\citet{hopkinsGIZMONewClass2015}, and \citet{hubberGANDALFGraphicalAstrophysics2018}. In this work,
this version will be referred to as the ``Hopkins version''. The other version, described in
\citet{ivanovaCommonEnvelopeEvolution2013} will be referred to as the ``Ivanova version''.

\section{Finite Volume Particle Methods Following \citet{hopkinsGIZMONewClass2015}}
\label{chap:meshless-hopkins}

Since the method is intended to be used with Lagrangian particles, i.e. particles co-moving with
the fluid, it is sensible to start with a co-moving description of a hyperbolic conservation law.
A Galilei-invariant hyperbolic conservation law can be written in a moving frame of reference as

\begin{align}
    \ddt \U + \ddxalpha \F_\alpha = 0
\end{align}

where the Einstein sum convention is applied over repeating Greek indices, which here denote
dimensional components of the terms. The components of the state vector $\U$ and the flux tensor
$\F$ are also taken to be in the moving frame.
The derivatives $\ddt$ and $\ddx$ denote the partial derivatives w.r.t. time and space,
respectively, in the moving frame.

Suppose the frame moves with a velocity $\V^{ref}$ w.r.t. the rest frame. The coordinate transform
to the moving frame is given by $\x' = \x - \V^{ref} t$, where $\x$ are the coordinates of the rest
frame. For an arbitrary differentiable function $f(x, t)$, the co-moving derivatives are

\begin{align*}
 \ddt f(x', t) &= \lim_{\Delta t \rightarrow 0} \frac{f(x', t+\Delta t) - f(x', t)}{\Delta t} \\
    &= \lim_{\Delta t \rightarrow 0}
        \frac{f(x - v^{ref}(t + \Delta t), t+\Delta t) - f(x - v^{ref}t, t)}{\Delta t} \\
    &= \lim_{\Delta t \rightarrow 0}
        \frac{f(x - v^{ref}t, t) -
                \DELDX{f}\big|_{(x-v^{ref}t, t)} v^{ref} \Delta t +
                \DELDT{f}\big|_{(x-v^{ref}t, t)} \Delta t +
                f(x - v^{ref}t, t)}{\Delta t} \\
    &= \DELDT{f} - v^{ref} \DELDX{f} \\[.5em]
\ddx f(x', t) &= \lim_{\Delta x \rightarrow 0} \frac{f(x', t+\Delta t) - f(x', t)}{\Delta x} \\
    &= \lim_{\Delta x \rightarrow 0}
        \frac{f(x - v^{ref}t + \Delta x, t) - f(x - v^{ref}t, t)}{\Delta x} \\
    &= \lim_{\Delta x \rightarrow 0}
        \frac{f(x - v^{ref}t, t) -
                \DELDX{f}\big|_{(x-v^{ref}t, t)} v^{ref} \Delta x +
                f(x - v^{ref}t, t)}{\Delta x} \\
    &= \DELDX{f}
\end{align*}

Hence in the rest frame, the conservation law reads as

\begin{align}
    \deldt{\U} + \deldxalpha \F_\alpha - \deldxalpha \V^{ref}_{\alpha} \U = 0 \ .
    \label{eq:conservation-law-reference-frame}
\end{align}

The additional term $\deldxalpha \V^{ref}_\alpha \U$ simply corresponds to the state $\U$ being
advected with the (constant) coefficient $-\V^{ref}$.

\subsubsection{Obtaining a Weak Solution}

To obtain a weak Galerkin-type solution, we multiply the conservation law in the moving frame with
an arbitrary function $\phi(\x, t)$:

\begin{align}
    \left(\DDT{\U} + \DDXALPHA{\F_\alpha} \right) \phi = 0 \ .
\end{align}

We add the following requirements for $\phi$:
\begin{itemize}
 \item $\phi$ is differentiable, and therefore smooth
 \item we demand $\phi$ to be Lagrangian, i.e. $\ddt \phi = 0$
 \item $\phi$ has compact support over the spatial domain $V$, i.e. $\phi(\x) = 0$ for $\x \notin
V$. Since $\phi$ is also required to be smooth, it follows that $\phi(\x) = 0$ along the domain
boundary $\del V$.
\end{itemize}

The spatial domain $V$ can be some arbitrary closed volume, and may be infinite. We now integrate
over the entire $V$:

\begin{align}
    \int_V \left(\DDT{\U} + \DDXALPHA{\F_\alpha} \right) \phi \ \de V & = 0 \\
    = \int_V \left(\DDT{\U} \phi + \DDXALPHA{\F_\alpha} \phi \right) \ \de V &
    \label{step:hopkins-integral-start}
\end{align}

To proceed, we make use of the product rule for each summand in the integrand:

\begin{align}
    \ddt ( \U \phi ) &= \DDT{\U}\phi + \U \underbrace{\DDT{\phi}}_{ = 0} = \DDT{\U} \phi \\
    \ddxalpha ( \F_\alpha \phi ) &= \DDXALPHA{\F_\alpha}\phi + \F_\alpha \DDXALPHA{\phi}
\end{align}

To simplify the integral further, we make use of Gauss divergence theorem for the flux term:

\begin{align}
    \int_V \ddxalpha (\F_\alpha \phi) \ \de V =
        \oint_{\delta V} \F_\alpha \phi \hat{n}_\alpha \de (\del V) = 0
\end{align}

where the final equality follows from our demands leading to $\phi = 0$ along the boundary $\del V$.
$\hat{n}_\alpha$ is the unit normal vector along the domain boundary surface $\del V$.

Inserting the found expressions

\begin{align}
    \DDT{\U} \phi &= \ddt ( \U \phi ) \\
    \int_V \DDXALPHA{\F_\alpha}\phi \ \de V &= - \int_V \F_\alpha \DDXALPHA{\phi} \de V
\end{align}

into eq.~\ref{step:hopkins-integral-start} leaves us with

\begin{align}
    \ddt \int_V \U \phi \ \de V - \int_V \F_\alpha \DDXALPHA{\phi} \ \de V = 0 \ .
\end{align}

\subsubsection{Using the Partition of Unity}

Now let's express the functions $(\U \phi)$ and $\left( \F_\alpha \DDXALPHA{\phi}  \right)$ using
the partition of unity interpolation (eq.~\ref{eq:psi-interpolation}) and re-arrange the equation
into a more convenient form:

\begin{align}
    & \ddt \int_V \U \phi \ \de V - \int_V \F_\alpha \DDXALPHA{\phi} \ \de V = 0 \\
    &= \ddt \int_V \sum_i \U_i \phi_i \psi_i(\x) \de V -
    \int_V \sum_i \left( \F_{\alpha,i} \DDXALPHA{\phi} \big|_{\x = \x_i} \right) \psi_i(\x) \ \de V
\\
    &= \ddt \left[\sum_i \U_i \phi_i \int_V \psi_i(\x) \de V \right] -
    \sum_i \left( \F_{\alpha,i} \DDXALPHA{\phi} \big|_{\x = \x_i} \right) \int_V \psi_i(\x) \ \de V
\\
    &= \ddt \left[\sum_i \U_i \phi_i V_i \right] -
    \sum_i \left( \F_{\alpha,i} \DDXALPHA{\phi} \big|_{\x = \x_i} \right) V_i \\
    &= \sum_i \left[ \phi_i \ddt (\U_i V_i) -
    V_i \F_{\alpha,i} \DDXALPHA{\phi} \big|_{\x = \x_i} \right] = 0
\label{step:intermediate-hopkins1}
\end{align}

where in the third line the definition of the particle associated volumes
(eq.~\ref{eq:psi-integral-volume}) was used, and the final equality made use of our demand for the
arbitrary function $\ddt \phi = 0$ to be Lagrangian.

We now express the gradient of the arbitrary test function $\DDXALPHA{\phi}$ using the second order
accurate least squares discrete gradient expression from
\cite{lansonRenormalizedMeshfreeSchemes2008a}:

\begin{align}
	\frac{\del}{\del x_{\alpha}} f(\x) \big{|}_{\x_i} &=
	\sum_j \left( f(\x_j) - f(\x_i) \right) \psitilde_j^\alpha (\x_i) + \order(h^2)
\label{eq:gradient} \\
	\psitilde_j^\alpha (\x_i) &= \sum_{\beta = 1}^{\beta=\nu} \mathcal{B}_i^{\alpha \beta}
	(\x_j - \x_i)^\beta \psi_j(\x_i) \label{eq:psitilde} \\
	\mathcal{B}_i &= \mathcal{E_i} ^ {-1} \label{eq:matrix_B}\\
	\mathcal{E}_i^{\alpha \beta} &= \sum_j (\x_j - \x_i)^\alpha (\x_j - \x_i)^\beta \psi_j(\x_i)
\label{eq:matrix_E}
\end{align}

where $\alpha$ and $\beta$ again represent the coordinate components for $\nu$ dimensions, and
$\mathcal{B}$ and $\mathcal{E}$ are symmetrical $\nu \times \nu$ matrices.
Inserting expression~\ref{eq:gradient} into the term involving the flux, we obtain

\begin{align}
& \sum_i V_i \F_{\alpha, i} \DDXALPHA{\phi}\big|_{\x = \x_i} =
    \sum_i V_i \F_{\alpha, i} \sum_j (\phi_j - \phi_i) \psitilde_j^\alpha (\x_i) = \\
&= \sum_{i,j} V_i \F_{\alpha, i} \phi_j \psitilde_j^\alpha (\x_i) -
    \sum_{i,j} V_i \F_{\alpha, i} \phi_i \psitilde_j^\alpha (\x_i) \\
&= \underbrace{\sum_{i,j} V_j \F_{\alpha, j} \phi_i \psitilde_i^\alpha (\x_j)}_{\text{switched } i
\leftrightarrow j} -
    \sum_{i,j} V_i \F_{\alpha, i} \phi_i \psitilde_j^\alpha (\x_i) \\
&= - \sum_{i,j} \phi_i \left[ V_j \F_{\alpha, j} \psitilde_i^\alpha (\x_j)
    - V_i \F_{\alpha, i} \psitilde_j^\alpha (\x_i) \right]
\end{align}

Inserting this into eq.~\ref{step:intermediate-hopkins1}, we obtain

\begin{align}
&\sum_i \left[ \phi_i \ddt (\U_i V_i) -
    V_i \F_{\alpha,i} \DDXALPHA{\phi} \big|_{\x = \x_i} \right] =  \\
&\sum_i \phi_i \left[ \ddt (\U_i V_i) + \sum_j
        \left( V_i \F_{\alpha, i} \psitilde_j^\alpha (\x_i)
        - V_j \F_{\alpha, j} \psitilde_i^\alpha (\x_j) \right)
    \right] = 0 \ .
\end{align}

This expression must hold for an arbitrary $\phi_i$ and for all $i$, therefore the expression
between the brackets must vanish:

\begin{align}
    \ddt (\U_i V_i) + \sum_j
        \left( V_i \F_{\alpha, i} \psitilde_j^\alpha (\x_i)
        - V_j \F_{\alpha, j} \psitilde_i^\alpha (\x_j) \right)
    = 0 \ .
\end{align}

So the time derivative of a conserved quantity $\U_i V_i$ of particle $i$ is described by the
exchange of fluxes with all its neighboring particles $j$. This flux exchange term however requires
a bit further discussion. While the fluxes $\F_i$ and $\F_j$ should in principle be the fluxes of
the corresponding states $\U_i$, $\U_j$ evaluated at the particle positions $\x_i$ and $\x_j$, i.e.
$\F_{i,j} = \F(\U(\x_{i,j}, t))$, this evaluation requires the evolution of the states $\U_{i}$,
$\U_j$ to be known at the particle positions, which in general is not available. Indeed the purpose
of the entire method is to give us a numerical scheme to find precisely this solution. A possibility
would of course be to take a cue from the approach that the MUSCL-Hancock method uses, and try and
find an approximation for the evolved states to use. This would however require yet another
interaction loop where every particle $i$ interacts with all its neighbors $j$ in order to find the
approximate expression first. A more viable solution is to ``move'' the problem: Instead of
evaluating the fluxes at the individual particle positions, we approximate the problem by treating
the particle positions $\x_{i}$, $\x_j$ as centers of (irregular) cells, and take the fluxes to be
the solution to the Riemann problem centered at the common cell boundary. The ``cell boundary'' is
positioned at some $\x_{ij}$, and the ``left'' and ``right'' states for the Riemann problem are
extrapolated states $\U_i$ and $\U_j$ at the position $\x_{ij}$. Since we already need to compute
the terms required for a general gradient (eq.~\ref{eq:gradient}),  the gradients of the states
$\U_{i}$, $\U_j$ at particle positions are readily available, and we can use them to extrapolate
the states at the ``cell interface'' $\x_{ij}$.

Taking the solution of the centered Riemann problem at the initial discontinuity between the left
and right state to be the solution for the fluxes $\F_{i}$, $\F_j$ at the ``cell interface'', then
the resulting fluxes for the ``left'' particle $i$ and the ``right'' particle $j$ are identical:

\begin{align}
    \F_i = \F_j
    = RP
    \left(\U_i + (\x_{i} - \x_{ij})_\alpha \DDXALPHA{\U_i}, \
    \U_j + (\x_{j} - \x_{ij})_\alpha \DDXALPHA{\U_j} \right)
    \equiv \F_{ij}
\end{align}

To obtain the actual solution of the Riemann problem, the either the exact Riemann solver (see Section~\ref{chap:exact-riemann-solver}) or an approximate one (see Section~\ref{chap:riemann-approximate}) can be used.
We can denote the ``effective surfaces'' \Aij as

\begin{align}
    \Aijm^\alpha \equiv V_i \psitilde_j^\alpha(\x_i) - V_j \psitilde_i^\alpha (\x_j) \ .
    \label{eq:HopkinsAij}
\end{align}

Note that the ``effective surfaces'' \Aij are vector quantities, since the $\psitilde$, which
are given by eq.~\ref{eq:psitilde}, are vector quantities as well. Furthermore, \Aij has the
dimension of a surface. Inserting the term into our previous result, we arrive at the update formula

\begin{align}
\boxed{
    \ddt (\U_i V_i) + \sum_j \F_{\alpha, ij} \Aijm^\alpha = 0 \label{eq:meshless-Hopkins}
}
\end{align}

This update formula tells us that we can evolve the system by exchanging fluxes between particles:
The rate of change of a state $\U_i$ (multiplied by the particle's volume $V_i$) of any particle
$i$ is given by the exchange of fluxes with other particles $j$ through ``effective'' surfaces
\Aij. In a method that is based on cells, the surfaces would be the cell surfaces. Here we were able
to obtain expressions which both ``look'' like surfaces, i.e. have a finite size and a direction,
and have the dimensionality of a surface. This is part of the magic of finite volume particle
methods: we were able to obtain surface-like quantities, while not constructing any cells
whatsoever. After all, all volume is shared between all particles in a weighted fashion. So in a
sense, we have obtained a finite volume method, where the volumes are overlapping.

The update formula can be written as an explicit update:

\begin{align}
    \U_i^{n+1} = \U_i^{n} + \frac{\Delta t}{V_i} \sum_j  \F_{\alpha, ij} \Aijm^\alpha
\label{eq:meshless-Hopkins-explicit}
\end{align}

Note that the explicit update formula assumed that the particle associated volumes $V_i$ are
constant w.r.t. time. This is correct for static particles, but in general is not valid for
Lagrangian particles that shift their positions. Practice shows however that this assumption is
sufficiently adequate, and errors can be kept at bay with a more restrictive choice of Courant
number.

\section{Finite Volume Particle Methods Following \citet{ivanovaCommonEnvelopeEvolution2013}}

We now derive the expression for the mesh-free finite volume particle method as describe in
\citet{ivanovaCommonEnvelopeEvolution2013}. To do so, we will make use of similar techniques as for
the Hopkins version, like using the Gauss divergence theorem, an arbitrary test function, and the
partition of unity. A central difference however is that we will not be using the discrete gradient
expressions given in eq.~\ref{eq:gradient} in order to derive the update formula, but keep the
analytical expressions throughout.

Once again we start the derivation by looking for a weak solution of a hyperbolic conservation law
by first multiplying it with an arbitrary test function $\phi(\x, t)$. Again we demand that

\begin{itemize}
 \item $\phi(\x, t)$ is differentiable, and therefore also smooth
 \item $\phi(\x, t)$ has compact support over the domain of interest, i.e. for a domain $T \times V
\in \mathds{R}_0^+ \times \mathds{R}^3$: $\phi(\x, t) = 0$ if $\x \notin V$ or $t \notin T$. Since
$\phi$ is required to be smooth, it follows that $\phi(\x, t) = 0$ at the boundaries of the
domain $\del T$ and $\del V$, i.e. if  $\x \in \del V$ or $t \in \del T$.
\end{itemize}

We integrate the conservation law multiplied by $\phi$ over the entire domain. However this time
around, we include both the spatial and the temporal domain:

\begin{align}
 \int_T \int_V \left[ \DELDT{\U} + \deldxalpha{\F_\alpha} \right] \phi(\x, t) \de t \de V = 0
 \label{step:ivanova1}
\end{align}

Since $\phi$ has compact support and evaluates to zero along the boundaries $t = 0, T$, we can write

\begin{align}
    \int_T \deldt(\U \phi) \de t &= \U \phi|_{t=0}^T = 0 \\
    &= \int_T \DELDT{\U} \phi \de t + \int_T \U \DELDT{\phi} \de t \\
    \Rightarrow \int_T \DELDT{\U} \phi \de t &= - \int_T \U \DELDT{\phi} \de t
\end{align}

Similarly we make use of the compact support of $\phi$ along $\x \in \del V$ for the flux term
after applying the Gauss divergence theorem:

\begin{align}
\int_V \deldxalpha (\F_\alpha \phi) \de V &=
    \oint_{\del V} (\F_\alpha \underbrace{\phi}_{= 0}) \hat{n}_\alpha \de (\del V) = 0 \\
&= \int_V \DELDXALPHA{\F_\alpha} \phi \de V + \int_V \F_\alpha \DELDXALPHA{\phi} \de V \\
\Rightarrow
\int_V \DELDXALPHA{\F_\alpha} \phi \de V &= - \int_V \F_\alpha \DELDXALPHA{\phi} \de V
\end{align}

Inserting these findings into eq.~\ref{step:ivanova1}, we get

\begin{align}
    \int_T \int_V \left[ \DELDT{\U} + \DELDXALPHA{\F_\alpha} \right] \phi(\x, t) \de t \de V &= \\
    \int_T \int_V \left[ \DELDT{\phi} \U + \DELDXALPHA{\phi} \F_\alpha \right] \de t \de V &= 0
\end{align}

\subsubsection{Using the Partition of Unity}

Using the partition of unity interpolation (eq.~\ref{eq:psi-interpolation}), we can write the term
of the integral containing the state vector $\U$ as

\begin{align}
\int_V \int_T \left[ \DELDT{\phi} \U \right] \de V \de T &=
    \int_V \int_T \left[ \sum_i \left( \DELDT{\phi_i} \U_i \right) \psi_i(\x) \right] \de V \de t \\
&= \int_T \sum_i \U_i \DELDT{\phi_i} \int_V \psi_i(\x) \de V \de t \\
&= \int_T \sum_i \U_i \DELDT{\phi_i} V_i \de t
\end{align}

We can once again make use of the compact support of $\phi$ to express

\begin{align}
\int_T \sum_i \deldt(\U_i V_i \phi_i) \de t &= \sum_i \U_i V_i \phi_i \big|_{t=0}^T = 0\\
&= \int_T \sum_i \left[ \deldt(\U_i V_i) \phi_i + \U_i V_i \DELDT{\phi_i} \right] \de t \\
\Rightarrow \int_T \sum_i \U_i V_i \DELDT{\phi_i} \de t &=
    - \int_T \sum_i \deldt(\U_i V_i) \phi_i \de t
\end{align}

giving us

\begin{align}
\int_V \int_T \left[ \DELDT{\phi} \U \right] \de V \de t &=
    \int_T \sum_i \U_i \DELDT{\phi_i} V_i \de t =
    - \int_T \sum_i \deldt(\U_i V_i) \phi_i \de t \label{step:ivanova-finished-U}
\end{align}

For the term containing the flux, we also express it using the partition of unity interpolation
(eq.~\ref{eq:psi-interpolation}) and keep modifying it into a convenient shape:

\begin{align}
\int_T \int_V \DELDXALPHA{\phi} \F_\alpha \de t \de V &=
    \int_T \int_V \F_\alpha \deldxalpha (\sum_i \phi_i \psi_i) \ \de t \ \de V
\\
&= \sum_i \int_T \int_V \F_\alpha \phi_i \DELDXALPHA{\psi_i} \ \de t \ \de V
\\
&= \sum_i \int_T \int_V \F_\alpha \phi_i \DELDXALPHA{\psi_i}
    \times \underbrace{ \sum_j \psi_j(\x)}_{\text{multiply by }1 = \sum_j \psi_j(\x)} \de t \de V
\\
&= \sum_{i,j} \int_T \int_V \F_\alpha \phi_i \psi_j \DELDXALPHA{\psi_i} \ \de t \ \de V
\\
&= \sum_{i,j} \int_T \int_V \left[
    \F_\alpha \phi_i \psi_j \DELDXALPHA{\psi_i} +
    \underbrace{0}_{\text{add zero}}
    \right] \ \de t \ \de V
\\
&= \sum_{i,j} \int_T \int_V \left[
\F_\alpha \phi_i \psi_j \DELDXALPHA{\psi_i} +
\underbrace{-\F_\alpha \phi_i \DELDXALPHA{\psi_j} \psi_i}_{=\text{0, will be shown later}}
\right] \ \de t \ \de V
\label{step:ivanova-add-zero}
\\
&= \sum_{i,j} \int_T \int_V \F_\alpha \phi_i \left[
\psi_j \DELDXALPHA{\psi_i} - \psi_i \DELDXALPHA{\psi_j}
\right] \ \de t \ \de V
\\
&= \sum_{i,j} \int_T \phi_i \int_V \F_\alpha \ \de \Aijm^\alpha \ \de t
\label{step:ivanova-intermediate-F}
\end{align}

where we introduced

\begin{align}
\de \Aijm^\alpha \equiv
    \left[ \psi_j \DELDXALPHA{\psi_i} - \psi_i \DELDXALPHA{\psi_j} \right] \de V \ .
\end{align}

It remains to be shown that the term added in eq.~\ref{step:ivanova-add-zero} is indeed zero:

\begin{align}
\sum_{i,j} \F_\alpha \phi_i \psi_i \DELDXALPHA{\psi_j} =
\sum_{i} \F_\alpha \phi_i \psi_i \sum_j \DELDXALPHA{\psi_j} =
\sum_{i} \F_\alpha \phi_i \psi_i
    \underbrace{\deldxalpha
        \underbrace{\left(\sum_j \psi_j \right)}_{= 1}
    }_{= 0} = 0
\end{align}

This term was added so that the resulting expression is anti-symmetric between particle $i$ and $j$, i.e. $\de \Aijm^\alpha = - \de \mathcal{A}_{ji}^\alpha$.

\subsubsection{Solving the Volume Integral}

To solve the volume integral in eq.~\ref{step:ivanova-intermediate-F}, further approximations are
necessary. In a first step, we may approximate the integral as a single point quadrature:

\begin{align}
    \int_V \F_\alpha \de \Aijm^\alpha \approx \F_{\alpha, ij} \Aijm^\alpha
\end{align}

where $\Aijm^\alpha$ is given by

\begin{align}
    \Aijm^\alpha = \int_V
    \left[ \psi_j \DELDXALPHA{\psi_i} - \psi_i \DELDXALPHA{\psi_j} \right] \de V
    \label{eq:IvanovaAij-analytical}
\end{align}

and $\F_{ij}$ is again the flux given by the solution of the Riemann problem centered at some
``cell interface'' located at some $\x_{ij}$ between particle $i$ and $j$ with left and right
states being gradient extrapolated states

\begin{align}
    \F_{ij}
    \equiv RP
    \left(\U_i + (\x_{i} - \x_{ij})_\alpha \DDXALPHA{\U_i}, \
    \U_j + (\x_{j} - \x_{ij})_\alpha \DDXALPHA{\U_j} \right) \ .
\end{align}

To obtain the actual solution of the Riemann problem, the either the exact Riemann solver (see Section~\ref{chap:exact-riemann-solver}) or an approximate one (see Section~\ref{chap:riemann-approximate}) can be used.
An approximate discretization for \Aij can be found by performing the integral over a Taylor
expansion of the $\DELDXALPHA{\psi_{i,j}}$ terms to second order. Let's show this only on one
integrand:

\begin{align}
    \int_V \psi_j \DELDXALPHA{\psi_i} \de V &=
    \int_V \psi_j \left[
        \DELDXALPHA{\psi_i(\x_i)}
        + (\x - \x_i) \frac{\del^2 \psi_i(\x_i)}{\del x_\alpha^2}
        + \order((\x - \x_i)^2)
    \right] \de V
\\
&= \DELDXALPHA{\psi_i(\x_i)} \int_V \psi_i(\x) \de x +
    \frac{\del^2 \psi_i(\x_i)}{\del x_\alpha^2}
        \underbrace{\int_V (\x - \x_i) \psi_i(\x) \de V}_{= 0}
    + \order((\x - \x_i)^2)
\\
&= \DELDXALPHA{\psi_i(\x_i)} V_i + \order(h^2)
\end{align}

The integral over the first order term being zero was already shown in
Section~\ref{chap:particle-volume}. This result allows us to write

\begin{align}
\Aijm^\alpha = V_i \DELDXALPHA{\psi_j(\x_i)} - V_j \DELDXALPHA{\psi_i(\x_j)} + \order(h^2)
\label{eq:IvanovaAij}
\end{align}

The full expressions for the gradients $\DELDXALPHA{\phi}$ assuming some kernel function $W(\x, h)$
are given in Appendix~\ref{app:psi-gradients-full}.
Combining all these results and inserting them into eq.~\ref{step:ivanova1}, we arrive at

\begin{align}
\int_T \int_V \left[ \DELDT{\U} + \DELDXALPHA{\F_\alpha} \right] \phi_i \de t \de V
= \int_T \sum_i \phi_i \left[ \deldt (\U_i V_i) + \sum_j \F_{\alpha,ij} \Aijm^\alpha \right] \de t =
0
\end{align}

This expression must hold for all $i$ and all arbitrary test functions $\phi$, and therefore

\begin{align}
\boxed{
    \deldt (\U_i V_i) + \sum_j \F_{\alpha,ij} \Aijm^\alpha = 0 \label{eq:meshless-Ivanova}
}
\end{align}

The method formally takes the same shape as the Hopkins version (eq.~\ref{eq:meshless-Hopkins}),
albeit with a different expression for the ``effective surfaces'' \Aij. The two expressions for \Aij and the resulting methods will be compared in Section~\ref{chap:meshless-comparison}.

\section{Physical Interpretation of the Effective Surfaces}

In this Section, the physical interpretation of the expressions of the \Aij in terms of physical
interpretation is discussed. The Ivanova \Aij, given in eq.~\ref{eq:IvanovaAij}, stem from an
approximate evaluation of the following integral:

\begin{align}
& \int_V \F_\alpha \ \de \Aijm^\alpha \\
\text{with } \quad &
\de \Aijm^\alpha \equiv \left[ \psi_j \DELDXALPHA{\psi_i} - \psi_i \DELDXALPHA{\psi_j} \right] \de
V
\end{align}

Effectively, this describes a local exchange of the divergences of fluxes between two particles $i$
and $j$ at every point in space, which is then summed through the volume integral over the entire
domain. The divergences of fluxes enter the equations from the very beginning, as they are a
component of the hyperbolic conservation law that is being solved. A part of the flux $\F(\x)$ at
any point $\x$ is associated with particle $i$ through its partition of unity at that point, which
is given by $\psi_i(\x)$. In the above equations, the divergences of \emph{weighted} fluxes have
simply been re-written as the flux components multiplied by the divergence of the respective
weights $\psi_i$ and $\psi_j$ through use of Gauss' theorem. They maintain their physical interpretation though: The term still describes the divergence of weighted fluxes at each point in space.

Contrary to finite volume methods using cells, where the domain is split exclusively between the
discretization elements, a core component of the finite volume particle methods is that the volume
is shared among particles at each point in space. While for cells we can only take into account the
divergence of fluxes across each cell face, we can't do that here, as there are no cell faces, and
there are no regions in space which are exclusively assigned to a single particle. So we need to
consider the flux exchanges, or more precisely the exchanges of divergences of fluxes between two
particles, at every point in space. The (divergence of) fluxes ``leaving'' particle $j$ act as a
source term for the fluxes associated with particle $i$. Hence the source term for particle $i$
will
be proportional to $\psi_j(\x) \F(\x)$, and vice versa.

In summary, the Ivanova surfaces describe the net sum of the exchanges in the weighted flux
divergences over all points in space. The fluxes are weighted due to the nature of the
discretization technique, which is the partition of unity. Similarly, due to the partition of unity
we need to take into account the exchange of flux divergences between two particles at each point
in
space, since each point in space is shared among particles.

As for the Hopkins version of the surfaces, the physical interpretation is a bit trickier. In fact,
I would argue that the terms for the surfaces don't have a physical representation. To arrive at
the expression for the surfaces given in eq.~\ref{eq:HopkinsAij}, the analytical gradients of the
arbitrary test function $\phi$ which is used to derive a weak solution are replaced by the
approximate discrete gradient expressions given in eq.~\ref{eq:gradient}. $\phi$ is arbitrary and
physically meaningless, and hence its derivatives are as well. The test functions $\phi$ later get
factored out, but the terms $\psitilde$, which encapsulate the approximate discrete gradients,
remain. So we are left with remnants of the approximate gradient expressions, stemming from the
originally analytical gradients of a physically meaningless function, along with volume integrated
fluxes, which can't be a physically meaningful quantity.

\section{Conservation Properties of the Methods}\label{chap:meshless-conservation-properties}

It is vital to examine the conservation properties of the underlying numerical methods in order to
understand how the solutions will behave, and whether the solutions will be correct in the first
place. In this Section, we'll look into the local conservation, the closure condition, and the
conservation of angular momentum of the methods.

\subsection{Local Conservation}\label{chap:meshless-conservation-local}

Consider for example a case where all fluid quantities are equal and in equilibrium throughout the
entire domain. The correct (analytical) solution is simply the fluid maintaining its original state
everywhere indefinitely. Considering that the resulting update formula for both the Hopkins and
Ivanova versions reads as

\begin{align}
    \U_i^{n+1} = \U_i^n + \frac{\Delta t}{V_i} \sum_j \F_{\alpha, ij} \Aijm^\alpha \ ,
\end{align}

in order for the numerical method to maintain the correct solution, i.e. $\U_i^{n+1} = \U_i^n$, we
must ensure that there are no net fluxes in the entire domain, i.e.

\begin{align}
    \sum_{ij}  \F_{\alpha, ij} \Aijm^\alpha = 0
\end{align}

This condition is easily satisfied by ensuring that everything that leaves a particle is exactly
received by another particle, i.e.

\begin{align}
    \F_{\alpha, ij} \Aijm^\alpha = -\F_{\alpha, ji} \mathcal{A}_{ji}^\alpha \ .
\end{align}

Since in both the Ivanova and the Hopkins version we approximated $\F_{ij} = \F_{ji}$ as the
solution to the centered Riemann problem between the particles $i$ and $j$, the condition
simplifies further for the effective surfaces \Aij to be anti-symmetric:

\begin{align}
    \Aijm^\alpha = -\mathcal{A}_{ji}^\alpha
\end{align}

which is satisfied in both versions of expressions for the effective surfaces.

\subsection{Closure Condition}\label{chap:meshless-conservation-closure}

\begin{figure}
 \centering
 \includegraphics[width=.5\textwidth]{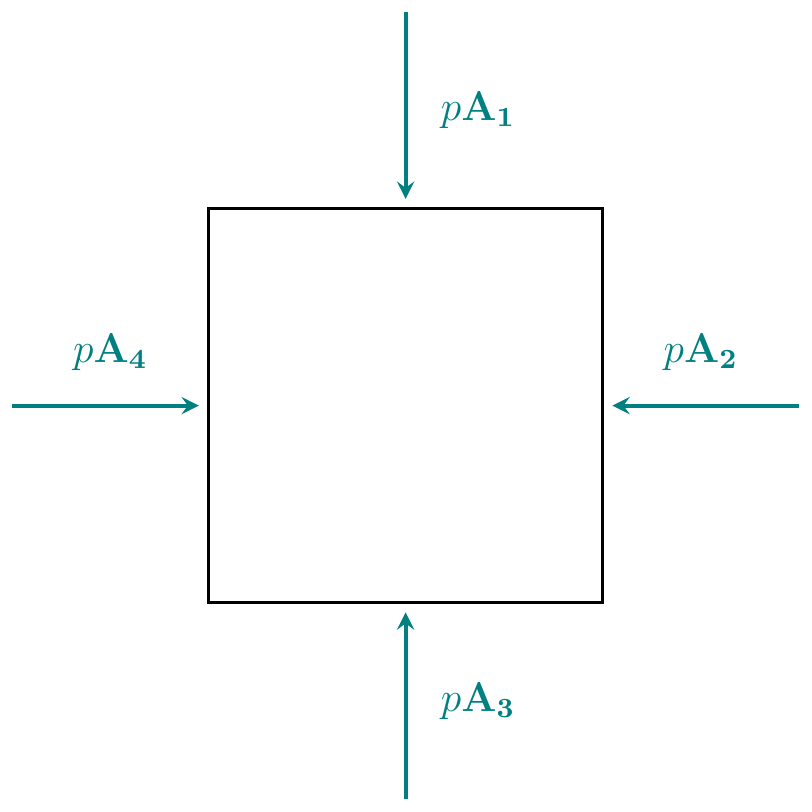}%
 \caption{
A single volume element surrounded by equal pressure $p$ on each side. The net force it experiences
is the sum of the pressure multiplied by the volumes, surfaces: $\mathbf{F}_{tot} = p \mathbf{A}_1
+
 p \mathbf{A}_2 +  p \mathbf{A}_3 +  p \mathbf{A}_4$. If the vector sum of the surfaces
$\mathbf{A}_j$ doesn't sum up to exactly zero, then the volume element experiences an unphysical
nonzero net force.
 }
 \label{fig:pressure-equilibrium}
\end{figure}

A second very important property of the numerical methods is to ensure that the vector sum of all
cell boundary areas, or in the case of FVPM the effective surfaces between all neighbors, is zero,
i.e.

\begin{align}
 \sum_j \Aijm  =  0 \ .
\end{align}

This condition ensures that no spurious forces and accelerations are being introduced. Consider for
example a fluid which is static and has equal pressure everywhere. Then in the case of a single
volume element, shown in two dimensions in Figure~\ref{fig:pressure-equilibrium}, the net force
experienced by the cell is the sum of the surrounding pressure multiplied by the surface upon which
it acts:

\begin{align}
 \mathbf{F}_{tot} = \sum_j^{\mathrm{all\ surfaces}} p \mathbf{A}_j
\end{align}

If the boundary areas don't have a vector sum of exactly zero, then the net force which the volume
element experiences will also be nonzero. In our example where pressure is equal everywhere, that
is incorrect and unphysical.

It can be shown that the analytical expression for the Ivanova surfaces,
eq.~\ref{eq:IvanovaAij-analytical}, satisfy the closure condition. Using the product rule, we can
write

\begin{align}
 \deldx (\psi_j \psi_i) = \DELDX{\psi_j} \psi_i + \psi_j \DELDX{\psi_i}
\end{align}

and use that to replace the first term in eq.~\ref{eq:IvanovaAij-analytical}:

\begin{align}
   \Aijm &= \int_V \left[ \psi_j \DELDX{\psi_i} - \psi_i \DELDX{\psi_j}  \right] \de V  \\
&= \int_V \left[ \deldx(\psi_j \psi_i) - 2 \psi_i \DELDX{\psi_j}  \right] \de V
\label{eq:closure-midstep}
\end{align}

We can get rid of the first term, $\deldx(\psi_j \psi_i)$, through use of the divergence theorem
once again. The volume integral is transformed into a surface integral:

\begin{align}
    \int_V  \deldx(\psi_j \psi_i) \de V = \int_{\del V} \psi_i \psi_j \hat{\mathbf{n}} \del V
\end{align}

Since we demanded that the partitions of unity have compact support, i.e. $\psi(\x, \x_i, H) = 0 \
\forall \ |\x - \x_i| > H$ and the surface integral is taken along the enclosing surface of the
entire domain, which in general is outside of the compact support radius, the integral evaluates to
zero.

To show that the closure condition holds, we need to show that for some particle $i$ the sum of the
surfaces with all the particles $j$ it interacts with is zero. We now write this sum using the
modifications from eq.~\ref{eq:closure-midstep} and the fact that the first term of
eq.~\ref{eq:closure-midstep} evaluates to zero:

\begin{align}
   \sum_j \Aijm &=
   \sum_j \int_V \left[ \psi_j \DELDX{\psi_i} - \psi_i \DELDX{\psi_j}  \right] \de V  \\
&= \sum_j \int_V \left[- 2 \psi_i \DELDX{\psi_j}  \right] \de V \\
&= \int_V \Bigg[- 2 \psi_i
    \underbrace{\deldx
        \bigg( \underbrace{\sum_j \psi_j}_{= 1} \bigg)
    }_{= 0}
    \Bigg] \de V  = 0
\end{align}

which confirms the closure condition. Note however that this is only valid for the analytical
expression for the Ivanova effective surfaces, not the approximate ones we actually use, given in
eq.~\ref{eq:IvanovaAij}. The approximate ones in general do not satisfy the closure condition
exactly. To demonstrate this, we again take the sum of all effective surfaces over all interacting
particles $j$:

\begin{align}
    \sum_j \Aijm &= \sum_j \left[ V_i \DELDX{\psi_j(\x_i)} - V_j \DELDX{\psi_i(\x_j)} \right] \\
    &=  V_i  \underbrace{
        \deldx \big( \underbrace{\sum_j \psi_j(\x_i)}_{=1} \big)
        }_{= 0} - \sum_j V_j \DELDX{\psi_i(\x_j)}
    = - \sum_j V_j \DELDX{\psi_i(\x_j)}
\end{align}

which in general is non-zero. For it to be exactly zero, we would need a perfectly regular particle
configuration: All volumes $V_j$ need to be equal everywhere, and for every neighboring particle
$j_1$ of particle $i$ there needs to be a neighboring particle $j_2$ which is at the exact same
distance from particle $i$ as $j_1$, but in the exact opposite direction, so that the directions of
the gradients cancel out exactly. This suggests that the method will perform worse when non-regular
particle configurations are present, and may be a hint towards the necessity to employ an
additional scheme to keep the particle configurations somewhat regular throughout the simulation.

As for the Hopkins expression for the effective surfaces \Aij, I am not aware of a general proof
that the closure condition is satisfied. The difficulty in this case is that contrary to the
Ivanova version, we don't have a fully analytical expression readily available, as the approximate
gradient (eq.~\ref{eq:gradient}) is used to obtain the expression. It is however possible to
demonstrate, just like for the Ivanova version, that in a perfectly regular particle configuration
the sum of surfaces is exactly zero.

Assuming a perfectly regular particle configuration leads to all geometry related quantities being
equal:
\begin{itemize}
\item The partition of unity of a particle at a different particle's position is exactly equal to
the second particle's partition of unity at the first particle's position: $\psi_j(\x_i) =
\psi_i(\x_j)$
\item Similarly, all the normalizations at each particle's position, $\omega(\x_i) = \omega$, must
be equal. As a consequence, the particle volumes $V_i = 1 / \omega(\x_i) = V$ are identical as well.
\item All the matrices $\mathcal{E}_i$ (given in eq.~\ref{eq:matrix_E}) are identical, as the sums
$\oldsum_j (\x_j - \x_i)^\alpha(\x_j - \x_i)^\beta$ must be equal everywhere, hence $\mathcal{E}_i
= \mathcal{E}$.
\item As a consequence, the matrices $\mathcal{B} = \mathcal{B}_i = \mathcal{E}_i^{-1}$ are
identical everywhere as well.
\end{itemize}

In this special case, the effective surfaces can be written as

\begin{align}
 \Aijm^\alpha &= V_i \psitilde_j^\alpha(\x_i) - V_j \psitilde_i^\alpha (\x_j) \\
 &= V_i \mathcal{B}_i^{\alpha\beta}(\x_j - \x_i)^\beta \psi_j(\x_i) -
    V_j \mathcal{B}_j^{\alpha\beta}(\x_i - \x_j)^\beta \psi_i(\x_j) \\
 &= V \mathcal{B}^{\alpha\beta}(\x_j - \x_i)^\beta \psi_j(\x_i) -
    V \mathcal{B}^{\alpha\beta}(\x_i - \x_j)^\beta \psi_j(\x_i) \\
 &= 2 V \mathcal{B}^{\alpha\beta}(\x_j - \x_i)^\beta \psi_j(\x_i)
\end{align}

The sum of all surfaces, $\oldsum_j \Aijm$, then vanishes in the case of a perfectly regular
particle configuration (just as was the case for the Ivanova surfaces) because for each neighbor
$j_1$ there is a neighbor $j_2$ for which $(\x_{j1} - \x_i) = -(\x_{j2} - \x_i)$, giving us
$\oldsum_j \Aijm = 0$.

\subsection{Conservation of Angular Momentum}\label{chap:meshless-conservation-angular-momentum}

To discuss the conservation of angular momentum of the schemes, we will make use of the momentum
equation of the Euler equations without any external forces. Suppose the particles are moving with
some velocity $\mathbf{w} = \DDT{\x}$ be the particle velocity. Then the momentum equation in the
particles' frame of reference is given by (compare eq.~\ref{eq:conservation-law-reference-frame}):

\begin{align}
\deldt (\rho V) + \nabla \cdot \left( \rho \V \otimes (\V - \mathbf{w}) + p \mathds{1} \right) = 0
\end{align}

which we can write component-wise as:

\begin{align}
\deldt (\rho v^\alpha) + \frac{\del}{\del x^\beta} \left( \rho v^\alpha (v^\beta - w^\beta) + p
\delta^{\alpha\beta} \right) = 0
\end{align}

for each spatial component $\alpha$, where the Einstein sum convention over same Greek indices is
assumed again and $\delta^{\alpha \beta}$ is the Kronecker delta. Applying the update formula we
found found for both the Ivanova version (eq.~\ref{eq:meshless-Ivanova}) and the Hopkins version
(eq.~\ref{eq:meshless-Hopkins}), we find for each particle $i$:

\begin{align}
\deldt (V_i \rho_i v_i^\alpha)
= - \sum_j \left( \rho v^\alpha  (v^\beta - w^\beta)  + p \delta^{\alpha\beta} \right)_{ij}
\Aijm^\beta
= \deldt(m_i v_i^\alpha)
\label{eq:angular-momentum-mid}
\end{align}

This expression will be useful to verify whether the angular momentum $\mathbf{L}$ of the
system,

\begin{align}
\mathbf{L} = \sum_i \mathbf{L}_i = \sum_i \x_i \times m_i \V_i
\end{align}

remains constant, where the sum is performed over all particles $i$. To demonstrate whether
$\mathbf{L}$ is constant w.r.t. time, we check that its derivative w.r.t. time is zero:

\begin{align}
\ddt \mathbf{L} &=
    \sum_i \ddt \mathbf{L}_i = \sum_i \ddt \left( \x_i \times m_i \V_i \right) \\
&= \sum_i \left( \DDT{\x_i} \times m_i \V_i + \x_i \times \ddt (m_i \V_i) \right)
\end{align}

To return to the component-wise formulation, we use the three-dimensional Levi-Civita symbol $\epsilon^{\alpha \beta \gamma}$ to express the vector product of the first term:

\begin{align}
\left(\DDT{\mathbf{x}_i} \times \V_i \right)^\alpha &=
    (\mathbf{w}_i \times \V_i)^\alpha =
    \epsilon^{\alpha \beta \gamma} w_i^\beta v_i^\gamma
\end{align}

Similarly, using expression~\ref{eq:angular-momentum-mid}, we can write the second term as:

\begin{align}
\left(\mathbf{x}_i \times \ddt (m_i \V_i)\right)^\alpha &=
\left(\mathbf{x}_i \times - \sum_j \left( \rho \V \otimes (\V - \mathbf{w}) + p
\mathds{1}\right)_{ij} \cdot \mathbf{\mathcal{A}}_{ij} \right)^\alpha \\
&= -\sum_j \epsilon^{\alpha \beta \gamma}  x^\beta
\left( \rho v^\gamma (v^\kappa - w^\kappa) + p \delta^{\gamma\kappa} \right)_{ij} \Aijm^\kappa
\end{align}

giving us:

\begin{align}
\ddt \mathbf{L}_i =
    m_i \epsilon^{\alpha \beta \gamma} w_i^\beta v_i^\gamma
    -\sum_j \epsilon^{\alpha \beta \gamma} x^\beta
        \left( \rho v^\gamma (v^\kappa - w^\kappa) + p \delta^{\gamma\kappa} \right)_{ij}
\Aijm^\kappa
\end{align}

for each particle $i$. Then the derivative w.r.t. time of the entire system is given by:

\begin{align}
\sum_i \ddt \mathbf{L}_i &=
\sum_i \left[
    m_i \epsilon^{\alpha \beta \gamma} w_i^\beta v_i^\gamma
    -\sum_j \epsilon^{\alpha \beta \gamma} x^\beta
        \left( \rho v^\gamma (v^\kappa - w^\kappa)   + p \delta^{\gamma\kappa} \right)_{ij}
\Aijm^\kappa
\right] \\
&=  \epsilon^{\alpha \beta \gamma} \sum_i m_i w_i^\beta v_i^\gamma
-  \epsilon^{\alpha \beta \gamma} \sum_i x_i^\beta \sum_j \left[
   \left( \rho v^\gamma (v^\kappa - w^\kappa)  + p \delta^{\gamma\kappa} \right)_{ij} \Aijm^\kappa
\right]
\end{align}

Making use of our assumption that the flux term $\fc_{ij} = \fc_{ji} = \left( \rho v^\gamma
(v^\kappa - w^\kappa) + p \delta^{\gamma\kappa} \right)_{ij} $ is symmetric in $i$ and $j$, while
the surfaces \Aij are anti-symmetric, and the fact that both summations over indices $i$ and $j$ are over all particles (where we make use of the compact support of the $\psi$ to limit the number of neighboring particles that have a non-zero contribution to the sum over $j$), we can express the
second term as follows:

\begin{flalign}
&\epsilon^{\alpha \beta \gamma} \sum_i x_i^\beta \sum_j \left[
   \left( \rho v^\gamma (v^\kappa - w^\kappa) + p \delta^{\gamma\kappa} \right)_{ij} \Aijm^\kappa
\right] = \\
&\epsilon^{\alpha \beta \gamma} \sum_{i,j} x_i^\beta \left[
   \left( \rho v^\gamma (v^\kappa - w^\kappa) + p \delta^{\gamma\kappa} \right)_{ij} \Aijm^\kappa
\right] = \\
&\frac{1}{2} \cdot 2
\epsilon^{\alpha \beta \gamma} \sum_{i,j} x_i^\beta \left[
   \left( \rho v^\gamma  (v^\kappa - w^\kappa) + p \delta^{\gamma\kappa} \right)_{ij} \Aijm^\kappa
\right] = \\
&\frac{1}{2} \left\{
\epsilon^{\alpha \beta \gamma} \sum_{i,j} x_i^\beta
    \left[ \left(
    \rho v^\gamma  (v^\kappa - w^\kappa) + p \delta^{\gamma\kappa}
    \right)_{ij} \Aijm^\kappa
    \right] +
\epsilon^{\alpha \beta \gamma} \sum_{i,j} x_j^\beta
    \left[ \left(
    \rho v^\gamma  (v^\kappa - w^\kappa) + p \delta^{\gamma\kappa}
    \right)_{ji} \mathcal{A}_{ji}^\kappa
    \right]
\right\} = \\
&\frac{1}{2} \left\{
\epsilon^{\alpha \beta \gamma} \sum_{i,j} x_i^\beta
    \left[ \left(
    \rho v^\gamma  (v^\kappa - w^\kappa) + p \delta^{\gamma\kappa}
    \right)_{ij} \Aijm^\kappa
    \right] -
\epsilon^{\alpha \beta \gamma} \sum_{i,j} x_j^\beta
    \left[ \left(
    \rho v^\gamma  (v^\kappa - w^\kappa) + p \delta^{\gamma\kappa}
    \right)_{ij} \mathcal{A}_{ij}^\kappa
    \right]
\right\} = \\
&\frac{1}{2}
\epsilon^{\alpha \beta \gamma} \sum_{i,j} (x_j^\beta - x_i^\beta)
    \left[ \left(
    \rho v^\gamma  (v^\kappa - w^\kappa) + p \delta^{\gamma\kappa}
    \right)_{ij} \Aijm^\kappa
    \right] = \\
&\frac{1}{2}
\epsilon^{\alpha \beta \gamma} \sum_{i,j} (x_j^\beta - x_i^\beta) (p \Aijm^\gamma) +
\frac{1}{2}
\epsilon^{\alpha \beta \gamma} \sum_{i,j} (x_j^\beta - x_i^\beta)
    \left( \rho v^\gamma  (v^\kappa - w^\kappa) \right)_{ij} \Aijm^\kappa
\end{flalign}

Inserting this expression into our previous findings, we obtain:

\begin{align}
\sum_i \ddt \mathbf{L}_i &=
\epsilon^{\alpha \beta \gamma} \sum_i m_i w_i^\beta v_i^\gamma
-  \epsilon^{\alpha \beta \gamma} \sum_i x_i^\beta \sum_j \left[
   \left( \rho v^\gamma (v^\kappa - w^\kappa)  + p \delta^{\gamma\kappa} \right)_{ij} \Aijm^\kappa
\right] && \\
&= \epsilon^{\alpha \beta \gamma} \sum_i m_i w_i^\beta v_i^\gamma \quad -
&& \text{(i)} \\ \nonumber
& \quad
\frac{1}{2} \epsilon^{\alpha \beta \gamma}
    \sum_{i,j} (x_j^\beta - x_i^\beta) (p \Aijm^\gamma) \quad -
    && \text{(ii)} \\ \nonumber
& \quad
\frac{1}{2} \epsilon^{\alpha \beta \gamma}
    \sum_{i,j} (x_j^\beta - x_i^\beta)
    \left( \rho v^\gamma  (v^\kappa - w^\kappa) \right)_{ij} \Aijm^\kappa
    && \text{(iii)}
\end{align}

For the methods to conserve angular momentum exactly, i.e. $\ddt \mathbf{L} = 0$, two conditions
need to be satisfied. Firstly, the particle velocity $\mathbf{w}$ must be exactly the fluid
velocity: $\mathbf{w} = \V$. This leads to the terms (i) being zero due to the cross product it
contains, as well as term (iii) vanishing.\footnote{
The cancellation of term (iii) due to the condition $\mathbf{w} = \V$ assumes that the expression
for the approximate flux $\fc_{ij} = \left( \rho v^\gamma  (v^\kappa - w^\kappa) \right)_{ij}$
also finds that the approximate factor $(v^\kappa - w^\kappa)_{ij}$ is zero.
That may in general not be the case. However, the second condition required for the total angular
momentum to be conserved, namely the surfaces \Aij to be exactly parallel to distance vector between the particles $i$ and $j$, ensures that the term zeroes out in any case.
}
Secondly, we need the normal vector $\hat{\mathbf{n}}$ of the surfaces $\Aijm =
|\Aijm|\hat{\mathbf{n}}$ to be exactly parallel to the distance vector $\x_j - \x_i$ between any
two particles $i$ and $j$, which leads to the cross product in term (ii) being zero as well. While
we are able to set the particle velocities $\mathbf{w}_i$ as we choose, it is in general not
possible to also enforce that the surfaces are parallel to the distance vector between two particles at all times. This may be the case for exactly regular particle configurations, but not for configurations containing irregularities, as will be shown in
Section~\ref{chap:meshless-comparison}. Hence we must conclude that the methods do not conserve
angular momentum in general.

\section{The Full Scheme Applied to the Euler Equations}\label{chap:meshless-full}

While the main update formulae for both the Hopkins and Ivanova version of the finite volume
particle methods have been derived, there are still a lot of other points that need to be addressed
in order to obtain a fully functioning method. In what follows, we focus on the application of the
finite volume particle methods on the Euler equations (\ref{eq:euler-equations}). The application
to the moments of the equation of radiative transfer will be discussed in Part~\ref{part:rt}.

Two major points that need to be discussed first are the exact time integration scheme and making
the particles co-move with the fluid. These two topics are intimately linked, as we will see. In
what follows, these topics will be discussed:

\begin{itemize}
 \item Choosing a frame of reference for the flux exchange between two particles
 \item Moving (``drifting'') particles
 \item The time integration scheme
 \item The neighbor search: Finding the smoothing length of particles
 \item Flux limiters for the finite volume particle method
 \item The CFL condition
\end{itemize}

Finally, the full algorithm of the method is presented.

\subsubsection{Choosing a Frame of Reference For the Flux Exchange Between Two Particles}

The equations that govern the time integration of the states traced by particles give us no
prescription on how to move particles. The Euler equations are Galilei-invariant, and hence are also
valid in co-moving form. So we are basically free to choose to move the particles however we want,
as long as the fluid quantities are correctly transformed into the corresponding frame of reference.
Specifically, the correct frames of reference need to be addressed when the inter-particle fluxes
$\F_{ij}$ of the finite volume particle methods are being computed, as we use the states carried by
the ``left'' and ``right'' particles $i$ and $j$ as left and right states for a Riemann problem. If
each particle's velocity is taken to be the velocity its reference frame, then their states need to
be boosted into a common frame of reference. An obvious choice for a common reference frame is to
take the frame of the ``effective surface'' \Aij. The ``surfaces'' are taken to be at position along
the line connecting particle $i$ and $j$ and at a distance weighted by the respective smoothing
lengths

\begin{align}
    \x_{ij} = \x_i + \frac{h_i}{h_i + h_j} (\x_j - \x_i)
\end{align}

The same weight is used to set the velocity of the surface w.r.t. the rest frame, and hence the
common frame of reference of the particles:

\begin{align}
    \V_{ij} = \V_i + \frac{h_i}{h_i + h_j} (\V_j - \V_i)
\end{align}

where $\V_{i,j}$ are \emph{particle} velocities, which may or may not correspond to the fluid
velocity.

Other choices for the weight to determine the position and velocity of the surfaces are possible,
as long as the resulting term remains anti-symmetric w.r.t. the particles $i$ and $j$. For example,
one could choose to set the surfaces exactly in the middle between both particles. In practice, it
makes little difference in the final results. The smoothing length however is a sensible choice, as
it's a quantity directly related to the local resolution and the local particle number density.

It still remains to determine how to choose the particle velocities. An obvious choice is to move
particles along with the fluid, i.e. to set the particle velocities to the local fluid velocity.
This way the particles, which are also the resolution elements in the scheme, naturally follow the
fluid and the spatial resolution is increased precisely in regions where we need it to be, whereas
``uninteresting'' regions that contain comparatively little material, have small densities, and
generally smooth flows are traced with lower resolution. Other choices are however possible as
well. For example, we could decide not to move the particles at all, which is a neat property of
the finite volume particle methods that we'll make use of to solve the equations of radiative
transfer simultaneously with the Euler equations in Part~\ref{part:rt}.

\subsubsection{Time Integration and Particle Drifts}

Moving the particles' position is an operation which is usually called a ``drift''. Drift
operations are also commonly used in higher order time integration schemes for particle systems
like the Leapfrog symplectic integrator. The Leapfrog integrator prescribes for a second order
differential equation of the form

\begin{align}
    \frac{\de ^2 x}{\de t ^2} = a && \DDT{x} = v
\end{align}

the following scheme:

\begin{align}
    v_{i+\half} &= v_i + a_i \frac{\Delta t}{2}  && \text{kick} \\
    x_{i+1} &= x_i + v_{i+\half} \Delta t  && \text{drift} \\
    v_{i+1} &= v_i + a_{i+1} \frac{\Delta t}{2}  && \text{kick}
\end{align}

Symplectic integrators like the Leapfrog are widely used for their conservational properties. In
particular, they are able to keep bound orbits in N-body systems bounded (i.e.  with a bounded
error in energy), whereas explicit integrators like the Euler integrator or the Runge-Kutta
integrators don't. This means that what physically should be bound orbits can (and will) severely
break energy conservation after a sufficiently high number of integration steps, making what should
be stable systems unstable. Hence the symplectic integrators are preferable.

However, due to the conservational properties of the finite volume particle methods, a full
kick-drift-kick scheme is not strictly necessary. The expressions for the ``effective surfaces''
\Aij (eq.~\ref{eq:HopkinsAij},\ref{eq:IvanovaAij}) are anti-symmetric in particles $i$ and $j$, and
the total flux exchange is thus always conserved. So in the case of hydrodynamics, a simpler
drift-kick scheme suffices, where the ``kick'' operation corresponds to the solution of the update
formulae~\ref{eq:meshless-Hopkins}~and~\ref{eq:meshless-Ivanova}. (Note however that in the case
where gravity is added to the system, the kick-drift-kick scheme becomes necessary once again to
treat the gravitational acceleration adequately.)

\subsubsection{Obtaining a Time-Centered Problem}

We can however make use of the drift operation to improve the flux estimates and the order of
accuracy of the scheme. Similarly to how the fluxes between two interacting cells are evaluated at
the midpoint in time between the start and the end of the step in the WAF and the MUSCL-Hancock
schemes (Sections~\ref{chap:WAF}~and~\ref{chap:MUSCL-Hancock}), we can do the same for the finite
volume particle schemes by obtaining the primitive variables (density $\rho$, velocity $v$, and
pressure $p$) at the midpoint in time with the drifts. The inter-particle fluxes in the finite
volume particle schemes are given by the solution to the Riemann problem positioned between two
particles. In turn, the Riemann solvers take the primitive variables as arguments to find the
solution. Hence providing the Riemann solvers with primitive variables advanced in time should allow
us to obtain a better estimate for the inter-particle fluxes throughout the step. To do so, we make
use of the primitive variable formulation of the Euler equations. It is straightforward to show that
the Euler equations can be written in the form

\begin{align}
    \deldt \W + A(\W) \deldx \W = 0 &&
\end{align}

with the primitive state vector $\W$ and Jacobi matrix $A(\W)$

\begin{align}
    \W = \begin{pmatrix}
          \rho \\ v \\ p
         \end{pmatrix}
&&
    A(\W) = \begin{pmatrix}
              v & \rho & 0 \\
              0 & v & 1/\rho \\
              0 & \rho c_s^2 & v
             \end{pmatrix}
\end{align}

Through the general gradient expression \ref{eq:gradient} the gradients of the primitive variables
$\ddx \W$ are available, and so the primitive state $\W^{n+\half}$ at the midpoint in time can be
obtained via

\begin{align}
    \W^{n + \half} &= \W^n - \frac{\Delta t}{2} A(\W) \deldx W
\end{align}

\subsubsection{The Neighbor Search: Determining the Smoothing Length}

Drifting particles has the consequence that the actual particle configuration keeps changing. If a
particle's smoothing length is defined via some requirement for the particle's kernel support radius
to contain an (approximate) number of neighboring particles, it means that after each drift the
smoothing length needs to be re-computed. This process is commonly referred to as a ``neighbor
search'', since determining the smoothing length of each particle also determines which other
particles it will exchange fluxes with. These interaction partners are called neighbors. The
neighbor search is a well known problem in smoothed particle hydrodynamics (see e.g.
\cite{priceSmoothedParticleHydrodynamics2012}) and needs to be solved iteratively using a root
finding technique like the Newton-Raphson or Bisection algorithms due to a circular dependence: If
the smoothing length is defined such that the compact support radius of a kernel includes some given
number of neighbors, then it is related to the local number density of the particles, i.e.

\begin{align}
    h(\x_i) \propto n_i^{1/\nu} \ .
\end{align}

The local number density $n_i = 1/V_i$ on the other hand depends on the associated particle volumes
$V_i$, which in turn require the smoothing length $h$ to be known (see eqs.~\ref{eq:omega}
and~\ref{eq:particle-volume}).

The (average) number of neighbors $N_{NGB}$ which is to be used for a given simulation is in
principle a free parameter. Following the approach of
\citet{dehnenImprovingConvergenceSmoothed2012c}, it can be defined via the parameter $\eta_{res}$:

\begin{align}
    N_{NGB} = V_{S,V} \left( \frac{H}{h} \eta_{res} \right)^\nu \label{eq:number-of-neighbors}
\end{align}

where $V_{S,\nu}$ is the volume of a $\nu$-dimensional sphere, i.e. $2 \pi$ in 2D and $4/3 \pi$ in
3D. The value of the ratio $H/h$ of the compact support radius $H$ to the smoothing length $h$
depends on the kernel choice and dimension. Some values of this ratio for various popular kernels
are given in Table~1 in \citet{dehnenImprovingConvergenceSmoothed2012c}. The default choice for
$\eta_{res}$ is

\begin{align}
    \eta_{res} = 1.2348
\end{align}

which for the cubic spline kernel (eq.~\ref{eq:cubic-spline-kernel}) leads to $\sim 48$ neighbors
in 3D, and $\sim 15$ neighbors in 2D.

\subsubsection{Computing the Gradients}

Once the neighbor search is complete and all smoothing lengths $h$ are known, the gradients
(eq.~\ref{eq:gradient}) can be computed via another sum over all neighboring particles. The
gradients are required to be known before we can proceed with the flux exchanges: While they are
necessary in both the Hopkins and Ivanova version to extrapolate the left and right states at the
``effective surface'' \Aij, the $\psitilde$ terms of \Aij itself in the Hopkins version requires
the gradient terms to be available first. With the gradients being available, the actual
computation of the fluxes and the ``kick'' time integration can be performed with a third sum over
all neighbors, where the sum $\oldsum_j \F_{\alpha, i} \Aijm^\alpha$ is accumulated.

\subsubsection{Flux and Slope Limiters}

Since the finite volume particle methods in this form are second order accurate, we need to employ
limiters in order to prevent spurious oscillations that are a consequence of higher order methods
(see Chapter~\ref{chap:higher-order-schemes}) and to make the scheme total variation diminishing. We
employ the same limiters described by \cite{hopkinsGIZMONewClass2015}. This limiter consists of two
steps: The first step limits the slopes of the individual gradients of each particle. The main idea
is to make sure that if we extrapolate the gradients of a particle to its neighboring particles'
positions, then the extrapolated value shouldn't be higher than the highest actual particle value,
nor lower than the lowest actual value. This prevents new extrema from forming, which is what the
TVD condition requires. So the ``true'' obtained gradient $\nabla \phi^i_{true}$ calculated by use
of eqn. \ref{eq:gradient} of any quantity $\phi^i$ needs to be limited by

\begin{align}
	\nabla \phi_{lim}^i &= \alpha_i \nabla \phi^i_{true} \label{eq:cell-limiter-first}\\
	\alpha_i &= \min\left[ 1, \frac{\phi_{ij\ ngb}^{max} - \phi_i}{\nabla \phi^i_{true} \cdot
r_{max}}, \frac{\phi_i - \phi_{ij\ ngb}^{min}}{\nabla \phi^i_{true} \cdot r_{max}} \right]
\label{eq:cell-limiter-last}
\end{align}

Secondly, to ensure more general stability, a pair-wise slope limiter from
\cite{hopkinsGIZMONewClass2015} is employed during each particle interaction. Unfortunately, only
slope limiting the gradients is not enough to make the method TVD. For a general quantity $Q_k$ of
particle $k$, the slope limited quantity $Q_k^{lim}$ at the intersection $\mathbf{x} =
\mathbf{x}_{kl}$ for the interaction between this particle $k$ and some other particle $l$ is

\begin{equation}
	Q_k^{lim} =
	\begin{cases}
		Q_k
						& \text{ if } Q_k = Q_l \\
		\max\left\{ Q_-,
			\min\{ \overline{Q}_{kl} + \delta_2, Q_k + \DELDX{Q_k} (\mathbf{x}_{kl} -
\mathbf{x}_k)\}
		\right\}
						& \text{ if } Q_k < Q_l \\
		\min\left\{ Q_+,
			\max\{ \overline{Q}_{kl} - \delta_2, Q_k + \DELDX{Q_k} (\mathbf{x}_{kl} -
\mathbf{x}_k)\}
		\right\}
						& \text{ if } Q_k > Q_l
	\end{cases}  \label{eq:face-limiter-first}
\end{equation}

where

\begin{align}
Q_- &=
	\begin{cases}
		Q_{min} - \delta_1
						& \text{ if } \mathrm{sign} (Q_{min} - \delta_1) = \mathrm{sign} (Q_{min})\\
		\frac{Q_{min}}{1 + \delta_1 / | Q_{min}|}
						& \text{ if } \mathrm{sign} (Q_{min} - \delta_1) \neq \mathrm{sign}
(Q_{min}) \\
	\end{cases}\\
Q_+ &=
	\begin{cases}
		Q_{max} + \delta_1
						& \text{ if } \mathrm{sign} (Q_{max} + \delta_1) = \mathrm{sign} (Q_{max})
\\
		\frac{Q_{max}}{1 + \delta_1 / | Q_{min}|}
						& \text{ if } \mathrm{sign} (Q_{max} + \delta_1) \neq \mathrm{sign}
(Q_{max}) \\
	\end{cases}\\
\overline{Q}_{kl} &=
	Q_k + \frac{|\mathbf{x}_{kl} - \mathbf{x}_{k}|}{|\mathbf{x}_{l} - \mathbf{x}_{k}|} (Q_l -
Q_k)\\
Q_{min} &= \min\{ Q_k, Q_l \} \\
Q_{max} &= \max\{ Q_k, Q_l \} \\
\delta_1 &= \gamma_1 | Q_k - Q_l | \\
\delta_2 &= \gamma_2 | Q_k - Q_l |  \label{eq:face-limiter-last}
\end{align}

where $\gamma_1$ and $\gamma_2$ are free parameters, the recommended choice being $\gamma_1 =
1/2$, $\gamma_2 = 1/4$.

\subsubsection{The CFL Condition}

The final point left to be discussed is the determination of the time step size of individual
particles. A ``cell size'' $\Delta x$ of  a particle is estimated using

\begin{align}
    \Delta x \approx \left(\frac{V_i}{V_{S,\nu}} \right)^{1/\nu}
\end{align}

where $V_{S,\nu}$ is the volume of a $\nu$-dimensional unit sphere, so $4/3 \pi$ in 3D. The signal
velocity is estimated as

\begin{align}
    v_{i,sig} = \max \{ |v_{i,fluid} - v_{i,particle}| + c_{s,i}, c_{s,i} + c_{s,j} \} \ .
\end{align}

We take the velocity difference between the fluid and the particle, since that is the velocity at
which the fluid in the co-moving frame with the particle would propagate, and the sound speed $c_s$
to estimate the propagation velocity of the emanating waves of the Riemann problem, similar to
the suggested estimate for Godunov's method, eq.~\ref{eq:wavespeed-estimate-godunov}. To ensure
that no problems arise in cases where two particles approach each other, the additional
upper boundary of the sum of the wave speeds $c_{s,i} + c_{s,j}$ for particle $i$ and all its
neighbors $j$ is added. This leads to the time step estimate

\begin{align}
    \Delta t = C_{CFL} \frac{\Delta x}{v_{sig}} \label{eq:meshless-cfl}
\end{align}

Typically a safe choice of $C_{CFL} \lesssim 0.4$ is recommended.

\subsubsection{The Full Algorithm}

To conclude this Section, let's explicitly write down the algorithm to solve the Euler Equations
from some $t_{start}$ to some $t_{end}$ using the full kick-drift-kick Leapfrog integration scheme:

\algo{Evolving The Euler Equations From $t_{start}$ to $t_{end}$ Using Finite Volume Particle
Methods}
{
To start, set $t_{current} = t^0 = t_{start}$ and set up the initial states $\U_i^0$ for
each particle $i$.\\[.5em]
While $t_{current} < t_{end}$, solve the $n$-th time step:\\[.5em]
\indent~~~~Find the maximal permissible time step $\Delta t$ using
eq.~\ref{eq:meshless-cfl}.\\[.5em]
\indent~~~~Kick the particles: Update the the particle velocity (not the fluid velocity) \\
\indent~~~~over the time step $\Delta t / 2$.\\[.5em]
\indent~~~~Drift the particles: Move their positions over the time step $\Delta t$.\\[.5em]
\indent~~~~Obtain the primitive variables at the midpoint in time, $\W_i^{n+\half}$.\\[.5em]
\indent~~~~Do a ``neighbor search loop'' over neighboring particles, and determine\\
\indent~~~~the smoothing length $h$ for all particles $i$.\\[.5em]
\indent~~~~Do a ``gradient loop'' over neighboring particles to determine the gradients\\
\indent~~~~of the primitive variables for each particle.\\[.5em]
\indent~~~~Do a ``flux exchange loop'' over neighboring particles, during which all\\
\indent~~~~neighboring particles $i$, $j$ interact with each other:\\[.5em]
\indent~~~~~~~~For each particle pair $i$, $j$, compute the ``effective surfaces'' \Aij.\\[.5em]
\indent~~~~~~~~Find the position $\x_{ij}$ and velocity $\V_{ij}$ of the \Aij.\\[.5em]
\indent~~~~~~~~Boost the primitive variables $\W^{n+\half}$ to the common reference frame \\
\indent~~~~~~~~with velocity $\V_{ij}$, and extrapolate them using gradients at the position
$\x_{ij}$.\\[.5em]
\indent~~~~~~~~Use the boosted and extrapolated states to find the inter-particle fluxes by \\
\indent~~~~~~~~solving the Riemann problem. Accumulate the sum of all fluxes for each \\
\indent~~~~~~~~particle, boosted back into the frame of reference of the respective
particle.\\[.5em]
\indent~~~~Obtain the updated states $\U_i^{n+1}$ using eq.~\ref{eq:meshless-Hopkins-explicit} over
the time step $\Delta t$.\\[.5em]
\indent~~~~Kick the particles again: update the particle velocities over the time step $\Delta
t/2$.\\[.5em]
\indent~~~~Update the current time: $t_{current} = t^{n+1} = t^n + \Delta t$ \\
}

Note that in the absence of external forces like gravity the first kick operation is unnecessary,
while the second kick operation consists of directly setting the particle velocities to the fluid
velocities (assuming the particles are treated as exactly Lagrangian, without any corrections or
modifications). Including external acceleration terms however makes the first kick operation
necessary again.

\section{Individual Timstepping}\label{chap:individual-timesteps}

\begin{figure}
 \centering
 \includegraphics[width=.5\textwidth]{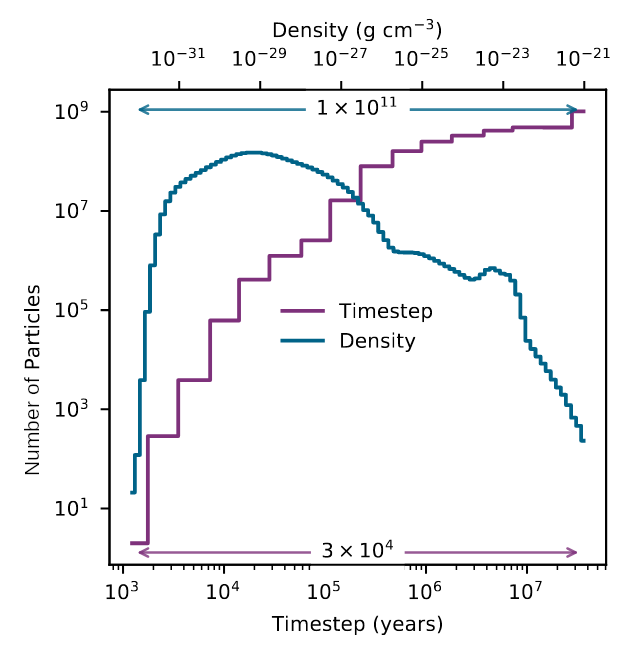}
 \caption{
Histogram of particle time steps and densities from the EAGLE simulation suite
\citep{schayeEAGLEProjectSimulating2015, schallerEAGLESimulationsGalaxy2015} at redshift $z = 0.1$,
where galaxies are fully formed.
With the density spanning 11 orders of magnitude, the resulting time step sizes for particles span
four orders of magnitude. Only a small minority of particles has very small time steps, motivating
the approach to allow particles to be integrated according to their individual time step sizes
rather than limiting all particles to the globally minimal time step size. \\
This figure is adapted from \cite{borrowSWIFTMaintainingWeakscalability2018} with permission from
the author.
}
 \label{fig:eagle-timesteps}
\end{figure}

A huge challenge in cosmological simulations is the extreme range of states that are present. For
example, the EAGLE simulation \citep{schallerEAGLESimulationsGalaxy2015}, shown in
Figure~\ref{fig:eagle-timesteps}, reports ranges in particle densities of $11$ orders of magnitude,
and $6$ orders of magnitudes in internal energies. We can make an estimate how that will affect
time step sizes: For the ``cell size'' $\Delta x$ (in 3D), we can use:

\begin{align}
\Delta x \sim V_i^{1/3} = (m_i/\rho_i)^{1/3} \propto \rho_i^{-1/3} \ .
\end{align}

For the propagation speed $v_{sig}$, we can use the approximation of the local sound speed, i.e.

\begin{align}
v_{sig} \sim c_{s} = \sqrt{p/\rho} = \sqrt{u (\gamma - 1)} \propto  u^{1/2} \ .
\end{align}

giving us the estimate

\begin{align}
\Delta t \propto \frac{\Delta x}{v_{sig}} \propto u^{-1/2} \rho^{1/3}
\end{align}

which leads to about 4 orders of magnitude in time step size differences. Additionally,
Figure~\ref{fig:eagle-timesteps} also shows that only a minority of particles will have small time
step sizes. If we were to use a global time step size throughout the simulation, i.e. limit the
time step of all particles by the globally minimal time step size, it would mean that the majority
of the particles is restricted to a much smaller time step size than their individual CFL-condition
would allow for. It would also mean that the simulation would require several orders of magnitude
more steps than strictly necessary if it weren't for the minority of particles with very small time
steps. This additional expense would make cosmological simulations prohibitively expensive.

To circumvent this issue, it is necessary to allow particles to have individual time step sizes,
albeit with some restrictions. Particles are given time step sizes based on their individual CFL
condition, rounded down to a power-of-two fraction of the maximal time step size of the system
$t_{max}$:

\begin{align}
    \Delta t_i = \frac{t_{max}}{2^n}
\end{align}

with $n \geq 0$. The maximal time step size of the system is typically the requested time from
beginning to end of the simulation. Allowing only specific values for time step sizes of particles
is akin to histogramming their time step sizes, and $n$ is typically referred to as the ``time
bin'' of the particles. The absolute minimal time step size is given by $t_{min} = t_{max} /
2^{N_{bins}}$, where $N_{bins}$ is the total number of available time bins. This method has been
used in SPH simulations since \citet{hernquistTREESPHUnificationSPH1989}.

\begin{figure}
 \centering
 \includegraphics[width=\textwidth]{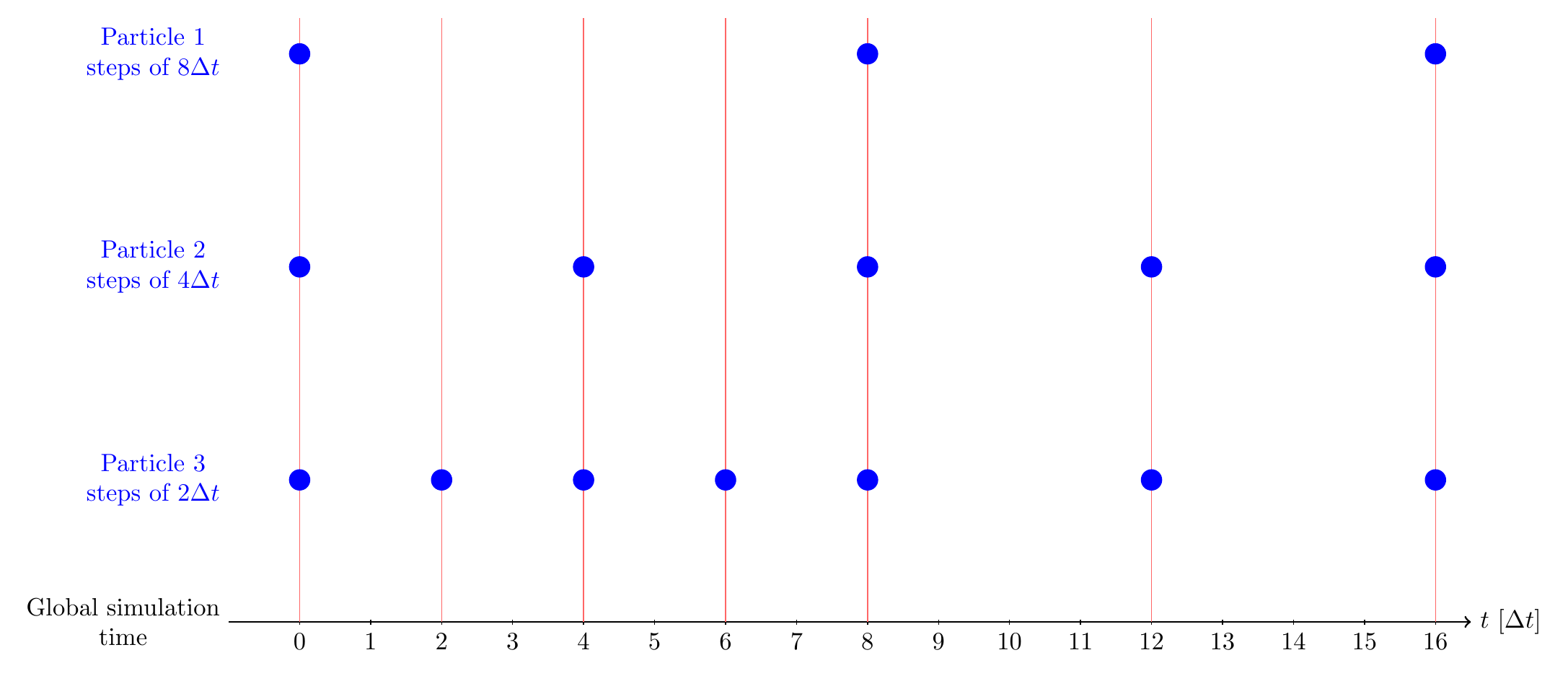}
 \caption{
Illustration of the individual time stepping scheme. Three particles with initially different time
step sizes $2 \Delta t$, $4 \Delta t$, and $8 \Delta t$, respectively, are shown, where $\Delta t$
denotes a minimal time steps size. The simulation progresses by advancing the time according to the
global minimal time step size, which in the beginning is $2 \Delta t$. Each simulation step is
marked as a red line. Since in this example there are no particles with time step size of $\Delta
t$, all times that are an odd integer multiple of $\Delta t$ are skipped over.
At each simulation step, only particles for which the current simulation time divided by the time
step size has a modulo of zero are being updated.\\
In this illustration, particle 3 changes its time step size at the at the simulation time $t = 8
\Delta t$ from $2 \Delta t$ to $4 \Delta t$ for reasons which are unimportant. From that point, the
simulation step can be increased to $4 \Delta t$, and fewer total number of simulation steps are
necessary.
}
 \label{fig:individual-timesteps}
\end{figure}

The entire simulation progresses by advancing the current simulation time by a time step size which
is a power-of-two multiple of $t_{min}$. The system time step size is still determined by the
minimal global time step, but only the particles that require such a small time step are also
updated. This is schematically shown in Figure~\ref{fig:individual-timesteps}. Whether a particle
needs to finish its time integration on the current simulation step is then determined by whether
the current simulation time divided by the particle's time step has a modulo of zero. Since
particles have time step sizes which are always a power-of-two apart from each other, this means
that at each simulation step, there is a maximal time bin $m$ for which the division of the current
system time with $m$ gives modulo of zero, and all particles with time bin $n \leq m$ will require
an update in that simulation step, as can be seen in Figure~\ref{fig:individual-timesteps}.

With particles being allowed to have individual time steps, the update formula for the finite volume
particle methods (eq.~\ref{eq:meshless-Hopkins-explicit}) needs to be slightly updated in order to
maintain its conservative properties. In particular, rather than exchanging fluxes, particles now
need to exchange time integrated fluxes:

\begin{align}
\U_i^{n+1} =
    \U_i^n + \frac{1}{V_i} \sum_j \min\{ \Delta t_i, \Delta t_j \} \F_{\alpha, ij} \Aijm^\alpha
\label{eq:meshless-explicit-individual-timesteps}
\end{align}

The time integration needs to be the smaller of the two time step sizes $\{\Delta t_i,\  \Delta
t_j\}$ since if one particle has a smaller time step size than the other, it also means that there
will be several interactions between that particle pair, each with the time step size of the smaller
of the two. This ensures that the total exchange of fluxes remains conservative, as the fluxes
``removed'' from particles $j$ remain exactly equal to the fluxed ``added'' to particle $i$. If some
neighboring particles $j$ have smaller time steps than particle $i$, then the net sum of the fluxes
is accumulated during the exchanges and applied only once at the point where particle $i$ is being
updated again.

\section{Comparing the ``Hopkins'' and ``Ivanova'' Versions}\label{chap:meshless-comparison}

With the full method presented in Section~\ref{chap:meshless-full}, we can now turn to comparing
the ``Hopkins'' and ``Ivanova'' versions of the methods.
The difference between the Hopkins version and the Ivanova version of the finite volume particle
method lies in the expression for the ``effective surface'' \Aij. These surfaces are rather abstract expressions, and it seemed worthwhile to invest time to investigate how they behave, how they can be interpreted, and to see whether there might be significant differences between the two versions.

\subsection{Behavior of the Effective Surfaces}

Let's begin with the behavior of the effective surfaces without looking into their application on
evolving problems in time just yet. To compute and visualize their behavior of these effective
surfaces, I have developed a dedicated Python package named \mbox{\codename{astro-meshless-surfaces}},
which is publicly available on
\href{https://github.com/mladenivkovic/astro-meshless-surfaces}{PyPI.org} and on
\url{https://github.com/mladenivkovic/astro-meshless-surfaces}.

\subsubsection{Differences in Direction and Magnitude of the \Aij}

\begin{figure}[htpb]
\centering
\includegraphics[width=\textwidth]{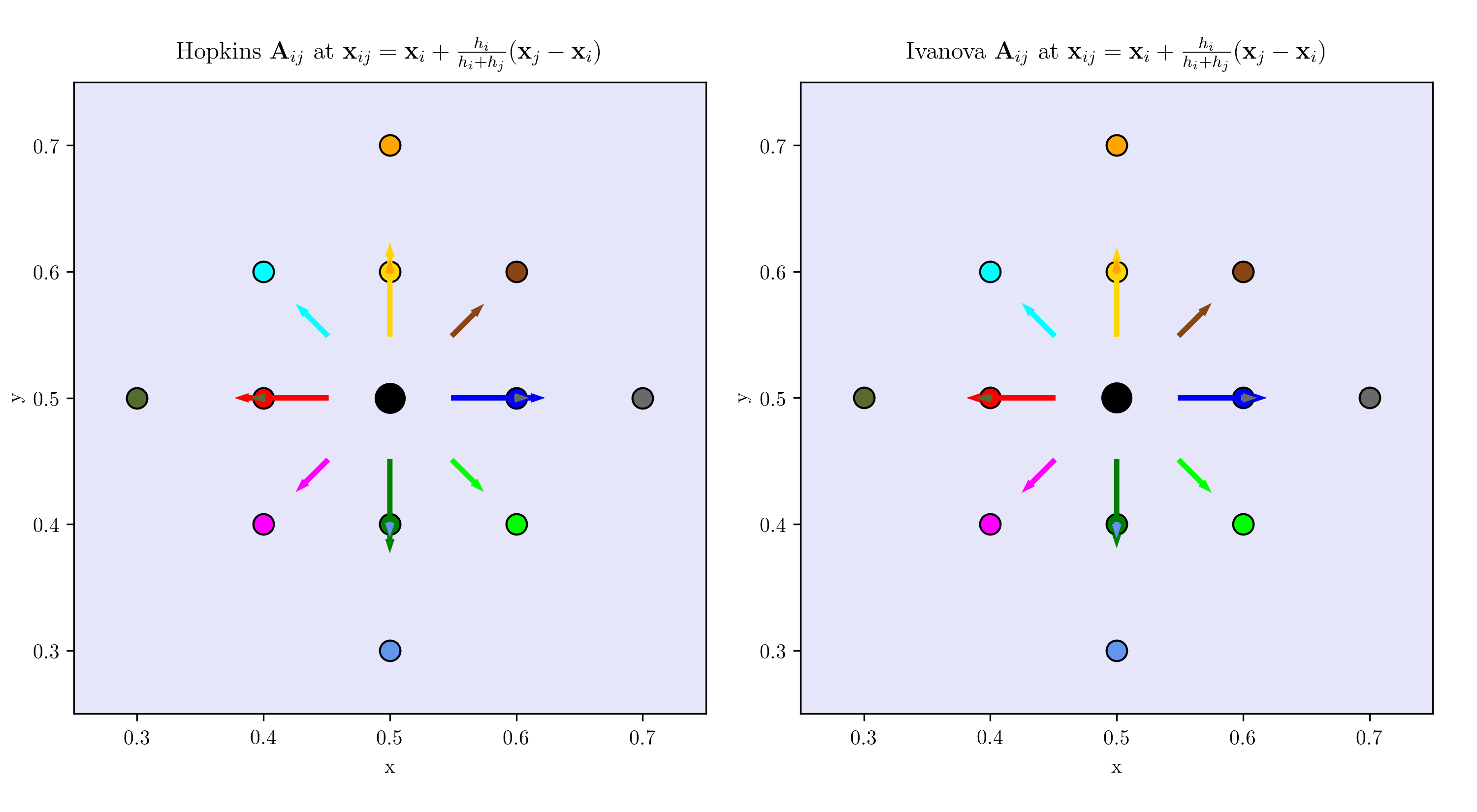}
\caption{\label{fig:uniform-arrows}
    The effective surfaces \Aij\ for the Hopkins and Ivanova expression for the black particle in a
    uniform particle distribution.  Only the particles within the compact support radius of the
    central particle are plotted.
}
\end{figure}

\begin{figure}[htpb]
\centering
\includegraphics[width=\textwidth]{
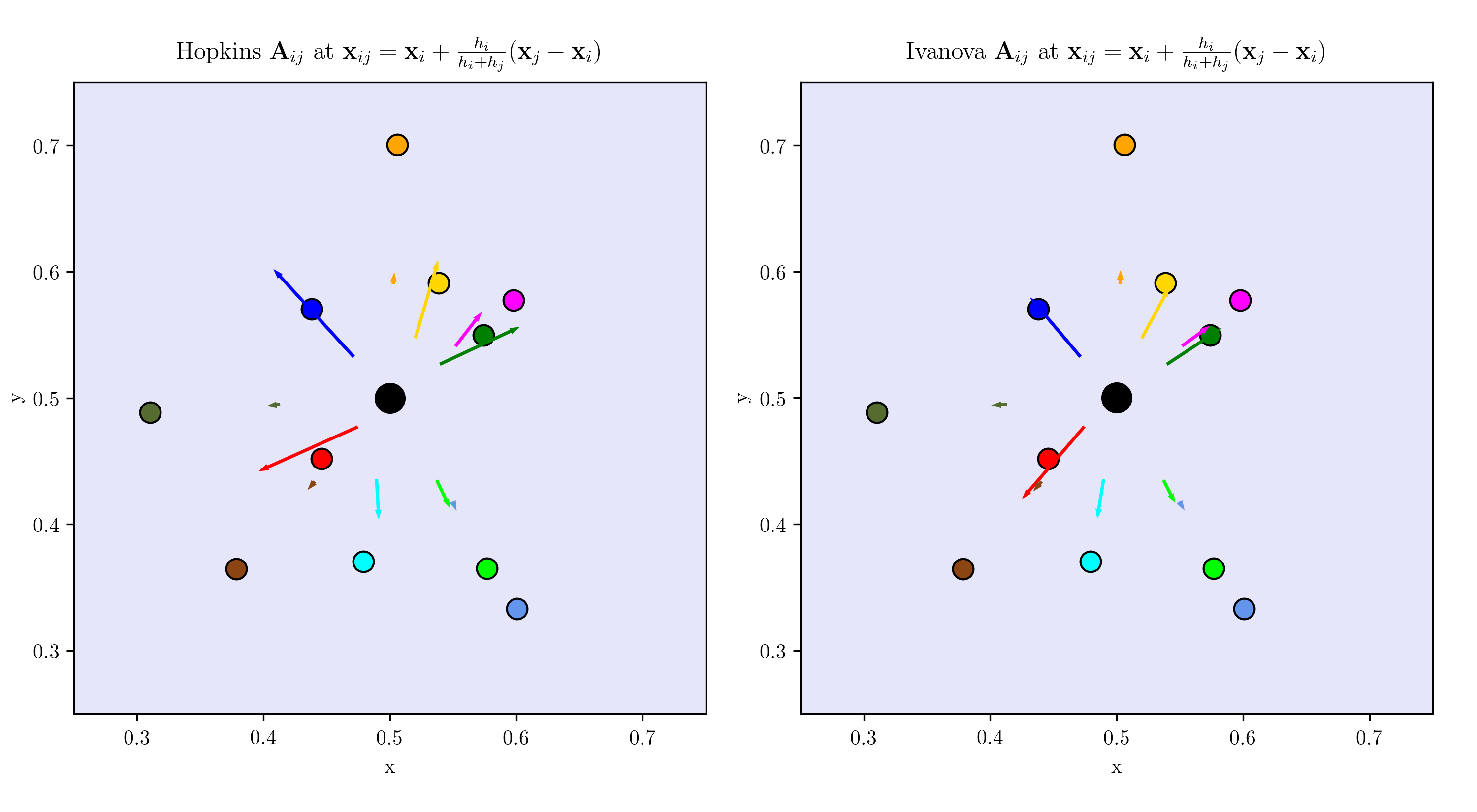}
\caption{\label{fig:perturbed-arrows}
    The effective surfaces \Aij\ for the Hopkins and Ivanova expression for the black particle
    in a non-uniform particle distribution. Only the particles within the compact support radius of
    the central particle are plotted.
}
\end{figure}

Figure \ref{fig:uniform-arrows} shows the effective surfaces of both versions for a central
particle in a uniform particle distribution, while Figure \ref{fig:perturbed-arrows} shows them for
a non-uniform particle configuration. The surfaces are drawn as (pseudo-)vectors starting at the
position

\begin{equation}
	\x_{ij} = \x_i + \frac{h_i}{h_i + h_j}(\x_j - \x_i)		\label{eq:xij}
\end{equation}

which is the position where the effective surfaces would be ``placed'' during the flux exchange of
the particles. Only the particles within the compact support radius of the central particle are
shown.

Comparing the norm of the \Aij\ between the Hopkins and the Ivanova version for the same
neighboring particles, the ratio of the Hopkins over Ivanova lengths varies between 1.06 and 0.33
for the uniform case and between 1.60 and 0.28 for the non-uniform case.
The ratio tends to lower values with increasing neighbor particle distance.

In the non-uniform particle distribution a striking feature is that the effective surfaces do not
point towards the neighboring particle which they are associated with.
In the Hopkins formulation, the reason for that is that the $\psitilde$ (eq. \ref{eq:psitilde}) in
the expression for \Aij have a non-zero contribution from other dimensions. These contributions
enter through the matrix multiplication, and occur even in the case where the two interacting
particles are positioned along a line parallel to the coordinate axes.
As for the Ivanova case, the direction of $\nabla \psi$ will in general not be radially symmetric,
as $\psi(\x)$ is also not radially symmetric in general (see Figure~\ref{fig:psi-of-x-contour}).

\subsubsection{Checking the Closure Condition}

A further check that can be made is to verify whether the closure condition (see
Section~\ref{chap:meshless-conservation-closure}), i.e. $\oldsum_j \Aijm = 0$, is satisfied. In the
uniform particle configuration this is satisfied to machine precision. In the non-uniform case
however, both expressions sum up to a value around the same order of magnitude of a single \Aij,
which tends to be $\approx 10^{-3} - 10^{-4}$. The effect remains for higher particle numbers and
higher neighbor numbers used, i.e. the magnitude of both the \Aij\ and the sum over all \Aij\ for an
individual particle decrease, but their ratio remains about the same. It should be noted that in any
case investigated, the $\oldsum_j \Aijm = 0$ for the Ivanova expression was closer to zero by about
one order of magnitude, but also the total sum of the norms of all \Aij, $\oldsum_j |\Aijm|$, of the
Ivanova version was in every case smaller than the one for the Hopkins version.

\subsubsection{\Aij as a Function of Neighbor Position}

\begin{figure}[htpb]
	\centering
	\includegraphics[width=\textwidth]{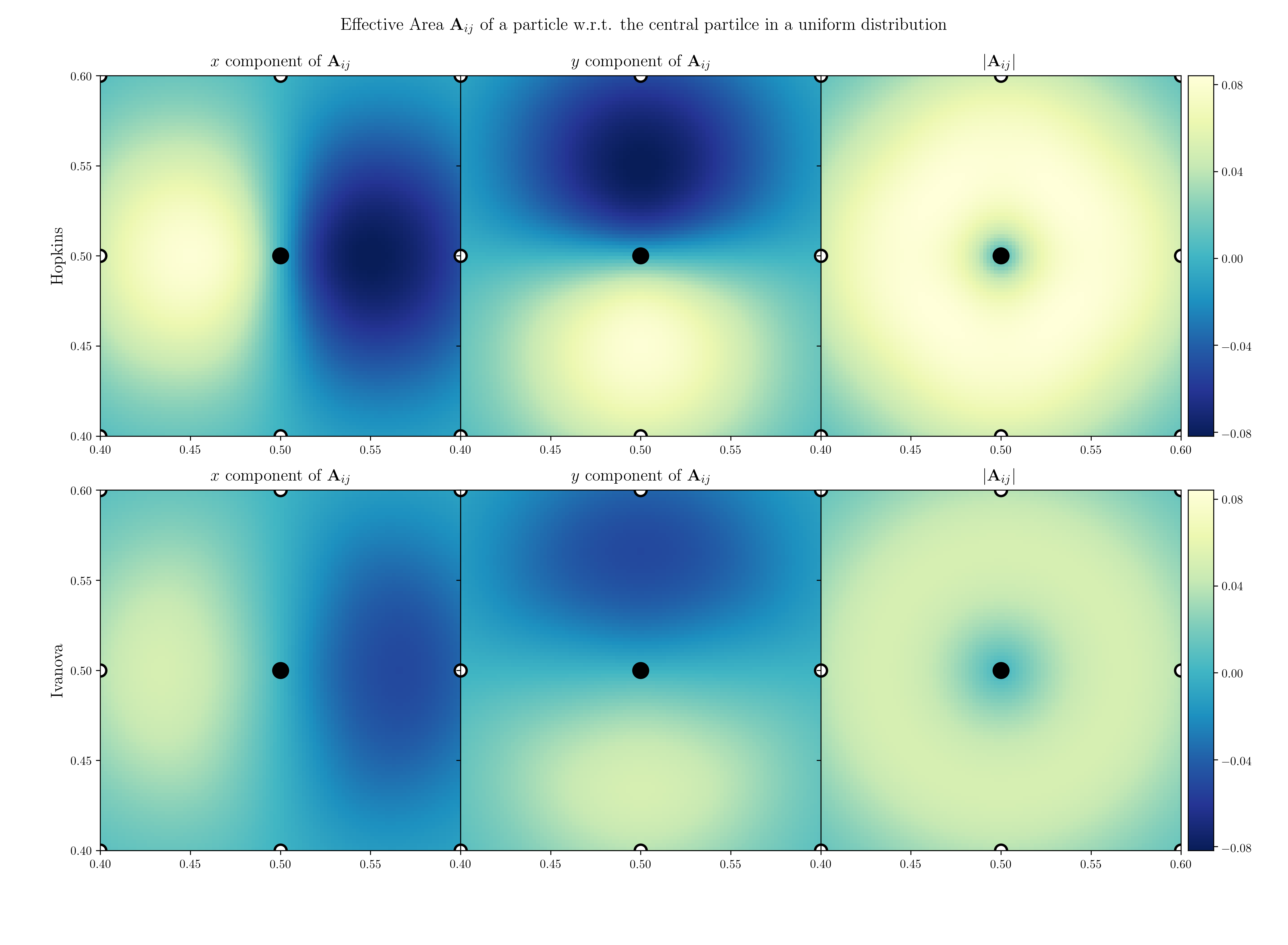}
	\caption{\label{fig:displaced-particle}
		x-component, y-component, and norm of \Aij\ for both the Hopkins and Ivanova expression for
a particle being displaced in an otherwise uniform particle configuration w.r.t the central (black)
particle.
		The white circles show the positions of the closest neighbor particles of the central one.
		The value of \Aij\ of the displaced particle at a certain point on the plotted plane
determines that point's color.
	}
\end{figure}

Let us now look into how the effective surfaces between two particles behave when one particle's
position is varied. To examine that, a particle is placed in an otherwise uniform configuration and
the \Aij\ are computed at that place with respect to the central particle. Figure
\ref{fig:displaced-particle} shows the x- and y- component of \Aij\ and $|\Aijm|$ for both
discretization methods. The obtained value is plotted at the particle's position. Note that in
during the actual flux exchanges the effective surface would be assumed to be in a different place
than is currently plotted, namely at the position specified by eq. \ref{eq:xij}.

As one would expect, the \Aij point towards the central particle and increase with distance until
another particle becomes too close and starts having a bigger contribution via the $\psi(\x)$ at
that position. The Hopkins \Aij reach higher peak values, and the contour shapes aren't perfect
circles, but a little ``boxy''. In accordance to the findings of the direct comparison in the
previous section, the Ivanova methods display a higher relative contribution with increasing
distance from the central particle. The same behavior of the effective surfaces pertains for varying
choices of neighbor number per particle, and for different kernels.

\subsubsection{Summary}

To summarize the findings, there are clear differences in the magnitudes of the \Aij\ obtained
using the Ivanova and Hopkins formulation, with factors up to $\sim 3$.
Not only the magnitudes differ, but the directions of the normal vectors to the surfaces \Aij\ do
as well. The total sum of the norms of \Aij\ of the Ivanova expression is almost always smaller
than the Hopkins version. This could lead to smaller total fluxes and possibly allow bigger time
step sizes when using the Ivanova version. Additionally, the Ivanova methods display a higher
relative contribution with increasing distance from the central particle than the Hopkins method.
This could lead to the method being more diffusive, but simultaneously could also mean that it's
more stable under extreme conditions like violent shocks.

A further advantage of the Ivanova formulation is that the effective surfaces are well-defined even
in troublesome particle configurations. The Hopkins version requires the inversion of a matrix (eq.
\ref{eq:matrix_B}), which can lead to problems if the matrix $\mathcal{B}$ is singular.
In that sense, the Ivanova version could make the hydrodynamics method more stable.

\subsection{Comparison in Hydrodynamical Applications}

\begin{figure}
    \centering
    \includegraphics[width=\textwidth]{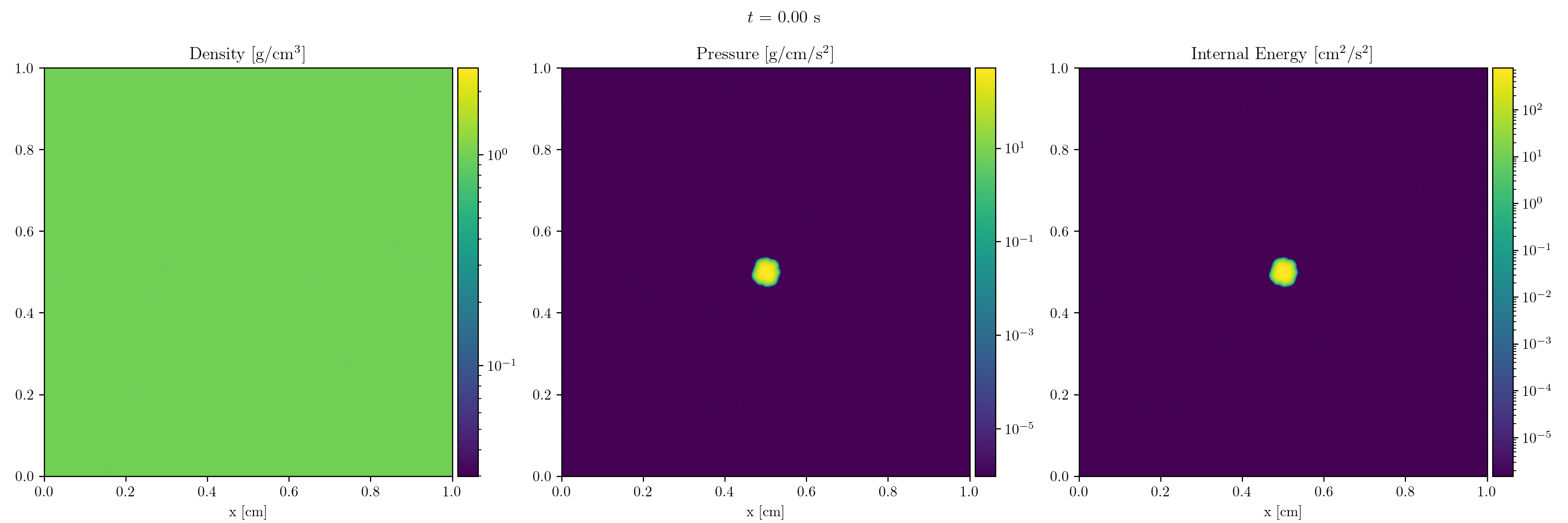}%
\\
    \includegraphics[width=\textwidth]{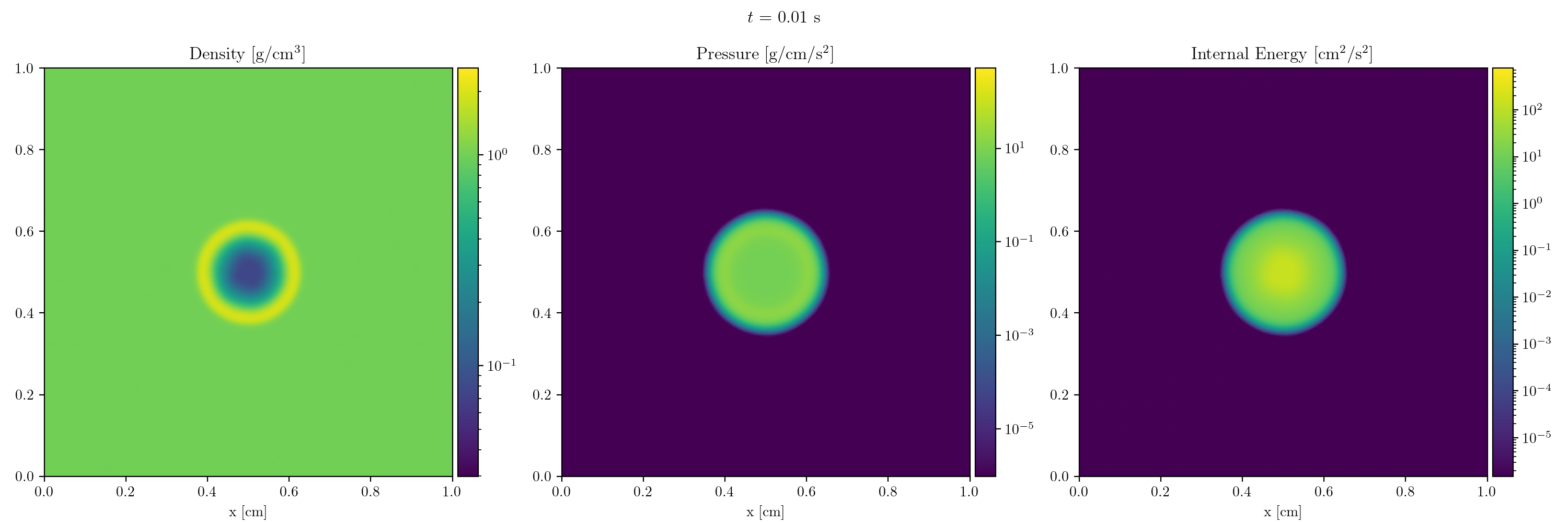}%
\\
    \includegraphics[width=\textwidth]{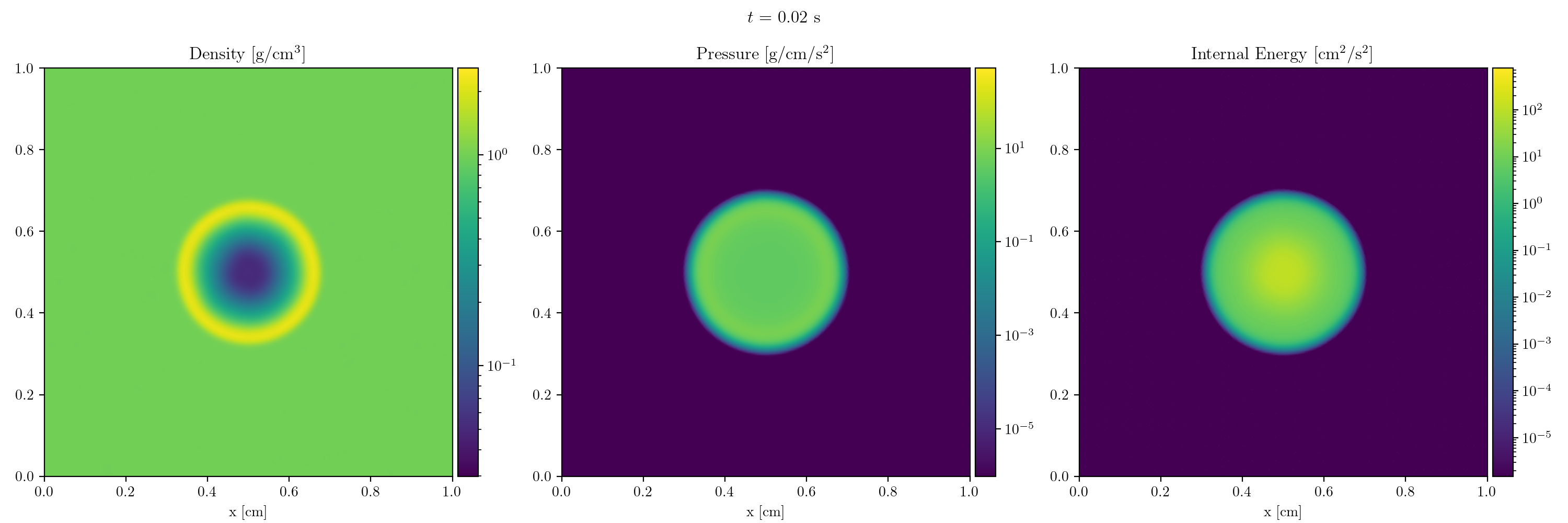}%
\\
    \includegraphics[width=\textwidth]{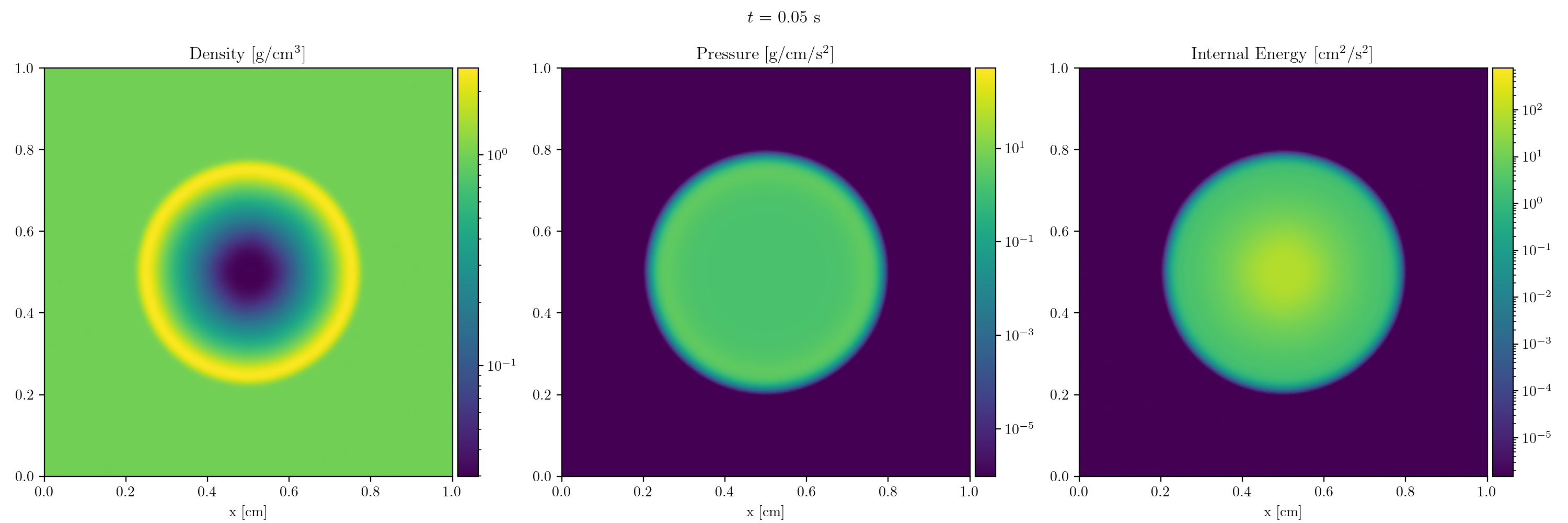}%
    \caption{
The results of the Sedov blast test in two dimensions at times $t = 0.0$s, $0.01$s, $0.02$s, and
$0.05$s for density, pressure, and internal energy. The initial conditions are set up such that a
few particles in the center of the box are injected with a high energy end pressure, which results
in a spherical explosion. The shown solution was obtained using the Hopkins \Aij and while keeping
particles static. Using the Ivanova \Aij with static particles results in a nearly identical
solution, which can be seen in the profiles shown in the top row of
Figure~\ref{fig:hopkins-ivanova-sedov}. These figures were created with the \swiftsimio python
library \citep{borrowSwiftsimioPythonLibrary2020}.
    }
    \label{fig:meshless-sedov-solution}
\end{figure}

\begin{figure}
    \centering
    \includegraphics[width=\textwidth]{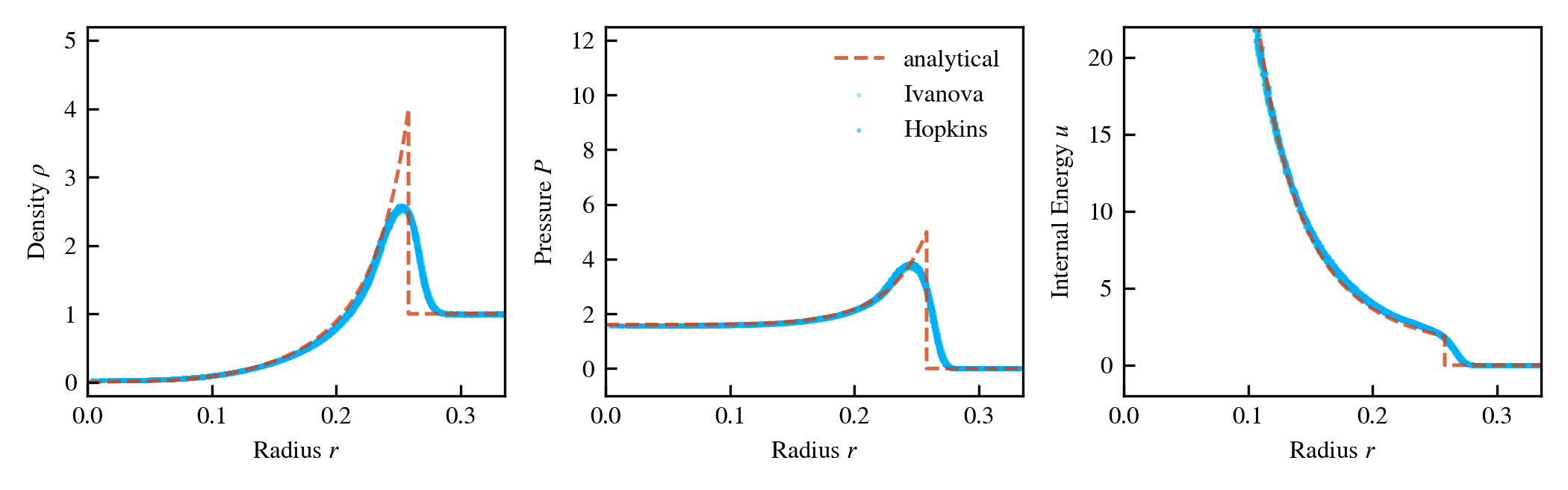}%
\\
    \includegraphics[width=\textwidth]{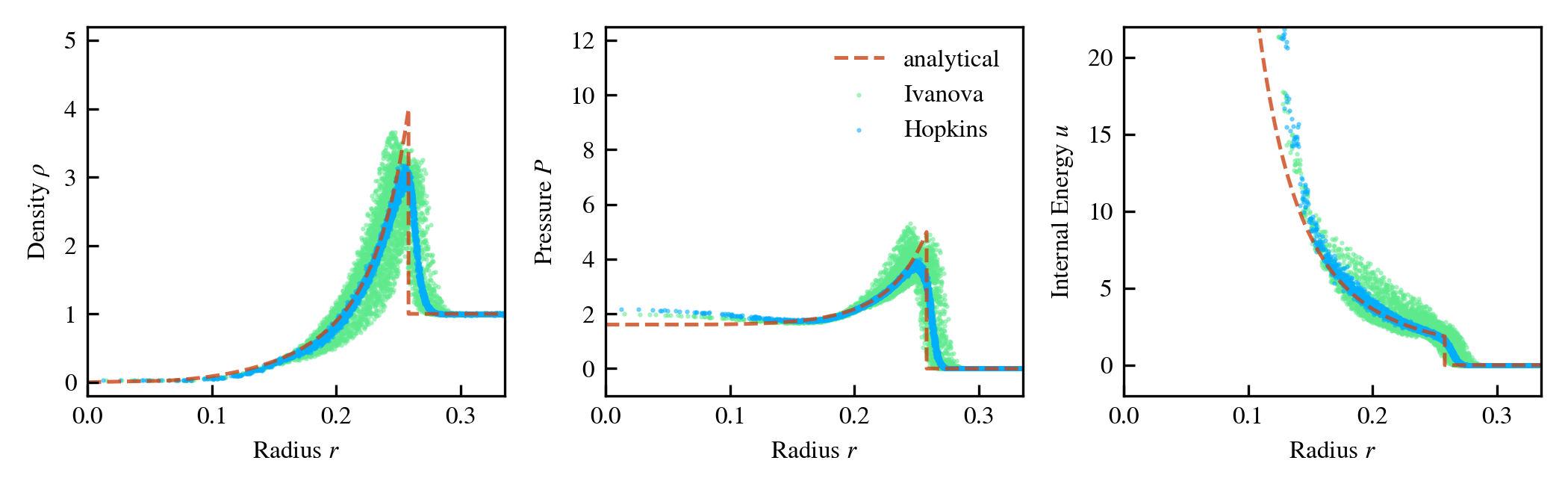}%
    \caption{
A Sedov blast test using both Hopkins and Ivanova effective surface. The blue points are the
solutions using the effective surfaces of \cite{hopkinsGIZMONewClass2015}, the green dots are the
solution using the \cite{ivanovaCommonEnvelopeEvolution2013} \Aij. The density, pressure, and
internal energy are shown as a function of radius from the center, in arbitrary units. The top row
shows the results when particles are kept static, i.e. are not being drifted. The results using
the Hopkins and Ivanova expressions for surfaces \Aij are virtually identical. The bottom row shows
the results for Lagrangian particles. The dashed line shows the analytical solution.
    }
    \label{fig:hopkins-ivanova-sedov}
\end{figure}

\begin{figure}
\minipage{0.33\textwidth}
	\centering
	Initial Conditions
  \includegraphics[width=\linewidth]{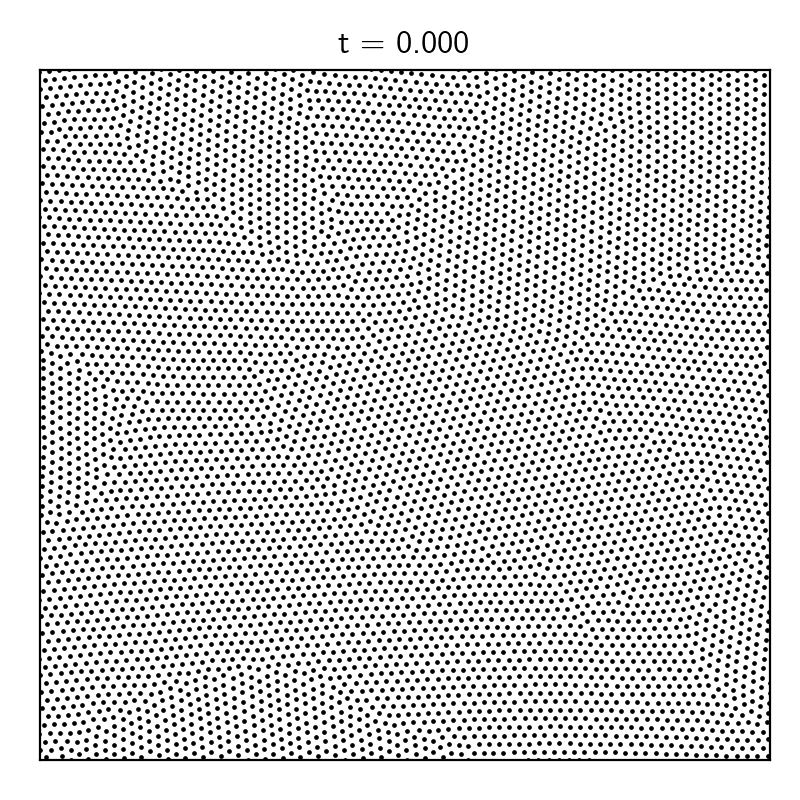}
\endminipage\hfill
\minipage{0.33\textwidth}
	\centering
	Hopkins Solution
  \includegraphics[width=\linewidth]{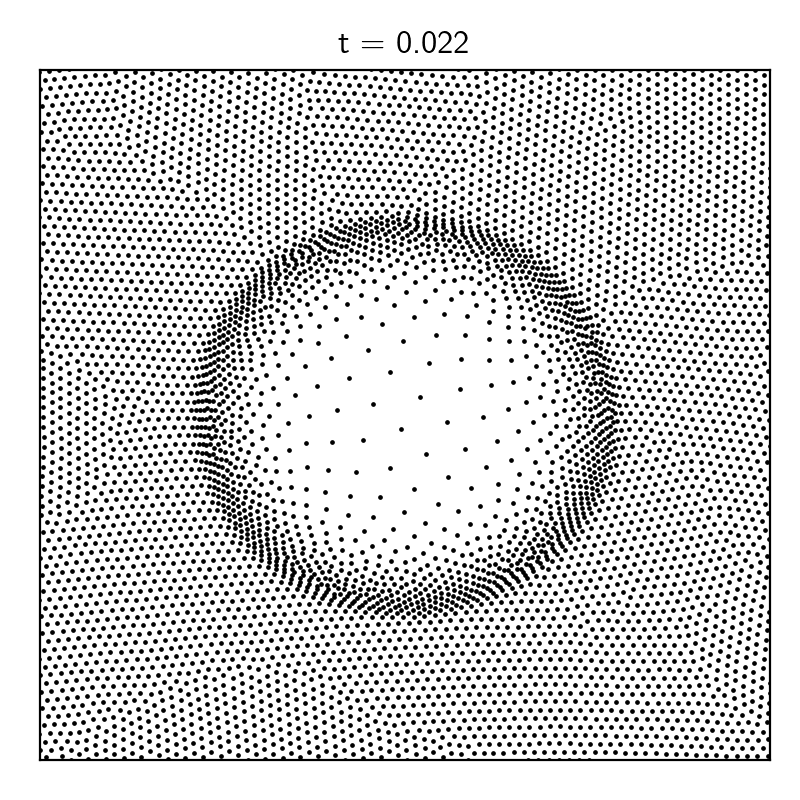}
\endminipage\hfill
\minipage{0.33\textwidth}%
	\centering
	Ivanova Solution
  \includegraphics[width=\linewidth]{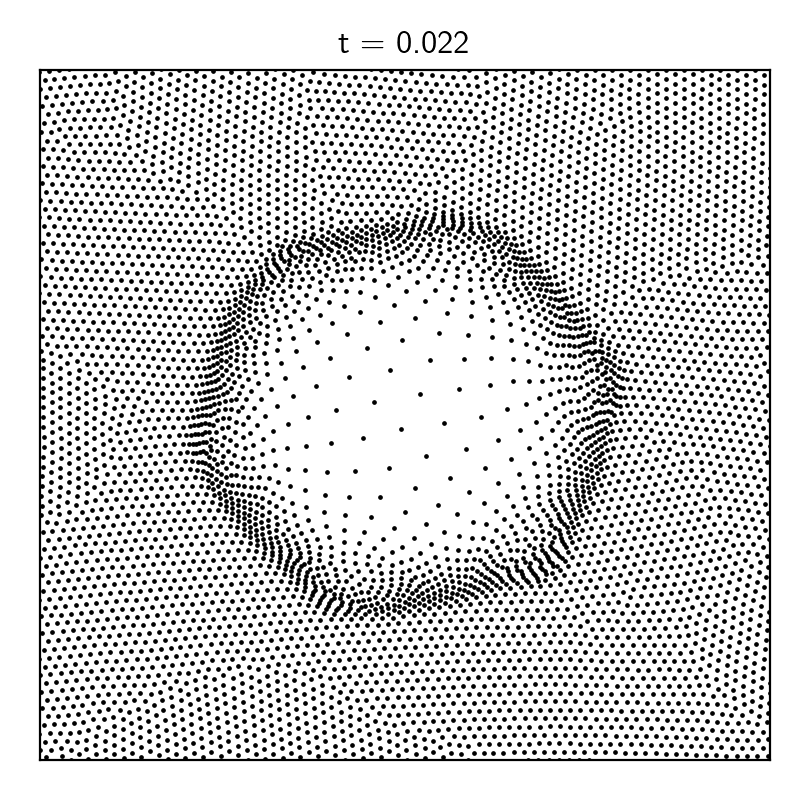}
\endminipage\\
%
%
\minipage{0.33\textwidth}
  \includegraphics[width=\linewidth]{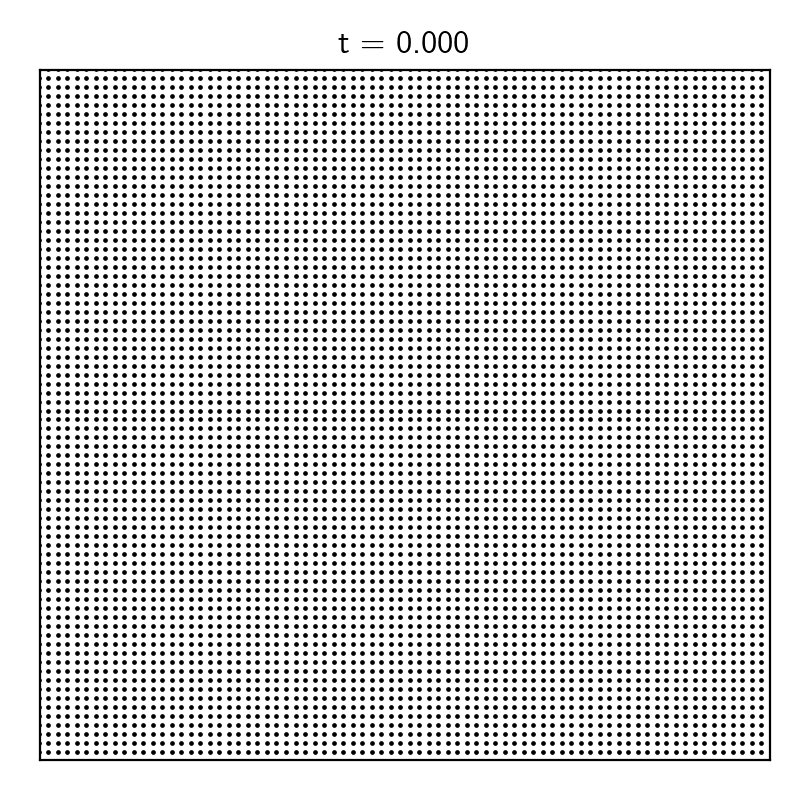}
\endminipage\hfill
\minipage{0.33\textwidth}
  \includegraphics[width=\linewidth]{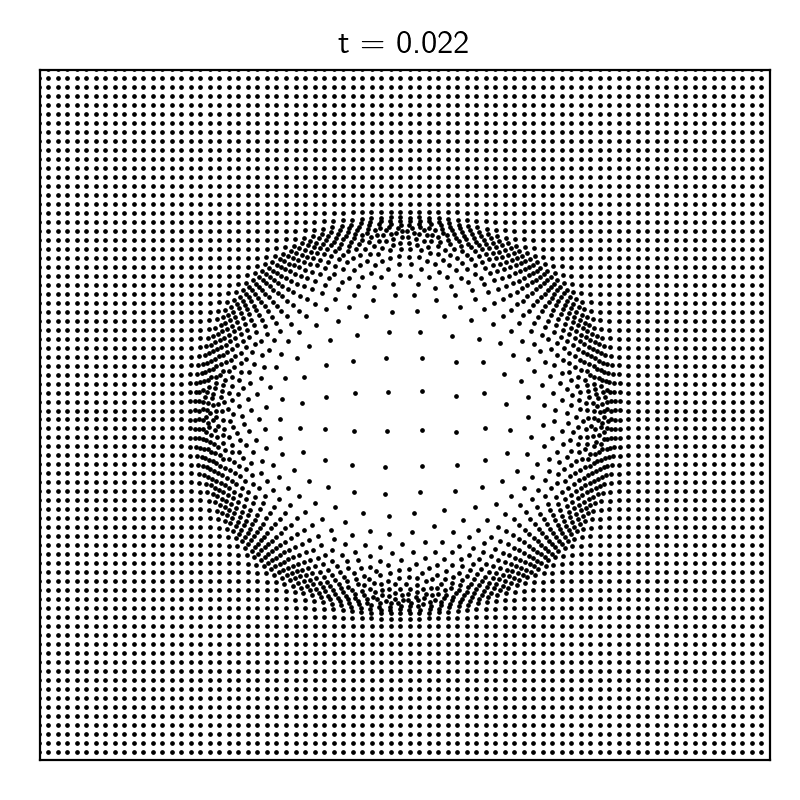}
\endminipage\hfill
\minipage{0.33\textwidth}%
  \includegraphics[width=\linewidth]{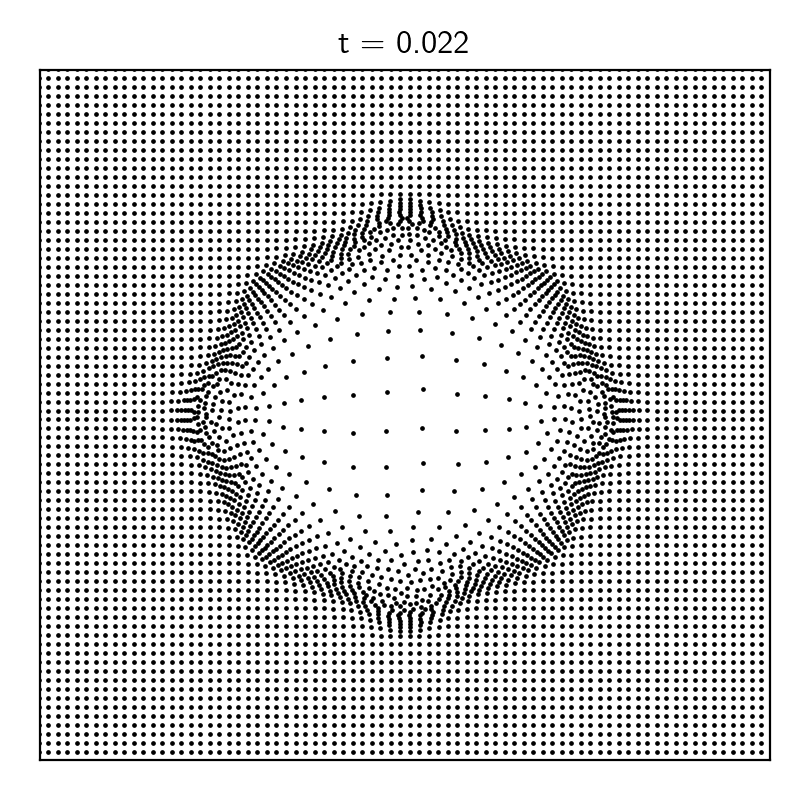}
\endminipage
\caption{
The influence of the initial conditions on the classical Sedov blast test. Plotted are the particle
positions of the initial conditions on the left, the solution provided by the
\citet{hopkinsGIZMONewClass2015} version in the middle and the
\citet{ivanovaCommonEnvelopeEvolution2013} version of the finite volume particle method on the right
for glass like initial particle positions (top) and initially uniformly placed particles (bottom)
for a Sedov blast, in arbitrary length and time units.
The blast should be a radially symmetric explosion from the center, which is well approximated with
the \citet{hopkinsGIZMONewClass2015} version, but not with the Ivanova version. Instead, the shape
of the blast wave is determined by the underlying particle configuration: hexagon-shaped for the
glass initial conditions, and octagon-like for the uniform particle configuration.
  }
\label{fig:sedov-particle-positions}
\end{figure}

\begin{figure}
    \centering
    \includegraphics[width=.8\textwidth]{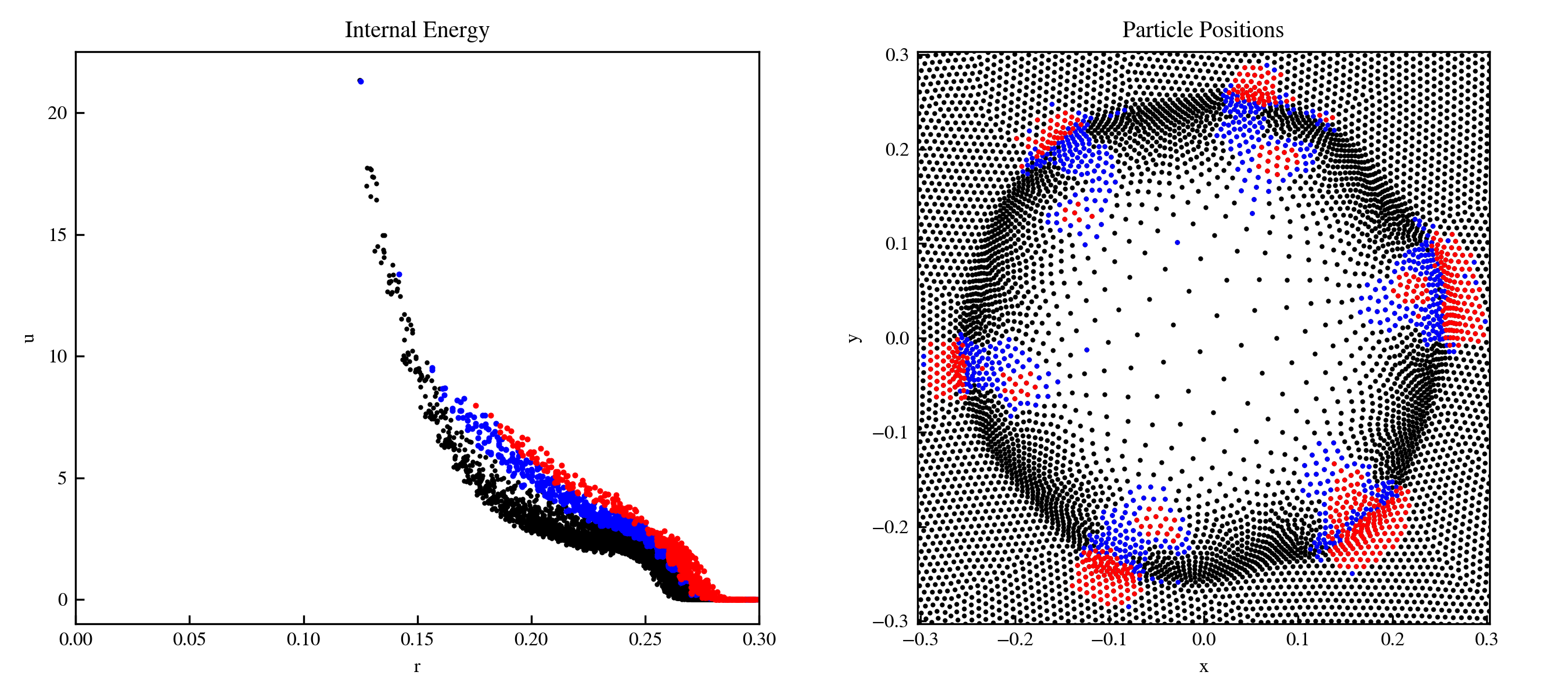}%
    \caption{
On the left, the internal energy $u$ as a function of radius $r$ from the center of the solution of
the Sedov blast test using the Ivanova \Aij and Lagrangian particles is shown. The output is
performed at the same time as in Figure~\ref{fig:hopkins-ivanova-sedov}. Particles with internal
energies above 1.2 and 1.5 times the average internal energy at that radius are marked blue and
red, respectively. On the right, the particle positions of the solution at the same time are shown.
The blue and red particles are the same particles marked correspondingly in the left plot,
highlighting that the scatter seen in the left plot and in Figure~\ref{fig:hopkins-ivanova-sedov}
are in regions where the particle configuration has become irregular and deformed following the
initial glass-like configuration.
    }
    \label{fig:sedov-ivanova-marked}
\end{figure}

To conclude the comparison between the Hopkins and Ivanova versions of FVPM, let's look at their
performance on actual hydrodynamical tests. The experiments are run using the simulation code \swift \citep{schallerSWIFTSPHInterdependent2018a}, which contained an implementation of the Hopkins version outlined in Section~\ref{chap:meshless-full} already, but not the Ivanova one yet. The details of the implementation will be discussed in detail later in Chapter~\ref{chap:meshless-implementation}.

Running actual hydrodynamics tests with both method revealed that the Ivanova formulation has some
big trouble when applied on Lagrangian particles. Figure~\ref{fig:hopkins-ivanova-sedov} shows the
solution to a classical test, the Sedov blast, in two dimensions. The initial conditions are set up
such that a few particles in the center of the box are injected with a high energy end pressure,
which results in a spherical explosion. When running the simulation as an Eulerian code, i.e.
keeping particles static, the results are nearly identical for both versions of effective surfaces
\Aij. The results are somewhat diffusive, which is to be expected since the interacting particle
pairs are spread out over a larger region compared to what a grid code that only interacts adjacent
cell pairs would require. Moving the particles along with the fluid however introduces a very strong scatter in the solution using the Ivanova \Aij, while the Hopkins version delivers adequate results.
The reason for the scatter appears to be a strong dependency of the Ivanova \Aij on the particle
configuration. Figure~\ref{fig:sedov-particle-positions} shows the particle positions for two
different underlying initial configurations: a glass-like ordering, and a uniform configuration. In
both cases the evolution of the blast wave with the Ivanova \Aij shows strong traces of the initial
particle positions. Figure~\ref{fig:sedov-ivanova-marked} shows the result of the Sedov blast with
the Ivanova \Aij where some particles are marked based on their internal energy $u$. It is clear
that the strong scatter is introduced by the deformities that follow the underlying initial particle configuration.

The same issues with the Ivanova surfaces appears for a wide variety of other tests. Unfortunately
neither a different choice of kernels, nor increasing the number of neighbors for particles to
interact with, nor very small Courant numbers, nor using much more restrictive limiters were able to alleviate the problem noticeably. This leads to the conclusion that in the presented form, the
Ivanova version of the finite volume particle method is not suitable for ``real life'' application
for hydrodynamics in the astrophysical and cosmological context. \footnote{
However, the Ivanova version appears to be an adequate choice for applications which don't require
co-moving particles, as is for example the case for the transport of radiation, which will be the
topic in Part~\ref{part:rt} of this thesis.}.
Having Lagrangian particles is absolutely necessary for cosmological simulations, and the Ivanova
\Aij do not work well with co-moving particles. A possible solution would be to determine the
particle velocities with a different method than the one used to evolve the fluid quantities. For
example, the particle velocities could be set using the results of a simple first order accurate
method, or using some basic SPH formulation, both of which typically aren't very sensitive to the
underlying particle configuration. Testing these solution attempts remains a subject for future
works.

%% file: main/Meshless/ML-3-implementation.tex
\chapter{Implementation in SWIFT}\label{chap:meshless-implementation}

To conclude the part on finite volume particle methods, their implementation in the cosmological
hydrodynamics code \swift \citep{schallerSWIFTUsingTaskbased2016,
schallerSWIFTSPHInterdependent2018a} is presented. It is fully open-source and publicly available on \url{https://gitlab.cosma.dur.ac.uk/swift/swiftsim} and on github under
\url{https://github.com/SWIFTSIM/swiftsim}, along with extensive documentation and a plethora of
ready-to-run examples.

\swift uses particles as the fundamental discretization elements, and has several flavors of
Smoothed Particle Hydrodynamics methods implemented. Gravity is solved using a fast multipole method \citep{chengFastAdaptiveMultipole1999, dehnenFastMultipoleMethod2014f} coupled to a particle mesh solver in Fourier space to deal with periodic volumes. In addition, several galaxy formation models are already implemented, most notably \codename{Eagle} \citep{schayeEAGLEProjectSimulating2015} and \codename{Gear} \citep{revazDynamicalChemicalEvolution2012}.

\swift is highly parallelized and makes use of a hybrid task-based parallelism strategy for shared
memory parallelism combined with MPI for distributed memory parallelism. The task-based parallelism
furthermore permits to exploit asynchronous MPI communications and a domain decomposition strategy
based on the work rather than data to efficiently utilize modern high performance computing
architectures. These technical aspects will be discussed in more detail in the subsequent sections,
as they are intimately tied to the manner in which the finite volume particle methods for the Euler
equations (and for the equations of radiative transfer in Part~\ref{part:rt}) need to be
implemented.

\section{Task Based Parallelism}

\begin{figure}
 \centering
 \includegraphics[width=\textwidth]{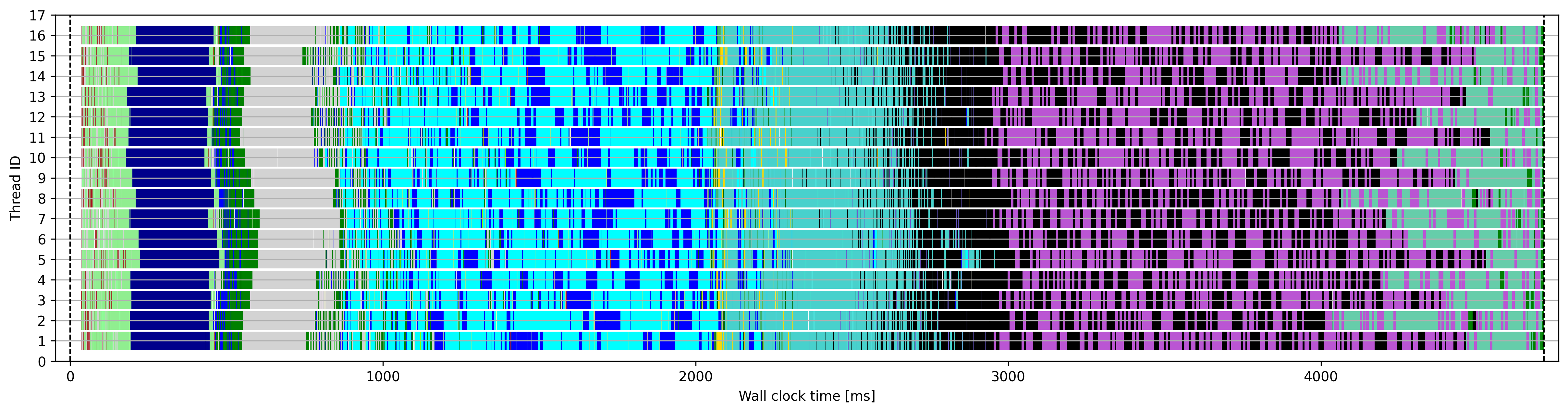}%
 \caption{
One time step of a simulation using \swift with 16 threads. The colored blocks represent various
tasks being solved. Each color represents a different task type. Note how there is no noteworthy
idle time (white blocks) and how the varying threads are solving different types of tasks
concurrently, and that all threads finish at nearly the identical time.
 }
 \label{fig:taskplot}
\end{figure}

\swift relies on a task-based parallelism scheme
\citep{gonnetSWIFTFastAlgorithms2013, gonnetQuickSchedTaskbasedParallelism2016b,
theunsSWIFTTaskbasedHydrodynamics2015}.
The task-based parallelism is a system in which a full computation is divided into a set of
individual however inter-dependent tasks. They are dynamically allocated to idle cores by a
dedicated task scheduler \citep{gonnetQuickSchedTaskbasedParallelism2016b}.
To illustrate the idea, consider the following scenario: ``\textit{If a construction worker requires
two weeks to pave a street, how long would it take ten construction workers?}'' In this analogy, the
paving of the street is the simulation that needs to be executed, while the construction workers are
computing cores.

In traditional parallelism schemes, every worker would execute the same sequence of tasks and
commands. Each construction worker would first clean up the place where the street needs to be
paved, then heat up their own portion of asphalt, pour it into place assigned for this worker, and
finally stomp the new asphalt in with a roller. Obviously employing more workers should lead to
finishing the job quicker, but there is room for improvement. For example, what if not all workers
can be assigned the same amount of street surface to work on? All workers that finished their share
of the job already would stand by idly waiting for the last one to complete their part. Having
workers just stand around with nothing to do is obviously a scenario we'd like to prevent.

This is one of the reasons why \swift (and other state-of-the-art codes like
\cite{stoneAthenaAdaptiveMesh2020}, \cite{wadsleyGasoline2ModernSmoothed2017},
\cite{menonAdaptiveTechniquesClustered2015})  moved towards a task-based parallelism scheme. This
means that instead of giving every worker exactly the same sequence of steps to do, we keep track of
all the individual jobs that need to be completed: Each section of the street needs to be cleaned
up, each section of the street requires its respective amount of asphalt heated up, then poured,
then rolled. Some of these jobs however can be completed concurrently: While some workers can clean
up the street, others can already start to heat up the asphalt. Once a section of the street is
cleaned up, the first asphalt can be poured. Simultaneously other workers can keep cleaning other
sections of the street and heating up more asphalt. This concurrent work is of prime interest for
the task-based parallelism scheme: As long as there is work to do, the workers shouldn't be idling.
To illustrate how the application of task-based parallelism can look like, Figure~\ref{fig:taskplot}
shows the tasks being executed during one time step of a simulation with \swift. The simulation was
run using 16 threads, and Figure~\ref{fig:taskplot} shows what each thread was working on during the
time step and how long it took. Different types of tasks are colored differently, and it clearly
shows how the varying threads are solving different types of tasks concurrently. Note how there is
no noteworthy idle time (white blocks) once the tasks start being executed and how all threads
finish at nearly the identical time.

This approach however requires us not only to define tasks, i.e. portions of work that a single
worker at a time could do, but also \emph{dependencies}. Some things need to be finished first
before others can be done: For example, one couldn't pour the asphalt before it's hot, nor could one
stomp it in before it's poured. Additionally, we need to also define \emph{conflicts}. Conflicts
between tasks arise in cases where there is work to be done for which the order of execution
doesn't matter, but it can't be done concurrently. This can for example be the case when two tasks
need to access and modify the same data, while the order of the access and modification is
unimportant. In the ``workers paving a road'' analogy, this could for example be the case when the
workers are finishing for the day and packing up their tools into their trucks. It doesn't really
matter which tools go in first, but only one tool can be put in the trunk at a time.

Task-based parallelism offers many benefits aside from reducing idle waiting time of CPUs. It can
be used to drastically decrease the times CPUs spend to fetch data from memory by ensuring that
tasks which act on a set of data are always executed by the same CPU. This way the memory remains
local, and the cache efficiency is increased. Furthermore, when dealing with distributed memory
architectures the tasks can be used as an adequate measure of the actual work that needs to be
performed. The domain decomposition between the nodes can then be performed based on sharing the
work equally rather than some proxy for the work like the number of cells or particles contained
within a region. Having other work available for threads to execute also enables \swift to use
asynchronous messages on distributed memory architectures: The threads don't need to perform
blocking messages, where both the sending and the receiving side do nothing but wait for each other
to complete the communication. Instead, the thread on the sending side can send the message and
continue doing other work, while the thread on the receiving side can start with the work once it's
notified that the message has been received and keep itself busy otherwise until the message
arrives.

However, task based parallelism also comes with several caveats. Firstly, it requires for tasks to
be generated during a simulation, and then for the scheduler to execute them, which constitutes
additional overhead work that needs to be done so the actual physics can be solved. The overhead is
noticeable for small problems, e.g. simulations that can be executed on a personal computer. With
increased problem size however, the benefits of the task-based parallelism make the overheads
negligible and lead to a significantly improved performance and scaling of the code in comparison
with traditional parallelization techniques (\cite{borrowSWIFTMaintainingWeakscalability2018}).
Secondly, the manual definition of tasks, dependencies, and conflicts necessary for the task-based
parallelization requires a considerable effort on the developing side. It doesn't suffice to write
code that solves the equations that govern the physics any longer. The entire framework of the
tasks, dependencies, and conflicts needs to be constructed first. The major difficulties with that
work is to identify and fix errors in the dependency logic. These difficulties arise because (a)
errors can be very hard to spot, since in real applications they manifest in certain variables being
accessed too early or too late without any warning, and (b) even when spotted, in general they
aren't reproducible because the execution order of the tasks is not fixed and may depend on
uncontrollable circumstances, e.g. which processing cores one gets access to for a given run, where
exactly in memory the processors need to fetch data from and how long that will take, etc. Matters
become even worse when distributed memory parallelization is included, as that adds an entire other
layer of unpredictability and irreproducibility.

\section[FVPM In \swift: Task Dependency Graph]{Hydrodynamics With Finite Volume Particle Methods
in SWIFT: Constructing the Task Dependency Graph}\label{chap:swift-hydro-tasks}

\begin{figure}[H]
\centering
\includegraphics[width=\linewidth]{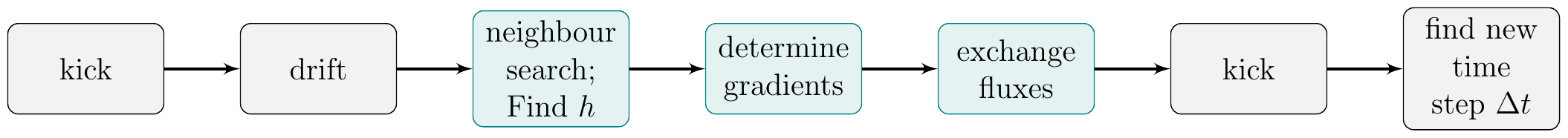}%
\caption{The order of operations for a single time step required for the finite volume particle
methods for each particle using a kick-drift-kick time integration scheme. The steps with a blue
background, namely the neighbor search, the computation of gradients, and the exchange of fluxes,
are the steps that involve loops over neighboring particles.}
\label{fig:meshless-order-of-operations}
\end{figure}

In this section, the tasks and dependencies required for the solution of hydrodynamics problems
using the finite volume particle methods in \swift is described. To begin with, let's remind
ourselves of the order of operations necessary for the finite volume particle hydrodynamics (see
Section~\ref{chap:meshless-full}), shown in Figure~\ref{fig:meshless-order-of-operations}. This
order of operations describes the order of work that needs to be done for each particle. In case
other physics are involved, in particular gravity, a second kick operation before the drift is
necessary for the integration to be sufficiently accurate. In what follows, we include that second
kick. The steps with a blue background, namely the neighbor search, the computation of gradients,
and the exchange of fluxes, are the steps that involve loops over neighboring particles.

\subsection{Constructing Tasks}

\subsubsection{Grouping Particles in Cells}

To develop the tasks and dependencies required for this order of operations to be executed on each
particle, we need to look into how the tasks and the particles are intertwined first.
As described in Section~\ref{chap:meshless-full}, a major component of the finite volume particle
schemes are the interactions between neighboring particles, and in particular the neighbor search
which is necessary each step. For the sake of optimization and quick access, particles are grouped
into cells whose size is determined by requiring that all neighbors of the particles within a cell
must be inside adjacent cells. This ensures that the number of adjacent cells that need to be
involved in an interaction are always known and predictable. Initially, the entire simulation domain
is divided into cells of equal size, called the ``\lingo{top level}'' cells. These \lingo{top
level} cells are then recursively split into cells with half the parent cell's size as much as
possible while the condition to have all neighbors of particles in adjacent cells is satisfied.
Having cells as small as possible and containing as few particles as possible is desirable as it
speeds up the neighbor search significantly by increasing the ratio of particles which are actually
neighbors of another particle to the number of total particles checked during the interactions.
However, we still
need to ensure that \emph{all} neighbors can be found in neighboring cells, which gives the lower
boundary of the cell size.

\subsubsection{Connecting Work and Data: Task Classes and Types}

\begin{figure}
 \centering
 \includegraphics[width=\linewidth]{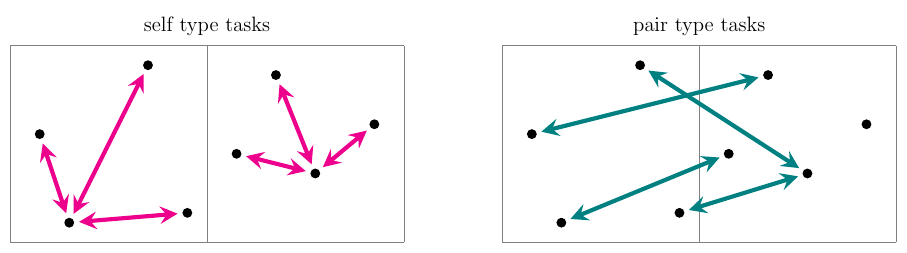}%
 \caption{
 Two adjacent cells containing some particles are shown. On the left, the interactions of
\lingo{self} type tasks are illustrated: They only interact particles with other particles that
share a cell with them. On the right, \lingo{pair} type tasks are shown. They only interact
particles form one cell with the particles from the other cell. Some interactions between particles
have not been drawn for clarity, but in principle for \lingo{self}-type tasks all particles within
a cell will interact with all other particles in that cell. Similarly all particles of a cell
should interact with all particles in the neighboring cell in \lingo{pair}-type tasks. }
 \label{fig:self-pair-tasks}
\end{figure}

Tasks are then attached to the cells, not the individual particles inside the cells. This means that
when a task is being executed, it operates on all the particles contained within a cell. At this
point, it's useful to begin differentiating between two classes of tasks. The first class would be
all the tasks which include interactions between particles, e.g. the neighbor search. Let's call
them ``\lingo{interaction}'' tasks. The other class of tasks conducts work exclusively on individual
particles, for example the drift and kick operations. The construction of these ``\lingo{plain}''
tasks is straightforward: All they need to do is go over all particles contained in a given cell to
which the task is attached to and do whatever it is set up to do on the particles individually. The
\lingo{interaction} tasks however are a little more contrived. They can again be split into two
different types of interactions. The first type, called ``\lingo{self}'' tasks, perform the
interactions of particles contained within a cell with all the other particles \emph{inside the same
cell}. \lingo{Self}-type tasks are always attached to a singe cell. The second type,
``\lingo{pair}'' tasks, perform the interactions of particles contained within a cell with the
particles contained within some neighboring cell. \lingo{Pair}-type tasks are always attached to a
pair of cells. The cells involved in the interaction are then always adjacent cells by construction.
Figure~\ref{fig:self-pair-tasks} illustrates the interactions involved with \lingo{self}- and
\lingo{pair}-type tasks. However, since cells are being split recursively into smaller ones until
they reach a lower limit, this means that not all leaf cells will have the same size, i.e. be at the
same level, and the tree structure of cells will have varying depths depending on the number of
particles currently situated there. This would mean that in some cases a task would need to interact
with cells of different sizes, and hence a variable number of \lingo{pair} type tasks needs to be
constructed between a cell and its potentially smaller sized adjacent cells. Rather than
constructing an individual task for each \lingo{pair} type interaction that might be necessary, we
pick some convenient cell level in the cell tree that we call the ``\lingo{super level}''.
All the tasks are then constructed and attached to cells at the \lingo{super level} in the tree.
The \lingo{super level} may or may not be the \lingo{top level}, i.e. the root of the cell tree. If
a \lingo{super level} cell is split and thus has children, \lingo{self}- and \lingo{pair}-tasks are
replaced by ``\lingo{sub-self}'' and ``\lingo{sub-pair}'' tasks which recursively descend to the
leaf cell level and perform all required interactions. Since each cell is split into eight children
cells of half the parent cell's dimensions, the recursion of a \lingo{sub-self} type tasks involves
calling a \lingo{sub-self} type task for each child cell as well as a \lingo{sub-pair} type task
for each pair of two child cells, such that all possible interactions are covered. The
\lingo{sub-pair} task recursion is handled analogously.

\subsection{A First Dependency Graph}

\begin{figure}
\centering
\includegraphics[width=\linewidth]{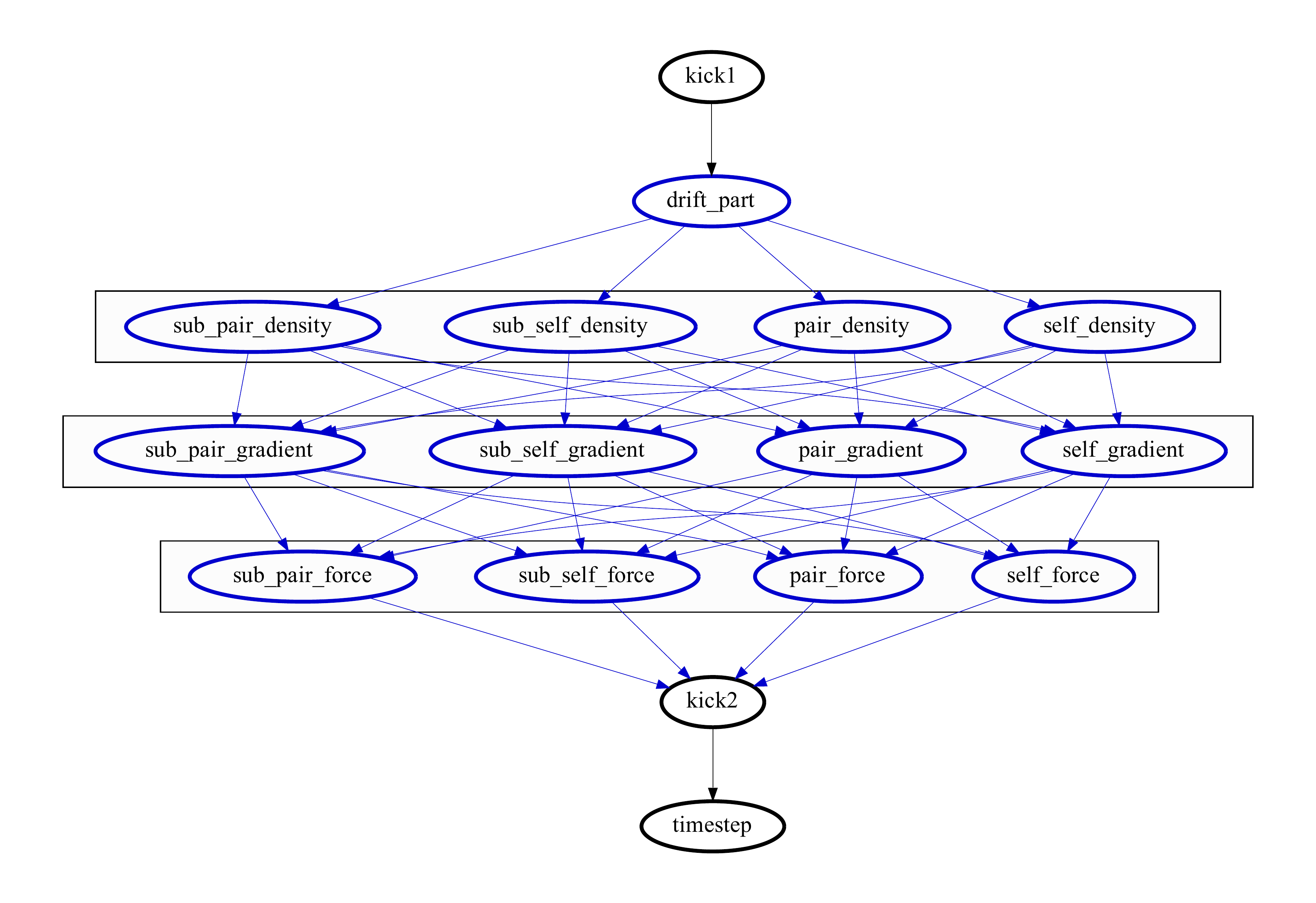}%
\caption{
A first dependency graph for tasks required for FVPM hydrodynamics with \swift based only on the
required order of operations.
}
\label{fig:dependency-graph-zeroth-order}
\end{figure}

Using the \lingo{plain} and \lingo{interaction} classes of tasks, we can now begin to construct the
dependency graph that executes the correct order of operations for the finite volume particle
hydrodynamics scheme. In a first approach, the dependency graph will look like the order of
operations shown previously: The drift, kick, and timestep tasks will be \lingo{plain} tasks, while
the neighbor search, the gradient, and the flux exchange tasks will be \lingo{interaction} type
tasks, i.e. will be \lingo{self}, \lingo{sub-self}, \lingo{pair}, and \lingo{sub-pair} type tasks.
The tasks and dependencies can be represented by a directed acyclic graph, where the nodes are the
tasks, and the edges are the dependencies. A simplified dependency graph based only on the required
order of operations is shown in Figure~\ref{fig:dependency-graph-zeroth-order}. For completeness,
the full kick-drift-kick integration scheme is shown, and hence two kick tasks are included. The
task names are slightly changed to the convention used for smooth particle hydrodynamics:

\begin{itemize}
\item the tasks that drift the particles are called ``\lingo{drift\_part}''
\item the neighbor search interactions are called ``\lingo{density}'' tasks following the SPH
convention, as the neighbor search determines the particle densities.
\item \lingo{self}, \lingo{pair}, \lingo{sub-self}, and \lingo{sub-pair} type tasks are marked as
such with the corresponding prefix. E.g. the sub-self neighbor search interaction task is called
``\lingo{sub\_self\_density}''.
\item the flux exchange interactions are called ``\lingo{force}'' tasks following
the SPH convention again, as the second interaction loop in SPH evaluates the forces between
particles.
\end{itemize}

\begin{figure}
\centering
\includegraphics[width=\linewidth]{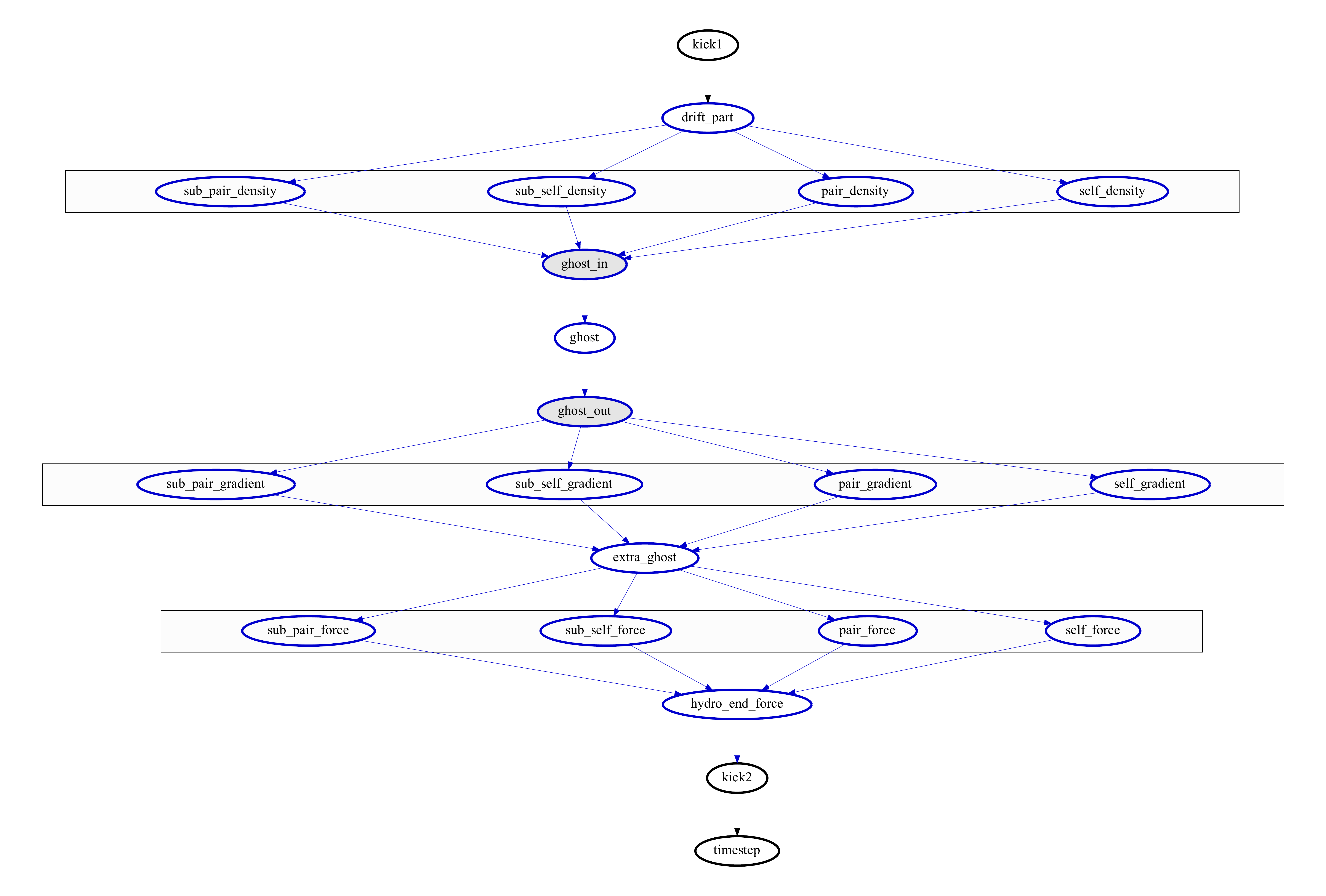}%
\caption{
A first modification of the dependency graph for tasks required for the hydrodynamics with \swift
with new \lingo{plain} tasks added sandwiched by the \lingo{interaction} type tasks. They are
necessary to complete required computations between the particle interaction operations.
Additionally implicit \lingo{ghost\_in} and \lingo{ghost\_out} tasks have been added (nodes with
gray background). They perform no actual work themselves, and exist only to collect dependencies and
make the development process easier.
}
\label{fig:dependency-graph-first-order}
\end{figure}

This simplified graph doesn't account for the fact that the required operations which involve
particle interactions can't be modeled exclusively by \lingo{interaction} tasks. Additional work on
individual particles needs to be performed. For example when computing the matrix $\mathcal{B}$
(eq.~\ref{eq:matrix_B}) for the gradients, first the matrix $\mathcal{E}$ (eq.~\ref{eq:matrix_E})
needs to be accumulated as a sum over neighboring particles, and then inverted. The inversion can
only occur \emph{after} $\mathcal{E}$ is accumulated in the gradient interaction tasks, but needs to
happen \emph{before} the flux exchanges in the \lingo{force} tasks. A second example is the fact
that the search for the smoothing length $h$ needs to be performed iteratively. Rather than
repeating all interactions between all neighboring cells, a list of neighbor candidates can be kept
for particles whose smoothing length hasn't converged yet. Only those neighbors can be checked
again. With the lists available after the first interaction task loop, the iteration can be modeled
as a \lingo{plain} task type instead of an \lingo{interaction} type task, which reduces the number
of conflicts in the algorithm. For this reason, new tasks between the density and the gradient, as
well as between the gradient and the force tasks are introduced, called ``\lingo{ghost}'' and
``\lingo{extra\_ghost}'', respectively. The \lingo{ghost} task is called this way because it
technically performs the work of an interaction type task while it isn't one, and hence does the
work in an ``invisible'' manner. These new tasks are shown in
Figure~\ref{fig:dependency-graph-first-order}.

Additionally, two ``\lingo{implicit}'' tasks are added before and after the \lingo{ghost} task,
named ``\lingo{ghost\_in}'' and ``\lingo{ghost\_out}''. \lingo{Implicit} tasks are a third class
of tasks that exists only to collect dependencies, while doing no actual work at all. They are added
an attempt to both reduce the total number of dependencies throughout the entire task graph as well
as to make the code development process easier.

\subsection{Optimizing \lingo{Pair}-Type Interactions: Sorting Cells}

\begin{figure}
\centering
\includegraphics[width=\linewidth]{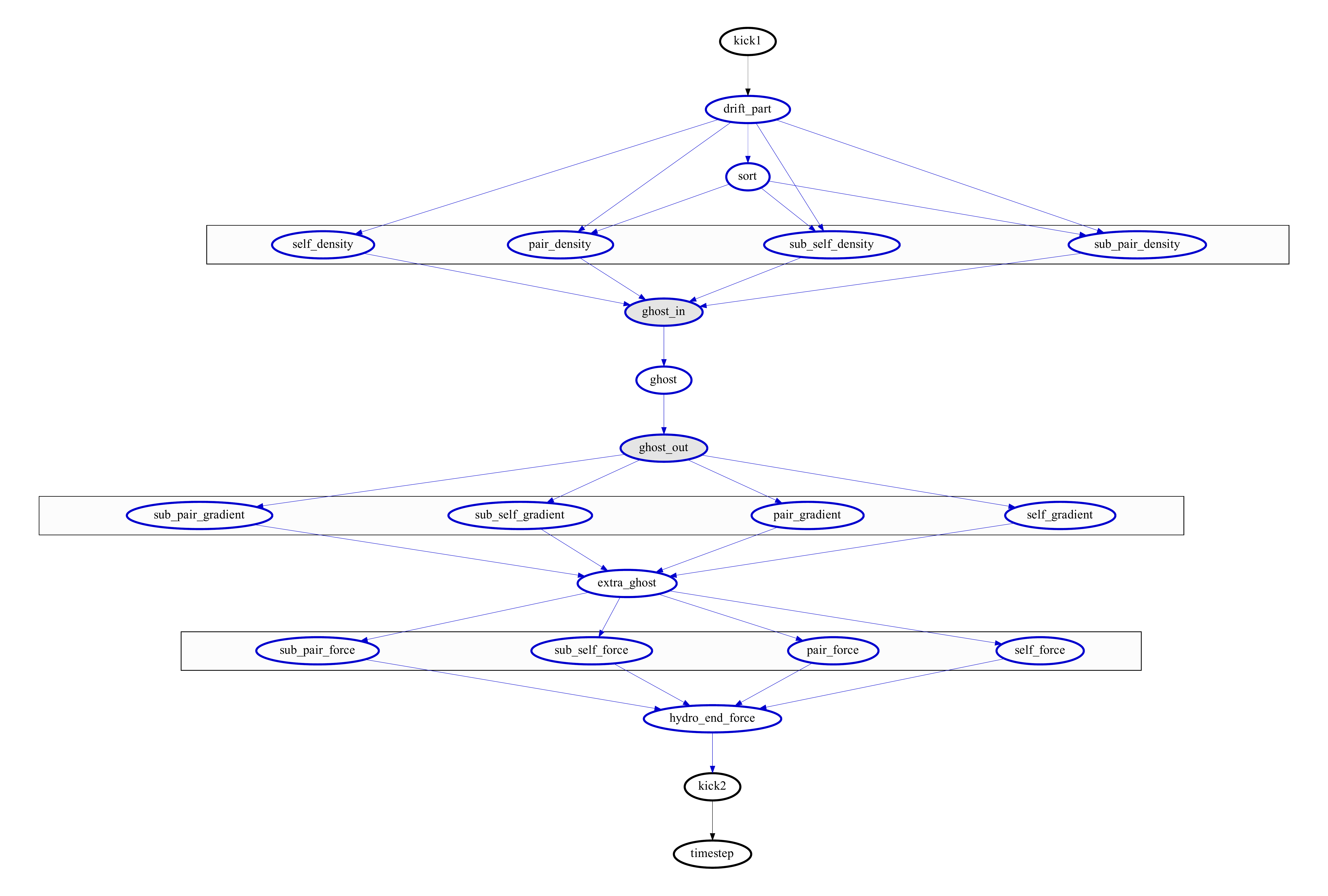}%
\caption{
The second modification of the dependency graph for tasks required for the hydrodynamics with
\swift, with the \lingo{sort} task and its dependencies added.
}
\label{fig:dependency-graph-second-order}
\end{figure}

The interactions between particles constitute a major part of the total workload each time step. For
the finite volume particle methods, we require three separate loops over neighboring particles.
Given this prevalence of the particle interactions, it is sensible to invest effort towards
optimizing the way particle interactions are performed. A good starting point is to limit the total
number of particles checked for possibly being neighbors even more than is already done by
constructing cells in a manner such that all neighbor particles must be situated within adjacent
cells. To this end, particles inside each cell are sorted along each of the 13 lines connecting a
cell center with an adjacent cell center. Any cell will have 26 adjacent cells of equal size, with
each of these 26 neighboring cells having a diagonally opposed counterpart. So it suffices to sort
the particles only along half of the 26 axes, and use the reverse order of the sorted list for the
diagonally opposed neighbor cells. The lists of sorted particles are then used in pair and sub-pair
type tasks as a means of reducing the number of viable possible neighbor particles during an
interaction loop by traversing the sorted lists rather than traversing all particles in the
neighboring cell. More precisely, instead of checking whether each particle in the first cell is
within range of each particle of the second cell, both cells' particles are traversed in the sorted
order along the axis that connects the cells' centers. The distance of the particles on the sorting
axis gives a lower limit on the actual distance between the two particles, since their positions
have been projected along the axis for the sorting. This way a significant amount of unnecessary
checks can be avoided. To facilitate the sorting, ``\lingo{sort}'' tasks must be added, which is
shown in Figure~\ref{fig:dependency-graph-second-order}. The sorting must take place after the
drifts, during which the particle positions are being modified, and before the \lingo{pair} and
\lingo{sub-pair} type tasks. Note that the \lingo{self} type tasks, which only interact particles
within a cell with other particles inside the same cell, do not require the cell itself to be
sorted, and hence have no dependency on the \lingo{sort} task. The \lingo{sub-self} tasks however
recursively interact particles of their child cells with each other, which again needs the particles
in the child cells to be sorted.

\subsection{Adapting to Individual Time Step Sizes of Particles}

\begin{figure}
 \centering
 \includegraphics[width=\textwidth]{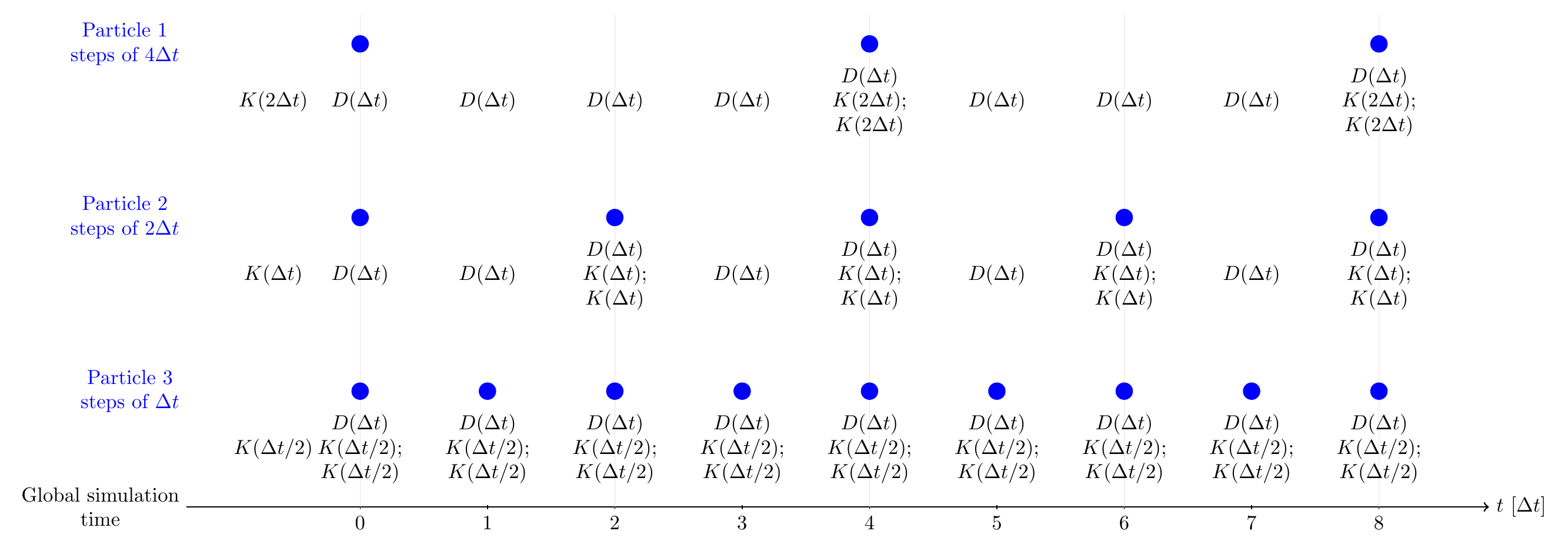}%
 \caption{
Illustration of how the kick and drift operations are performed for particles with individual time
step sizes. A full time integration for a particle with time step size $\Delta t$ is given by
$K(\Delta t / 2) \circ D(\Delta t) \circ K(\Delta t/2)$, where $K$ denotes a kick operator, and $D$
denotes a drift operator. Three particles with time step sizes $\Delta t$, $2 \Delta t$, and $4
\Delta t$, respectively, are depicted. The blue dots represent the points in the simulation time
(shown on the x-axis) at which the respective particles undergo a full hydrodynamics integration
update, i.e. complete the kick-drift-kick integration. Between the full updates, particles with
higher time step sizes compared to their neighbors are drifted to the current simulation time so
their neighbors, which finish their integration step in this simulation step, have access to the
correct particle positions. In order for all particles to be in the correct state to be drifted any
number of times, the first kick operator is applied to all particles before the first simulation
step. At the end of a simulation step, the particles aren't left in the state where the integration
step finishes, i.e. after the second kick operator has been applied, but are kicked again with their
new time step size. This way, they can again be drifted any number of times until the simulation
reaches the step where the particle completes its own time integration step.
 }
 \label{fig:individual-time-steps-drifts}
\end{figure}

The next required modification is necessary to facilitate the individual time steps for particles
(see Section~\ref{chap:individual-timesteps}). Consider a case where some
particle $i$ has a time step of size $\Delta t$ with a neighboring particle $j$ which has a time
step of size $2 \Delta t$. Allowing the particles to have individual time steps would mean that
while particle $i$ does two time step integrations following its own time step size $\Delta t$,
particle $j$ does only one. During the second update of particle $i$, while particle $j$ is
skipped, particle $i$ needs to go through its entire required order of operations, which includes
the neighbor search to determine the smoothing length $h$. However $h$ depends on the neighboring
particle positions, so for $h$ to be accurate, all neighboring particle positions must be correct,
i.e. drifted to the correct time. This means that particle $j$ would need to be drifted to a time at
 which it is not being integrated itself because particle $i$ requires it (see
Figure~\ref{fig:individual-time-steps-drifts}). Luckily drifting the particle positions is a linear
operation, which can be exploited. Let $K(\Delta t)$ denote the kick operation for a particle, and
$D(\Delta t)$ denote the drift operation. For particle $j$, which has a time step size of $2 \Delta
t$, a kick-drift-kick time integration can then be written as
\begin{align}
    K(\Delta t) \circ D(2 \Delta t) \circ K(\Delta t)
\end{align}

However, since the drift operation is linear, we can just as well do

\begin{align}
    K(\Delta t) \circ D( \Delta t ) \circ D(\Delta t) \circ K(\Delta t) \\
    = K(\Delta t) \circ D( \Delta t / n ) \circ \hdots \circ D(\Delta t / n) \circ K(\Delta t)
\end{align}

\begin{figure}
 \centering
 \includegraphics[width=\textwidth]{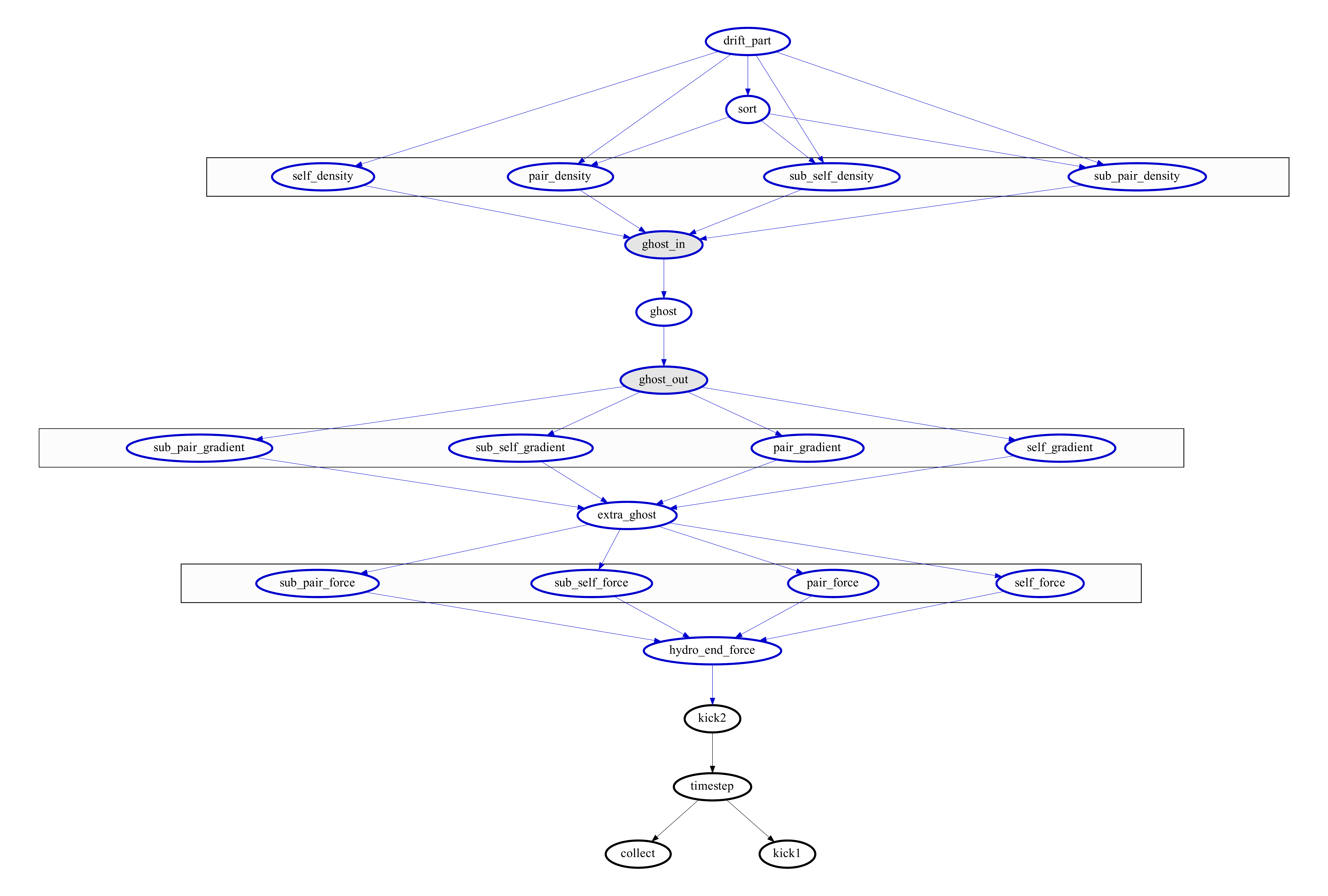}%
 \caption{
The dependency graph for tasks required for the hydrodynamics with \swift for shared memory
architectures. The \lingo{kick1} task has been moved to the bottom of the dependency graph, and the
\lingo{collect} task has been added.
}
 \label{fig:dependency-graph-nompi}
\end{figure}

So we can concatenate any number $n$ of drift operations on a particle between the two kick
operators. This means that as long as the first kick operation on particle $j$ has been performed
already, we can keep drifting the particle $j$ even in time steps where it is skipped so the
neighboring active particle $i$ has access to the correct positions of its neighbor. To this end,
we perform the following modifications: When a simulation starts, the first kick operation is
performed on all particles. This way, we can begin each subsequent simulation time step with the
drift operation. If a particle is being skipped in the current time step due to its larger
individual time step size, then it will be in the correct state to be drifted to the current time.
These drifts can be repeated an arbitrary number of times until the time is reached where the
particle undergoes a full hydrodynamics step itself. The first kick operation is then moved in the
task dependency graph from being at the root of the dependency graph (like in
Figure~\ref{fig:dependency-graph-second-order}) to the end (see
Figure~\ref{fig:dependency-graph-nompi}), after the new time step size for a
particle has been computed in the timestep tasks. Once the particle has done its first kick
operation, it is once again in the correct state to be drifted any required number of times until
its own next time step begins. Figure~\ref{fig:individual-time-steps-drifts} shows how the kick and
drift operations are performed with three interacting particles with different time step sizes.

\subsection{Facilitating Task Activation}

There is one more additional task required before we can turn our attention towards how to deal
with distributed memory architectures. This task is related to the \emph{task activation}. The
tasks themselves as structures that contain information what work is to be performed on a cell are
created when the cells themselves are created. Before a simulation step is executed, tasks are
activated depending on whether the work they represent needs to be performed in the following
simulation step. This is not always the case: For example, while one cell may contain particles
that finish their time step in this simulation step, there may be cells somewhere else in the
simulation which contain no particles that require an update in this simulation step. A cell which
contains particles that need to be updated in the current simulation step is called
``\lingo{active}'', otherwise ``\lingo{inactive}''. Only tasks which perform work on an
\lingo{active} cell are also activated, i.e. passed to the scheduler for subsequent execution.

In order to facilitate this task activation, cells need to store the information whether they
are \lingo{active} at the current simulation time. For split cells, this information needs to be
available recursively for their child cells as well. This information is gathered and stored for
each cell during the \lingo{timestep} tasks. \lingo{Timestep} tasks are the ones where the new time
steps for particles that reside in that cell are computed after the integration step of the
particles is complete. Similarly, the information also needs to be available to the parent cells of
the cells to which tasks are attached to. Recall that tasks are attached to the \lingo{super level}
in the tree, which may or may not be the \lingo{top level}. This work is performed by the
``\lingo{collect}'' task, whose only purpose is to recursively pass on the activity information
from the \lingo{super level} to the \lingo{top level}. The \lingo{top level} cells require the
activity information because the task activation is performed recursively starting at the
\lingo{top level}. When other physics (or more precisely other particle types than particles
representing the fluid) like gravity are included in the simulation, the corresponding tasks can
have a \lingo{super level} which is different from the \lingo{super level} of the hydrodynamics
tasks. To ensure that the task activation process is performed correctly, it needs to begin at the
root of the cell tree, which is the \lingo{top level}.

\subsection{Distributed Memory Parallelism: Domain Decomposition and MPI}

This concludes the construction of the dependency graph for finite volume particle hydrodynamics
with \swift on shared memory architectures, where all threads have access to all data at all times.
Modern high performance computing software however needs to go beyond shared memory architectures to
be able to solve sizable problems: There are limits to how many CPUs can have access to the same
blocks of memory before the architecture becomes too expensive, too inefficient, or even impossible
to create. In order to be able to increase the computing power beyond the capabilities of one node,
or shared memory region, these nodes can instead be connected in a network and form a distributed
memory architecture. However, the CPUs between different nodes (individual regions of shared
memory) have no access to each others respective memory, and necessary data needs to be communicated
between the nodes explicitly. The industry standard for such communications is the ``Message
Passing Interface'' MPI, which defines a standard for the communications and has several available
implementations. In MPI terminology, each separated memory region with associated CPUs is called a
``\lingo{rank}''. While the main benefit of MPI is the possibility to pass messages between
different nodes, MPI is perfectly capable of running several \lingo{ranks} on a single node as
well. However, each \lingo{rank} still only has access to its own data, even when being executed on
the same node, where in principle the memory can be shared between CPUs.

\subsubsection{Domain Decomposition}

\begin{figure}
 \centering
 \includegraphics{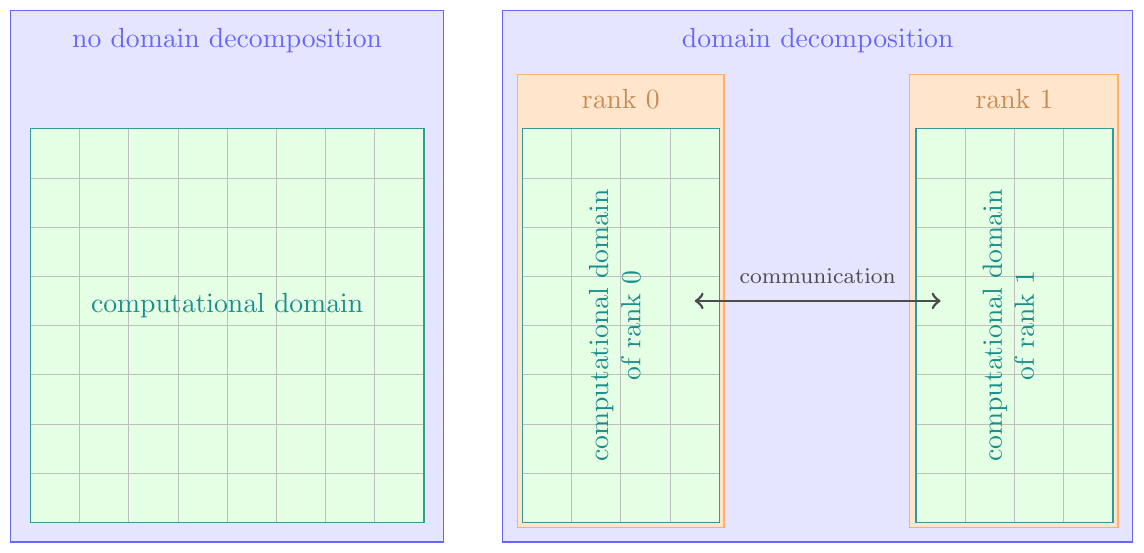}%
 \caption{
Schematic illustration of the domain decomposition strategy for distributed memory architectures:
The initial total volume is split between the two \lingo{ranks} in this example. Splitting the
domain requires communications of necessary data between the \lingo{ranks}. However, thanks to the
possibility to solve the problem on separate \lingo{ranks}, a much bigger problem can be solved. In
this example we can use the total memory of two computing nodes instead of only one.
 }
 \label{fig:domain-decomposition}
\end{figure}

At its core, the strategy to run simulations on distributed memory architectures is to split the
entire problem into separate regions and assign a region to each MPI rank. This process is called
``domain decomposition''. To illustrate this process, suppose that the simulation you'd like to run
consists of a square. A very simple domain decomposition would be to split the square in half and
give each half of the total square to an individual MPI \lingo{rank} to work with, as is shown in
Figure~\ref{fig:domain-decomposition}. Communications become necessary when a \lingo{rank} requires
data which is stored on a different \lingo{rank}: For example, consider a particle-particle
interaction loop along the edge of a the halves of the square. Particle data along the edges of each
half needs to be sent to the other \lingo{rank}, and vice versa. While the communications are extra
work that needs to be performed, and hence constitute an overhead, a significant advantage is that
through MPI the total amount of memory available for a problem increases.\footnote{
Additionally, since MPI is capable of running multiple \lingo{ranks} on shared memory architectures
just as if it were a distributed memory architecture, it can be used as a method to parallelize
serial software on shared memory architectures as well. An MPI only parallelization for shared
memory architectures will almost certainly not be as efficient as a dedicated shared memory
parallelization method due to the overheads it involves. However, it would still constitute an
improvement over strictly serial code.}
For example if we use two MPI \lingo{ranks}, like in Figure~\ref{fig:domain-decomposition}, and
assign one \lingo{rank} per computing node, then we can use the entire memory available on two
nodes, rather then one, and fill the memory with twice\footnote{Not accounting for additional
overheads that arise due to the addition
of MPI and domain decomposition.}  as many particles than before.

The separation of the work into individual tasks required for the task-based parallelism in
\swift allows us to make more sophisticated choices for the domain decomposition than a simple
geometrical partition of the volume. The computational load for each MPI \lingo{rank} can be
modeled much more accurately using tasks rather than basing the decomposition on equal volumes or
particle counts. This allows for better load balancing between each MPI \lingo{rank}, and hence
reduces idle times of entire \lingo{ranks} which are waiting for other \lingo{ranks} to finish
their respective work. Furthermore, the number of communications can be drastically reduced by
splitting the domain along regions with large time step sizes of particles. For example, it is much
more preferable to split the domain in regions largely void of particles rather than splitting the
center of a galaxy, where particles tend to have very short time steps. Large time step sizes means
that fewer time integrations for these particles are necessary, and hence fewer total communications
and associated overheads are required. So using the data available through the existence of tasks
can both optimize the load balancing as well as minimize the number of necessary communications.
This is achieved by modeling the domain, the work associated with it, and the communications as a
graph, and making use of the graph partition library \codename{Metis}
\citep{karypisFastHighQuality1998} to optimize the domain decomposition.

\subsubsection{MPI Communications}

\begin{figure}
 \centering
 \includegraphics[height=.8\textheight]{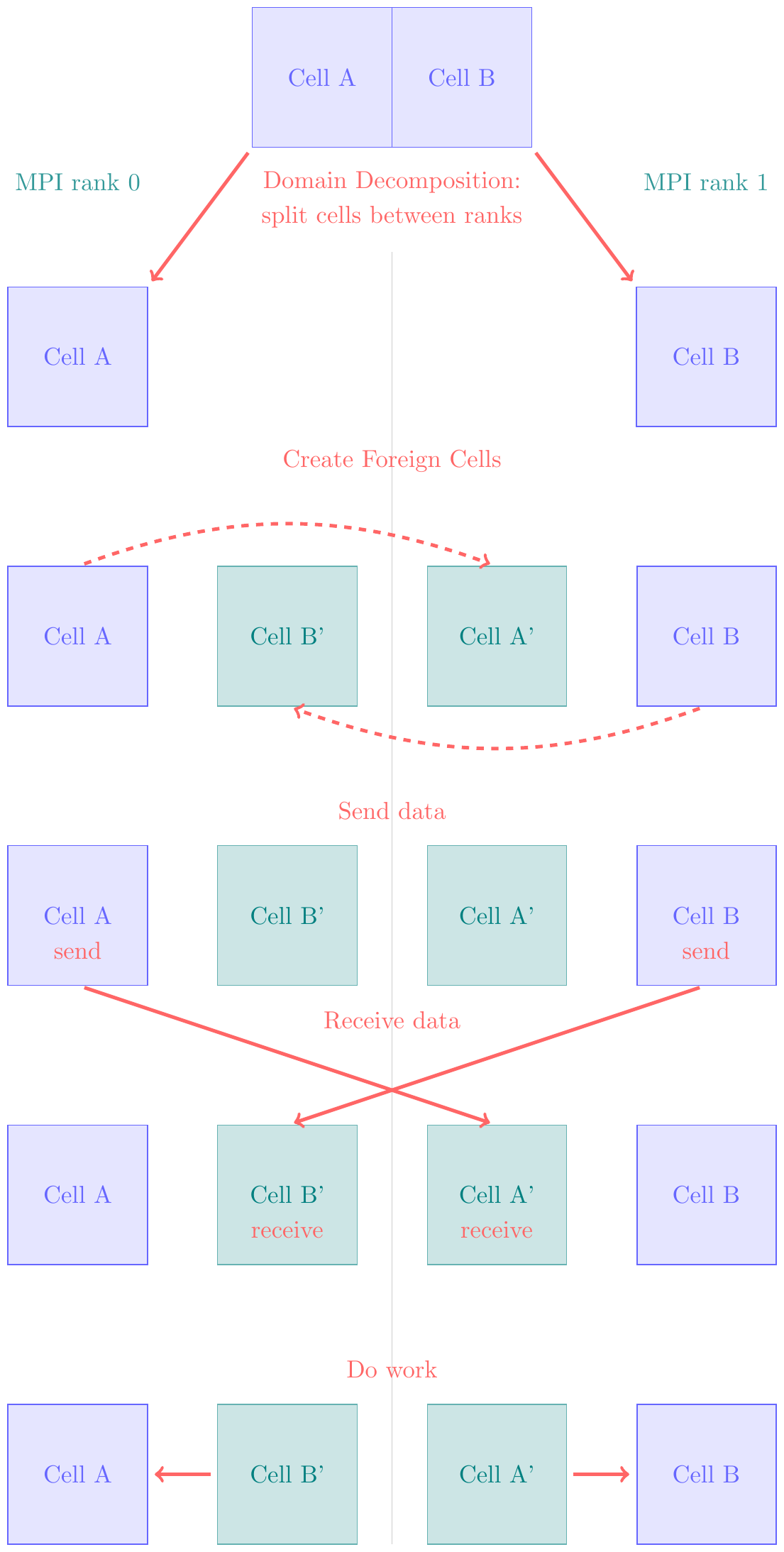}
 \caption{
Sketch on how the MPI communications with a decomposed domain work. Two cells, $A$ and $B$, first
get assigned to two different MPI \lingo{ranks} during the domain decomposition, and a
\lingo{foreign} counterpart cell $A'$ and $B'$ is created on the other \lingo{rank}, respectively.
A
communication then consists of the following order of operations: First the \lingo{real} cells $A$
and $B$ send their data to the foreign counterparts $A'$ and $B'$, respectively. The
\lingo{foreign}
cells $A'$ and $B'$ receive data when the corresponding \lingo{receive} tasks are executed. Finally
the work can proceed, and the \lingo{real} cells $A$ and $B$ can updated correctly. Note that
\lingo{foreign} cells are never updated on foreign \lingo{ranks}, only \lingo{real} cells are
updated. \lingo{Foreign} cells' only purpose is to make the necessary data available for other
\lingo{real} cells on their respective \lingo{ranks}. Also note that despite what this sketch might
suggest, the send and receive operations (and tasks) of cells do not need to occur simultaneously,
and in general will not be executed concurrently.
 }
 \label{fig:mpi-comm}
\end{figure}

\begin{figure}
 \centering
 \includegraphics[height=.75\textheight]{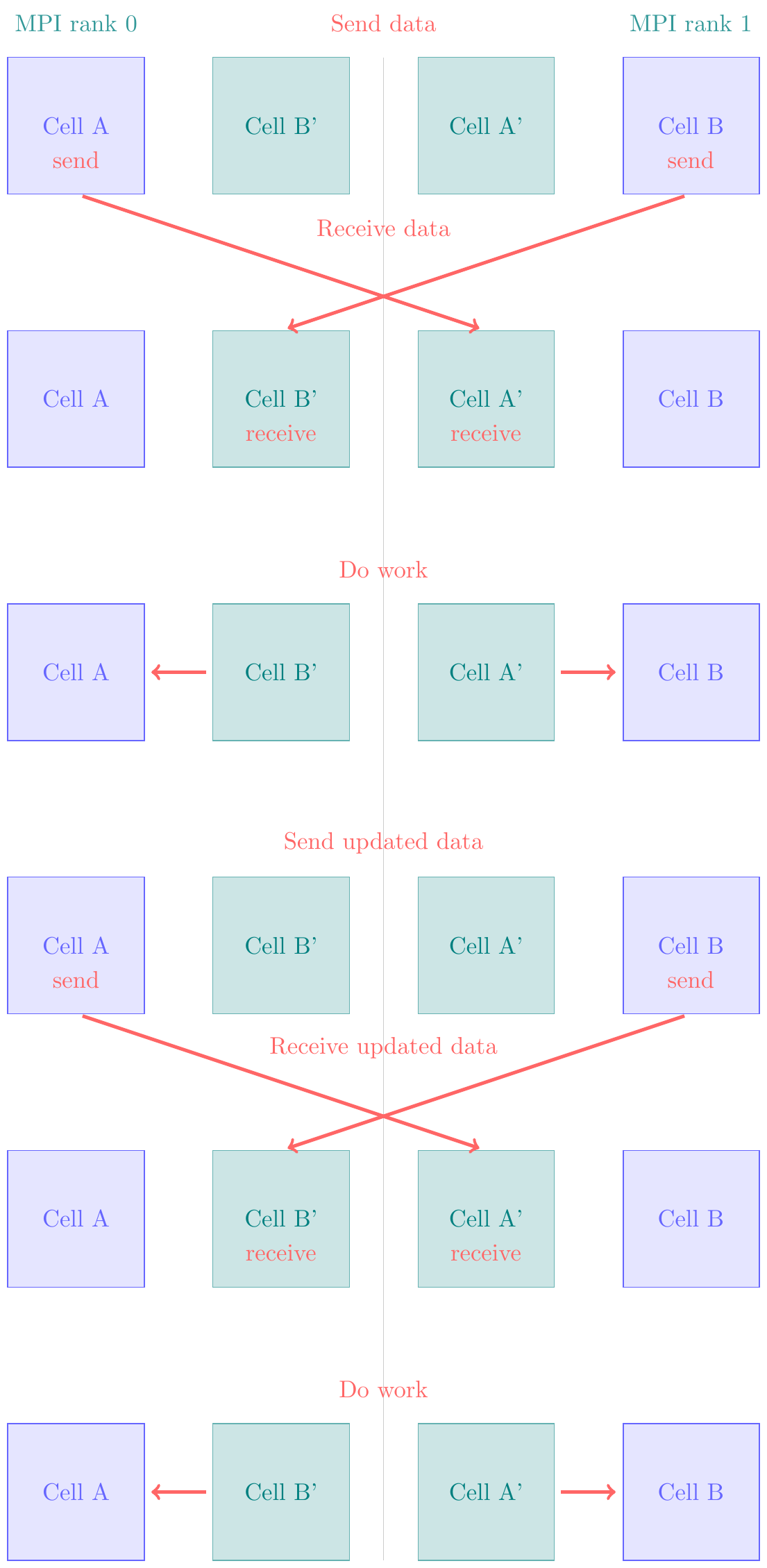}
 \caption{
Illustration of how the communication logic is set up in \swift for situations where updated data
of
\lingo{foreign} cells are necessary. It starts with \lingo{real} cells $A$ and $B$ sending their
current state of data to their \lingo{foreign} counterparts, $A'$ and $B'$, respectively. Once $A'$
and $B'$ have received the data, the individual ranks can proceed with the work necessary for their
\lingo{real} cells $A$ and $B$. In the depicted situation, cells $A$ and $B$ then undergo a
second interaction with each other, and hence the \lingo{foreign} cells $A'$ and $B'$ need to be
updated to the current state of $A$ and $B$. To this end, the cells $A$ and $B$ once again send
their data, and the second interaction can take place once $A'$ and $B'$ have received the data.
Note that \lingo{foreign} cells are never updated on foreign ranks, only \lingo{real} cells are
updated. \lingo{Foreign} cells' only purpose is to make the necessary data available for other
\lingo{real} cells on their respective \lingo{ranks}. This restriction allows to never need to send
data back from foreign cells $A'$ and $B'$ to their \lingo{real} counterparts $A$ and $B$. Also
note that despite what this sketch might suggest, the send and receive operations (and tasks) of
cells do not need to occur simultaneously, and in general will not be executed concurrently.
}
 \label{fig:mpi-comm-repeated}
\end{figure}

MPI provides developers with a programming interface to transfer data between individual
\lingo{ranks}, or collectively among all \lingo{ranks} involved in a process. The communications
between individual \lingo{ranks} consist of explicit calls to send and to receive certain data. If
for example the MPI \lingo{rank} 0 needed to send data to MPI \lingo{rank} 1, \lingo{rank} 0 would
send a message addressed to \lingo{rank} 1, while \lingo{rank} 1 would need to expect and receive
that message. Traditionally, this data exchange would occur with so-called ``\lingo{blocking}'' or
``\lingo{synchronous}'' calls to the send and receive functions. The message is sent and received
only when both \lingo{ranks} arrive at the point where a message needs to be exchanged, and then
both \lingo{ranks} would proceed with their respective work. Until each \lingo{rank} involved in
the communication reaches the synchronization point at which the message is exchanged, the others
are blocked from proceeding any further, and wait for all \lingo{ranks} involved in this
communication to reach the synchronization point. However, the task-based parallelism of \swift
allows for the communications to be completely \lingo{asynchronous}: Any \lingo{rank} can send the
data it needs to send whenever it is ready to send it, and then proceed to work on other tasks
without waiting at a synchronization point until the message is received. Conversely, the receiving
\lingo{ranks} can keep checking whether the necessary data has arrived. While the data hasn't been
received yet, the receiving \lingo{rank} can also simply proceed to work on other tasks without
wasting time waiting at synchronization points, and check again whether data has arrived after it
completed a different task.

More concretely, the data that needs to be communicated consists of data contained within cells.
This includes all particles within a cell, as well as other cell data such as the minimal time step
size within the cell. A local copy of cells assigned to some given MPI \lingo{rank}'s domain is
created on the \lingo{ranks} which require that particular cell's data. These cells are called
``\lingo{foreign}'' cells. The required communications then consist of transmitting the up-to-date
data from the ``\lingo{real}'' cell into the corresponding \lingo{foreign} cell. This way, the
other \lingo{ranks} can do the necessary work as if the \lingo{foreign} cell is a normal one and
part of their domain, with the exception that they don't need to do work on the particles in the
\lingo{foreign} cell itself. The actual work on the \lingo{foreign} cells is performed by the
\lingo{rank} where their corresponding \lingo{real} counterpart situated. Figure~\ref{fig:mpi-comm}
shows the working principle how two cells $A$ and $B$ first get assigned to two different MPI
\lingo{ranks} during the domain decomposition, and a \lingo{foreign} counterpart cell $A'$ and $B'$
is created on the other \lingo{rank}, respectively. It then goes on to show how a first
communication takes place: \lingo{Real} cells $A$ and $B$ send their data to the foreign
counterparts $A'$ and $B'$, respectively. The \lingo{foreign} cells $A'$ and $B'$ receive data when
the corresponding receive tasks are executed. Finally the work can proceed, and the ``real'' cells
$A$ and $B$ can updated correctly.

In cases where several updates of \lingo{foreign} cells are required during a single simulation
step, the updated data of the \lingo{real} cells is being sent and received again. Let us stress
once again that data of \lingo{foreign} cells is never updated by the ranks where the cells are
\lingo{foreign}, they are merely used to make necessary data for the \lingo{real} cells of the
\lingo{rank} available. \lingo{Foreign} cells are only updated by receiving the updated data from
the \lingo{real} cells. While this may seem like a harsh restriction, it allows to make the
algorithm work without ever needing to send data back from \lingo{foreign} cells to their
\lingo{real} counterparts. So in total, it reduces the required number of communications.
Figure~\ref{fig:mpi-comm-repeated} shows how the algorithm works when several updates and hence MPI
communications are required.

\subsubsection{MPI Tasks}

\begin{figure}
 \centering
 \includegraphics[width=\textwidth]{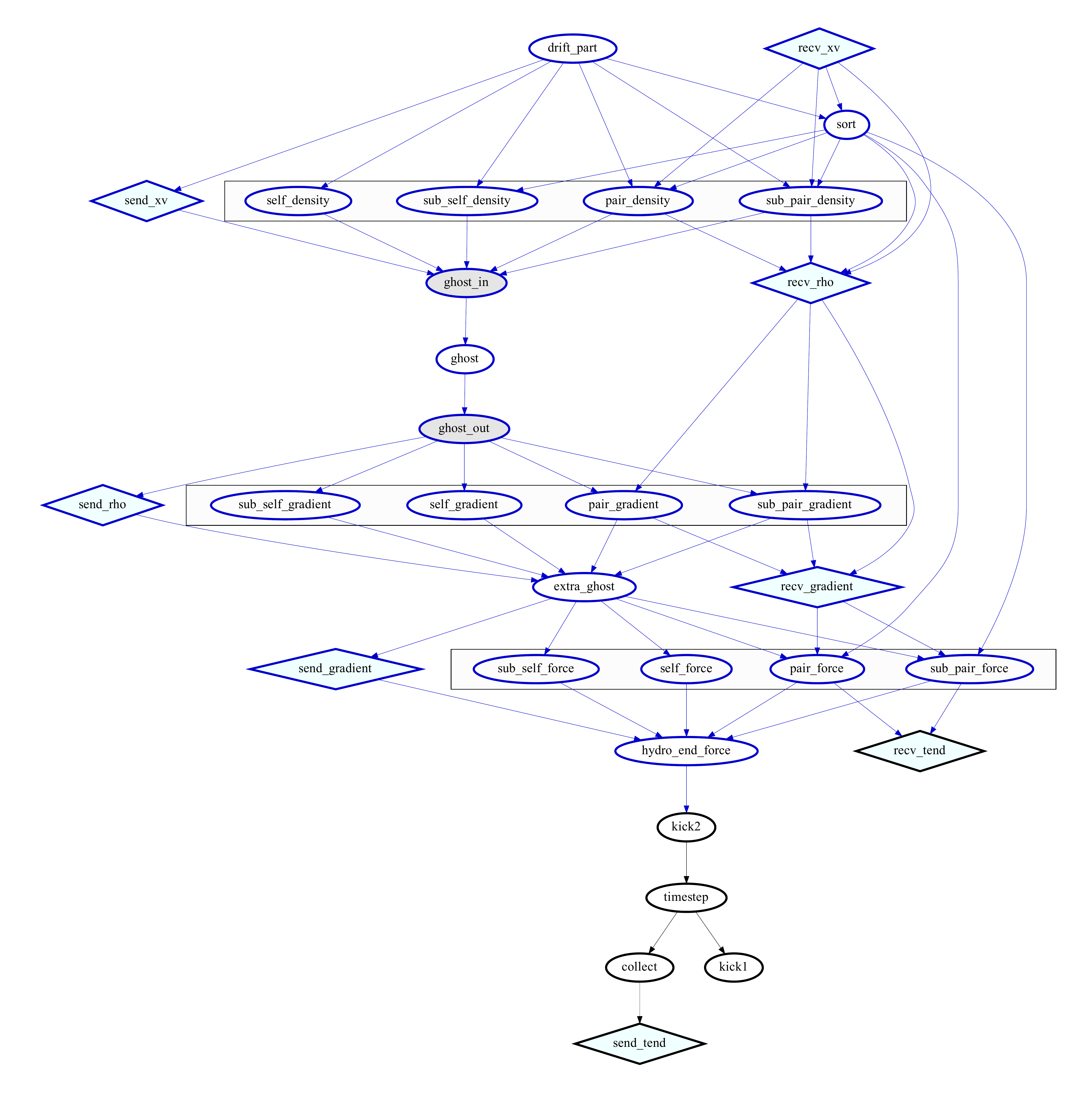}
 \caption{
The full dependency graph required for finite volume particle hydrodynamics with \swift and MPI.
Before each
interaction loop (density, gradient, and force) an updated state of foreign cells is necessary.
After the timestep tasks have finished, an additional update with the current minimal time step
sizes inside the foreign cells is necessary to facilitate the correct task activation at the
beginning of the subsequent simulation step.
}
 \label{fig:dependency-graph-hydro}
\end{figure}

With the basic working principle of asynchronous MPI communications having been discussed, we can
now proceed to add the corresponding required tasks and dependencies to the dependency graph for
finite volume particle hydrodynamics in \swift, whose final form is shown in
Figure~\ref{fig:dependency-graph-hydro}. The \lingo{foreign} cells need to be up-to-date before each
particle-particle interaction, i.e. they need to receive data before the \lingo{density} tasks are
run, then again before the \lingo{gradient} tasks are run, and finally again before the fluxes are
exchanged in \lingo{force} tasks. The first receiving task on the graph, called
``\lingo{recv\_xv}'', is the task that is used to receive the data from the corresponding
\lingo{real} cell before the neighbor search (\lingo{density}) interaction can take place. It is
named with the suffix ``\lingo{xv}'' as its purpose is primarily to receive the current positions
and velocities of particles. The particles are first drifted in the corresponding \lingo{real}
cell before the data is sent, hence there will be a dependency from the \lingo{drift\_part} to the
\lingo{send\_xv} tasks, and the data \lingo{recv\_xv} tasks receive will contain particles drifted
to the correct time.  Once the data is received, the particles are sorted along each neighboring
cells' axes for an optimized access to particles during interactions, as was previously mentioned.
(Note that this doesn't modify any particle values, but only creates lists of sorted particles for
the cells). The \lingo{send\_xv} task has an outgoing dependency to the first task after the
interaction loop, which in this case is the \lingo{ghost\_in} task. The reason this dependency is
needed is because we can't allow the cell to be updated any further before the data has been sent,
which is what this dependency ensures. Explicitly, this dependency ensures that the
\lingo{ghost\_in} tasks can't be executed before the \lingo{send\_xv} task is finished.

On the receiving side, dependencies between \lingo{recv\_xv} and \lingo{pair}-type tasks of the
\lingo{density} loop are necessary: The interactions must not take place before the data has
arrived. Recall that actual work is never done on \lingo{foreign} cells, which are the only cell
types that have an associated \lingo{receive}-type task. So cells which have \lingo{receive}-type
tasks will never have \lingo{self}-type tasks, which are the ones that interact particles within a
cell with each other. Therefore no dependencies between \lingo{receive}-type tasks and
\lingo{self}-type tasks are necessary. For the same reason, only dependencies from \lingo{pair}-type
(and \lingo{sub-pair}) type tasks are required for the subsequent \lingo{receive}-type task, which
is the \lingo{recv\_density} task. Additionally, it is necessary to add a dependency between two
\lingo{receive}-type tasks: MPI doesn't guarantee in which order messages will arrive. Allowing two
\lingo{receive}-type tasks of a single \lingo{foreign} cell to be ready for execution concurrently
permits the message that should be arriving later to be received first. The remaining message, which
should have arrived first, will then overwrite the already received message's data with data which
is incorrect at that point. Therefore it is necessary to forbid \lingo{receive}-type tasks to run
concurrently, which is achieved by adding a dependency between them. The same logic applies to the
\lingo{send}-type and \lingo{receive}-type tasks in the subsequent particle-particle interaction
loops.

To summarize, the required sending and receiving types of tasks for a hydrodynamics step are as
follows:

On the sending side:

\begin{itemize}
 \item After drifting particles, send over the updated particle positions using \lingo{send\_xv}.
 \item After doing the neighbor search loop with the \lingo{density} tasks, send the updated
particle data over for the computation of gradients with the \lingo{send\_rho} tasks before you
continue modifying the particle data.
 \item After finishing the gradient computations in the \lingo{gradient} tasks, send the updated
particle data over with the \lingo{send\_gradient} tasks for the computation of fluxes before you
continue modifying the particle data.
\end{itemize}

On the receiving side:

\begin{itemize}
 \item Before doing a neighbor search loop, receive the updated particle positions using
\lingo{recv\_xv} and sort the particles.
 \item Before doing the gradient loop, receive the updated particle ``densities'' with
\lingo{recv\_rho}.
 \item Before doing the flux exchange (\lingo{force}) loop, receive the updated particle
gradients.
\end{itemize}

Finally, once the new time step sizes for particles have been computed with the \lingo{timestep}
tasks once the time integration in this simulation step has been completed, a final update needs to
take place. The \lingo{foreign} cells need to have the correct cell information available which is
required for the task activation process in the subsequent simulation step. Concretely, this is the
smallest time step size contained within the cell, and the last time the cell was updated. Rather
than sending all particle data again, it suffices to only send this cell metadata. The full updated
particle data will be sent over with the \lingo{send\_xv} task at the beginning of the next
simulation step anyway. The cell metadata is being sent using the \lingo{send\_tend} tasks, and
correspondingly received by the \lingo{recv\_tend} tasks.

%% file: main/RHD/RHD-1-introduction.tex
\chapter{Introduction}

Having discussed solution strategies for hyperbolic conservation laws in
Part~\ref{part:finite-volume} (in particular the Euler equations, which describe ideal gases),
Finite Volume Particle Methods (which are used to solve hyperbolic conservation laws) and their
implementation in \swift in Part~\ref{part:meshless}, we can now turn our attention to the
\emph{K\"onigsdisziplin} in this thesis - radiation hydrodynamics.

Radiation hydrodynamics is tricky on many, if not all, fronts. Part of the complexity is the sheer
dimensionality of the problem: The radiation fields in principle need to keep track of both the
direction and the intensity of the radiation for each photon frequency individually and for each
point in space. The equation of radiative transfer, discussed in more detail later, also needs in
principle to be solved for each photon frequency individually, for each direction, and for each
point in space. The photon frequency dependency is due to the frequency dependency of the
interaction rates between particles and photons. A different part of the intricacy is the fact the
interactions between radiation and particles come with their own set of complexities: Each chemical
species in the gas has its own set of reactions it undergoes, which are temperature dependent, but
also depend on the abundances of other elements present. For example, ionized atoms and free
electrons can recombine, or neutral atoms can excite and ionize each other through collisions, which naturally depend on the number of other species currently present. Hence
the chemistry of the gas is described through a network of interconnected differential equations
between each present chemical species, and to make matters worse, they are typically stiff
equations. The interactions with radiation constitute additional terms in the chemistry equations
for the creation and destruction of some chemical species, as for example a photo-ionization event of an atom is equivalent to the destruction of the atom, and the creation of an ion and a free
electron.

To begin with, let's focus on radiative transfer (RT) and leave the thermochemistry for later. Over the years, many different approaches have been developed in order to solve radiative transfer. They can broadly be separated into two classes of methods: (i) ray-tracing, or ray-casting methods, and (ii) moment based methods.

Ray-tracing methods cast photon beams from radiative sources over the volume, and solve the equation
of radiative transfer along the ray as a one dimensional problem and as a function of the optical
depth $\tau$, which encodes the probability of the photons to be absorbed or scattered. The optical depth needs to be accumulated along the ray, traversing a multitude of cells. Methods which cast such long rays
are called ``Long Characteristics'' methods
\citep[e.g.][]{baczynskiFerventChemistrycoupledIonizing2015, grondTREVRGeneralLog2019,
susaSmoothedParticleHydrodynamics2006}.
While generally being very accurate, in parts due to explicitly handling radiation from far-away
sources, Long Characteristics ray-casting methods' main caveat is arguably the associated
computational expense. Firstly, the expense scales linearly with the number of radiating sources in
the simulation volume. Secondly, the non-locality of the method makes parallelization over
distributed memory architectures tricky and inefficient. Finally, they typically need to be
performed iteratively. As the radiation photo-ionizes the gas and changes its chemical composition, its optical depth changes as well, whereas the optical depth is required to be accumulated along the
ray to infer how much radiation is placed at the destination. This process needs to be repeated
until the optical depths converge.

An approach to reduce the computational cost of Long Characteristics methods is used by Monte
Carlo schemes \citep[e.g.][]{vandenbrouckeMonteCarloPhotoionization2018,
smithAREPOMCRTMonteCarlo2020, michel-dansacRASCASRAdiationSCattering2020, baekSimulated21Cm2009,
molaroARTISTFastRadiative2019, campsSKIRTAdvancedDust2015}, which randomly sample the radiation
field from radiating sources by emitting a number of photon packets. The sampling is typically
performed in both frequency and angular direction. The tracing of photon packets allows to also
trace scatterings of photons, making Monte Carlo methods ideal for line radiation transfer.
However, due to the sampling nature of the method, statistical noise is introduced, which only
decreases proportional to the square root of the number of emitted photon packets per source, and
the expense remains proportional to the number of sources in the simulated volume.

A different approach to reduce the expense of Long Characteristics methods is to try to avoid
performing the same accumulation of the optical depth of Long Characteristics methods over and over
again. This is for example the case for cells (or volume elements) close to a radiation source
which are repeatedly transversed by many rays. To this end, ``Short Characteristics'' methods were
developed \citep[e.g.][]{mellemaPhotoevaporationClumpsPlanetary1998,
shapiroPhotoevaporationCosmologicalMinihaloes2004,mellema2RayNewMethod2006,
sarkarNewIonizationNetwork2021,jauraSPRAICouplingRadiative2018,
jauraSPRAIIIMultifrequencyRadiative2020,peterSweepMethodRadiative2023}. Rather than casting the
rays from a source to a destination cell, Short Characteristics methods propagate the radiation and
accumulate the optical depths on a cell-by-cell basis, and in an ordered fashion. This removes the
redundancy of the repeated summation over the optical depths, and in some cases removes the linear
dependency of the number of sources in the simulation volume, but makes the method inherently
serial, as the cell sweep needs to be performed in a specific order. The casting of Long
Characteristics rays on the other hand is an ``embarrassingly parallel'' problem, as all castings
can be performed independently from each other. Hybrid Characteristics methods
\citep[e.g.][]{rijkhorstHybridCharacteristics3D2006}, which combine elements of Long and Short
Characteristics to make efficient use of adaptive meshes, have also been proposed.

Other approximate ray-based methods include Adaptive Ray Tracing
\citep[e.g.][]{abelAdaptiveRayTracing2002, kimModelingUVRadiation2017}, where rays are created at
point sources and successively split as they are traced outward, adapting to the angular resolution
depending on the distance from the source. In a similar spirit, some methods
\citep[e.g.][]{petkovaNovelApproachAccurate2011, pawlikTRAPHICRadiativeTransfer2008,
pawlikMultifrequencyThermallyCoupled2011} follow the propagation of radiation through cones from
radiating sources.

The second class of RT methods, the moment-based methods, take a fundamentally different approach
to solving radiative transfer. To reduce the dimensionality of the problem, rather than solving the equation of radiative transfer in all directions, angular moments of the equation of RT are taken, which removes the angular directionality component from the equations. In addition, frequencies are typically discretized into discrete intervals, or groups, and frequency dependent quantities are integral-averaged over the interval.

The removal of directionality through moments is at the same time a great advantage and a great
disadvantage. The equations to be solved take the shape of a purely local hyperbolic conservation
law, which, as discussed before, is a well-known and well-studied problem in physics, and many known methods to solve them exist. Furthermore, a local problem can be parallelized more efficiently, even across shared memory domains. It also removes the dependency of the number of radiating sources in the simulation volume.

At the same time, the loss of directionality leads moment-based methods to predict unphysical
solutions for radiation. The radiation is behaving more akin to a fluid, rather than radiation. For
example, two colliding photon beams should in reality just pass through each other, whereas
moment-based methods predict a collision akin to the impact of two fluid waves. Furthermore, the
formation of sharp shadows is severely limited, as the fluid-like representation of radiation will
diffuse around corners (see e.g. \citet{ramses-rt13}).

Several versions of moment-based methods have been described to date in literature. The ``Flux Limited Diffusion'' \citep[e.g.][]{commerconRadiationHydrodynamicsAdaptive2011,
normanSimulatingCosmologicalEvolution2007a}, perhaps the simplest form of a moment-based method, only uses the zeroth moment, and provides a closure in the form of a local diffusion relation that
diffuses radiation along the local energy gradient. This is a reasonable approximation in optically thick regimes, but not for optically thin ones.

Other approaches use both the zeroth and the first moment w.r.t. angular direction. This results in
two equations (in 1D), one for the energy density and one for the photon flux. However, these
equations are not closed, and an approximate closure relation is required. In particular, an
expression for the pressure tensor, which serves as the hyperbolic flux $\fc$ for the photon flux
$\mathbf{F}$, is missing. The ``OTVET'' \citep[``Optically Thin Variable Eddington Tensor'',
e.g.][]{gnedinMultidimensionalCosmologicalRadiative2001, petkovaImplementationRadiativeTransfer2009} closure
approximates the pressure tensor by gathering the directional components from sources of radiation
under the assumption of the optically thin limit. This re-introduces some non-locality in the
radiation fields again, but also re-introduces the dependency on the number of radiating sources in the simulation volume. Other variable Eddington tensor methods have also been proposed
\citep[e.g.][]{finlatorNewMomentMethod2009,menonVETTAMSchemeRadiation2022}, but usually come at an
even higher computational expense.

A different approach, the so-called ``M1 Closure''
\citep[e.g.][]{gonzalezHERACLESThreedimensionalRadiation2007, ramses-rt13,
kannanAREPORTRadiationHydrodynamics2019, fuksmanRadiativeTransferModule2019,
chanSmoothedParticleRadiation2021} approximates the pressure tensor based on local quantities only, and keeps the locality, and hence the gains in computational expense. The pressure tensor is then estimated as an interpolation between the optically thin and optically thick limits based on the local photon energy density and flux.

The M1 Closure method has already been used in a variety of simulations of the Epoch of Reionization
\citep[e.g.][]{rosdahlSPHINXCosmologicalSimulations2018,trebitschObeliskSimulationGalaxies2021,
xuTHESANProjectLymanalpha2022,borrowTHESANHRHowDoes2022, katzInterpretingALMAObservations2017} and has been demonstrated to be an adequate method for simulations of Cosmic Reionization, albeit with caveats \citep[see][]{wuAccuracyCommonMomentbased2021,ocvirkImpactReducedSpeed2019}. Motivated by the superior computational efficiency of the method, \GEARRT, the novel radiative transfer solver in \swift which will be described in this Part of my thesis, also employs a moment-based method with the M1 Closure. The core idea of the numerical solution for the radiative transfer equations follows the strategy of \cite{ramses-rt13} closely. In particular, I adapt their technique of discretizing frequencies and the operator splitting approach used to evolve moments of the equations of radiative transfer in time. In Section~\ref{chap:rt-equations}, the equations of radiative transfer and the M1 Closure are described. Section~\ref{chap:rt-numerical-strategy} discusses the numerical solution strategies to solve the equations, while Section~\ref{chap:rt-implementation} describes the implementation of \GEARRT in \swift. A series of tests and validations is presented in Section~\ref{chap:rt-validation}.

\GEARRT is open source software and available under \url{https://github.com/swiftsim/swiftsim}. It
is extensively documented and comes along with several prepared example problems which are ready to be run. Additional test examples and many peripheral RT related tools are available under URL
\url{https://github.com/swiftsim/swiftsim-rt-tools}, including the validation tests used in
Section~\ref{chap:rt-validation}.

%% file: main/RHD/RHD-2-rt-equations.tex
\chapter{The Equations of Moment-Based Radiative Transfer}\label{chap:rt-equations}

\section{The Equations of Radiative Transfer and the M1 Closure}

Before we begin with the introduction of the equation of radiative transfer and its associated
equations, let's take a quick aside and note that they contain a plethora of variables and
coefficients. For clarity, an overview of the relevant quantities and coefficients is given in
Table~\ref{tab:rt-variables} along with their respective units.

Returning to the topic at hand, radiation is described by a quantity called the specific intensity
$I_\nu$ \citep[e.g.][]{mihalasFoundationsRadiationHydrodynamics1984,
teyssierTheoreticalAstrophysics2021}, which depends on the frequency $\nu$ of the radiation and is
defined as:

\begin{equation}
	\de E = I_\nu(\x, \mathbf{n}, t) \de A \de \Omega \de \nu \de t \  \label{eq:specific-intensity}
\end{equation}

where $\de E$ is the energy absorbed by the surface element $\de A$ of a detector per unit time
$\de t$ in the frequency range $\de \nu$ coming from a beam projected along the normal of
the surface element $\de A$,\footnote{
The projection along the normal unit vector $\mathbf{n}_A$ of the surface element $\de A$ happens
as a scalar product of the normal unit vector and the direction of propagation of the radiation,
i.e. $\mathbf{n}_A \cdot \mathbf{n}$, which can also be written as an additional factor
$\cos(\theta_A)$, where $\theta_A$ is the angle between $\mathbf{n}$ and $\mathbf{n}_A$. So
eq.~\ref{eq:specific-intensity} may as well be written as
\begin{equation*}
	\de E = I_\nu(\x, \mathbf{n}, t) \cos(\theta_A) \de A \de \Omega \de \nu \de t
\end{equation*} to include the projection explicitly, rather than assume $I_\nu$ is already
projected along the normal vector $\mathbf{n}_A$. This additional factor $\cos(\theta_A)$ is
however the reason why the units of the specific intensity are per radians instead of per
steradians.}
pointing in direction $\mathbf{n}$, and with (solid angle) size $\de \Omega$. $I_\nu$ has units of
erg cm$^{-3}$ rad$^{-1}$ Hz$^{-1}$ s$^{-1}$.

The equation of radiative transfer (RT) is given by:

\begin{align}
    \frac{1}{c} \DELDT{I_\nu} + \mathbf{n} \cdot \nabla I_\nu
        &= \eta_\nu - \alpha_\nu I_\nu \label{eq:RT} \\
        &= \eta_\nu - \sum_j^{\text{photo-absorbing\ species}} \sigma_{j,\nu} n_j I_\nu \ .
\label{eq:RT-sigma}
\end{align}

$\eta_\nu$ is a source function of radiation, i.e. the term describing radiation being added along
the (dimensionless) direction $\mathbf{n}$ due to yet unspecified processes, and has units of erg
cm$^{-3}$ rad$^{-1}$ Hz$^{-1}$ s$^{-1}$, which is the same as the units of the specific intensity
$I_\nu$ per cm. $\alpha_\nu$ is an absorption coefficient, describes how much radiation is being
removed, and has units of cm$^{-1}$. Naturally only as much radiation as is currently present can be
removed, and so the sink term must be proportional to the local specific intensity $I_\nu$.

The equation holds for any photon frequency $\nu$ individually. In eq.~\ref{eq:RT-sigma} we split
the absorption coefficient $\alpha_\nu$ into the sum over the photo-absorbing species $j$, which
throughout this work will only be the main constituents of primordial gas, namely hydrogen, helium,
and singly ionized helium. The photo-absorption process is modeled as binary collisions between
radiation and the photo-absorbing species, where the photo-absorbing species are treated as targets
which collide with the incoming photons. The probability of a collision that leads to a photon being
absorbed is described by an interaction cross section $\sigma_{j,\nu}$, which has units of cm$^2$,
while $n_j$ represents the number density of photo-absorbing species $j$ in cm$^{-3}$.

\input{tables/RHD/rt_variables.tex}


The equation of radiative transfer already takes the shape of a hyperbolic conservation law, but it
describes how the specific intensity behaves \emph{along a given direction} $\mathbf{n}$. So in
order to solve the equation of radiative transfer for an entire field, we would need to solve it for
all possible directions. This is something we would like to avoid due to the incredible associated
computational expense, and in order to do so, we take the zeroth and first angular moment of
\ref{eq:RT}. Specifically, for the zeroth moment, this means integrating the equation once over all
solid angles $\Omega$, i.e. $\int_\Omega \de \Omega$. For the first moment, the equation is first
multiplied by the direction unit vector $\mathbf{n}$ and then integrated over the entire solid
angle, i.e. $\int_\Omega \mathbf{n} \de \Omega$. Additionally, we make use of the following
quantities:

\begin{align}
	E_\nu (\x, t) &= \int_{4 \pi} \frac{I_\nu}{c} \de \Omega
			&& \text{total energy density }
			&& [E_\nu] = \frac{\text{erg}}{\text{cm}^3 \text{ Hz}}\\
	\Fbf_\nu(\x, t) &= \int_{4 \pi}  I_\nu \mathbf{n} \de \Omega
			&& \text{radiation flux }
			&& [\Fbf_\nu] = \frac{\text{erg}}{\text{cm}^2 \text{ s Hz}}\\
	\mathds{P}_\nu (\x, t) &= \int_{4 \pi} \frac{I_\nu}{c} \mathbf{n} \otimes \mathbf{n} \de \Omega
			&& \text{radiation pressure tensor }
			&& [\mathds{P}_\nu ] = \frac{\text{erg}}{\text{cm}^3 \text{ Hz}}
\end{align}

where $\mathbf{n} \otimes \mathbf{n}$ denotes the outer product, which in components $k$, $l$ gives
\begin{align*}
 (\mathbf{n} \otimes \mathbf{n})_{kl} = \mathbf{n}_k \mathbf{n}_l \ .
\end{align*}

This gives us the following equations:

\begin{align}
	\DELDT{E_\nu} + \nabla \cdot \Fbf_\nu &=
		- \sum\limits_{j}^{\absorbers} n_j \sigma_{\nu j} c E_\nu + \dot{E}_\nu
		\label{eq:dEdt-freq} \\
	\DELDT{\Fbf_\nu} + c^2 \ \nabla \cdot \mathds{P}_\nu &=
		- \sum\limits_{j}^{\absorbers} n_j \sigma_{\nu j} c \Fbf_\nu
		\label{eq:dFdt-freq}
\end{align}

which again take the shape of a hyperbolic conservation law with a state vector $\U$, flux tensor
$\F$, and source term $\mathcal{S}$:

\begin{align}
&&
    \DELDT{\U} + \nabla \cdot \F = \mathcal{S}
&&
\\
    \U = \begin{pmatrix}
          E_\nu \\ \Fbf_\nu
         \end{pmatrix}
&&
    \F = \begin{pmatrix}
            \Fbf_\nu \\ c^2 \mathds{P}_\nu
         \end{pmatrix}
&&
    \mathcal{S} = \begin{pmatrix}
		- \sum\limits_{j}^{\absorbers} n_j \sigma_{\nu j} c E_\nu + \dot{E}_\nu \\
		- \sum\limits_{j}^{\absorbers} n_j \sigma_{\nu j} c \Fbf_\nu
         \end{pmatrix}
\end{align}

Note that $E_\nu$ (and $\dot{E}_\nu$) is the radiation energy \emph{density} (and \emph{density}
injection rate) in the frequency interval between frequency $\nu$ and $\nu + \de \nu$ and has
units of $\text{erg / cm}^3 \text{ / Hz}$ (and $\text{erg / cm}^3 \text{ / Hz / s}$).
$\Fbf$ is the radiation flux, and has units of $\text{erg / cm}^2 \text{ / Hz / s}$, i.e.
dimensions of energy per area per frequency per time.

Furthermore, it is assumed that the source term $\dot{E}_\nu$ stems from point sources which radiate
isotropically. This assumption has the consequence that the vector net flux $\Fbf_\nu$ must sum up
to zero, and hence the corresponding source terms in eq.~\ref{eq:dFdt-freq} are zero.

To close this set of equations, a model for the pressure tensor $\mathds{P}_\nu$ is necessary. In
the case of the Euler equations, which can also be derived as moments of the Boltzmann equation,
the closure is provided by the equation of state, which relates the pressure, the temperature, and
the internal energy of the gas. For the moments of the equation of radiative transfer, we use the
so-called ``M1 closure'' \citep{levermoreRelatingEddingtonFactors1984} where we describe the
pressure tensor via the Eddington tensor $\mathds{D}_\nu$:

\begin{equation}
	\mathds{P}_\nu = \mathds{D}_\nu E_\nu
\end{equation}

The Eddington tensor is a dimensionless quantity that encapsulates the local radiation field
geometry and its effect in the radiation flux conservation equation. The M1 closure sets the
Eddington tensor to have the form:

\begin{align}
\mathds{D}_\nu &=
    \frac{1- \chi_\nu}{2} \mathds{I} + \frac{3 \chi_\nu - 1}{2} \mathbf{n}_\nu \otimes
    \mathbf{n}_\nu
    \label{eq:eddington-freq} \\
\mathbf{n}_\nu &=
    \frac{\Fbf_\nu}{|\Fbf_\nu|} \\
\chi_\nu &=
    \frac{3 + 4 f_\nu ^2}{5 + 2 \sqrt{4 - 3 f_\nu^2}} \\
f_\nu &=
    \frac{|\Fbf_\nu|}{c E_\nu}
\end{align}

\begin{figure}
 \centering
 \includegraphics[width=\textwidth]{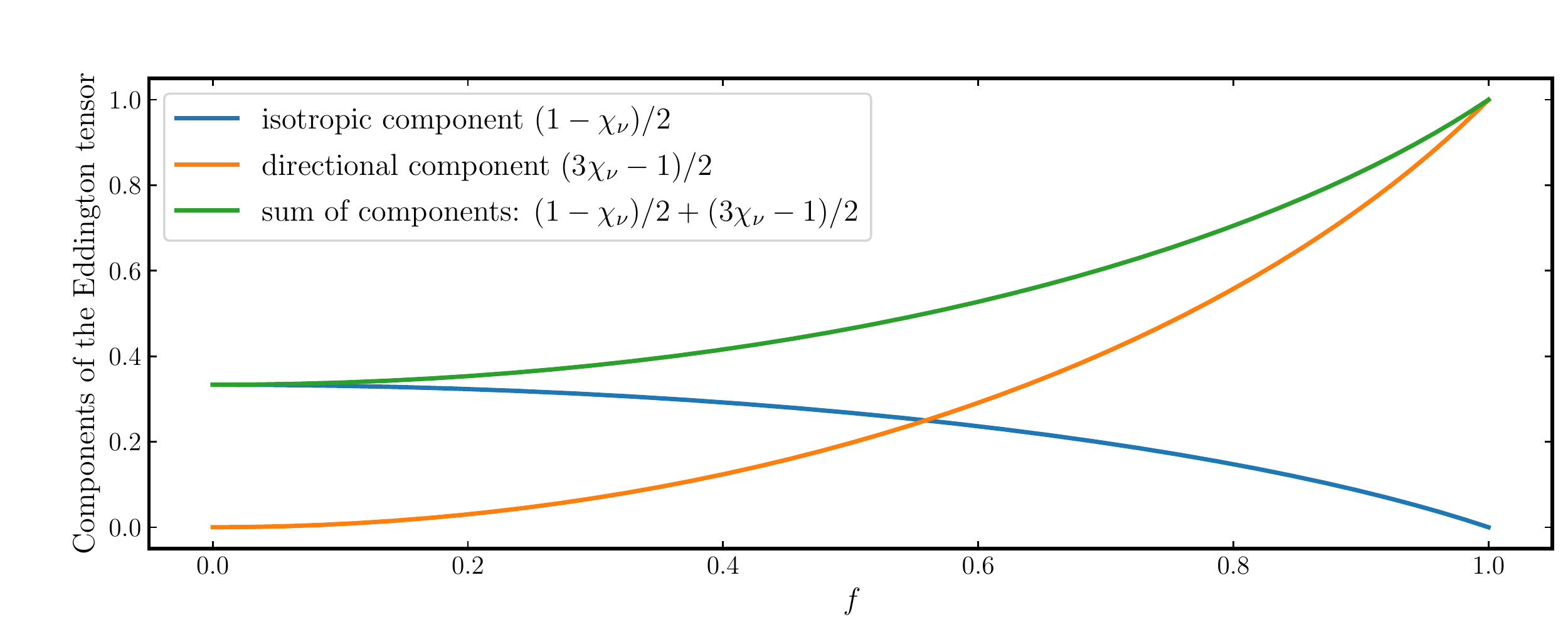}%
 \caption{Components of the Eddington tensor (eq.~\ref{eq:eddington-freq}) depending on the reduced
flux $f_\nu$.}
 \label{fig:eddington-chi}
\end{figure}

The behavior of the Eddington tensor depending on the ``reduced flux'' $f_\nu$ is shown in
Figure~\ref{fig:eddington-chi}. The M1 Closure is an interpolation between extreme cases of fully
isotropic radiation (like blackbody radiation), which is typical for optically thick regimes, and
the free streaming limit which is the case for optically thin regimes. A low value of $f_\nu$
corresponds to predominantly isotropic radiation, while a high value means it is predominantly
flowing in one direction.

This asymptotic behavior can be readily verified, which we shall do now. In the free streaming
limit, the specific intensity of a single point source may be described using a Dirac delta
function:

\begin{equation}
	I_\nu = I_\nu^* \delta(\mathbf{n} - \mathbf{n}_0)
\end{equation}

which gives:

\begin{align}
	\Fbf_\nu &= \int\limits_{4 \pi} I_\nu \mathbf{n} \ \de \Omega = I^*_\nu \mathbf{n}_0 = c E_\nu
\mathbf{n}_0 \\
	\mathds{P}_\nu &= \int\limits_{4 \pi} \frac{I_\nu}{c} \mathbf{n} \otimes \mathbf{n} \de \Omega
=
E_\nu \mathbf{n}_0 \otimes \mathbf{n}_0
\end{align}

and hence

\begin{align}
	|\Fbf_\nu| = c E_\nu && f = \frac{|\Fbf_\nu|}{c E_\nu} = 1 \ .
\end{align}

So for $f_\nu = 1$, we also have $\chi_\nu = 1$, which leads to the correct $\mathds{P}_\nu = E_\nu
\mathbf{n} \otimes \mathbf{n}$ with the M1 Closure.

In the fully isotropic, optically thick case (like blackbody radiation), which constitutes the other
asymptotic case, the pressure tensor is also isotropic, i.e. $\mathds{P}_{11} = \mathds{P}_{22} =
\mathds{P}_{33}$, and using the relation

\begin{align}
	\mathrm{Tr}\ \mathds{P} = \int\limits_{4 \pi} \frac{I_\nu}{c} \de \Omega = E_\nu =
\mathds{P}_{11} + \mathds{P}_{22} + \mathds{P}_{33}
\end{align}

it follows that

\begin{align}
	\mathds{P}_\nu = \frac{E_\nu}{3} \mathds{I} && \mathds{D}_\nu = \frac{1}{3} \mathds{I}
\end{align}

which the M1 Closure correctly gives for $\chi_\nu = 1/3$, or equivalently for $f_\nu = 0$.

\section{Interactions Between Radiation and Gas}\label{chap:coupling-to-hydrodynamics}

The interactions between gas and radiation manifests in a variety of effects. Radiation can heat
the gas, ionize it, and dissociate molecules. If the radiation field is directed, as opposed to
isotropic, the continuous transfer of momentum from photons onto the gas results in an acceleration
of the gas, which is an effect referred to as ``radiation pressure''. Conversely, the gas can both
absorb and scatter the radiation, as well as emit new radiation. Processes that emit radiation are
for example recombination, which describes a positively charged ion capturing a free electron to
form a neutral atom under emission of a photon, where the photon's energy corresponds to the binding
energy of the newly captured electron. Other examples include the radiation emitted by charged
particles being accelerated, like Bremsstrahlung and synchrotron radiation.

While these interactions can be quite contrived on a microscopic level, their macroscopic
description, particularly in the context of hyperbolic conservation laws, is quite straightforward:
Processes which remove energy and momentum from the radiation fields, i.e. act as sink terms, will
be added as energy and momentum to the gas, i.e. act as source terms in the Euler equations. The
inverse is also true: Gas emitting radiation will remove energy from the gas and add it to the
radiation field, and act as source terms in the moments of the equations of radiative transfer.

In the context of interactions between ionizing radiation and gas, the gas emitting recombination
radiation is particularly common, as it makes an entrance as soon as ionized particles and free
electrons are present. It is then convenient to split up the source term of radiation ($\dot{E}_\nu$
in eq.~\ref{eq:dEdt-freq}) into two individual terms: One containing radiation from radiating
sources like stars, $\dot{E}_\nu^*$, and another that contains the recombination radiation emitted
by the gas, $\dot{E}_\nu^{rec}$, i.e.

\begin{align}
 \dot{E}_\nu = \dot{E}_\nu^* + \dot{E}_\nu^{rec} \ . \label{eq:split-injection-terms}
\end{align}

In this work, I focus only on the heating and ionization of the gas. Specifically, the effects of
radiation pressure and the explicit emission of recombination radiation are omitted. Instead,
recombination is modeled as ``Case B recombination'', where emitted recombination photons are
assumed to be directly re-absorbed by the surroundings, leading to a net lower recombination rate.
This approach is called the ``On The Spot Approximation'' (OTSA), and is generally valid in
optically thick regimes.

\subsection{Modeling Interactions as Binary Collisions}

Most of these interactions between gas and radiation can be modeled on a macroscopic scale as
collisions. For example, when a photon and a particle collide, the photon can either be absorbed or
scattered. The scattering can be elastic, and the photon only changes direction, while the photon
energy and the particle's kinetic energy remain constant. This type of scattering is known as
``Thomson scattering''. The scattering can also be inelastic, where the photon and the particle
exchange energies as a consequence of the collision. More precisely, the photon changes both its
direction and its frequency, which is directly proportional to its energy. This process is called
``Compton  Scattering''. Similarly, a photon may be absorbed entirely as a consequence of a
collision.

Binary collisions are commonly described using interaction cross sections $\sigma$, like in
eq.~\ref{eq:RT-sigma}. The basic underlying concept is to have some ``projectiles'' being thrown
with some velocity at some ``targets'' which are at rest. The net collision rates are then
described using probabilities for projectiles and targets to collide. The probability increases
with increasing number of targets, as more targets can be hit. Similarly the probability increases
with increasing number of projectiles for the same reasons. Additionally, the collision \emph{rate}
increases with a higher projectile velocity: A higher velocity means that the projectiles cover a
greater volume over some fixed time interval $\Delta t$. A bigger volume being covered means in turn
that there are more chances for a collision to occur, as the bigger volume would contain more
targets. So some interaction rate $R$ described by the model of binary collisions would need to
satisfy

\begin{align}
    R \propto v_{projectile} \ n_{projectile} \ n_{target}
\end{align}

where $v_{projectile}$ is the projectile velocity, $n_{projectile}$ is the number (density) of
projectiles, and $n_{target}$ is the number (density) of targets. The proportionality can be
resolved into an equality by adding the cross sections $\sigma_{projectile,\ target}$ as the
proportionality constants which encode the interaction probability and have units of cm$^{-2}$:

\begin{align}
    R = \sigma_{projectile,\ target} \ v_{projectile} \ n_{projectile} \ n_{target}
\end{align}

\subsection{Photo-ionization and Photo-heating Rates}

\begin{figure}
 \centering
 \includegraphics[width=\linewidth]{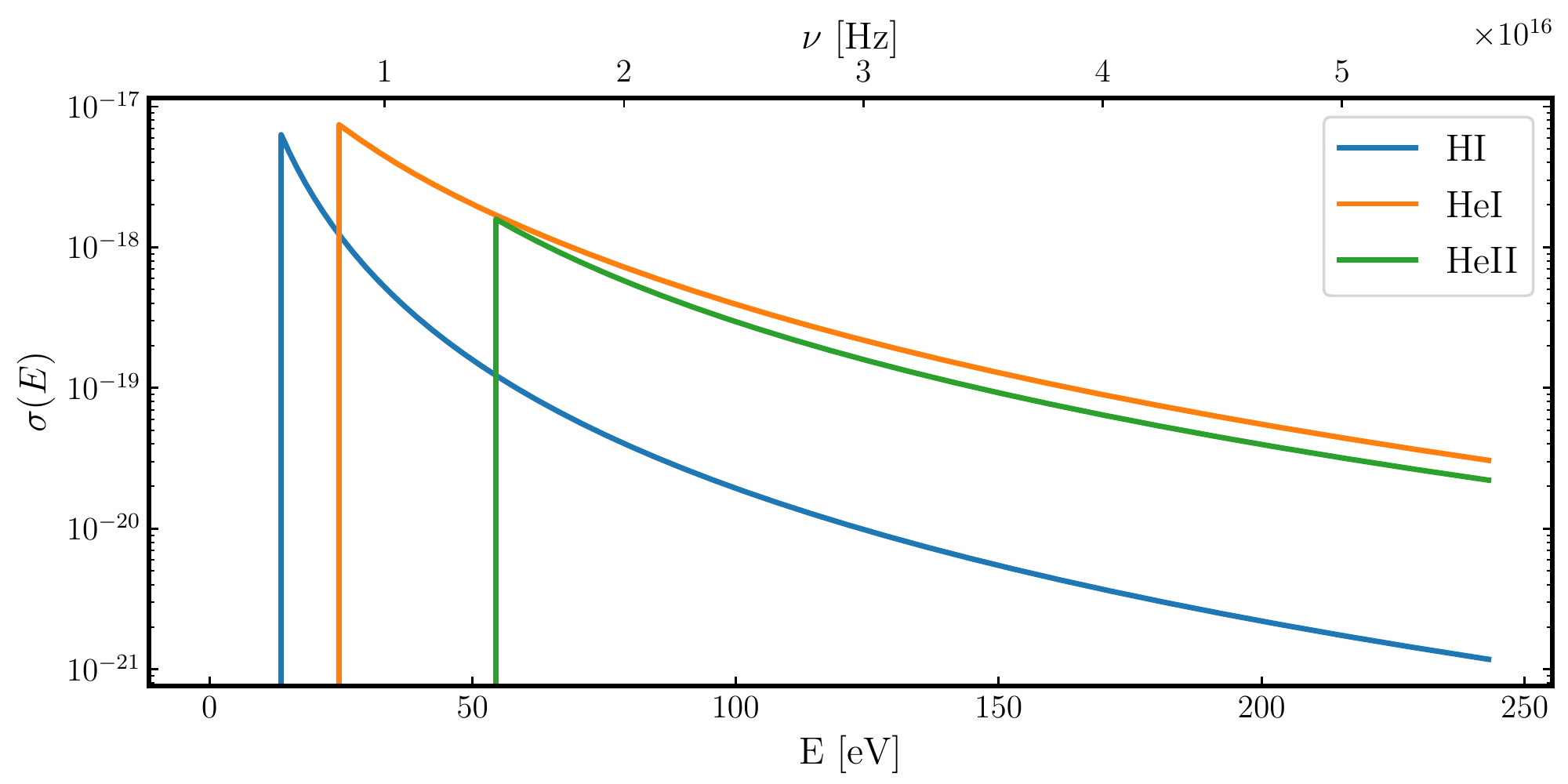}%
\caption{The ionization cross sections parametrizations given by eq.~\ref{eq:sigma-parametrizaiton}.
}
\label{fig:cross-sections}
\end{figure}

In the context of radiative transfer and photo-ionization, the photo-ionization rate $\Gamma_{\nu,
j}$ in units of s$^{-1}$ for photons with frequency $\nu$ and a photo-ionizing particle species $j$
is then given by

\begin{align}
   \DELDT{n_j} = -\Gamma_{\nu, j} \ n_j = - c \ \sigma_{\nu j} \ N_\nu \ n_j
\end{align}

where $N_\nu = E_\nu / (h \nu)$ is the photon number density. Note that the interaction cross
sections are specific to a frequency $\nu$ and the photo-ionizing species $j$ as well. In this work,
I use the analytic fits for the photo-ionization cross sections from
\cite{vernerAtomicDataAstrophysics1996} (via \cite{ramses-rt13}), which are given by

\begin{align}
\sigma(E) &= \sigma_0 F(y) \times 10^{-18} \text{ cm}^2  \label{eq:sigma-parametrizaiton}
\\
F(y) &= \left[(x - 1)^2 + y_w^2 \right] y ^{0.5 P - 5.5} \left( 1 + \sqrt{y / y_a} \right)^{-P}
\\
x &= \frac{E}{E_0} - y_0 \\
y &= \sqrt{x^2 + y_1^2}
\end{align}

where $E$ is the photon energy $E = h \nu$ in eV, and $\sigma_0$, $E_0$, $y_w$, $y_a$, $P$, $y_0$,
and $y_1$ are fitting parameters. The fitting parameter values for hydrogen, helium, and singly
ionized helium are given in Table~\ref{tab:cross-sections}. Figure~\ref{fig:cross-sections} shows
the frequency dependency of the cross sections for these three ionizing species. This fit is valid
for photon energies $E$ which are above the ionization thresholds for the corresponding particle
species. The thresholds are given as frequencies in eqs.~\ref{eq:nuIonHI}-\ref{eq:nuIonHeII}. Below
this threshold, no ionization can take place, and hence the cross sections are zero.

\input{tables/RHD/fitting_parameters.tex}

Conversely, the rate at which photons are absorbed, i.e. ``destroyed'',  must be equal to the
photo-ionization rate, which means

\begin{align}
\DELDT{N_\nu} &= -c \ \sigma_{\nu j} n_j N_\nu
\end{align}

or in terms of photon energy density $E_\nu$:

\begin{align}
\DELDT{E_\nu} &= h \ \nu \DELDT{N_\nu}  = -h\ \nu \ c \ \sigma_{\nu j} n_j N_\nu
\end{align}

Finally, the photo-heating rate is modeled as the rate of excess energy absorbed by the gas during
photo-ionizing collisions. To ionize an atom, the photons must carry a minimal energy corresponding to the ionizing frequency  $\nu_{ion,j}$ for a photo-ionizing species $j$. In the case of hydrogen and helium, their values are

\begin{align}
    \nu_{\text{ion,HI}} &= 2.179 \times 10^{-11} \text{ erg} = 13.60 \text{ eV} \label{eq:nuIonHI}\\
    \nu_{\text{ion,HeI}} &= 3.940 \times 10^{-11} \text{ erg} = 24.59 \text{ eV}
\label{eq:nuIonHeI}\\
    \nu_{\text{ion,HeII}} &= 8.719 \times 10^{-11} \text{ erg} = 54.42 \text{ eV}
\label{eq:nuIonHeII}
\end{align}

All excess energy is added to the gas in the form of internal energy, and the heating rate
$\mathcal{H}$ (in units of erg cm$^{-3}$ s$^{-1}$ ) for photons of some frequency $\nu$ and some
photo-ionizing species $j$ is described by

\begin{align}
\mathcal{H}_{\nu, j} = (h \nu - h \nu_{ion,j}) \ c \ \sigma_{\nu j} \ n_j \ N_\nu
\end{align}

%% file: tables/RHD/rt_variables.tex
\begin{table}
\begin{center}
\begin{small}
\begin{tabular}{p{0.32\textwidth} p{0.42\textwidth} p{0.26\textwidth}}

variable  & name & units (cgs) \\[.5em]
\hline \\

$I_\nu (\x, \mathbf{n}, t)$ &
        specific intensity &
        $[I_\nu] = \frac{\text{erg}}{\text{cm}^2 \text{ rad Hz s}}$ 
\\[.5em]
$u_\nu (\x, \mathbf{n}, t) = \frac{I_\nu}{c}$ &
        radiation energy density &
        $[u_\nu] = \frac{\text{erg}}{\text{cm}^3 \text{ rad Hz}}$
\\[.5em]
$E_\nu (\x, t) = \int_{4 \pi} \frac{I_\nu}{c} \de \Omega$ &
        total energy density &
        $[E_\nu] = \frac{\text{erg}}{\text{cm}^3 \text{ Hz}}$
\\[.5em]
$N_\nu (\x, t) = E_\nu / h$ &
        photon number density &
        $[N_\nu] = \frac{1}{\text{cm}^3 \text{ Hz}}$
\\[.5em]
$E_{rad} (\x, t) = \int_{0}^{\infty} E_\nu \de \nu$ &
        total integrated energy density &
        $[E_{rad}] = \frac{\text{erg}}{\text{cm}^3}$ 
\\[.5em]
$N_{rad} (\x, t) = E_{rad} / h$ &
        integrated photon number density &
        $[N_{rad}] = \frac{1}{\text{cm}^3}$
\\[.5em]
$\mathbf{F}_\nu(\x, t) = \int_{4 \pi}  I_\nu \mathbf{n} \de \Omega$ &
        radiation flux &
        $[\mathbf{F}_\nu] = \frac{\text{erg}}{\text{cm}^2 \text{ s Hz}}$
\\[.5em]
$\mathbf{P}_\nu (\x, t) = \frac{\F_\nu}{c^2}$ & 
        radiation momentum density &
        $[\mathbf{P}_\nu] = \frac{\text{erg}}{ \text{cm}^4 \text{ s}^{-1} \text{ Hz}}$
\\[.5em]
$\mathds{P}_\nu (\x, t) = \int_{4 \pi} \frac{I_\nu}{c} \mathbf{n} \otimes \mathbf{n} \de \Omega$ &
        radiation pressure tensor &
        $[\mathds{P}_\nu ] = \frac{\text{erg}}{\text{cm}^3 \text{ Hz}}$

\\
\hline\\

$\eta_\nu(\x, t)$ &
        source function (of $I_\nu$)&
        $[ \eta_\nu ] = \frac{\text{erg}}{\text{cm}^3 \text{ rad Hz s}}$ 
\\[.5em]
$n(\x, t)$ &
        number density &
        $[ n ] = \text{cm}^{-3}$
\\[.5em]
$\mathcal{J}_\nu(T)$ &
        (specific intensity of a) spectrum &
        $[\mathcal{J}_\nu (T)] = \frac{\text{erg}}{\text{cm}^2 \text{ rad Hz s}}$ 
\\[.5em]
$\Gamma_{\nu, j}$ &
        photoionization rate of species $j$ &
        $[\Gamma] = \text{s}^{-1}$
\\[.5em]
$\mathcal{H}_{\nu,j}$ &
        (photo)heating rate of species $j$ &
        $[\mathcal{H}_{\nu,j}] = \frac{\text{erg}}{s\ cm}^{3}$
\\[.5em]
$\alpha_\nu = \frac{1}{\lambda_\nu} = \rho \kappa_\nu = n \sigma_\nu$ &
        absorption coefficient &
        $[\alpha_\nu] = \text{cm}^{-1}$
\\[.5em]
$\lambda_\nu $ &
        mean free path &
        $[\lambda_\nu] = \text{cm}$
\\[.5em]
$\kappa_\nu $ &
        opacity &
        $[\kappa_\nu] = \frac{\text{cm}^2}{\text{g}}$
\\[.5em]
$\tau_\nu(s) = \int_0^s \alpha_\nu(x) \de x$ &
        optical depth &
        $[\tau_\nu] = 1$
\\[.5em]
$\sigma_{j\nu}$ &
        interacton cross section (of species $j$)&
        $[\sigma_{j\nu}] = \text{cm}^2$
\\[.5em]
$\sigma_{ij}^N = \int_{\nu_{i}} N_\nu \ \sigma_{j\nu} \de \nu / \int_{\nu_i} N_\nu \de \nu$ &
        number weighted average cross section &
        $[\sigma_{ij}^N] = \text{cm}^2$
\\[.5em]
$\sigma_{ij}^E = \int_{\nu i} E_\nu \ \sigma_{j\nu} \de \nu / \int_{\nu i} E_\nu \de \nu$ &
        energy weighted average cross section &
        $[\sigma_{ij}^E] = \text{cm}^2$

\end{tabular}
\end{small}
\end{center}
\caption{Common quantities appearing in the context of radiative transfer, and their units.}
\label{tab:rt-variables}
\end{table}

%% file: tables/RHD/fitting_parameters.tex
\begin{table}
\centering
\begin{tabular}{l | c c c c c c c }
    & $E_0$ [eV]  & $\sigma_0$ [cm$^2$] & $y_a$ & $P$ & $y_w$ & $y_0$ & $y_1$ \\
\hline\\[-0.5em]
H$^0$  & 0.4298 & 5.475$\times 10^{-14}$ & 32.88 & 2.963 & 0     & 0 & 0 \\
He$^0$ & 13.61  & 9.492$\times 10^{-16}$ & 1.469 & 3.188 & 2.039 & 0.4434 & 2.136 \\
He$^+$ & 1.720  & 1.369$\times 10^{-14}$ & 32.88 & 2.963 & 0     & 0 & 0 \\
 \end{tabular}
\caption{Fitting parameters used for the ionizing cross section parametrizations given in 
eq.~\ref{eq:sigma-parametrizaiton}. The parametrization and these parameters are taken from 
\cite{vernerAtomicDataAstrophysics1996}.}
\label{tab:cross-sections}
\end{table}

%% file: main/RHD/RHD-3-rt-numerical-method.tex
\chapter{Numerical Solution Strategy}\label{chap:rt-numerical-strategy}

\section{Discretization of Frequencies}

The moments of the equation of radiative transfer, given in
eqns.~\ref{eq:dEdt-freq}~-~\ref{eq:dFdt-freq}, are still frequency specific, and would need to be
solved for each frequency between 0 Hz and infinity individually. This is obviously not a feasible
approach. Instead, these equations need to be discretized in frequency first. I follow the approach
of \cite{ramses-rt13} and for a rough approximation of multi-frequency, a relevant frequency range
is split into a number $M$ of mutually exclusive groups of frequency ranges:

\begin{align}
	& [\nu_{00}, \nu_{01}\ : \ \nu_{10}, \nu_{11}\ : ... \ : \ \nu_{M0}, \nu_{M1}\ ] =
        [\nu_{0}, \infty [
\end{align}

The frequency ranges are chosen to be convenient for us. Specifically, since the interaction cross
sections of ionizing species are zero below the ionizing frequency of their corresponding species,
it makes sense to use the various ionizing frequencies as the boundaries for the frequency groups.
Since the prime interest in the cosmological context is to heat and ionize the gas, it makes little
sense to keep track of photons with frequencies below the smallest ionizing frequency. So in
practice the lower threshold $\nu_0$ is usually the hydrogen ionization frequency given in
eq.~\ref{eq:nuIonHI}. (The interaction cross sections are all zero below that threshold anyway.)
However if effects like radiation pressure are included, then the relevant frequency range is in
the infrared spectrum, and the $\nu_0$ is lower than the hydrogen ionization frequency, which is in
the ultraviolet frequency range. Additionally, frequencies below the lowest interaction threshold
are of interest when Doppler effects due to the relative motion of the gas are accounted for, which
we also neglect for now.

Rather than treating the photon energy densities and fluxes for individual frequencies, we now use
their integrated averages over a frequency group. For any frequency group $i$, we have

\begin{align}
	E_i &=
        \int\limits_{\nu_{i0}}^{\nu_{i1}} E_\nu d\nu \\
	\Fbf_i &=
       \int\limits_{\nu_{i0}}^{\nu_{i1}} \Fbf_\nu d\nu
\end{align}

giving us the following discretized equations to solve:

\begin{align}
	\DELDT{E_i} + \nabla \cdot \Fbf_i &=
		- \sum\limits_{j}^{\absorbers} n_j \sigma_{i j}^N c E_i + \dot{E}_i^* + \dot{E}_i ^{rec}
		\label{eq:dEdt-group} \\
	\DELDT{\Fbf_i} + c^2 \ \nabla \cdot \mathds{P}_i &=
		- \sum\limits_{j}^{\absorbers} n_j \sigma_{i j}^N c \Fbf_i
		\label{eq:dFdt-group}
\end{align}

The expression for the number weighted average cross section $\sigma_{ij}^N$ is given in eq.
\ref{eq:sigma_N}. The radiation pressure tensor is discretized in the same manner:

\begin{align}
	\mathds{P}_i &=
        \mathds{D}_i E_i \label{eq:pressure-tensor-group-start}\\
	\mathds{D}_i &=
        \frac{1- \chi_i}{2} \mathds{I} + \frac{3 \chi_i - 1}{2} \mathbf{n}_i \otimes \mathbf{n}_i
        \label{eq:eddington-group} \\
	\mathbf{n}_i &=
        \frac{\Fbf_i}{|\Fbf_i|} \\
	\chi_i &=
        \frac{3 + 4 f_i ^2}{5 + 2 \sqrt{4 - 3 f_i^2}} \\
	f_i &=
        \frac{|\Fbf_i|}{c E_i} \label{eq:pressure-tensor-group-end}
\end{align}

For the computation of the  photo-heating and photo-ionization rates, we need to introduce the mean
photon energy $\overline{\epsilon}_i$ of the frequency bin $i$. In order to estimate the average
energy over a frequency interval, the distribution of the radiation energy over the frequencies,
i.e. the photon spectrum, needs to be known. Unfortunately, the exact spectrum of the radiation
field can't be known without tracing the frequencies individually, as the initially emitted spectrum
changes over time due to frequency dependent interactions with the gas. In addition, if radiation
sources with different spectra are present, the superposition of the radiation stemming from these
varying sources would change the local spectrum of the radiation as well. So the exact spectrum
needs to be approximated with a guess. \cite{ramses-rt13} recommend taking an average value among
all stellar radiation sources. In this work, I assume a blackbody spectrum, which for a temperature
$T$ is given by:

\begin{align}
    \mathcal{J}_\nu(T) = \frac{2 \nu^2}{c^2} \frac{h \nu}{\exp\left(h\nu/k_B T\right) - 1} \ .
    \label{eq:blackbody}
\end{align}

The temperature $T$ in this case would be some effective temperature for stellar sources, not the
local temperature of the gas. With an assumed spectral shape, the mean photon energy
$\overline{\epsilon}_i$ in each frequency group $i$ can be estimated as

\begin{align}
\overline{\epsilon}_i \equiv
    \frac{E_i}{N_i} =
    \frac{
        \int\limits_{\nu_{i0}}^{\nu_{i1}} E_\nu \ d\nu
        }{
        \int\limits_{\nu_{i0}}^{\nu_{i1}} N_\nu \ d\nu
        }
    \approx
    \frac{
        \int\limits_{\nu_{i0}}^{\nu_{i1}} \mathcal{J}_\nu \ d\nu
        }{
        \int\limits_{\nu_{i0}}^{\nu_{i1}} \mathcal{J}_\nu / h\nu \ d\nu
        }
\end{align}

Using the mean photon energy $\overline{\epsilon}_i$, the photo-heating rate $\mathcal{H}_{i,j}$
of the gas for the photon group $i$ and ionizing species $j$ then becomes:

\begin{align}
\mathcal{H}_{i, j} &=
\left[
		\int\limits_{\nu_{i0}}^{\nu_{i1}}\de \nu h \nu N_\nu  \sigma_{j\nu} -
	h \nu_{ion,j}\
		\int\limits_{\nu_{i0}}^{\nu_{i1}}\de \nu N_\nu \sigma_{j\nu} \
\right] \ c \ n_j \\
&=
\left[
	\sigma_{ij}^E E_i - h \nu_{ion,j}\ \sigma_{ij}^N N_i
\right]  c \ n_j \\
&=
\left[
	\sigma_{ij}^E \overline{\epsilon}_i - h \nu_{ion,j}\ \sigma_{ij}^N
\right]  N_i\ c \ n_j \\
&=
\left[
	\sigma_{ij}^E  - h \nu_{ion,j}\ \sigma_{ij}^N /\ \overline{\epsilon}_i
\right]  E_i\ c \ n_j
\label{eq:photoheating-group}
\end{align}

And the photo-ionization rate can be written as

\begin{align}
\Gamma_{i, j}
	&=
		c \ \int\limits_{\nu_{i0}}^{\nu_{i1}}\de \nu \ \sigma_{\nu j} N_\nu \\
	&= c \ \sigma_{ij}^N N_i
	= c \ \sigma_{ij}^N E_i / \overline{\epsilon}_i \label{eq:photoionization-group}
\end{align}

Finally, the photon absorption (destruction) rates for a frequency bin $i$ and ionizing species $j$
are given by

\begin{align}
\DELDT{E_i} &=
    \deldt{(N_i \ \overline{\epsilon}_i)} =
    \overline{\epsilon}_i \DELDT{N_i} =
    -\overline{\epsilon}_i c \ \sigma_{i j} ^ N n_j
\end{align}

Here we have introduced the number- and energy-weighted average cross sections:
\begin{align}
\sigma_{ij}^N &=
		\frac{
			\int\limits_{\nu_{i0}}^{\nu_{i1}}\de \nu \ N_\nu \ \sigma_{j\nu}
		} {
		  \int\limits_{\nu_{i0}}^{\nu_{i1}}\de \nu \ N_\nu
		}
		\approx
		\frac{
			\int\limits_{\nu_{i0}}^{\nu_{i1}}\de \nu \ \mathcal{J}(\nu) / (h \ \nu) \sigma_{j\nu}
		} {
  		\int\limits_{\nu_{i0}}^{\nu_{i1}}\de \nu \ \mathcal{J}(\nu) / (h \ \nu)
		} \label{eq:sigma_N}\\
\sigma_{ij}^E &=
		\frac{
			\int\limits_{\nu_{i0}}^{\nu_{i1}}\de \nu \ h \nu N_\nu \ \sigma_{j\nu}
		}	{
			\int\limits_{\nu_{i0}}^{\nu_{i1}}\de \nu \ h \nu N_\nu
		}
		\approx
		\frac{
			\int\limits_{\nu_{i0}}^{\nu_{i1}}\de \nu \ \mathcal{J}(\nu) \  \sigma_{j\nu}
		}	{
			\int\limits_{\nu_{i0}}^{\nu_{i1}}\de \nu \ \mathcal{J}(\nu)
		} \label{eq:sigma_E}
\end{align}

\section{One RT Time Step}

\subsection{Outline}\label{chap:rt-numerics-outline}

For each photon frequency group, the equations \ref{eq:dEdt-group} and \ref{eq:dFdt-group} are
solved with an operator-splitting strategy. Following the approach of \cite{ramses-rt13}, the
equations are decomposed into three steps that are executed in sequence over the same time step
$\Delta t$:

\begin{enumerate}

\item \textbf{Photon injection step}: the radiation from radiative sources is injected into the
grid.

\item \textbf{Photon transport step}: Photons are transported in space by solving the homogenized
equations \ref{eq:dEdt-group} and \ref{eq:dFdt-group}, i.e. the right hand side of these equations
is set to zero.

\item \textbf{Thermochemistry step}: The rest of the source terms (the right hand side) of the
equations \ref{eq:dEdt-group} and \ref{eq:dFdt-group} is solved.
\end{enumerate}

In what follows, the numerical and algorithmic aspects of these three steps are discussed
individually. But first, we should look at the broader picture and discuss some fundamentals upon
which the RT scheme will be built.

\subsubsection{Coupling to Hydrodynamics}

As mentioned before, the effects of radiation on the gas act as source terms for the Euler equations
(eq.~\ref{eq:euler-equations}). Let $\mathcal{S}_{RT}$ denote the source term stemming from the
radiation. Then, neglecting all other source terms, the Euler equations can be written as

\begin{align}
    \DELDT{\U} + \nabla \cdot \F = \mathcal{S}_{RT}
\end{align}

Let $H(\Delta t)$ denote the operator that solves the homogenized (i.e. source-free) part of the
equation over some time step $\Delta t$,

\begin{align}
  H(\Delta t) [\U]: \quad \text{Get }\U(t+\Delta t) \text{ according to } \DELDT{\U} + \nabla \cdot
\F = 0
\end{align}

Similarly, let $S(\Delta t)$ denote the operator that solves only the source part of the equation
over some time step $\Delta t$:

\begin{align}
  S(\Delta t) [\U]: \quad \text{Get }\U(t+\Delta t) \text{ according to } \DELDT{\U} =
\mathcal{S}_{RT}
\end{align}

The operator splitting approach consists of solving the homogenized and the source part of the
equation in sequence rather than concurrently, and the full solution of the system at time $t +
\Delta t$ is given by

\begin{align}
    \U(t+\Delta t) = S(\Delta t)\left[ H(\Delta t) \left[ \U(t) \right] \right] = S(\Delta t) \circ
H(\Delta t) [\U(t)] \label{eq:operator-splitting-hydro-rt}
\end{align}

The solution is first order accurate in time as long as both $H$ and $S$ are also at least first
order accurate operators. In terms of accuracy, the order in which the operators are applied is
inconsequential\footnote{
The analysis of the order of accuracy and interchangeability of operators in operator splitting
techniques, also known as ``fractional-step methods'', relies on the analysis of how the truncation
errors of the used methods behave when the operators are applied in sequence instead of
simultaneously, and what the \emph{numerical} result would look like when using split operators. The
application of a discrete approximate numerical method inevitably leads to error terms being
introduced anyway (see Section~\ref{chap:numerical_diffusion}). The important point is to take into
account whether, and if it does: how the leading order of error terms change when the operators are
split and solved in succession. See \citet{levequeFiniteVolumeMethods2002} for more details.
}, so we can set the order according to our convenience.

\subsubsection{Using a Particle Based Method: Selecting Reference Frames}

In contrast to \cite{ramses-rt13}, who use a finite volume method and adaptive mesh refinement to
solve the moments of the equation of radiative transfer, \GEARRT uses a finite volume particle
method, whose derivation and application to the Euler equations was previously discussed in
Part~\ref{part:meshless}. Rather than cells, which are traditionally employed in finite volume
methods, \GEARRT uses particles as discretization elements. The particles, which represent the gas,
are Lagrangian, i.e. are being moved along with the fluid, and have individual time step sizes (see
Section~\ref{chap:individual-timesteps}).
This approach is maintained with \GEARRT: the motion (drifts) of particles is exclusively determined
by the hydrodynamics. The radiative transfer is solved using a different approach by exploiting the
fact that finite volume particle methods are arbitrary Lagrangian-Eulerian, i.e. are valid in both
co-moving and static frames of reference. In the context of RT, the particles are treated as static
interpolation points with respect to the simulated volume, rather than elements co-moving with the
radiation. As such, the particles carry the fluid quantities stored as a co-moving quantity, while
the radiation fields are taken to be in a static frame of reference. This means however that when
particles are drifted, the radiation fields they carry are drifted along as well, which is wrong.
Hence corrections in the radiation fields are necessary when particles are drifted. The exact form
of these corrections will be discussed in Section~\ref{chap:rt-drift}.

\subsubsection{Setting the Order of Operators to Re-Use Neighbors of Hydrodynamics}

An essential part of finite volume particle methods is the (repeated) interaction of every particle
with its neighboring particles. To facilitate these interactions, first a neighbor search must be
conducted to determine which particles are ``neighbors'' of each other (see
Section~\ref{chap:meshless-full} for details). This needs to be done each time particle positions
change, i.e. each time particles are drifted. Both the hydrodynamics and the radiative transfer
require the neighbor search to be completed before the actual interactions between particles can
take place. Since the neighbor search is already implemented for the hydrodynamics, and the
particles aren't being drifted in the middle of a simulation step, it is practically convenient to
solve the radiative transfer \emph{after} the hydrodynamics and re-use the existing known neighbors.
Using the operators $S$ and $H$, this means that the solution strategy can be written as:

\begin{align}
    \U(t + \Delta t) = S(\Delta t) \circ H(\Delta t) [ \U (t)] \ .
\end{align}

This equation is the same as eq.~\ref{eq:operator-splitting-hydro-rt}, but here it's intended to
\emph{define} the exact order in which the operators will be solved in \GEARRT.

\subsubsection{On the Choice to Treat Radiation in a Static Reference Frame}

The approach to treat the radiation in a static frame of reference with respect to the simulated
volume and the hydrodynamics in a Lagrangian one has caveats. For example, sharp discontinuities in
radiation energy densities and photon fluxes will be more diffusive then they could be, as the
particle positions aren't compressed along the front of the discontinuity. A second caveat is that
corrections to the radiation fields are necessary when particles are drifted, as outlined above and
discussed in Section~\ref{chap:rt-drift}. These corrections may in general not be strictly
conservative, as will be shown in Section~\ref{chap:validation-drift-corrections}.

However, the caveats of the approach to treat the radiation in an Eulerian frame of reference and
the hydrodynamics in a Lagrangian one are outweighed by the many advantages it offers. Firstly, the
hydrodynamics remain self-consistent, as the approach can be basically formulated as ``let the
hydrodynamics determine how the hydrodynamics is solved''. The particle positions will trace the
fluid rather than the radiation fields. Secondly, as discussed in the previous section, the
radiation fields are separated into several photon frequency groups. Suppose we decided to make the
particles co-moving with the radiation instead of with the gas. Then the first question would be:
Which frequency group should we choose to determine the particle motion? By treating the particles
as static interpolation points in the context of RT, all frequency groups are treated equally.
Thirdly, it allows us to save a tremendous workload when a sub-cycling approach is applied. The main
idea behind sub-cycling is to somewhat decouple the time integration of the hydrodynamics and the
radiative transfer from each other. In general, radiation, whose signal velocity is always the speed
of light, will require much shorter time step sizes compared to the fluid, with a difference of
several orders of magnitude. The idea is to allow the RT to progress over several (hundreds) of time
steps for each particle individually using its own radiation time step size, while the hydrodynamics
is updated according to its own hydrodynamics time step size. In terms of operators $S$ and $H$, the
sub-cycling can be written as:

\begin{align}
    \U(t + \Delta t) =
        S(\Delta t/n) \circ S(\Delta t/n) \circ ... \circ S(\Delta t/n) \circ H(\Delta t)[\U(t)]
\end{align}

where $n$ sub-cycles with time steps $\Delta t/n$ have been computed for a single homogenized
hydrodynamics operator $H$. Furthermore, just as is the case for hydrodynamics, particles are also
given individual radiation time step sizes (see Section~\ref{chap:individual-timesteps}),
independently of the particles' hydrodynamics time step sizes.

The alternative would be needing to restrict the hydrodynamics time step size to the radiation time
step size, and hence having to perform many hydrodynamics time steps which are in principle not
strictly necessary. Additionally, if particle drifts are performed only according to what the
hydrodynamics prescribe, it means that each RT sub-cycle can proceed without drifting particles, and
hence there is no need for neighbor search interaction loops during RT sub-cycles as well. This
saves a tremendous amount of work, as will be demonstrated in Section~\ref{chap:subcycling-results}. The
sub-cycling approach is discussed in more details in Sections~\ref{chap:dynamic-sybcycling} and \ref{chap:subcycling}.

\subsubsection{Summary}

Before we continue with the description of each of the operator splitting steps required
for the solution of the radiative transfer described at the beginning of this section, let's
summarize the fundamental approach upon which \GEARRT is built:

\begin{itemize}
\item The basic discretization elements are particles.
\item The underlying method to solve the hyperbolic conservation laws for hydrodynamics and for
radiative transfer is a finite volume particle method, as described in Part~\ref{part:meshless}.
\item The particles are co-moving with the gas, and the motion of the particles is determined by
the hydrodynamics.
\item Radiation fields at particle positions are treated in a static frame of reference.
\item Particles are given individual time step sizes, for both the hydrodynamics and the radiative
transfer.
\item In a simulation step, the hydrodynamics are solved before the radiative transfer.
\end{itemize}

\subsection{First step: Injection} \label{chap:injection-step}

\subsubsection{Injecting the Energy Density}

In the injection step, the radiation is gathered from radiating sources and injected into the gas.
Radiation emitting sources are taken to be stars or entire stellar populations, which in \swift are
represented by individual ``star particles''. The equation to be solved in this step is

\begin{equation}
    \DELDT{E_i} = \dot{E}_i^* \label{eq:solve-dEdt}
\end{equation}

where $\dot{E}_i^*$ is the total energy density in the frequency group $i$ emitted by all stars $k$
that are within compact support radius of the corresponding particle. If each star $k$ deposits
some fraction $\xi_k$ of its respective total radiation energy density to be injected,
then the total radiation energy density injected into a single gas particle $i$ is given by:

\begin{align}
    \dot{E_i}^* = \sum_k E_{i,k}^* \xi_k \label{eq:injection_onto_particle} \ .
\end{align}

The exact choice how the fraction $\xi_k$ is determined will be discussed in the subsequent
subsection. Eq.~\ref{eq:injection_onto_particle} is solved using a simple finite difference
discretization and first order forward Euler integration:

\begin{equation}
    E_i(t + \Delta t) = E_i(t) + \Delta t \dot{E}_i^*
\end{equation}

for each particle.

The number of neighbors a star particle injects energy into is a free parameter. It is determined
in the same manner as for the number of neighbors for the hydrodynamics interactions, given in
eq.~\ref{eq:number-of-neighbors}, by specifying a dimensionless parameter $\eta_{res}$. The default
choice is $\eta_{res} = 1.2348$, which leads to approximately 48 neighbors for the cubic spline
kernel (eq.~\ref{eq:cubic-spline-kernel}). Note that the number of neighbors for star particles and
for gas particles can be specified using two independent parameters $\eta_{res}$, one for stars,
and one for the gas.

Star particles also have individual time step sizes, just like gas particles (see
Section~\ref{chap:individual-timesteps}). Because both star and gas particles have individual time
step sizes, the way the energy density $\dot{E}_i^*$ is deposited from stars onto particles is
determined by the star particles' time steps. Each time a star particle is ``\lingo{active}'', i.e.
the simulation is at a step where the star particle finishes its time integration, the star particle
``pushes'' radiation onto both active and inactive gas particles. With the known time step size of a
star, the correct amount of energy density can be found for given stellar luminosities and
subsequently deposited on neighboring gas particles. Other stellar feedback is performed in the
same manner in \swift.

Doing things the other way around, i.e. injecting energy by having the gas ``pull'' the radiation
from stars instead is too restrictive and inefficient. It would require star particles to perform a
neighbor search at a rate determined by the smallest neighboring gas particle time step\footnote{
In fact, first we would need additional neighbor search during which gas particles identify their
neighboring \emph{star} particles in order to identify the star neighbors they need to ``pull''
radiation from. Such a neighbor loop, where gas particles search for neighboring stars, is not
necessary to solve the hydrodynamics nor the propagation of radiation, and hece would be a new
additional neighbor search loop. \\
Then the star particles would need a neighbor search to identify which (and
in particular: how many) gas particles will ``pull'' radiation from them, so each star particle $k$
has the information available how to determine the fraction $\xi_k$ of radiation it will deposit on
each of the gas particles it interacts with. This is necessary to ensure that the total radiation
``pulled'' from star particles is the correct amount. The frequency of this second neighbor search,
where the star particles search for neighboring gas particles, will be determined by the smallest
neighboring gas particle time step size.
}.
Secondly, since gas particles have individual time step sizes, it's difficult to ensure that the
total emitted energy from the stars is the correct amount. At a given step, gas particles with
bigger individual time step sizes should receive a higher amount of energy, which is supposed to
represent the energy injected into them over their respective time step size. However at the next
simulation step, when particles with smaller time step sizes are active again, while the ones with
bigger time step sizes are inactive, we would have to account for all the energy already injected
into the particles with the bigger time step size in the previous step. Generally, this information
is not available. It would only be available if the particles with bigger time step sizes are also
accounted for during the neighbor search for the energy injection requested by the gas particles
with smaller time step sizes, which would mean that they'd have to be \lingo{active} despite having
a bigger time step size. This effectively means that for an energy injection scheme where the gas
particles ``pull'' radiation from star particles, individual time step sizes for the gas can't be
used. So that is not a viable scheme.

For similar reasons, stars need to inject energy density rather than luminosities, i.e. energy
density \emph{rates}, onto neighboring gas particles. Problems arise when there are multiple stars
injecting energy into the same gas particle: Since star particles have individual time step sizes,
and the injected rates are valid for the duration of the respective star's time step length, we
would then need to keep track of the injected energy density rates of each source for each gas
particle individually. Additionally, we would also have to keep track of the time step size of each
source for each gas particle. Given how well the alternative, where the energy density rather than
luminosities is injected, works, all this additional expense and effort that would be required is
not worthwhile.

\subsubsection{Weights For Distribution Of Energy Density of a Star onto
Particles}\label{chap:injection-weights}

We now define the weights $\xi_s$ used in eq.~\ref{eq:injection_onto_particle} to distribute the
total radiation energy ejected by a star $s$ over a time step $\Delta t$ onto the surrounding gas
particles $p$.
A natural way of depositing the energy density from the star on a neighboring gas particle is to
make use of the already present partition of unity, $\psi_p(\x_p - \x_s)$ (see eqs. \ref{eq:psi} -
\ref{eq:omega}), which is a fundamental building block for finite volume particle methods. It
guarantees to sum up to unity, and additionally is taking into account the particle configuration
due to its normalization.
%
%
%
%
%
%
Then the total emitted energy of the star $s$, $E^*_i(\x_s)$, is fully and self-consistently
distributed on the gas (here gas particles have the index $g$):

\begin{align}
	\sum_g \psi_g(\x_s)\  E_i^*(\x_s) = \sum_g \psi(\x_s - \x_g, h_s) \ E_i^*(\x_s) = E_i^*(\x_s)
\end{align}

For the actual distribution of the radiation from stars to gas particles, the sum over neighboring
stars from the gas particle's point of view looks like:

\begin{align}
	E^*_i (\x = \x_{gas}) = \sum_{stars} E^*_i(\x_{star}) \ \psi(\x_{star} - \x_{gas}, h(\x_{star}))
\end{align}

which ultimately gives us

\begin{align}
    \xi_s = \psi(\x_{s} - \x_{g}, h(\x_{s})) \ .
\end{align}

\begin{figure}
	\centering
	\includegraphics[width=.6\textwidth]{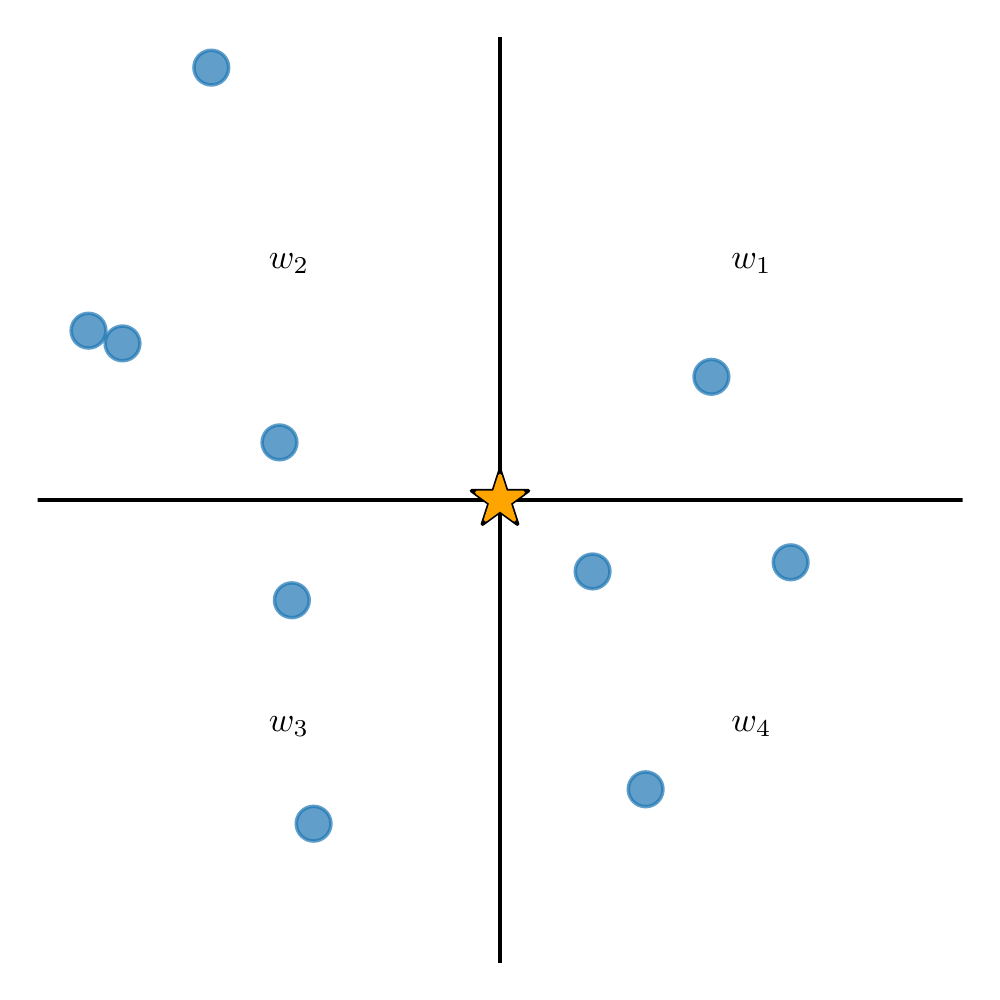}%
	\caption[Sketch of the weight correcting scheme for radiation injection]{
Sketch of the weight correcting scheme for radiation injection for a two-dimensional case. The
space
around a star which is surrounded by gas particles (blue circles) is divided into four quadrants.
The sum of all weights within a quadrant, $w_1$, $w_2$, $w_3$, $w_4$, is then used to improve the
isotropy of the injected energy and flux.
    }
	\label{fig:flux-injection-correction-method}
\end{figure}

A potential problem here is that unless the particle distribution is somewhat symmetric, the
resulting injected energy density (and flux) will not be isotropic, which is what stellar radiation
is typically assumed to be. So instead, we ``correct'' the weight $\psi_p$ in an attempt to improve
the isotropy of the deposited net energy and flux. We are limited in the manner we can compute the
correction term - we don't want the additional expense of a new star-particle interaction loop, nor
can we store all the neighbor data for every star individually in an attempt to determine isotropy.
So instead, we determine a correction term as follows.
For simplicity, let's consider the problem in 2D. We divide the space around a star into four
quadrants like in Figure~\ref{fig:flux-injection-correction-method}, and sum up the total weights
$\xi_s$ of the neighboring gas particles $g$ in each quadrant. This sum is denoted as $w_1$,
$w_2$, $w_3$, and $w_4$, as shown in Figure~\ref{fig:flux-injection-correction-method}. We now add
a correction factor $\mu_1$,  $\mu_2$,  $\mu_3$,  $\mu_4$ to each quadrant in an attempt to improve
the isotropy by applying two constraints:

\begin{enumerate}
    \item The total weight must remain constant:

\begin{equation}
\mu_1 w_1 + \mu_2 w_2 + \mu_3 w_3 + \mu_4 w_4 = w_1 + w_2 + w_3 + w_4
\end{equation}

    \item The modified weight of each quadrant should be equal, such that at least along these four
    directions, the injected energy and flux are isotropic:

\begin{equation}
\mu_1 w_1 = \mu_2 w_2 = \mu_3 w_3 = \mu_4 w_4
\end{equation}

\end{enumerate}

After some simple algebra, we arrive at

\begin{equation}
\mu_a = \frac{\oldsum_b w_b}{4 w_a} \label{eq:isotropy-correction-general}
\end{equation}

The extension to three dimensions is trivial: the factor 1/4 is just turned into 1/8, as we now
have octants instead of quadrants. The correction term~\ref{eq:isotropy-correction-general} needs a
small modification for cases where there is a octant which contains zero particles and hence zero
weight: Let $q_{nz}$ be the number of octants with non-zero weight, i.e. with non-zero particles
in them. Then

\begin{equation}
	\mu_a = \frac{\oldsum_b w_b}{q_{nz} w_a} \label{eq:isotropy-correction-with-zero}
\end{equation}

this ultimately gives us

\begin{align}
    \xi_s = \mu_a \ \psi(\x_{s} - \x_{g}, h(\x_{s}))
\end{align}

where $a$ is the octant in which the gas particle resides.

Since the weights in each octant are being collected for the anisotropy correction terms $\mu_a$,
we can in principle use any weight for particles instead of $\psi_i(\x)$ and use the total weight
to normalize the net injected energy and flux. Indeed, I tested particle weights $\xi_i \propto
r^{-\alpha}$ with $\alpha \in [0, 1, 2, 3]$ instead of the partition of unity $\psi_i(\x)$, but
none of these choices was able to out-perform the partitions of unity in terms of isotropy. The
reason is that the $\psi(\x)$ already contain information about neighboring particle positions
through the normalizations $\omega(\x)$ (eq.~\ref{eq:omega}). Additionally, the correction terms
$\mu_a$ can only improve isotropy, but not guarantee it, as they only take into account inequalities
between the octants around the star particle. For this reason, the choice of $\psi(\x)$ as the
individual weights is selected.

%
%
%
%
%
%
%
%
%
%
%
%

\subsubsection{Injecting Photon Fluxes From Radiative Sources}

A further topic with regards to injection of radiation from sources which requires a bit of
discussion is whether, and if so, in which manner to inject photon fluxes. For grid codes, the net
injected flux into a cell will automatically be zero since we assume that stars are points that
radiate isotropically, and the vector sum of an isotropic flux is zero. However for particle
codes, we don't have that luxury given automatically, as the total flux isn't injected into a
single cell, but into several particles. So as long as we can ensure that the vector sum of all
injected fluxes remains zero, we have some freedom to choose how (and how much) photon fluxes to
inject. However, if we inject the flux carelessly, we can end up with anisotropic fluxes.


To inject fluxes, we could in principle use the same approach as for the energy density, and
inject the flux $\Fbf_{i,p,s}$ of photon group $i$ injected from star $s$ that we assign to a
particle $p$ according to

\begin{align}
	\Fbf_{i,p,s} &= \xi_s c E_{i,s} \mathbf{n}_{ps}  \\
    \mathbf{n}_{ps} &=  \frac{\x_p - \x_s}{|\x_p - \x_s|}
\end{align}

by making use of the energy density-flux relation for the optically thin limit $|\Fbf| = c E$, or
some other relation. What remains unclear is to determine (i) whether flux should be injected at
all, and (ii) if so, what energy density-flux relation to use. To this end, I tested three
different flux injection models:

\begin{itemize}
 \item The ``no flux injection'' model injects no radiation flux $\Fbf$. This corresponds to the
optically thick limit.
 \item The ``full flux injection'' model injects flux assuming an optically thin limit, i.e. $\Fbf
= c E$
 \item The ``modified flux injection'' model injects a linearly increasing amount of flux starting
with no flux and ending in the optically thin limit $\Fbf = c E$ at the distance $0.5H$, where $H$
is the maximal distance at which energy and flux is injected, i.e. the star's compact support
radius. For distances above $0.5H$, again the value of the optically thin limit is taken.
\end{itemize}

Note that (at least in ideal cases) the net injected flux, i.e. the vector sum of all injected
fluxes, still remains zero just as is the case for grid codes, but split among several gas
particles. The results are shown in Section~\ref{chap:results-injection}, where the ``no flux
injection'' model is shown to perform best.

\subsection{Second Step: Transport}\label{chap:transport-step}

In this step, we solve
\begin{align}
    \DELDT{E_i} + \nabla \cdot \Fbf_i &= 0 \label{eq:homogenized-dEdt}\\
    \DELDT{\Fbf_i} + c^2 \nabla \cdot \mathds{P}_i &= 0 \label{eq:homogenized-dFdt}
\end{align}

They are both in the form of a hyperbolic conservation law
(\ref{eq:conservation-law-1D-introduction}), so we can use the finite volume particle method:

\begin{align}
	\deldt(\U_k V_k) + \sum\limits_{l} \mathcal F_{\alpha, kl} \mathcal{A}^\alpha_{kl} = 0
\label{eq:rt-meshless}
\end{align}

where $k$, $l$ are particle indices, $V_k$ the associated particle volumes
(eq.~\ref{eq:particle-volume}), $\mathcal{A}_{kl}$ the effective surfaces given
in eq.~\ref{eq:HopkinsAij}, and we have the state vector $\U$ and flux tensor $\F$

\begin{align}
	\U =
		\begin{pmatrix}
			E_i \\
			\Fbf_i
		\end{pmatrix}
	\quad , \quad
	\F =
		\begin{pmatrix}
			\Fbf_i \\
			c^2 \mathds{P}_i
		\end{pmatrix}
\end{align}

The total sum $\oldsum_l \F_{kl} A_{kl}$ for each particle $k$ is collected during a neighbor
interaction loop. Once the sum is accumulated, the final explicit update of the state is given by

\begin{align}
\U_k(t + \Delta t_k) =
	\U_k (t) - \frac{1}{V_k} \sum\limits_{l} \min\{\Delta t_k, \Delta t_l\}
    \F_{\alpha,kl} \mathcal{A}_{kl}^\alpha
\label{eq:transport-update-explicit}
\end{align}

To facilitate the individual time step sizes, the minimum between individual particle time steps
$\Delta t_k$ and $\Delta t_l$ is taken again, and the exchanged quantities between particles are
time integrated fluxes rather than just fluxes. The time integration needs to be the smaller of the
two time step sizes $\{\Delta t_k,\  \Delta t_l\}$ since if one particle has a smaller time step
size than the other, it also means that there will be several interactions between that particle
pair, each with the time step size of the smaller of the two. This ensures that the total exchange
of fluxes remains conservative, as the fluxes ``removed'' from particles $l$ remain exactly equal
to the fluxed ``added'' to particle $k$. If some neighboring particles $l$ have smaller time steps
than particle $k$, then the net sum of the fluxes is accumulated during the exchanges and applied
only at the point where particle $k$ is being updated again. More details on individual
timestepping is discussed in Section~\ref{chap:individual-timesteps}.

The number of neighbors each particle interacts with, or more precisely each particle's smoothing
length $h$, is the same as for hydrodynamics. This is necessary in order to be able to re-use all
quantities required to compute the effective surface terms \Aij and gradients of all radiation
fields. The precise number of neighbors is a free parameter, and is determined by specifying a
dimensionless parameter $\eta_{res}$ (see eq.~\ref{eq:number-of-neighbors}). The default choice is
$\eta_{res} = 1.2348$, which leads to approximately 48 neighbors for the cubic spline kernel
(eq.~\ref{eq:cubic-spline-kernel}). Note that the number of neighbors for star particles and for gas
particles can be specified using two independent parameters $\eta_{res}$, one for stars, and one for
the gas.

The time step size of the particles is computed in the same manner as we do it for hydrodynamics.
A ``cell size'' $\Delta x$ of a particle is estimated using

\begin{align}
    \Delta x \approx \left(\frac{V_i}{V_{S,\nu}} \right)^{1/\nu}
\end{align}

where $V_{S,\nu}$ is the volume of a $\nu$-dimensional unit sphere, i.e. $4/3 \pi$ in 3D. The signal
velocity in case of radiation is simply the speed of light $c$, which leads to the time step
estimate

\begin{align}
    \Delta t = C_{CFL} \frac{\Delta x}{c} \label{eq:rt-cfl}
\end{align}

A safe choice of $C_{CFL} \lesssim 0.6$ is recommended.

To get the inter-particle fluxes $\F_{kl}$, we also follow the scheme outlined by the finite volume
particle method in Section~\ref{chap:FVPM}. We use the ``Hopkins'' formulation of the method
(Section~\ref{chap:meshless-hopkins}), since the ``Ivanova'' version was shown to still have some
trouble in Section~\ref{chap:meshless-comparison}. Even though the radiative transfer is solved
assuming an Eulerian frame of reference, for which the ``Ivanova'' version is working adequately,
the ``Hopkins'' formulation seems like a safer and therefore better choice, at least for the time
being.

As it is done for the hydrodynamics in Section~\ref{chap:meshless-full}, we approximate the problem
by defining an ``interface'' at the position

\begin{equation}
	\mathbf{x}_{kl} = \mathbf{x}_k + \frac{h_k}{h_k + h_l} ( \mathbf{x}_l - \mathbf{x}_k )
\end{equation}

and extrapolate the value of the conserved variables at the position:

\begin{equation}
	\U_{k, l}(\mathbf{x} =
        \mathbf{x}_{kl}) \approx \U_{k, l} + \DELDX{ \U_{k, l}}\ (\mathbf{x}_{kl}
- \mathbf{x}_{k,l})
\end{equation}

such that the fluxes $\F_{kl}$ in the update formula~\ref{eq:rt-meshless} between particles is
estimated as the solution of the Riemann problem with ``left'' state $\U_k(\x_{kl})$ and ``right''
state $\U_l(\x_{kl})$

\begin{align}
    \F_{kl}
    = RP
    \left(\U_k + (\x_{k} - \x_{kl})_\alpha \DELDXALPHA{\U_k}, \
    \U_l + (\x_{l} - \x_{kl})_\alpha \DELDXALPHA{\U_l} \right) \label{eq:rt-riemann-problem}
\end{align}

The gradients $\DELDX{\U_{k,l}}$ are estimated again using the least-squares second order accurate
gradient expression given in eqn. \ref{eq:gradient}, for which a particle-particle interaction loop
is required. Since the RT is solved after the hydrodynamics and the particle positions haven't
changed since the last hydrodynamics drift, the particles' smoothing lengths are still accurate, and
we don't require an additional density loop for the radiative transfer. For that reason the gradient
particle loop constitutes the first particle interaction loop in the RT scheme.

\subsubsection{Flux Limiters}

The gradient extrapolation is equivalent from going from a piece-wise constant reconstruction of the radiation field to a piece-wise linear one, which makes the method second order accurate.
For each quantity $Q_{k,l} \in \U = (E_{k,l}, \Fbf_{k,l})^T$, we extrapolate the value at the
interface $\x_{kl}$ using the estimated gradient $dQ_{k,l}$. To prevent oscillations and
instabilities from occurring, we need to employ a flux limiter at this point, which effectively
reduces the gradient slopes $dQ_k$ and $dQ_l$ whenever necessary  (see
Section~\ref{chap:higher-order-schemes}). The limiters are implemented in such a way that they
return a factor $\alpha$, with which both slopes $dQ_k$ and $dQ_l$ are to be multiplied in order to
obtain the limited slopes, i.e:

\begin{align}
	\alpha &= \text{limiter}(dQ_k, dQ_l)\\
	dQ_{k, limited} &= \alpha\ dQ_k\\
	dQ_{l, limited} &= \alpha\ dQ_l
\end{align}

The two limiters that worked well in tests are the \emph{minmod limiter}

\begin{align}
    \alpha_{\text{minmod}}(dQ_k, dQ_l) =
    \begin{cases}
      dQ_k      & \text{ if } |dQ_k| < |dQ_l| \text{ and } dQ_k dQ_l > 0 \\
      dQ_l      & \text{ if } |dQ_k| > |dQ_l| \text{ and } dQ_k dQ_l > 0 \\
      0         & \text{ if }  dQ_k dQ_l \leq 0
    \end{cases} \label{eq:rt-minmod}
\end{align}

and the \emph{monotonized central difference (MC) limiter}:
\begin{align}
	r &= dQ_k / dQ_l \\
	\alpha_{\text{MC}}(r) &= \max \left\{ 0, \min\left[\frac{1 + r}{2}, 2, 2r \right] \right\}
	\label{eq:rt-MC}
\end{align}

The minmod limiter appears to be the most robust one with regards to ensuring stability for
radiative transfer, and for this reason is the recommended choice. The effects of the various
limiters on the transport of radiation will be shown in Section~\ref{chap:rt-riemann-limiters}.
Other possible choices are the \emph{superbee} limiter:

\begin{align}
	r &= dQ_k / dQ_l \\
	\alpha_{\text{superbee}}(r) &= \max \left\{0,  \min (1, 2r), \min(2, r) \right\}
	\label{eq:rt-superbee}
\end{align}

And the \emph{van Leer} limiter:

\begin{align}
	r &= dQ_k / dQ_l \\
	\alpha_{\text{vanLeer}}(r) &= \frac{r + |r|}{1 + |r|}
	\label{eq:rt-vanLeer}
\end{align}

While the slope and flux limiter combination described in \cite{hopkinsGIZMONewClass2015} for
hydrodynamics, whose full form is given in
eqns.~\ref{eq:face-limiter-first}-\ref{eq:face-limiter-last}, is a viable option, tests (shown in
Section~\ref{chap:rt-riemann-limiters}) have shown that it is more diffusive than the previously
described limiters for the application on radiation transport, and in some cases instabilities arise
that are seemingly connected to the first gradient limiting process. As such, I do not recommend to
use this sophisticated limiter in the context of RT. Note however that this limiter is used for
hydrodynamics, and that different flux limiting methods are used for hydrodynamics and radiative
transfer.

%
%
%
%
%
%
%

\subsubsection{Riemann Solvers for the Moments of the RT equation}\label{chap:riemann-rt}

It remains to find a solution for the Riemann problem~\ref{eq:rt-riemann-problem} which gives us
the inter-particle (``inter-cell'') fluxes $\F_{kl}$.
To find $\F_{kl}$, we use the extrapolated and flux limited states $E_{k,l}$ and $\Fbf_{k,l}$ at the
interface position $\x_{kl}$ to compute the states and fluxes of the conservation law,
$\U_{k,l}(\x_{kl})$ and $\F_{k,l}(\x_{kl})$. The components of the vector valued photon flux density
$\Fbf_{k,l}$ are projected along the unit vector pointing from the particle towards the surface,
allowing us to treat each component individually as a one dimensional problem. We then adapt the
convention that particle $k$ is the left state of the Riemann problem, while particle $l$ is the
right state.

The \emph{Global Lax Friedrich (GLF)} Riemann solver \citep{ramses-rt13} then gives the
following solution for the flux $\F_{i+\half}$ between a cell $i$ and $i+1$:
\begin{equation}
	\F_{1/2}(\U_L, \U_R) =
		\frac{\F_{L} + \F_{R}}{2} -
		\frac{c}{2} \left(\U_R - \U_L \right) \label{eq:riemann-GLF}
\end{equation}

The GLF Riemann solver can be derived from the Lax-Friedrichs scheme, which for cells with sizes
$\Delta x$ takes the solution $\U_i^{n+1}$ to be the integral average of the solution
$\tilde{\U}(x, \Delta t/2)$ at the midpoint in time, $\Delta t/2$, of the Riemann problem which
uses the adjacent cells $i-1$ and $i+1$ as the left and right states, i.e.

\begin{align}
\U^{n+1}_i &=
    \frac{1}{\Delta x}
    \int_{-x_{i-\half \Delta x}}^{x_{i+\half \Delta x}} \tilde{\U}(x, \Delta t/2) \de x \\
\tilde{\U} &= RP(\U_{i-1}, \U_{i+1})
\end{align}

To proceed, we make use of the integral form of the conservation law
(eq.~\ref{eq:conservation-law-integral-form}) over the volume $[-\Delta x /2, \Delta x/2]$ and the
time interval $[0, \Delta t]$

\begin{align}
 \int\limits_{-\Delta x/2}^{\Delta x/2} \tilde{\U}(x, \Delta t/2) \de x  =
 \int\limits_{-\Delta x/2}^{\Delta x/2} \tilde{\U}(x, 0) \de x  +
 \int\limits_{0}^{\Delta t/2} \F(\tilde{\U}(-\Delta x /2, t)) \de t -
 \int\limits_{0}^{\Delta t/2} \F(\tilde{\U}(\Delta x/2, t)) \de t
\end{align}

The solution of the Riemann problem at $t = 0$ is simply the initial conditions, which in this
case is $\U_{i-1}$ on the left side, and $\U_{i+1}$ on the right side. The first integral on
the right hand side evaluates to

\begin{align}
\int\limits_{-\Delta x/2}^{\Delta x/2} \tilde{\U}(x, 0) \de x  =
\frac{\Delta x}{2} (\U_{i+1} + \U_{i-1})
\end{align}

The fluxes in the integrals are evaluated at the cell boundaries $-\Delta x/2$ and $\Delta x/2$.
Assuming an appropriate CFL condition is chosen, the resulting waves of the Riemann problem
situated at the cell center, which in this case is placed at $x = 0$, are not allowed to reach the
cell boundaries, i.e. travel a distance of $\Delta x/2$, by the time $\Delta t/2$. This means that
the states $\tilde{\U}$ at the cell boundaries $\pm \Delta x/2$ are constant in the time interval
$[0, \Delta t/2]$, and are also simply $\tilde{\U}(\pm\Delta x/2) = \U_{i\pm1}$. So the integrals
involving the fluxes evaluate to

\begin{align}
\int\limits_{0}^{\Delta t/2} \F(\tilde{\U}(\pm \Delta x /2, t)) \de t &=
    \int\limits_{0}^{\Delta t/2} \F(\U_{i\pm1}) \de t = \frac{\Delta t}{2} \F(\U_{i\pm1})
\end{align}

This leads to the scheme

\begin{align}
    \U_i^{n+1} = \frac{1}{2} (\U_{i-1} + \U_{i+1}) + \frac{1}{2} \frac{\Delta t}{\Delta x}
(\F^n_{i-1} - \F^n_{i+1})
\end{align}

which can be re-written as

\begin{align}
    \U_i^{n+1} =
        \U_i^n + \frac{\Delta t}{\Delta x} \left(\F_{i-\half}^{LF} - \F_{i+\half}^{LF} \right)
\end{align}

with the Lax-Friedrichs fluxes

\begin{align}
\F_{i+\half}^{LF} = \frac{1}{2} (\F^n_i + \F_{i+1}^n) + \frac{1}{2}\frac{\Delta x}{\Delta t} (
\U_i^n - \U_{i+1}) \label{eq:lax-friedrichs-flux}
\end{align}

By setting the Courant number to the maximally allowable value, $C_{cfl} = \frac{\Delta t}{\Delta x
c} = 1$, we can replace the term $\Delta t / \Delta x = c$ in eq.~\ref{eq:lax-friedrichs-flux},
and we obtain the GLF flux from eq.~\ref{eq:riemann-GLF}, where $c$ is the speed of light.

An alternative is the \emph{Harten - Lax - van Leer (HLL)} Riemann solver
(\cite{gonzalezHERACLESThreedimensionalRadiation2007}), which can be derived by assuming that the
solution of the Riemann problem consists of two discontinuous waves, and applying the
Rankine-Hugeniot jump conditions (eq.~\ref{eq:rankine-hugeniot}) across these waves. Like the
approximate HLLC solver for the Euler equations (discussed in Section~\ref{chap:riemann-hllc}), the
actual speed of the velocities needs to be estimated, for which the Eigenvalues of the conservation
law are used, as they describe the velocities of the characteristics (see
Chapter~\ref{chap:riemann}). Explicitly, the solver is given by

\begin{align}
	\F(\U_L, \U_R) &=
		\frac{ \lambda^{+} \F_{L} - \lambda^{-} \F_{R} +  \lambda^{+}
\lambda^{-} (\U_R - \U_L)}{ \lambda^{+} - \lambda^{-} }
\label{eq:riemann-HLL} \\
    \lambda^+ &= \max(0, \lambda_{max})\\
    \lambda^- &= \min(0, \lambda_{min})
\end{align}

Here, $\lambda^{max}$ and $\lambda^{min}$ are the minimum and the maximum of the Eigenvalues of the
Jacobian matrix $\frac{\del \F(\U)}{\del \U}$. It turns out that the Eigenvalues can be described
using only two parameters, the angle between the flux and the interaction interface, $\theta$, and
the reduced flux $\mathbf{f} = \Fbf / (cE)$. Rather then continuously computing them on-the-fly for
every interaction, they are tabulated\footnote{
A program to produce the tables of the Eigenvalues depending on $\theta$ and $\mathbf{f}$ is
provided on \url{https://github.com/SWIFTSIM/swiftsim-rt-tools}.
}
and interpolated. Note that the HLL solver predicts the same solution of the Riemann problem as the
GLF solver for $\lambda^{max} = c$, $\lambda^{min} = -c$ (compare eqs.~\ref{eq:riemann-GLF} and
\ref{eq:riemann-HLL}). The interest in the HLL solver is mainly because while it is more expensive
than the GLF solver, it was shown \citep{ramses-rt13, gonzalezHERACLESThreedimensionalRadiation2007}
to be less diffusive and is thus better suited to form shadows more correctly than the GLF solver.
The influence of the Riemann solvers are discussed in Sections~\ref{chap:results-transport} and
\ref{chap:Iliev3}.

\subsection{Third step: Thermochemistry}

In this final step, we solve for the interaction between photons and the gas. The equations to be
solved are:

\begin{align}
	\DELDT{E_i}  &=
		- \sum\limits_{j}^{\absorbers} n_j \sigma_{i j} c E_i + \dot E_i ^{rec}
\label{eq:thermochemistry-E} \\
	\DELDT{\Fbf_i} &=
		- \sum\limits_{j}^{\absorbers} n_j \sigma_{i j} c \Fbf_i \label{eq:thermochemistry-F}
\end{align}

The source term for recombination radiation, $\dot{E_i}^{rec}$, is added in
eq.~\ref{eq:thermochemistry-E} for completeness. In this work, the emission of recombination
radiation is neglected, and I use the approximate ``Case B recombination'' rates instead, which
assumes that all emitted recombination radiation is immediately re-absorbed by the gas, even in
optically thin regimes. However, once the emission of recombination radiation is treated explicitly
in the future, it will be added and solved for in this operator splitting step.

The actual thermochemistry is solved using the \grackle library
\citep{smithGrackleChemistryCooling2017}. This involves the evolution of the individual species
number densities, as well as the heating of the gas. The number densities of individual species
change over time due to processes like photo-ionization, collisional ionization, and recombinations.
These processes depend on the current gas temperature as well as the photo-heating and photo-ionizing rates which are determined by the present radiation field. Currently \GEARRT supports the ``6 species network'', composed of H$^0$, H$^+$, He$^0$, He$^+$, He$^{++}$, and $e^{-}$, provided by \grackle.

\grackle requires us to provide it with the (total) photo-heating rate $\mathcal{H} =
\oldsum_{i,j} \mathcal{H}_{ij}$ and the individual photo-ionization rates $\Gamma_{j} = \oldsum_i
\Gamma_{i,j}$ for each species $j$ and photon group $i$. It then solves the network of equations of
thermochemistry for the 6 species and evolves the internal energy of the gas over some time step
$\Delta t$. More precisely, \grackle takes into account the following interactions:

\begin{itemize}
 \item Collisional ionization of ionizing species H$^0$, He$^0$, and He$^+$
 \item Collisional excitation cooling of ionizing species H$^0$, He$^0$, and He$^+$
 \item Recombination of ions H$^+$, He$^+$, and He$^{++}$
 \item Bremsstrahlung cooling of all ionized species
 \item Compton cooling/heating off the Cosmic Microwave Background
 \item Photo-ionization and photo-heating.
\end{itemize}

While \grackle evolves the state of the gas, we need to take care of the removal of the absorbed
radiation energy over the thermochemistry time step manually.
We solve this again using a simple first order forward Euler integration:

\begin{align}
\frac{E_i (t + \Delta t) - E_i(t)}{\Delta t} =
    -\overline{\epsilon}_i N_i(t) c \ \sum_j \sigma_{i j} ^ N n_j
\end{align}

A minor complication is that the number density of the ionizing species $n_j$ is not constant over
the time step $\Delta t$, as we are currently in the process of ionizing the gas. This may lead to
scenarios where the absorption rate is overestimated by not accounting for the ionization during
the thermochemistry time step. To account for this effect, we take the average absorption rate over
the time step instead:

\begin{align}
    \frac{E_i (t + \Delta t) - E_i}{\Delta t}
        = -\overline{\epsilon}_i \ N_i(t) c \ \sum_j \sigma_{i j} ^ N  \
    \frac{n_j(t + \Delta t) + n_j(t)}{2}
\end{align}

The final equation used to remove absorbed photons from the radiation field is:

\begin{align}
E_i (t + \Delta t)
    &= E_i(t) - \overline{\epsilon}_i N_i(t) c \ \sum_j \sigma_{i j} ^ N  \
    \frac{n_j(t + \Delta t) + n_j(t)}{2} \ \Delta t \\
    &= E_i(t) \left(1 - c \ \sum_j \sigma_{i j} ^ N  \
    \frac{n_j(t + \Delta t) + n_j(t)}{2} \ \Delta t \right)
\end{align}

The same is applied to the photon fluxes:

\begin{align}
\Fbf_i (t + \Delta t)
    &= \Fbf_i(t) \left(1 - c \ \sum_j \sigma_{i j} ^ N  \
    \frac{n_j(t + \Delta t) + n_j(t)}{2} \ \Delta t \right)
\end{align}

\section{Dynamic Sub-Cycling}\label{chap:dynamic-sybcycling}

Many radiation hydrodynamics solver codes \citep[e.g.][]{rosdahlSPHINXCosmologicalSimulations2018,
kannanAREPORTRadiationHydrodynamics2019} use a sub-cycling approach to solve the radiation transport
and thermochemistry using several dozens or hundreds of time steps over the duration of a single
hydrodynamics time step. This approach is motivated by the vastly different typical time step sizes
of hydrodynamics and radiation. The maximal permissible time step size is determined by their
respective CFL conditions (eq.~\ref{eq:rt-cfl} for radiation, and eq.~\ref{eq:meshless-cfl} for
hydrodynamics), which in turn each depend on the signal velocities of the respective conserved
quantities. Radiation, traveling at the speed of light of $\sim 3 \times 10^5$ km/s, usually has a
propagation velocity which is several orders of magnitude higher than fluid velocities, which in
the context of galaxies can reach orders of $\sim 10^2 - 10^3$ km/s, but is typically lower. This
also leads to the radiation time step sizes being similarly orders of magnitude smaller than
hydrodynamics time step sizes.

Enforcing the hydrodynamics and the radiative transfer to be solved using equal time step sizes
restricts the hydrodynamics time steps to much smaller sizes, which are not strictly necessary. So
the basic idea behind sub-cycling is, written in terms of the homogenized hydrodynamics operator $H$
and the source terms coming from radiation $S$, to modify the solution using equal time step sizes
$\Delta t$

\begin{align}
    \U(t + \Delta t) = S(\Delta t) \circ H(\Delta t)[\U(t)]
\end{align}

and to instead solve

\begin{align}
    \U(t + \Delta t) =
        S(\Delta t/n) \circ S(\Delta t/n) \circ ... \circ S(\Delta t/n) \circ H(\Delta t)[\U(t)]
\end{align}

where $n$ sub-cycles with time steps $\Delta t/n$ have been computed for a single homogenized
hydrodynamics operator $H$.

\begin{figure}
 \centering
 \includegraphics[width=\textwidth]{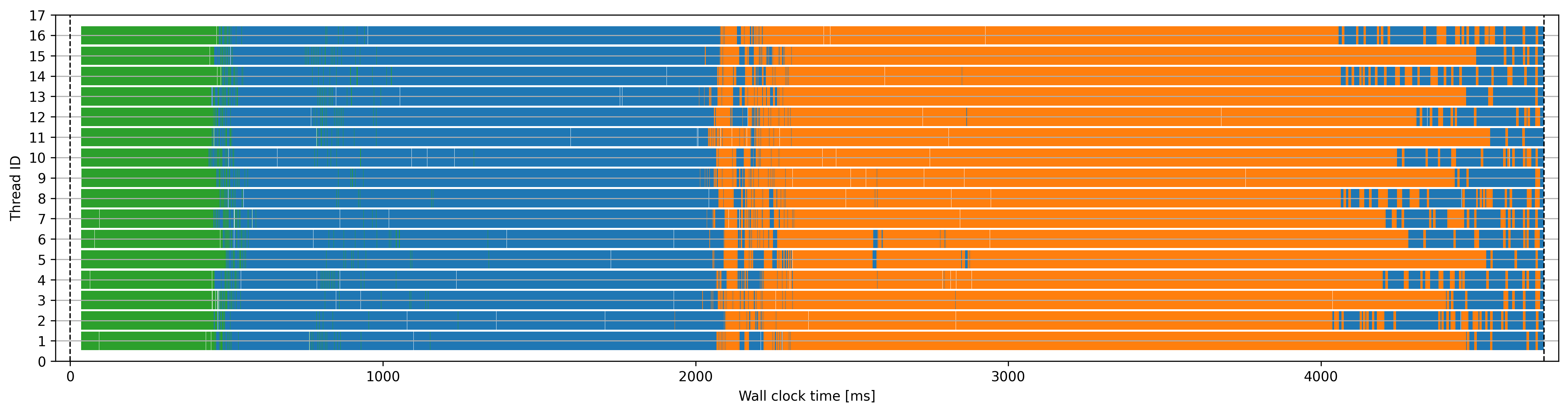}%
 \caption{
One time step of a simulation using \swift with 16 threads. This is the same plot as in
Figure~\ref{fig:taskplot}, but tasks have been colored in differently. Orange blocks show tasks
related to radiative transfer only, blue tasks show the tasks related to hydrodynamics. The
sub-cycling approach allows us to only solve the radiative transfer during each sub-cycle, saving
about half of the total runtime per sub-cycle through omission of the hydrodynamics. Green tasks
are tasks related to the neighbor search, which can also be omitted during RT sub-cycles due to our
choice to treat radiation in a static frame of reference w.r.t. the simulation box (see
Section~\ref{chap:rt-numerics-outline}).
 }
 \label{fig:taskplot-rt-marked}
\end{figure}

\begin{figure}
 \centering
 \includegraphics[width=\textwidth]{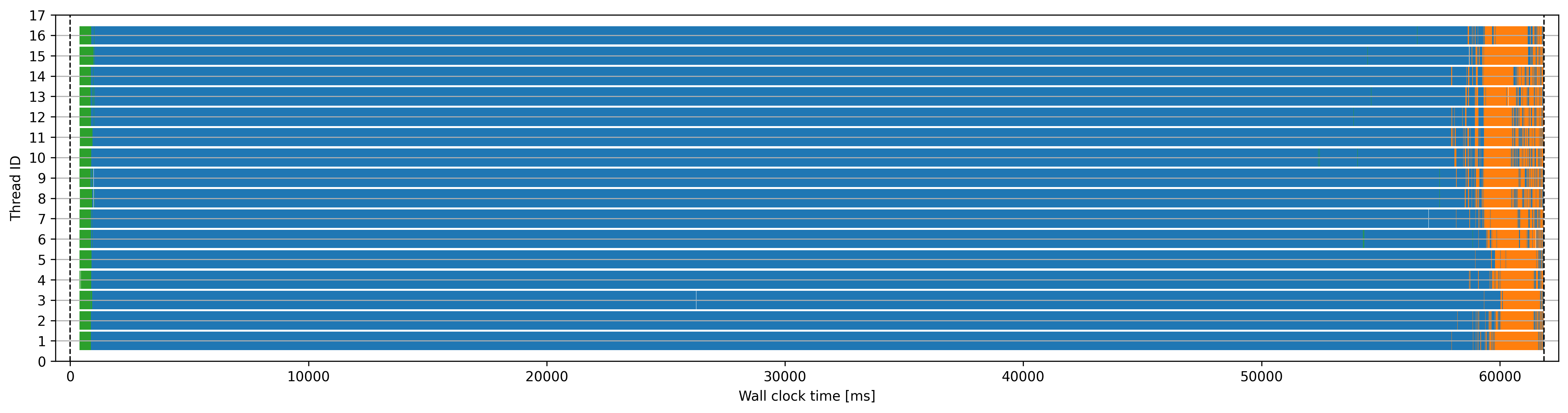}%
 \caption{
One time step of a simulation using \swift with 16 threads. This task plot includes (self-)gravity
in addition to radiation hydrodynamics. Orange blocks show tasks related to radiative transfer
only, blue tasks show the tasks related to hydrodynamics and gravity. Green tasks are tasks related
to the neighbor search, which can also be omitted during RT sub-cycles due to our choice to treat
radiation in a static frame of reference w.r.t. the simulation box (see
Section~\ref{chap:rt-numerics-outline}). The sub-cycling approach allows us to only solve the
radiative transfer during each sub-cycle, saving a huge fraction (the blue and green blocks) of the
total runtime per sub-cycle through omission of the hydrodynamics and gravity.
 }
 \label{fig:taskplot-rt-marked-gravity}
\end{figure}

The sub-cycling approach allows us to save a tremendous workload. To motivate the possible benefits,
Figure~\ref{fig:taskplot-rt-marked} shows the same as in Figure~\ref{fig:taskplot}, tasks being
executed by 16 threads during one step solving radiation hydrodynamics with \swift, with the only
difference that the tasks have been given other colors: Orange blocks show tasks related to
radiative transfer only, blue tasks show the tasks related to hydrodynamics. Green tasks are tasks
related to the neighbor search, which can also be omitted during RT sub-cycles due to our choice to
treat radiation in a static frame of reference w.r.t. the simulation box (see
Section~\ref{chap:rt-numerics-outline}). Figure~\ref{fig:taskplot-rt-marked} shows that by skipping
unnecessary hydrodynamics updates by using sub-cycles, we could save roughly half the computational
expense per RT time step. The gains increase even further when other physics are involved.
Figure~\ref{fig:taskplot-rt-marked-gravity} shows the same radiation hydrodynamics problem being
solved with the addition of gravity. In this case, the fraction of the runtime occupied by RT tasks
is even smaller, meaning that by skipping needless updates of gravity and hydrodynamics (i.e.
forcibly reducing their time step sizes to match the RT time step sizes), we could save up a
significant amount of runtime. The actual reduction of runtime through the sub-cycling approach will
be de demonstrated in Section~\ref{chap:subcycling-results}.

\begin{figure}
 \centering
 \includegraphics[width=\textwidth]{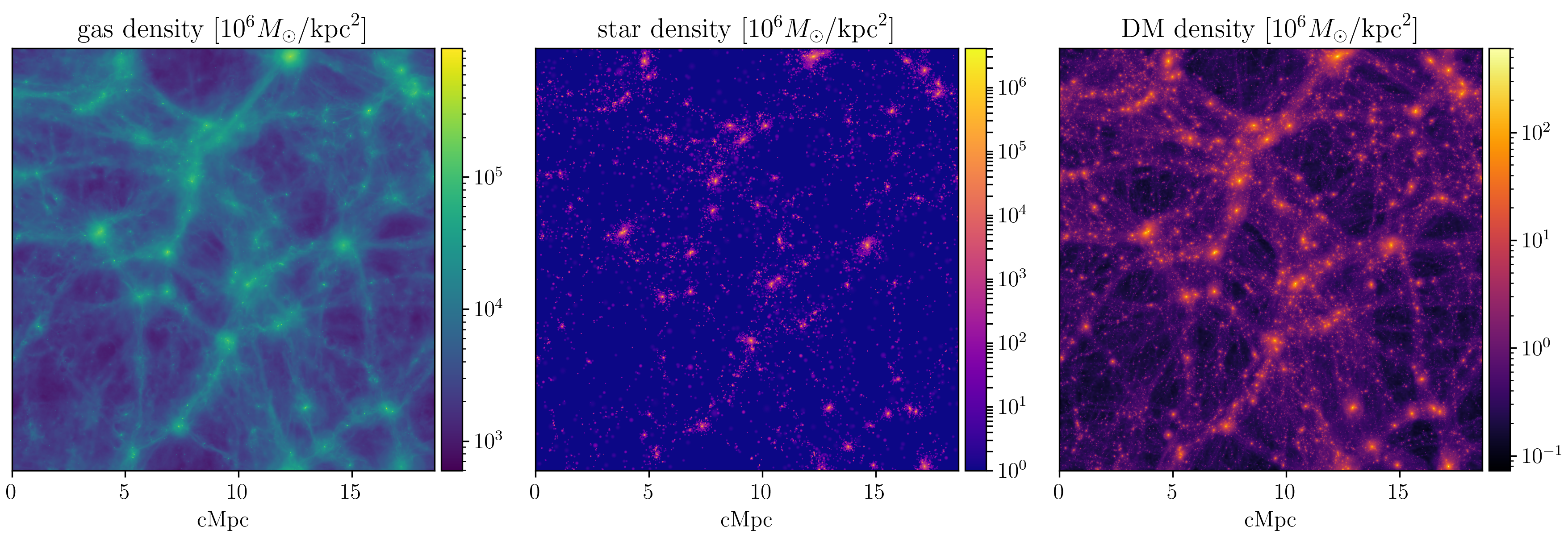}%
 \caption{
Projections along the $z$-axis of the gas mass (left), the stellar mass (center), and dark matter
mass (right) of the simulation used to extract particles' radiation and hydrodynamics time bins
which are shown in Figure~\ref{fig:eagle-timebins-rt}. The simulation uses initial conditions
generated from the redshift $z = 0.1$  snapshots of the EAGLE suite of simulations
\citep{schayeEAGLEProjectSimulating2015}. It has a box size of $\sim 25$ co-moving Mpc and contains
$\sim 52 \times 10^6$ dark matter particles, $\sim 50 \times 10^6$ gas particles, and $\sim 2 \times
10^6$ star particles.\\
The initial conditions are publicly available via the \swift repository under
\url{https://github.com/SWIFTSIM/swiftsim}.
 }
 \label{fig:eagle-25}
\end{figure}

\begin{figure}
 \centering
 \includegraphics[width=\textwidth]{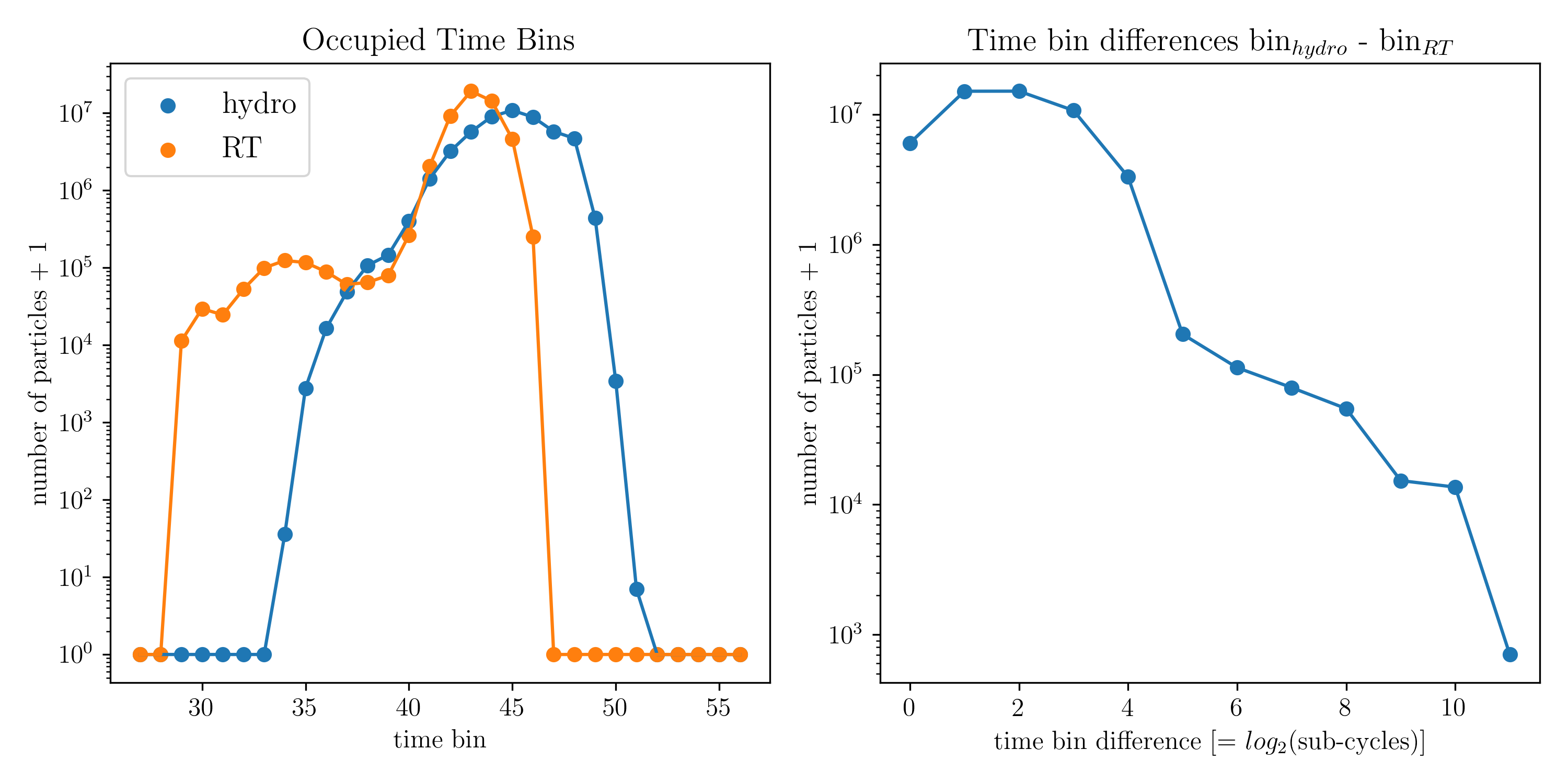}%
 \caption{
\emph{Left}: The distribution of particle time bins for hydrodynamics and radiative transfer at a
single point in time in the simulation described in Figure~\ref{fig:eagle-25}. Particles with a
time bin $n + 1$ have a time step twice the size of particles with time bin $n$.\\
\emph{Right}: The distribution of time bin differences between the hydrodynamics time bin and the
RT time bin of each particle. This difference corresponds to the base 2 logarithm of the number of
sub-cycles each particle would perform. The upper limit of $2^{11} = 2048$ sub-cycles was imposed.
 }
 \label{fig:eagle-timebins-rt}
\end{figure}

An additional point that needs to be discussed with regards to the sub-cycling is how the number of
RT sub-cycles per hydrodynamics step is to be determined. On one hand, it is desirable to have as
many sub-cycles as possible to decrease the total runtime as far as possible. However, forcing
\GEARRT to perform \emph{too many} RT sub-cycles also introduces unnecessary work. For example, say
a particle has a hydrodynamics time step $\Delta t_h = 100$ in arbitrary units, while its RT time
step is $\Delta t_{rt} = 1$. Forcing \GEARRT to use 1000 sub-cycles leads to solving the radiative
transfer with unnecessarily reduced time steps of $\Delta t_{rt} = 0.1$. Clearly that's something
to be avoided. Instead, we allow each particle to set its number of sub-cycles individually
depending on its local conditions, and to change it dynamically after each completed hydrodynamics
step according to its own requirements. For example, a particle may perform 32 sub-cycles during one
hydrodynamics time step, during which the particle get heated and increases its internal energy,
which in turn decreases its hydrodynamics time step size due to the increased sound speed $c_s$
which enters the CFL condition~\ref{eq:meshless-cfl}. To continue the example, in the subsequent
hydrodynamics time step the particle may only require 16 or 8 RT sub-cycles. To further motivate
this approach, the time bins occupied for hydrodynamics and for radiative transfer extracted from a
cosmological simulation are shown in Figure~\ref{fig:eagle-timebins-rt}. (Details on the simulation
are given in the caption of Figure~\ref{fig:eagle-25}.) In order to facilitate a synchronous
integration in time alongside individual particle time step sizes, the time steps are histogrammed
into power-of-two multiples of some minimal system time step size, i.e. $\Delta t_i = 2^n t_{min}$,
where $n$ is usually referred to as the ``time bin'' of the particle (as described in
Section~\ref{chap:individual-timesteps}). Figure~\ref{fig:eagle-timebins-rt} illustrates that there
is a wide range of time step sizes, or rather time bins, for both hydrodynamics and radiative
transfer occurring. Figure~\ref{fig:eagle-timebins-rt} furthermore shows that there is a wide range of
RT sub-cycles occurring, including the maximal permitted value in that example of $2^{11} = 2048$,
which corresponds to a time bin difference of $11$.\footnote{
Arguably a maximal permitted number of RT sub-cycles of $2048$ is a bit unreasonably high. It was
used in Figure~\ref{fig:eagle-timebins-rt} for illustrative purposes. The maximal number of
sub-cycles is a free parameter, and should be determined depending on the actual use case at hand.
It should however probably not exceed 512 or 1024.
}
Notably, there is also a significant amount of particles with equal RT and hydrodynamics time bins,
validating the approach to let particles decide for themselves how many sub-cycles they require,
rather than enforcing a fixed number of sub-cycles for all particles. It should however be noted
that a time step limiter is being employed as well. Following the findings of
\citet{saitohNecessaryConditionIndividual2009} and \citet{durierImplementationFeedbackSmoothed2012},
any two particles that interact with each other are enforced to have a maximal time step difference
of a factor of 4 (or equivalently, a difference in time bins of 2) in order to preserve energy
conservation. This limiter is applied to both the hydrodynamics and to the radiative transfer
solvers individually.

While the fundamental approach of sub-cycling the radiative transfer is fairly straightforward, its
implementation while accounting for individual particle time step sizes and task-based parallelism
used in \swift were a considerable challenge. The details of the implementation are presented in
Section~\ref{chap:subcycling}.

\section{Additional Topics}

\subsection{Dealing With Particle Drifts}\label{chap:rt-drift}

As outlined in Section~\ref{chap:rt-numerics-outline}, while we assume particles are static w.r.t.
the simulated volume in the context of RT, they are drifted for the purposes of hydrodynamics. As a
consequence, we need to make corrections to the radiation fields when a particle is drifted for
hydrodynamics purposes. We do that by simply extrapolating the particle value at the new point given
its current state and the gradients $\nabla \U$ we obtained using the general gradient expression
given in eq~\ref{eq:gradient}. Explicitly, for a particle $i$ and for each component of the
radiation state vector $\U^k$, we first obtain the drift distance over the time $\Delta t_{i,
drift}$ as

\begin{align}
\Delta \x_{i,\text{drift}} &= \V_{i} \Delta t_{i,\text{drift}}
\end{align}

where $\V_i$ is the particle velocity. We then obtain the value corrected at the drifted
location

\begin{align}
\U_{i}^k (\x + \Delta \x_{\text{drift}}) \approx \U_{i}^k (\x) + \nabla \U_{i}^k (\x) \cdot \Delta
x_{i,\nu,\text{drift}}
\end{align}

A caveat for these corrections is that they are  no longer strictly conservative for obvious
reasons: Firstly, we use gradients to extrapolate what the radiation field values at the drifted
position should be. Secondly, the gradients themselves are only approximate and $\mathcal{O}(h^2)$
accurate. Thirdly, there are no further mechanisms to ensure that the resulting radiation field
quantities are actually conserved in total. However, the error introduced is acceptably small, as
will be shown in Section~\ref{chap:validation-drift-corrections}, where the application of these
corrections is validated.


\subsection{Reducing The Speed of Light}

A huge challenge related to radiative transfer is the very high speed of light compared to the
fluid velocities. Since the signal velocity determines the maximal time step size, the speed of
light leads to minute time step sizes for the radiative transfer, several orders of magnitude
smaller than those for the hydrodynamics. Indeed this discrepancy is the main motivation to solve
radiative transfer with in a sub-cycling manner, which is discussed in
Section~\ref{chap:subcycling}.

In an attempt to reduce the computational expense associated with these tiny time step sizes,
\cite{ramses-rt13} employ the Reduced Speed of Light Approximation (RSLA), which was introduced by
\cite{gnedinMultidimensionalCosmologicalRadiative2001}. I also adapt the RSLA with \GEARRT. The
approach consists of globally reducing the speed of light everywhere by some factor $f_c$, i.e.
replace the speed of light $c$ in all equations involved with radiative transfer by

\begin{align}
    \tilde{c} = f_c c
\end{align}

Naturally, this means that the radiation will propagate (much) slower than it actually should.
However, that does not necessarily mean that the propagation of ionization fronts will be slowed
down as well. \cite{gnedinMultidimensionalCosmologicalRadiative2001} argue that as long as the
reduced speed of light remains much higher than the gas velocity, the Newtonian limit (in which the
speed of light is infinite) is maintained, and the evolution of ionized fronts is not hindered.
\cite{ramses-rt13} provide a framework to estimate the smallest permissible $f_c$ based on the
light crossing time in spherical ionized regions around a single ionizing source, also known as
``Str\"omgren spheres''. For example, they estimate that for the interstellar medium modeled with
a number density of $n = 10^{-1}$ cm$^-3$ and an ionizing source of $2 \times 10^{50}$ ionizing
photons per second, $f_c$ may be $\sim 10^{-2}$.

Note that the reduced speed of light $\tilde{c}$ is not only used in the photon transport, but also
to compute the photo-heating and photo-ionization rates (eq.~\ref{eq:photoheating-group} and
\ref{eq:photoionization-group}). This can be understood when considering the photo-heating and
photo-ionizations as binary collisions between photons and photo-absorbing particles, which are
treated as targets in this scenario. Now consider a photon packet, like a top hat function,
traveling at the reduced speed $\tilde{c}$. In the binary collision scenario, the photons are
projectiles being propelled at the targets with their velocity $\tilde{c}$, and as discussed in
Section~\ref{chap:coupling-to-hydrodynamics}, the interaction rates are directly proportional to
that velocity. For $\tilde{c} < c$, this means that the interaction rates, and thus the
photo-ionization rates, will also be decreased. However, since the photon packet travels at a
reduced
speed, it will spend a longer time passing through the location of the targets, which in turn again
increases the total number of photo-ionization events. So keeping the speed of light at its actual
physical value in the interaction rates while decreasing it for the photon transport would lead to
an over-ionization over time, which is what \cite{ocvirkImpactReducedSpeed2019} have also found.

\subsection{Ion Mass Fluxes}

Hydrodynamics with Finite Volume Particle Methods exchange mass fluxes $\Delta m$ between particles.
This means that we need to pay attention to the individual ionizing and ionized species' mass
fractions being exchanged. The mass of each individual species being exchanged needs to be traced
individually, and depending on the direction: If a particle loses mass, then it loses mass and
species mass fractions according to its own mass fractions. If it gains mass during an exchange,
then it gains species mass fractions according to the interaction partner's mass fractions.

The usual convention for a mass flux between particles ``$i$'' and ``$j$'' is to treat $i$ as the
``left'' particle and $j$ as the ``right'' particle in an analogy to cells. Let $X_k$ denote the
mass fraction of each species $k$. During each \emph{hydrodynamics} flux exchange, for each species
$k$ the total mass flux $\Delta x_{i,k}$ is being accumulated:

\begin{align}
\Delta x_{i,k} &= \sum_j \Delta m_{ij} X_{ij,k}\\
\text{where }\quad X_{ij,k} &=
	\begin{cases}
		X_{i,k} \quad \text{ if } \quad \frac{\Delta m_{ij}}{\Delta t} > 0 \\
		X_{j,k} \quad \text{ if } \quad \frac{\Delta m_{ij}}{\Delta t} < 0
	\end{cases}
\end{align}

The masses which particles carry are updated during the kick operation, during which the individual
species' masses $x_{i,k}$ are also evolved. For a particle $i$ with mass $m_i$, the masses of
individual species at the beginning of a step are given by:

\begin{align}
    x_{i,k}^{t} &= m_i X_{i,k}^{t}
\end{align}

During the kick operation, they get updated as follows:

\begin{align}
    x_{i,k}^{t + \Delta t} &= x_{i,k}^{t} + \Delta x_{i,k}
\end{align}

Finally, after the update, the new mass fractions of the species can be obtained using

\begin{align}
X_{i,k}^{t + \Delta t} &= \frac{x_{i,k}^{t + \Delta t}}{\sum_k x_{i,k}^{t + \Delta t}}
\end{align}

\subsection{Creating Collisional Ionization Equilibrium Initial Conditions}

At the beginning of a simulation, \GEARRT offers the possibility to generate the ionization mass
fractions of the gas particles assuming the gas is in collisional ionization equilibrium, composed
of hydrogen and helium, and that there is no radiation present. In order to determine the
ionization mass fractions of all species (H$^0$, H$^+$, He$^0$, He$^+$, He$^{++}$) for a given
specific internal energy $u$, an iterative procedure needs to be applied because the gas variables
are interconnected in a circular manner. To explain this circular dependence, we make use of the
(unitless) mean molecular weight $\mu$, which is commonly used to express the average particle mass
$\overline{m}$ of a gas:

\begin{align}
    \overline{m} = \mu m_u
\end{align}

where $m_u$ is the atomic mass unit. To begin, we note that the ionization state of the gas
influences the mean molecular weight $\mu$. Consider a medium with $j$ different elements of atomic
mass $A_j$ with a mass fraction $X_j$ and a number of free electrons $E_j$. Then the number density
of the gas is given by

\begin{align}
n = \frac{\rho}{\mu m_u}
    = \sum_j
    \underbrace{\rho X_j}_{\text{density fraction of species } j} \times
    \underbrace{\frac{1}{A_j m_u}}_{\text{mass per particle of species}} \times \underbrace{(1 +
    E_j)}_{\text{number of particles per species}} \label{eq:number-density-MMW}
\end{align}

where we neglect the mass contribution of electrons. In the case for hydrogen and helium, the values
of $A_j$ and $E_j$ are shown in Table \ref{tab:mass-and-electron-numbers}.
Eq.~\ref{eq:number-density-MMW} simplifies to

\begin{align}
\frac{1}{\mu} = \sum_j \frac{X_j}{A_j} (1 + E_j)
\end{align}

Specifically, ionization changes the mass fractions $X_j$ of the species, and therefore also the
mean molecular weight $\mu$. In turn, the mean molecular weight determines the gas temperature at a
given specific internal energy. Using the equation of state for ideal gases using the pressure $p$,

\begin{align}
    p = n k T = \frac{\rho}{\mu m_u} k T
\end{align}

and the expression for the specific internal energy  $u$,

\begin{align}
    u = \frac{1}{\gamma - 1} \frac{p}{\rho}
\end{align}

the gas temperature $T$ = $T(u, \mu)$ is given by

\begin{align}
    T = u (\gamma - 1) \mu \frac{m_u}{k}
\end{align}

Lastly, the gas temperature determines the collisional ionization and recombination rates, which
need to be balanced out by the correct number density of the individual species in order to be in
ionization equilibrium, i.e. at a state where the ionization and recombination rates exactly cancel
each other out. We take the ionization and recombination rates from
\citet{katzCosmologicalSimulationsTreeSPH1996}, which are given in Table
\ref{tab:coll-ion-rates-katz}. For a gas with density $\rho$, hydrogen mass fraction $X_H$ and
helium mass fraction $X_{He} = 1 - X_H$, the total number densities of all hydrogen and helium
species are

\begin{align}
    n_H &= X_H \frac{\rho}{m_u} \\
    n_{He} &= X_{He}  \frac{\rho}{4 m_u}
\end{align}

and in equilibrium, the number densities of the individual species are given by

\begin{align}
n_{H^0} &= n_H \frac{A_{H^+}}{A_{H^+} + \Gamma_{H^0}} \\
n_{H^+} &= n_H - n_{H^0} \\
n_{He^+} &= n_{He} \frac{1}{1 + (A_{He^+} + A_d) / \Gamma_{He^0} + \Gamma_{He^+} / A_{He^{++}}} \\
n_{He^0} &= n_{He^+} \frac{A_{He^+} + A_d}{\Gamma_{He^0}} \\
n_{He^{++}} &= n_{He^+} \frac{\Gamma_{He^+}}{A_{He^+}}
\end{align}

To summarize, the tricky bit here is that the number densities determine the mean molecular weight,
the mean molecular weight determines the temperature of the gas for a given density and specific
internal energy, while the temperature determines the number densities of the species through the
ionization rates. To find the correct mass fractions, the iterative Newton-Raphson root finding
method is used. Specifically, using some initial guesses for temperature and mean molecular
weights, $T_{guess}$ and $\mu_{guess}$, in each iteration step we determine the resulting specific
internal energy

\begin{align}
u_{guess} = k T_{guess} / (\gamma - 1) / (\mu_{guess} m_u)
\end{align}

The function whose root we're looking for is

\begin{align}
    f(T) = u - u_{guess}(T)  = 0
\end{align}

which has the derivative

\begin{align}
\frac{\del f}{\del T} =
    - \frac{\del u}{\del T} (T = T_{guess}) =
    \frac{k}{(\gamma - 1) / (\mu_{guess} m_u )}
\end{align}

The specific internal energy of the gas $u$ is fixed and provided by the initial conditions. We now
look for the $T$ at which $f(T) = 0$. The Newton-Raphson method prescribes to find the $n+1$th
$T_{guess}$ using

\begin{align}
    T_{n+1} = T_n + \frac{f(T_n)}{\frac{\del f}{\del T}(T_n)} \ .
\end{align}

\begin{figure}
 \centering
 \includegraphics[width=.8\textwidth]{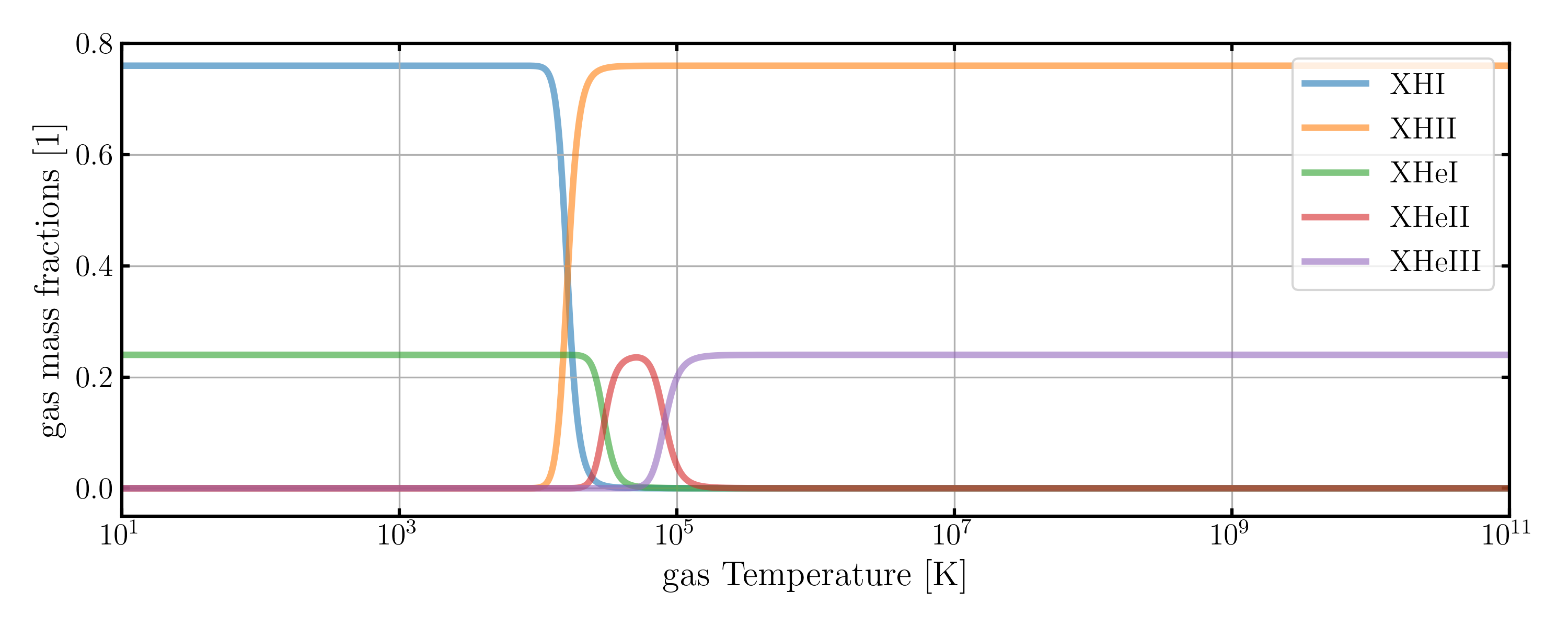}
 \caption{
Mass fractions of hydrogen and helium ion species in collisional ionization equilibrium  as
computed by \GEARRT for a wide range of temperatures $T$, corresponding to specific internal
energies of the gas between $10^9$ erg/g and $10^{20}$ erg/g, specified by the initial conditions.
The initial composition of the gas consists of 76\% hydrogen and 24\% helium by
mass.
 }
 \label{fig:ionization-equilibrium}
\end{figure}

During each iteration, the new mass fractions and the resulting mean molecular weight given the
latest guess for the temperature are computed. At the start, the first guess for the temperature
$T_{guess}$ is computed assuming a fully neutral gas. Should that gas temperature be above $10^5$
K, the first guess is changed to a fully ionized gas. The iteration is concluded once $f(T) \leq
\epsilon = 10^{-4}$. The results \GEARRT provides for a wide range of temperatures is shown in
Figure~\ref{fig:ionization-equilibrium}.

\input{tables/RHD/thermochemistry_rates}

\input{tables/RHD/mass_and_electron_numbers}

\section{Implementation in SWIFT}\label{chap:rt-implementation}

\subsection{Task Dependency Graph for Radiative Transfer}

Just as for the hydrodynamics in Chapter~\ref{chap:meshless-implementation}, the task-based
parallelism used by \swift requires the manual definition of the individual tasks and dependencies
that solve the numerical scheme. In this section, we do this for radiative transfer with \GEARRT.

\begin{figure}
 \centering
 \includegraphics[width=.8\textwidth]{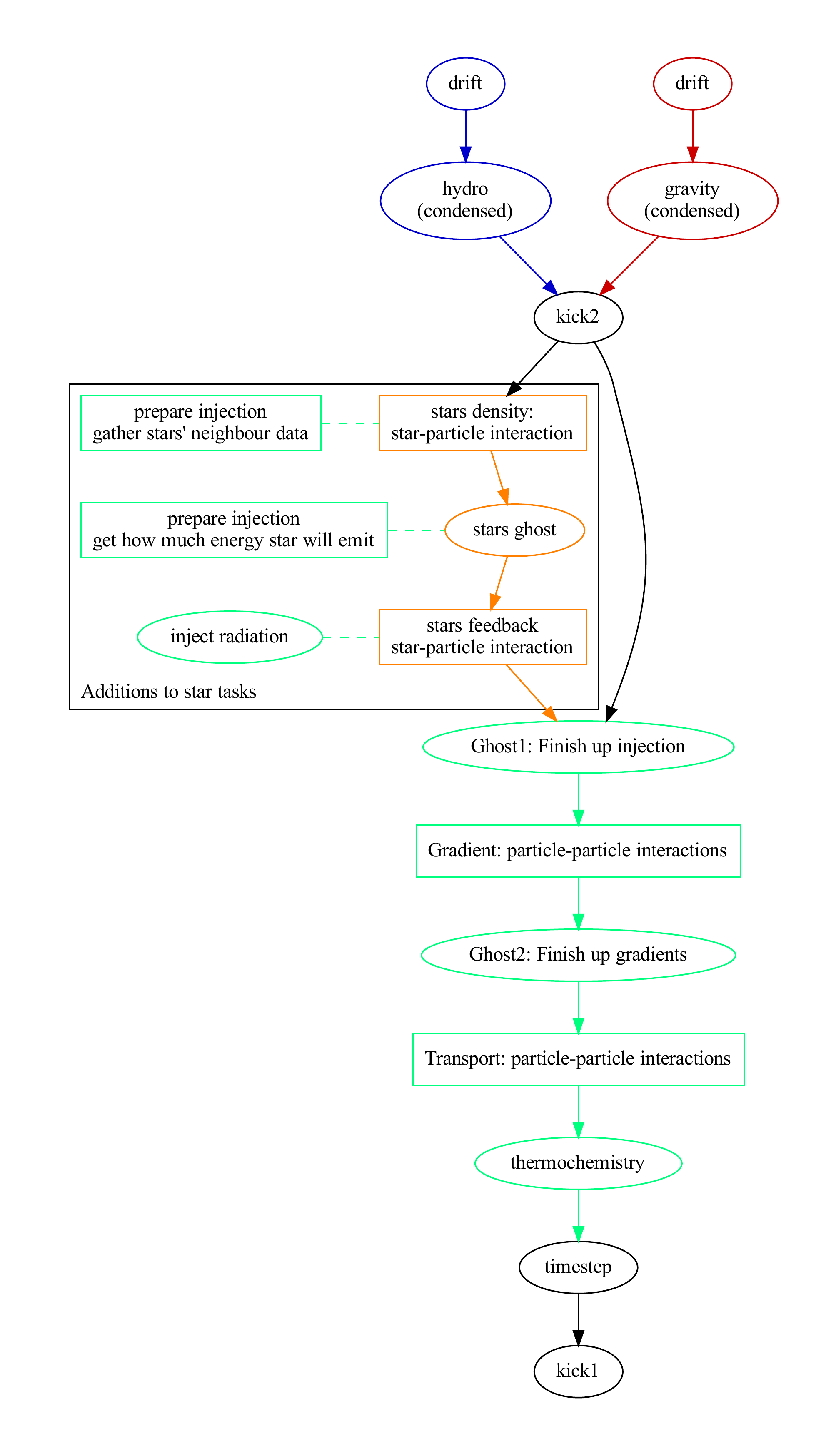}%
 \caption{
A simplified task dependency graph of the tasks required for RT in \swift. For the sake of clarity,
the tasks required for the hydrodynamics and gravity have been condensed into a single node each.
Nodes with round boundaries represent \lingo{plain} type tasks, while nodes with rectangular
edges represent \lingo{interaction} type tasks.
 }
 \label{fig:RTtaskplot-simplified}
\end{figure}

The tasks required for RT will be added to the already existing tasks which are required for the
hydrodynamics. A simplified sketch is shown in Figure~\ref{fig:RTtaskplot-simplified}. The full
task dependency graph for hydrodynamics is shown in Figure~\ref{fig:dependency-graph-hydro}. As
mentioned before, the radiative transfer will take place \emph{after} the hydrodynamics have been
solved, or explicitly after the ``\lingo{kick2}'' task in Figure~\ref{fig:dependency-graph-hydro}.

The first step of RT, namely the injection of radiation energy density from stellar sources into
particles, requires two interaction loops for star particles. The first loop constitutes a neighbor
search for stars, during which the stars' smoothing lengths and neighbor lists of gas particles are
established. We make use of this first loop to accumulate the octant weights $w_a$ (which are
discussed in Section~\ref{chap:injection-step}) in order to be able to compute the weight
corrections $\mu_a$ (eq.~\ref{eq:isotropy-correction-with-zero}) during the second star-gas particle
interaction loop. The actual injection of energy density into particles then occurs during this
second star-gas particle interaction loop.

Other stellar feedback processes in \swift require the same two interaction loops for the exactly
same reasons. As the injection of radiation energy density is in principle independent from these
feedback processes, we can conveniently re-use these tasks, i.e. add the injection of radiation to
them. Additionally, the \lingo{star ghost} task, which is required between the two interaction
loops for stars for the same reasons it is necessary after the neighbor search (``\lingo{density}'')
loop for hydrodynamics, can be used to compute the amount of energy each star needs to inject this
step based on their current luminosity and time step sizes.

After the injection through the star-gas particle interaction loops, a first RT ghost task
(``\lingo{rt\_ghost1}'') is added to finish any necessary work after the accumulation of injected
radiation, and to collect all dependencies before proceeding further down the dependency graph.
Since not all gas particles need to have a neighboring star particle, some gas particles may not
undergo the injection process. For this reason, an additional dependency from \lingo{kick2} to the
\lingo{rt\_ghost1} task is necessary (see Figure~\ref{fig:RTtaskplot-simplified}).

The next step in the RT scheme consists of the photon transport step, which needs to be done in two
parts. First, a gas-gas particle interaction loop is necessary to determine the gradients of the
radiation quantities, which are required for the flux exchanges. The flux exchanges are then
performed in a second gas-gas particle interaction loop, named the ``\lingo{transport}'' loop.
Between the \lingo{gradient} and the \lingo{transport} loop, a second ghost task,
``\lingo{rt\_ghost2}'', is required to finish the gradient computations after the accumulation of
sums in the \lingo{gradient} loop is completed.

The ``\lingo{thermochemistry}'' task solves the final step in the RT scheme, which consists of the
actual thermochemistry. All the required work is on individual particles, which can be modeled as a
\lingo{plain} type task.

\begin{figure}
 \centering
 \includegraphics[angle=90,height=.9\textheight]{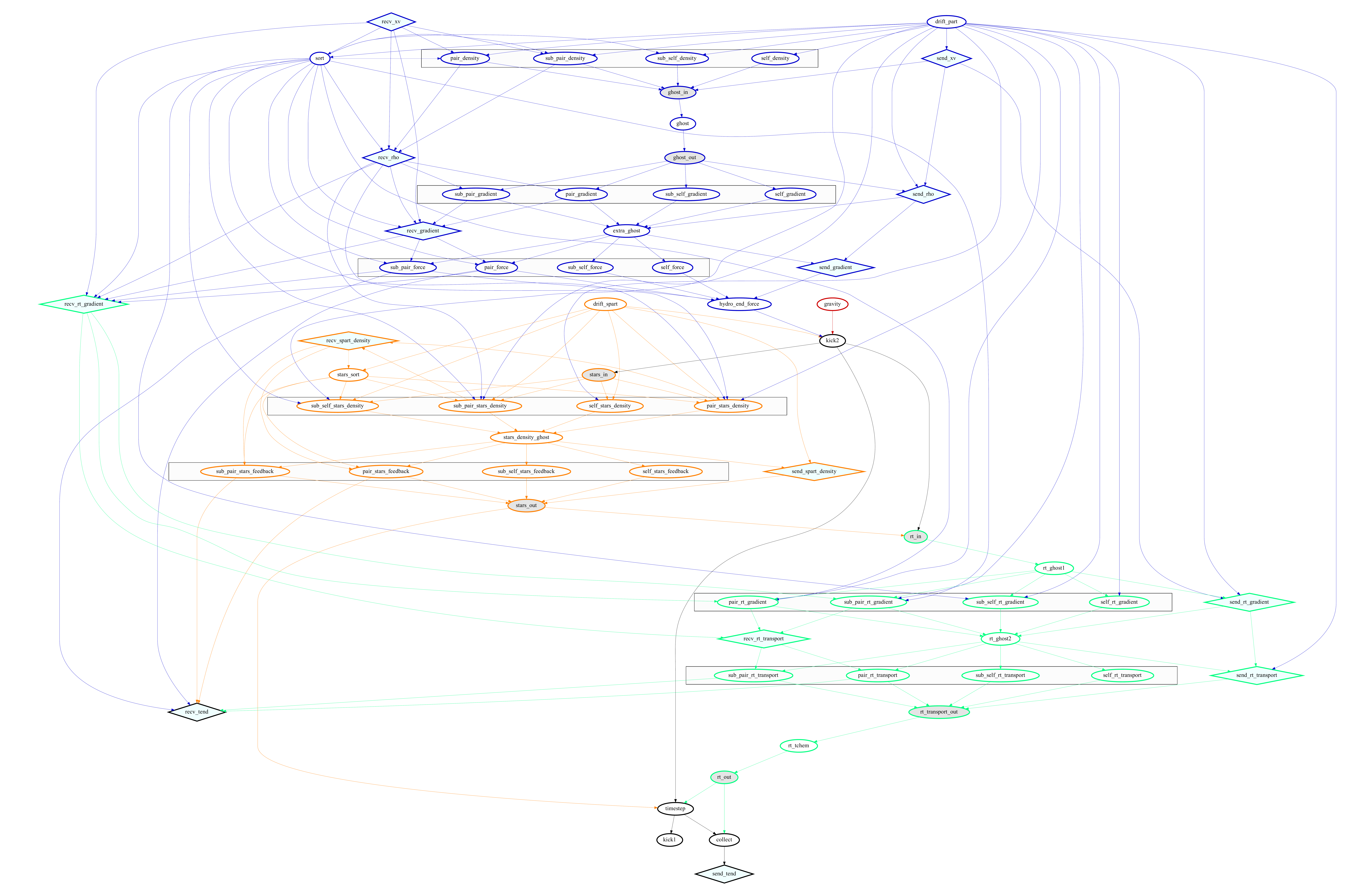}%
 \caption{
The full task dependency graph required for radiative hydrodynamics with \swift. Gravity has been
condensed into a single task.
 }
 \label{fig:RTtaskplot-nosubcycling}
\end{figure}

Figure~\ref{fig:RTtaskplot-nosubcycling} shows the full task dependency graph for radiation
hydrodynamics with \GEARRT and \swift. The interaction loops have been expanded to depict all four
types of \lingo{interaction} type tasks, which are the \lingo{self}, \lingo{sub\_self},
\lingo{pair}, and \lingo{sub\_pair} tasks, for both the star interaction tasks
(``\lingo{stars\_density}'' and ``\lingo{stars\_feedback}'', in orange) and the RT interaction tasks
(``\lingo{rt\_gradient}'' and ``\lingo{rt\_transport}'', in green). Additionally, the MPI
communication tasks have been added. Just as was the case for hydrodynamics, the current state of
the particles needs to be sent and received before each interaction loop, using the respective
``\lingo{send\_rt\_gradient}'', ``\lingo{recv\_rt\_gradient}'' and ``\lingo{send\_rt\_transport}'',
``\lingo{recv\_rt\_transport}'' tasks. Dependencies between the sending tasks and the receiving
tasks are necessary, respectively, in order to assure the order of messages arriving being correct,
and to prevent race conditions since the order at which messages arrive are not guaranteed by MPI.
Finally, the first interaction loop of both the star related task block (orange) and the RT related
task block (green) each need explicit dependencies on the \lingo{drift} and \lingo{sort} tasks of
the hydrodynamics block. The reason is that both \lingo{active} and \lingo{inactive} gas particles
are drifted, after which they need to be sorted along the neighboring cells' axes for an optimized
access during interaction loops. \lingo{Inactive} particles however will not go through any of the
other hydrodynamics related (blue) tasks in that step, and all the dependencies inherited from the
hydrodynamics related task will not be available. Therefore the additional dependencies are
necessary to ensure particles will already be drifted and sorted before they are accessed in the
star and the RT block.

\subsection{Sub-Cycling}\label{chap:subcycling}

\begin{figure}
 \centering
 \includegraphics[width=.6\textwidth]{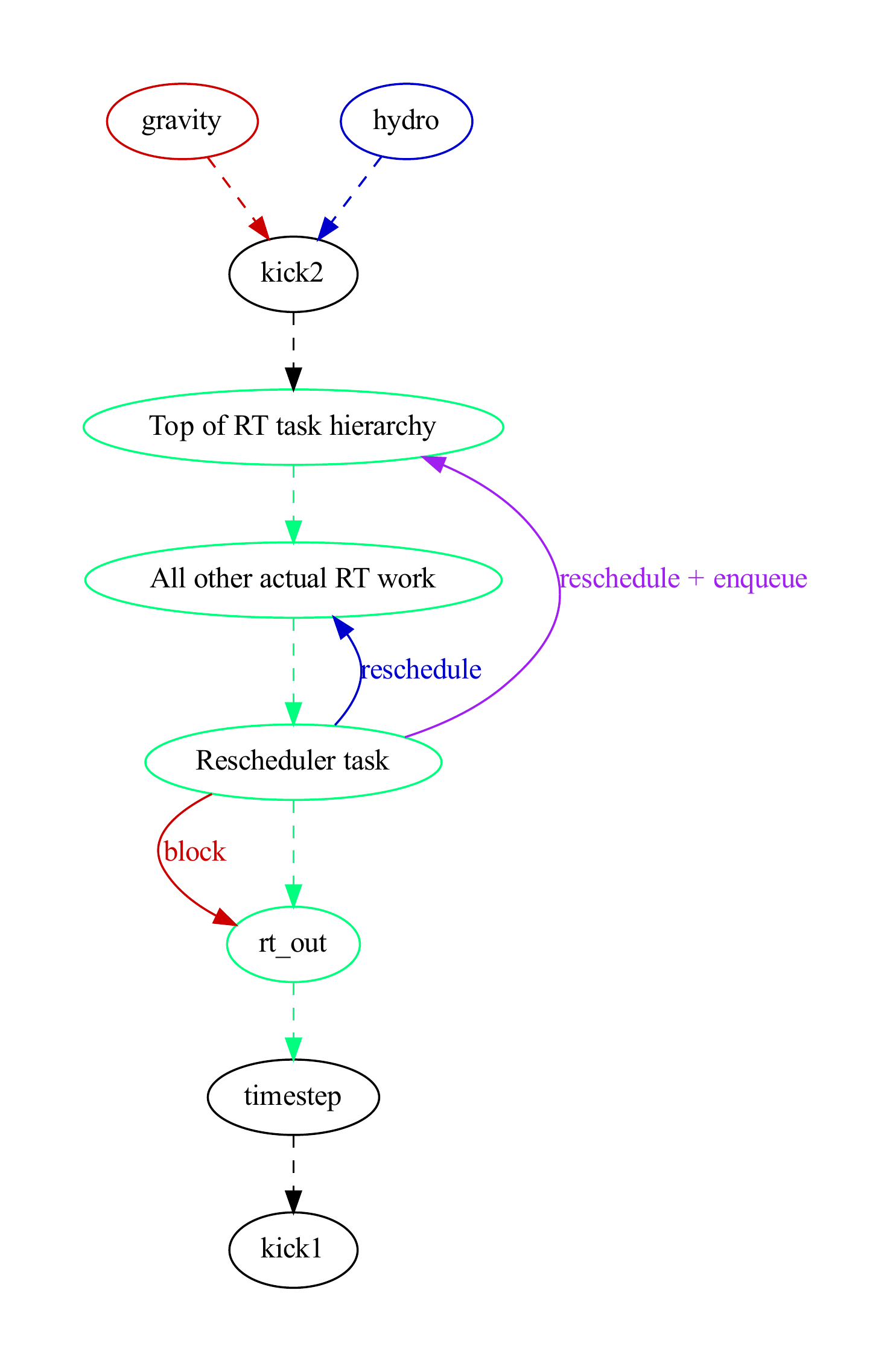}%
 \caption{
 Illustration of the rescheduling approach for sub-cycling.  A new ``\lingo{rescheduler}'' task is
added, which blocks tasks beyond the scope of the RT block to proceed until the correct number of
sub-cycles has been completed. Instead, it sets the entire RT block of tasks to the correct state
to be executed again, and directly enqueues the task at the root of the RT hierarchy so the
subsequent sub-cycle can immediately start.
 }
 \label{fig:rescheduling}
\end{figure}

On the algorithmic side, sub-cycling (previously introduced in
Sections~\ref{chap:rt-numerics-outline} and \ref{chap:dynamic-sybcycling}) essentially consists of
repeating the entire RT block of tasks over and over again. This can be achieved in several ways.
The simplest solution is probably to repeatedly create the entire RT block of tasks, and concatenate
them correctly such that they are executed in sequence. By generating the entire RT block of tasks
$N$ times, up to $N$ sub-cycles can be executed. This is however not a very efficient solution. On
one hand, the scheduler would need to spend a lot of time fetching tasks that do the same thing over
and over again and adding them to the queues. Secondly, this approach can get quite expensive in
memory. A single RT block of tasks consists of two interaction loops with other cells, which
averages on 13 \lingo{pair} type tasks per loop per cell. In addition, there is a \lingo{self}-type
task per interaction loop, there are the two \lingo{ghost} tasks, the \lingo{thermochemistry} task,
and the MPI communication tasks, totaling 33 tasks per cell per RT block. If we needed several
hundred sub-cycles for thousands of cells, the memory expense and the associated overhead would rise
prohibitively high.

Since each sub-cycle requires the same tasks to be executed in the same order, an obvious
improvement would be to re-use the existing tasks of a singe RT block over and over again. This
could be done in the following manner, which is illustrated in Figure~\ref{fig:rescheduling}. A new
``\lingo{rescheduler}'' task is added, which has three functions. Firstly, it must block tasks
beyond the scope of the RT task block to be enqueued and proceed until the correct number of
sub-cycles has been completed. Secondly, it needs to set the entire RT block of tasks to the correct
state to be executed again. Finally, it must and directly enqueue the task at the root of the RT
hierarchy so the subsequent sub-cycle can start.

\begin{figure}
 \centering
 \includegraphics[width=\textwidth]{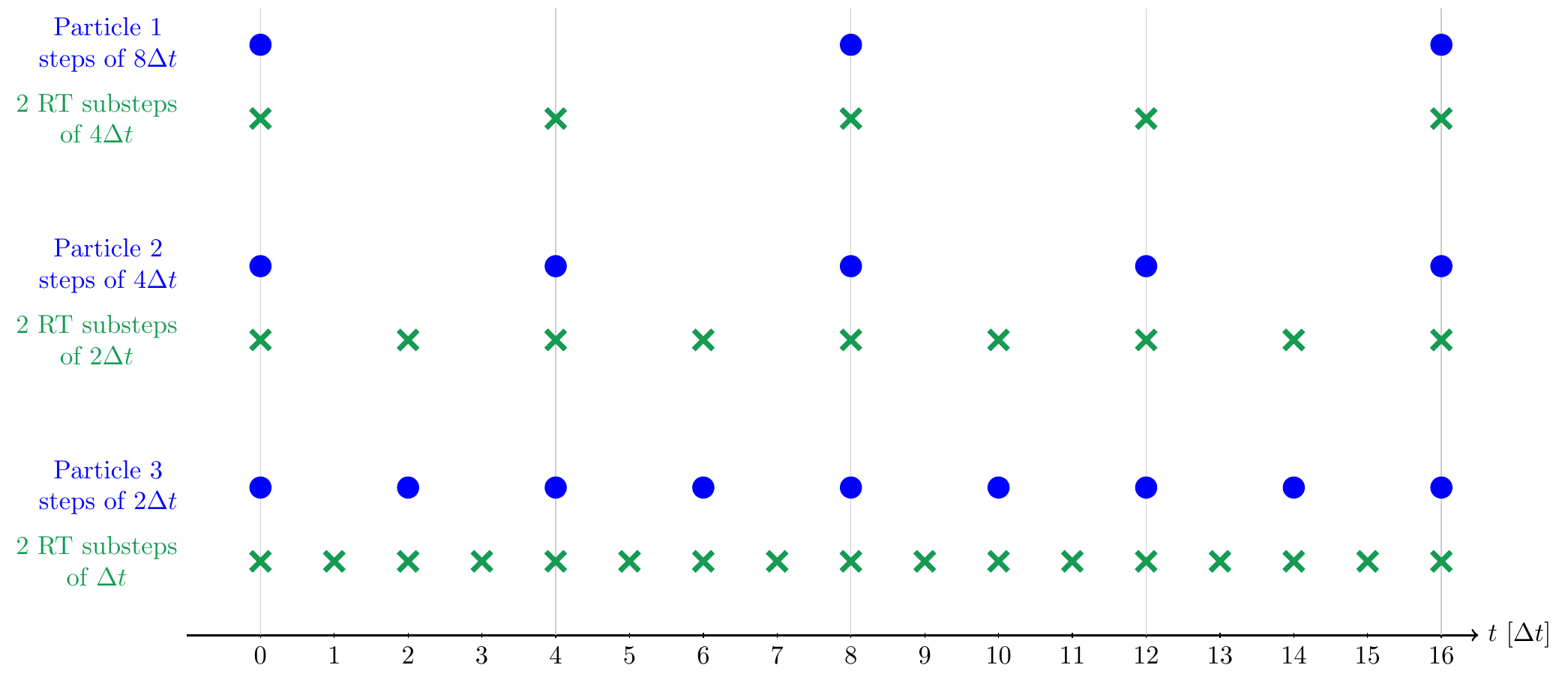}%
 \caption{
The evolution of three particles with different individual time step sizes and two RT sub-cycles
each. Along the $x$-axis, the time at which the entire simulation has been advances is shown in
units of some minimal time step size $\Delta t$. Blue circles represent times at which the
particles do a hydrodynamics update. Green crosses represent the times of RT updates. Note how some
RT updates during sub-cycles overlap with global simulation times where other particles have their
regular hydrodynamics update.
 }
 \label{fig:two-subcycles-per-step}
\end{figure}

\begin{figure}
 \centering
 \includegraphics[width=\textwidth]{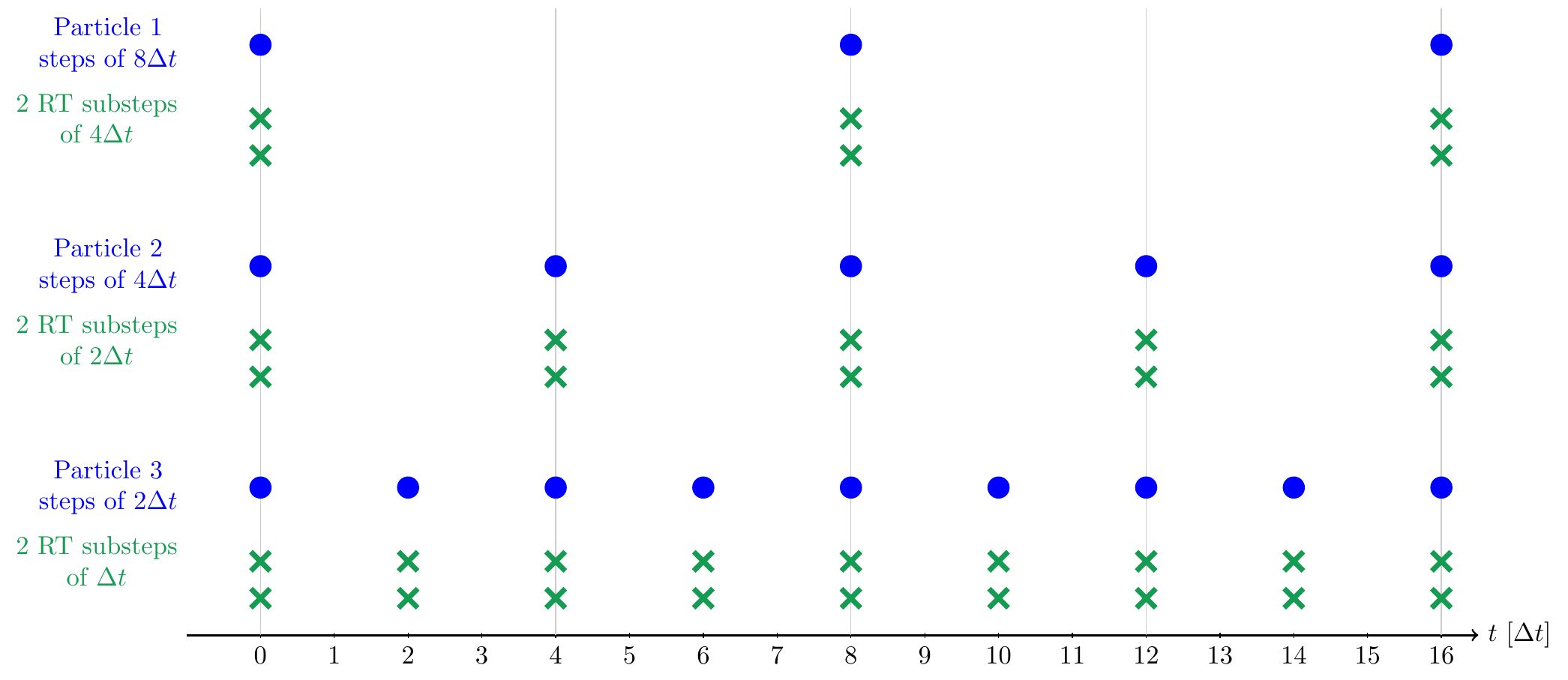}%
 \caption{
The evolution of three particles with different individual time step sizes and two RT sub-cycles
each using the rescheduling approach. Along the $x$-axis, the time at which the entire simulation
has been advances is shown in units of some minimal time step size $\Delta t$.  Blue circles
represent times at which the particles do a hydrodynamics update. Green crosses represent the times
of RT updates. Due to the nature of the re-scheduling approach, all sub-cycles of any particle must
be executed in the same global simulation step as the hydrodynamics update, even if they get
integrated in time further than the global simulation time at which they are updated. This leads to
synchronicity issues, which are shown more clearly in Figure~\ref{fig:rescheduling-problem}.
 }
 \label{fig:rescheduling-reality}
\end{figure}

A working proof of concept version of the sub-cycling through a \lingo{rescheduler} task has
successfully been developed. Unfortunately, due to the individual time step sizes of particles a
sub-cycling method using a \lingo{rescheduling} task is severely limited. To demonstrate the issues
that arise, consider the simple case where for each hydrodynamics update we do only two RT
sub-cycles. Each RT sub-cycle is then performed over exactly half the hydrodynamics time step size.
The evolution of three particles with different individual time step sizes and two RT sub-cycles
each is shown in Figure~\ref{fig:two-subcycles-per-step}. The case illustrated in
Figure~\ref{fig:two-subcycles-per-step} is actually how we want the sub-cycling to proceed: The RT
updates are performed at the appropriate global simulation times $t$. Most importantly, the RT
updates which coincide with the hydrodynamics updates of other particles with smaller hydrodynamics
time step sizes on the global simulation time axis are solved simultaneously.
The comparison of the time at which RT and hydrodynamics of different particles is solved is
important since the hydrodynamics for all particles needs to be solved \emph{before} the particle
can proceed with its RT updates. So the hydrodynamics update constitutes a barrier to proceed with
further RT updates. Consider for example the second RT update of Particle 2 and the second
hydrodynamics update of Particle 3 in Figure~\ref{fig:two-subcycles-per-step} should both occur on
the global simulation time $t = 2 \Delta t$. The second RT update of Particle 2 and the third RT
update of Particle 3 need to happen at the same time. For this to be possible, Particle 3 needs to
have completed its second hydrodynamics update.
However, this is not what a re-scheduling approach can offer. How the RT sub-cycles would be
executed with a rescheduling approach is shown in Figure~\ref{fig:rescheduling-reality}. The problem
is the fact that the rescheduling approach requires all sub-cycles to be completed in the same
global simulation step as the hydrodynamics. The rescheduling approach can only repeat all the tasks
which have been run in the current simulation step, it can not and must not advance the global
simulation time. Otherwise, it would introduce anachronisms: Some particles would be integrated
further in time than their neighbors without correctly accounting for the interactions with the
neighbor particles. This leads to very incorrect results. For example, radiation wouldn't propagate
past the particles which are \lingo{active} in that particular simulation step.

\begin{figure}
 \centering
 \includegraphics[width=\textwidth]{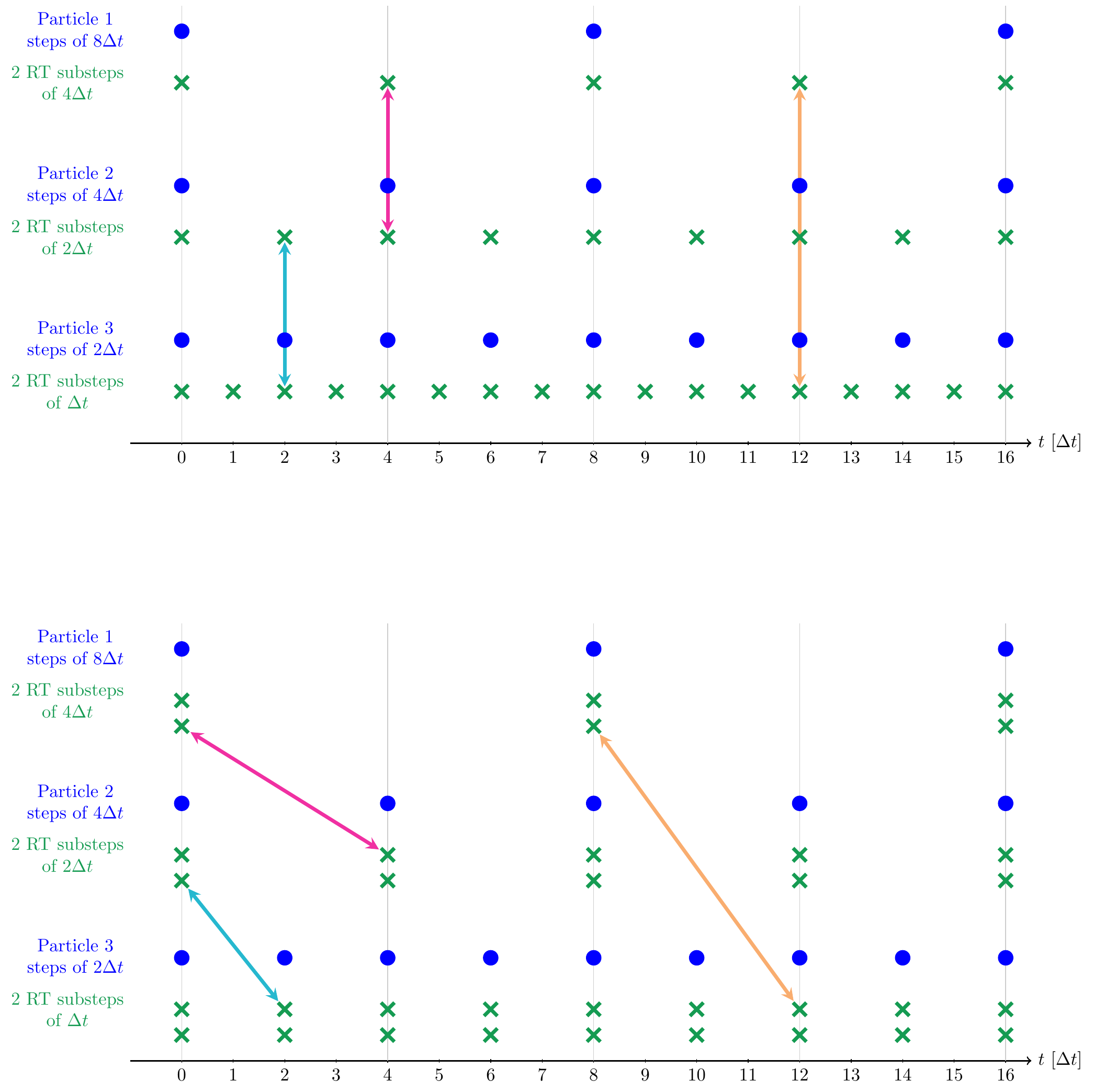}%
 \caption{
The evolution of three particles with different individual time step sizes and two RT sub-cycles
each using the rescheduling approach. Along the $x$-axis, the time at which the entire simulation
has been advances is shown in units of some minimal time step size $\Delta t$. Blue circles
represent times at which the particles do a hydrodynamics update. Green crosses represent the times
of RT updates. \\
On the top, the intended way of sub-cycling RT updates w.r.t hydrodynamics steps is shown. This is
the same as Figure~\ref{fig:two-subcycles-per-step}, except that three RT updates which should
occur simultaneously are highlighted through arrows.\\
Bottom: How the sub-cycling with a rescheduling approach would proceed. This is the same as in
Figure~\ref{fig:rescheduling-reality}, except that the same updates as in the top row are
highlighted with arrows of the same color as above. For example, the magenta arrow connects the
second RT update of Particle 1 with the third RT update of Particle 2 in both plots. The problem
with the rescheduler approach is that RT updates like the highlighted ones cannot be performed
simultaneously. On one hand, all sub-cycles must be completed before the simulation step is
finished, as in the bottom plot. But the data these highlighted RT updates require won't be
available until the simulation progresses to a further step.
 }
 \label{fig:rescheduling-problem}
\end{figure}

This has the consequence that data which a sub-cycling RT update requires will not be available at
the time it needs to be executed. To make use of the previous example, the second RT update of
Particle 2 in Figure~\ref{fig:two-subcycles-per-step} would need to happen at global simulation time
$t = 0$ (right after the first RT update), while the third RT update of Particle 3 is blocked by the
second hydrodynamics update, which would still occur at global simulation time $t = 2 \Delta t$.
However, these two RT updates should be happening simultaneously, as illustrated in
Figure~\ref{fig:two-subcycles-per-step}.
To illustrate the issue, both the case of how sub-cycling is supposed to work as well as how the
rescheduling approach would execute it are shown in Figure~\ref{fig:rescheduling-problem}. On the
top of Figure~\ref{fig:rescheduling-problem}, the intended way of sub-cycling RT updates w.r.t
hydrodynamics steps is shown. This is the same as Figure~\ref{fig:two-subcycles-per-step}, except
that three RT updates which should occur simultaneously are highlighted through arrows. At the
bottom is how the sub-cycling with a rescheduling approach would proceed. This is the same as in
Figure~\ref{fig:rescheduling-reality}, except that the same updates as in the top row are
highlighted with arrows of the same color as above. For example, the magenta arrow connects the
second RT update of Particle 1 with the third RT update of Particle 2 in both plots. The problem
with the rescheduler approach is that RT updates like the highlighted ones cannot be performed
simultaneously. On one hand, all sub-cycles must be completed before the simulation step is
finished, as in the bottom plot. But the data these highlighted RT updates require won't be
available until the simulation progresses to a further step.

The two requirements to finish all sub-cycles during a single simulation step and to have all data
available updated to the correct time unfortunately contradict each other. This can only be resolved
by abandoning individual time step sizes for particles, which is a necessity in order to be able to
run cosmological simulations to completion. As such, the rescheduling approach can't be used for
adequate sub-cycling, and we need to turn to a different solution.

In order to facilitate sub-cycling alongside the use of individual time step sizes, a secondary
time marching scheme needed to be developed. To be more precise, a regular simulation step in
\swift always performs the following operations:

\begin{itemize}
 \item Activate all tasks that are attached to a cell which is active in this step
 \item Launch the scheduler and execute all tasks according to the task dependency graph
\end{itemize}

Each RT sub-cycle needs to perform these operations as well. However, during each sub-cycle, only
tasks from the RT block are \lingo{activated} and executed. In
Figure~\ref{fig:two-subcycles-per-step}, such a simulation sub-cycle step corresponds to e.g. the
update of Particle 3 at $t = \Delta t$ and $t = 3 \Delta t$, where nothing besides an RT step is
required. In addition, the sub-cycling scheme needs to be able to work concurrently alongside a
normal step. Take for example the case shown in Figure~\ref{fig:two-subcycles-per-step} at $t = 2
\Delta t$. While Particle 3 has a full hydrodynamics and RT update at that time, Particle 2 only
does an RT update. So the regular simulation step and the sub-cycling step need to be intertwined in
order to facilitate this functionality. This means that there are two classes of problems to solve:
First, sub-cycling steps between regular simulation steps need to be enabled. Second, a sub-cycling
step needs to be able to be performed during a regular simulation step.

Let's look at the sub-cycling steps between regular steps first. To facilitate this functionality,
we can exploit some peculiarities of the radiative transfer. For example, the RT doesn't drift
particles, and the signal velocity is always the (reduced) speed of light. This has the consequence
that during all sub-cycling steps, the particle's RT time step sizes always remain constant. In
other words, we know exactly how many sub-cycling steps we need to perform between two regular
steps, and can simply perform all the required sub-cycling steps in a loop. Each loop consists of
first \lingo{activating} the tasks attached to cells that contain particles which require an RT
update in that sub-cycle, and then launching the scheduler to execute all the \lingo{active} tasks
according to their respective dependencies. This launching procedure is essentially repeatedly
running regular steps for time steps determined by the sub-cycling steps, where only RT related
tasks are being \lingo{activated} and executed. One difference is a modified task dependency graph
(the modifications will be discussed further below) so the tasks related to radiative transfer can
be executed without any other tasks, in particular without hydrodynamics tasks and without the first
MPI communication in the dependency graph (\lingo{send\_xv} and \lingo{recv\_xv}). A second
difference is that cells containing particles now need to keep track of additional time and time
integration related variables for radiative transfer in order to perform the time marching and
integration correctly. This will also be discussed in more detail further below. The launching
procedure by itself however is essentially the same as for a regular step.

Before looking into the specific case where a sub-cycle's time coincides with a regular step's
time, one minor addition to the task dependency graph and to the tracking of time related variables
of cells are already necessary, as mentioned above. In order to facilitate the correct task
\lingo{activation}, cells need to store both the minimal time step size of any particle they
contain, as well as the next smallest time any particle within them will be active. Cells having
this data readily available is a general requirement for any physics, not something specific to RT
or sub-cycling. However, in order to decouple the integration times and time steps of radiative
transfer from other physics, the RT time steps and time step sizes need to be traced separately
from other now.

\begin{figure}
 \centering
 \includegraphics[angle=90,height=.9\textheight]{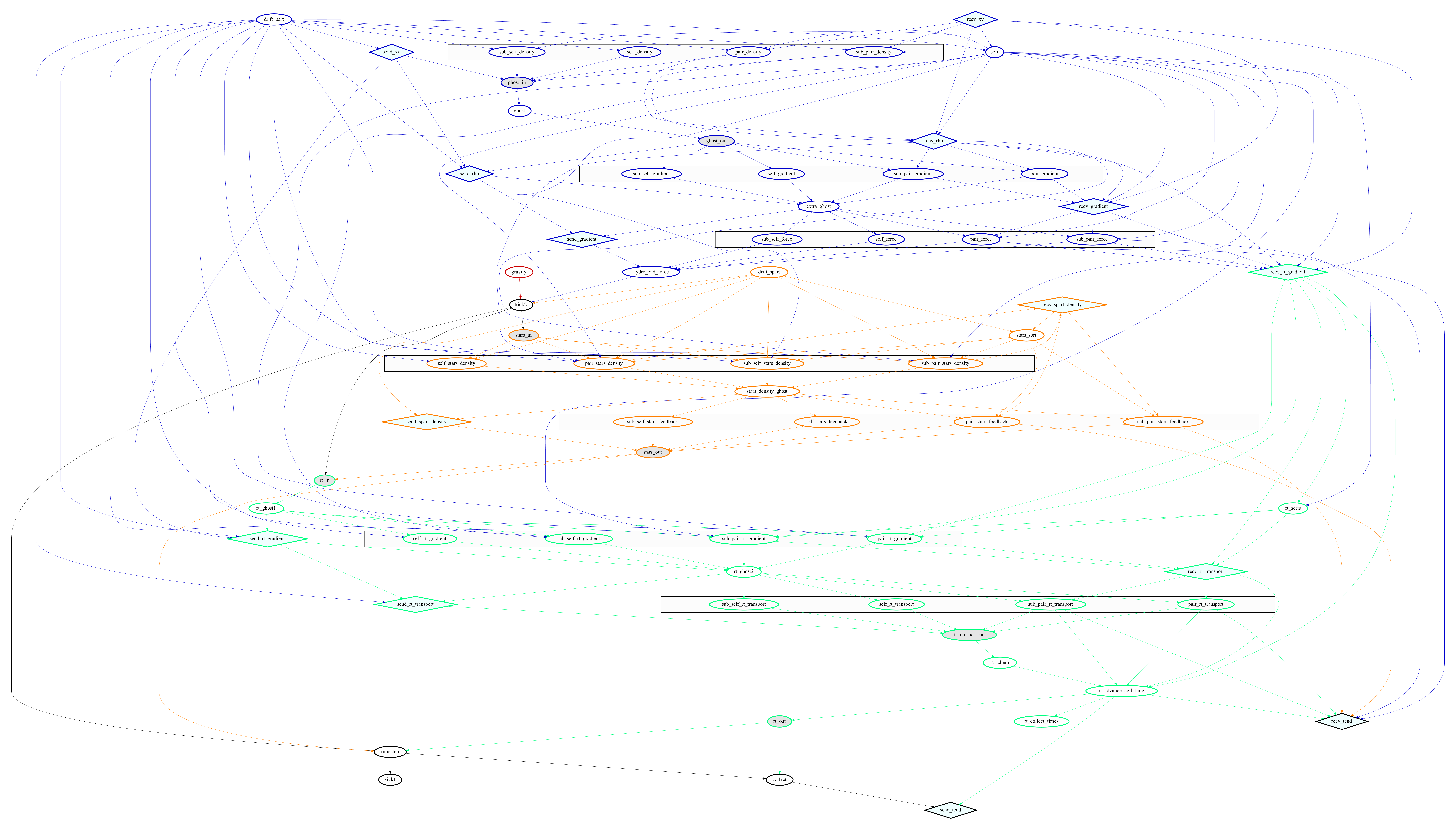}%
 \caption{
The full task dependency graph required for radiative hydrodynamics and sub-cycling with \swift.
Gravity has been condensed into a single task. To update cell times correctly, the
\lingo{rt\_advance\_cell\_time} and \lingo{rt\_collect} tasks needed to be added. Since with the
sub-cycling a cell may be drifted during a main step for a variety of reasons, the particles may not
have been sorted due to no particle being active in that cell in the current step. For this reason,
an additional \lingo{rt\_sorts} task has been added, which is only activated if no regular
\lingo{sort} task has been run in the current regular step.
 }
 \label{fig:RTtaskplot}
\end{figure}

At the end of a step (both a regular step and a sub-cycle), i.e. after all tasks have been executed
and all active particles have been updated, the cell information needs to be updated accordingly as
well to store the up-to-date minimal time step size within them and the next smallest time any
particle within them will be \lingo{active}. In regular simulation steps, this is done in
\lingo{timestep} tasks, where the new particle time step sizes are also being computed according to
the CFL condition (eq.~\ref{eq:meshless-cfl} for hydrodynamics, and eq.~\ref{eq:rt-cfl} for RT).
Since the \lingo{timestep} tasks aren't executed during sub-cycles, a replacement task named
``\lingo{rt\_advance\_cell\_time}'' needs to be added. Note that \lingo{rt\_advance\_cell\_time}
tasks only update the cell metadata on the time step sizes they contain, not the particles within
the cells, as the time step sizes of particles don't change during sub-cycles. Similarly, a
replacement for \lingo{collect} tasks, which propagate the same time step metadata of a cell from
the \lingo{super level} to the \lingo{top level} in the cell tree hierarchy, is necessary, and has
been added as ``\lingo{rt\_collect}'' tasks. The  \lingo{rt\_advance\_cell\_time} and
\lingo{rt\_collect} tasks are shown in the final task dependency plot in
Figure~\ref{fig:RTtaskplot}.

Now let's turn to the case where a sub-cycling step overlaps with a regular simulation step.
Specifically, this means that during a regular simulation step some cells only execute the RT block
of tasks, while others do other physics as well. Complications arise here because some key
assumptions in \swift's task dependency scheme are violated. In particular, it is assumed that a
regular step in the context of hydrodynamics always starts with particle drifts, after which
particles are being sorted along the axes of active neighbor cells. If a cell is foreign, its root
of the task dependency graph starts with a \lingo{recv\_xv} task, where the (drifted) particle
positions are being received from the ``\lingo{real}'' cell. After the particle data is received,
the cell is sorted along the axes of active neighbor cells. All this doesn't take place if the cell
in question is currently doing a sub-cycle, not a regular simulation step. As particles are not
drifted in RT sub-cycles, there is no need to sort them. In the case of a sub-cycle being executed
during a regular step, the root of the task dependency graph for a cell starts with either RT tasks,
or star tasks, provided there are neighboring cells which contain star particles. In other words,
the crucial \lingo{recv\_xv} and \lingo{sort} tasks are skipped for \lingo{foreign} cells.

The missing \lingo{recv\_xv} tasks are actually not a very big problem. Since communications in
\swift always send all particle data, the first communication in the RT block, the
\lingo{recv\_rt\_gradient} tasks, will ensure that up-to-date data is available. However, the
missing \lingo{sort} task needs to be compensated with a new \lingo{rt\_sort} task which runs
after the \lingo{recv\_rt\_gradient} task is completed. Furthermore, the \lingo{rt\_sort} task  is
only \lingo{activated} if the regular \lingo{sort} task isn't. Note that since particles aren't
drifted outside of regular simulation steps, there is no need for sorts outside of sub-cycles which
coincide with regular simulation steps. The full task dependency graph required for \GEARRT which
includes sub-cycling is shown in Figure~\ref{fig:RTtaskplot}.

A final constraint that should be noted is that since the sub-cycling effectively works like a
secondary time stepping machinery, the same restriction for individual time step sizes apply. In
particular, the radiation time step sizes must be a power-of-two fraction of the maximally
permitted time step size. This requirement can be translated to demanding that the number of
RT sub-cycles a particle performs must be a power-of-two fraction of its hydrodynamics time step
size.

%% file: tables/RHD/thermochemistry_rates.tex
\begin{table}
\begin{center}
\begin{tabular}{lll}
\hline\\
$A_{H^+}         $ & 
    $8.4 \times 10^{-11} T^{-1/2} T_3^{-0.2} (1 + T_6^{0.7})^{-1}$ & 
    RR for H$^{+}$ 
\\[.5em]
$A_{d}           $ &
    $1.5 \times 10^{-10} T^{-0.6353}$    & 
    dielectronic RR for He$^{+}$ 
\\[.5em]
$A_{He^+}        $ & 
    $1.9 \times 10^{-3} T^{-1.5} \mathrm{e}^{-470000/T} (1 + 0.3 \mathrm{e}^{-94000/T})$    & 
    RR for He$^{+}$
\\[.5em]
$A_{He^{++}}     $ & 
    $3.36 \times 10^{-10} T^{-1/2} T_3^{-0.2} (1 + T_6^{0.7})^{-1}$    & 
    RR for He$^{++}$
\\[.5em]
\hline\\
$\Gamma_{H^{0}}  $ & 
    $5.85 \times 10^{-11} T^{1/2} \mathrm{e}^{-157809.1/T} ( 1 + T_5^{1/2})^{-1}$& 
    CIR for H$^{0}$
\\[.5em]
$\Gamma_{He^{0}} $ & 
    $2.38 \times 10^{-11} T^{1/2} \mathrm{e}^{-285335.4.1/T} ( 1 + T_5^{1/2})^{-1}$& 
    CIR for He$^{0}$
\\[.5em]
$\Gamma_{He^{+}} $ & 
    $5.68 \times 10^{-12} T^{1/2} \mathrm{e}^{-631515/T} ( 1 + T_5^{1/2})^{-1}$& 
    CIR for He$^{+}$
\\[.5em]
\hline
\end{tabular}
\end{center}
\caption{Temperature ($T$) dependent recombination rates (RR) and collisional ionization rates 
(CIR) for Hydrogen and Helium species, adapted from \citet{katzCosmologicalSimulationsTreeSPH1996}. 
All rates are in units of cm$^3$ s$^{-1}$. $T_n$ is shorthand for $T / 10^n$K.}
\label{tab:coll-ion-rates-katz}
\end{table}

%% file: tables/RHD/mass_and_electron_numbers.tex
\begin{table}
\begin{center}
\begin{tabular}{l|ll}
species  & $A$ & $E$ \\
\hline
H$^0$    & 1 & 0 \\
H$^+$    & 1 & 1 \\
He$^0$   & 4 & 0 \\
He$^+$   & 4 & 1 \\
He$^{++}$  & 4 & 2 \\
\end{tabular}
\end{center}
\caption{Atomic mass numbers $A$ and free electron numbers $E$ for ionization species of Hydrogen 
and Helium.}
\label{tab:mass-and-electron-numbers}
\end{table}

%% file: main/RHD/RHD-4-validation.tex
\chapter{Tests and Validation}\label{chap:rt-validation}

This chapter presents a series of tests of \GEARRT. Section~\ref{chap:results-transport} is centered
around pure radiation transport using the Finite Volume Particle Methods, without any interactions
between radiation and the gas. In section~\ref{chap:results-injection} the various photon flux
injection methods previously outlined in Section~\ref{chap:rt-numerics-outline} are tested.
Sections~\ref{chap:IL6} and \ref{chap:IL9} reproduce standard tests of radiation hydrodynamics set
by the \citet{ilievCosmologicalRadiativeTransfer2006} and
\citet{ilievCosmologicalRadiativeTransfer2009} comparison projects, and the results of \GEARRT are
compared with results obtained by other radiation hydrodynamics codes which participated in the
comparison projects. More precisely, in Section~\ref{chap:IL6}, the radiation transport is tested
alongside the thermochemistry on static gas density fields, where, gas is only permitted to heat and
ionize due to the influence of radiation and any hydrodynamics are neglected. Section~\ref{chap:IL9}
tests the full radiation hydrodynamics. Lastly, section~\ref{chap:subcycling-results} presents the
perfomance gains though use of the sub-cycling.

Many initial conditions used in this Section as well as nearly all presented figures made heavy use
of the \codename{swiftsimio} \citep{borrowSwiftsimioPythonLibrary2020} visualization and analysis
python library.

\section{Radiation Transport}\label{chap:results-transport}

\subsection{Testing Riemann Solvers and Limiters}\label{chap:rt-riemann-limiters}

\begin{figure}
 \centering
 \includegraphics[width=\textwidth]{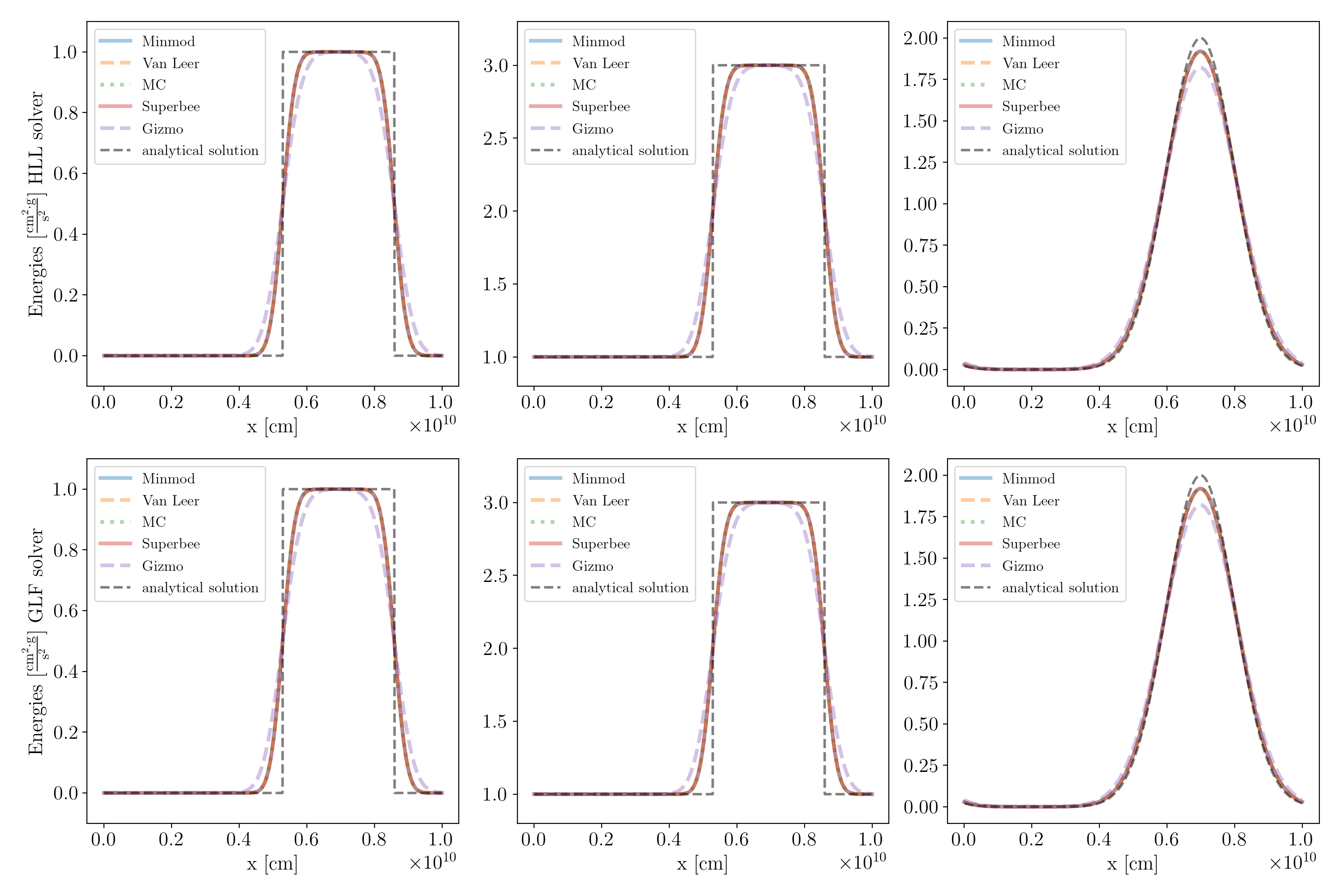}%
 \caption{
Left: a top hat function; Center: a top hat function where the lower value is nonzero; And Right: a
smooth Gaussian initial energy density being advected using the HLL Riemann solver (top row) and
GLF Riemann solver (bottom row) using various flux limiters.
 }
 \label{fig:rt-riemann-limiter-1D}
\end{figure}

\begin{figure}
 \centering
 \fbox{\includegraphics[width=\textwidth]{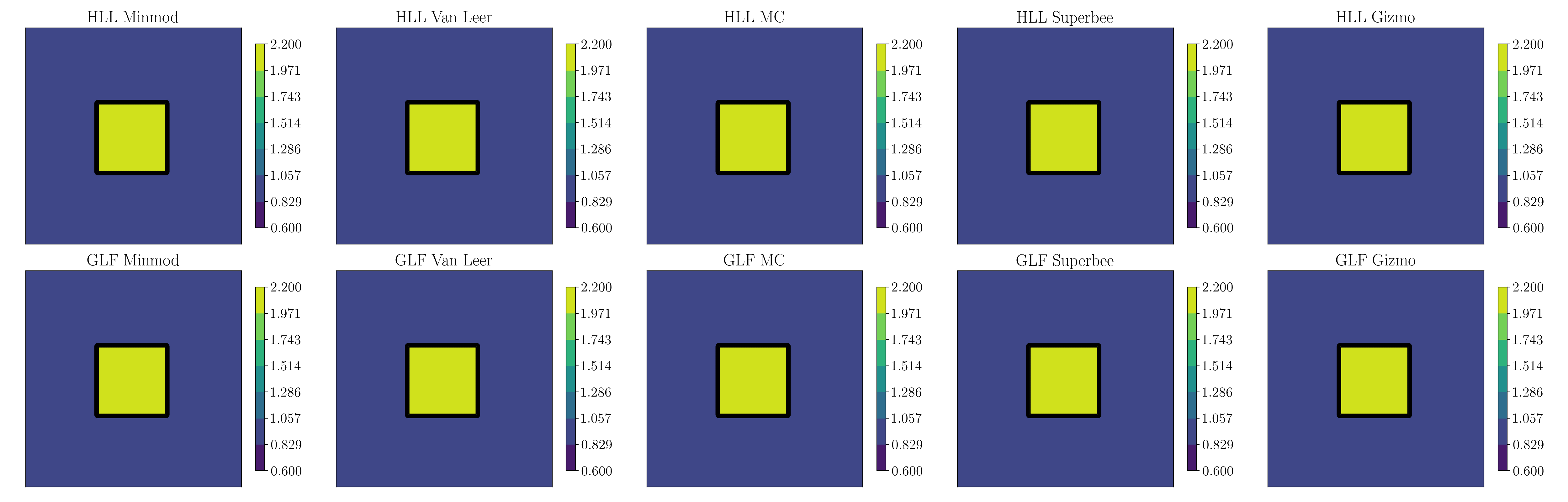}}\\
 \fbox{\includegraphics[width=\textwidth]{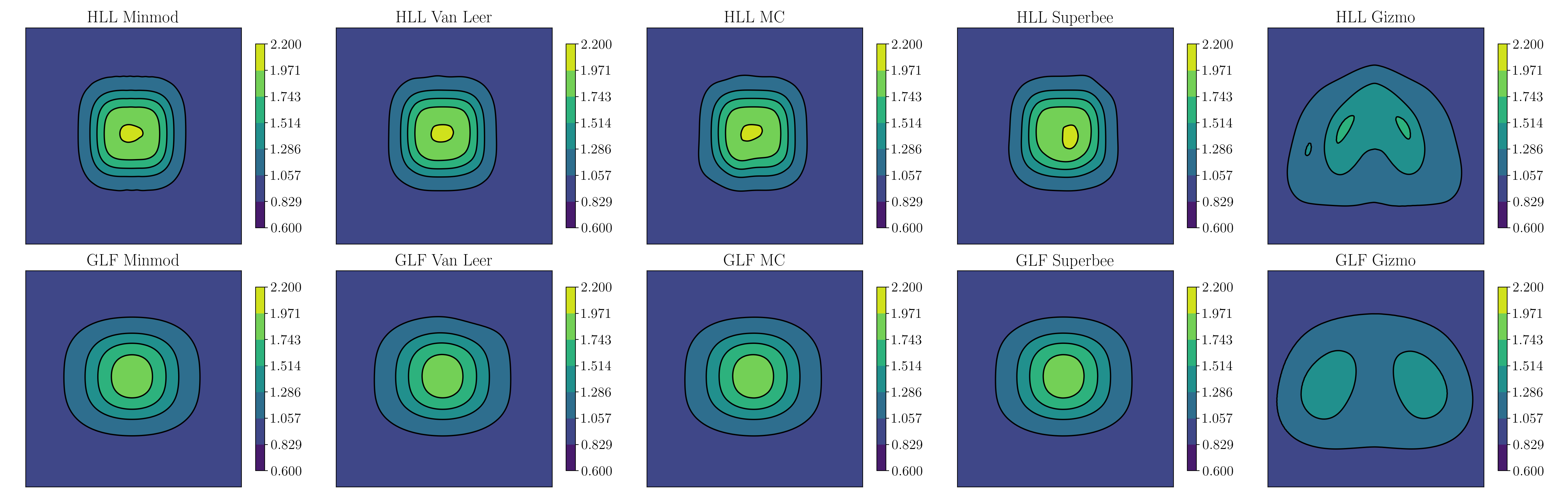}}%
 \caption{
Propagation of a square with non-zero background energy density along the $y$-axis using the GLF
and HLL Riemann solver and various flux limiters. The top figure shows (contours of) the initial
conditions, the bottom figure shows the (contours of the) results after the square propagated a
full box length. The underlying particles were distributed along a uniform grid.
 }
 \label{fig:rt-riemann-limiter-2D-Group2-uniform}
\end{figure}

\begin{figure}
 \centering
 \fbox{\includegraphics[width=\textwidth]{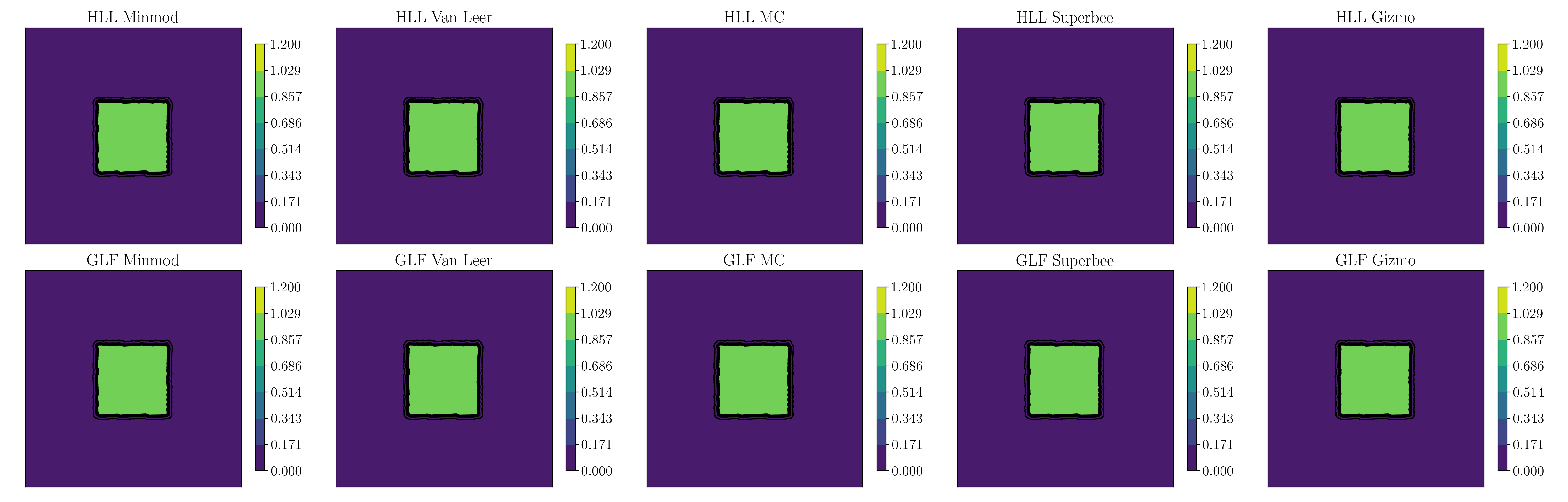}}\\
 \fbox{\includegraphics[width=\textwidth]{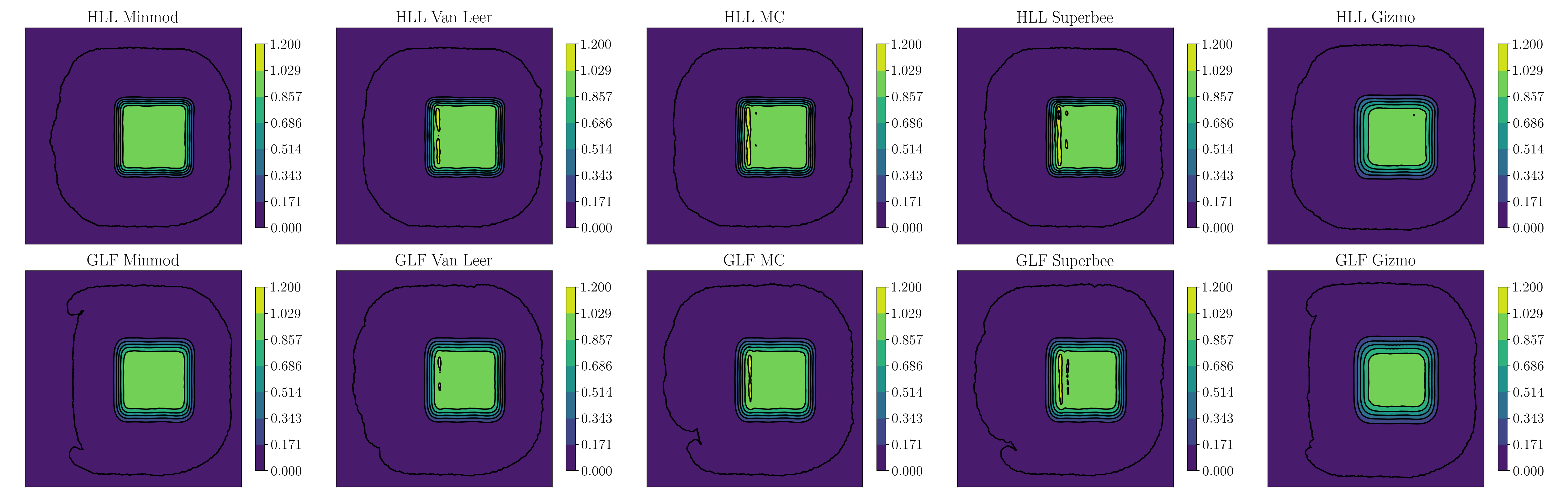}}\\
 \fbox{\includegraphics[width=\textwidth]{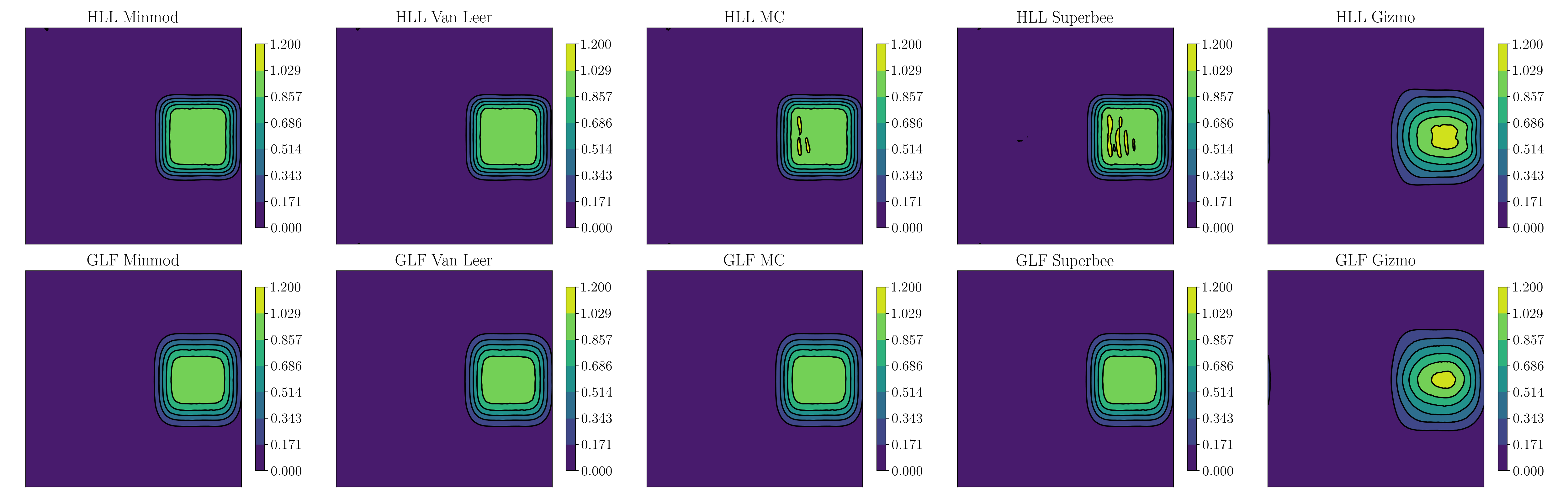}}\\
 \fbox{\includegraphics[width=\textwidth]{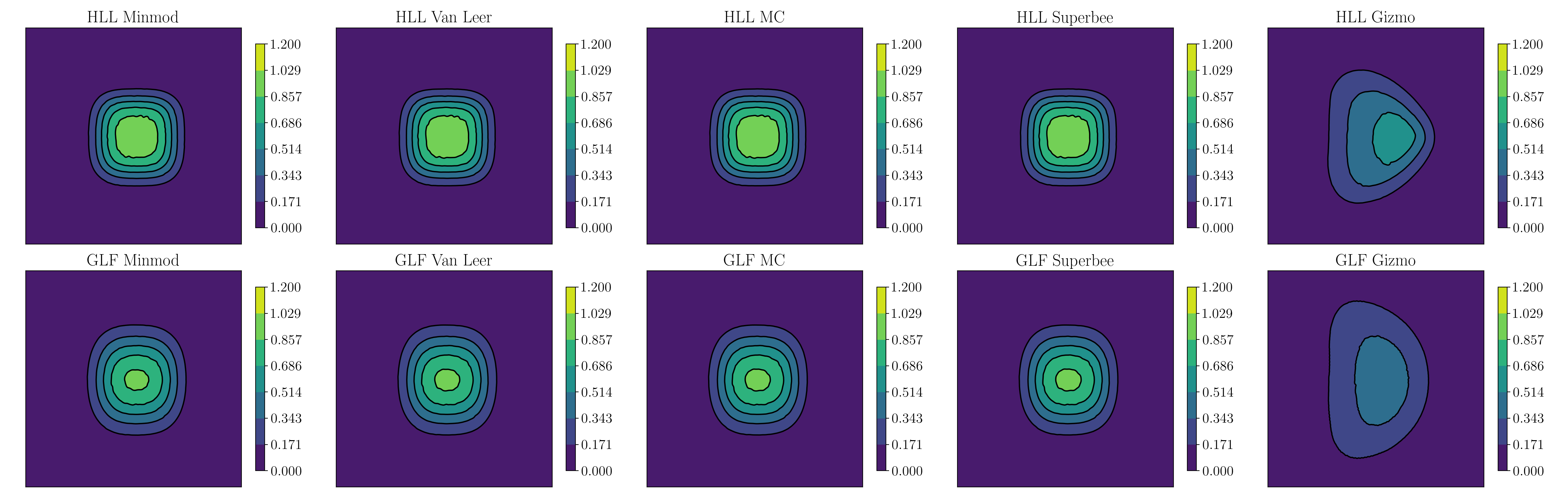}}%
 \caption{
Propagation of a square with non-zero background energy density along the $x$-axis using the GLF
and HLL Riemann solver and various flux limiters. The top figure shows (contours of) the initial
conditions, the bottom figure shows the (contours of the) results after the square propagated a
full box length. The second and third figure show intermediate steps, where the van Leer, MC, and
superbee limiters develop local maxima. The underlying particles were placed in a glass-like
distribution.
 }
 \label{fig:rt-riemann-limiter-2D-Group1-glass}
\end{figure}

\begin{figure}
 \centering
 \fbox{\includegraphics[width=\textwidth]{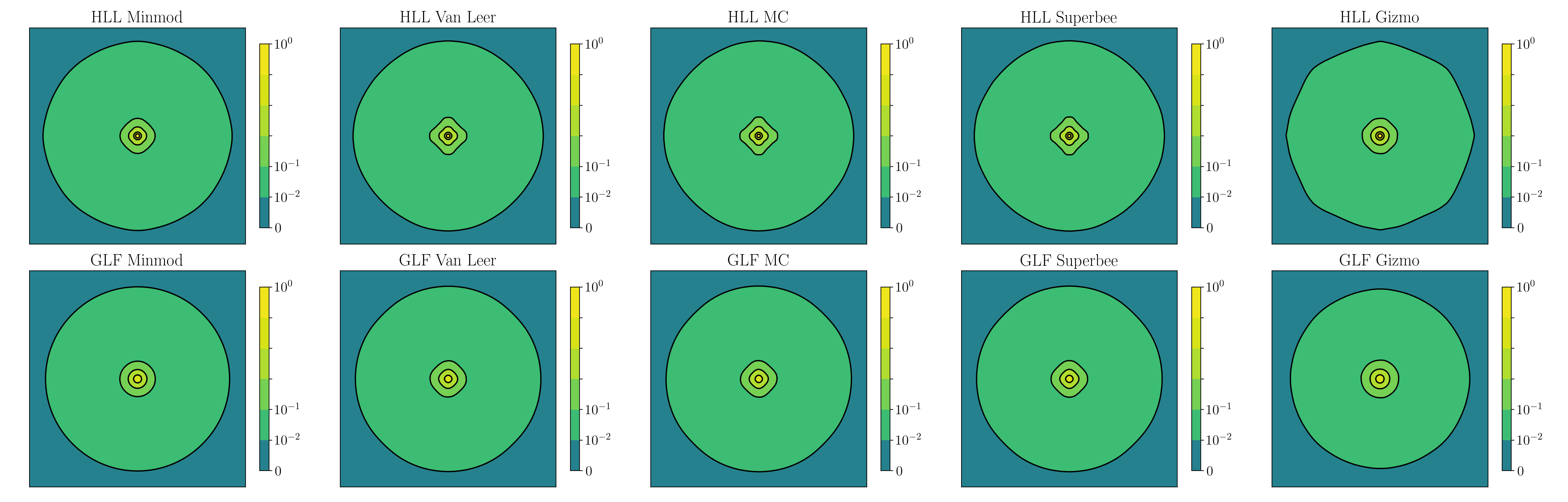}}\\
 \fbox{\includegraphics[width=\textwidth]{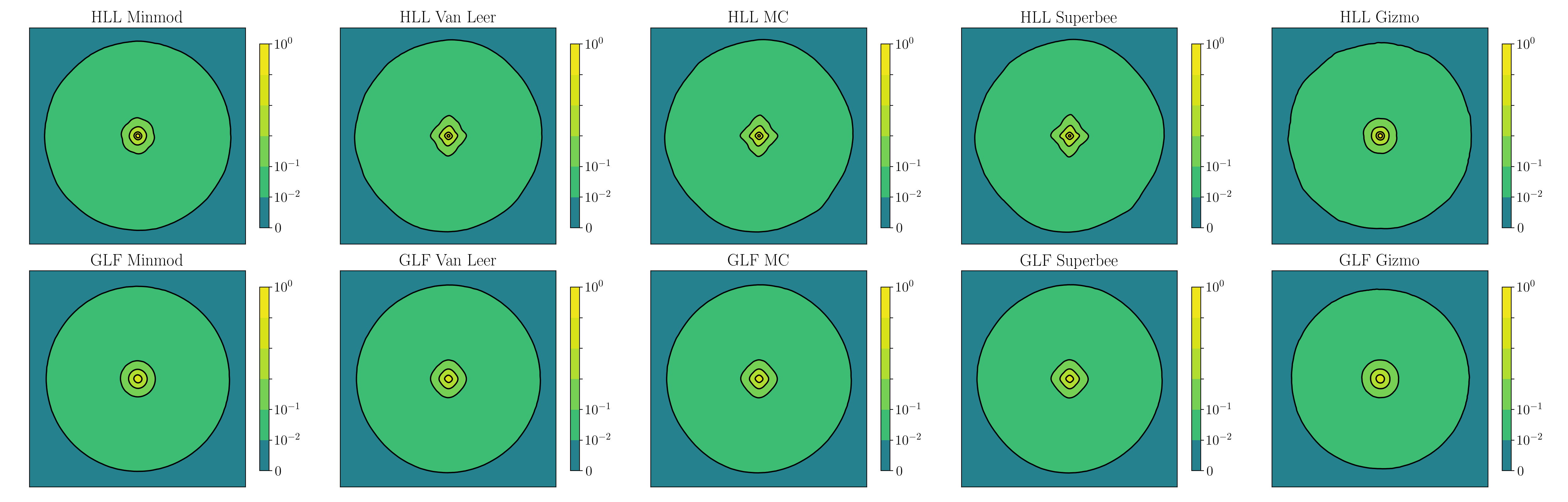}}%
 \caption{
Isotropic injection of radiation energy density from a central source evolved over time using
various limiters and both Riemann solvers. The top figure shows the results using a regular
particle distribution, the bottom figure shows the solution on a glass-like particle configuration.
 }
 \label{fig:rt-riemann-limiter-2D-injection}
\end{figure}

In this section, results of simple photon transport tests using varying Riemann solvers and flux
limiters are shown. The experiments are set up such that the radiation is only advected, and
doesn't interact with the gas at all.

To test the photon transport in 1D, the following initial conditions are set up: (i) a top hat
function, (ii) a top hat function where the lower value is nonzero, and (iii) a smooth Gaussian. The
second top hat function with a nonzero lower value is used to test for cases where the solution
develops oscillations that would predict negative energy densities if the lower limit were zero. In
such cases \GEARRT would correct them, i.e. set the energy density to zero. By having a nonzero
lower value, these instabilities will show up.

Figure~\ref{fig:rt-riemann-limiter-1D} shows the results using various limiters for both the HLL and GLF Riemann solvers (see Section \ref{chap:riemann-rt} for details). The results are virtually
indistinguishable, with the only exception of the ``Gizmo'' limiter
(eqns.~\ref{eq:face-limiter-first}-\ref{eq:face-limiter-last}) being somewhat more diffusive.

Figure~\ref{fig:rt-riemann-limiter-2D-Group2-uniform} shows the propagation of a square along the
$y$-axis with a non-zero background energy density using the two Riemann solvers and various
limiters. In general, the HLL solver indeed delivers less diffusive results, as it promised to do.
The ``Gizmo'' flux limiter however delivers much too diffusive results, despite being a more
sophisticated model. Due to its additional expense yet worse results, it shouldn't be used for the
purposes radiative transfer.

Figure~\ref{fig:rt-riemann-limiter-2D-Group2-uniform} used uniformly distributed particles as the
underlying configuration. Using a glass-like file instead leads to similar results, which are shown
in Figure~\ref{fig:rt-riemann-limiter-2D-Group1-glass} for a square being advected along the
$x$-axis. Additionally, in the test shown in Figure~\ref{fig:rt-riemann-limiter-2D-Group1-glass}, a
zero background energy density is assumed. Looking at intermediate steps, i.e. before the square
transverses a full box length, some local maxima develop for the van Leer, superbee and MC flux
limiters, indicating that they do not make this method truly TVD, and as such are not
unconditionally stable. Hence I advise strongly against the use of them, and to opt for the minmod
limiter instead.\footnote{
It should be noted that due to these findings, the choice of the limiter is not given as a free
parameter to users, as it could lead to unstable results and crashes. The minmod limiter is taken
by default. However, changing the flux limiter is a simple exercise, and the functions of the other
limiters described in this work remain in the source code of \GEARRT. The choice of the Riemann
solver remains a free compile time parameter.
}

Lastly, Figure~\ref{fig:rt-riemann-limiter-2D-injection} shows the results of radiation energy
density being injected isotropically from a single central source and evolved over time using
various flux limiters and both Riemann solvers, using both a glass-like and a uniform underlying
particle configuration. In nearly all cases small deformities, i.e. deviations from perfect circles,
can be seen close to the source, but they even out with increasing distance. Notably the HLL Riemann
solver doesn't show deformities nowhere nearly as strong as reported in \citet{ramses-rt13} even
when using the uniform particle configuration. The reason is that contrary to grid codes, which in
2D would only interact any cell with its eight adjacent neighbors, \GEARRT uses more neighboring
particles ($\sim 15$ in this example for $\eta_{res} = 1.2348$ and the cubic spline kernel), leading
to the improved isotropy of the solution. The use of more neighbors compared to grid codes was
indeed a motivation to make use of the HLL solver, which should perform better in terms of creating
shadows compared to the GLF solver, while the particle nature of \GEARRT and the use of more
neighbors can sidestep the anisotropies seen in grid codes.

\subsection{Order of Accuracy}

\begin{figure}
    \centering
    \includegraphics[width=\textwidth]{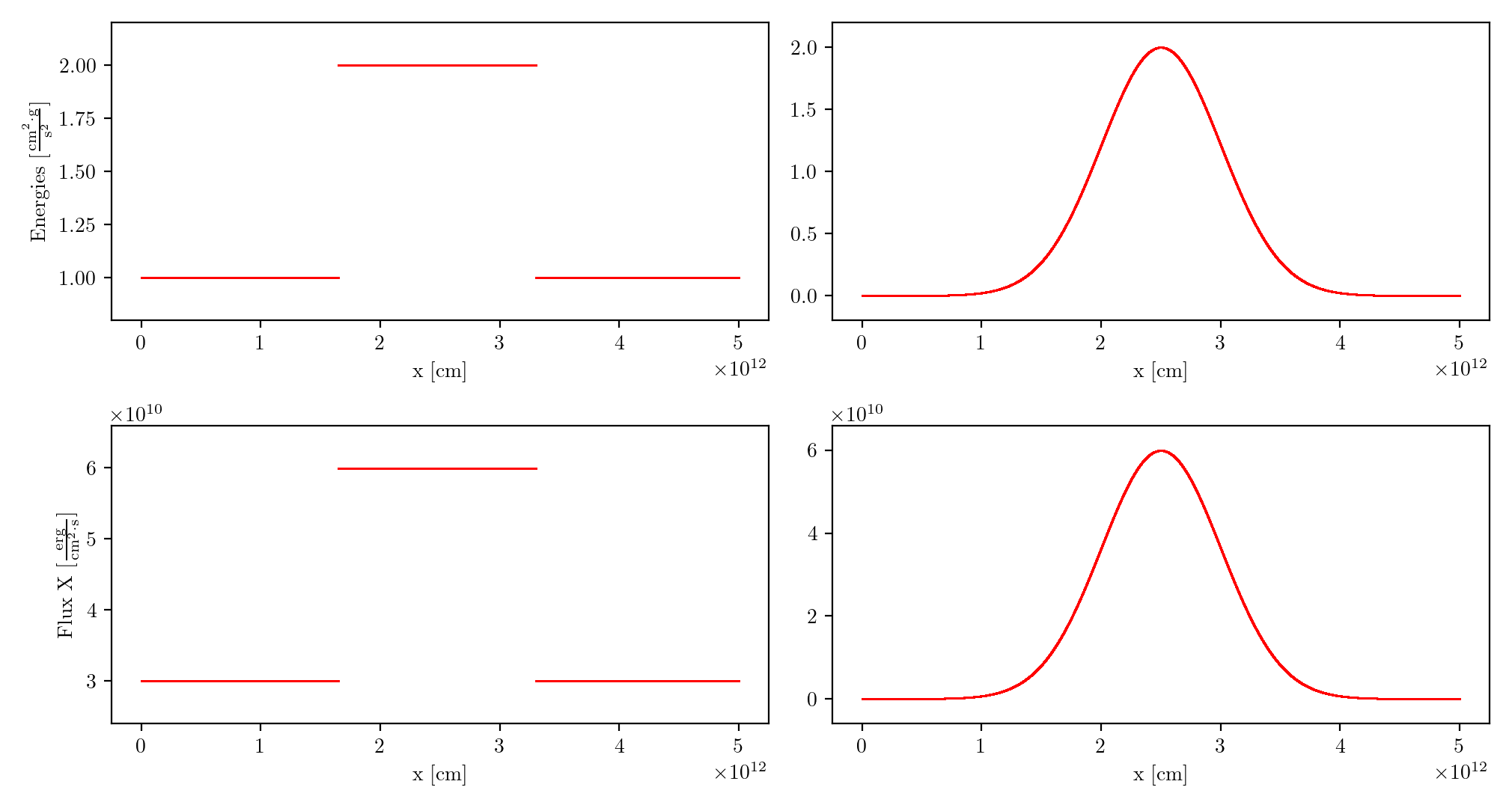}%
    \caption{Initial conditions for the order of accuracy test: A smooth Gaussian and a top hat
function.}
    \label{fig:result-convergence-IC}
\end{figure}

\begin{figure}
    \centering
    \includegraphics[width=\textwidth]{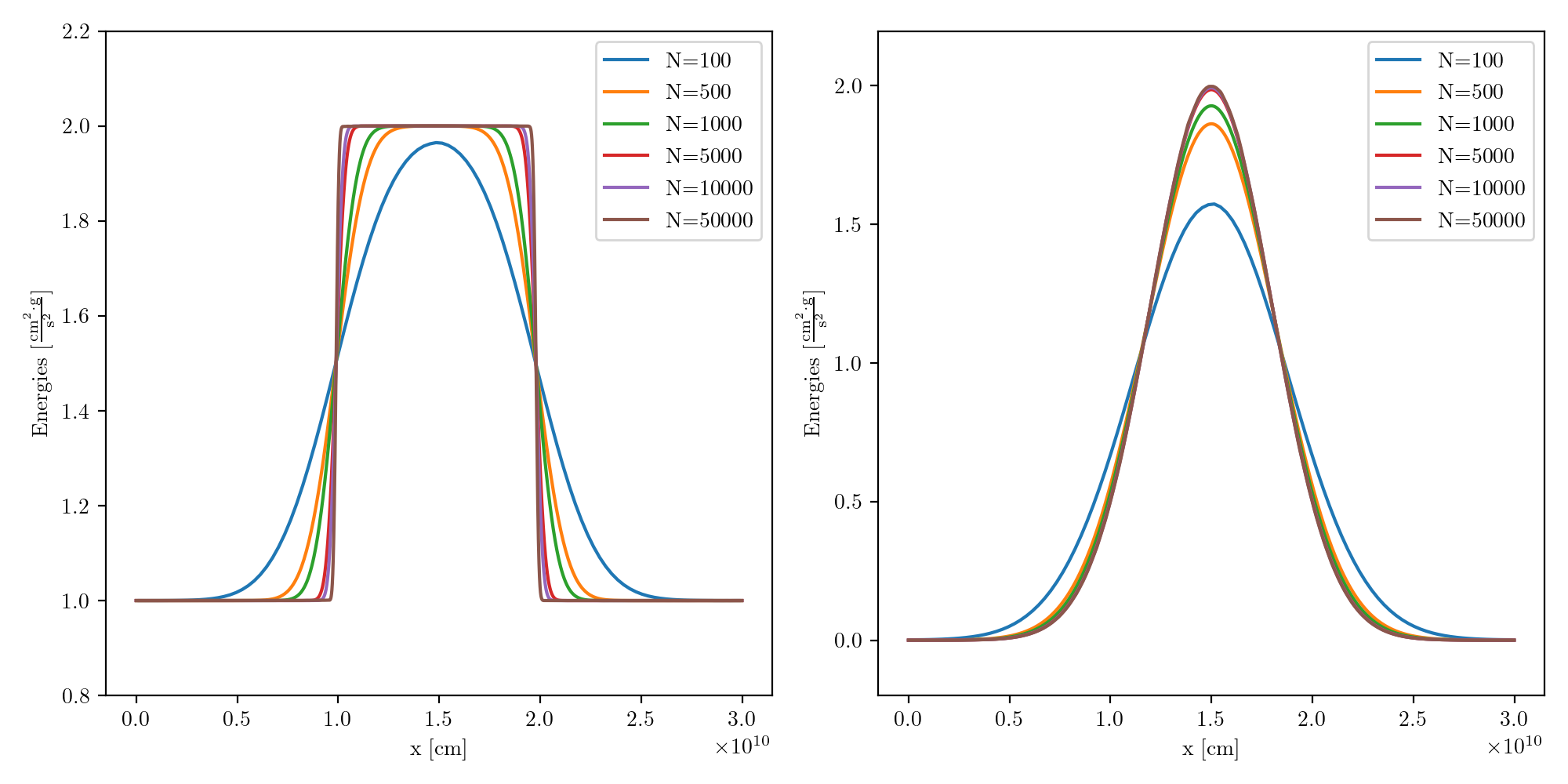}%
    \caption{
The results of the order of accuracy test at $t = 1$s for varying number of particles $N$ used, as
indicated in the legend, for a fixed box size $L = c \times 1$s.
    }
    \label{fig:convergence-energy-fixed-boxsize}
\end{figure}

\begin{figure}
    \centering
    \includegraphics[width=.75\textwidth]{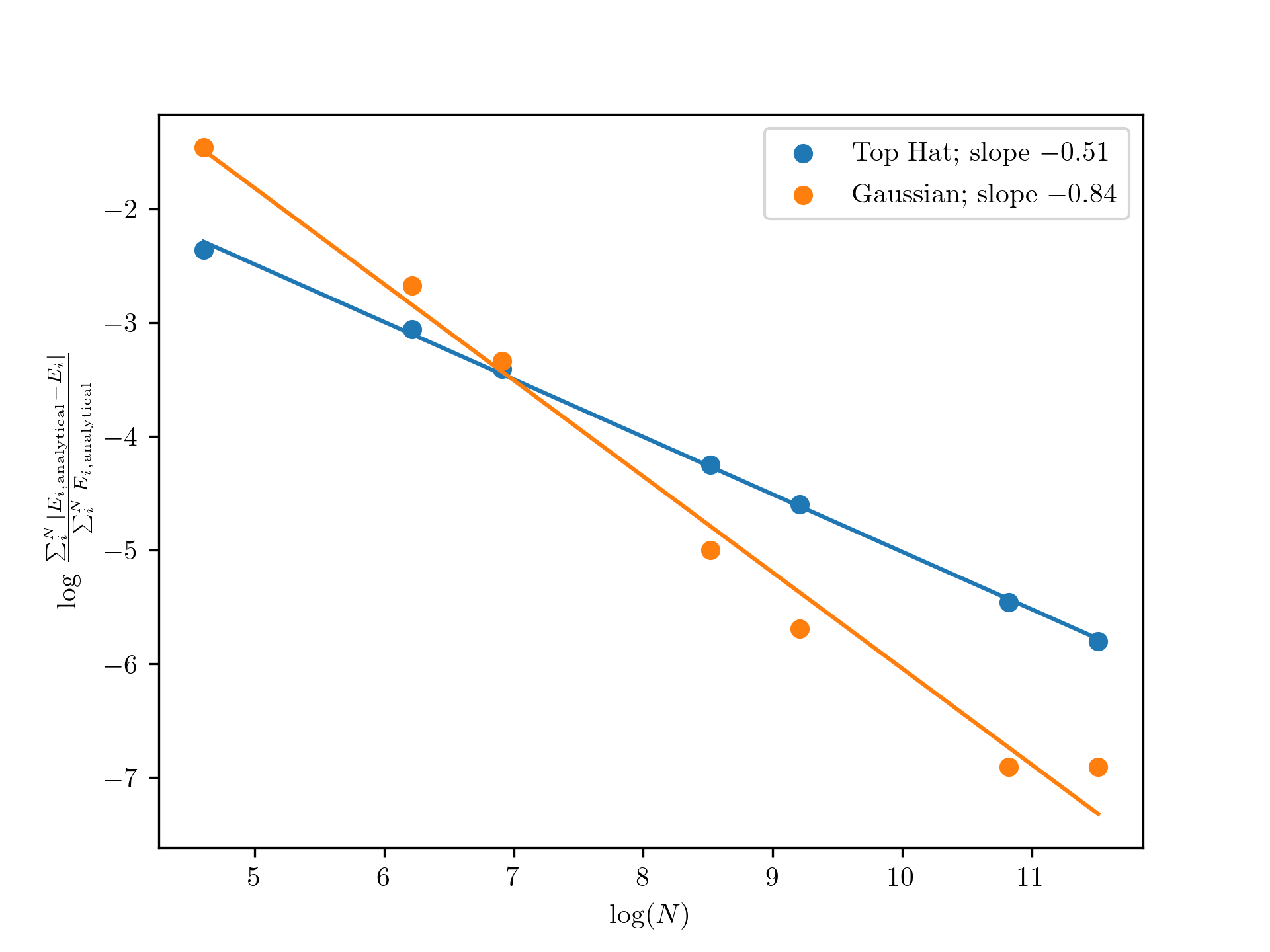}%
    \caption{
Order of accuracy for \GEARRT with increasing particle numbers $N$, as indicated in the legend,
for simulations using the identical box size length $L = c \times 1$s and end time $t = 1$s.
According to expectations for a second order accurate scheme, the photon transport for the smooth
Gaussian converges with a power of nearly $-1$, while the exponent is reduced to $-1/2$ for the top
hat function due to its discontinuities. The slopes are best fits to the measurements (circles).
Note that the slopes aren't $-2$ and $-1$, respectively, because in this setup, the simulations
with
higher particle numbers $N$ also need to perform more time steps to reach the same end time $t$,
and thus accumulating more one step errors over the course of the run (compare with
Section~\ref{chap:numerical_diffusion}). The expected slopes of $-2$ and $-1$ are achieved when
keeping the number of time steps fixed as well, which is shown in
Figure~\ref{fig:result-convergence}.
    }
    \label{fig:result-convergence-fixed-boxsize}
\end{figure}

\begin{figure}
    \centering
    \includegraphics[width=.75\textwidth]{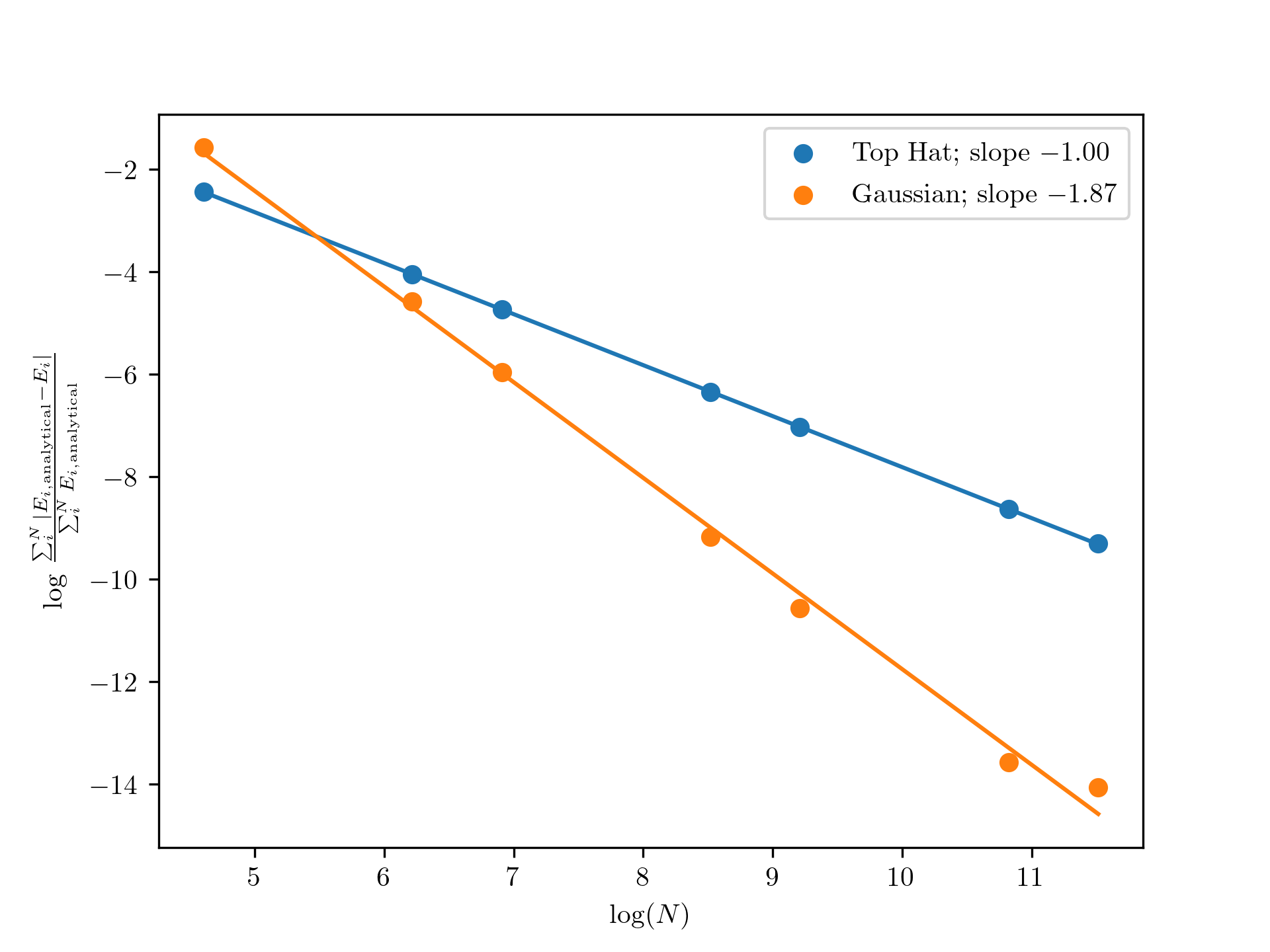}%
    \caption{
Order of accuracy for \GEARRT with increasing particle numbers. According to expectations for a
second order accurate scheme, the photon transport for the smooth Gaussian converges with a power
of
nearly $-2$, while the exponent is reduced to $-1$ for the top hat function due to its
discontinuities. The slopes are best fits to the measurements (circles).
    }
    \label{fig:result-convergence}
\end{figure}

To demonstrate the order of accuracy of radiation transport solved by the finite volume particle
method, the same experiments as for the finite volume methods in Part~\ref{part:finite-volume} are
conducted. Two test cases are set up: One where the initial conditions are a top hat function for
the photon energy density, and one where the initial conditions are a smooth Gaussian. The photon
fluxes are assigned assuming the free streaming limit $\Fbf = c E$, such that the analytical
solution consists of the equivalent of linear advection of the initial functions to the right. The
gas isn't allowed to interact with the radiation. The initial setup is shown in
Figure~\ref{fig:result-convergence-IC}. The experiments are conducted in one dimension, and with an
increasing number of particles, as indicated in the plots depicting the results.

A first test, intended to facilitate visual inspection of the results alongside the order of
accuracy of the method, keeps the simulation box size and end time constant, while varying the
number of particles, and hence the mean inter-particle distance. The simulation box size is set up
to be $L = c \times 1$s, and the simulation runs until $t = 1$s is reached.
Figure~\ref{fig:convergence-energy-fixed-boxsize} shows the resulting energy densities at $t = 1$s.
In agreement with the findings in Section~\ref{chap:numerical_diffusion}, the results improve, i.e.
are less diffusive and maintain the original shapes better, with increasing particle numbers used.
The resulting order of accuracy along with a fit for the slopes are shown in
Figure~\ref{fig:result-convergence-fixed-boxsize} of this setup, where the errors in each simulation
is estimated as

\begin{align}
 Err = \frac{\sum_{i=1}^N |E_{i, \text{analytical}} - E_i|}{\sum_{i=1}^N E_{i, \text{analytical}}}
\ .
\end{align}

Here $i$ denotes the index of particles, and $N$ is the total number of particles. The slopes don't
reach the optimal values of $-2$ for the smooth Gaussian and $-1$ for the discontinuous top hat
function, respectively, because the runs with more particles also require more time steps to reach
the same end time, thus accumulating more one step errors (compare with
Section~\ref{chap:numerical_diffusion}). Comparing the order of accuracy after a fixed number of
time steps for each number of particles $N$ used leads to the anticipated slopes of (nearly) $-2$
and $-1$, respectively, which is shown in Figure~\ref{fig:result-convergence}.

\subsection{Testing the Drift Corrections}\label{chap:validation-drift-corrections}

\begin{figure}
 \centering
 \includegraphics[width=\textwidth]{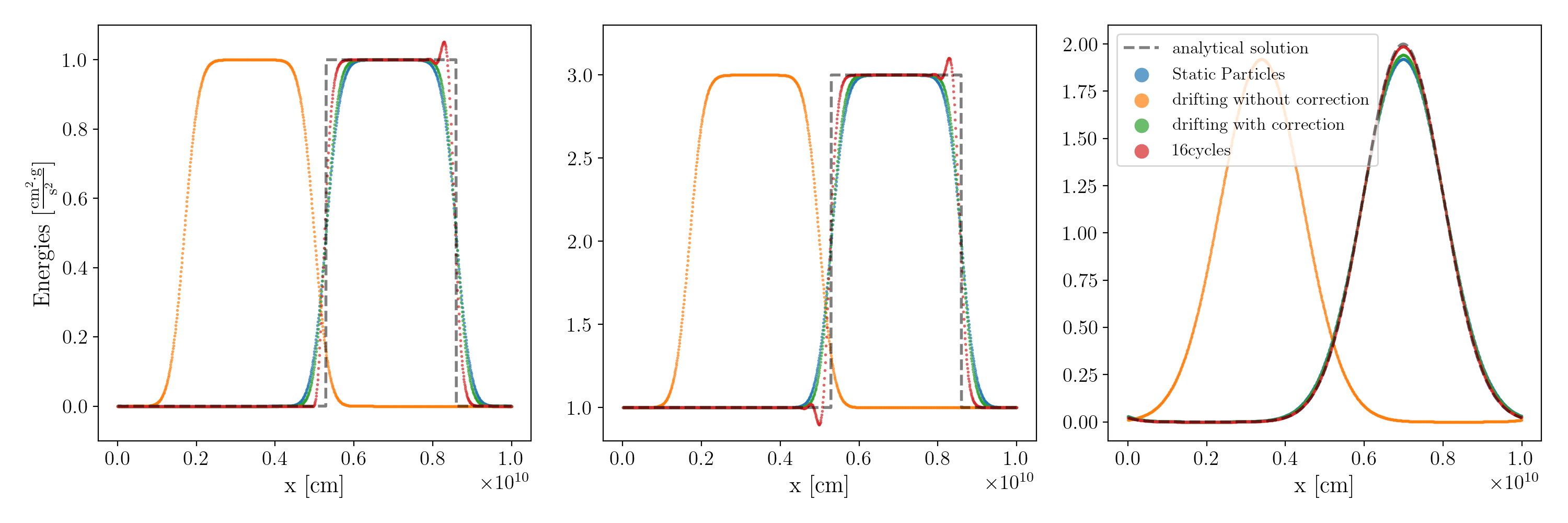}%
 \caption{
Results of the photon transport in a scenario where the fluid has a velocity of $-0.3c$ in
cases where particles are kept static (blue dots), when particles are drifted, but no drift
correction terms are applied (orange dots), when the drift correction terms are applied (green
dots), and when additionally 16 sub-cycles have been used (red dots).
 }
 \label{fig:drift-0.3c}
\end{figure}

\begin{figure}
 \centering
 \includegraphics[width=\textwidth]{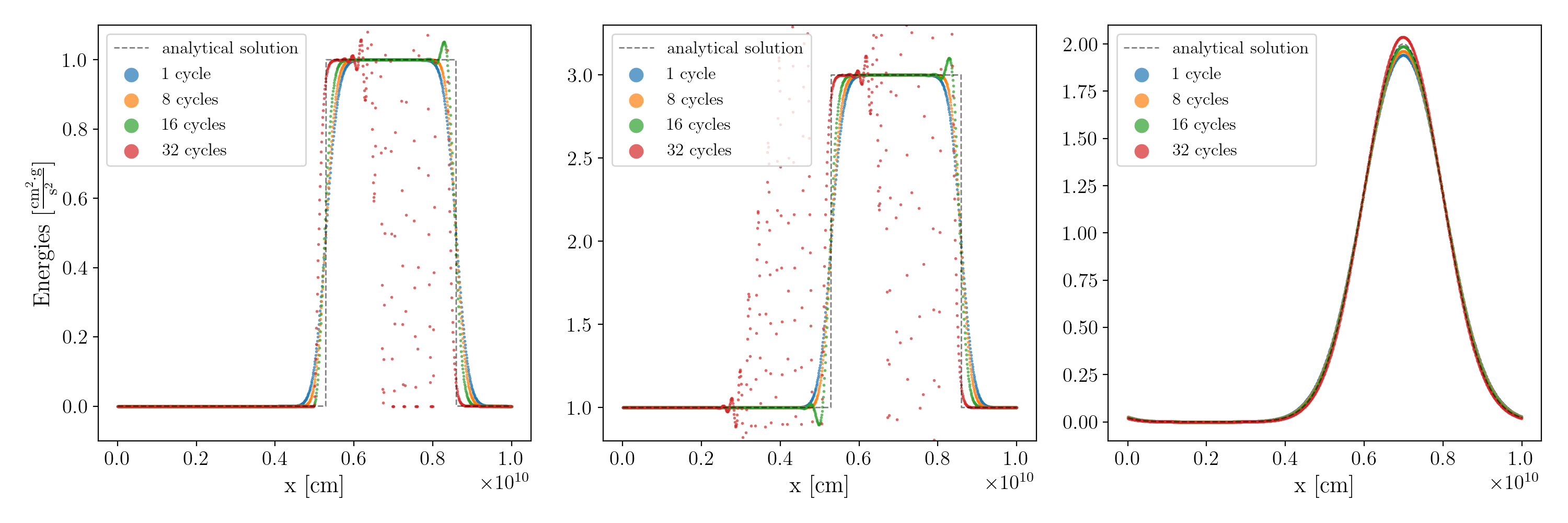}%
 \caption{
Same as Figure~\ref{fig:drift-0.3c}, but testing various numbers of sub-cycles along with the drift
correction. While higher sub-cycle numbers give catastrophic results, 8 sub-cycles still give
adequate results, even though they are above the realistic limit of 4 sub-cycles for a fluid
velocity of $-0.3c$.
 }
 \label{fig:drift-0.3c-subcycles}
\end{figure}

\begin{figure}
 \centering
 \includegraphics[width=\textwidth]{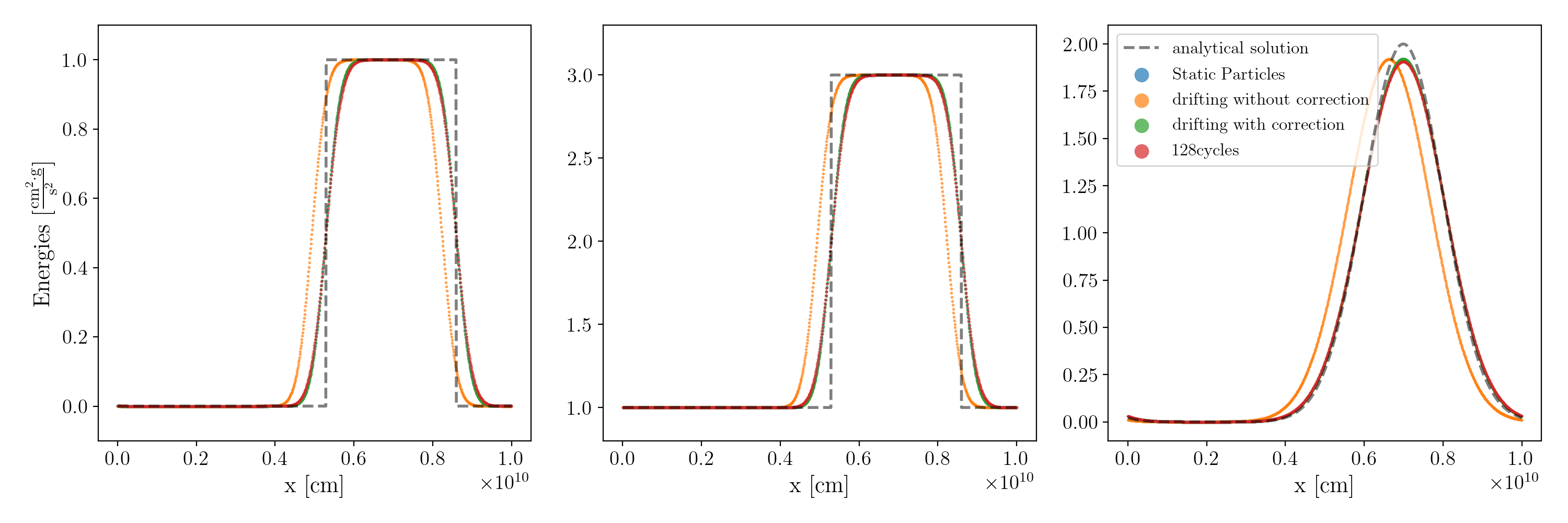}%
 \caption{
Same as Figure~\ref{fig:drift-0.3c}, but with a fluid velocity of $-0.03c$ instead of $-0.3c$. In
this case, even 128 sub-cycles don't develop instabilities.
 }
 \label{fig:drift-0.03c-subcycles}
\end{figure}

\begin{figure}
 \centering
 \includegraphics[width=\textwidth]{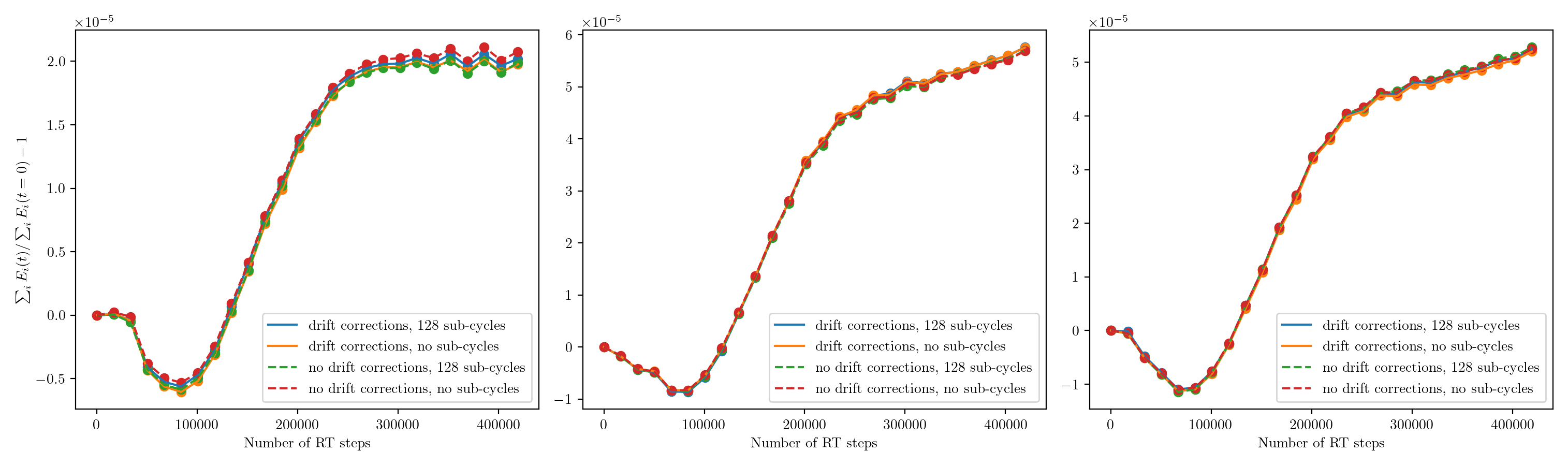}%
 \caption{
The evolution of the total energy densities of slightly adapted initial conditions compared to
what is shown in Figure~\ref{fig:drift-0.3c}, as indicated in the main text, where the widths and
amplitudes of the initial functions have been tweaked in an attempt to introduce larger errors due
to the approximate drift correction scheme. The errors introduced by the approximate and not strictly conservative drift correction scheme remain within the order of the floating point precision, although they accumulate systematically over the course of hundreds of thousands of performed time steps, while remaining acceptably low.
 }
 \label{fig:drift-correction-conservation}
\end{figure}

As explained in Section~\ref{chap:rt-numerics-outline}, since the radiative transfer is solved as if
the particles were static interpolation points, but the particles are moved along with the fluid, a
correction term needs to be applied when particles are drifted. In this section, I test the validity
of that approach.

To this end, first an experiment in one dimension is set up as follows. As before, the photons
aren't allowed to interact with the gas, and are only transported. The initial conditions are (i) a
top hat function, (ii) a top hat function where the lower value is nonzero, and (iii) a smooth
Gaussian. The second top hat function with a nonzero lower value is used to test for cases where the
solution develops oscillations that would predict negative energy densities if the lower limit were
zero. In such cases \GEARRT would correct them, i.e. set the energy density to zero. By having a
nonzero lower value, these instabilities will show up.
The initial conditions are set up such that the radiation has fluxes of $\Fbf = c E$, i.e. the free
streaming limit, and the analytical solution is equal to a linear advection with constant
coefficients. The particles are given a velocity in negative $x$ direction, i.e. in the opposite
direction of the photon flux.

Figure~\ref{fig:drift-0.3c} shows the result for a particle velocity of $0.3c$. The following
results are shown:

\begin{itemize}
 \item The blue dots show the results when particles are kept static.
 \item The orange dots show the results when particles are drifted, but no correction terms are
applied. Clearly the radiation has been carried away by the particle drifts to the wrong position.
 \item The green dots show the results when particles are drifted along with the fluid, and the
correction terms discussed in Section~\ref{chap:rt-numerics-outline} are applied. The results agree
very well with the static particle case (blue dots).
 \item Finally, the red dots show the results for when the drift corrections are applied, and
additionally 16 RT sub-cycles have been used. The discontinuities develop an instability.
\end{itemize}

While the development of the instability with the sub-cycling may seem alarming at first, it is in
fact not a real problem. It develops because the sub-cycling allows for a larger hydrodynamics time
step, and hence a larger drifting distance. Because of the larger drift distance, the gradient
extrapolation then introduces new extrema which the limiters aren't able to keep in check any more.
However, 16 sub-cycles are not a realistic scenario, as the fixed number of 16 sub-cycles in
this case was enforced. In a real application, the number of sub-cycles would instead dynamically
adjust itself and be limited by the ratio of the speed of light to the particle velocity, in this
case 4. Indeed, Figure~\ref{fig:drift-0.3c-subcycles} compares the results for 1, 8, 16, and 32
sub-cycles, and the results for up to 8 sub-cycles are perfectly adequate.
Figure~\ref{fig:drift-0.03c-subcycles} shows the results with a ten times lower fluid velocity of
$-0.03c$, where even 128 sub-cycles show no trace of an instability developing. Tests with similar
setups in 2D confirm these findings, and I conclude that using gradients of the radiation quantities
to extrapolate their values in order to deal with particle drifts is an adequate approach.

A caveat of the correction scheme for particle drifts is that the method to solve radiation
transport is no longer strictly conservative. However, testing the radiation energy density
conservation explicitly by comparing the current total energy densities with the initial ones shows
that the introduced errors are negligibly small and within the precision limits of the single
precision floating point variables used to carry the radiation fields.
Figure~\ref{fig:drift-correction-conservation} shows the results of such a comparison for a test
similar to the experiment setup previously described in this section: We transport two top hat
functions, one of which has a nonzero lower value, and a smooth Gaussian function. The amplitude of
the Gaussian and the upper value of the top hat function have been increased by a factor of 5,
while
their width has been reduced to take one fifth of the box size in an attempt to produce sharper
discontinuities which will require more time for the diffusivity of the method to smooths them out
significantly, therefore keeping steeper gradients for longer. Additionally, in an attempt to
increase the distances of each drift, the ratio of the speed of light to the gas (drift) velocity
was set to -40. Varying these parameters (amplitudes and widths of initial conditions,
ratio of speed of light to drift velocity), as well as the number of particles used all had
negligible effect on the outcome with regards to total energy conservation: The errors remain
within the precision limits of the floating point numbers (although they accumulate systematically
over the course of hundreds of thousands of performed time steps, while remaining acceptably low).
As such, I deem the drift correction scheme as appropriate for use without any further corrections
or adaptations.

\section{Testing Injection Models From Radiation Sources}

\subsection{Testing Flux Injection Models From Radiation Sources}\label{chap:results-injection}

\begin{figure}
 \centering

\includegraphics[width=\textwidth]
{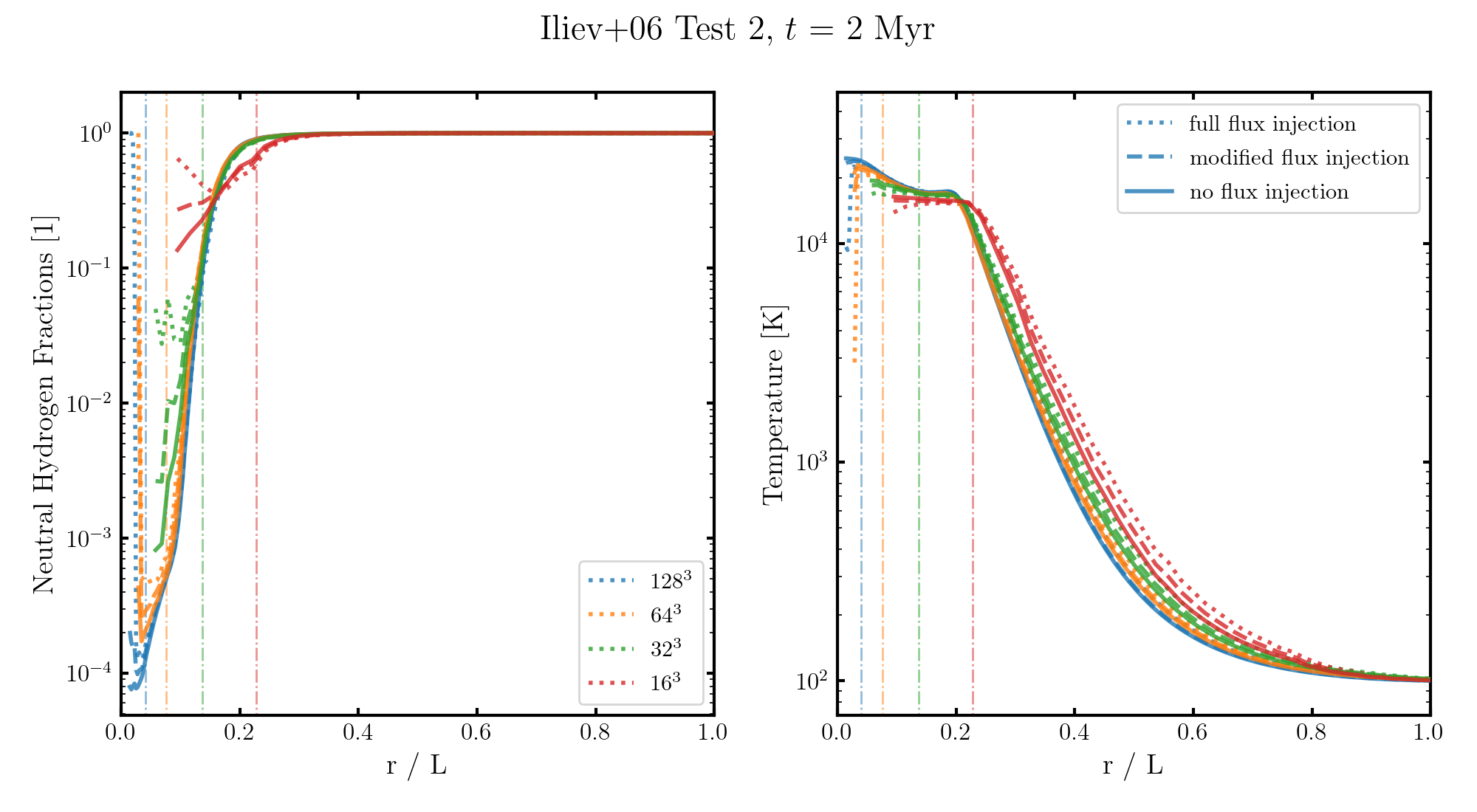}
\\
\includegraphics[width=\textwidth]
{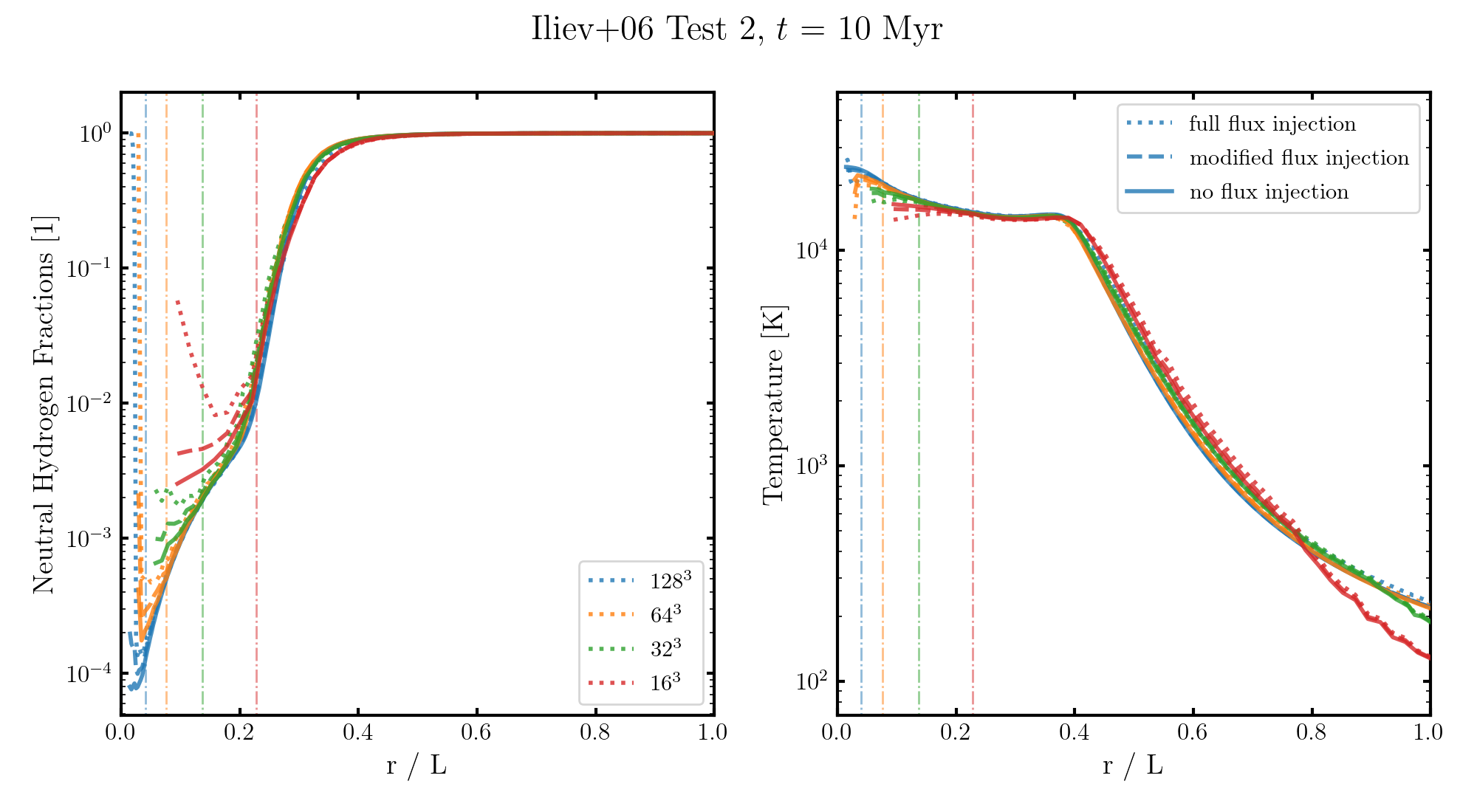}
\caption{
Radial profiles of the results of the Iliev 2 test (see Section~\ref{chap:Iliev2} for details) for
varying flux injection methods and resolutions, as indicated in the legends, at $t = 2$Myr (top)
and $t = 10$Myr (bottom). The vertical dash-dotted lines show the maximal radius at which energy
and fluxes are injected from a source, which is located at $r = 0$, for the various resolutions.
 }
 \label{fig:injection_convergence}
\end{figure}

\begin{figure}
\centering
\includegraphics[width=\textwidth]
{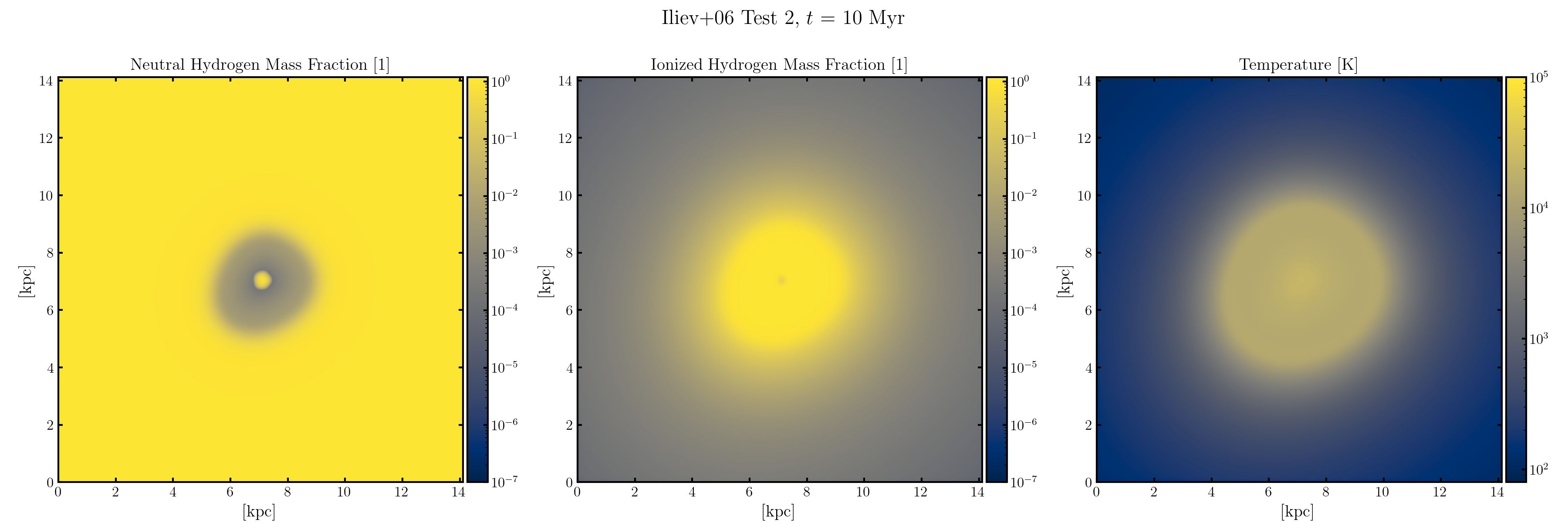}\\
\includegraphics[width=\textwidth]
{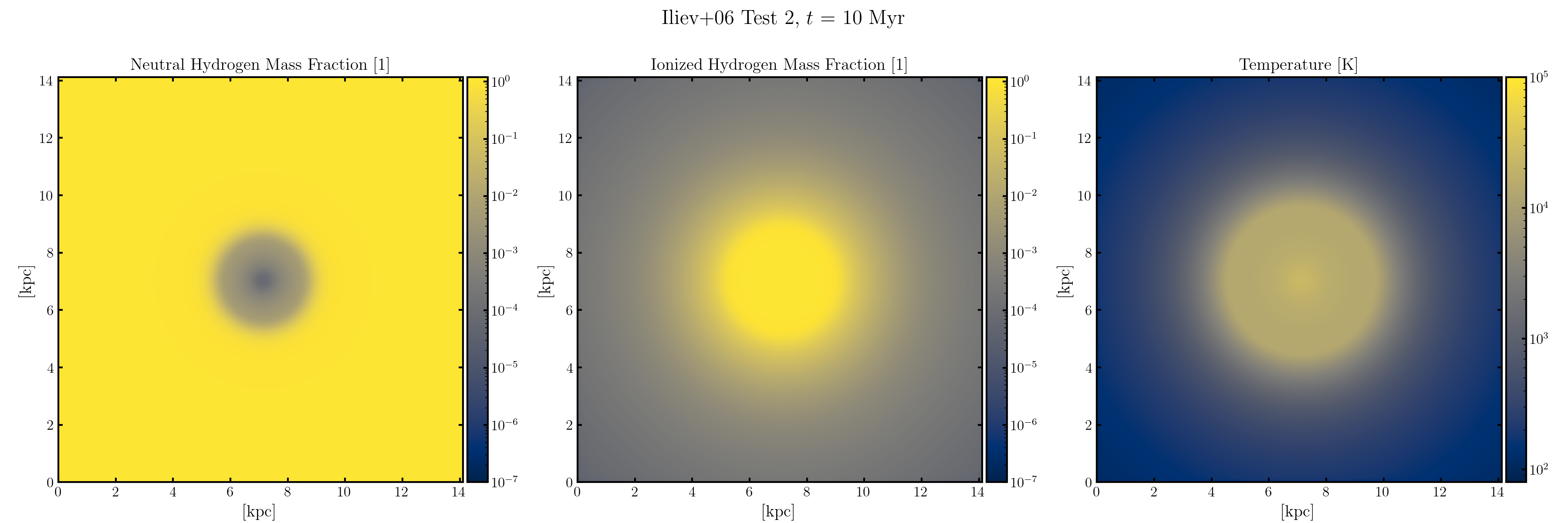}
\caption{
Slice along the mid-plane of the results of the Iliev 2 test (see Section~\ref{chap:Iliev2} for
details) for the ``full flux injection'' model (top) and the ``no flux injection'' model (bottom)
in the $128^3$ particle simulation at $t = 10$Myr. The ``full flux injection'' model transports
too much radiation energy density away from the innermost regions of the box, leaving the immediate
vicinity of the radiation source under-ionized.
 }
 \label{fig:flux-injection-slices}
\end{figure}

In this section, three different ways of injecting radiation \emph{flux} are tested, as mentioned in
Section~\ref{chap:injection-step}. Contrary to mesh based codes, where the net injected flux from a
point source radiating isotropically inside a single cell always neatly sums up to zero, \GEARRT is
a particle based code, and radiation is injected into a group of particles. This leaves us with
some freedom to choose how exactly to inject the photon fluxes onto the particles, while still
maintaining that the vector sum of the injected fluxes should remain zero to preserve isotropy.
Three flux injection models are tested:

\begin{itemize}
 \item The ``no flux injection'' model injects no radiation flux $\Fbf$, i.e. only energy density
is injected onto particles.
 \item The ``full flux injection'' model injects flux assuming an optically thin limit, i.e. $\Fbf
= c E$
 \item The ``modified flux injection'' model injects a linearly increasing amount of flux
with increasing distance starting with no flux and ending in the optically thin limit $\Fbf = c E$
at the distance $0.5H$, where $H$ is the maximal distance at which energy and flux is injected. For
distances above $0.5H$, again the value of the optically thin limit is taken.
\end{itemize}

Note that the anisotropy correction terms for the injection of energy density described in
Section~\ref{chap:injection-weights} are also applied here for all flux injection models.

The test setup is identical to the Iliev 2 test, which is described in detail in
Section~\ref{chap:Iliev2}. For the scope of this section however, it suffices to note that the test
consists of a single ionizing source injecting photons into its uniform surroundings. The
temperature and ionization state of the gas is only modified through the influence of the radiation,
actual hydrodynamics are turned off.

Figure~\ref{fig:injection_convergence} shows the results at 2 Myr and at 10 Myr using various
resolutions, ranging from $16^3$ to $128^3$ particles. The strongest deviations are within the
region where energy density is injected, i.e. within the compact support radius $H$ of the source,
beyond which the solutions for all resolutions and flux injection methods converge quickly.

The ``full flux injection'' model and the ``modified flux injection'' model are somewhat
problematic, as they leave a peak in neutral hydrogen fraction and a dip in temperature very close
to the source (which is located at $r = 0$), whereas the solution is expected to be monotonously
increasing in the neutral hydrogen fraction and decreasing in the temperature, respectively.
Essentially the immediate vicinity around the source remains under-heated and under-ionized. The
reason for this phenomenon lies in the order of operators used in the operator splitting scheme to
solve the equations of radiative transfer: The photon transport is done before the thermochemistry.
With a flux corresponding to the free streaming limit, too much energy is being transported away
before it can heat and ionize the gas sufficiently in the thermochemistry step. This is supported by
the fact that the ``no flux'' model indeed delivers the best results. The lack of ionization and
heating in around the source can be seen very clearly in Figure~\ref{fig:flux-injection-slices},
where the a slice through the mid-plane of the box is shown for the solution of the ``full flux'' and
the ``no flux'' models. Additionally, Figure~\ref{fig:flux-injection-slices} shows that the ``full
flux'' model exacerbated anisotropies in the ionized region, adding a further reason to prefer to
avoid it.

With an increasing amount of flux injected into the gas, the gas heats up a little more at large
distances. The difference is negligible though, and decreases with increasing resolution. As such,
it is safe to conclude that the amount of flux injected from a single source is not really relevant
for regions far from the source. Indeed, even when injecting only energy density and no flux at all
particles ``develop'' a flux in the subsequent time step (see
Appendix~\ref{app:zero-flux-nonzero-energy}). Given how injecting zero flux leads to best results
close to the source, and to negligible differences far from the source, I recommend that model to
be used.\footnote{
In fact, the flux injection method is not left to users as a free parameter for \GEARRT, but the
``no flux'' model is used as the default and only choice.
}

\subsection{Testing Different Time Step Sizes for Injection and Radiation Transport}
\label{chap:results-star-timesteps}

\begin{figure}
\centering
\includegraphics[width=\textwidth]{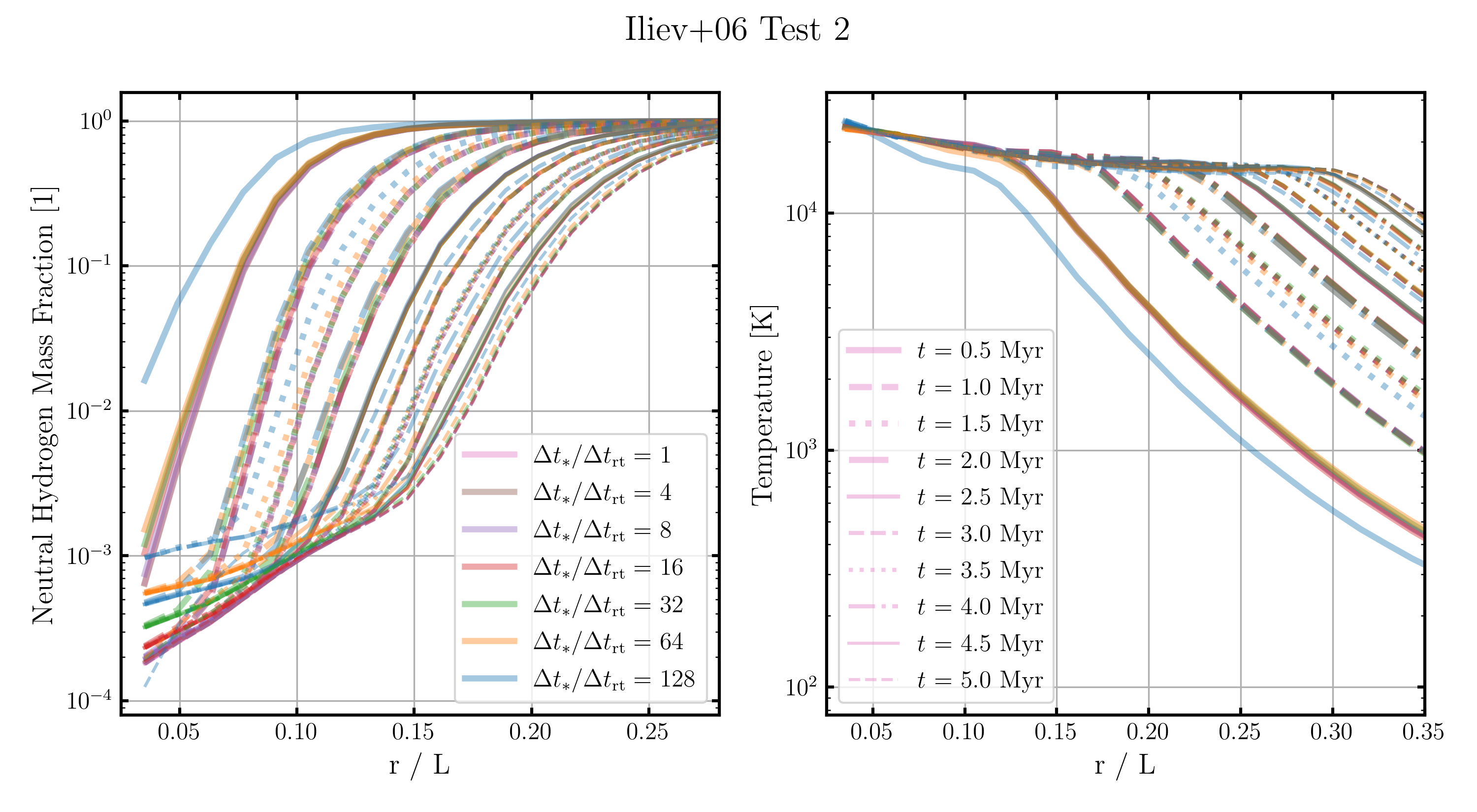}%
\caption{
Profiles of the neutral hydrogen mass fractions and the gas temperature over time for increasing
power-of-two multiples of star time step sizes $\Delta t_*$ compared to the radiative transfer time
step size $\Delta t_{\mathrm{rt}}$ for initial conditions identical to the Iliev Test 2 (see
Section~\ref{chap:Iliev2} for details). The colors of the lines represent a specific ratio $\Delta
t_* / \Delta t_{\mathrm{rt}}$, while the line widths and styles depict the results at different
times, as indicated in the legends.
}
\label{fig:injection-timesteps-profiles}
\end{figure}

\begin{figure}
\centering
\includegraphics[width=\textwidth]{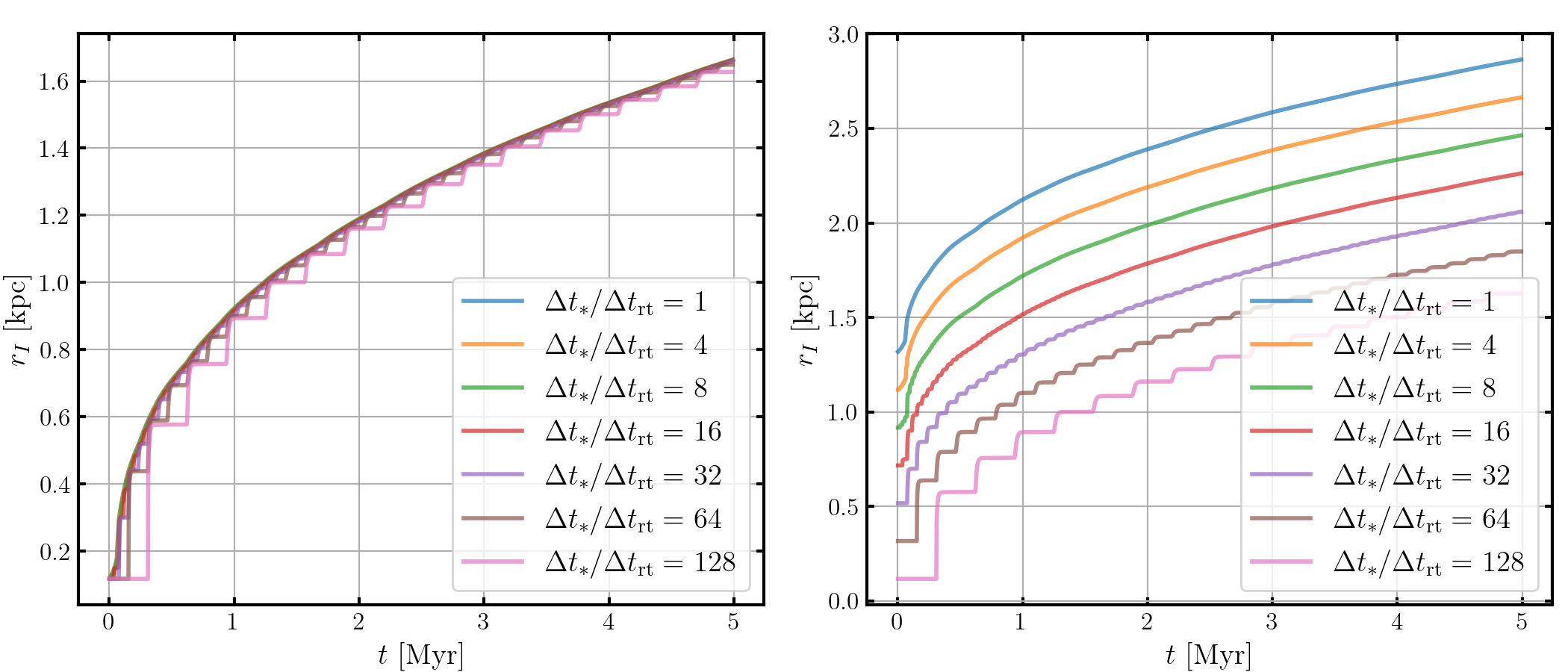}%
\caption{
Position of the radius of the ionization front, defined as the point where the ionized hydrogen mass
fraction is one half, over time for increasing power-of-two multiples of star time step sizes
$\Delta t_*$ compared to the radiative transfer time step size $\Delta t_{\mathrm{rt}}$ for initial
conditions identical to the Iliev Test 2 (see Section~\ref{chap:Iliev2} for details). The right
plot is the same as the left plot, but with a displacement of $0.2$kpc added between each ratio
$\Delta t_* / \Delta t_{\mathrm{rt}}$ so each individual line can be seen clearly.
}
\label{fig:injection-timesteps-ionization-front-radius}
\end{figure}

As mentioned before, particles are permitted to have individual time step sizes (which must be
multiples-of-two of some global minimal time step size). While this is true for gas particles
compared to each other, it is also true between star particles and gas particles. This means that
the photon injection step, which is determined by the star particles' time step sizes, can and will
be performed less frequently\footnote{In principle, the injection step can also occur \emph{more}
frequently than the transport step, but this is not a situation we'd need to be concerned about at
all. Star particles having time step sizes smaller than radiative transfer time step sizes
is not really a realistic scenario, and as such is not expected to occur at all. If it does occur
at any point, the injected energy density will just be added to the neighboring gas particles
several times instead of only once, where the total sum of the injected energy density remains
identical. So the injection of radiation energy density in the situation where a star particle has
a smaller time step size than its neighboring gas particles results in the identical outcome as if
it had the same time step size as the gas particles.}
than the photon transport steps, who in turn are determined by the gas particles' RT time step sizes
(see Section~\ref{chap:injection-step}).

To test the influence of increasing time step sizes of star particles, I use the same setup as in
the previous section, which is also identical to the Iliev 2 test, described in detail in
Section~\ref{chap:Iliev2}. For the scope of this section however, it suffices to note that the test
consists of a single ionizing source injecting photons into its uniform surroundings. The
temperature and ionization state of the gas is only modified through the influence of the radiation,
actual hydrodynamics are turned off.

Figure~\ref{fig:injection-timesteps-profiles} shows the resulting profiles of the neutral hydrogen
mass fractions and the gas temperature over time for increasing powers-of-two of star time step
sizes compared to the radiative transfer time step size. The profiles with time step ratios $\Delta
t_* / \Delta t_{\mathrm{rt}} \leq 32$ agree quite well with each other save for the innermost
region $r / L \lesssim 0.075$, where smaller ratios find an increasingly lower neutral hydrogen
mass fraction. This is also the region closest to the radiation source, and it is to be expected
that it would react the most sensitively to a change in the frequency of energy density injection.
The differences in the neutral hydrogen mass fraction for the ratios $\Delta t_* / \Delta
t_{\mathrm{rt}} \leq 32$ are rather small, with a maximum of a factor of $\sim 2$ closest to the
source for $\Delta t_* / \Delta t_{\mathrm{rt}} = 32$, and the profiles become virtually identical
at $r / L \gtrsim 0.075$. This distance is also roughly the compact support radius of the star
particle, within which particles are injected with radiation energy density, and in the context of
radial profiles a rather poorly resolved region. As such, these differences are deemed
unproblematic.

The situation is different for higher ratios $\Delta t_* / \Delta t_{\mathrm{rt}} = 64$ and $128$.
They show clear differences to other profiles. In particular the ratio 128: At 0.5 Myr, it lags
behind other profiles. At 1 Myr, it has caught up with them again, only to lag behind again at 1.5
Myr. This cycle repeats for later times as well. The situation becomes more clear when we look at
the evolution of the ionization front radius over this time span. The ionization front radius is
defined as the radius at which the ionized hydrogen mass fraction is exactly half. It's evolution
through time is shown in Figure~\ref{fig:injection-timesteps-ionization-front-radius} for the same
ratios $\Delta t_* / \Delta t_{\mathrm{rt}}$ as shown before. Here, the influence of the increasing
time intervals between two energy density injections is obvious. For larger time step size ratios,
the ionization front radius expands almost instantaneously after each injection, and then remains
static until the next injection occurs, leading to the step-like curves that can be seen in
Figure~\ref{fig:injection-timesteps-ionization-front-radius}. The step sizes decrease proportionally
to the decreasing time step size ratios, and also decrease with the distance of the ionization
front radius from the source. For  $\Delta t_* / \Delta t_{\mathrm{rt}} = 8$, the steps are only
detectable at early times $t < 0.5$ Myr, while for  $\Delta t_* / \Delta t_{\mathrm{rt}} = 16$,
they are visible until $t \sim 2$ Myr.

The step-wise ionization for higher time step size ratios (save for perhaps $\Delta t_* / \Delta
t_{\mathrm{rt}} = 128$ in the vicinity of the center of the box) shows no signs of affecting the
profiles on small spatial scales, as evidenced by the smooth profiles shown in
Figure~\ref{fig:injection-timesteps-profiles}, i.e. they show no evidence of a wave-like propagation
of the ionization front. However, they clearly have an effect on larger scales. For example for the
ratio $\Delta t_* / \Delta t_{\mathrm{rt}} = 128$, after the first injection the ionization front
jumps nearly instantly over a region of $\sim 0.5$kpc, so essentially a region with diameter of
1kpc was ionized instantly. Clearly that is a problem, and there is a need to restrict the star
time step sizes according to the local radiative transfer time step sizes. Following the results of
Figure~\ref{fig:injection-timesteps-ionization-front-radius}, the upper threshold shouldn't exceed
$\Delta t_* / \Delta t_{\mathrm{rt}} = 16$, and probably needs to be lower for more luminous sources
of radiation.  Throughout this work, it was sufficient to restrict the time step sizes using a
global upper limit for star particles time step sizes, such that the injection is usually performed
at the same rate as the radiation transport. A more flexible way, where star particles also collect
neighboring gas particles' radiative transfer time step sizes and limit theirs according to the
neighbors' values is left for future work.

\section{Radiation Transport And Thermochemistry}\label{chap:IL6}

This section tests the radiative transfer and the thermochemistry following the standard tests set
by the comparison paper \cite{ilievCosmologicalRadiativeTransfer2006} (hereafter IL6). They
prescribe a series of tests for static gas fields, i.e. all hydrodynamics is turned off. Radiation
hydrodynamics will be discussed in the subsequent section. The gas temperature and internal energy
is allowed to evolve due to present radiation fields (with the exception of Test 1, where the
temperature is kept constant).

Unless noted otherwise, the default parameters used with \GEARRT are to use the GLF Riemann
solver, the minmod limiter, and the ``no flux'' flux injection model.\footnote{
Note that in the current implementation of \GEARRT, the ``no flux'' flux injection model and the
minmod flux limiter are the default choices, and are not actually provided as free parameters to
users. Tests described in Sections~\ref{chap:rt-riemann-limiters} and \ref{chap:results-injection}
revealed possible problems with other choices in certain circumstances, in particular possible
instabilities for flux limiters other than the minmod limiter, and as such could be dangerous to use
carelessly. To prevent conceivable catastrophic failures, the choice has been taken from users.
} Furthermore, the underlying particle distribution is glass-like. For both star and gas particles,
the smoothing length is determined by the default parameter $\eta_{res} = 1.2348$
(eq.~\ref{eq:number-of-neighbors}). This choice results in $\sim 48$ neighbors in 3D for the cubic
spline kernel, which was used for all tests without exception.

The reference solutions shown in this section are the solutions of the codes which participated in
IL6. Most of the participating codes, namely
\codename{C2Ray} \citep{mellema2RayNewMethod2006},
\codename{CRASH} \citep{ciardiCosmologicalReionizationFirst2001, maselliCRASHRadiativeTransfer2003},
\codename{RSPH} \citep{susaSmoothedParticleHydrodynamics2006},
\codename{Zeus-MP} \citep{whalenMultistepAlgorithmRadiation2006},
\codename{IFT} \citep{alvarezIIRegionFirst2006},
\codename{ART} \citep{nakamotoEffectsRadiativeTransfer2001},
\codename{FTTE} \citep{razoumovFullyThreadedTransport2005}, and
\codename{FLASH-HC} \citep{rijkhorstHybridCharacteristics3D2006},
used some variant of a ray tracing method to solve the radiative transfer.
\codename{OTVET} \citep{gnedinMultidimensionalCosmologicalRadiative2001} used a moment based method,
but unlike \GEARRT, which uses the M1 closure,  it used the ``Optically Thin Variable Eddington
Tensor'' (OTVET) closure instead.
Lastly, \codename{SimpleX} \citep{ritzerveldTriangulatingRadiationRadiative2004} solved the equation
of radiative transfer along the connecting lines (Delauney tessellations) of sampling points
placed on a irregular, deformed mesh.

The reference solutions and prescribed initial conditions for the test are publicly available under
\url{https://astronomy.sussex.ac.uk/~iti20/RT_comparison_project/index.html}.
%
The initial conditions to run all the tests described and shown in the subsequent sections with
\GEARRT are publicly available under \url{https://github.com/SWIFTSIM/swiftsim-rt-tools}.
In figures showing (radial) profiles of solutions, where a comparison with a multitude of reference
solutions is sensible and manageable, all the reference solutions will be shown as gray lines. One
particular reference solution, the one of the \codename{C2Ray} code, will be highlighted
throughout. \codename{C2Ray} was selected for two reasons: Firstly, \codename{C2Ray} was one of the
few codes whose reference solutions were available for every presented test in both the
\citet{ilievCosmologicalRadiativeTransfer2006} and \citet{ilievCosmologicalRadiativeTransfer2009}
papers. Secondly, \codename{C2Ray} was also used as a reference in \citet{ramses-rt13}, and thus
presenting the solutions along with the same reference facilitates easier comparison by eye between
\GEARRT and \codename{Ramses-RT} as well.

\subsection{Iliev Test 0}\label{chap:Iliev0}

\begin{figure}
 \centering
 \includegraphics[width=\textwidth]{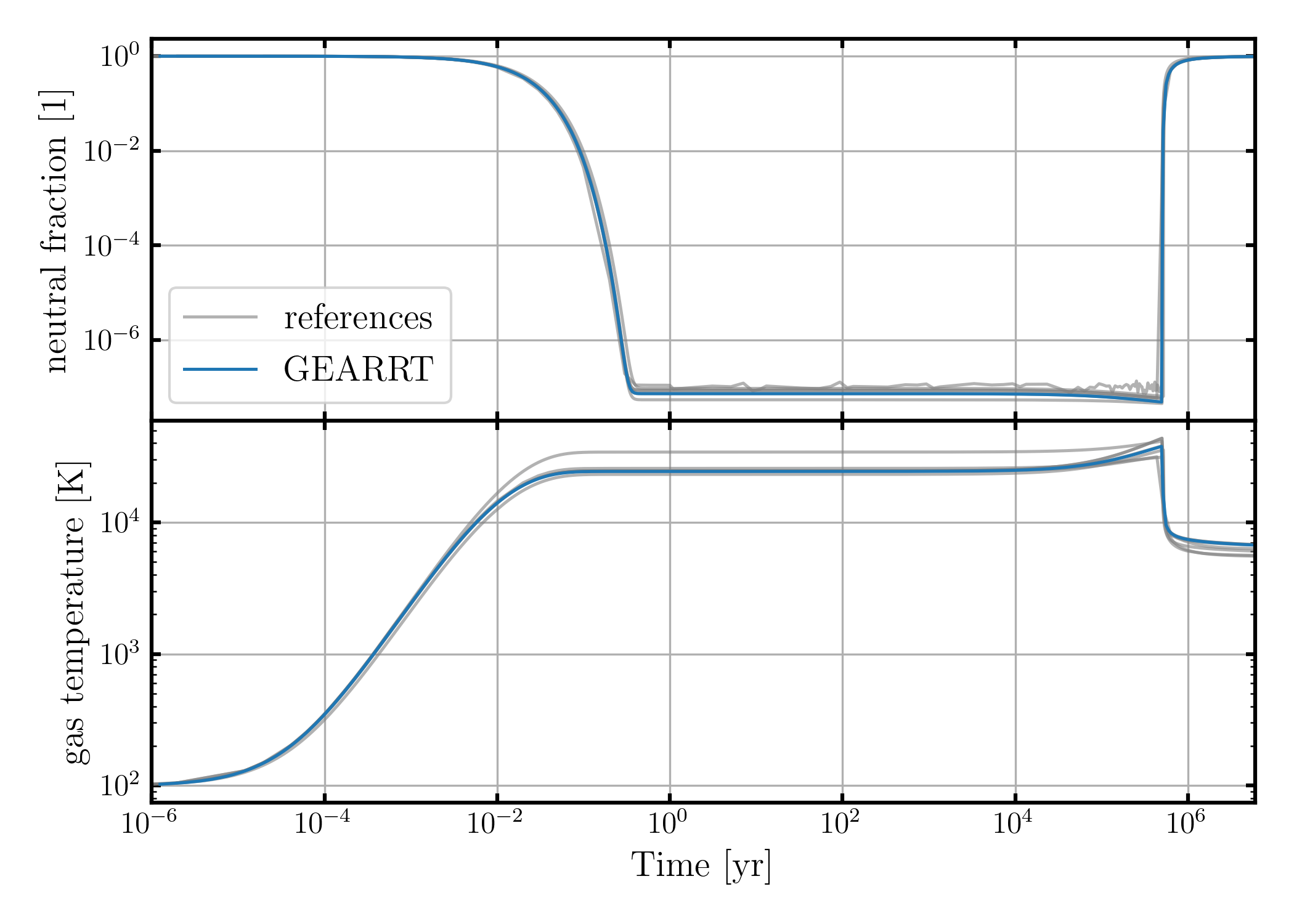}
 \caption{
 Test 0: Single zone photo-heating and ionization with subsequent cooling and recombination.
 }
 \label{fig:iliev0}
\end{figure}

Test 0 doesn't involve radiative transfer, but tests only the photo-ionization and photo-heating for
a given radiation field. A single cell (or in the case of \GEARRT, particle) containing only
hydrogen gas with number density $n = 1$cm$^{-3}$ and temperature $T = 100K$ is subjected to a
fixed photo-ionizing flux of $F = 10^{12}$ photons/s/cm$^2$ with a blackbody spectrum for 0.5 Myr.
During this time, the gas heats up and becomes ionized. After the initial 0.5 Myr, the ionizing flux
is turned off and the particle is let to cool and recombine for further 5 Myr. Admittedly, this
is a more of a test of \grackle rather than of \GEARRT, but I show it nonetheless for completeness.
The resulting neutral fraction of hydrogen and the gas temperature over time are shown in
Figure~\ref{fig:iliev0} and show good agreement with the reference solutions.

\subsection{Iliev Test 1}\label{chap:Iliev1}

\begin{figure}
 \centering
 \includegraphics[width=.49\textwidth]{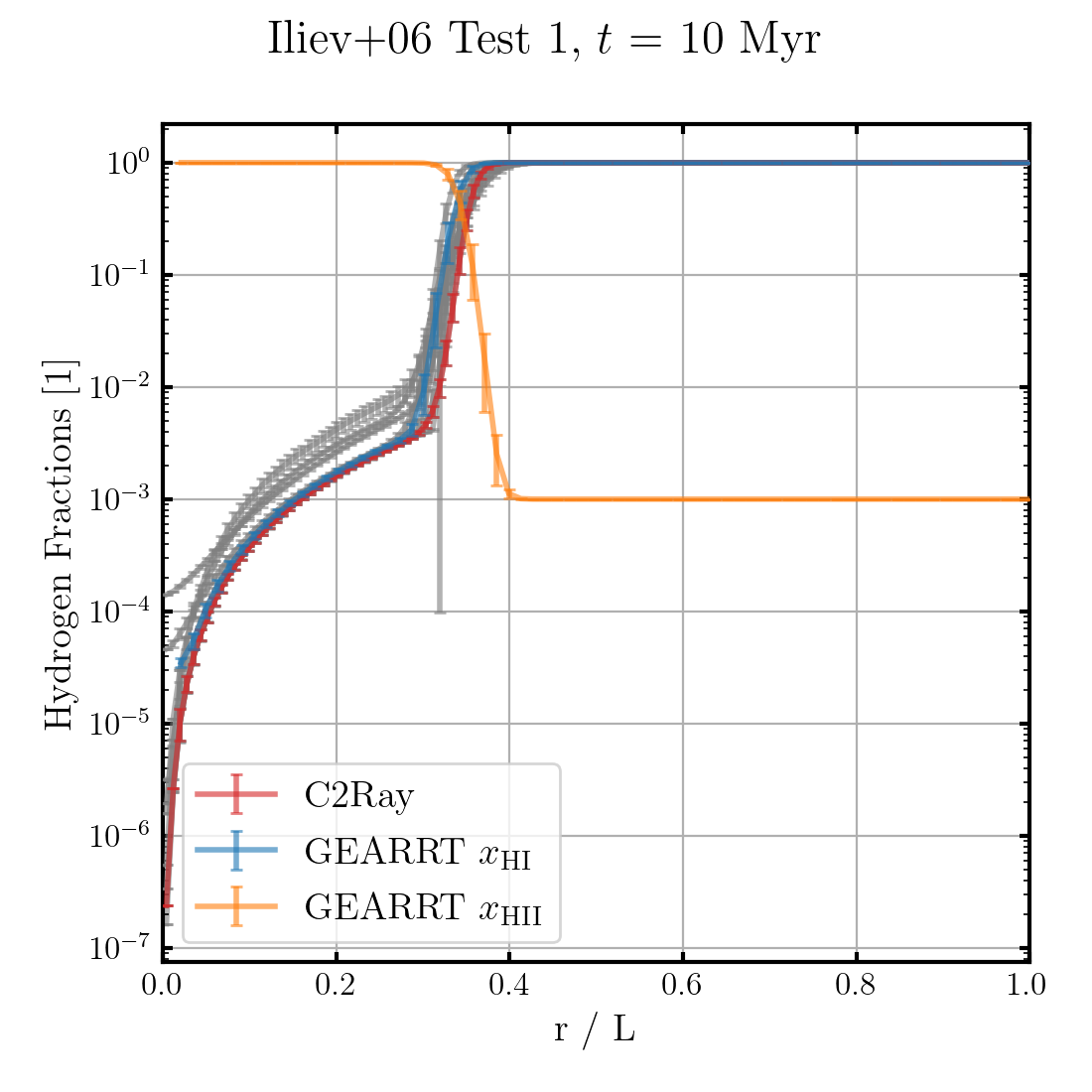}%
 \includegraphics[width=.49\textwidth]{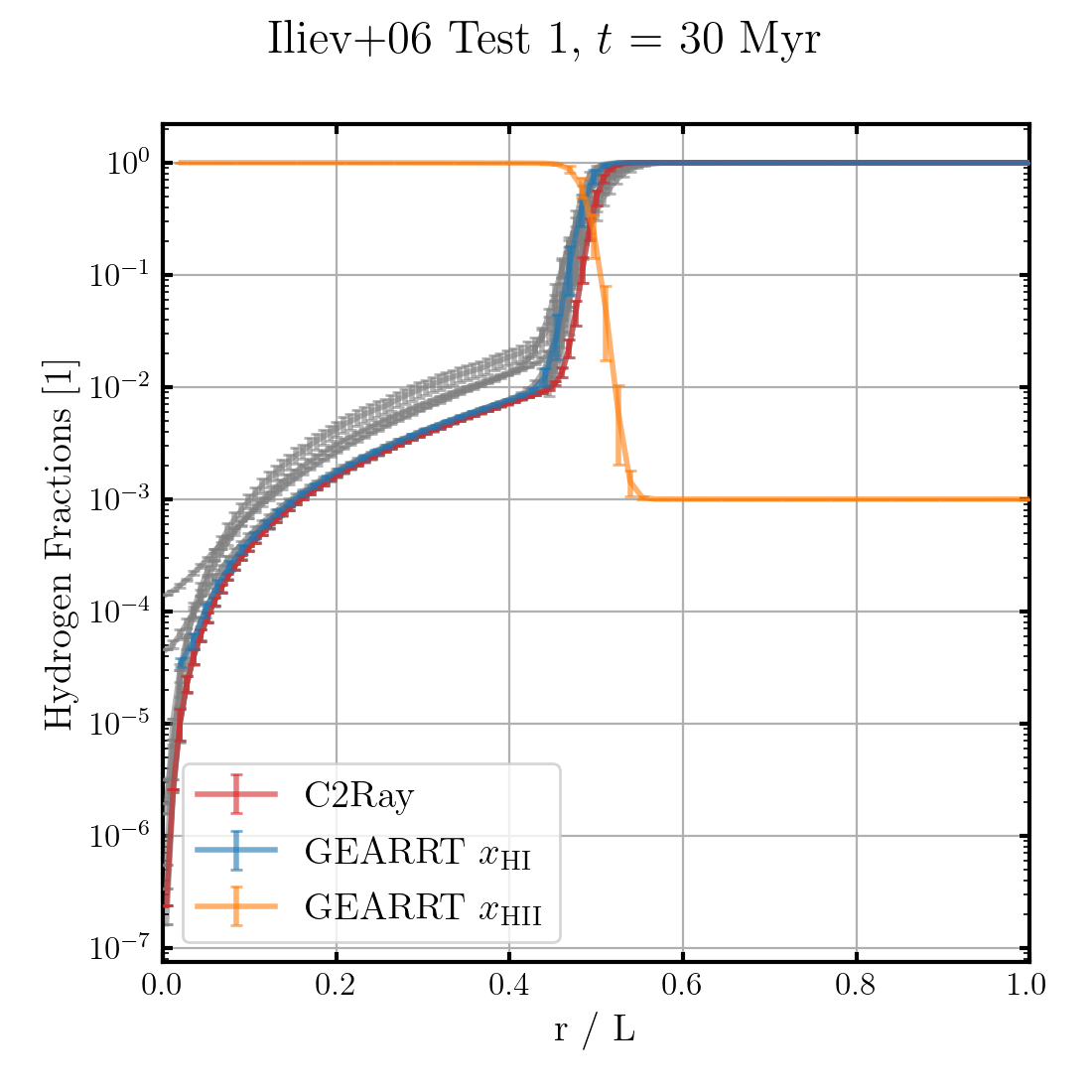}\\%
 \includegraphics[width=.49\textwidth]{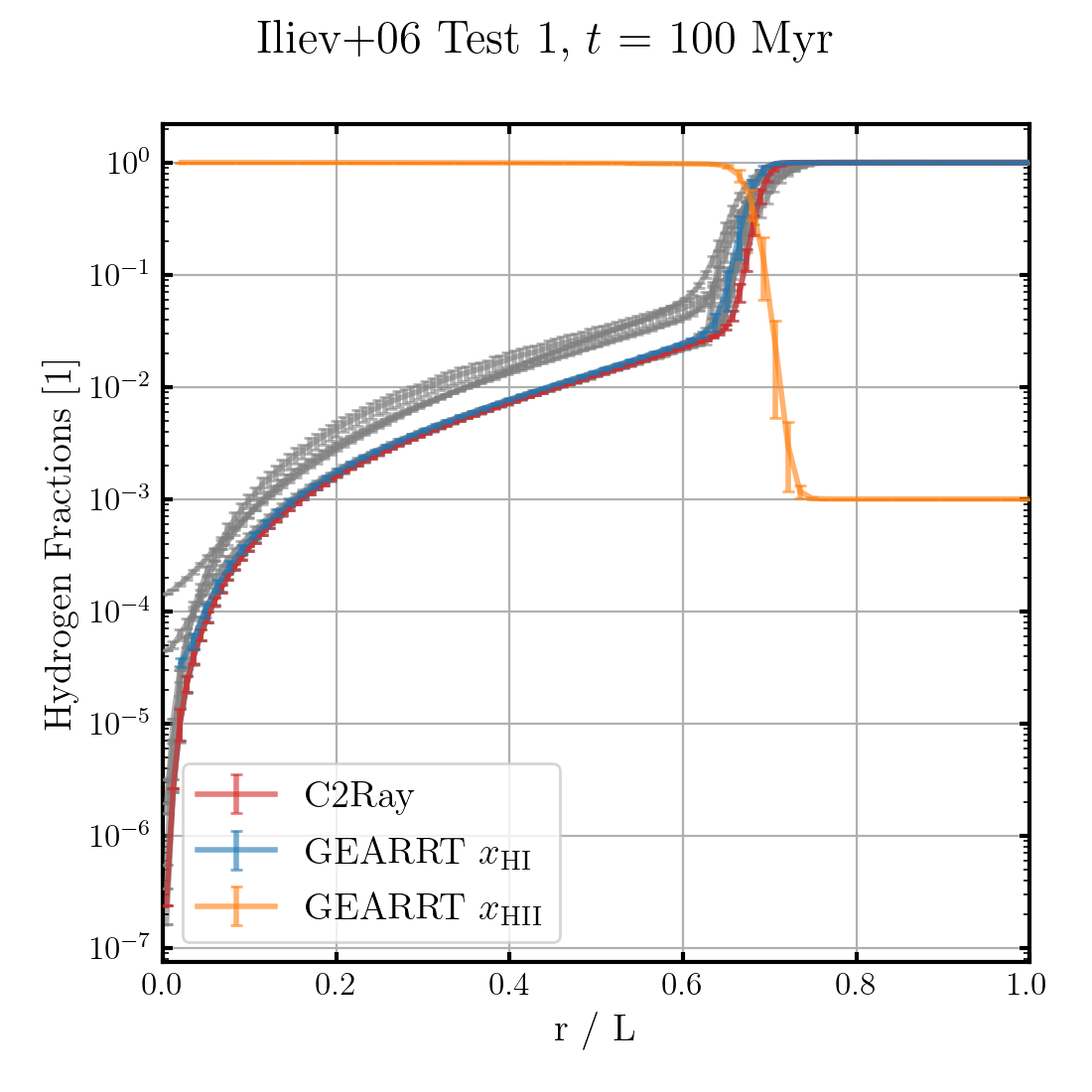}%
 \includegraphics[width=.49\textwidth]{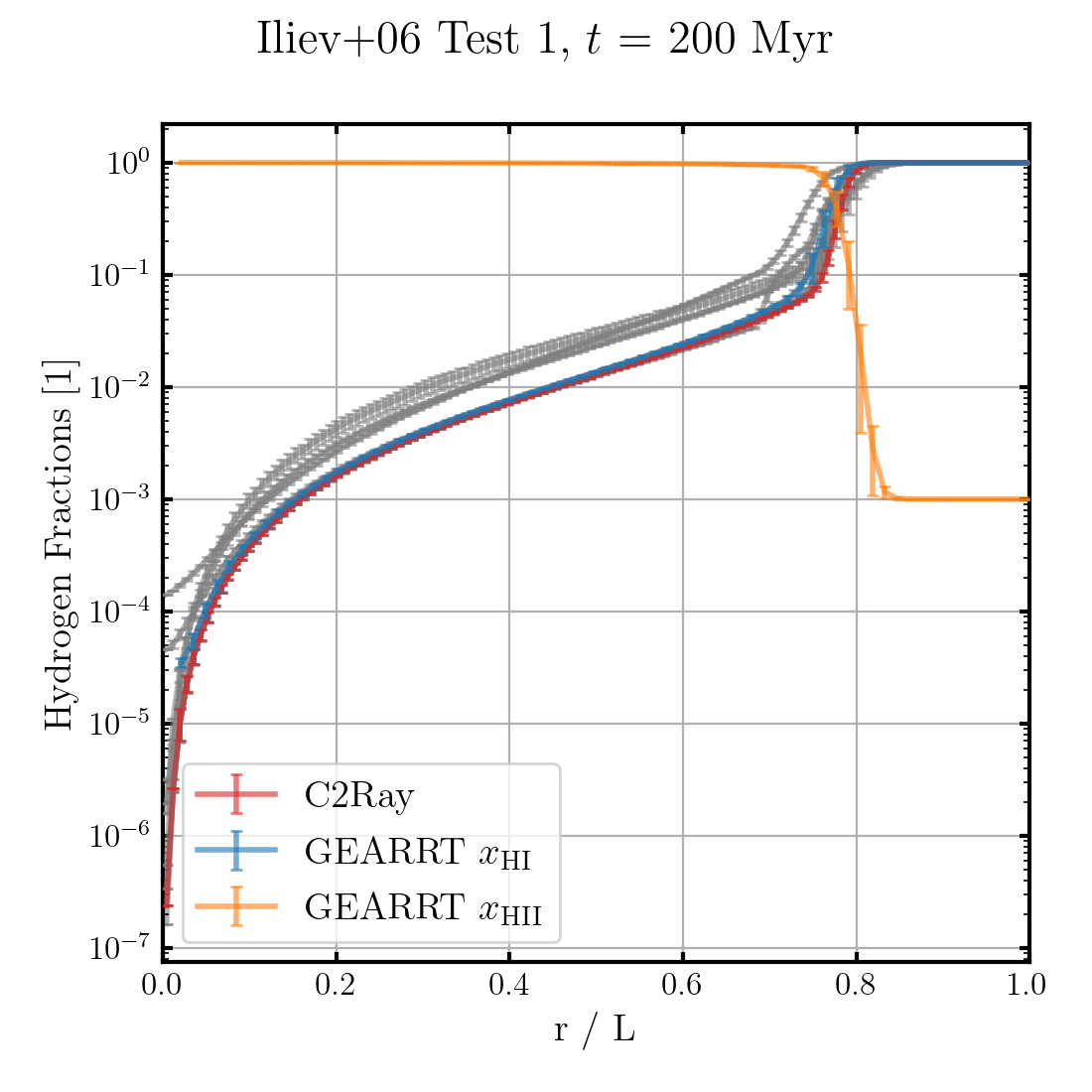}%
 \caption{
Spherically averaged ionized and neutral fractions of hydrogen in Test 1, where a single source
emits monochromatic ionizing radiation with frequency $h \nu = 13.6$eV at 10, 30, 100, and 200 Myr,
while the temperature is artificially held fixed at $T = 10^4$K. The error bars are standard
deviations.
 }
 \label{fig:iliev1}
\end{figure}

\begin{figure}
 \centering
 \includegraphics[width=.49\textwidth]{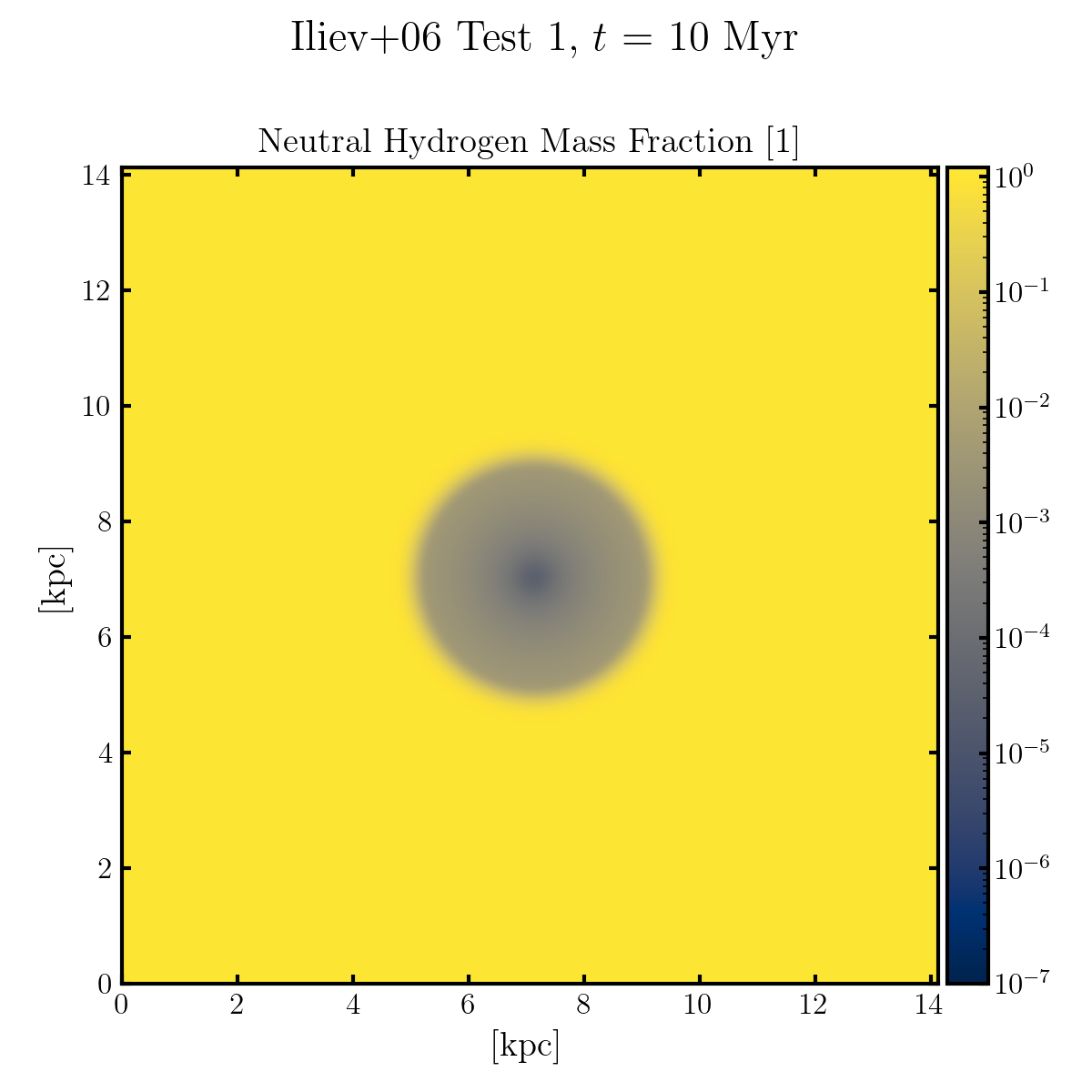}%
 \includegraphics[width=.49\textwidth]{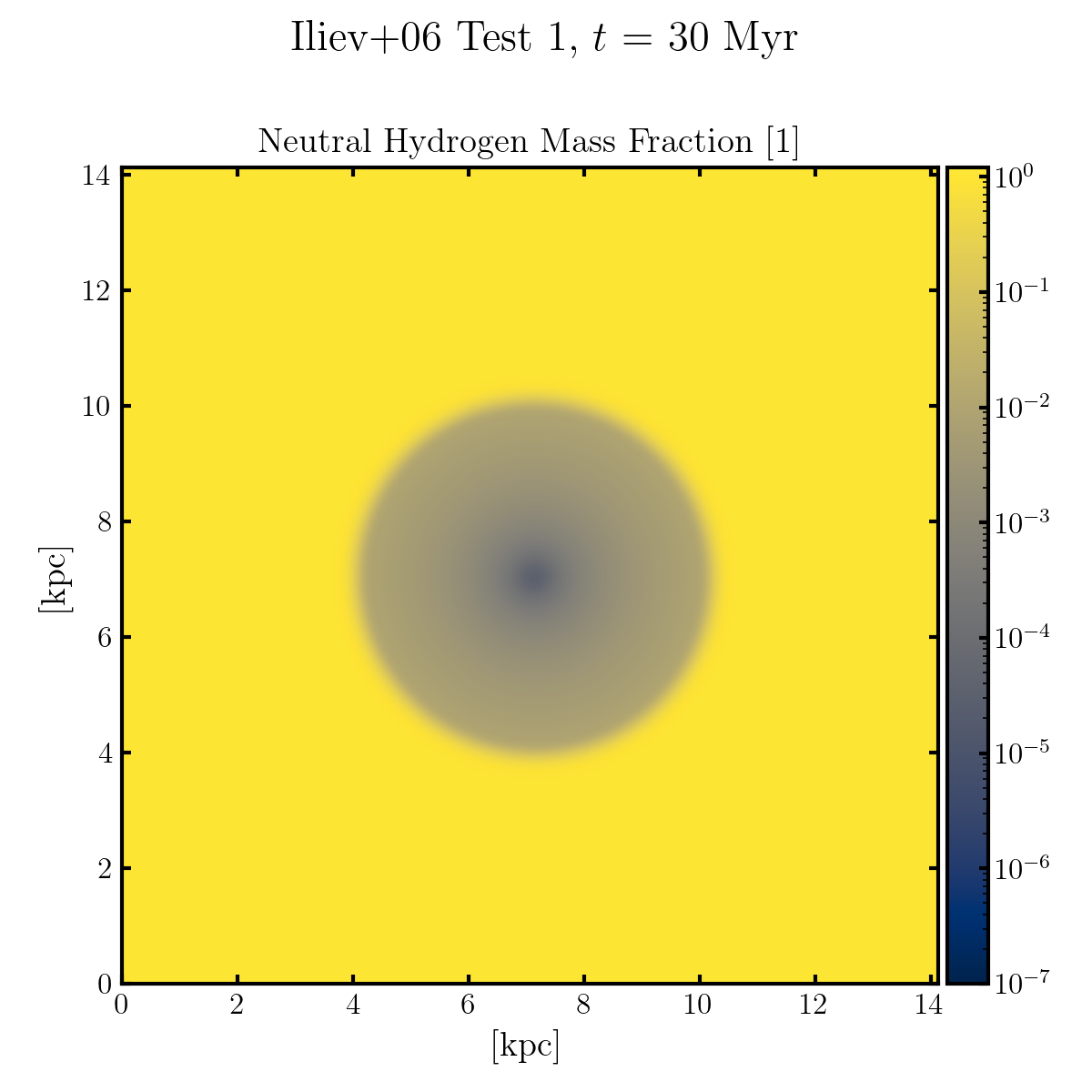}\\%
 \includegraphics[width=.49\textwidth]{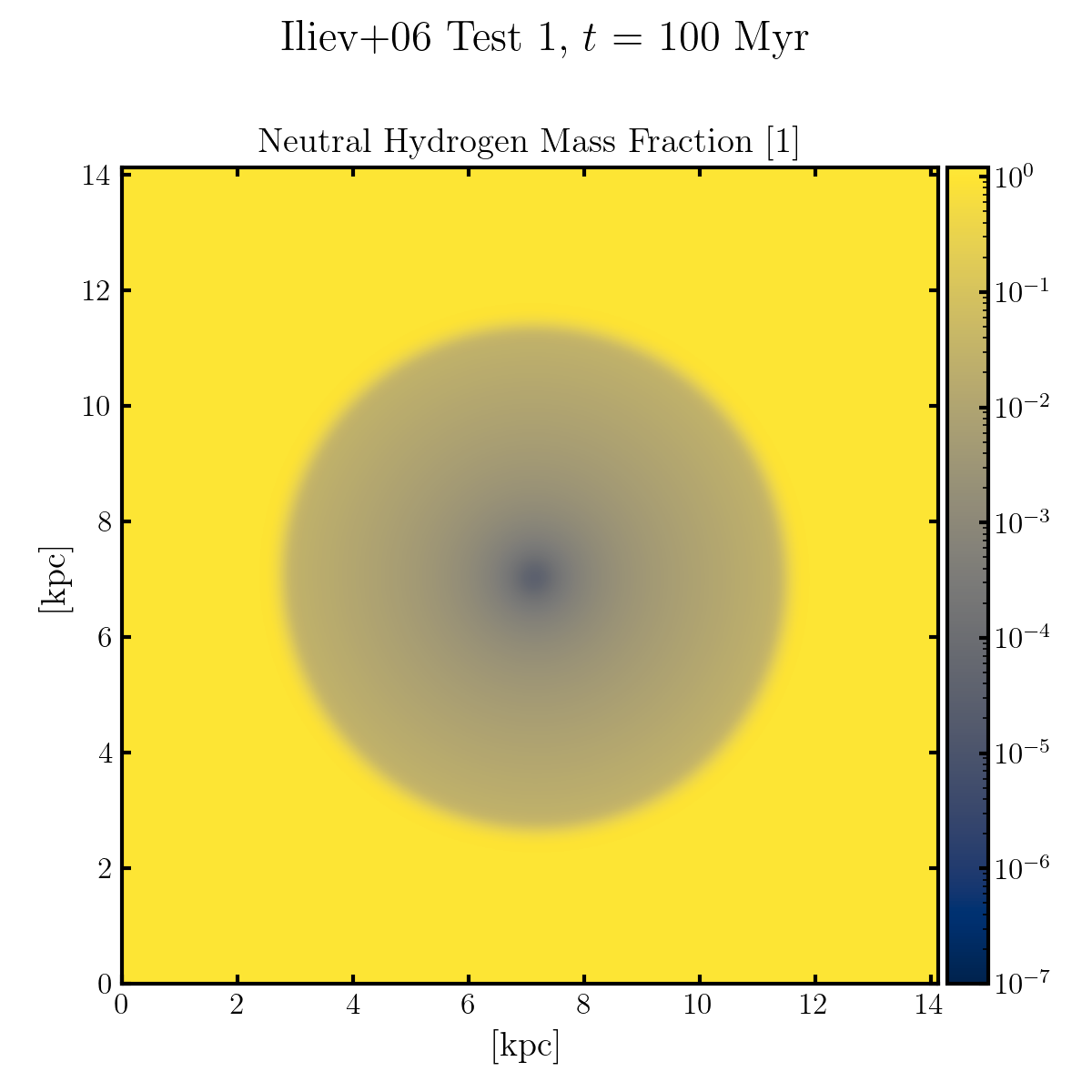}%
 \includegraphics[width=.49\textwidth]{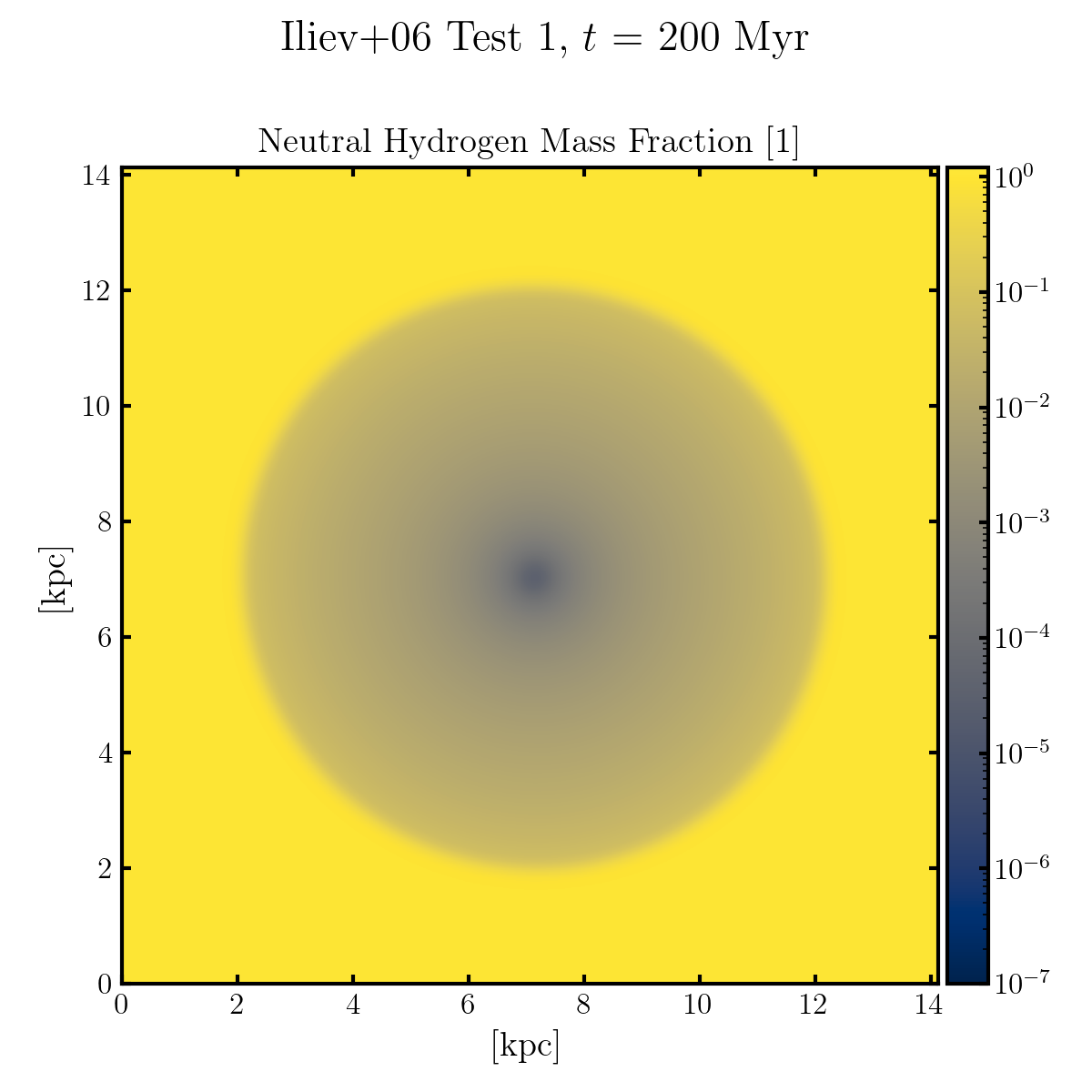}%
 \caption{
Slices through the mid-plane of the box of the ionized hydrogen mass fraction in Test 1, where a
single source emits monochromatic ionizing radiation with frequency $h \nu = 13.6$eV at 10, 30, 100,
and 200 Myr.
 }
 \label{fig:iliev1-slices}
\end{figure}

\begin{figure}
 \centering
 \includegraphics[width=.7\textwidth]{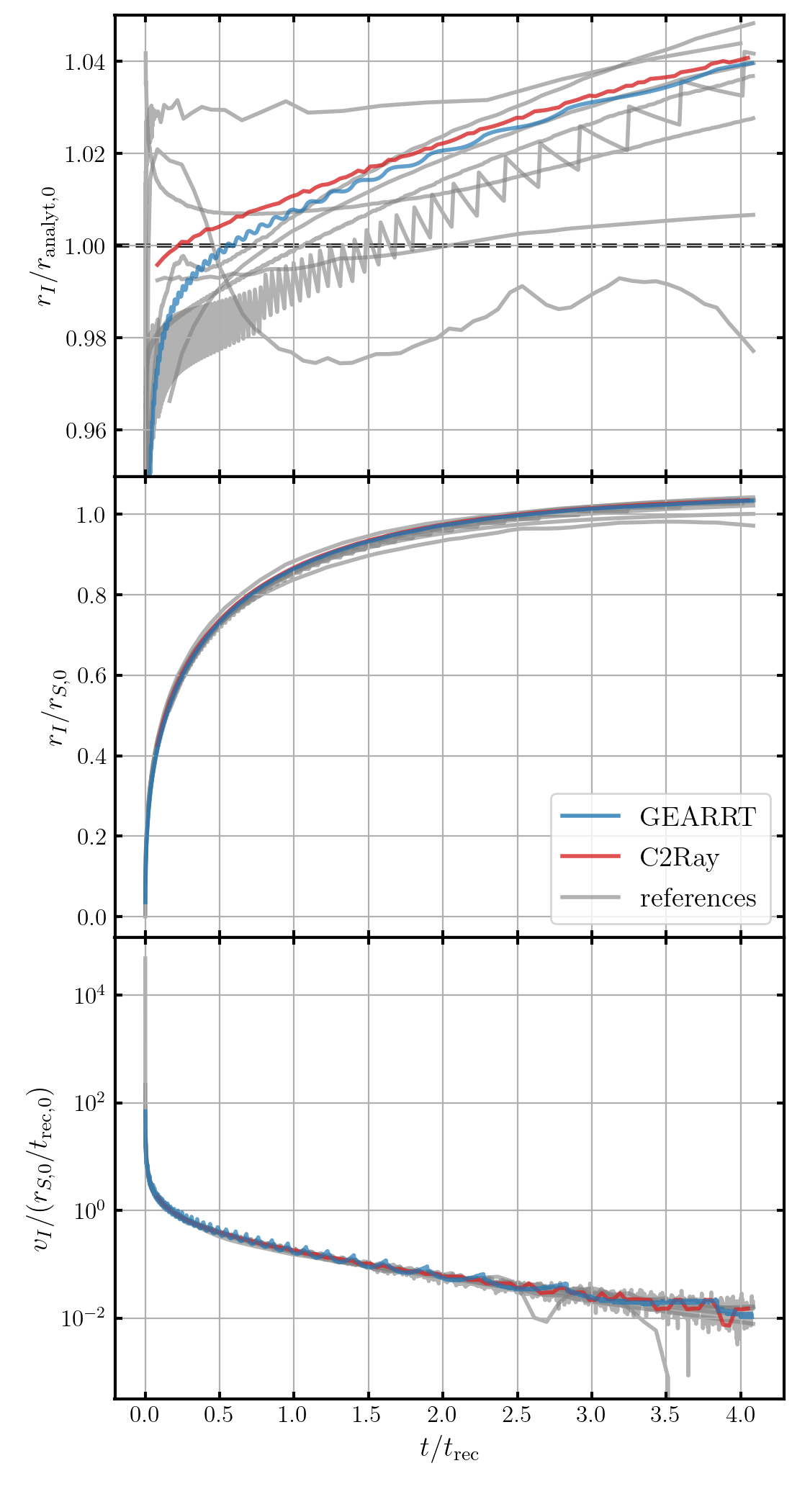}%
 \caption{
 Evolution of the I-front position and velocity over time for Test 1.
 Top: Evolution of the I-front position over time, compared to the analytical position
(eq.~\ref{eq:rI}) and reference solutions from other codes.
 Middle: Evolution of the I-front position over time, compared to the final Str\"omgren radius
(eq.~\ref{eq:rS}).
 Bottom: Evolution of the I-front velocity.
 }
 \label{fig:iliev1-Ifront}
\end{figure}

Test 1 consists of a single source emitting monochromatic ($h\nu = 13.6$eV) radiation into an
initially neutral uniform hydrogen gas with number density $n = 10^{-3}$cm$^{-3}$. The temperature
of the gas is being kept constant at $T = 10^4$K throughout the entire simulation. The test in IL6
prescribe to put the ionizing source in the lower left corner of the simulation box, and to use
reflective boundary conditions along the three box sides adjacent to the source. However, in order
to avoid having to construct reflective boundary conditions for \GEARRT, which is not a trivial
matter with particles, I put the source in the center of the box, and make the box have twice the
size of what is prescribed in IL6, which results in a box size of $L = 13.2$ kpc. I use a glass file
consisting of $128^3$ particles for the particle positions of the uniform gas. IL6 use $128^3$
cells, but with half the box size compared to the one used here, so the results presented here will
be at half the spatial resolution compared to the results in IL6.\footnote{
While in principle it would be possible to run the tests at the same resolution as in IL6, the
results obtained with the reduced resolution are perfectly sufficient to demonstrate the validity of
\GEARRT, and I opted to avoid the otherwise increased computational load of a factor of eight.}
The ionizing photon rate emitted by the source is $\dot{N}_\gamma = 5 \times 10^{48}$ photons/s. I
reduce the speed of light by a factor of 100. (The validity of this value for the reduced speed of
light will be verified in Test 2).

The expected solution should be an expanding HII region, known as a Str\"omgren sphere. Assuming
the ionization front (I-front) is infinitely sharp, the I-front radius has an analytical solution
given by

\begin{align}
    r_I = r_S ( 1 - \exp(-t/t_{rec}))^{1/3} \label{eq:rI}
\end{align}

and its velocity is

\begin{align}
    v_I = \deldt{r_I} = \frac{r_S}{3 t_{rec}} \frac{\exp(-t/t_{rec})}{( 1 -
\exp(-t/t_{rec}))^{2/3}} \label{eq:vI}
\end{align}

where

\begin{align}
    r_S &= \left[ \frac{3 \dot{N}_\gamma}{4 \pi \alpha(T) n_H^2} \right]^{1/3} \label{eq:rS} \\
    t_{rec} &= \frac{1}{\alpha_B(T) n_H}
\end{align}

are the Str\"omgren radius and the recombination time, respectively.
$\alpha_B(T)$ is the Case B recombination rate of hydrogen. For $T = 10^4$K, $\alpha_B =
2.59 \times 10^{-13}$cm$^3/$s and $\dot{N_\gamma} = 5 \times 10^{48}$ photons/s, we have $r_S =
5.4$ kpc and $t_{rec} = 122.4$ Myr.

Figure~\ref{fig:iliev1} shows the spherically averaged profiles of neutral fractions of hydrogen at
10, 30, 100, and 200 Myr, while Figure~\ref{fig:iliev1-slices} shows the slices through the
mid-plane of the box at the same output times. \GEARRT shows again good agreement with the reference
solutions. The position of the I-front radius, defined as the radius at which the ion mass fraction
is exactly 0.5, and velocity is shown in Figure~\ref{fig:iliev1-Ifront}, and also agrees well with
the reference solutions. At early times ($t \lesssim t_{rec}$) the ionization front radius lags a
little behind compared to the results of \codename{C2Ray}, which is due to the reduced speed of
light used by \GEARRT. At later times, the I-front radius is ahead of the analytical solution for
most reference codes as well as for \GEARRT due to the assumption of a sharp I-front used to derive
the analytical solution. More precisely, the analytical solution assumed that the I-front is a sharp
discontinuity, and that the HII region is fully ionized, which is not exactly the case.
Indeed \citet{pawlikTRAPHICRadiativeTransfer2008} have derived that the analytical equilibrium
solution for the position of the I-front defined as the radius at which the ionization is exactly
half results in 1.05$r_S$, which is in good agreement with the results by \GEARRT and other codes
shown in Figure~\ref{fig:iliev1-Ifront}. In order to reproduce figures close to those presented in
IL6 and \citet{ramses-rt13}, I chose to keep the reference values as specified by IL6.

\subsection{Iliev Test 2}\label{chap:Iliev2}

\begin{figure}
 \centering
 \includegraphics[width=.85\textwidth]{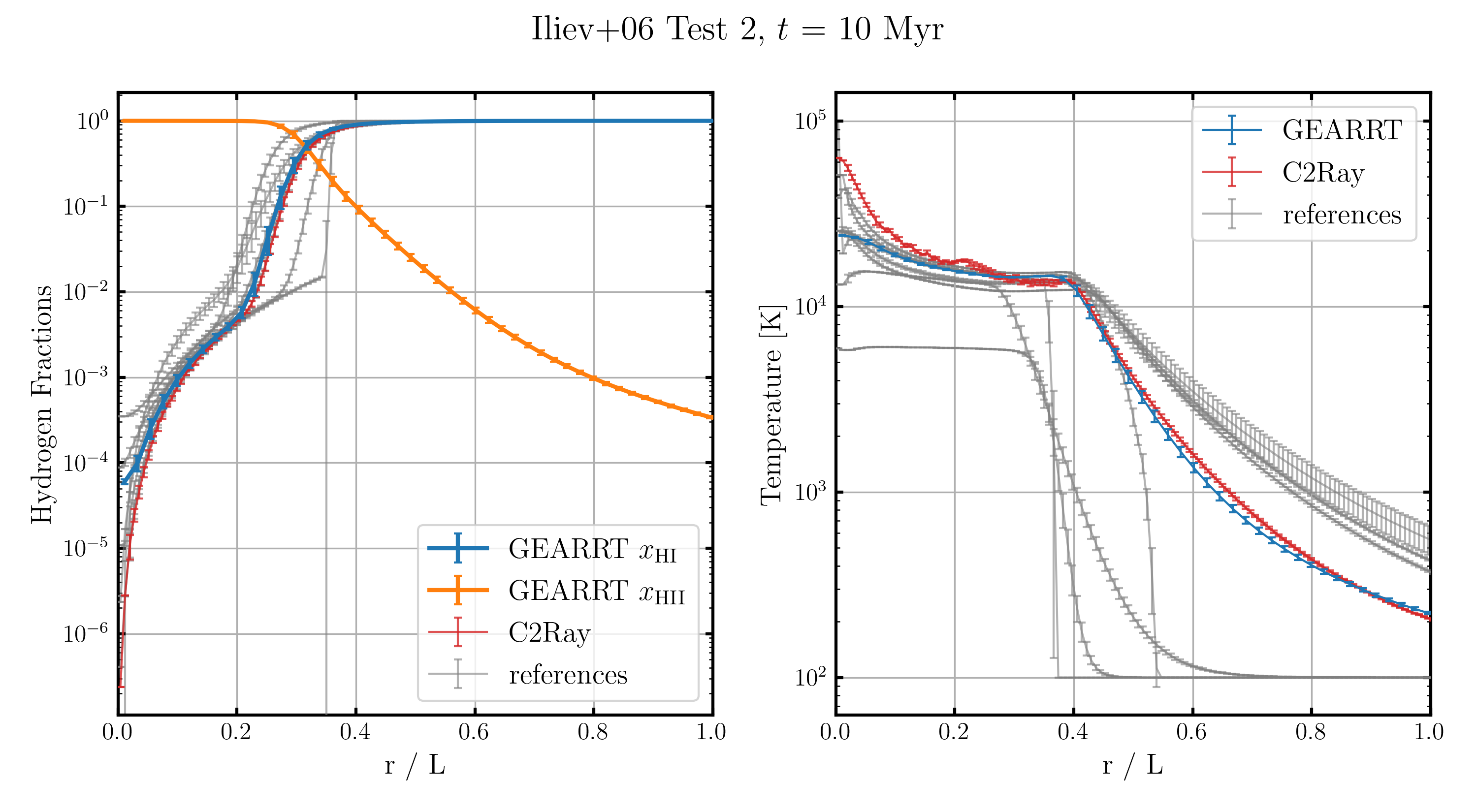}\\%
 \includegraphics[width=.85\textwidth]{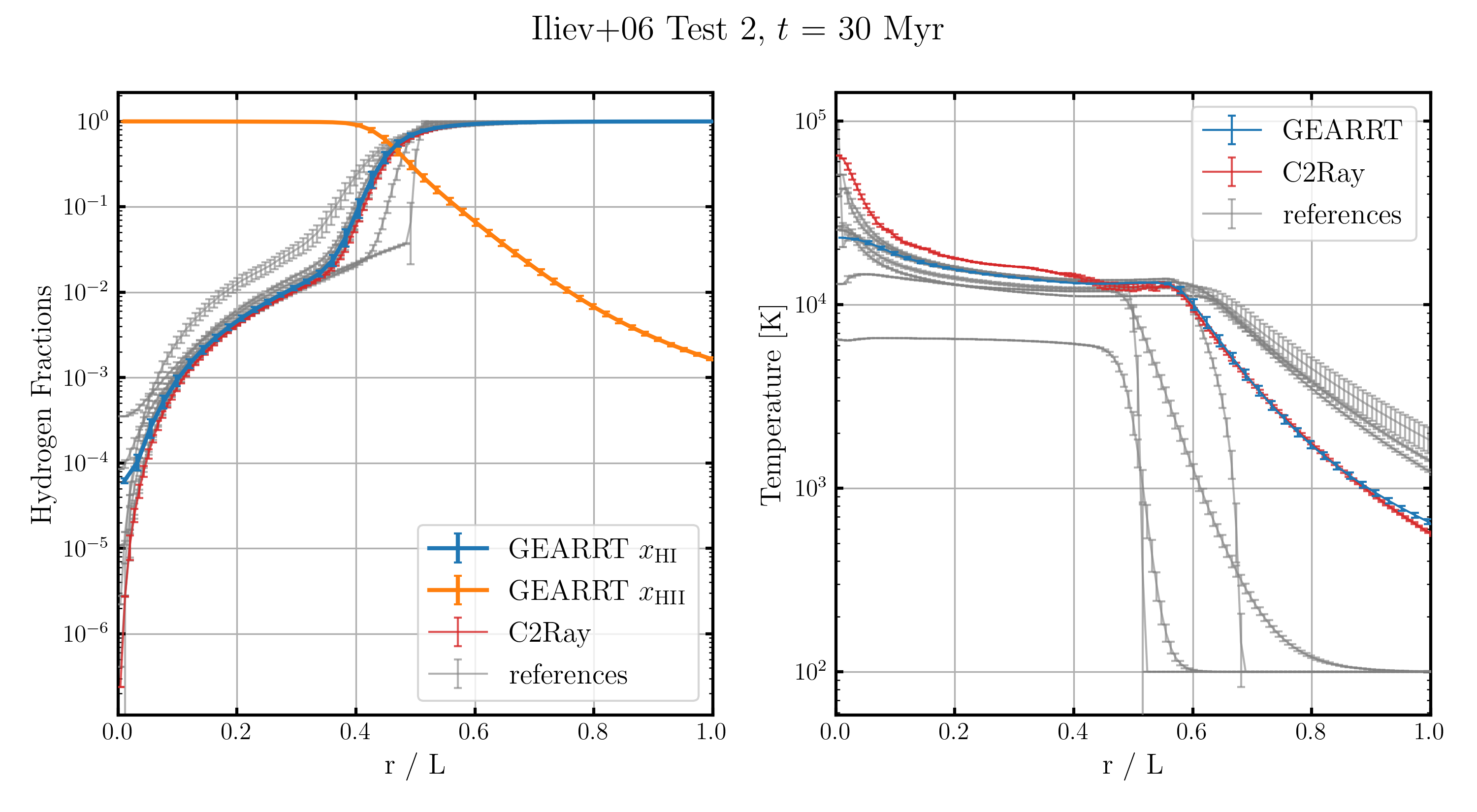}\\%
 \includegraphics[width=.85\textwidth]{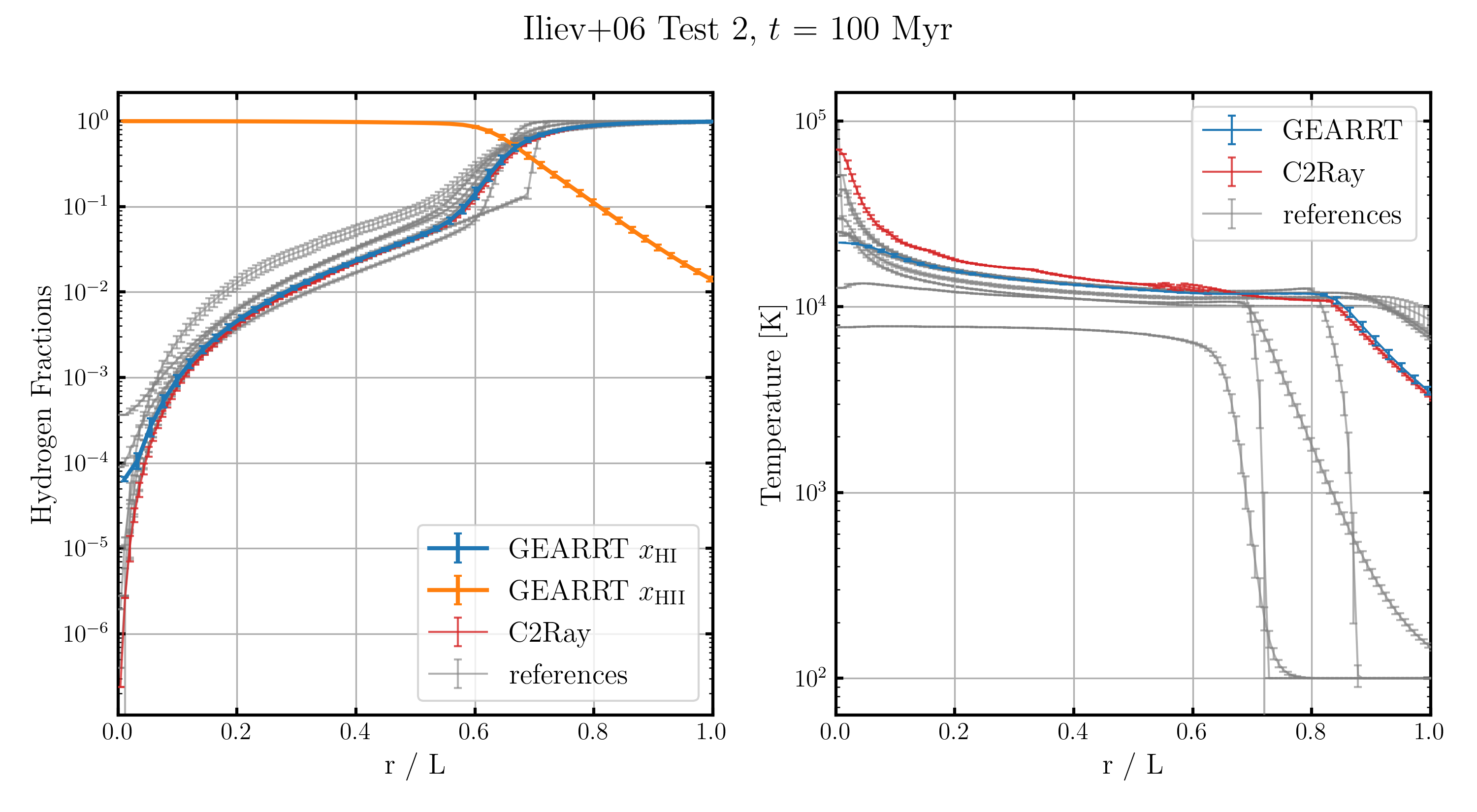}%
 \caption{
Spherically averaged Ionized and neutral fractions of hydrogen and temperatures in  Test 2 at 10,
30, and 100 Myr solved with \GEARRT and the reference solutions from IL6, where the solution of
\codename{C2Ray} has been highlighted. The error bars are standard deviations.
 }
 \label{fig:iliev2}
\end{figure}

\begin{figure}
 \centering
 \includegraphics[width=\textwidth]{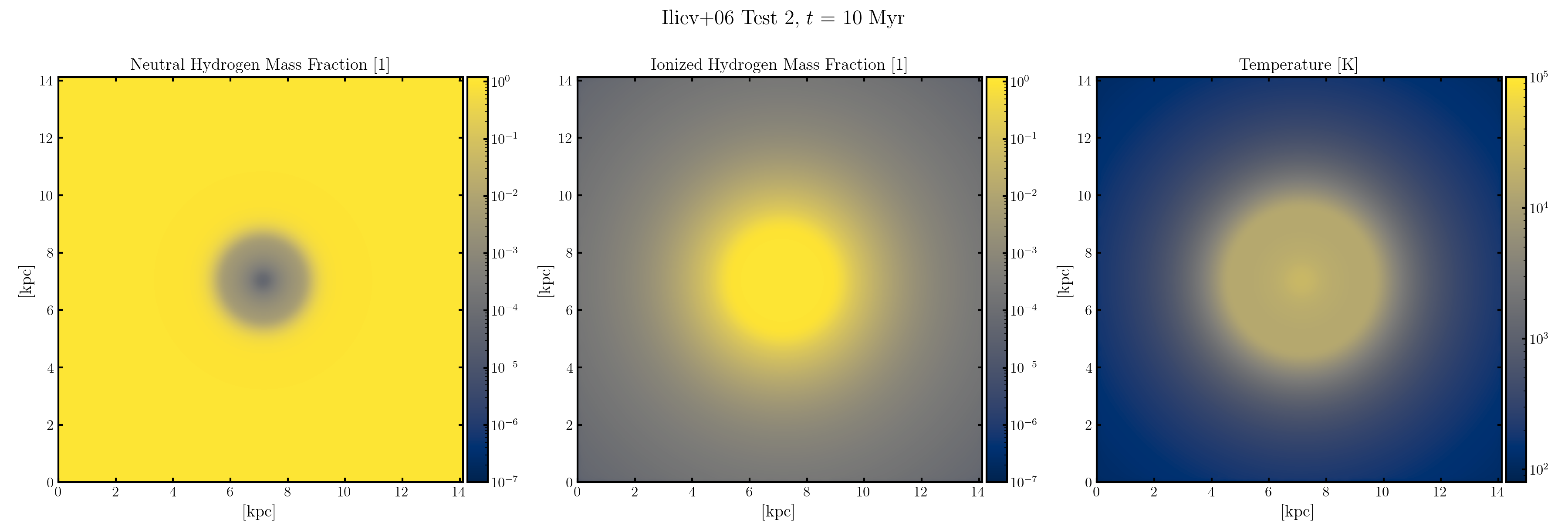}\\%
 \includegraphics[width=\textwidth]{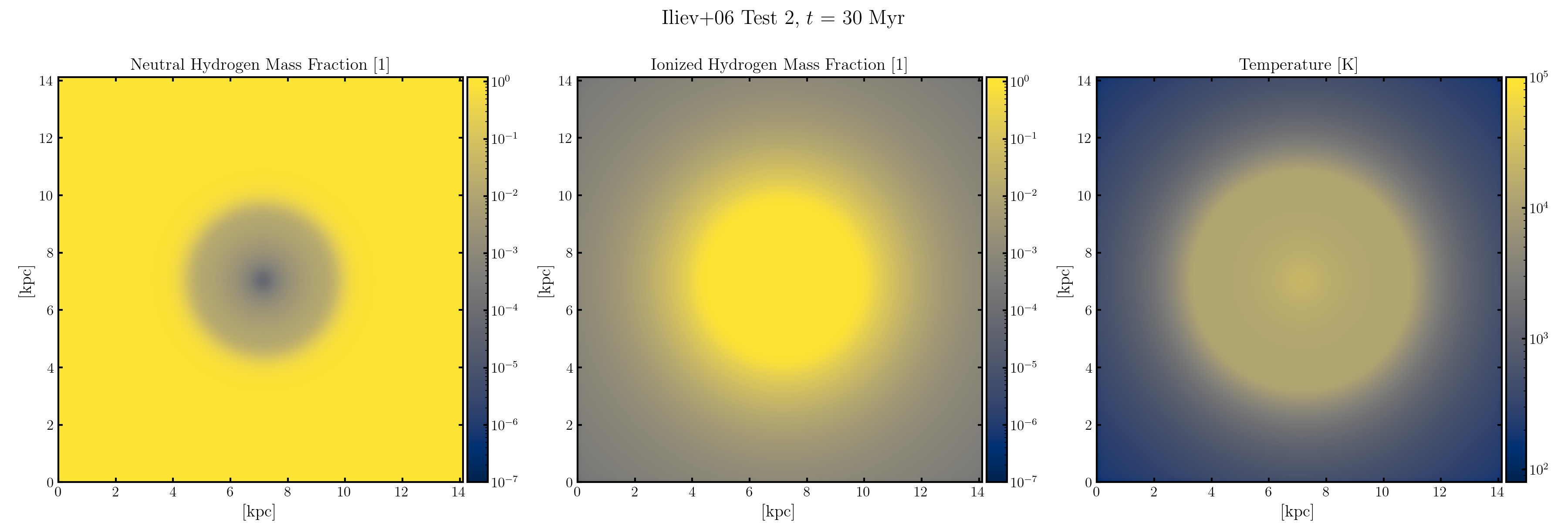}\\%
 \includegraphics[width=\textwidth]{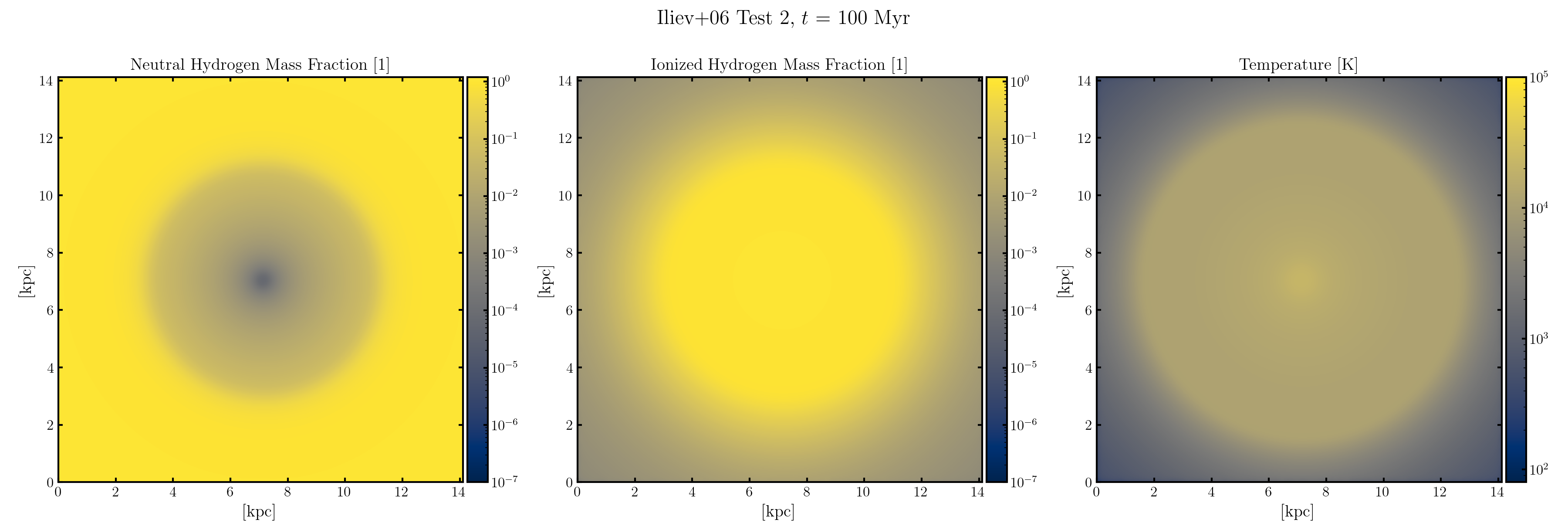}\\%
 \includegraphics[width=\textwidth]{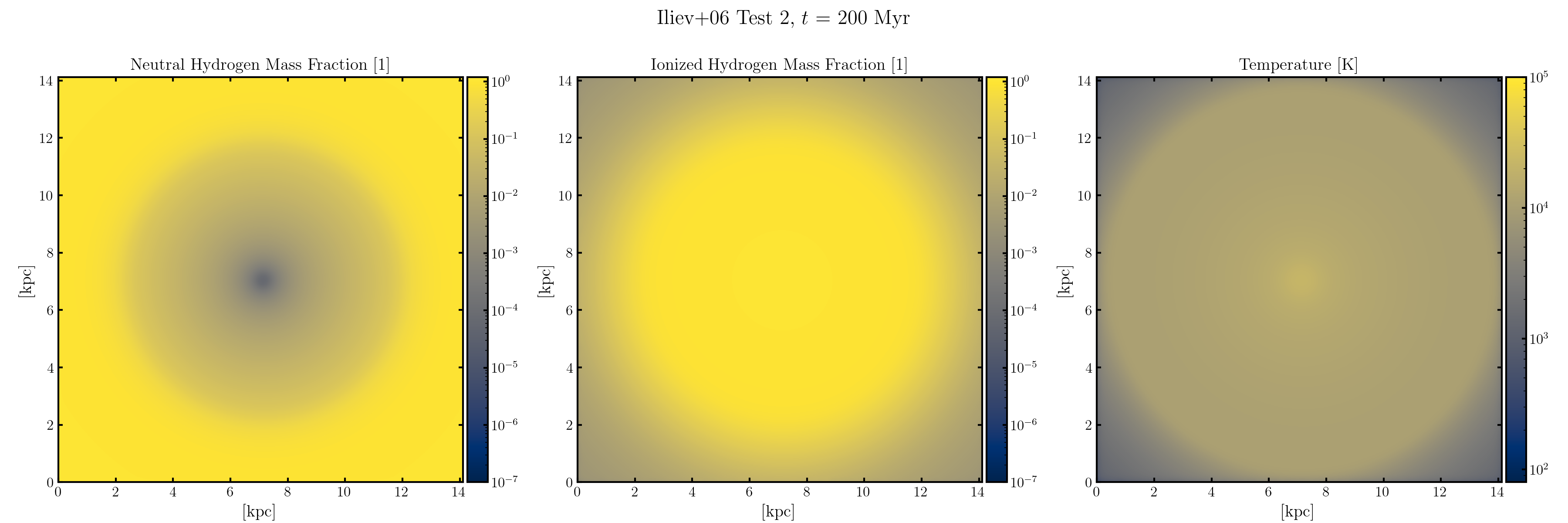}%
 \caption{
Slices through the mid-plane of the box of the ionized and neutral fractions of hydrogen and
temperatures in Test 2 at 10, 30, 100, and 200 Myr solved with \GEARRT.
 }
 \label{fig:iliev2-slices}
\end{figure}

\begin{figure}
 \centering
 \includegraphics[width=.7\textwidth]{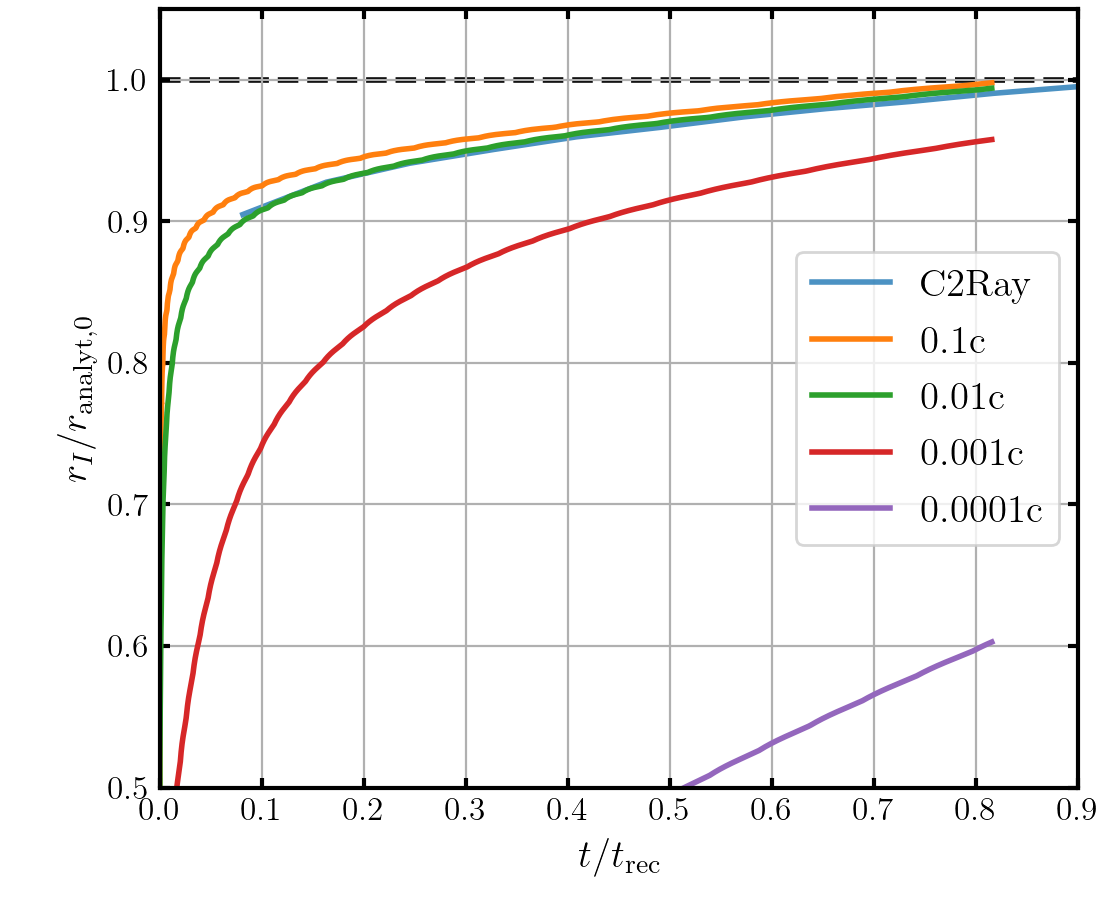}%
 \caption{
 The position of the ionization front for the Test 2 over 100 Myr with different values for the
reduced speed of light, as indicated in the legend, compared to the expected analytical position
given in eq.~\ref{eq:rI}, along with the solution of \codename{C2Ray}, which assumed an infinite
speed of light.
 }
 \label{fig:iliev2-compare-c}
\end{figure}

\begin{figure}
 \centering
 \includegraphics[width=.7\textwidth]{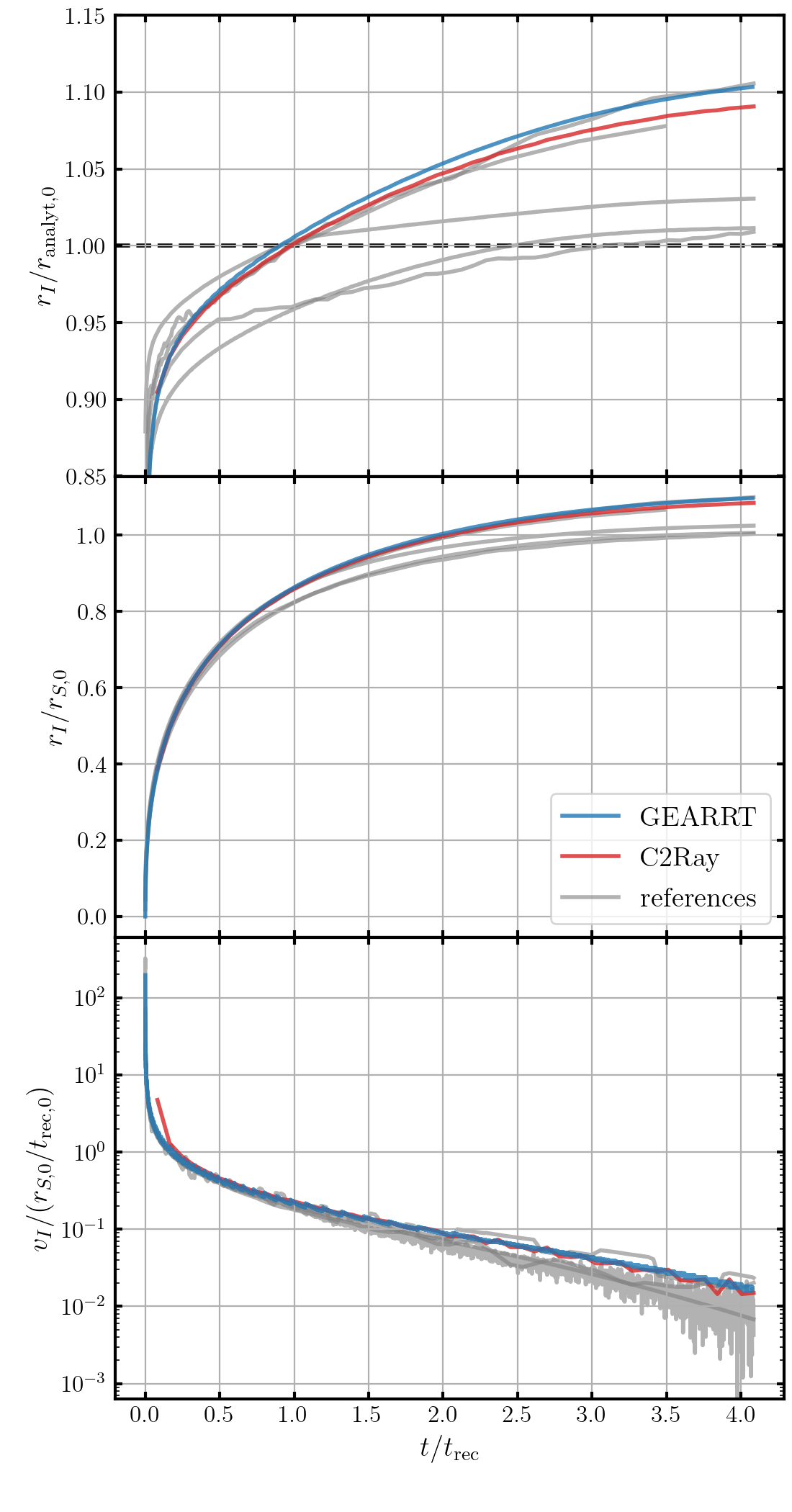}%
 \caption{
 Evolution of the I-front position and velocity over time for Test 2.
 Top: Evolution of the I-front position over time, compared to the analytical position of Test 1
(eq.~\ref{eq:rI}).
 Middle: Evolution of the I-front position over time, compared to the final Str\"omgren radius
(eq.~\ref{eq:rS}).
 Bottom: Evolution of the I-front velocity.
 }
 \label{fig:iliev2-Ifront}
\end{figure}

\begin{figure}
 \centering
 \includegraphics[width=\textwidth]{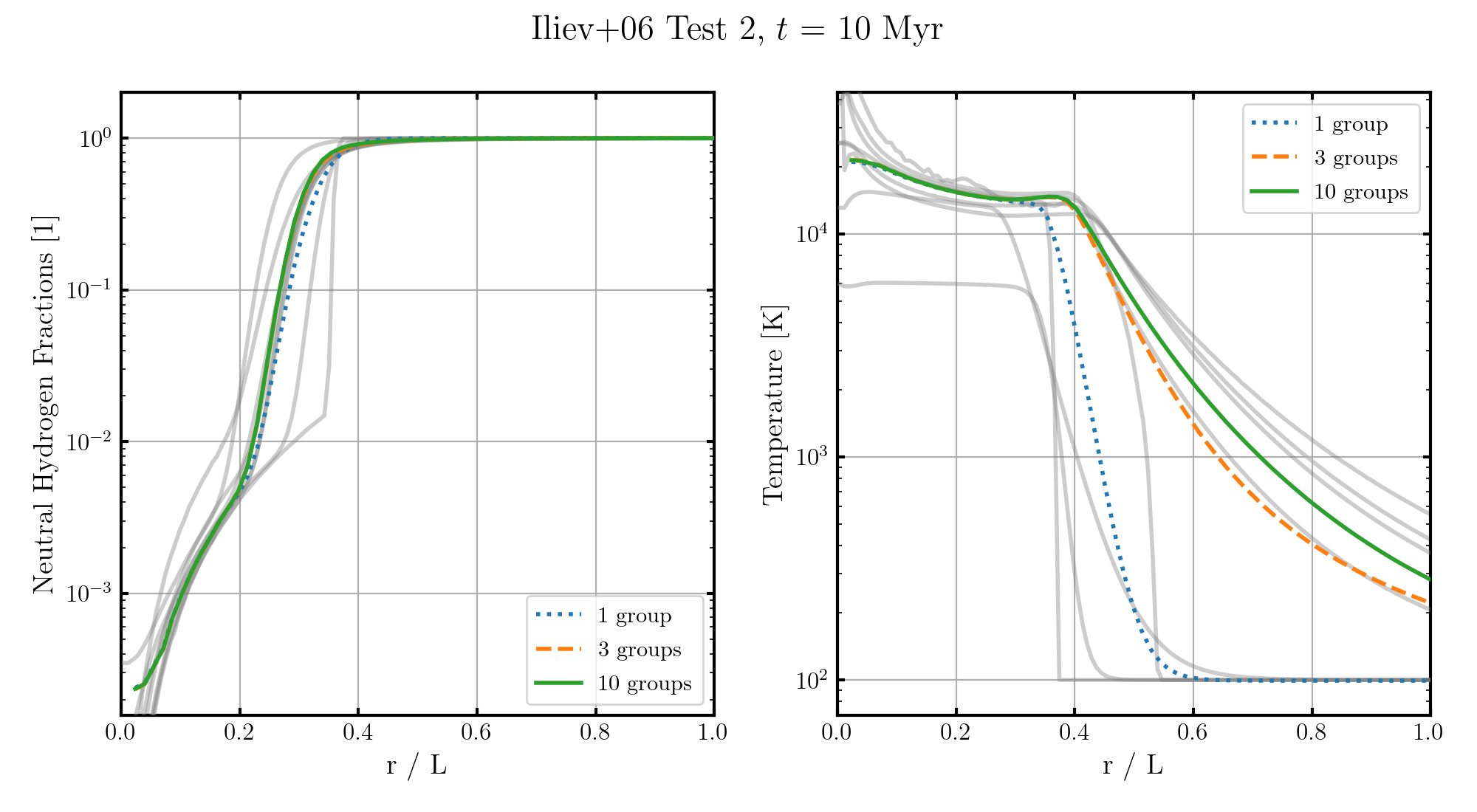}%
 \caption{
Spherically averaged ionized and neutral fractions of hydrogen and temperatures of Test 2 at 10
Myr for 1, 3, and 10 photon frequency groups used. With an increasing number of groups, the
approximation becomes more accurate and closer to treating frequencies individually. The average
ionization cross sections are treated more accurately, and high energy photons have lower
interaction rates, as they should. The high energy photons are then able to reach regions further
from the source, where they can heat the gas.
 }
 \label{fig:iliev2-photon-groups}
\end{figure}

\begin{figure}
 \centering
 \includegraphics[width=\textwidth]{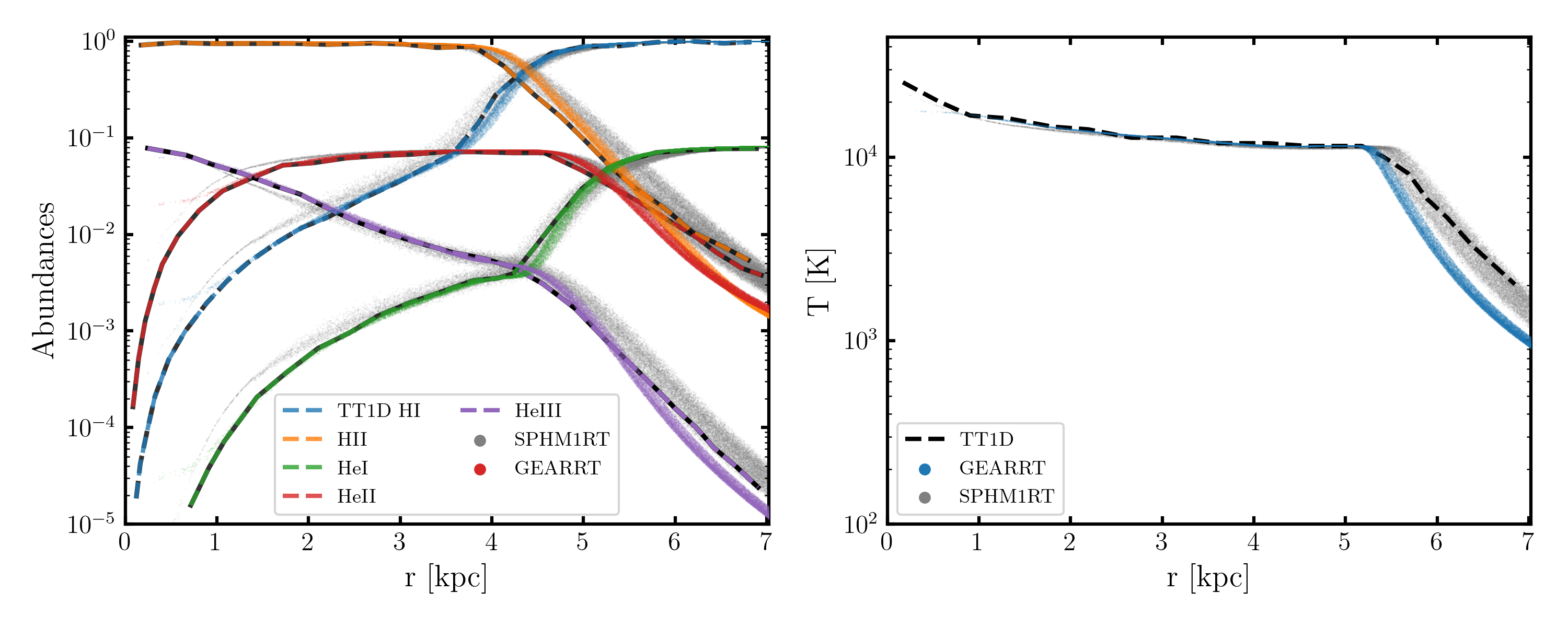}%
 \caption{
Spherically averaged ionized and neutral fractions of hydrogen and helium of Test 2 at 10 Myr for
a gas composed of 75\% hydrogen and 25\% helium. The solution of \GEARRT are colored points. The
reference solutions are from the \codename{TT1D} code
\citep[][dashed lines]{pawlikMultifrequencyThermallyCoupled2011} and \codename{SPHM1RT}
\citep[][gray points]{chanSmoothedParticleRadiation2021}.
}
 \label{fig:iliev2-HHe}
\end{figure}

Test 2 is very similar to Test 1, with the exception that gas is also allowed to heat up due to the
effects of radiation, whereas the gas temperature was held constant in Test 1, and the source
doesn't emit monochromatic radiation, but follows a blackbody spectrum (eq.~\ref{eq:blackbody}) with temperature $T_{bb} = 10^5$K. The initially neutral uniform hydrogen gas with number density $n = 10^{-3}$cm$^{-3}$ is given the initial temperature $T = 100$K. I again use a box size twice the size prescribed by IL6 in order to avoid using reflective boundary conditions, $128^3$ particles in a glass-like configuration, and reduce the speed of light by a factor of 100. The ionizing flux is prescribed to be $\dot{N}_\gamma = 5 \times 10^{48}$ ionizing photons/s. For three photon groups divided at the ionizing frequencies (eq.~\ref{eq:nuIonHI}-eq.~\ref{eq:nuIonHeII}), this translates to the group
luminosities $L_i$ (see Appendix~\ref{app:number-to-luminosity}):

\begin{align}
& \text{Group 1 } &&
    \nu \in [3.288\times 10^{15}, 5.945 \times 10^{15}] \text{ Hz} &&
    L_1 = 1.764 \times 10^4 \Lsol \label{eq:group-luminisoties-1} \\
& \text{Group 2 } &&
    \nu \in [5.945 \times 10^{15}, 13.157 \times 10^{15}] \text{ Hz} &&
    L_2 = 3.631 \times 10^4 \Lsol \\
& \text{Group 3 } &&
    \nu \in [13.157 \times 10^{15}, \infty] \text{ Hz} &&
    L_3 = 8.037 \times 10^3 \Lsol \label{eq:group-luminosities-3}
\end{align}

Figure~\ref{fig:iliev2} shows spherically averaged profiles of the neutral fraction and the
temperatures at 10, 30, and 100 Myr along with reference solutions, while
Figure~\ref{fig:iliev2-slices} shows slices through the mid-plane of the box of the \GEARRT solution
only. The results agree well with the reference solutions on the overall size of the ionized region
and its internal structure. Even for the reference solutions, there is a significant scatter in the
temperature profiles outside the HII region though. The reason behind this phenomenon is the way the
different codes handle multi-frequency, and in particular spectral hardening. Photons at higher
frequencies also deposit more energy during an ionization event. However, the photo-ionization cross
sections also decrease with increasing frequency (see Figure~\ref{fig:cross-sections}). Since the
photon energies in \GEARRT are averaged over the frequency bin, frequencies past the peak of the
blackbody spectrum will tend to be grouped along with higher averaged ionization cross sections.
This results in too much energy of the high frequency photons being absorbed too early compared with
what should happen when frequencies are treated individually. To illustrate this effect,
Figure~\ref{fig:iliev2-photon-groups} shows the result at 10 Myr for 1, 3, and 10 photon frequency
groups used. Specifically, for the simulation with a single photon group, the following luminosity
was used:

\begin{align}
 \text{Group 1 } &&
    \nu \in [3.288 \times 10^{15}, \infty] \text{ Hz} &&
    L_1 = 6.198 \times 10^4 \Lsol
\end{align}

For the 10 photon group run, the following values were used:

\begin{flalign}
& \text{Group 1 } &&
    \nu \in [3.288 \times 10^{15}, 6.576 \times 10^{15}] \text{ Hz} &&
    L_1 = 2.221 \times 10^4 \Lsol && \\
& \text{Group 2 } &&
    \nu \in [ 6.576 \times 10^{15}, 9.864 \times 10^{15}] \text{ Hz} &&
    L_2 = 2.020 \times 10^4 \Lsol && \\
& \text{Group 3 } &&
    \nu \in [ 9.864  \times 10^{15}, 13.152 \times 10^{15} ] \text{ Hz} &&
    L_3 = 1.153 \times 10^4 \Lsol && \\
& \text{Group 4 } &&
    \nu \in [13.152  \times 10^{15}, 16.440 \times 10^{15}] \text{ Hz} &&
    L_4 = 5.122 \times 10^3 \Lsol && \\
& \text{Group 5 } &&
    \nu \in [16.440 \times 10^{15}, 19.728 \times 10^{15}] \text{ Hz} &&
    L_5 = 1.952 \times 10^3 \Lsol && \\
& \text{Group 6 } &&
    \nu \in [19.728 \times 10^{15}, 23.016 \times 10^{15}] \text{ Hz} &&
    L_6 = 6.705 \times 10^2 \Lsol && \\
& \text{Group 7 } &&
    \nu \in [23.016 \times 10^{15}, 26.304 \times 10^{15}] \text{ Hz} &&
    L_7 = 2.140 \times 10^2 \Lsol && \\
& \text{Group 8 } &&
    \nu \in [26.304 \times 10^{15}, 29.592 \times 10^{15}] \text{ Hz} &&
    L_8 = 6.461 \times 10^1 \Lsol && \\
& \text{Group 9 } &&
    \nu \in [29.592 \times 10^{15}, 32.880 \times 10^{15}] \text{ Hz} &&
    L_9 = 1.869 \times 10^1 \Lsol && \\
& \text{Group 10 } &&
    \nu \in [32.880 \times 10^{15}, \infty] \text{ Hz} &&
    L_{10} = 7.158 \Lsol &&
\end{flalign}

With an increasing number of
groups, the approximation becomes more accurate and closer to treating frequencies individually.
The averaged ionization cross sections are treated more accurately, and high energy photons are
able to reach regions further from the source, where they can heat the gas.

Figure~\ref{fig:iliev2-Ifront} shows the evolution of the I-front position and velocity compared
to analytical values taken from Test 1. \GEARRT's results agree with the reference solutions,
although both the I-front position and velocity tend to be towards the higher end, especially at
late times.

This test setup is also convenient to demonstrate the validity of the reduced speed of light
approach. Figure~\ref{fig:iliev2-compare-c} shows the propagation of the I-front radius using
decreasing values for the speed of light along with the solution of \codename{C2Ray}, which assumed
an infinite speed of light. The I-front radius using a factor of 100 to reduce the speed of light,
which was used in Test 2 and Test 1, is nearly identical with the radius of the I-front when the
factor 10 was used as well as with the solution of \codename{C2Ray}, although it lags a little
behind in early times, which is to be expected. A factor of 1000 however is shown to be too large
of a reduction.

Finally, to demonstrate the thermochemistry involving helium as well, Figure~\ref{fig:iliev2-HHe}
shows the solution of the same test with a gas composed of 75\% hydrogen and 25\% helium at 100 Myr,
with reference solutions of the \codename{TT1D} code
\citep{pawlikMultifrequencyThermallyCoupled2011} and \codename{SPHM1RT}
\citep{chanSmoothedParticleRadiation2021}.

\subsection{Iliev Test 3}\label{chap:Iliev3}

\begin{figure}
 \centering
 \includegraphics[width=.95\textwidth]{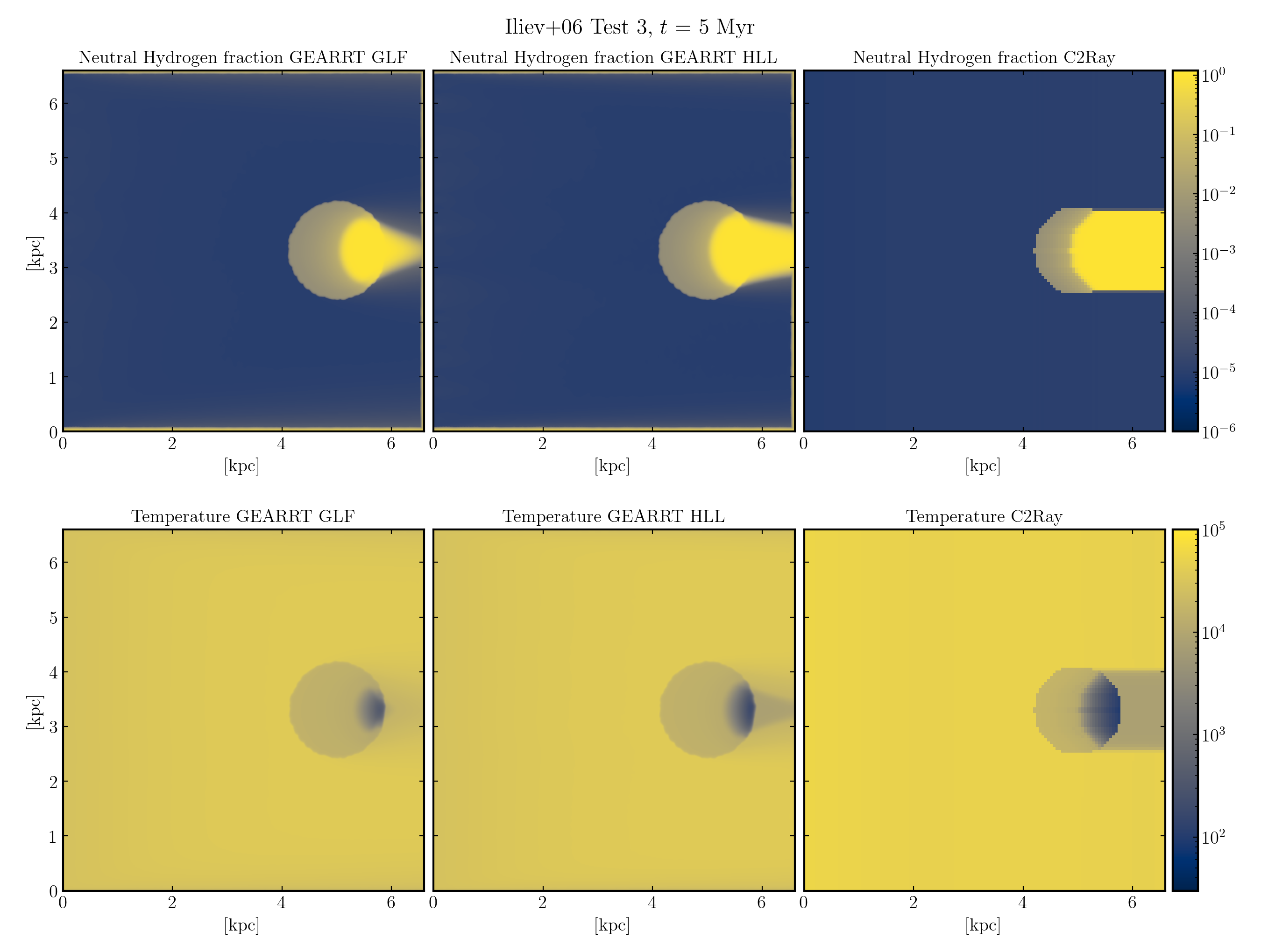}\\
 \includegraphics[width=.95\textwidth]{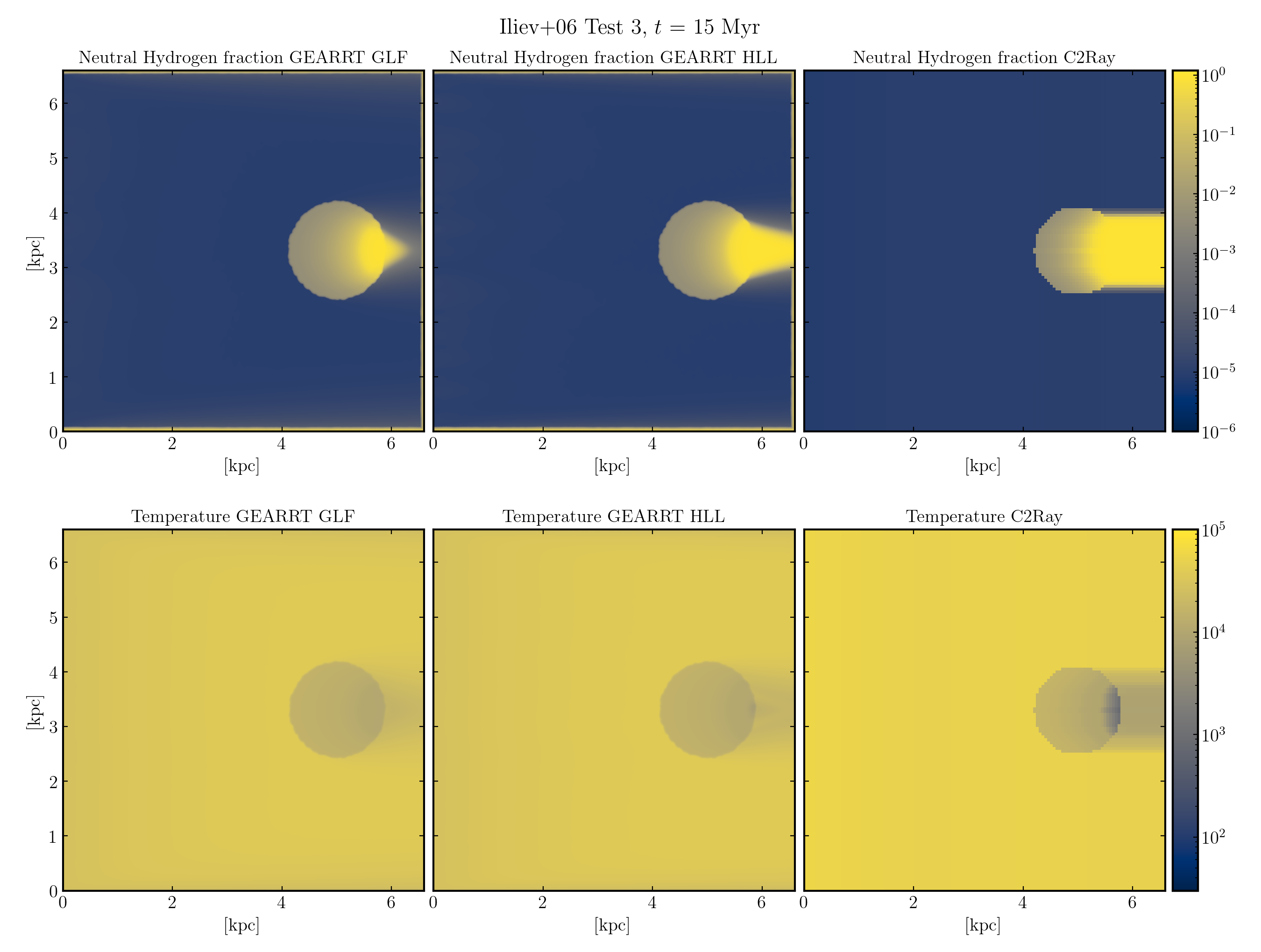}
 \caption{Slice along the mid-plane of Test 3 at 5 Myr (top) and 15 Myr (bottom) using both the GLF
and the HLL Riemann solver, compared with the reference solution of the \codename{C2Ray} code.
}
 \label{fig:iliev3-slices}
\end{figure}

\begin{figure}
 \centering
 \includegraphics[width=\textwidth]{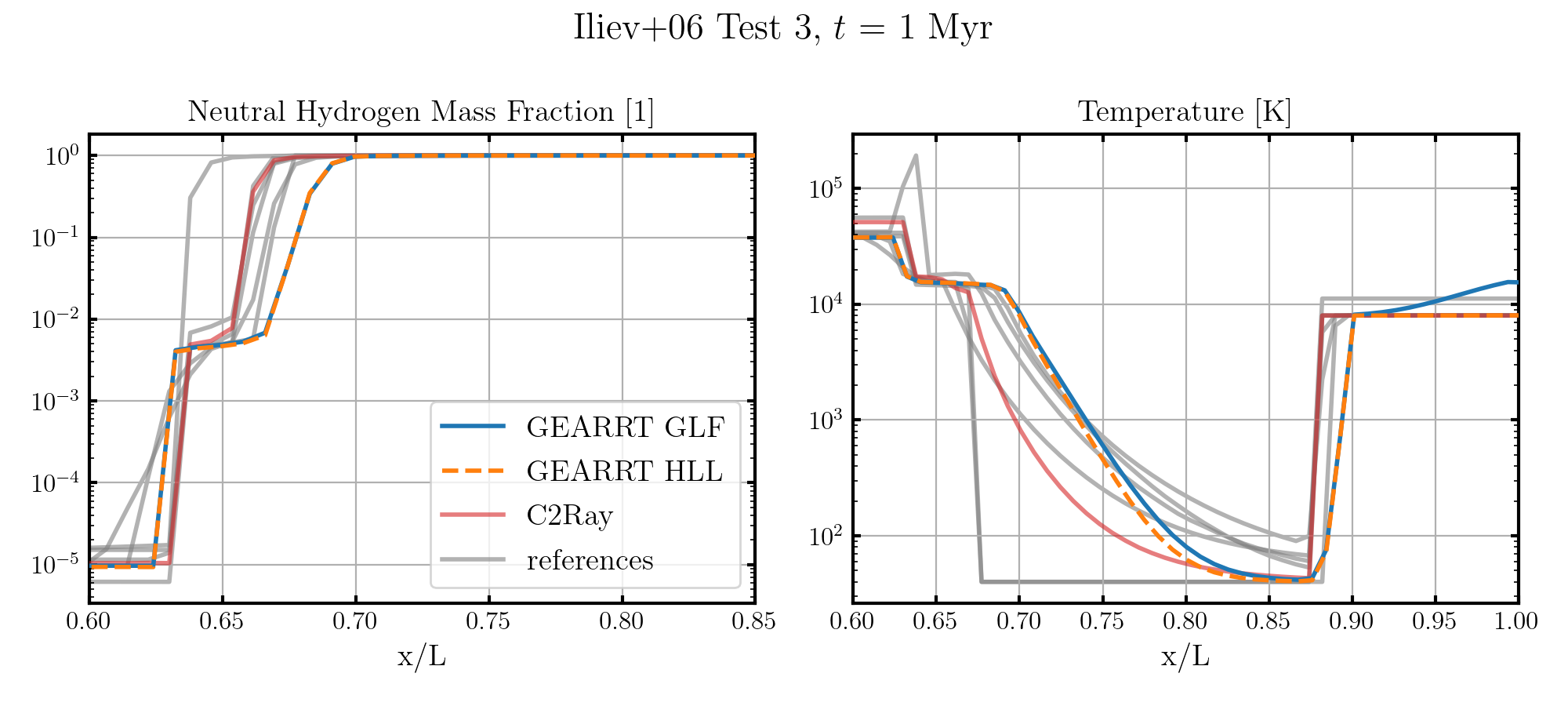}\\%
 \includegraphics[width=\textwidth]{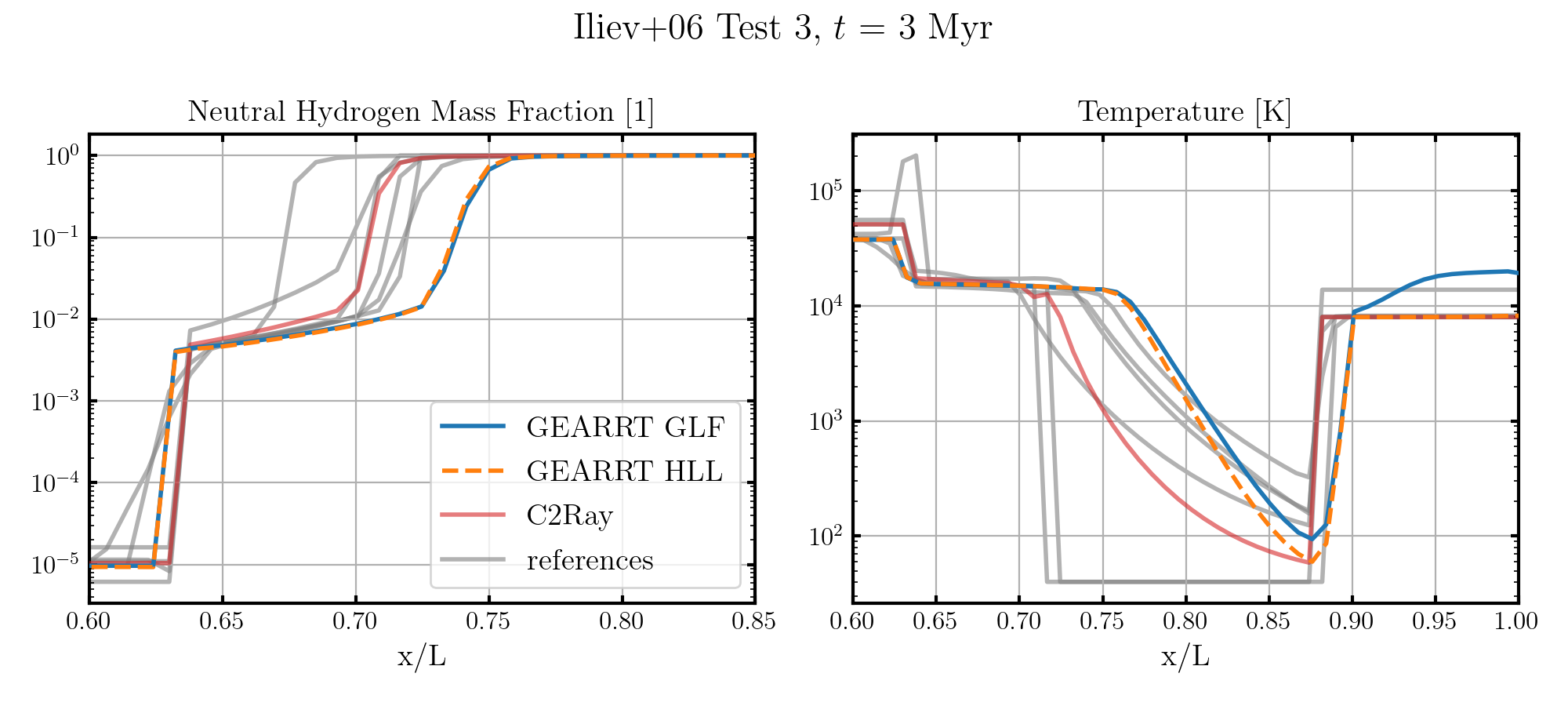}\\%
 \includegraphics[width=\textwidth]{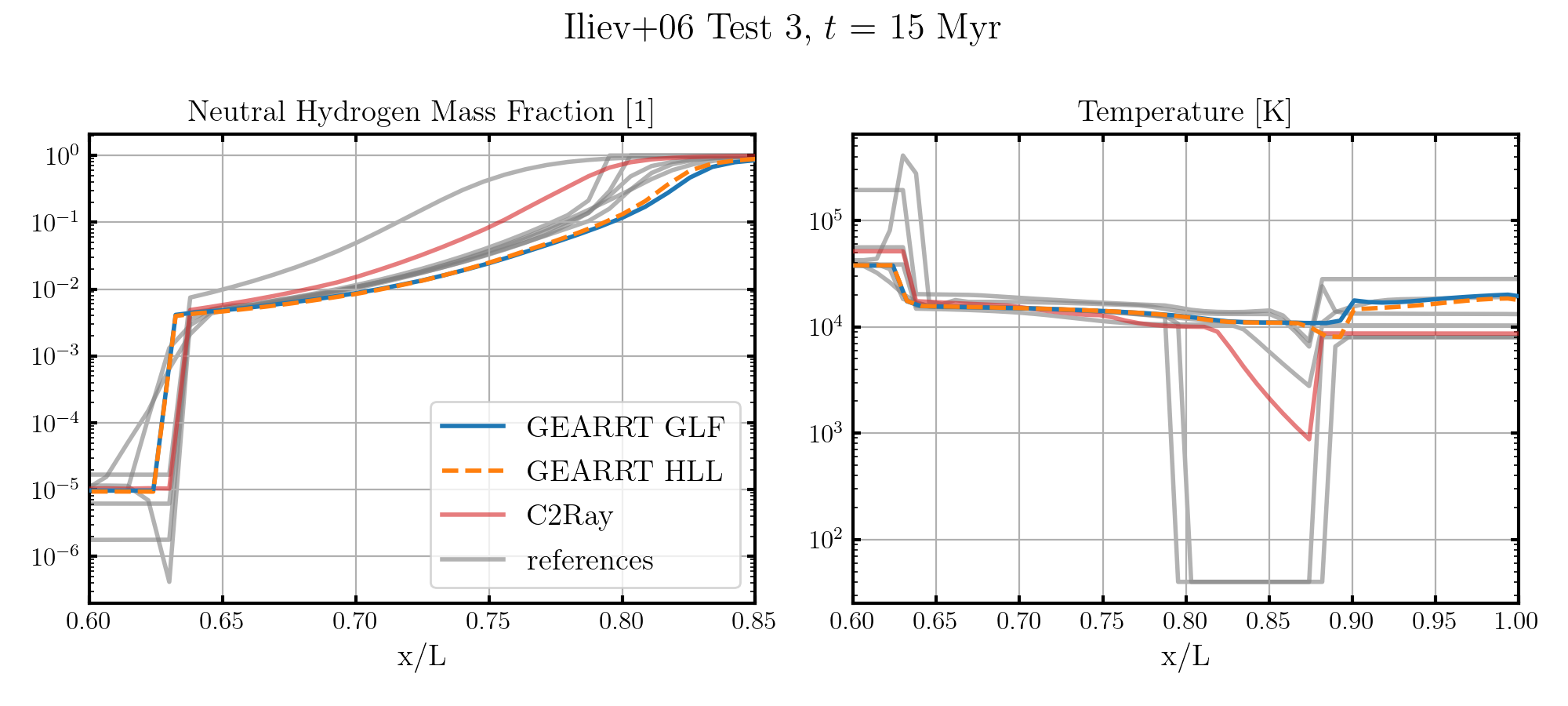}%
 \caption{
Left: Line cuts of the ionized and neutral fraction along the axis of symmetry through the center
of the clump. Right: Line cuts of the temperature along the axis of symmetry through the center of
the clump. Results are shown at times $t=$ 1, 3, and 15 Myr for \GEARRT using the HLL and the GFL
Riemann solver, for \codename{C2Ray}, and for other references.
 }
\label{fig:iliev3-profiles}
\end{figure}

\begin{figure}
 \centering
\includegraphics[
width=.85\textwidth
]{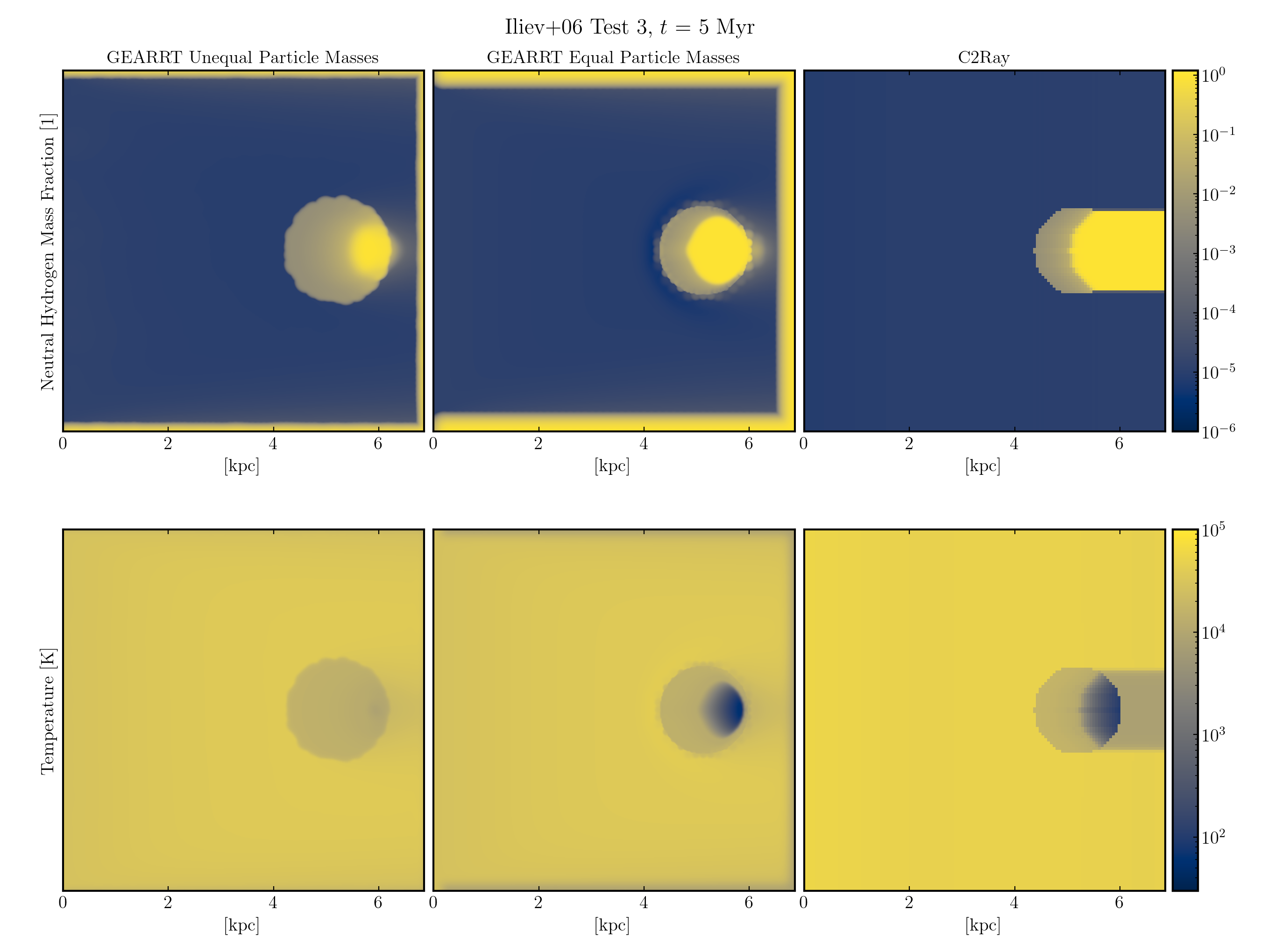} \\
\includegraphics[
width=.85\textwidth
]{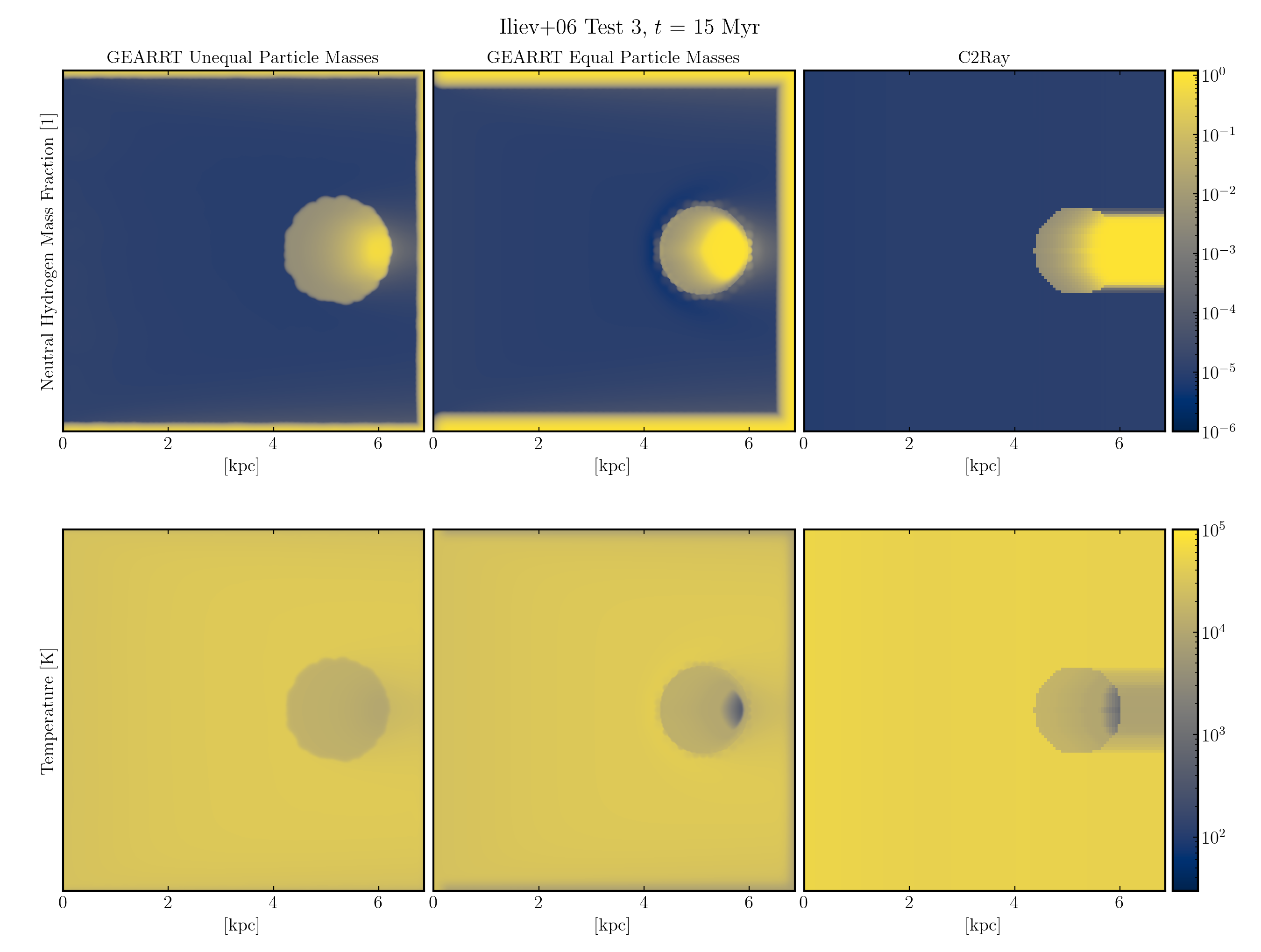}
 \caption{
 \small{
Slice along the mid-plane of Test 3 at 5 Myr and 15 Myr for initial conditions where
particles have equal masses, and hence a higher number of particles are placed inside the clump than
outside. This is a more adequate setup for particle simulations than setting particles with regular
distances, as is done in Figure~\ref{fig:iliev3-slices}. This simulation was however produced using a
lower resolution of $\sim 64^3$ particles. The neutral gas along the edges in the solution of
\GEARRT arise due to projection effects of the boundary particles, which form an additional layer
around the box and ``swallow'' all radiation that would escape the box. The reference solution on
the right is from the \codename{C2Ray} code.}
}
 \label{fig:iliev3-equal-mass}
\end{figure}

\begin{figure}
 \centering
 \includegraphics[width=\textwidth]{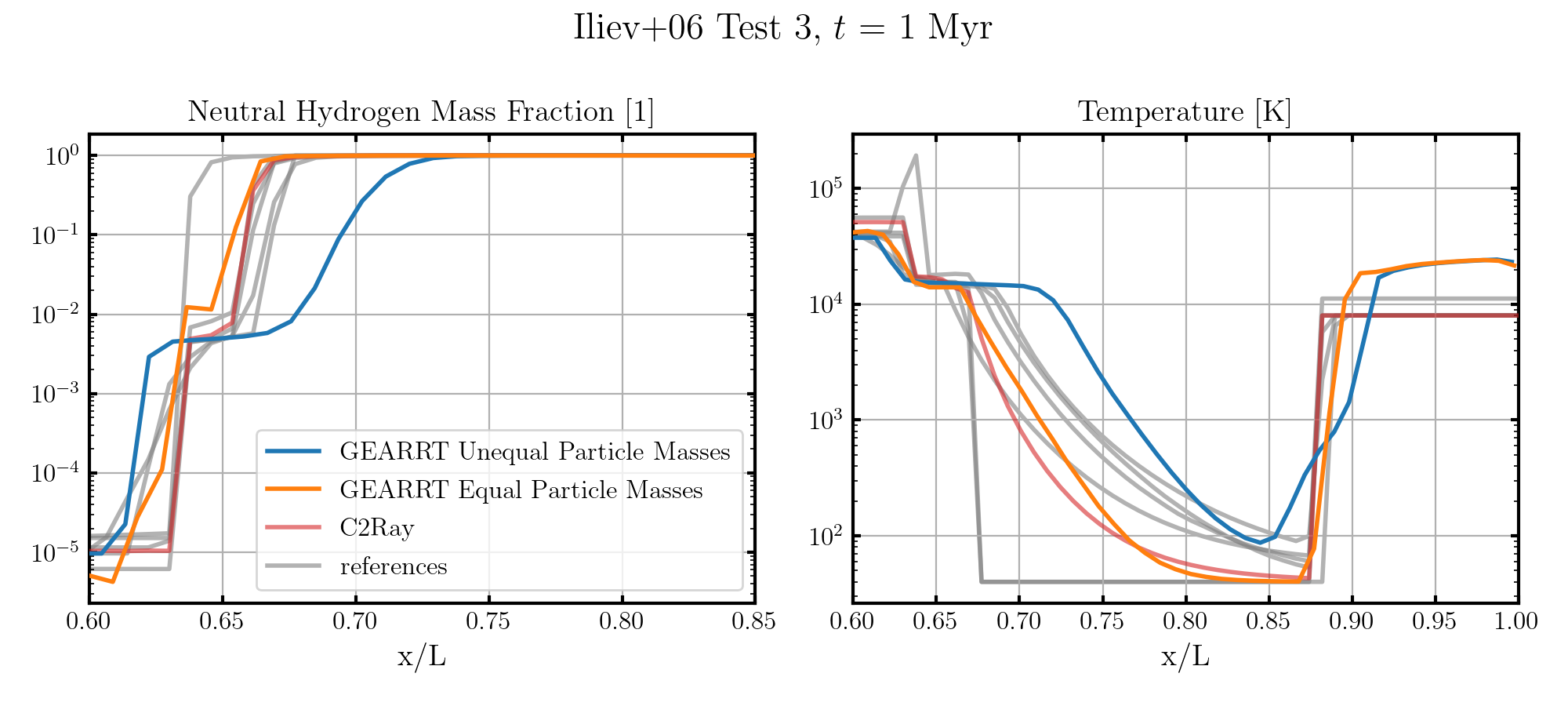}\\%
 \includegraphics[width=\textwidth]{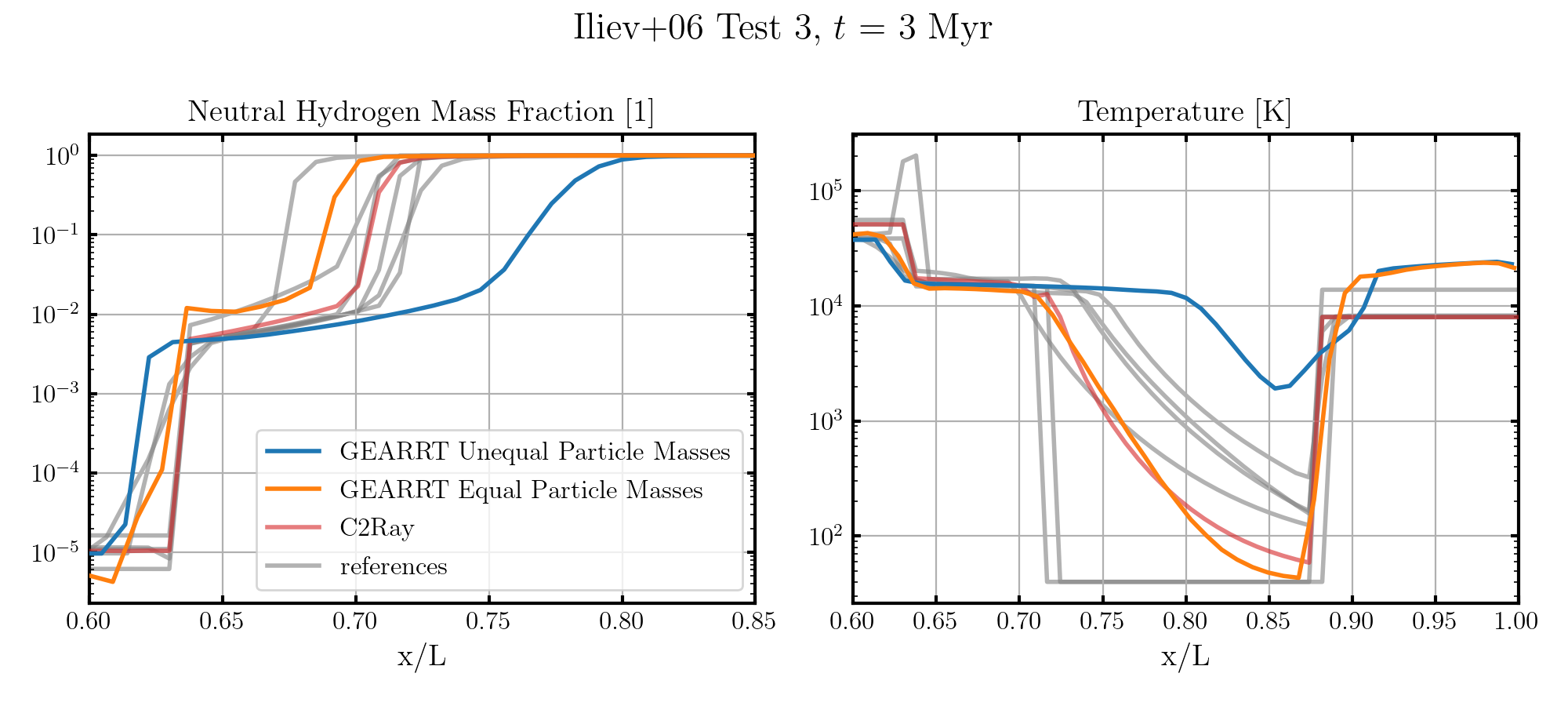}\\%
 \includegraphics[width=\textwidth]{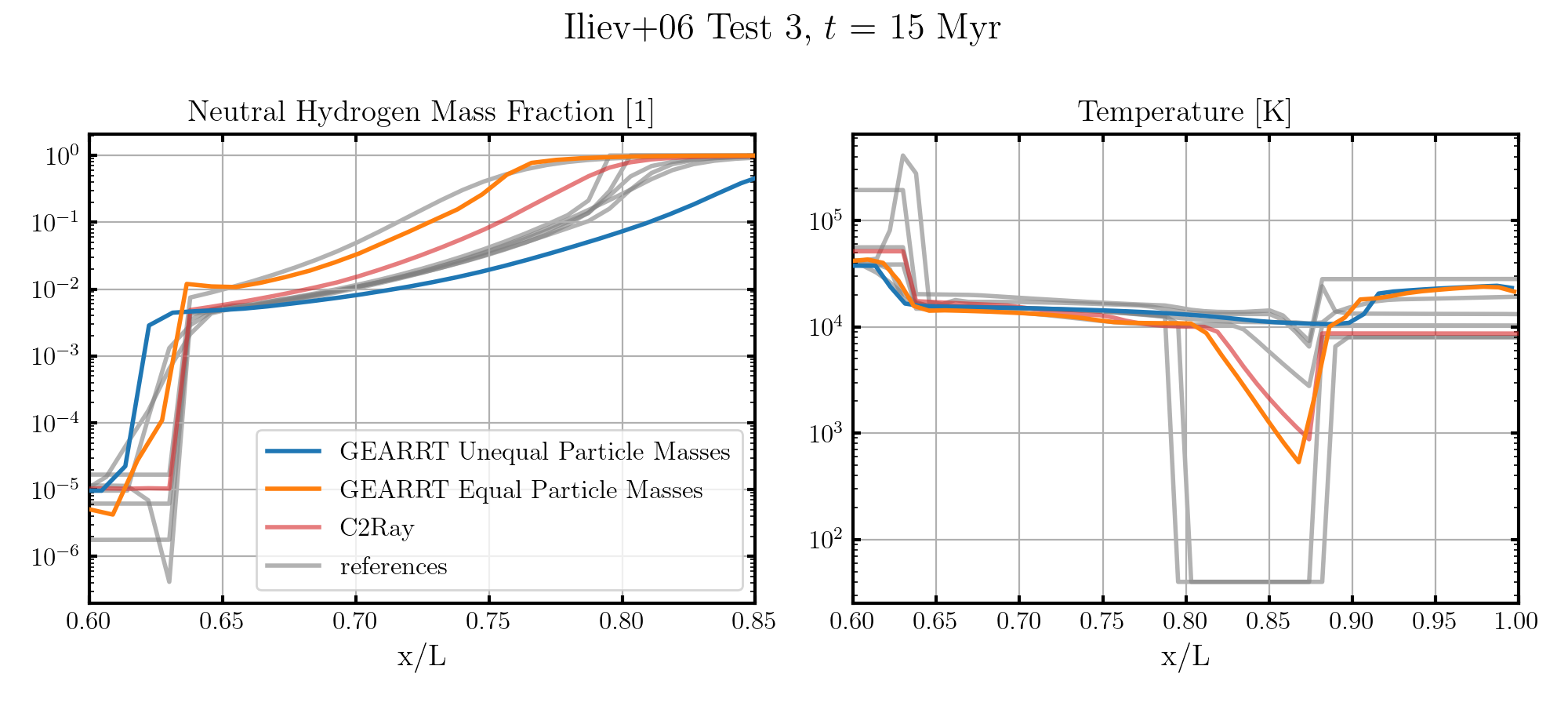}\\%
 \caption{
Left: Line cuts of the ionized and neutral fraction along the axis of symmetry through the center
of the clump. Right: Line cuts of the temperature along the axis of symmetry through the center of
the clump. Results are shown at times $t=$ 1, 3, and 15 Myr for initial conditions where particles
have equal masses, and hence a higher number of particles are placed inside the clump than outside,
using $\sim 64^3$ particles, as well as when using unequal masses, but regular distances. Using
equal particle masses, which is a more adequate choice for particle simulations, significantly
improves the self-shielding of the dense clump.
 }
\label{fig:iliev3-equal-mass-profiles}
\end{figure}

\begin{figure}
 \centering
\includegraphics[
width=.85\textwidth
]{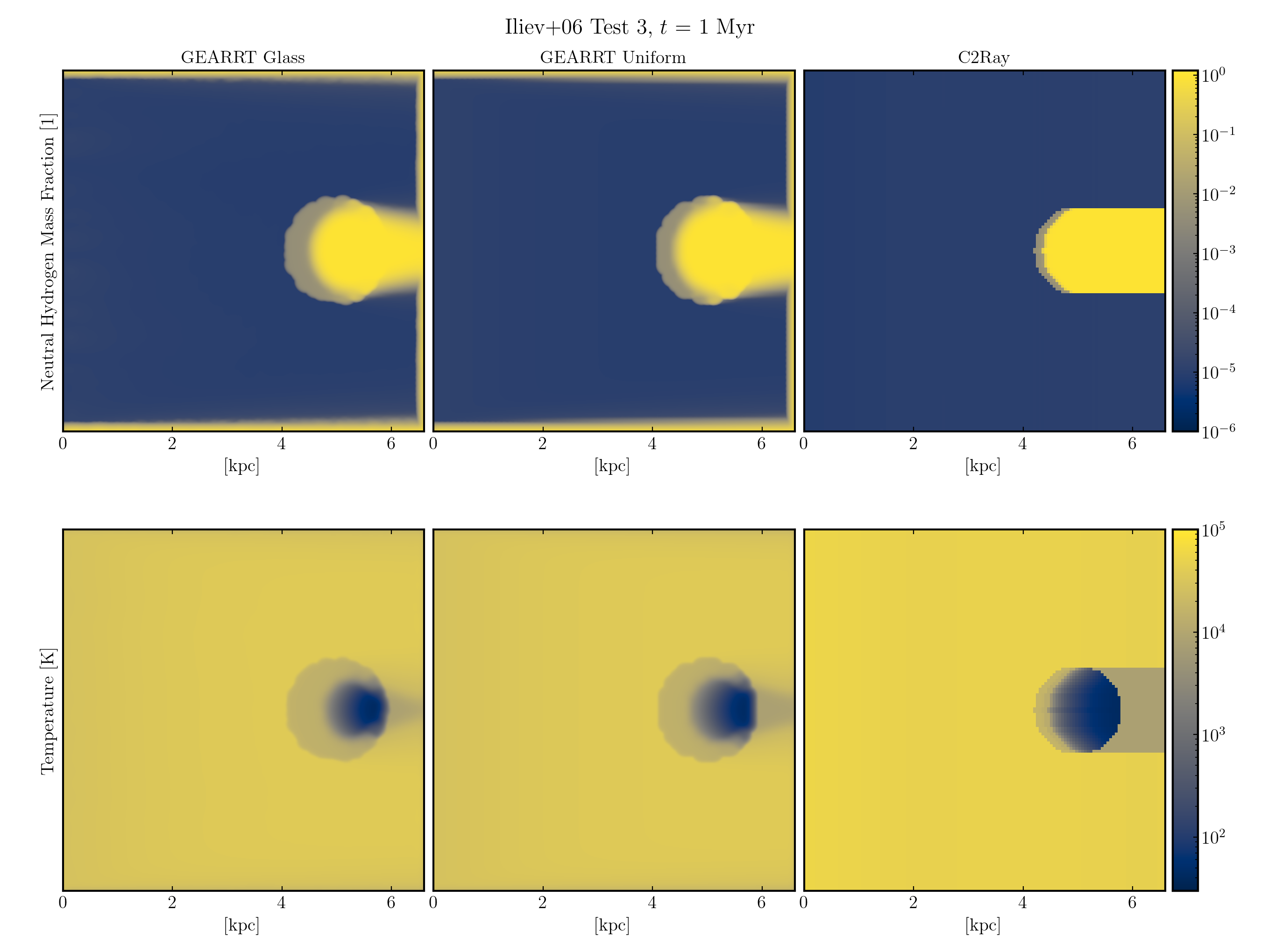} \\
\includegraphics[
width=.85\textwidth
]{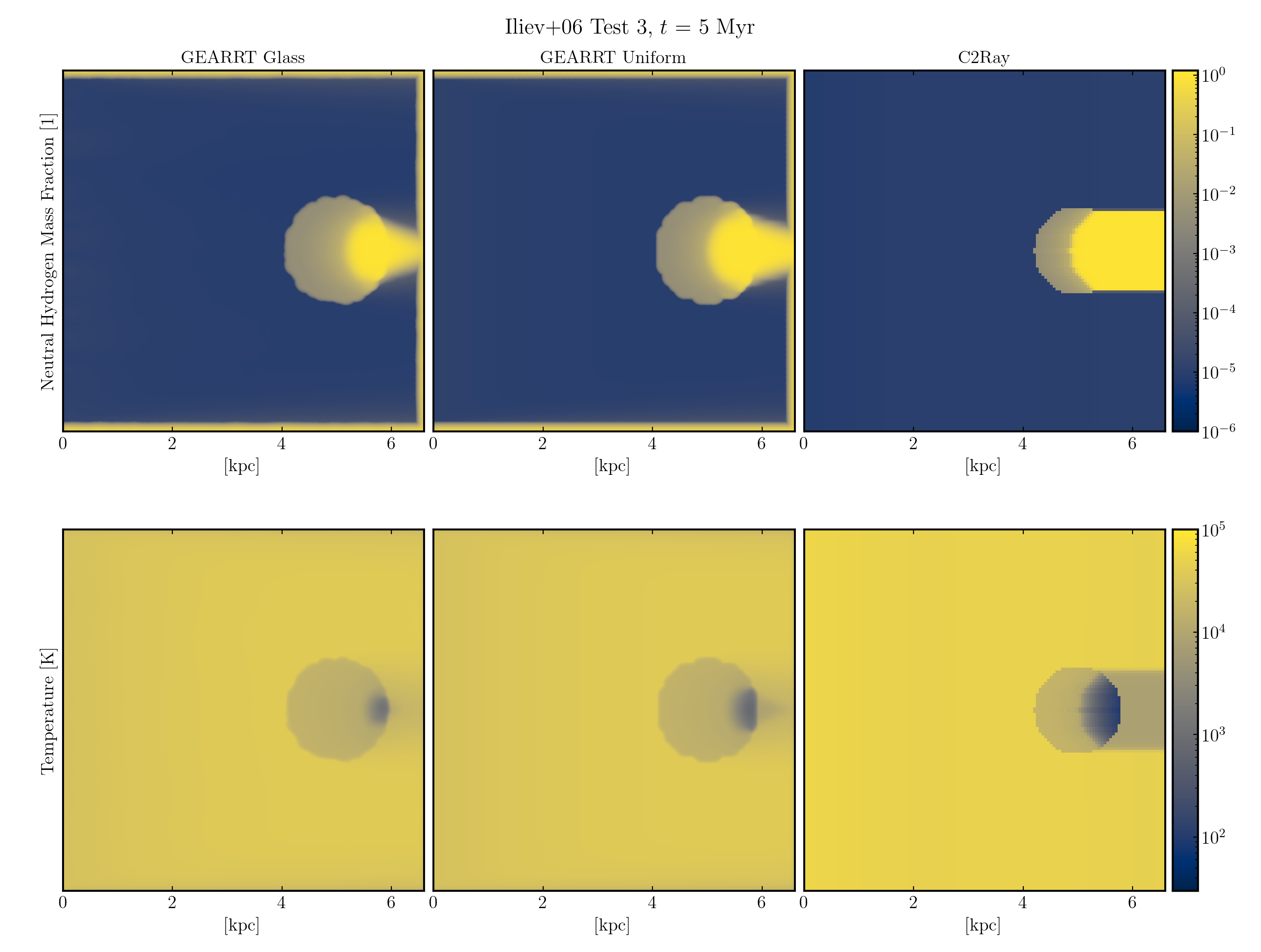}
 \caption{
Slice along the mid-plane of Test 3 at 1 Myr and 5 Myr for initial conditions where particles are
placed in a glass-like distribution (left) and on a uniform grid (middle) using the HLL Riemann
solver. The reference solution on the right is from the \codename{C2Ray} code. Particles being
placed on a uniform grid improves the formation of the shadow behind the dense clump, since the
perfectly perpendicular particle placement w.r.t. to the direction of the radiation minimizes the
diffusion in the perpendicular direction.
}
 \label{fig:iliev3-uniform}
\end{figure}


This test examines the self-shielding of a dense gas clump and the formation of a shadow. The
simulation box has a size of $6.6$ kpc. A spherical cloud of gas with radius $r_{cloud} = 0.8$ kpc
is placed centered at $(5, 3.3, 3.3)$ kpc. The surrounding hydrogen gas has an number density of
$n_{out} = 2 \times 10^{-4}$cm$^{-3}$ and temperature $T_{out} = 8000$K. The cloud is given a
number density of $T_{cloud} = 40$K and number density of $n_{cloud} = 200 n_{out}$. A constant
flux of $F = 10^6$ photons / s / cm$^2$ following a blackbody spectrum with temperature $T_{bb} =
10^{5}$K is injected from the $x = 0$ plane of the box. I again use three photon frequency
intervals split at the photo-ionizing frequencies of hydrogen and helium, leading to the same source radiation luminosities as given in eqs.~\ref{eq:group-luminisoties-1}-\ref{eq:group-luminosities-3}.
The speed of light was reduced by a factor of 8 (compared to 100 in Test 1 and 2) so that the
radiation emitted at the $x = 0$ plane would reach the clump before the first snapshot at 1 Myr as
prescribed by IL6. The particle distribution remains glass-like; In order to achieve the correct
densities in- and outside of the clump, the particle masses were modified accordingly.

This test displays a known shortcoming of moment based radiative transfer, which is its inability
to form shadows correctly. While the clump in this test is still composed of neutral hydrogen, a
shadow should form behind it. The formation of shadows is limited because the radiation, which is
being treated similar to a fluid, can diffuse into regions where light rays wouldn't. We thus test
the two different Riemann solvers described in Section~\ref{chap:riemann-rt}: The inexpensive GLF
solver, and the HLL solver, which promises to be less diffusive than the GLF solver
\citep{ramses-rt13, gonzalezHERACLESThreedimensionalRadiation2007}.

Figure~\ref{fig:iliev3-slices} shows a slice of the mid-plane of the box at 5 Myr and at 15 Myr for
both the GLF and the HLL Riemann solver, along with a reference solution of the \codename{C2Ray}
code. The HLL solver indeed improves the formation of the shadow. Figure~\ref{fig:iliev3-profiles}
shows line cuts of the neutral hydrogen mass fraction and the temperature along the axis of symmetry
of the clump at 1, 3, and 15 Myr. The improved shadow formation with the HLL solver can be seen in
the temperature profile remaining at the initial $8000$K behind the clump, i.e. at $x/L \geq 0.9$.

The line cuts in Figure~\ref{fig:iliev3-profiles} clearly reveal a further issue with the solution
obtained by \GEARRT: The ionization front penetrates much further into the dense clump than the
reference solutions. The reason for this phenomenon lies in particle nature of \GEARRT, and the
number of neighbors used.
While mesh-based codes would typically only exchange fluxes between adjacent cells, FVPM exchange
fluxes between $\sim 40-50$ neighboring particles. If these $40-50$ neighbors were arranged on a
regular grid around the particle, this would correspond to about 2 neighboring particles on each
side, and the diffusion can propagate further compared to grid codes. The diffusivity of the method
in this case can be expected to be much worse than with a mesh-based code, and the ionization front
can propagate too fast.

However, the setup in Figure~\ref{fig:iliev3-slices} and Figures~\ref{fig:iliev3-profiles} is
sub-optimal for particle codes like \GEARRT. To mimic the prescribed test setup, particles were
distributed uniformly (in a glass-like manner), and, more importantly with regard to the currently
discussed issue, assigned different masses such that the correct densities are reproduced. A more
adequate representation would be to use particles of equal masses, and adapt their positions to
reproduce the correct densities. So in this case, much more particles would be positioned inside the
clump compared to the outside region.

The results using such a setup with the GLF solver are shown in Figure~\ref{fig:iliev3-equal-mass},
albeit with a resolution of only $\sim 64^3$ particles in total.
Figure~\ref{fig:iliev3-equal-mass-profiles} shows the line cuts of the neutral hydrogen mass
fraction and the temperature along the axis of symmetry of the clump at 1, 3, and 15 Myr, comparing
the results to the ones using unequal particle masses (but somewhat regular inter-particle
distances). The improvement in the self-shielding when using particle number densities rather than
masses to represent the dense clump is evident. The formation of the shadow however isn't improved,
and is in fact worsened, since in order to keep roughly the correct total number of particles, the
rarefied region behind the clump is now sampled through fewer particles.

A contributing factor to the inadequate shadow formation with \GEARRT is the fact that it uses
particles as the underlying discretization method. In general, the particle distribution will not be regular. The exchange of fluxes (in the hyperbolic sense) between particles not perfectly aligned with the direction the radiation propagates in will lead to radiation diffusing perpendicularly to the direction of propagation, leading to reduced shadows as seen in
Figures~\ref{fig:iliev3-slices}-\ref{fig:iliev3-equal-mass-profiles}. This can be readily verified by repeating the test using a particle configuration distributed along a uniform grid. The results of such a setup are shown in Figure~\ref{fig:iliev3-uniform}, where the shadow is indeed improved.
However, the diffusivity of the method will never be able to be fully eliminated, due to several
reasons. Firstly, there will always be neighboring particles which aren't perfectly aligned with
the direction of propagation of the radiation, even on a perfectly uniform particle grid. Secondly,
the diffusivity of the method is also due to the nature of the M1 closure, and as such can't be
fully eliminated. And finally, numerical diffusion is a necessary component of the underlying
numerical method, which for FVPM enters the equation in the form of discretization errors and flux
limiters.

\subsection{Iliev Test 4}\label{chap:Iliev4}

\begin{figure}
 \centering
 \includegraphics[width=\textwidth]{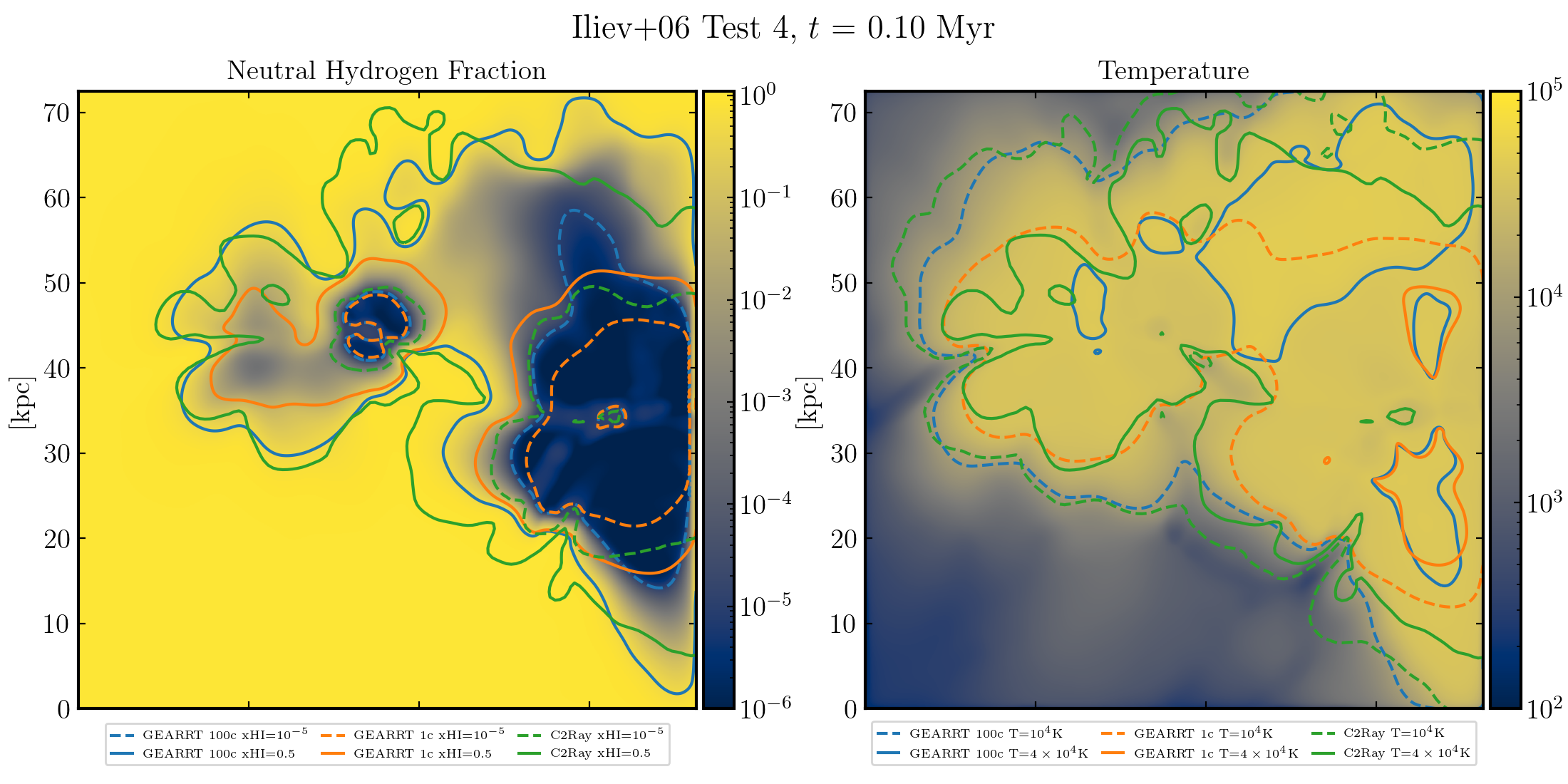}
 \caption{
Slice through the mid-plane of the box of the solution to Test 4 at 0.1 Myr.
Contour lines for neutral hydrogen fractions of 0.5 (solid lines) and $10^{-5}$ (dashed lines) are
shown on the left. On the right, contour lines for temperatures of $10^4$K (dashed lines) and $4
\times 10^4$K (solid lines) are shown. The solutions of \GEARRT (orange lines), \GEARRT with an
increased speed of light by a factor of 100 (blue lines) are plotted, along with the reference
solution of the \codename{C2Ray} code (green lines).
}
 \label{fig:iliev4-0.1Myr}
\end{figure}

\begin{figure}
 \centering
 \includegraphics[width=\textwidth]{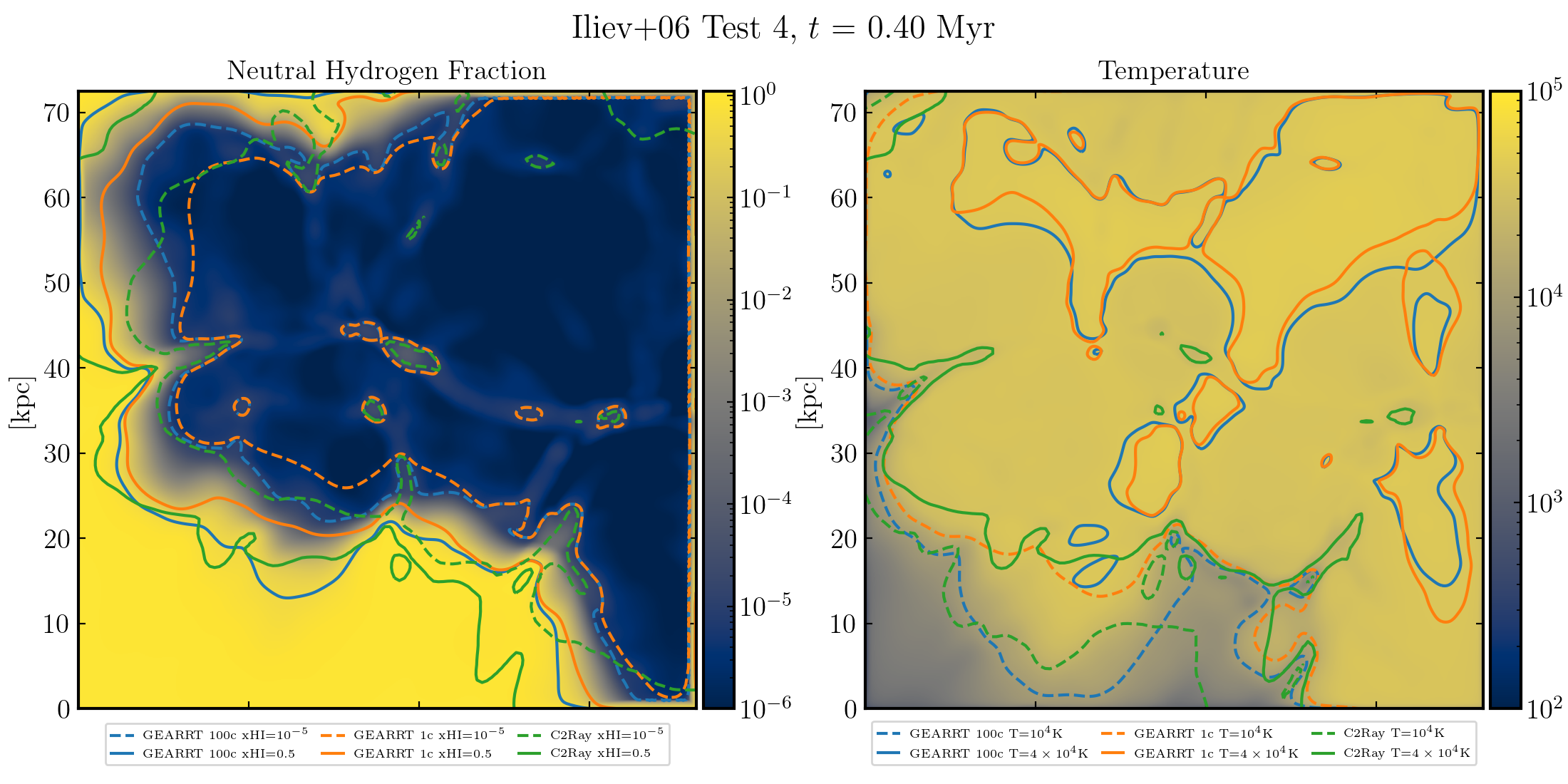}
\caption{
Slice through the mid-plane of the box of the solution to Test 4 at 0.8 Myr.
Contour lines for neutral hydrogen fractions of 0.5 (solid lines) and $10^{-5}$ (dashed lines) are
shown on the left. On the right, contour lines for temperatures of $10^4$K (dashed lines) and $4
\times 10^4$K (solid lines) are shown. The solutions of \GEARRT (orange lines), \GEARRT with an
increased speed of light by a factor of 100 (blue lines) are plotted, along with the reference
solution of the \codename{C2Ray} code (green lines).
}
 \label{fig:iliev4-0.4Myr}
\end{figure}

\begin{figure}
 \centering
 \includegraphics[width=\textwidth]{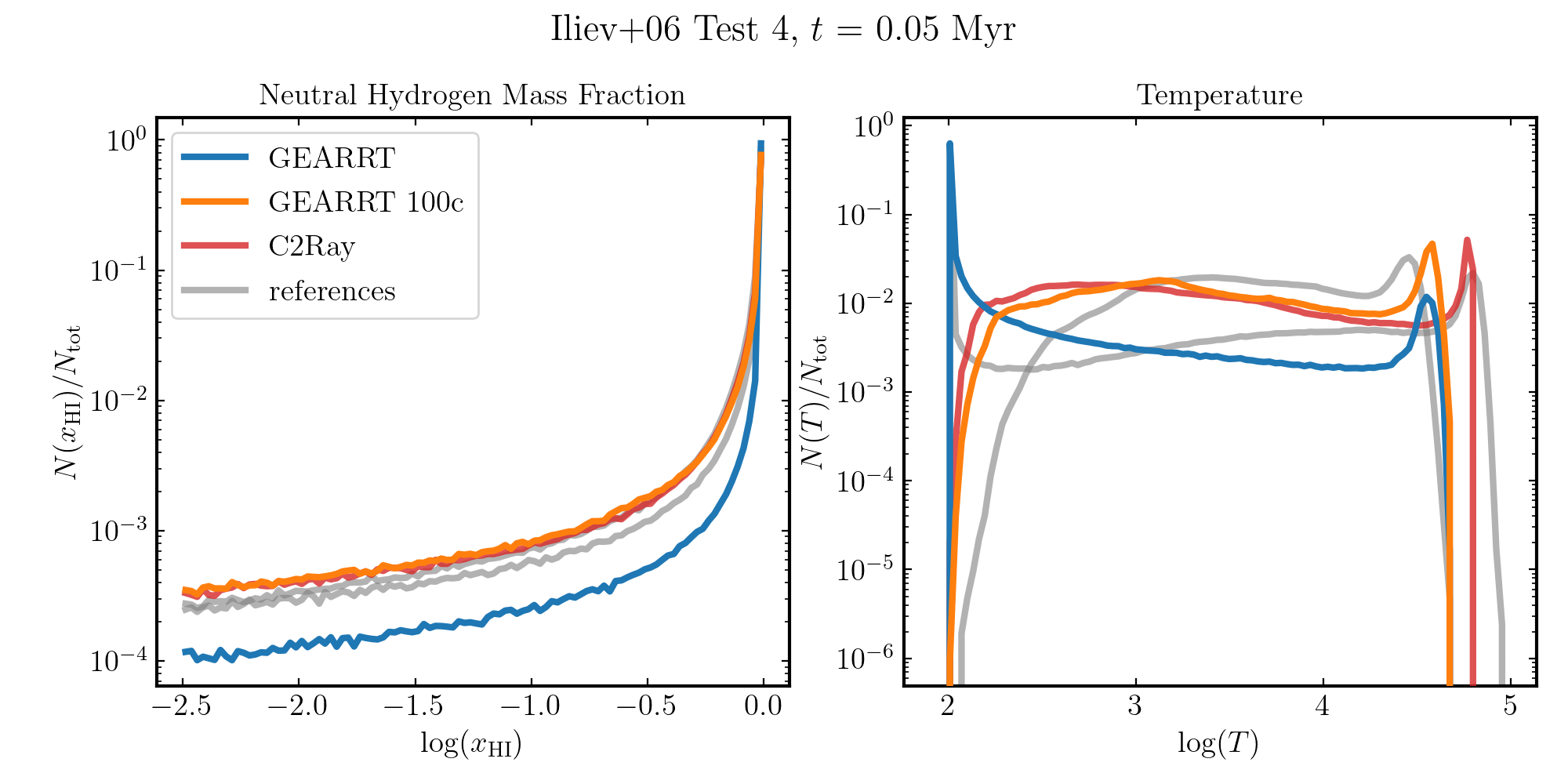}\\
 \includegraphics[width=\textwidth]{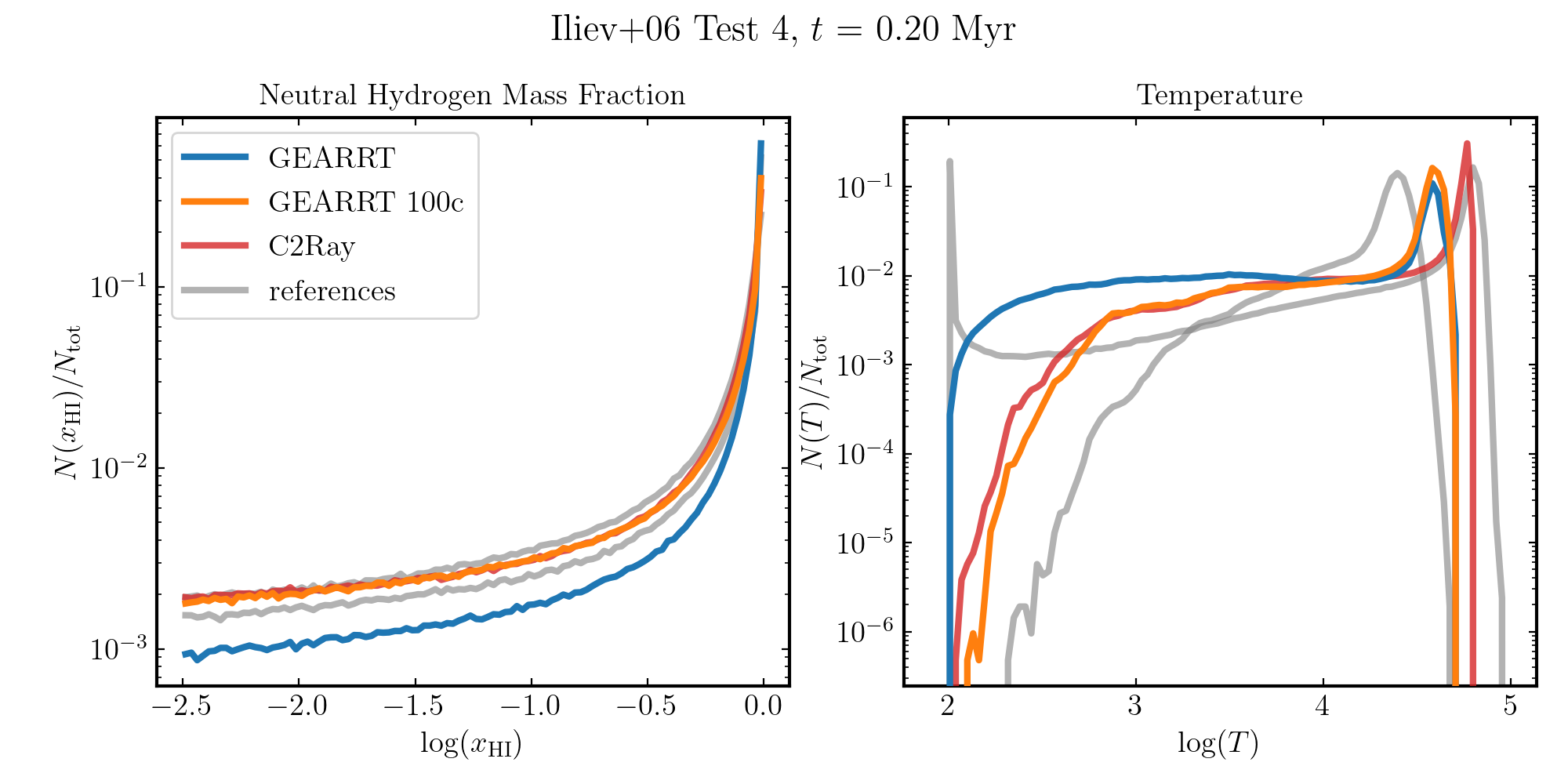}\\
 \includegraphics[width=\textwidth]{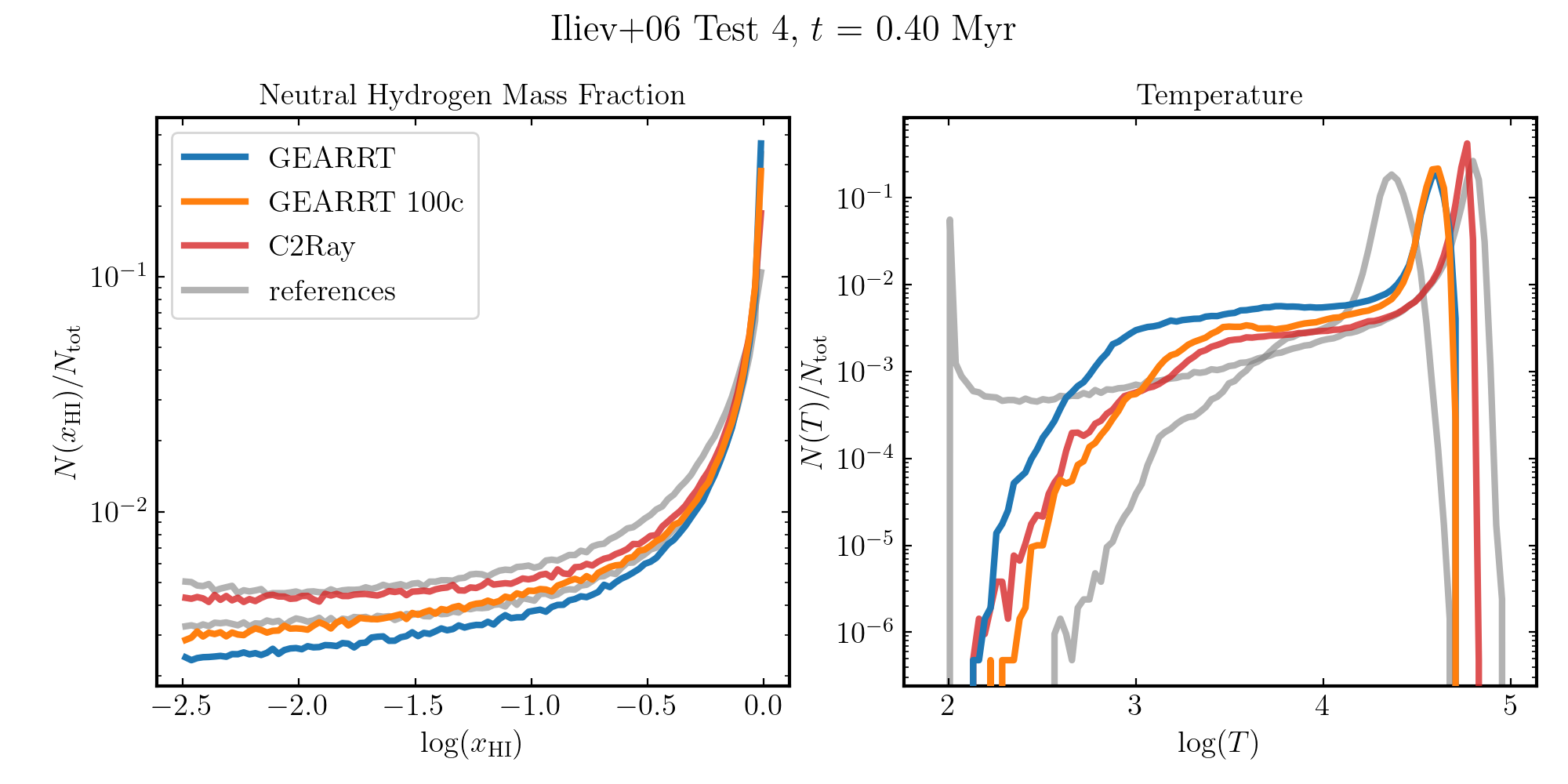}
\caption{
Histograms of the neutral hydrogen fraction and the temperature of the entire simulation box at
times $t = $ 0.05 Myr, 0.2 Myr, and 0.4 Myr solved with the normal speed of light, and the speed of
light increased by a factor of 100.
}
 \label{fig:iliev4-histograms}
\end{figure}

In this test, a cosmological density field extracted from a simulation is being heated and ionized
by 16 sources placed at the positions of the most massive halos, and given a luminosity
proportional to the halo mass $M$. More precisely, the luminosity is assigned assuming that each
source lives for $t_s = 3$ Myr and emits $f_\gamma = 250$ ionizing photons per atom during its
lifetime, resulting in

\begin{align}
\dot{N}_\gamma = f_\gamma \frac{\Omega_b}{\Omega_0} \frac{M}{m_p} \frac{1}{t_s}
\end{align}

where $m_p$ is the proton mass, and the cosmological density parameters $\Omega_0 = 0.27$ and
$\Omega_b = 0.043$ were used. The spectrum of the emitted radiation source is again taken to be a
blackbody spectrum with effective temperature $T_{bb} = 10^5$K. I split the spectrum into three
frequency intervals at the photo-ionizing frequencies of hydrogen and helium, as given in
eqns.~\ref{eq:group-luminisoties-1}-\ref{eq:group-luminosities-3}. The simulation was extracted on a
$128^3$ grid at high redshift $z \sim 9$ for a box size of 500 $h^{-1}$ co-moving kpc. For this test, particles are also arranged on a uniform grid, and given the appropriate mass to reproduce the density field prescribed by IL6. I again use three photon frequency intervals split at the photo-ionizing frequencies of hydrogen and helium, leading to the same source
radiation luminosities as given in eqns.~\ref{eq:group-luminisoties-1}-\ref{eq:group-luminosities-3}.

Figure~\ref{fig:iliev4-0.1Myr} shows the results of \GEARRT compared with the results of
\codename{C2Ray} results at 0.1 Myr, while Figure~\ref{fig:iliev4-0.4Myr} shows the results at 0.4
Myr. In agreement to the findings in \cite{ramses-rt13}, the ionization fronts appear to propagate
slower for \GEARRT, which is likely due to the assumption of infinite speed of light used by the
ray tracing reference code \codename{C2Ray}. Following the approach of \cite{ramses-rt13}, I also
re-run the test with the speed of light \emph{increased} by a factor of 100 to approximate the
infinite speed of light for a comparison with the reference solution of \codename{C2Ray}, the
results of which are also shown in Figure~\ref{fig:iliev4-0.1Myr} and \ref{fig:iliev4-0.4Myr}.

Especially at the early times like in Figure~\ref{fig:iliev4-0.1Myr} increasing the speed of light
by a factor of 100 matches the results of \codename{C2Ray} quite well. \codename{C2Ray} however
develops more distinct features in the ionization front, which is likely due to the diffusivity of
the FVPM and the M1 closure's difficulties to capture shadows.

Figure~\ref{fig:iliev4-histograms} shows histograms of the neutral hydrogen fraction and the gas
temperature at 0.05 Myr, 0.2 Myr, and at 0.4 Myr. They confirm the previous findings: Increasing
the speed of light by a factor of 100 leads to results very similar to the reference solutions, and
nearly identical histograms of neutral hydrogen mass fractions at early times. The peak
of the temperatures however vary between \GEARRT and reference solutions due to the individual
treatment of multi-frequency and the resulting differences in the treatment of spectral hardening.
Nonetheless, the solution obtained with \GEARRT agrees fairly well with reference solutions.

\section{Radiation Hydrodynamics}\label{chap:IL9}

This section tests the radiative transfer and the thermochemistry following the standard tests set
by the comparison paper \cite{ilievCosmologicalRadiativeTransfer2009} (hereafter IL9). These tests
now involve hydrodynamics in addition to the radiative transfer and thermochemistry.
The reference solutions shown in this section are the solutions of the codes which participated in
the comparison project, whose features are summarized in Table 1 of IL9. Most of these codes, namely
\codename{C2Ray+Capreole} \citep{mellema2RayNewMethod2006, mellemaDynamicalIiRegion2006},
\codename{C2Ray+TVD} \citep{mellema2RayNewMethod2006, tracMovingFrameAlgorithm2004},
\codename{HART} \citep{kravtsovAdaptiveRefinementTree1997,
kravtsovConstrainedSimulationsReal2002,gnedinModelingMolecularHydrogen2009},
\codename{Zeus-MP} \citep{whalenMultistepAlgorithmRadiation2006},
\codename{Coral} \citep{mellemaPhotoevaporationClumpsPlanetary1998,
shapiroPhotoevaporationCosmologicalMinihaloes2004},
\codename{Flash-HC} \citep{rijkhorstHybridCharacteristics3D2006}, and
\codename{Enzo-RT} \citep{normanSimulatingCosmologicalEvolution2007a}
solved both radiative transfer and the hydrodynamics on meshes, either on uniform or adaptive ones.
\codename{RH1D} \citep{ahnDoesRadiativeFeedback2007} uses Lagrangian spherical shells instead, and
was created to solve spherically symmetric problems.
\codename{Licorice} \citep{baekSimulated21Cm2009} used a grid for radiative transfer, but SPH for
hydrodynamics, while \codename{RSPH} \citep{susaSmoothedParticleHydrodynamics2006}
used SPH for hydrodynamics and a particle-based ray-tracing method for the
radiation transport. Furthermore, with the exception of \codename{HART}, which used the variable
eddington tensor closure for the moments of the equation of radiative transfer, and
\codename{Enzo-RT}, which used the flux limited diffusion method (which only uses the zeroth
moment of the equation of radiative transfer), all other participating codes used some form of
ray-tracing to solve the radiation transport.

Unless noted otherwise, the default parameters used with \GEARRT are again to use the GLF Riemann
solver, the minmod limiter, and the ``no flux'' flux injection model. The underlying particle
distribution is glass-like. For both star and gas particles, the smoothing length is determined by
the default parameter $\eta_{res} = 1.2348$ (eq.~\ref{eq:number-of-neighbors}). This choice results
in $\sim 48$ neighbors in 3D for the cubic spline kernel, which was used for all tests without
exception. The default reduced speed of light was $\tilde{c}_r = 0.01 c$.

\subsection{Iliev Test 5}\label{chap:Iliev5}

\begin{figure}
\centering
\includegraphics[width=.8\textwidth]{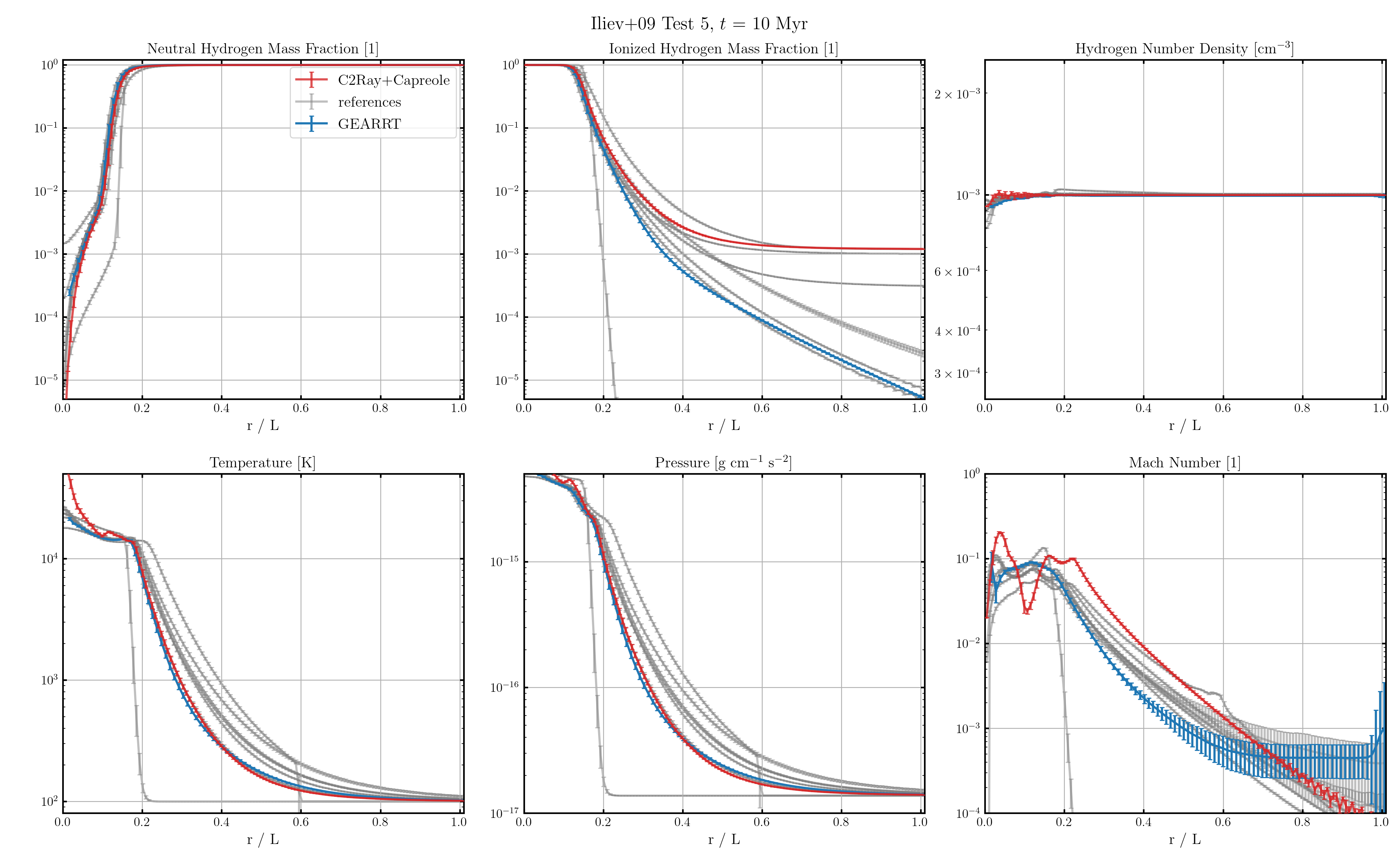}\\
\includegraphics[width=.8\textwidth]{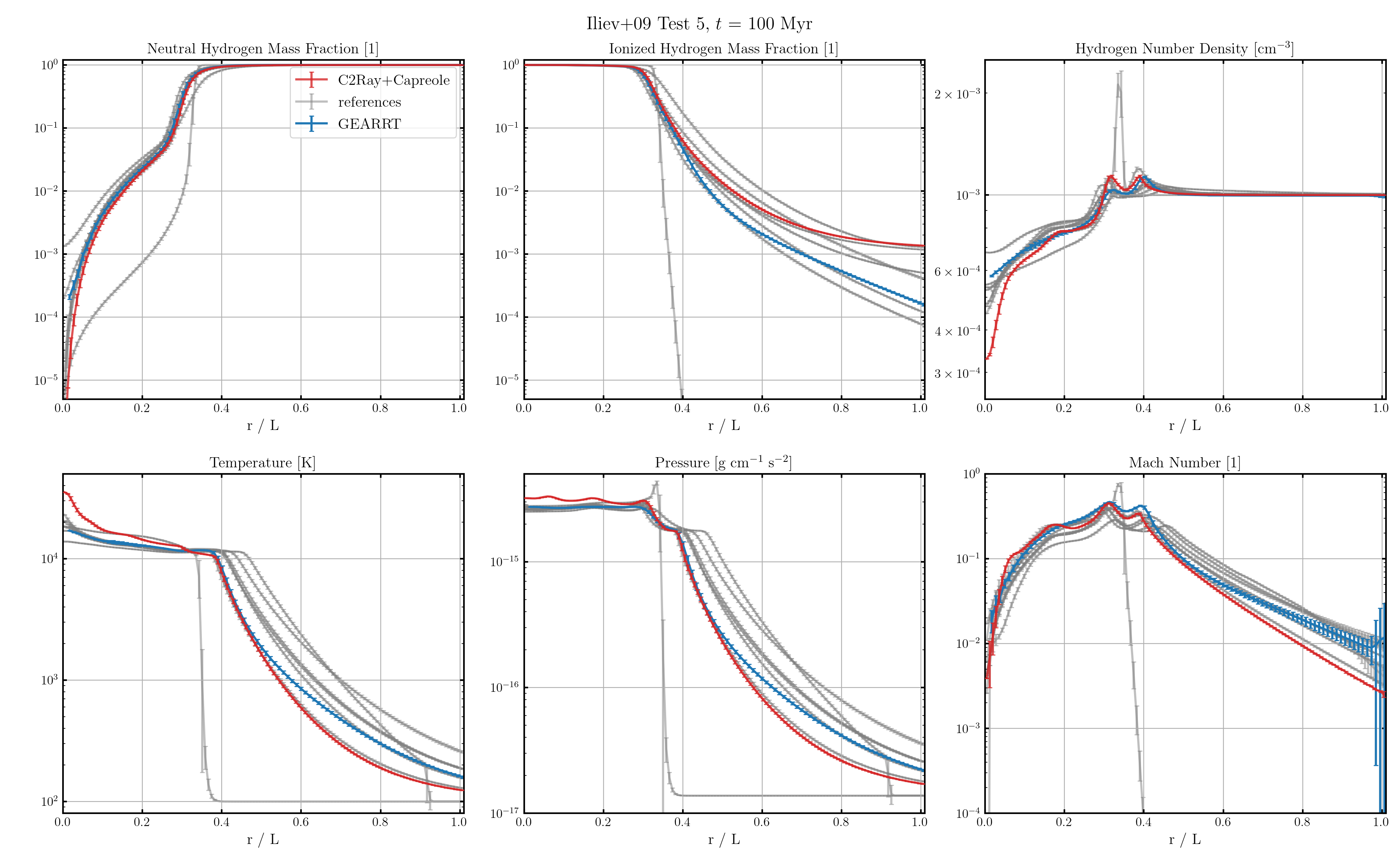}\\
\includegraphics[width=.8\textwidth]{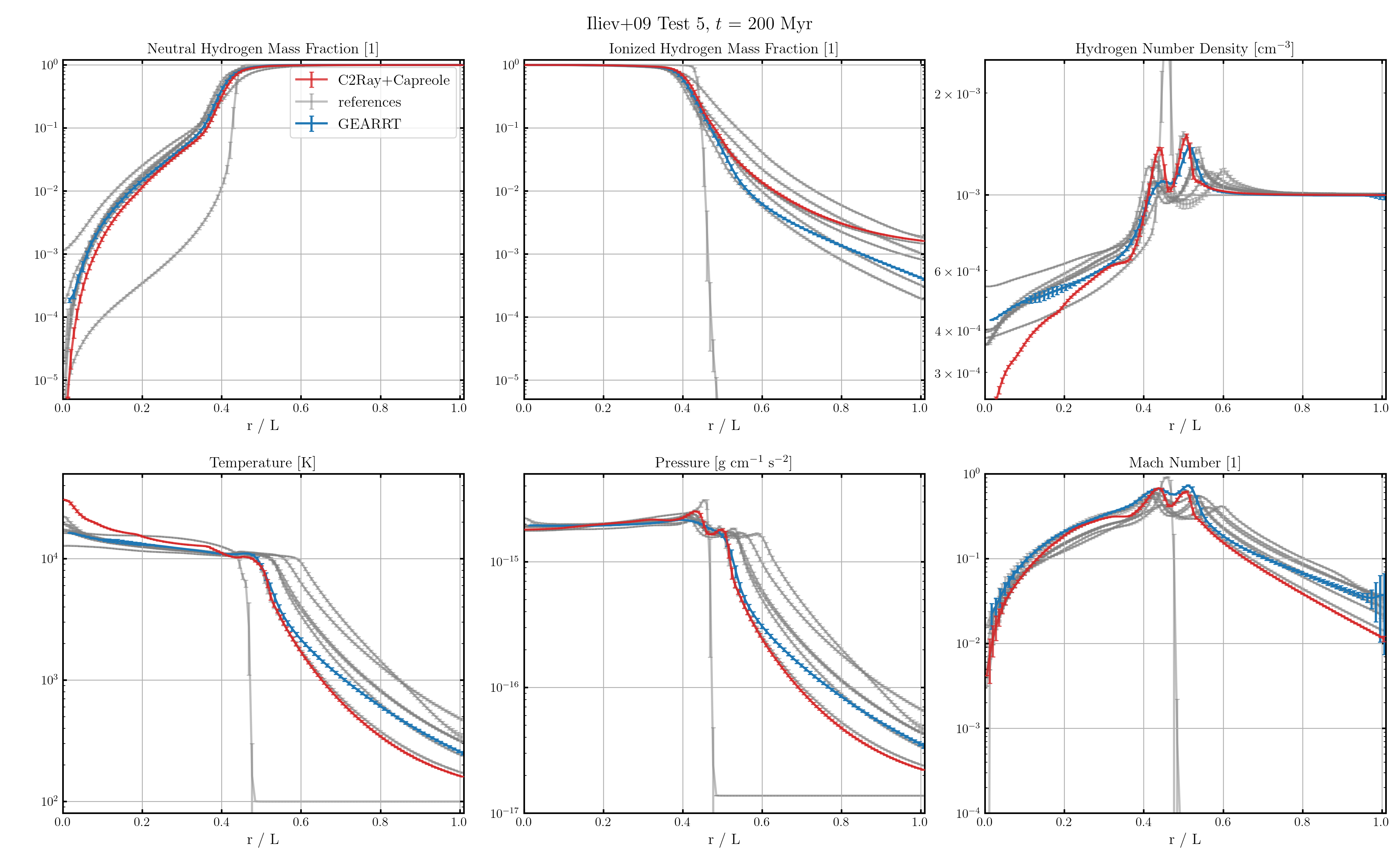}
\caption{
Spherically averaged profiles of the Test 5 problem at 10 Myr, 100 Myr, and 200 Myr. Shown are the
neutral hydrogen fraction, the ionized hydrogen fraction, the temperature, pressure, and local Mach
number of the gas. Shown are the solution of \GEARRT and reference solutions from the IL9 paper,
with \codename{C2Ray+Capreole} highlighted. The error bars are standard deviations.
}
\label{fig:iliev5}
\end{figure}

\begin{figure}
\centering
\includegraphics[width=0.7\textwidth]{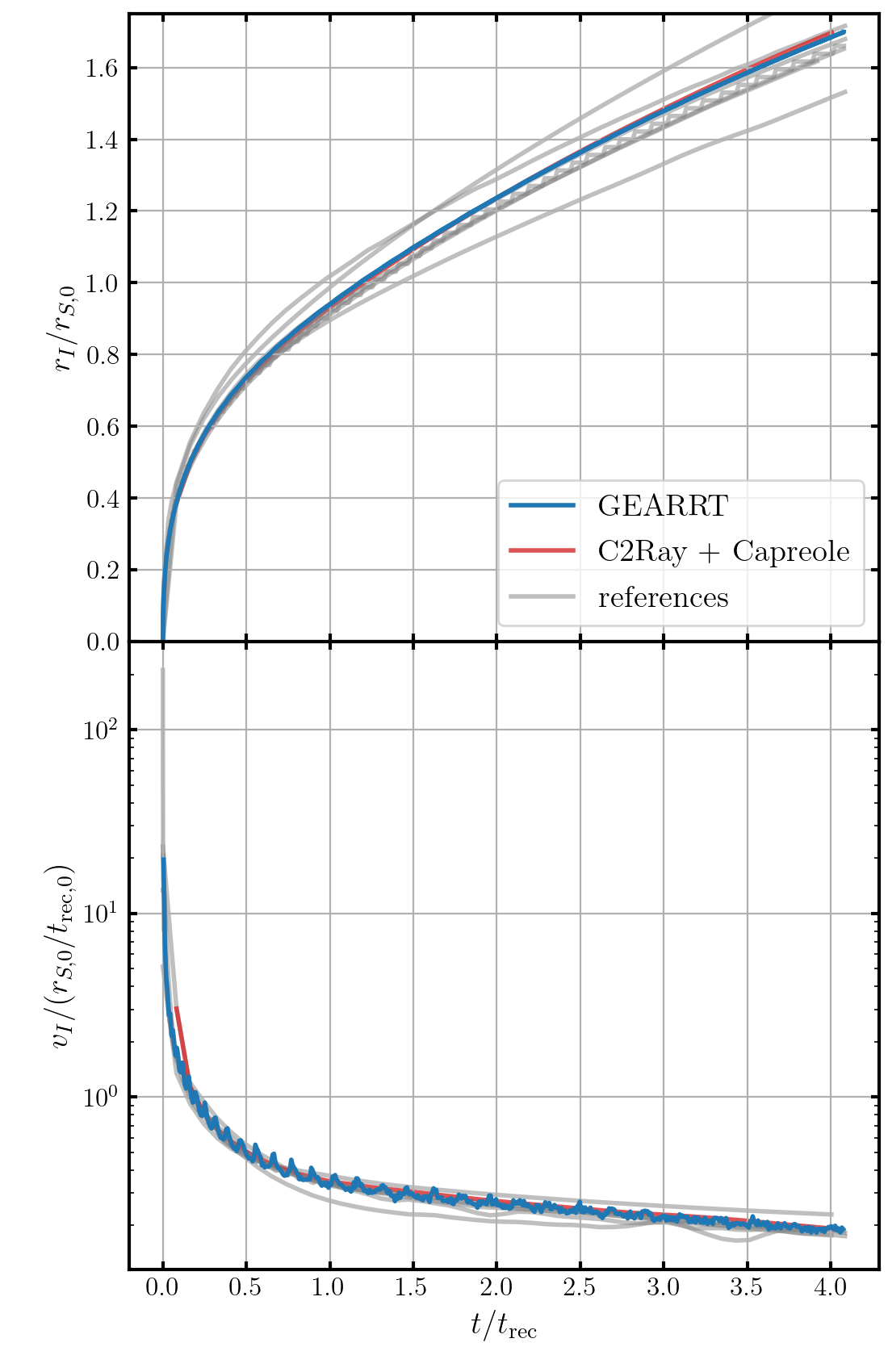}
\caption{
 Evolution of the I-front position and velocity over time for Test 5.
 Top: Evolution of the I-front position over time, compared to the analytical position
 Bottom: Evolution of the I-front velocity. Shown are the solution of \GEARRT and reference
 solutions from the IL9 paper, with \codename{C2Ray+Capreole} highlighted.
}
\label{fig:iliev5-Ifront}
\end{figure}

\begin{figure}
\centering
\includegraphics[width=\textwidth]{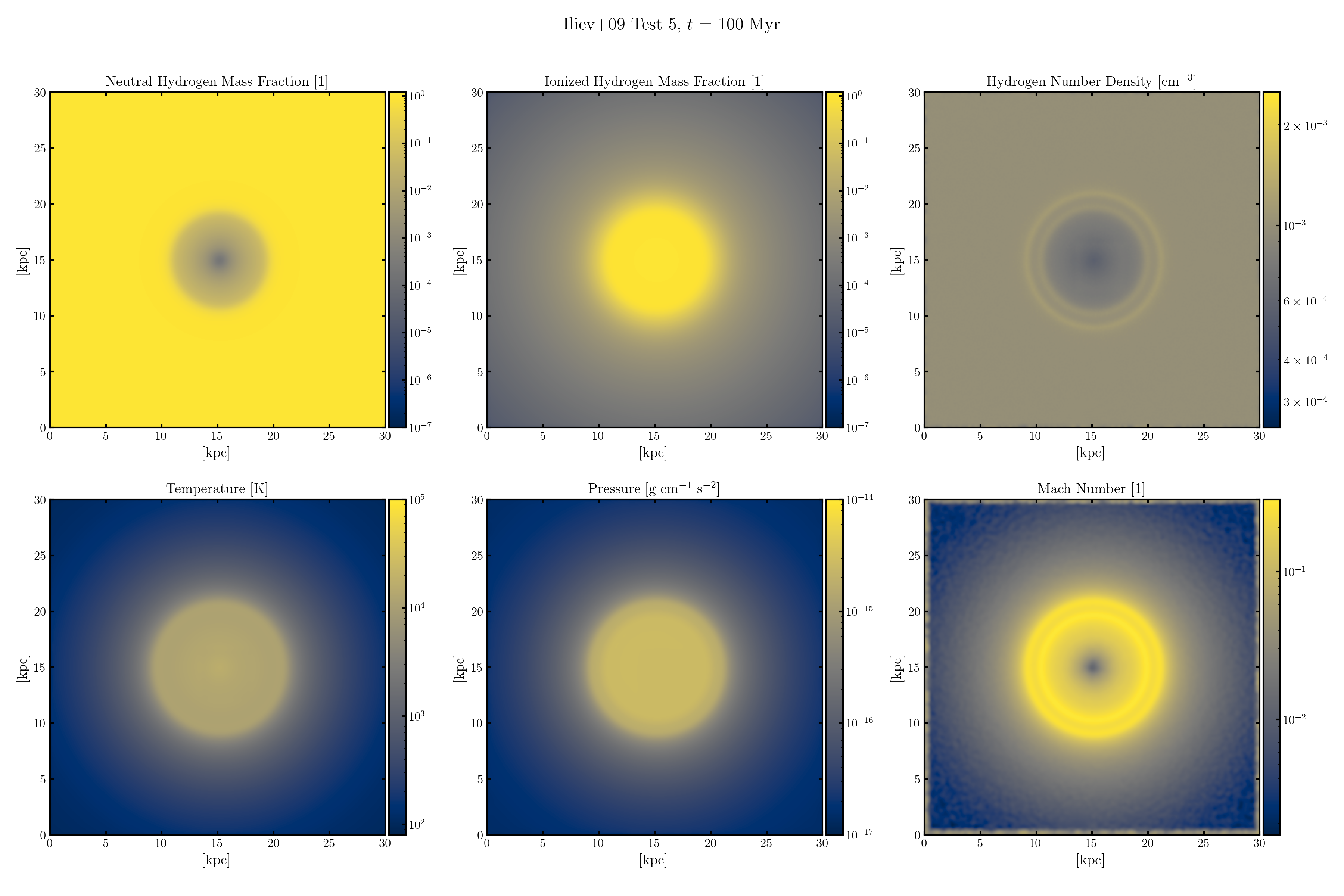}\\
\includegraphics[width=\textwidth]{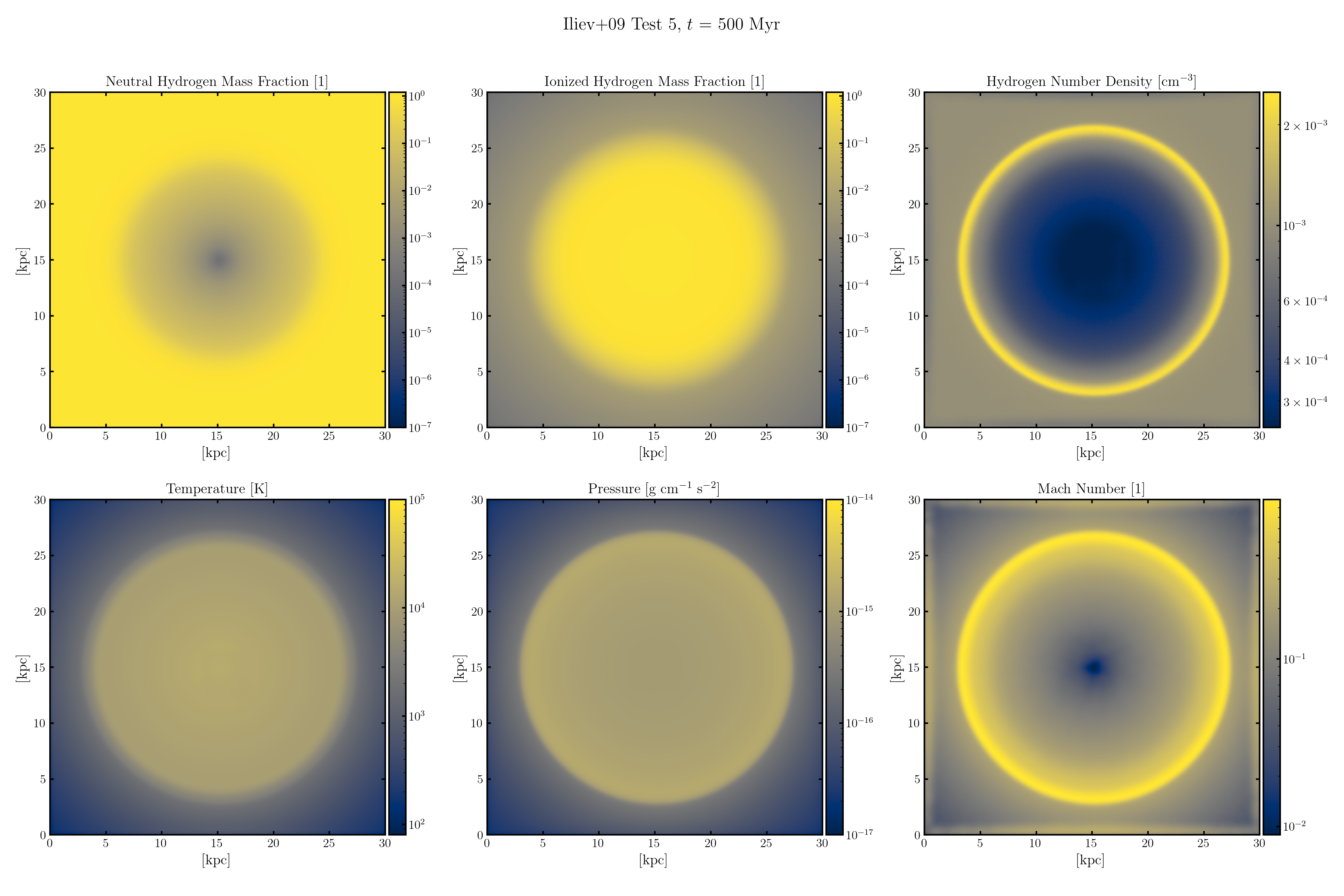}
\caption{
Slice across the mid-plane for the Iliev Test 5 with \GEARRT at 100 Myr (top) and 500 Myr (bottom).
Shown are the neutral hydrogen fraction, the ionized hydrogen fraction, the temperature, pressure,
and local Mach number of the gas.
}
\label{fig:iliev5-slice}
\end{figure}

\begin{figure}
\centering
\includegraphics[width=\textwidth]{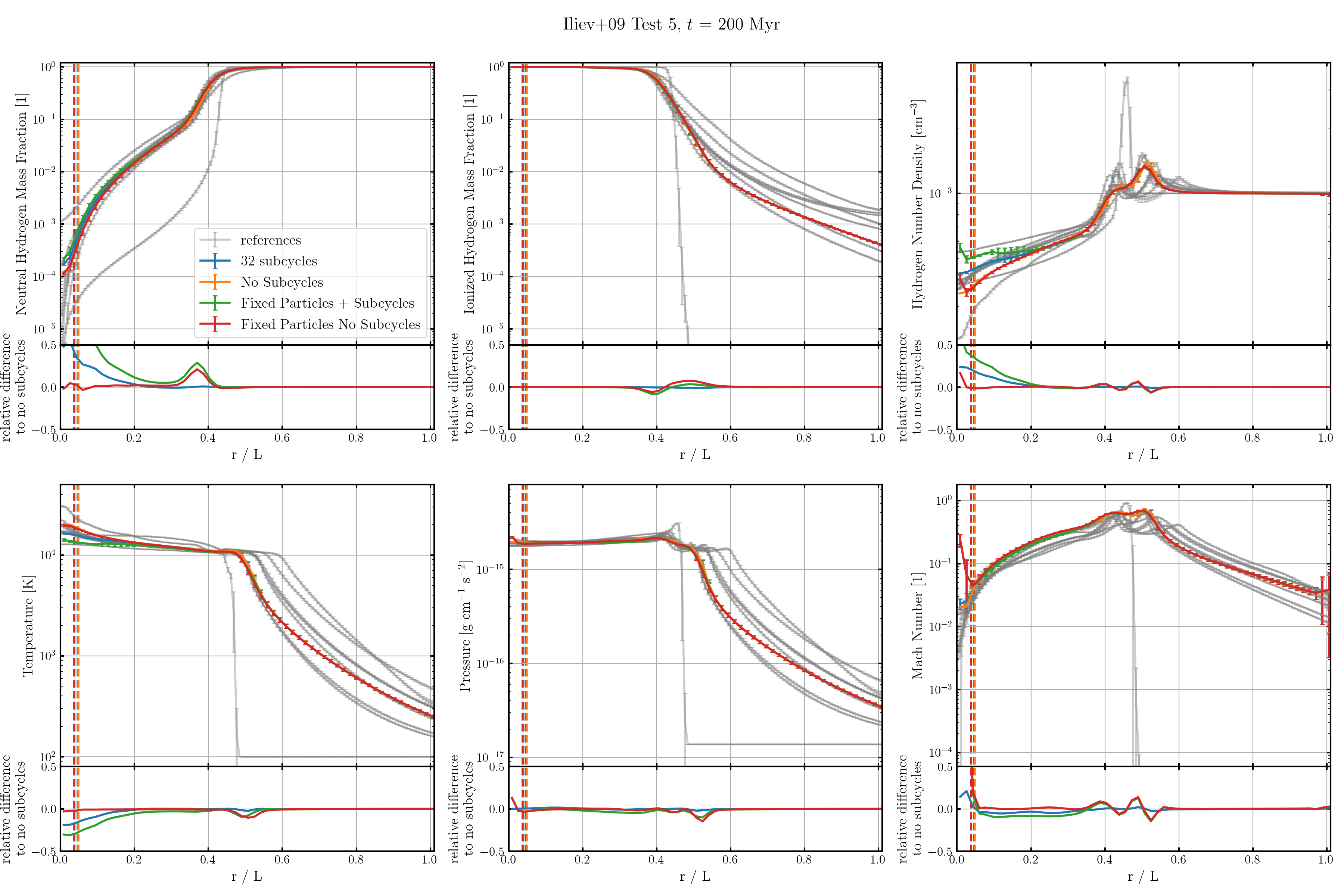}
\caption{
Spherically averaged profiles of the Test 5 problem at 200 Myr both with and without sub-cycling,
and with and without drifting particles. Shown are the neutral hydrogen fraction, the ionized
hydrogen fraction, the temperature, pressure, and local Mach number of the gas, as well as the
relative difference of the results when comparing with the run without sub-cycles.
}
\label{fig:iliev5-subcycling}
\end{figure}

This test prescribes the classical HII region expansion in an initially uniform gas. Like
in Tests 1 and 2, radiation is injected from a single source, which I place in the center of the
box to avoid needing reflective boundary conditions along the box faces, and take a box size twice
the prescribed size to a total of $30$ kpc. The gas is hydrogen only with a number density of $n =
10^{-3}$cm$^{-3}$ and initial temperature of $100$K. The resolution is $128^3$ particles, which
compared to the $128^3$ cells used in IL9 leads to half the spatial resolution due to the doubled
box size in my simulation. The spectrum of the source is taken to be a blackbody spectrum with
temperature $T_{bb} = 10^5$K, which I again split into three frequency intervals following
eqs.~\ref{eq:group-luminisoties-1}-\ref{eq:group-luminosities-3}.

Figure~\ref{fig:iliev5} shows spherically averaged profiles of the neutral hydrogen mass fraction,
the ionized hydrogen mass fraction, the gas number density, the gas temperature, pressure, and Mach
number at different times. Figure~\ref{fig:iliev5-Ifront} shows the position and velocity of
the ionization fronts over time. They agree quite well with the reference solutions.
Figure~\ref{fig:iliev5-slice} shows the slices through the mid-plane at 100 Myr and at 500 Myr, and
demonstrates that \GEARRT is able to grow nicely symmetrical Str\"omgren spheres.

As the first test which includes hydrodynamics, it offers a good opportunity to validate the
sub-cycling approach. Figure~\ref{fig:iliev5-subcycling} shows the solution to this test problem at
200 Myr performed in four different ways: The particles can either be forced to remain static, or
be used in a Lagrangian manner, which is the default behavior. Both these approaches are tested
with and without sub-cycling, and the results and their relative differences compared to the
solution with Lagrangian particles and without sub-cycling are plotted.

The sub-cycling indeed introduces a difference in regions very close to the source, which is
located at $r = 0$. This is to be expected, as in terms of spherical averaged profiles, this is a
poorly resolved region that contains only a few particles. However, comparing to reference
solutions, these differences are all in an acceptable range, and as such not deemed as an issue. A
second set of differences can be seen around the region where the shock is currently located, which
in Figure~\ref{fig:iliev5-subcycling} is located around $r/L \sim 0.5$. However, the sub-cycling
has virtually no effect there. Instead, the differences are dominated by whether the particles are
Lagrangian or static. This is to be expected, as the Lagrangian mode should allow the shock to be
better resolved, as the particles' positions will also be compressed close to the shock front.
Indeed, looking at the density profile, the two peaks around $r/L \sim 0.5$ are less pronounced
when the particles are forced to be static.

\subsection{Iliev Test 6}\label{chap:Iliev6}

\begin{figure}
\centering
\includegraphics[width=\textwidth]{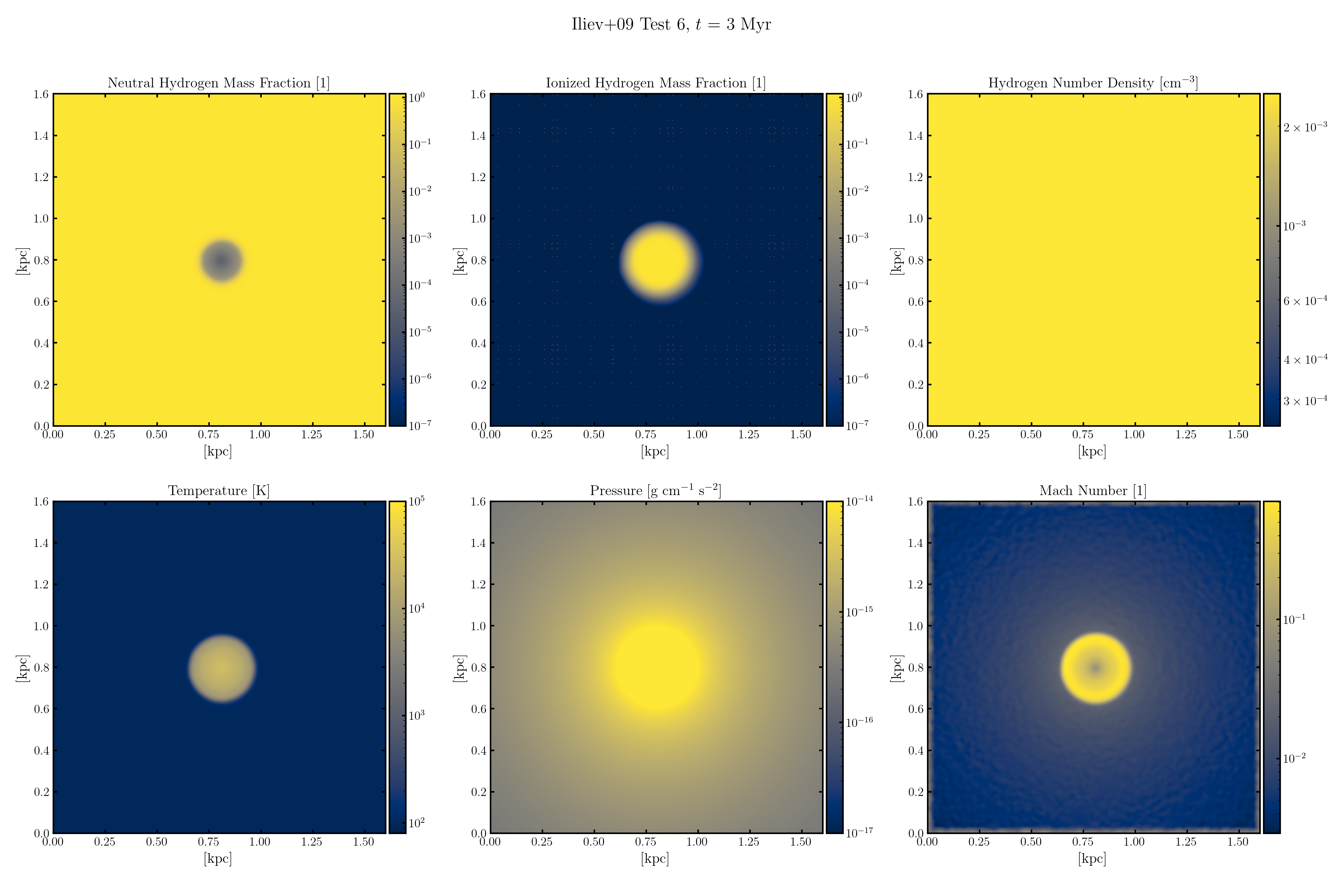}\\
\includegraphics[width=\textwidth]{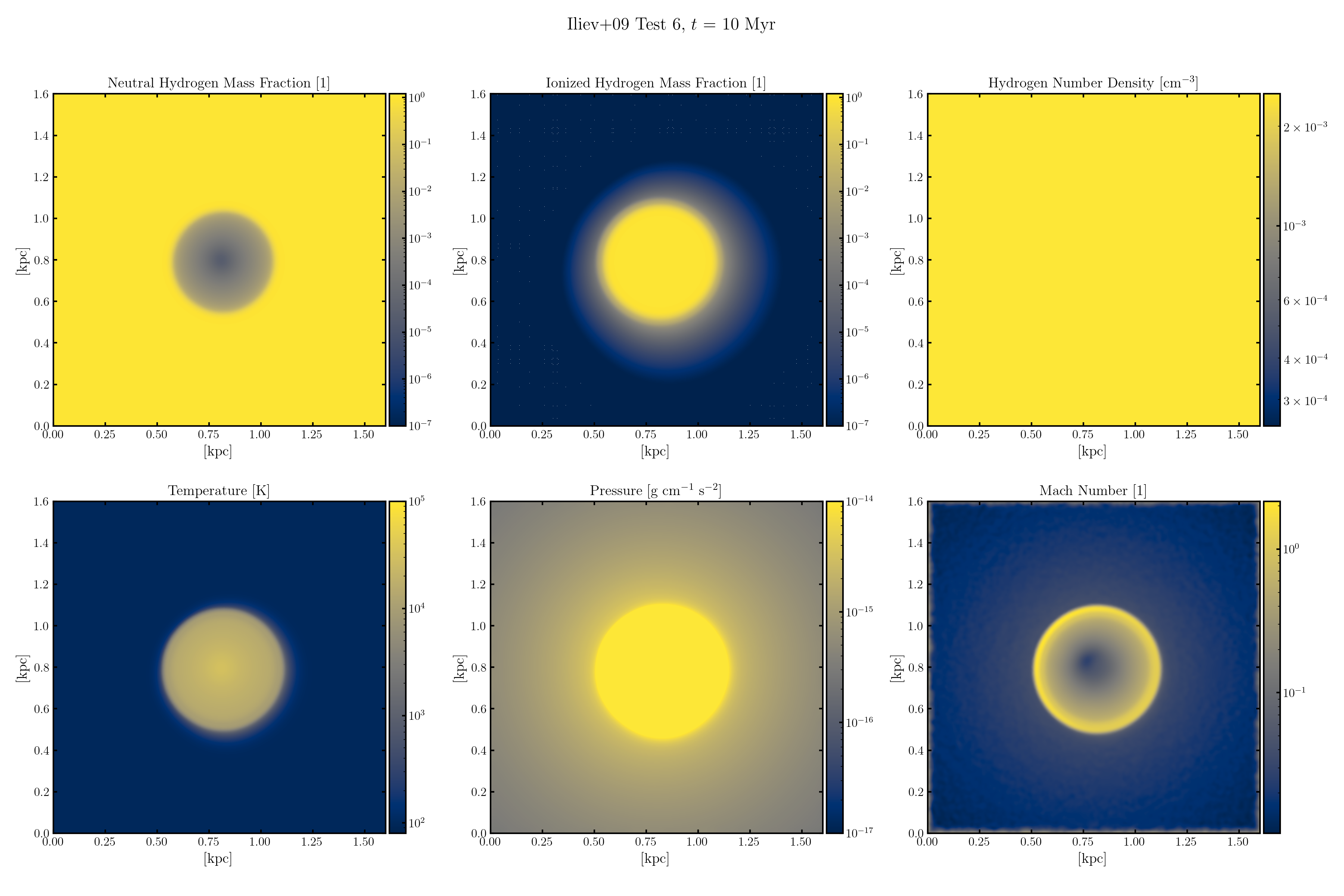}
\caption{
Slice across the mid-plane for the Iliev Test 6 with \GEARRT at 3 Myr (top) and 10 Myr (bottom).
Shown are the neutral hydrogen fraction, the ionized hydrogen fraction, the temperature, pressure,
and local Mach number of the gas. Some anisotropies develop at later times because the stronger
shock (compared to the previous tests) from the photo-heating advects too many particles away from
the inner region, leaving the central regions poorly resolved. The anisotropies are reflected in
the comparatively much bigger error bars in the profiles of these quantities in
Figure~\ref{fig:iliev6}.
}
\label{fig:iliev6-slices}
\end{figure}

\begin{figure}
\centering
\includegraphics[width=.8\textwidth]{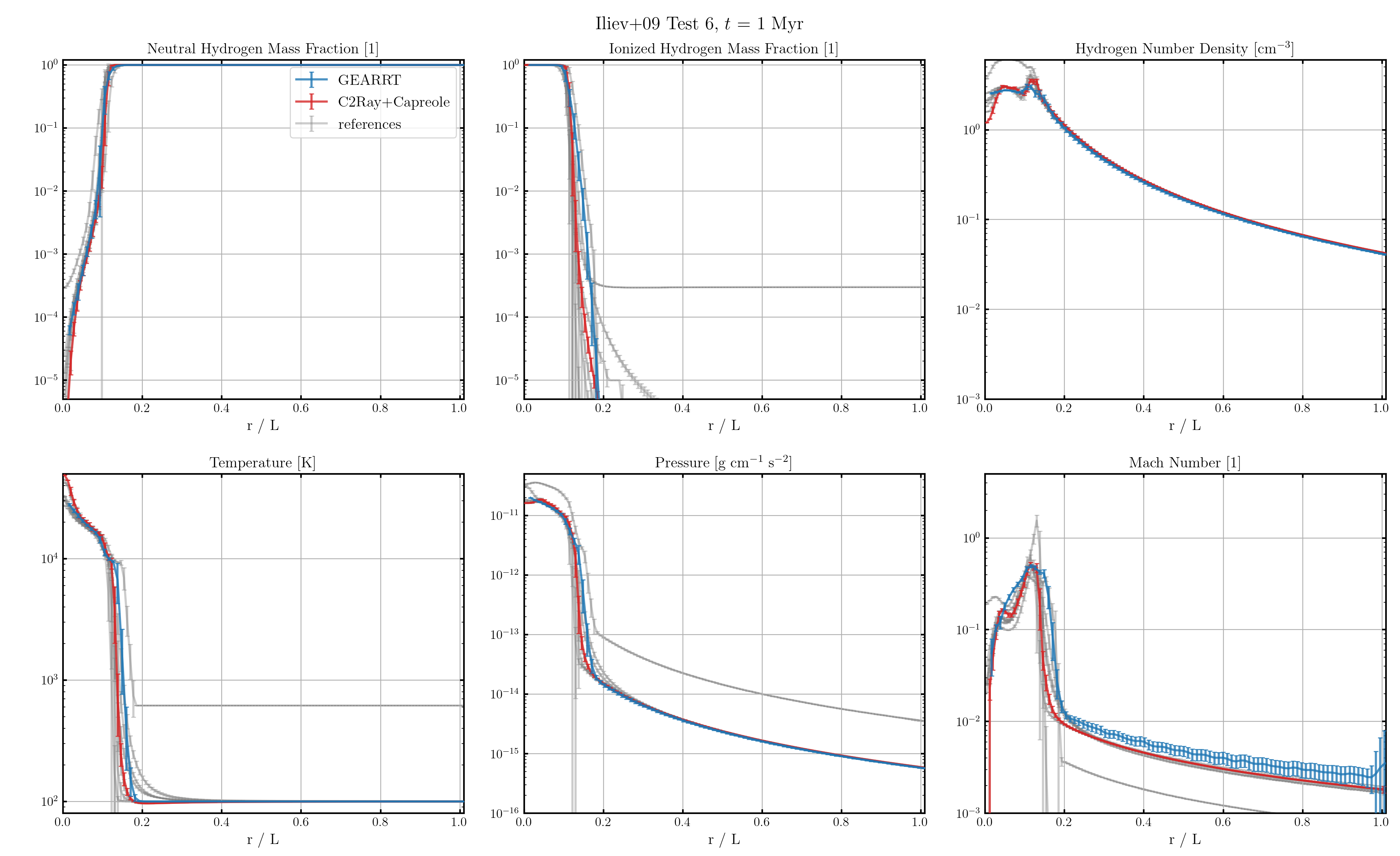}\\
\includegraphics[width=.8\textwidth]{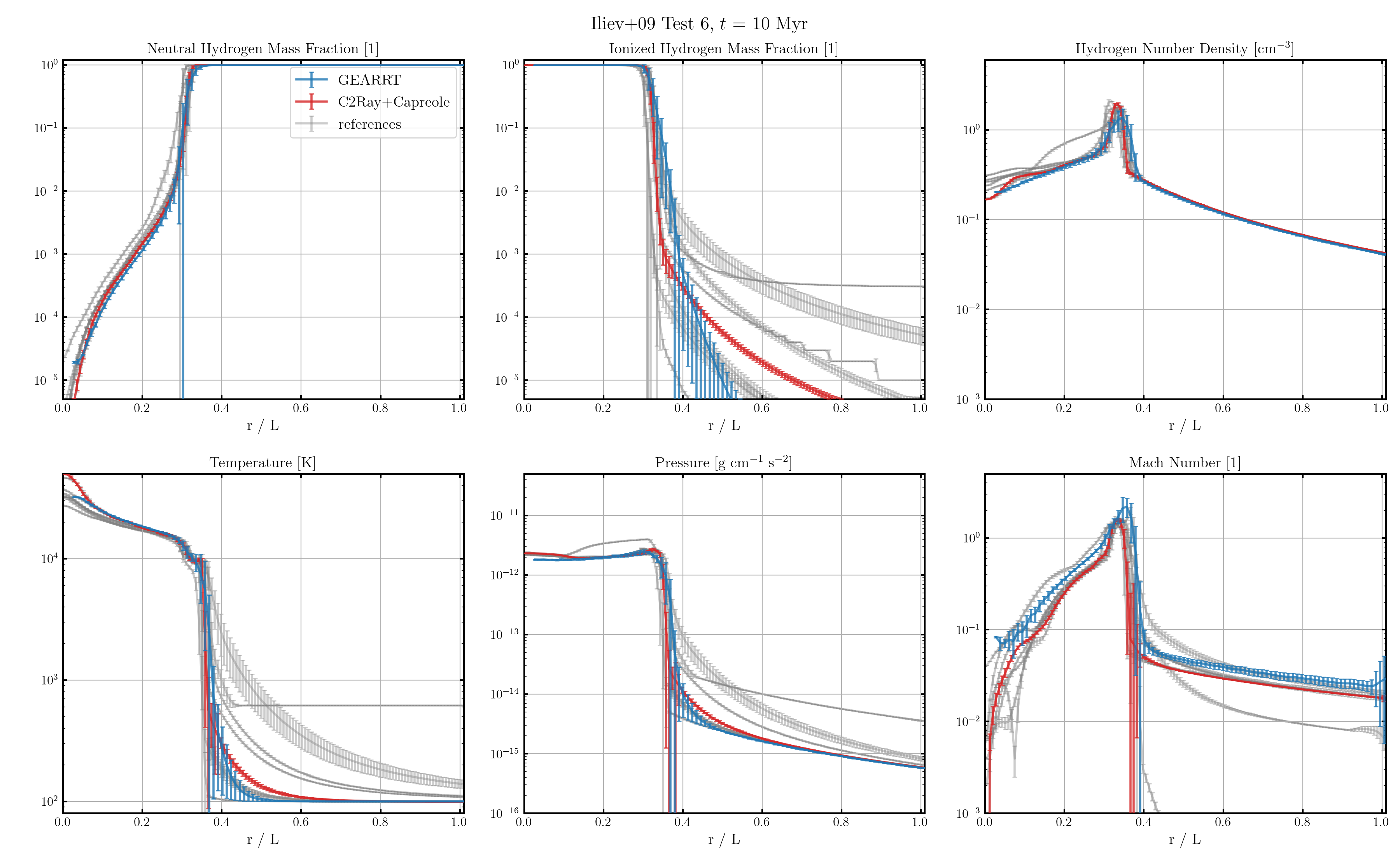}\\
\includegraphics[width=.8\textwidth]{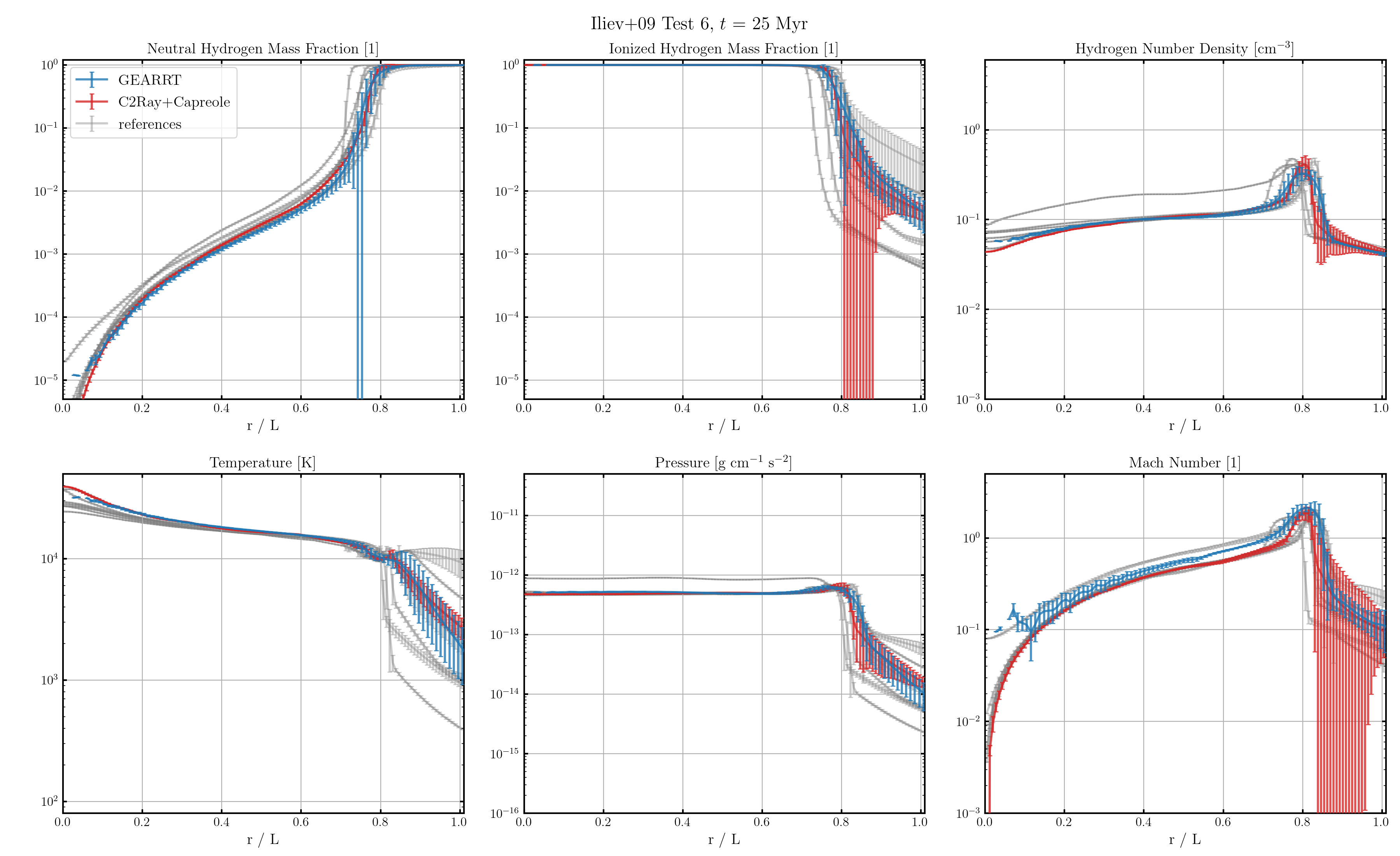}
\caption{
Spherically averaged profiles of the Test 6 problem at 1 Myr, 10 Myr, and 25 Myr. Shown are the
neutral hydrogen fraction, the ionized hydrogen fraction, the temperature, pressure, and local
Mach number of the gas. The error bars are standard deviations.
}
\label{fig:iliev6}
\end{figure}

\begin{figure}
\centering
\includegraphics[width=0.7\textwidth]{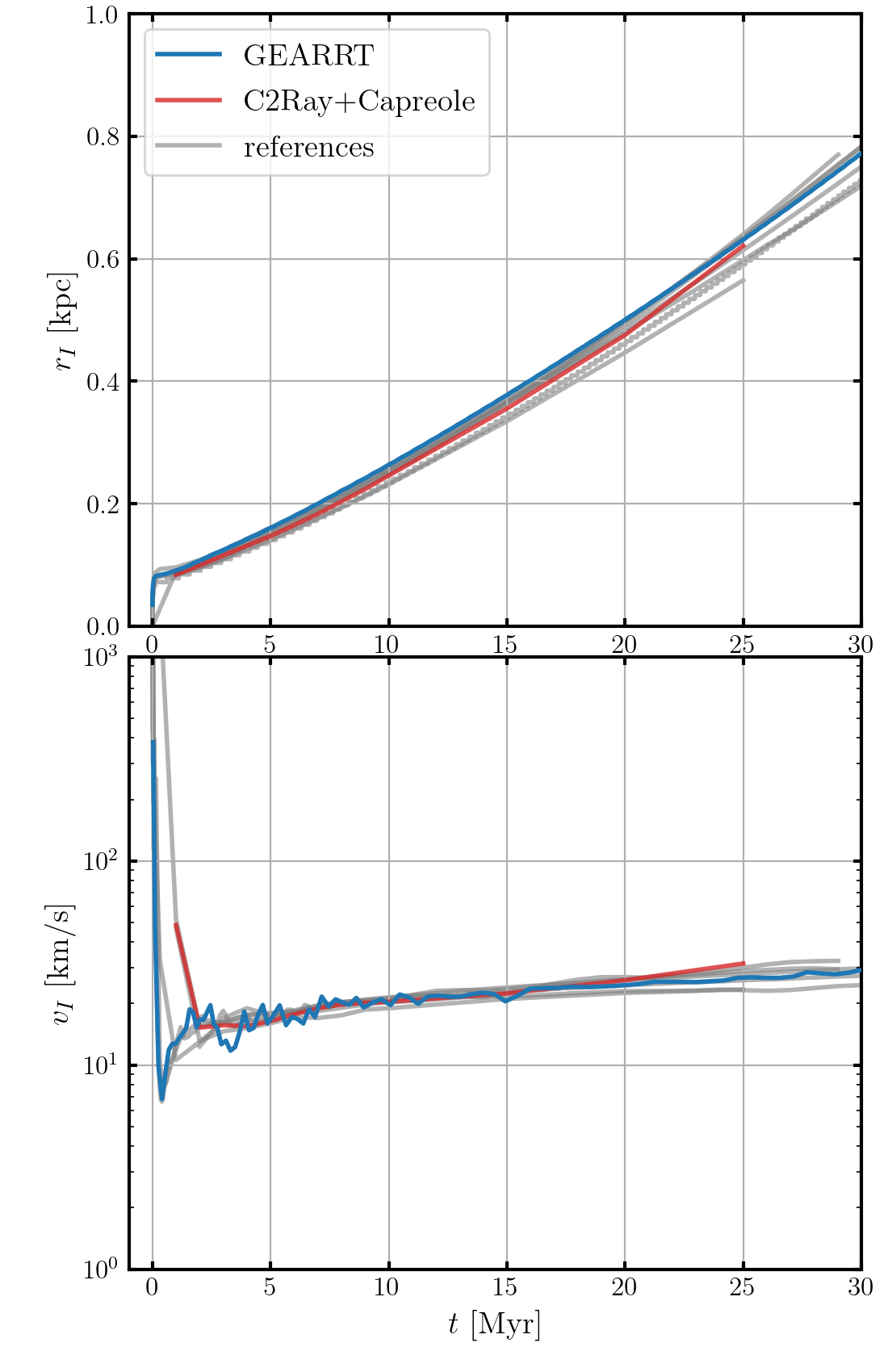}
\caption{
 Evolution of the I-front position and velocity over time for Test 6.
 Top: Evolution of the I-front position over time, compared to the analytical position
 Bottom: Evolution of the I-front velocity.
}
\label{fig:iliev6-Ifront}
\end{figure}

This test is similar to the Test 5, but uses an inhomogeneous initial gas density of

\begin{align}
    n(r) = \begin{cases}
            n_0 & \text{ if } r \leq r_0 \\
            n_0 \frac{r_0}{r^2} & \text{ if } r \geq r_0
           \end{cases}
\end{align}

with $n_0 = 3.2$cm$^{-3}$ and $r_0 = 91.5$ pc. The box size is 0.8 kpc, which is again doubled in
my simulation. The gas has an initial temperature of $T = 100$K. The radiation source has a
blackbody spectrum with temperature $T_{bb} = 10^5$K and emits $10^{50}$ ionizing photons per
second, which is a larger emission compared to the previous tests by a factor of 20. The aim of this
test is to study the initial transition of the I-front from R-type, where the I-front propagates
much faster than the gas' response to it, to D-type, in which case the I-front propagates at about
the same velocity as the gas, and back to R-type. The initial transition from R-type to D-type
occurs during the growth of the Str\"omgren sphere, which in this test's setup occurs at $\sim
70$pc, i.e. inside the core with a flat density profile. The acceleration back to an R-type I-front
occurs due to the decreasing density after the I-front has reached the region of the density profile
with the steep gradient.

Figure~\ref{fig:iliev6-slices} shows slices through the mid-plane of the box of the solution at 1
Myr, 10 Myr, and 25 Myr. The neutral hydrogen fraction, the ionized hydrogen fraction, the
temperature, pressure, and local Mach number of the gas are shown. In this test, \GEARRT was unable
to maintain a perfectly symmetrical spherical expansion of the I-front at later times, which can be
seen particularly well in the slice at 25 Myr for the ionized hydrogen mass fraction and the mach
number. Figure~\ref{fig:iliev6} shows the spherically averaged profiles of the same quantities and
at the same times as Figure~\ref{fig:iliev6-slices}. The anisotropies can be seen in the profiles in
Figure~\ref{fig:iliev6} as comparatively much larger error bars close to the ionization front,
which show the standard deviation, compared to the results from previous tests. The reason for the
anisotropies forming is that compared to the previous tests, the heating and the resulting shock in
this test was much stronger, which lead to most particles being drifted away from the central region
behind the I-front, leaving the region poorly resolved. Indeed at 25 Myr, where the I-front is at
$r/L \sim 0.75$, only about 6.8\% of the total number of particles remained within $r/L \leq 0.75$,
whereas the initial fraction of particles within the same radius was 22.1\%. The low sampling can be
seen particularly well in the noisy Mach numbers at 10 and 25 Myr. At 25 Myr, only $\sim 2500$
particles, or 0.1\% of the total number of particles, remained within the region $r/L \leq 0.4$,
leading to the noisy results. Aside from that, the results of \GEARRT once again agree fairly well
with the reference solutions. This is also the case for the positions and velocities of the
ionization front over time, which are shown in Figure~\ref{fig:iliev6-Ifront}.

\subsection{Iliev Test 7}\label{chap:Iliev7}

\begin{figure}
\centering
\includegraphics[width=.75\textwidth]{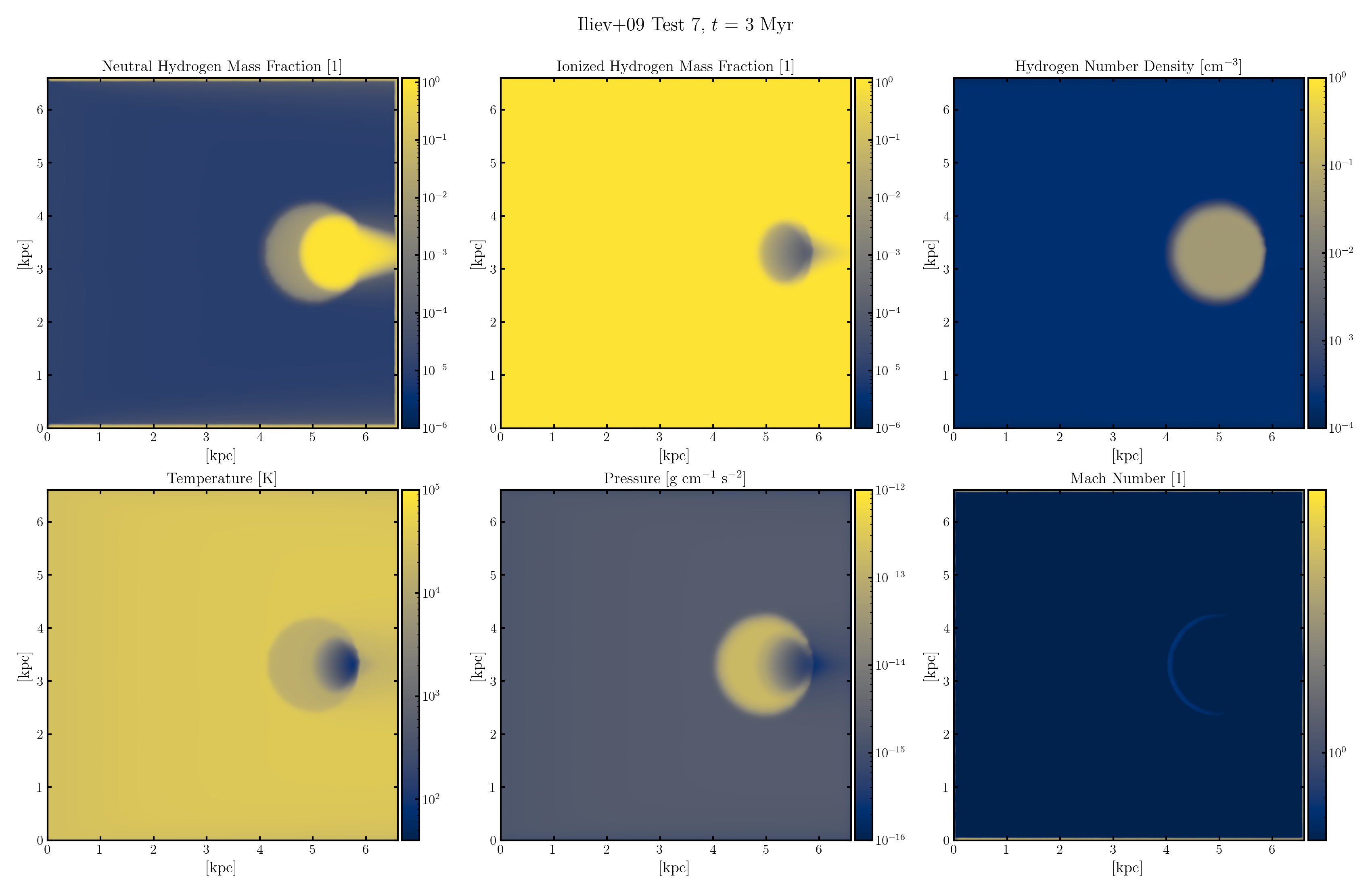}\\
\includegraphics[width=.75\textwidth]{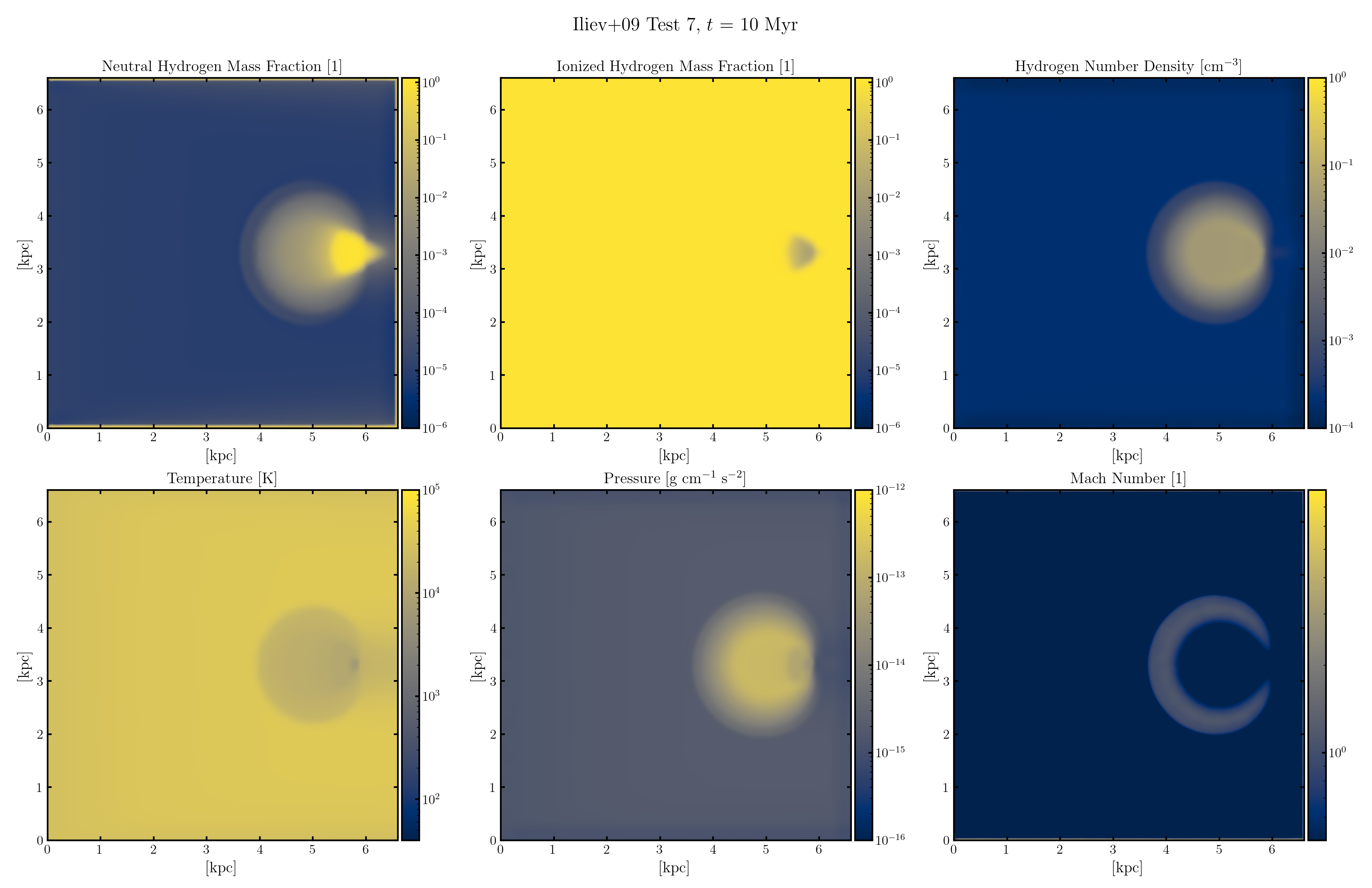}\\
\includegraphics[width=.75\textwidth]{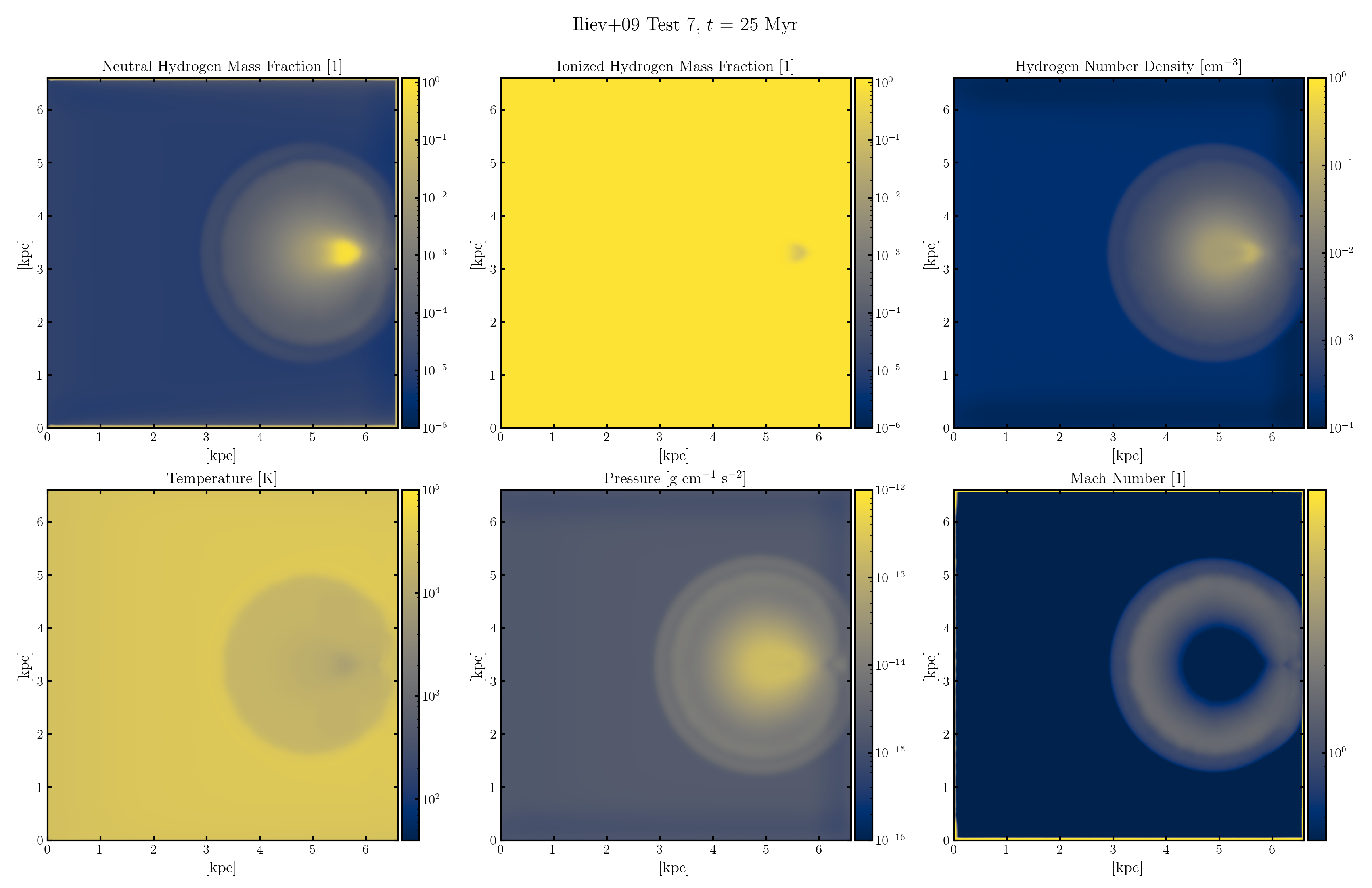}
\caption{
Slices through the midplane of the box for the Iliev 7 test at 3, 10, and 25 Myr using the GLF
Riemann solver. Shown are the neutral hydrogen fraction, the ionized hydrogen fraction, the
temperature, pressure, and local Mach number of the gas.
}
\label{fig:iliev7-slices-GLF}
\end{figure}

\begin{figure}
\centering
\includegraphics[width=.75\textwidth]{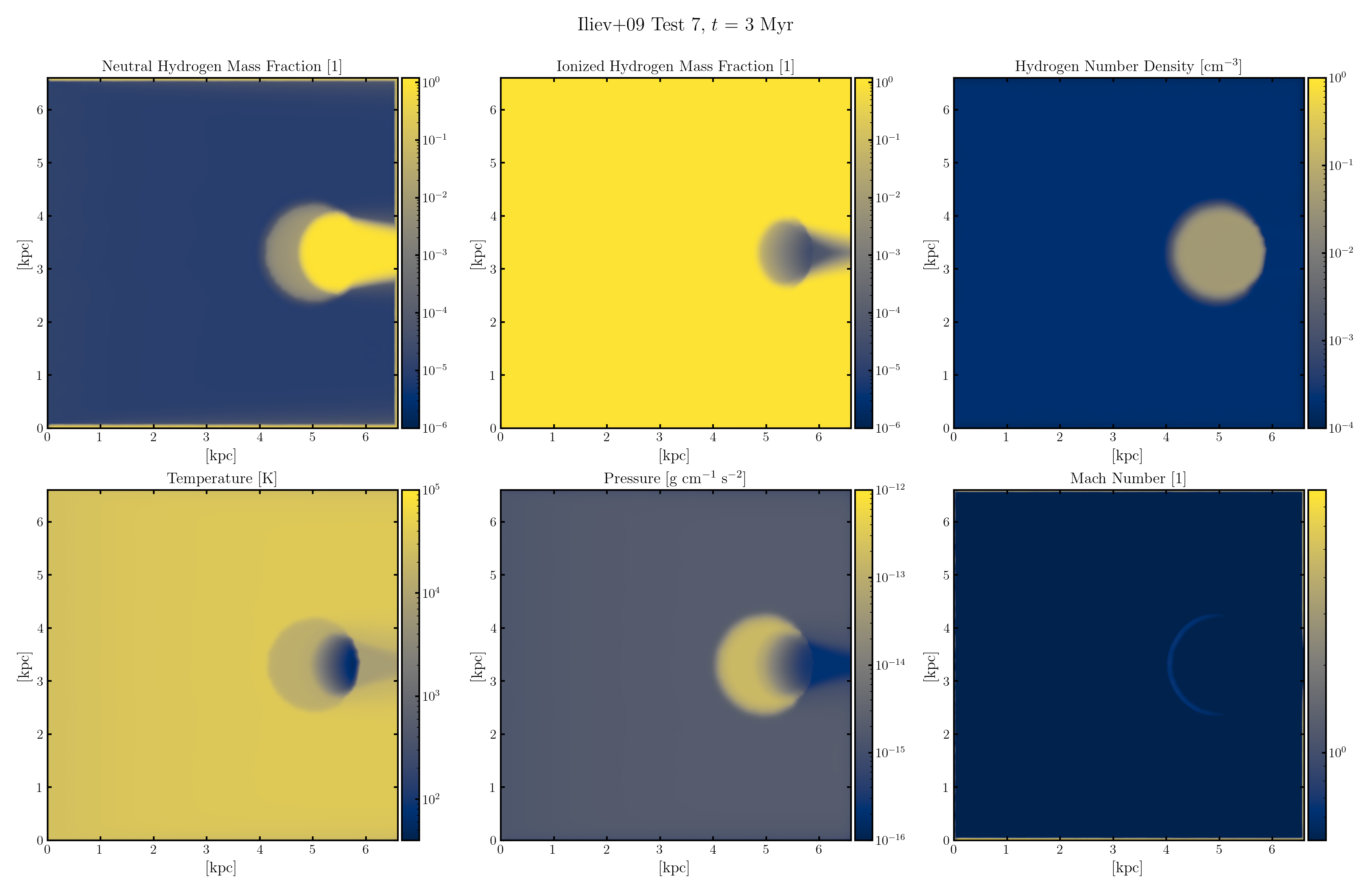}\\
\includegraphics[width=.75\textwidth]{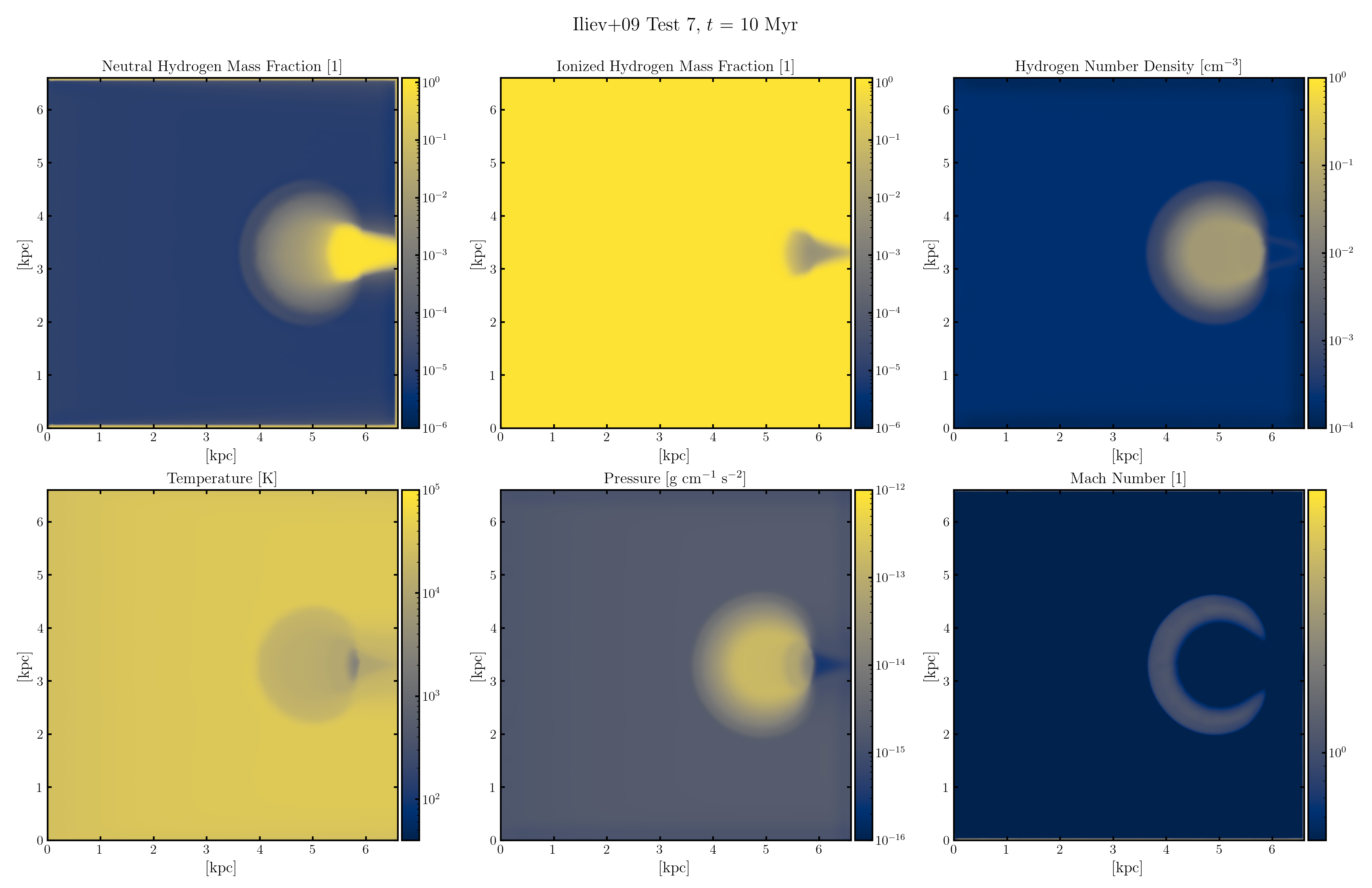}\\
\includegraphics[width=.75\textwidth]{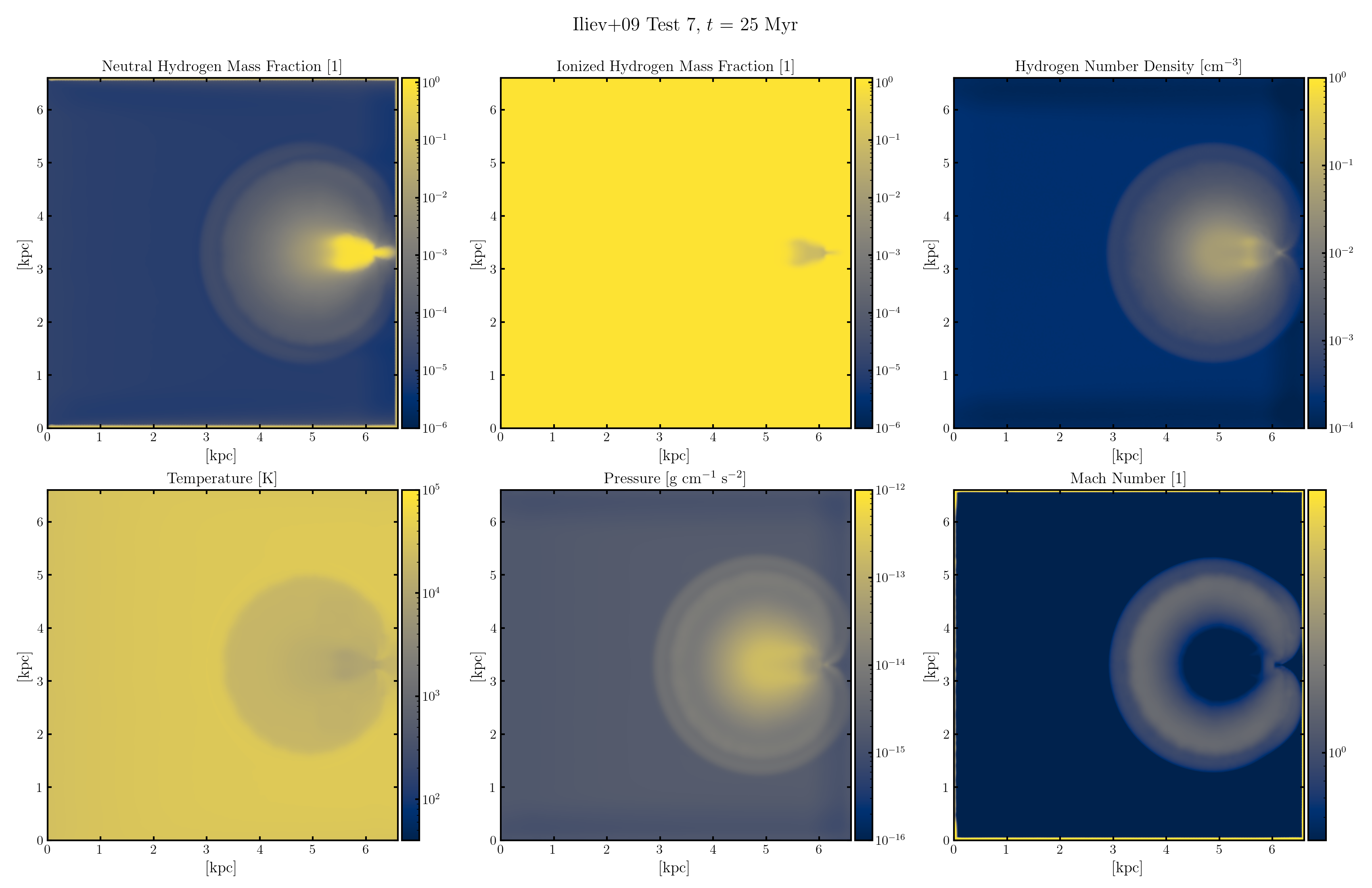}
\caption{
Slices through the midplane of the box for the Iliev 7 test at 1, 10, and 50 Myr using the HLL
Riemann solver. Shown are the neutral hydrogen fraction, the ionized hydrogen fraction, the
temperature, pressure, and local Mach number of the gas.
}
\label{fig:iliev7-slices-HLL}
\end{figure}

\begin{figure}
\centering
\includegraphics[width=.8\textwidth]{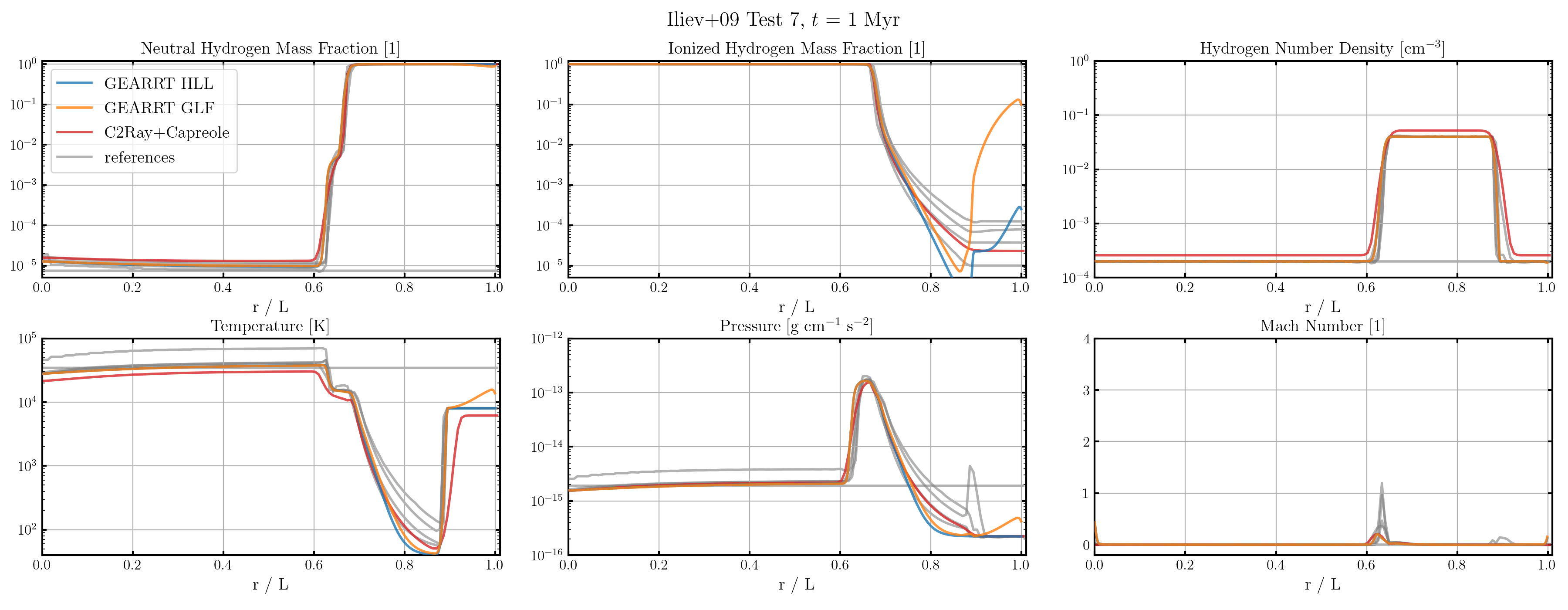}\\
\includegraphics[width=.8\textwidth]{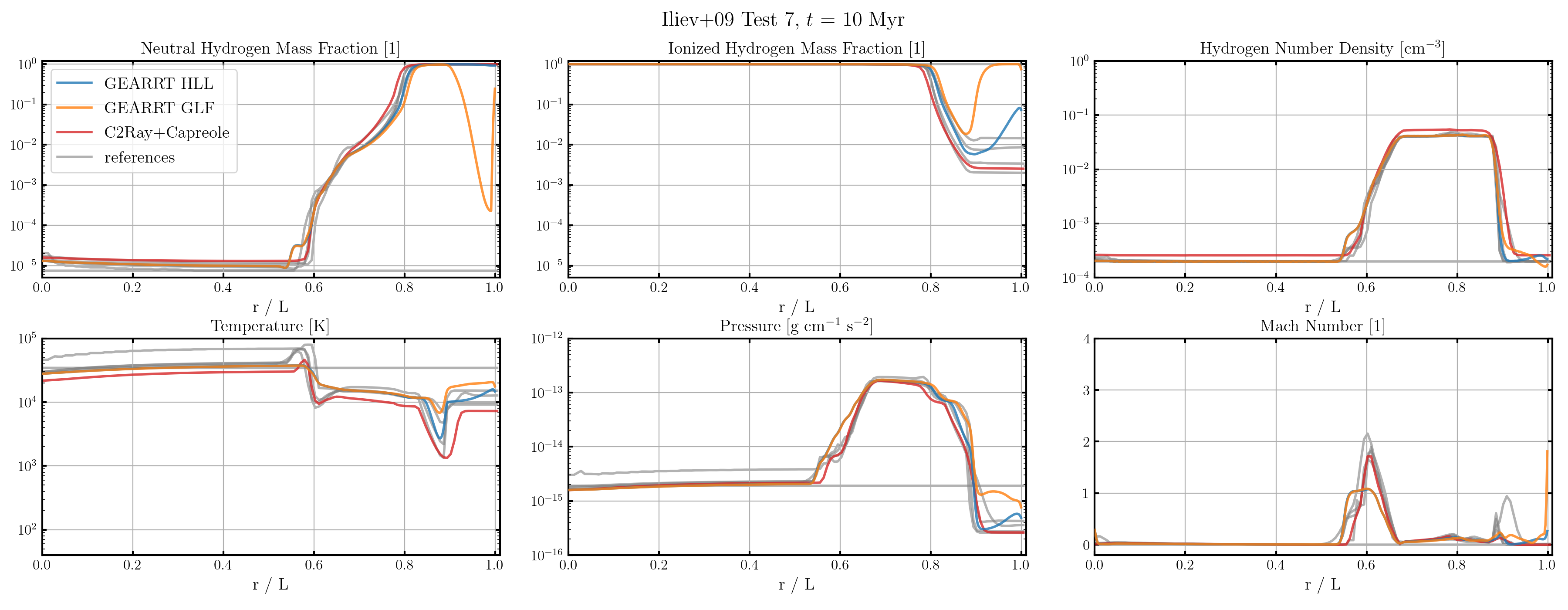}\\
\includegraphics[width=.8\textwidth]{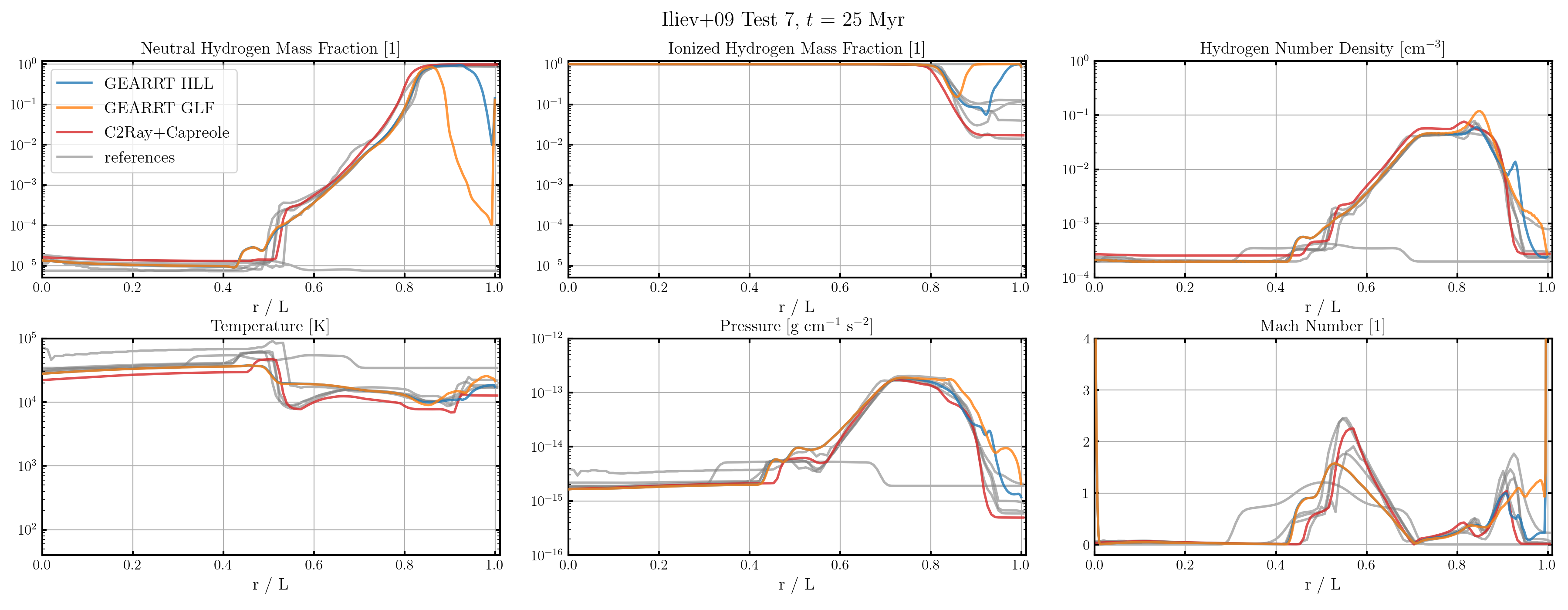}\\
\includegraphics[width=.8\textwidth]{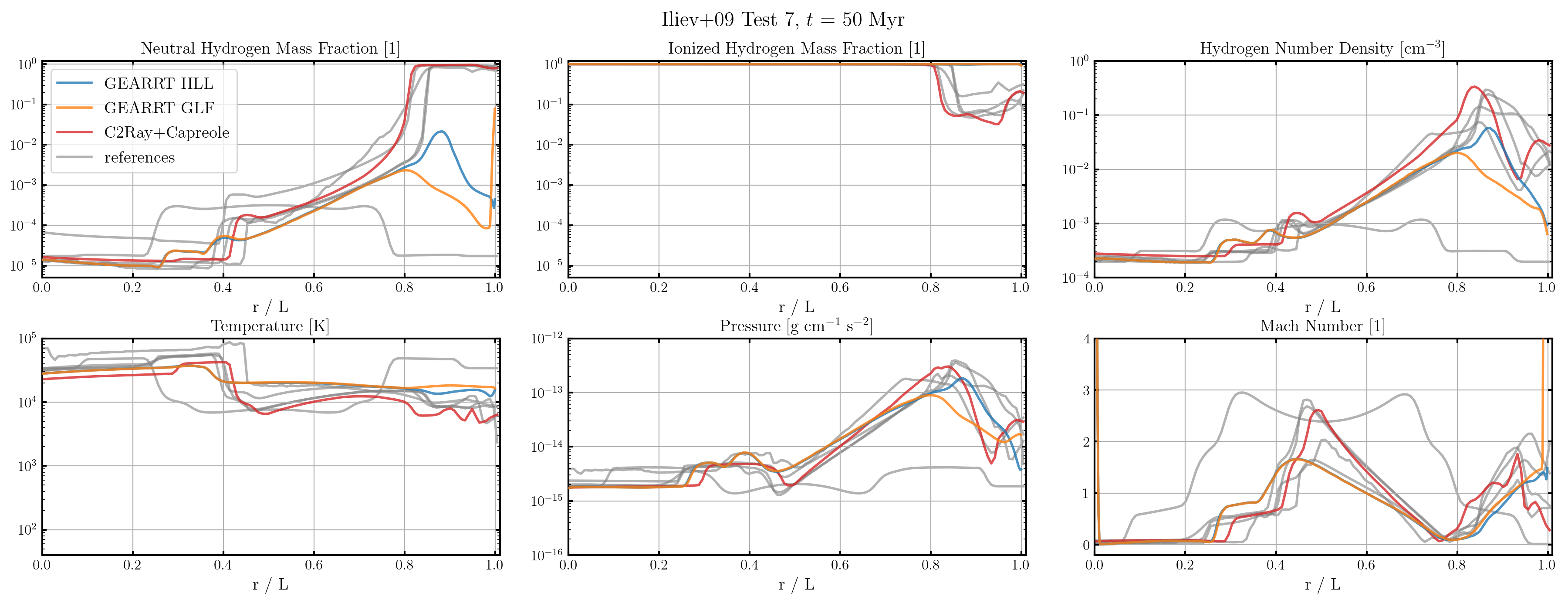}
\caption{
Line cuts along the axis of symmetry of the neutral hydrogen fraction, the ionized hydrogen
fraction, the temperature, the pressure, and the local Mach number for the Iliev 7 test at 1, 10,
25, and 50 Myr for \GEARRT using the HLL and the GLF Riemann solver, for \codename{C2Ray+Capreole},
and for other references.
}
\label{fig:iliev7-profiles}
\end{figure}

This test is set up the same way as Test 3 (Section~\ref{chap:Iliev3}): A uniform overdense clump
is placed in an otherwise uniform background density, and a plane-parallel front of radiation is
emitted from the $x = 0$ plane of the box. As in Test 3, the simulation box has a size of $6.6$ kpc.
A spherical cloud of gas with radius $r_{cloud} = 0.8$ kpc is placed centered at $(5, 3.3, 3.3)$
kpc. The surrounding hydrogen gas has an number density of $n_{out} = 2 \times 10^{-4}$cm$^{-3}$ and
temperature $T_{out} = 8000$K. The cloud is given a number density of $T_{cloud} = 40$K and number
density of $n_{cloud} = 200 n_{out}$. A constant flux of $F = 10^6$ photons / s / cm$^2$ following a
blackbody spectrum with temperature $T_{bb} = 10^{5}$K is injected from the $x = 0$ plane of the
box. The reduced speed of light used was $\tilde{c}_r = 0.125 c$ so that the incoming radiation can
reach the clump before the first required snapshot at 1 Myr.

In contrast to test 3, this test allows for the evolution of the hydrodynamics. As a consequence,
the initially dense cold clump is slowly evaporated due to the incoming radiation. As was done in
Test 3, I present results for \GEARRT using both the HLL and the GLF Riemann solvers.
Figure~\ref{fig:iliev7-slices-GLF} shows slices through the mid-plane of the box at 3, 10, and 25
Myr using the GLF solver, while Figure~\ref{fig:iliev7-slices-HLL} shows the same for the HLL
solver. A more quantitative comparison is shown in Figure~\ref{fig:iliev7-profiles}, where a line
cut along the axis of symmetry of the neutral hydrogen mass fraction, the ionized hydrogen mass
fraction, the hydrogen number density, the gas temperature, pressure, and the Mach number are shown
and compared to reference solutions at 1 Myr, 10 Myr, 25 Myr, and 50 Myr. Just as was found in Test
3, the moment based method which \GEARRT uses fails to produce sharp shadows and the region behind
the clump ($r/L \gtrsim 0.9$ at 1 Myr) heats up and ionizes, where reference solutions find the
background gas to be still in the initial state. Again the HLL solver is confirmed to perform better
at maintaining the direction of the radiation, and in providing less diffusive results in that
region. As for the hydrodynamics, the results of \GEARRT largely agree with the reference solutions.
\GEARRT finds somewhat broader peaks in the Mach number profiles at early times, but with smaller
maximal values, and can be attributed to both the more diffusive nature of the FVPM and the moment
based RT method used. The broadening on the gas velocities results in the front of the cloud
material being transported towards $r/L = 0$ a bit faster than for the reference solutions, as can
be seen in the profiles of later times. Aside from that, \GEARRT is shown to perform reasonably well
again.

\section{Sub-Cycling Performance}\label{chap:subcycling-results}

\begin{figure}
    \centering
    \includegraphics[width=\textwidth]{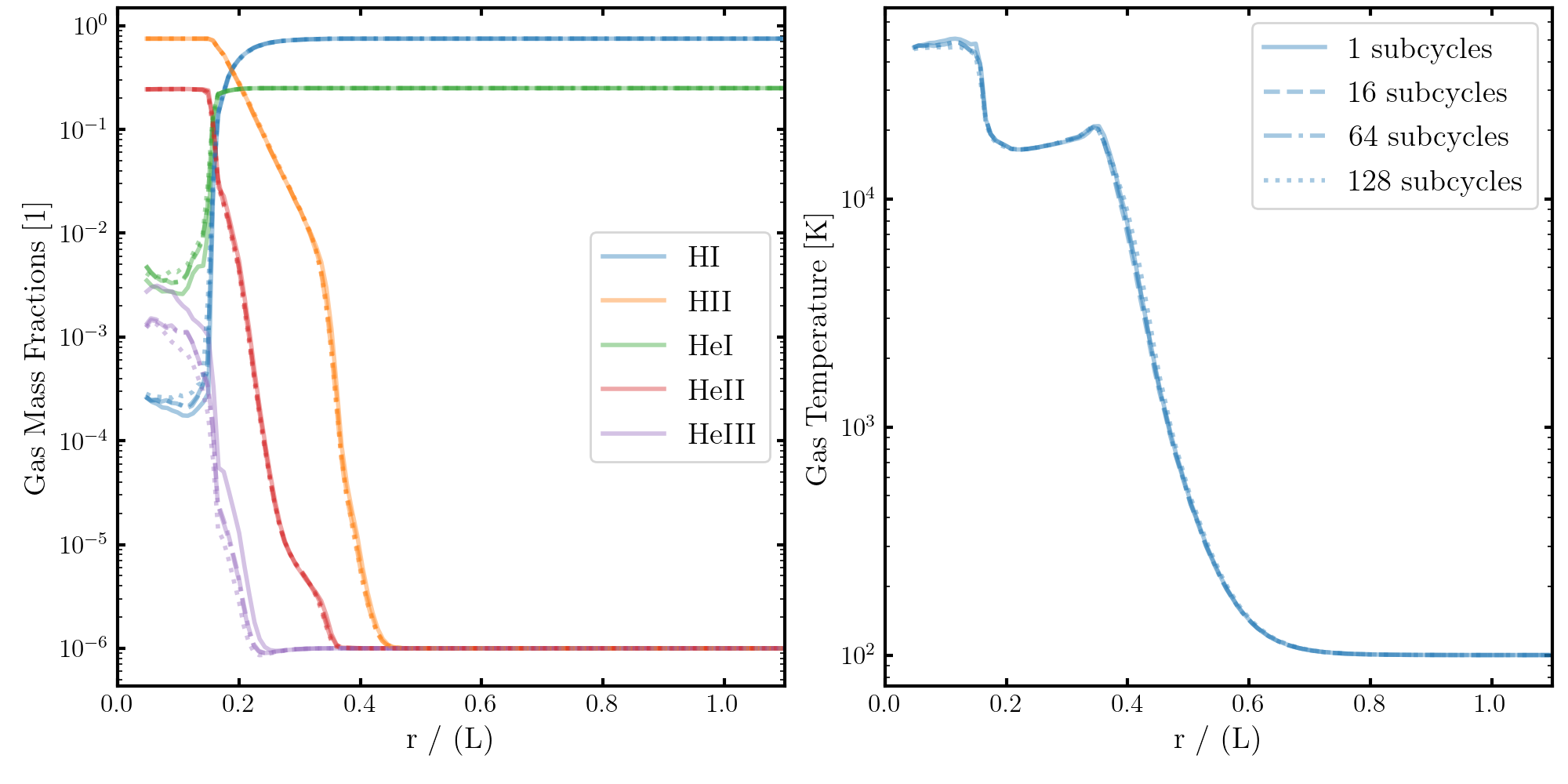}
    \caption{
The gas mass fractions and temperature of the Iliev5 test (Section~\ref{chap:Iliev5}) for a gas
composed of 75\% hydrogen and 25\% helium by mass at 10 Myr when different number of sub-cycles
are used.
    }
    \label{fig:subcycling-gas}
\end{figure}

To demonstrate the benefits of the sub-cycling scheme, the same setup as the Iliev 5 test
(Section~\ref{chap:Iliev5}) is used and evolved to 10 Myrs with varying numbers of sub-cycles.
However, instead of a gas consisting only of hydrogen, a gas composed of  75\% hydrogen and 25\%
helium by mass fractions is used. Figure~\ref{fig:subcycling-gas} shows the results of the gas mass
fractions and gas temperatures for different numbers of sub-cycles. The differences are small, and
mainly situated in the poorly resolved region close to the radiation source at $r = 0$, where the
influence of delayed particle drifts due to an increased number of sub-cycles is most influential.

\begin{figure}
    \centering
    \includegraphics[width=.6\textwidth]{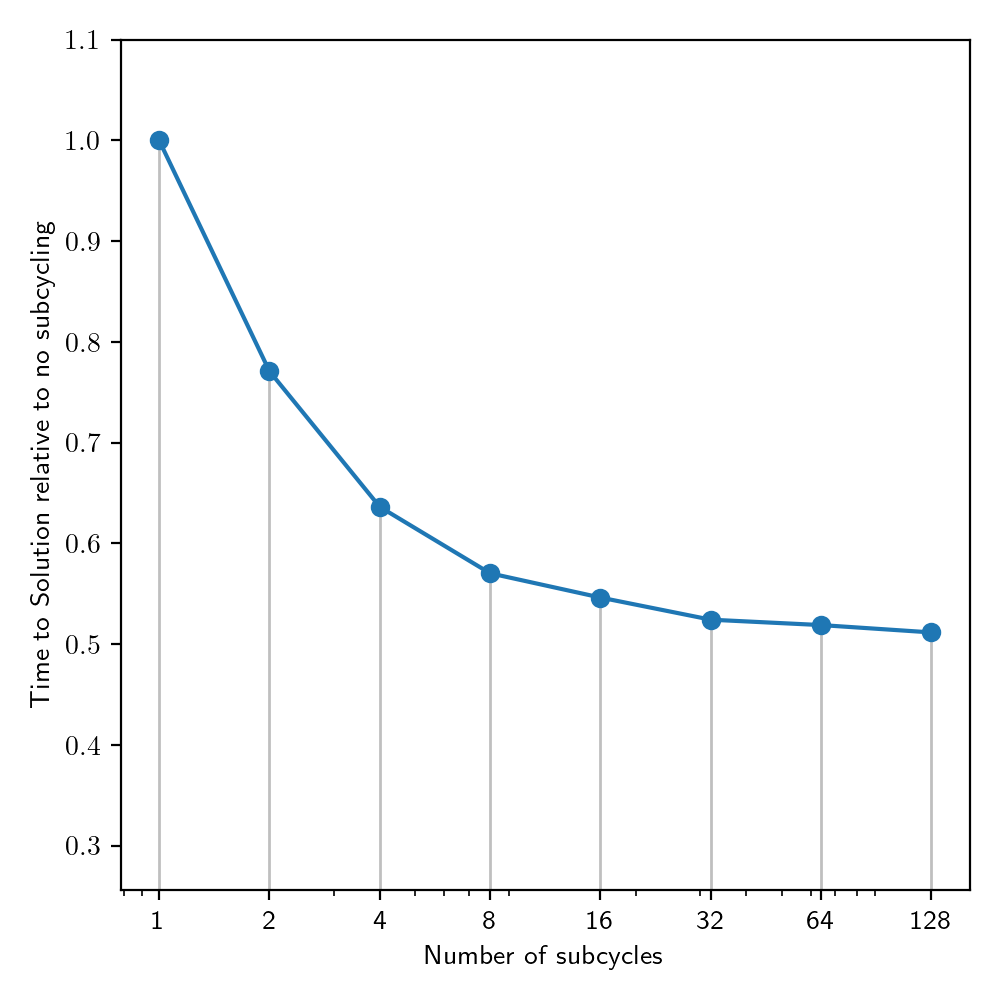}
    \caption{
The time-to-solution for varying numbers of sub-cycles compared to a simulation run without
sub-cycling.
The problem being solved is to evolve the Iliev5 test (Section~\ref{chap:Iliev5}) for a gas
composed of 75\% hydrogen and 25\% helium by mass to 10 Myr.
Note that this is not a scaling plot: The number of processors used is always the same.
    }
    \label{fig:subcycling-speedup}
\end{figure}

\begin{figure}
    \centering
\includegraphics[width=.6\textwidth]{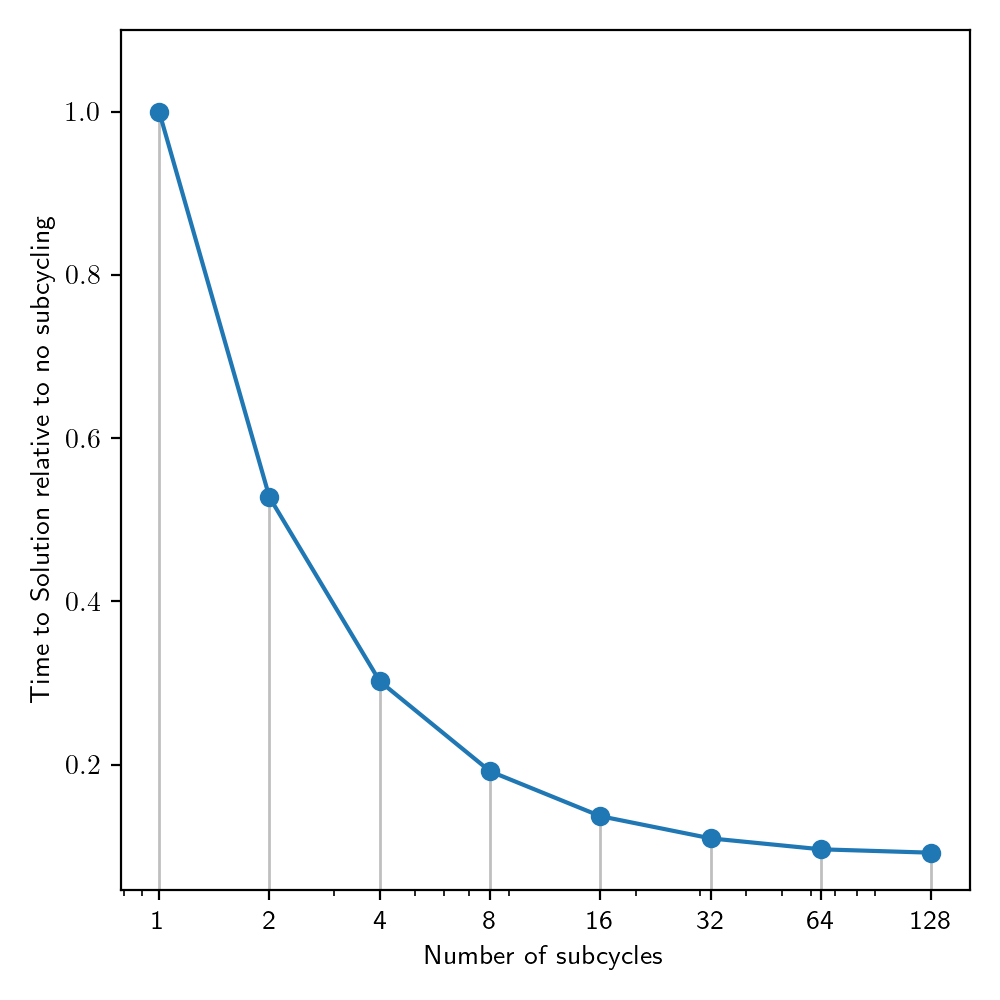}
    \caption{
Same as Figure~\ref{fig:subcycling-speedup}, but with self-gravity included. Note that this is not a
scaling plot: The number of processors used is always the same.
    }
    \label{fig:subcycling-speedup-gravity}
\end{figure}

Figure~\ref{fig:subcycling-speedup} shows the time-to-solution for an increasing number of
sub-cycles used compared to the case without sub-cycling. The initially steep decrease of the
required time to complete the simulation decreases and eventually nearly plateaus for 128 sub-cycles
at a time-to-solution reduction of about 50\% compared to a run without sub-cycling. The eventual
plateau is to be expected, as the sub-cycling's purpose is to allow us to omit work that may not be
strictly necessary. However, there will always be a minimal amount of work that needs to be done. In
particular, the total amount of RT time steps remains constant in all simulations regardless of how
many sub-cycles were used; Only the additional amount of hydrodynamics steps is decreased, which
can't be decreased endlessly. Eventually the minimal amount of required hydrodynamics updates will
be reached, and no further time gains will be possible by means of sub-cycling. However, more
relative improvements, i.e. relative reductions of the time-to-solution compared to a run without
sub-cycling, are expected when more physics are involved. For example,
Figure~\ref{fig:subcycling-speedup-gravity} shows the reduction in time-to-solution with increasing
numbers of sub-cycles for the same test as before, but with gravity included. Using 128 sub-cycles,
\GEARRT was able to reduce the time-to-solution by over 90\%.

%% file: main/RHD/RHD-5-conclusion.tex
\chapter{Conclusion and Outlook}

In this Part of my thesis, the implementation of \GEARRT, a novel radiation hydrodynamics solver for
astrophysical and cosmological applications into the task-based parallelized code \swift was
presented. The equations of radiation hydrodynamics, namely the moments of the equation of radiative
transfer together with the M1 Closure and the Euler equations for hydrodynamics, take the shape of
hyperbolic conservation laws. Hyperbolic conservation laws are a common problem in physics, and
solution strategies and associated difficulties and caveats were introduced previously in
Part~\ref{part:finite-volume} of this thesis. \GEARRT uses a particle based approach to discretize
the equations to be solved and uses a Finite Volume Particle Method, which was introduced and
discussed in Part~\ref{part:meshless} of this thesis, along with a description of the task-based
parallelism of \swift and a discussion of the implementation.

In Chapter~\ref{chap:rt-validation} a series of tests and validations of \GEARRT are presented. The
method is demonstrated to be second order accurate, and the minmod flux limiter
(eq.~\ref{eq:rt-minmod}) is determined to be the recommended choice, as for the other choices some
cases have been found where new minima and maxima develop in the radiation fields, making the method
not TVD, and therefore possibly unstable. Additionally, the drift correction method which \GEARRT
uses is shown to introduce only insignificant errors. The drift corrections are required because the
particles are being moved along (``drifted'') with the fluid in a Lagrangian manner, while the
radiative transfer is being solved by treating the particles as static interpolation points. Since
the particles move along with the fluid, the radiation field quantities are corrected during a drift
by extrapolating the local gradients to the new position. These corrections are found not to
introduce errors above the floating point representation precision limit, and are as deemed to be unconcerningly small, even
though the method is not strictly photon conserving any longer due to these corrections.

Chapter~\ref{chap:rt-validation} also shows \GEARRT's solution of the tests prescribed by the
Cosmological Radiative Transfer Comparison Project \citep{ilievCosmologicalRadiativeTransfer2006,
ilievCosmologicalRadiativeTransfer2009}, and \GEARRT is demonstrated to produce comparable results
to those of the codes which originally participated in the Comparison Project. Some caveats of the
moment-based approach which \GEARRT uses to solve radiative transport are apparent, like the
inability to form and maintain sharp shadows or the unphysical fluid-like collisions of the
radiation. They are however well-known caveats of the method, not an issue particular to \GEARRT,
and appear with other simulation codes which use the moment-based method and the M1 closure
too, like e.g. \citet{ramses-rt13} and \citet{kannanAREPORTRadiationHydrodynamics2019}. \GEARRT is
also shown to be slightly more diffusive compared to grid codes due to the larger number of
neighboring particles ($\sim 48$) used for interactions and flux exchanges, compared to 26
neighboring cells in mesh codes. While this is an unfortunate necessity, it should also not
constitute a problem with regards to the propagation of the ionization front. In real applications
particles to follow the flow of the fluid, and overdense regions should be resolved through a higher
particle number density as well, leading to smaller distances between neighboring particles, and
adequately slowing down the propagation of the I-front, as was demonstrated in
Section~\ref{chap:Iliev3}.

The task-based parallelism of \swift, while offering numerous benefits, constituted a major
challenge in the development process due to its complexity and intricacy, and unpredictable,
irreproducible nature of the execution of the program. In combination with the individual time step
sizes of particles, both for hydrodynamics and for radiative transfer independently, the
implementation of a dynamic sub-cycling feature, where a number of radiative transfer time steps is
being solved during a single hydrodynamics time step depending on local conditions for any particle
individually, was particularly challenging and required modifications deep inside \swift's core
functionalities. The sub-cycling is a completely novel feature in \swift, and both the task-based
algorithm to solve radiation hydrodynamics as well as the sub-cycling have been rigorously tested
for their correctness. As such, the development of \GEARRT provides \swift not only with a method to
solve radiation hydrodynamics, but also with the core algorithms to solve moment-based radiative
transfer and for dynamic sub-cycling which other methods, like \codename{SPHM1RT}
\citep{chanSmoothedParticleRadiation2021}, can take advantage of.

On the algorithmic side, two further features would improve the stability and flexibility of \GEARRT
(and other possible RT schemes in \swift) for future projects. Firstly, an option to ``wake up''
particles whose hydrodynamics time step size has decreased due to the increased specific internal
energy through the photo-heating is necessary to ensure that the hydrodynamics remain accurate
enough. This would require a global check after each sub-cycling simulation step for possible
changes in particles hydrodynamics time bins, and to abort the sequence of the RT sub-cycling
simulation steps, if necessary, in order to perform a global simulation step which includes
hydrodynamics updates as well. Secondly, a feature similar to the time step limiting required
between hydrodynamics time step sizes for all particles would be required between gas particles and
star particles as well. The time step limiting for gas particles is used to impose an upper
threshold of a factor of four between any two interacting gas particles' time step sizes. Similarly,
an upper limit between star and gas particles' time step sizes for strong sources of ionizing
radiation would ensure that not too large regions of space are ionized instantly after a single
injection step, since the injection of radiation into the gas is determined by the stars' time step
sizes. Findings in Section~\ref{chap:results-star-timesteps} suggest that an upper threshold of
eight can be permitted, but it might be more sensible to impose physically motivated upper thresholds as well.

Aside from these two useful features, the algorithmic side of the development of moment-based
radiative transfer for \swift is complete with the presented implementation of \GEARRT. This sets
the stage for future projects to focus on the improvement and extension of physical models and to
run simulations with scientific goals in mind. Specifically, the following improvements and
extensions are planned:

\begin{itemize}

\item Adding the explicit treatment of radiation pressure, i.e. transfer of photon momentum onto
the gas, following the approach of \citet{ramses-rt15} and
\citet{hopkinsNumericalProblemsCoupling2019}

\item Extending the thermochemistry module to treat more chemical species and to include metal line
radiative cooling. Aside from the current ``six species network'' containing H$^0$, H$^+$, He$^0$,
He$^+$, He$^{++}$, and $e^-$, the currently used \grackle \citep{smithGrackleChemistryCooling2017}
library offers the option to solve non-equilibrium thermochemistry for the ``nine species network'',
which additionally includes H$_2$, H$^-$, and H$_2^+$. The ``twelve species network'' furthermore
adds D, D$^+$, and HD. For the metal heating and cooling, \grackle is able to include the
corresponding rates from pre-computed tables. These options have not been explored yet in
coordination with the explicit treatment of radiative transfer. For an explicit treatment of
non-equilibrium thermochemistry including even more species and metals, the use of the \grackle
library needs to be replaced with some solver which is able to perform these vast networks of
thermochemistry equations. This can be in the form of a different already existing library like
\codename{Krome} \citep{grassiKROMEPackageEmbed2014}. Alternatively, \GEARRT may also follow the
approach of many other codes like e.g. \citet{katzRAMSESRTZNonEquilibriumMetal2022,
richingsNonequilibriumChemistryCooling2014, baczynskiFerventChemistrycoupledIonizing2015,
sarkarNewIonizationNetwork2021} and implement a non-equilibrium thermochemistry solver tailored
towards the specific needs at hand.

\item Move from a global reduced speed of light approximation to a local variable speed of light
approximation, similar in spirit to the work in \citet{katzInterpretingALMAObservations2017}. By
how much the speed of light may be reduced without impacting the propagation of ionization fronts
is limited by the condition that the propagation velocity of the ionization front must remain much
smaller than the reduced speed of light. The upper limit depends on the local gas conditions, in
particular the gas density (see discussion in \citet{ramses-rt13}). As such, simulations which
entail the underdense intergalactic medium will require a high value for the speed of
light, whereas overdense regions like the interstellar medium could be treated reasonably well with
a much smaller value for the speed of light. The approach would be to determine the local speed of
light of each particle individually depending on its local conditions such as density and smoothing
length.

\item The addition of treatment of Doppler effect for the interactions between
radiation and the moving gas. This is currently neglected.

\item Treatment of cosmological redshifting of the radiation energy density after each time step
for each particle. This is currently neglected.

\item Use more sophisticated stellar luminosity spectra, which take into account mass, age, and
metallicities of stars and stellar populations
\citep[e.g.][]{bruzualStellarPopulationSynthesis2003, leithererStarburst99SynthesisModels1999,
stanwayReevaluatingOldStellar2018}. Currently only a blackbody and a constant spectrum are
supported.

\item Permit for other sources of radiation aside from stars and stellar populations, like active
galactic nuclei \citep[e.g.][]{costaDrivingGasShells2018, barnesRadiativeAGNFeedback2020}.

\item Use an advanced or modified closure for the moments of the equation of radiative transfer
\citep[e.g.][]{chanSmoothedParticleRadiation2021}

\item Explicitly account for and trace recombination radiation. This is currently neglected.

\end{itemize}

Lastly, it should be noted that in principle \GEARRT can be coupled to SPH hydrodynamics as well.
The quantities required for the computation of the effective surfaces \Aij and the gradients, in
particular the matrix $\mathcal{B}$ (eq.~\ref{eq:matrix_B}), can be added to the second SPH neighbor
interaction loop (``\lingo{force}'' loop) once the smoothing lengths have been determined in the
first SPH neighbor interaction loop (``\lingo{density}'' loop), thus making it available for the
subsequent radiative transfer operations. This would not only enable the use of an entirely
different class of hydrodynamics methods to use, but also permit to couple \GEARRT to a variety of
sophisticated sub-grid models which have been developed for use with SPH, such as EAGLE
\citep{schayeEAGLEProjectSimulating2015} and GEAR \citep{revazDynamicalChemicalEvolution2012}.

The long term goals for \GEARRT are to study dwarf galaxies during the Epoch of Reionization. It is
still debated whether the numerous but fainter dwarf galaxies are the main drivers of cosmic
reionization in the early universe, or whether massive and brighter but less common galaxies are the
main actors responsible. We intend to run simulations of both isolated dwarf galaxies using a
zoom-in technique as well as cosmological volumes in search to determine and constrain the role of
dwarf galaxies with regards to cosmic reionization. A first milestone for future work will be to
repeat the simulations of \citet{revazPushingBackLimits2018} using \GEARRT to additionally account
for the explicit treatment of radiative transfer for the UV background, for the self-shielding of
the gas, and for stellar emission of radiation, which thus far have been treated as sub-grid models only, or in the case of photo-ionizing radiation from stellar sources, have been neglected. It is our intention to run these zoom-in simulations of dwarf galaxies up to redshift zero, and to verify that the GEAR model of galaxy formation and evolution is able to produce realistic dwarf galaxies which agree well with observations even with the explicit treatment of radiative feedback.

%% file: main/ACACIA/ACA0-introduction.tex
\chapter{Introduction}

Mock galaxy catalogues generated using N-body or hydrodynamical
simulations are important tools for extragalactic astronomy and
cosmology.  They are used to test current theories of galaxy
formation, to explore systematic and statistical errors in large scale
galaxy surveys and to prepare analysis codes for future dark energy
missions such as Euclid or LSST.  There is a large variety of methods
to generate such mock galaxy catalogues.  The most ambitious line of
products is based on full hydrodynamical simulations, where dark
matter, gas, and star formation are directly simulated
\citep[e.g.][]{duboisDancingDarkGalactic2014,
  khandaiMassiveBlackIISimulationEvolution2015,
  vogelsbergerPropertiesGalaxiesReproduced2014,
  schayeEAGLEProjectSimulating2015}.
The intermediate approach is based on semi-analytic modelling
(hereafter SAM) \citep[e.g.][]{whiteGalaxyFormationHierarchical1991,
  bowerBreakingHierarchyGalaxy2006, somervilleSemianalyticModellingGalaxy1999,
    kauffmannFormationEvolutionGalaxies1993,
  kangSemianalyticalModelGalaxy2005,crotonManyLivesActive2006} for
which galaxy formation physics, although simplified, is still at the
origin of the mock galaxy properties. Finally, the simplest and most
flexible approach is based on a purely empirical modelling of galaxy
properties, sometimes called Halo Occupation Density (HOD hereafter)
\citep[e.g.][]{seljakAnalyticModelGalaxy2000, berlindHaloOccupationDistribution2002,
  peacockHaloOccupationNumbers2000,
  bensonNatureGalaxyBias2000,wechslerGalaxyFormationConstraints2001,
  scoccimarroHowManyGalaxies2001}.
The last two techniques (SAM and HOD) both require the complete
formation history of dark matter
haloes, and possibly their sub-haloes. This formation history is
described by halo `\emph{merger trees}'
\citep{roukemaSpectralEvolutionMerging1993,
  roukemaFailureSimpleMerging1993, laceyMergerRatesHierarchical1993}.
Accurate merger trees are essential to obtain realistic mock galaxy
catalogues, and constitute the backbone of SAM and HOD models.

The advantage of using SAM and HOD techniques to generate mock galaxy
catalogues is that one does not need to model explicitly the gas
component, but only the dark matter component.  The corresponding
N-body simulations are commonly referred to as `\emph{dark matter
only}' (DMO) simulations.  With growing processing power, improved
algorithms and the use of parallel computing tools and architectures,
larger and better resolved DMO simulations are becoming possible.  The
current state-of-the-art is the Flagship simulation performed for the
preparation of the Euclid mission \citep{potterPKDGRAV3} and featured 2
trillion dark matter particles.  Such extreme simulations make
post-processing analysis tool such as merger tree algorithms
increasingly difficult to develop and to use, mostly because of the
sheer size of the data to store on disk and to load up later from the
same disk back into the processing unit memory.  In some extreme
cases, the amount of data that needs to be stored to perform a merger
tree analysis in post-processing is simply too large.  Storing just
particle positions and velocities in single precision for trillions of
particles requires dozens of terabytes per snapshot.  Another issue is
that most modern astrophysical simulations are executed on large
supercomputers which offer large distributed memory.  Post-processing
the data they produce may also require just as much memory, so that
the analysis will also have to be executed on the distributed memory
infrastructures as well.  The reading and writing of such vast amount
of data to a permanent storage remains a considerable bottleneck,
particularly so if the data needs to be read and written multiple
times.  One way to reduce the computational cost is to include
analysis tools like halo-finding and the generation of merger trees in
the simulations and run them ``\textit{on the fly}'', i.e. run them
during the simulation, while the necessary data is already in memory.

The main motivation for this work is precisely the necessity for such
a merger tree tool for future ``beyond trillion particle''
simulations. While many state-of-the-art N-body simulation codes include
structure finders that are run on-the-fly, codes like \codename{Gadget4}
\cite{springelSimulatingCosmicStructure2021} who are able to build
merger trees on-the-fly are still a relative rarity. It is crucial that
multiple, distinct, codes have the capacity to do this to provide the
possibility to cross-check results and their convergence.
To this end, a new algorithm that we named
\acacia was designed to work on the fly within the parallel
AMR code \ramses.  One novel aspect of this work is the use of the
halo finder \phew\ \citep{bleulerPHEWParallelSegmentation2015} for the parent halo catalogue.
Different halo finders have been shown to have a strong impact on the
quality of the resulting merger trees \citep{SUSSING_HALOFINDER}.
\phew\ falls into the category of ``watershed'' algorithms that are
not so common in the cosmological halo finding literature.  This type
of algorithm assigns particles (or grid cells) to density peaks above
a prescribed density threshold and according to the so-called
``watershed segmentation'' of the negative density field.

This Part of the work is a slightly modified version of the corresponding published article
\citep{ivkovicACACIANewMethod2022} to fit the format of a thesis better. While the bulk of the
algorithm development and implementation precedes the starting point of my thesis, a substantial
effort during my thesis was directed towards finishing the project, fixing a handful of bugs,
extending the performance analysis of the algorithm, and validating the quality of the resulting
merger trees by generating mock galaxy catalogues and comparing them to observational data.

This Part is structured as follows. In Chapter \ref{chap:phew}, a brief description of the \phew\
halo finder and its new particle unbinding method is given. Chapter \ref{chap:making_trees}
describes some common difficulties that arise when making merger trees and the way we address them
in our merger algorithm \acacia, which is ultimately described in detail in Chapter
\ref{chap:my_code}. Chapter \ref{chap:tests} shows test results to determine what parameters give
the best results.  Using the halo catalogue and its corresponding merger tree generated on the fly
by a cosmological N-body simulation, we use the stellar-mass-to-halo-mass (SMHM) relation from
\cite{behrooziAVERAGESTARFORMATION2013} to produce a mock galaxy catalogue. We analyse in
Section~\ref{chap:mock_catalogues} the properties of our mock galaxy catalogue and show that the
introduction of orphan galaxies improve the comparison to observations considerably.

The \ramses\ code is  publicly available and  can be  downloaded from
\url{https://bitbucket.org/rteyssie/ramses/}. Instructions  on how to use \acacia\ and \phew\ during
a simulation can be found under \url{https://bitbucket.org/rteyssie/ramses/wiki/Content}.

%% file: main/ACACIA/ACA1-from-halofinding-to-mocks.tex
\chapter{Halo Finding and Particle Unbinding}\label{chap:phew}

Halo finding plays a central role in the exploitation of N-body
simulations. Paradoxically, a unique definition of what is a halo or a
sub-halo has never been adopted so far.  The current state of affairs
in the halo finding business is quite the opposite, with a multitude
of definitions emerging over the last decades, each definition
corresponding to a different halo finding algorithm.  The Halo Finder
Comparison Project \citep{knebeHaloesGoneMAD2011} lists 29 different codes and roughly
divides them into two distinct groups:
\begin{enumerate}
\item Percolation algorithms, for which particles are linked together
  if closer to each other than some specified linking length. The
  typical example is the algorithm ``\emph{friends-of-friends}''
  (thereafter FOF) \citep{davisEvolutionLargescaleStructure1985}.
\item Segmentation algorithms, for which space is segmented into
  separate regions around local peaks of the density field. Particles
  within these regions are then collected and assigned to the same
  halo or sub-halo. The typical example is the ``\emph{Spherical
  Overdensity}'' method (thereafter SOD) \citep{pressFormationGalaxiesClusters1974}.
\end{enumerate}
The outer boundary of the haloes are defined in both method by a
density iso-surface, whose exact value determines the properties of
the resulting halo statistics.  Halo catalogues derived from FOF and
SOD and their corresponding merger trees have been studied quite
extensively in the literature \citep[see e.g.][]{SUSSING_HALOFINDER}.

\section{The PHEW halo finder}

In this work, we extend these earlier studies to the \phew\ halo
finder \citep{bleulerPHEWParallelSegmentation2015} developed specifically
for the \ramses\ code \citep{teyssierCosmologicalHydrodynamicsAdaptive2002}.
The \phew\ algorithm belongs to the category of
segmentation methods.  Particle masses are first deposited to the AMR
grid using the ``\emph{cloud-in-cell}'' technique. All density maxima
are then marked as potential sites for a \emph{clump}. Clumps are what
we call any structure, haloes and sub-haloes, in contexts where we
don't need to differentiate between them. The volume is then segmented
into peak patches by assigning each cell of the grid to the closest
density maximum in the direction of the steepest density gradient.

This segmentation method provides well defined regions separated by
density saddle surfaces. The minimum density in the saddle surface
between two adjacent peaks marks the saddle point between the two
peaks.  This well known method is often called `\emph{watershed
segmentation}`. In order to define proper halo boundaries, \phew\ uses
an outer density isosurface, like most methods described
above. Subhaloes, on the other hand, are just the ensemble of all
peak patches within the halo boundaries. This allows to identify
haloes and sub-haloes without the assumption of spherical symmetry,
unlike other popular methods such as SOD.

Subhaloes can be organised into a hierarchy of sub-structures based
on the same steepest gradient technique, for which individual clumps
can be assigned to the closest densest peak.  After this first pass,
only a few sub-haloes survived the merging process, which is then
repeated a second time, assigning these surviving sub-haloes to their
densest neighbours.  Ultimately, all sub-haloes will be collected into
a single peak that corresponds to the main halo. Each pass defines a
level in the hierarchy of sub-haloes. More details can be found in the
original \phew\ paper \citep{bleulerPHEWParallelSegmentation2015}.
As a consequence, a halo can have
a number of sub-haloes, each one of them containing subsub-haloes and
so on.  This well defined hierarchy is a very important feature for us
to uniquely assign particles to haloes and sub-haloes based on a
binding energy criterion.

There are four parameters that \phew\ requires a user to choose in
order to identify clumps and haloes.  Firstly, a ``relevance
threshold'' needs to be defined.  If for any given peak patch the
ratio of the peak's density to the maximal density of the entire
saddle surface of the respective peak patch is smaller than the chosen
relevance threshold, then the peak patch is considered to be noise,
not a genuine structure.  The peak patch is then merged into a
neighbour.  Secondly, a density threshold determines the minimal
density a cell needs to have to be part of any peak patch.  Thirdly, a
``saddle threshold'' defines the maximal density for a saddle surface
between two peak patches for the two patches to be considered parts of
two different haloes.  If the saddle surface density is above the
threshold, then the peak patches will be parts of the same halo (but
different sub-haloes within the host halo).  Finally, a mass threshold
determines the minimal mass a peak patch needs to have to be kept.  We
list the parameters that we used throughout this paper in Table~\ref{tab:phew-parameters}.

\begin{table}
\centering
\caption{Parameters used for the \phew\ clump finder throughout this
  paper.  The numerical values given can be directly used in the
  namelist file that \ramses\ uses to read in runtime parameters.
  Here $\bar \rho$ here is the mean background density and $m_p$ is
  the particle mass.}
\label{tab:phew-parameters}
\begin{tabular}[c]{l l l}
  parameter				&	value		& units \\
  \hline
  relevance threshold			&	3		& 1		\\
  density threshold			& 80			& $\bar \rho$ 	\\
  saddle threshold			& 200			& $\bar \rho$	\\
  mass threshold			& 10			& $m_p$		\\
  \hline
\end{tabular}
\end{table}



\section{Particle unbinding}

We now describe how we assign each dark matter particle to a given
sub-halo, a process that has not been implemented so far in the
\phew\ code.  For this, we follow a physically motivated criterion,
quite common in the halo finding literature, based on the binding
energy of the particle \citep[e.g.][]{knollmannAHFAmigaHalo2009,
springelPopulatingClusterGalaxies2001, stadelCosmologicalNbodySimulations2001}.  If a
particle is not bound to the first sub-halo of the hierarchy, it is
then passed recursively to the next sub-halo in the hierarchy, where
the binding energy is checked again and so on.  If the particle is not
bound to any sub-halo, it is assigned to the main halo.

In the previous hierarchical unbinding process, the key component is
the criterion adopted for deciding whether a particle is bound to a
sub-halo or not.  Traditionally, this is done using the \emph{static}
gravitational potential, since we are dealing with a single time step
and we have to assume that the N-body system is stationary.  In this
case, a particle at position $\mathbf{r}$ is considered as unbound if its
velocity $\mathbf{v}$ exceeds the escape velocity given by

\begin{equation}
v_{\mathrm{esc}} = \sqrt{- 2\Phi(\mathbf{r})}
\label{eq:boundv}
\end{equation}

\begin{figure}
  \centering
  \fbox{\includegraphics[width=.95\linewidth, keepaspectratio]
	{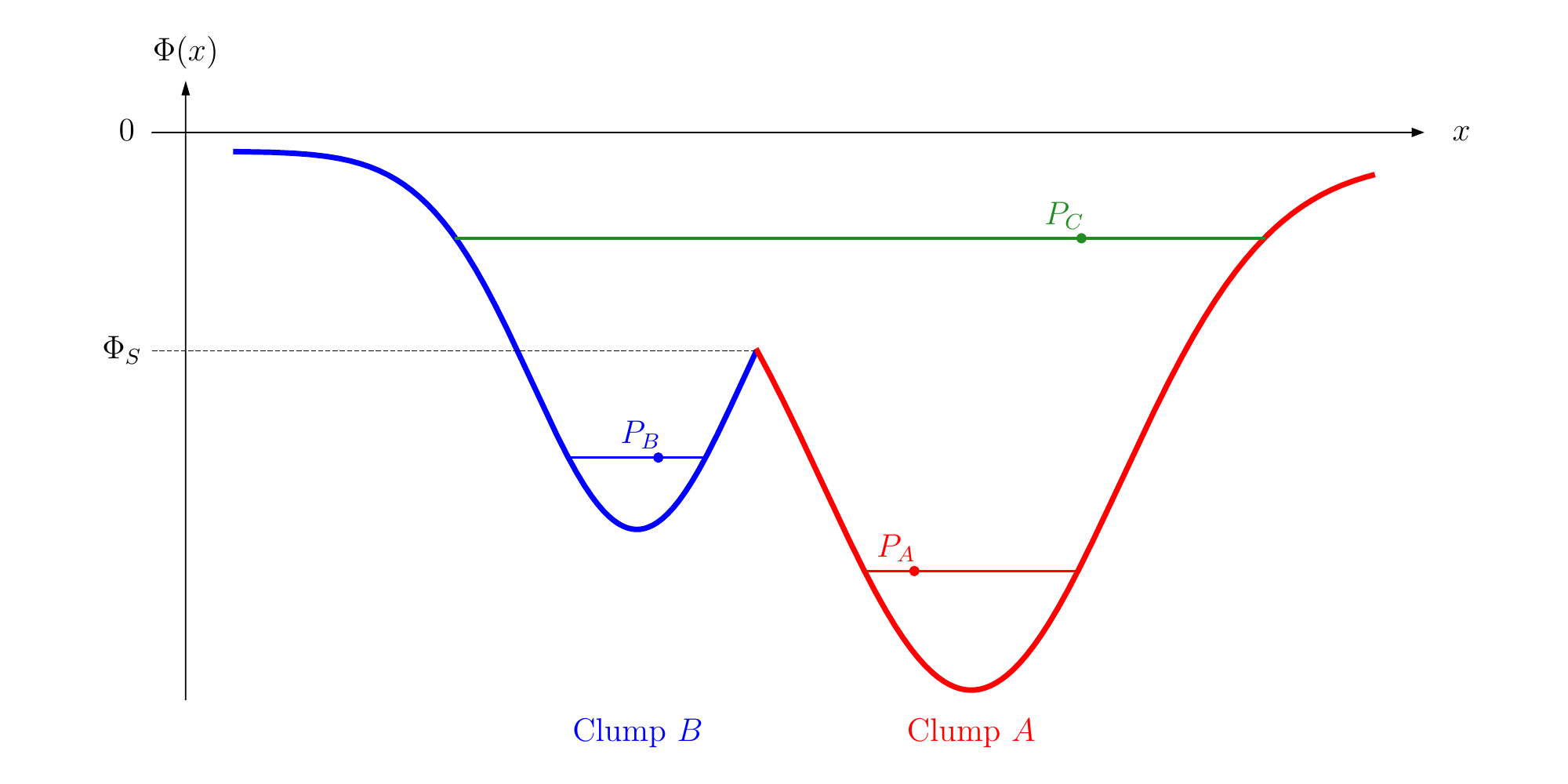}}%
  \caption{Simple sketch of the gravitational potential of a halo that
    consists of two clumps, $A$ and $B$.  The position of the
    horizontal lines marks the energies of three example particles,
    $P_A$, $P_B$, and $P_C$, while the length of the lines shows the
    spatial extent of the orbits.  We call particles like $P_A$ and
    $P_B$ ``strictly bound'', as their predicted orbit boundaries
    don't allow them to escape from the clump they are assigned to.
    Particle $P_C$ however, although energetically bound to clump $A$,
    can wander off deep into clump $B$, and for that reason is called
    ``loosely bound''. To discriminate between these two types of
    particles, we use the potential of the saddle point between clump
    $A$ and $B$ marked as $\Phi_S$.
  }%
  \label{fig:potentials}
\end{figure}

More precisely, this means that the particle will be able to travel to
infinity where the potential goes to  zero. If the velocity is smaller
than the escape  velocity, the particle will follow a  bound orbit and
come back to its current location.  Note that this orbit can leave the
boundaries of the  sub-halo.  The particle will stay for  some time in
the sub-halo,  but can visit at  a later time a  neighbouring sub-halo
and then come back along the  same bound orbit.  This kind of particle
does not \emph{exclusively} belong  to its original sub-halo. It should
be in fact assigned to the parent sub-halo in the hierarchy.  In order
to identify particles as more  strictly bound, we re-define the escape
velocity using the potential of the closest saddle point $\Phi_S$.
\begin{equation}
v_{\mathrm{esc}} = \sqrt{- 2(\Phi(\mathbf{r})-\Phi_S)}
\label{eq:boundv_corr}
\end{equation}
This new  escape velocity is  smaller than the previous  one, allowing
more particle to  exceed it and leak out of  the current sub-halo into
neighbouring ones.  In what  follows, we will  use these two unbinding
criteria, calling the first method the ``loosely bound'' criterion and
calling   the   second   one   the   ``strictly   bound''   criterion.
Figure~\ref{fig:potentials} illustrates  the difference  between these
two  criteria.   The  gravitational  potential  of   two  neighbouring
sub-haloes labelled A  and B is represented.  We show  an example of a
strictly bound particle in each sub-halo,  and an example of a loosely
bound particle that can wander from one sub-halo to the other one.  If
one uses the first binding criterion, this loosely bound particle
will be  assigned to sub-halo A,  because this is where  it is located at
the present time.  If one uses the second, stricter binding criterion,
then the loosely bound particle will be assigned to the parent halo, but
not to sub-halo A nor B.



%
%

When computing the velocity of the particle, it is important to use
the velocity \emph{relative} to the velocity of the sub-halo centre of
mass (also called the bulk velocity of the sub-halo).  Because of this
requirement, the unbinding process has to be performed iteratively,
since removing a particle that is unbound requires to recompute the
centre of mass velocity.  Let us finally repeat that the particle
unbinding is performed recursively, following the sub-halo hierarchy
from the bottom up.  Starting with the lowest (finest) level of
sub-haloes, unbound particles are assigned to the higher (coarser)
level of parent sub-halo for unbinding, and so on following the
structure hierarchy.  Particles that are unbound from all sub-haloes
are collected into the main halo, marking the end of the hierarchical
unbinding process.

%% file: main/ACACIA/ACA2-mergertree_algorithm.tex
\chapter{Merger Trees: Basic Principles}\label{chap:making_trees}

In this work, we adopt the terminology set by the ``Sussing Merger
Tree Comparison Project'' \citep{SUSSING_COMPARISON,
  SUSSING_CONVERGENCE, SUSSING_HALOFINDER,leeSussingMergerTrees2014}.
For sake of clarity, we repeat here some important definitions:

\begin{itemize}

\item For two snapshots at different times, a halo from the first one
  (i.e. higher redshift) is always referred to using the capital
  letter $A$ and a halo from the second one (i.e. lower redshift)
  using $B$.

\item Recursively, $A$ itself and progenitors of $A$ are all
  \emph{progenitors} of $B$.  When it is necessary to distinguish $A$
  from earlier progenitors, the term \emph{direct progenitor} will be
  used.

\item Recursively, $B$ itself and descendants of $B$ are all
  \emph{descendants} of $A$.  When it is necessary to distinguish $B$
  from later descendants, the term \emph{direct descendant} will be
  used.

\item In this work, we restrict ourselves to merger trees for which
  there is \emph{precisely one direct descendant for every halo}.

\item When there are multiple direct progenitors, it is required that
  one of these is identified as the \emph{main progenitor}.

\item The \emph{main branch} of a halo is a complete list of main
  progenitors tracing back along its cosmic history.

\end{itemize}
We finally define an important convention we use here: when no
distinction between sub-haloes and main haloes is necessary, they are
collectively referred to as \emph{clumps}.

\section{Linking Clumps Across Snapshots}

\begin{figure}
	\centering
  \includegraphics[height=.8\textheight]{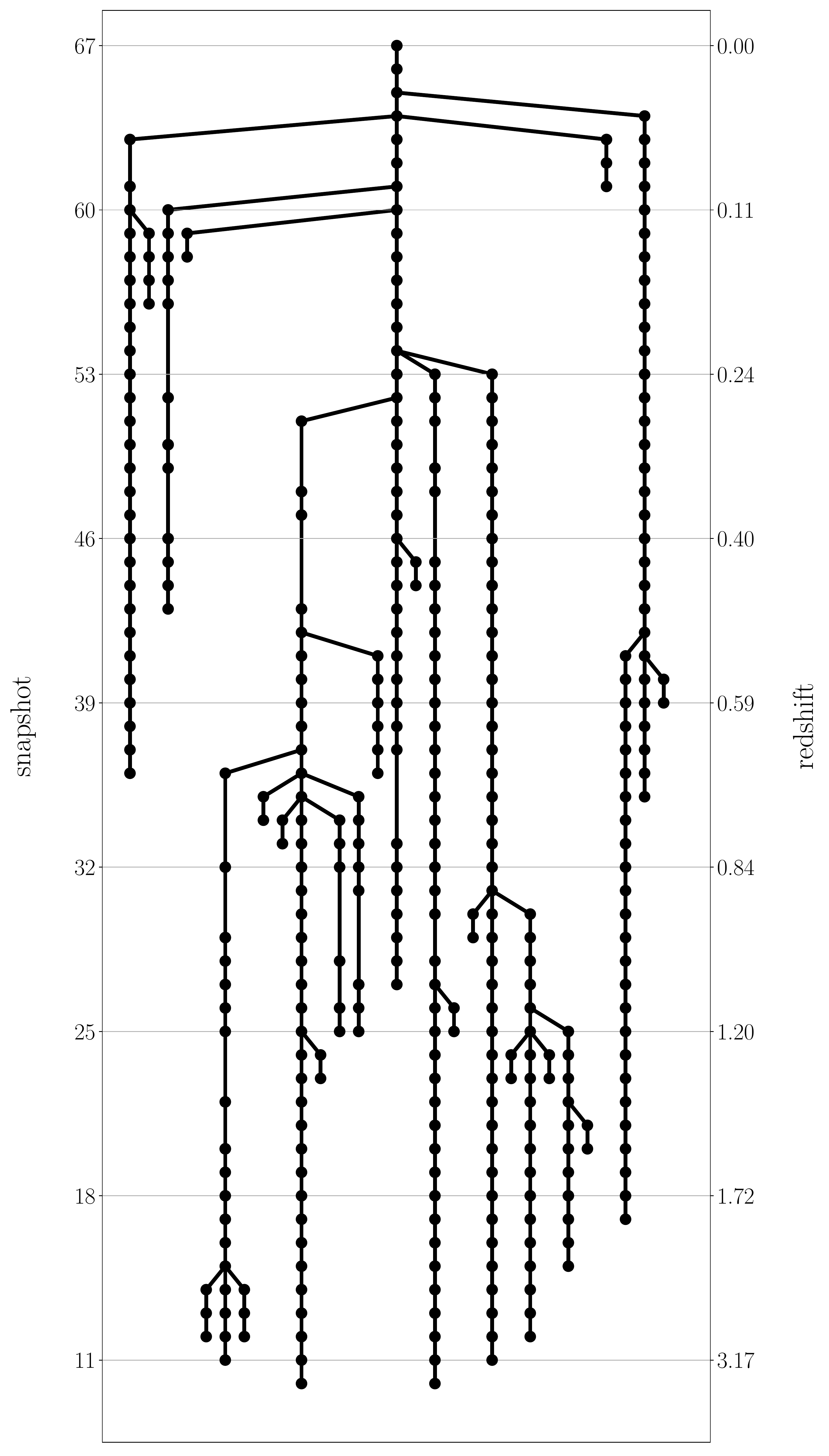}%
  \caption{The merger tree of a main halo at redshift zero as found by
    \texttt{ACACIA}.  This tree was extracted from a low resolution
    cosmological DMO simulation containing $64^3$ particles for
    illustrative purposes. Using higher resolutions quickly leads to
    hundreds and thousands of branches, resulting in a rather messy
    plot. On the $y$ axis, the snapshot numbers and their
    corresponding redshifts are given.  The $x$ axis has no physical
    meaning.  Each dot represents a clump identified at the given
    snapshot.  Two dots connected by a vertical line represent clumps
    that have been linked as main progenitor and main descendant.
    Diagonal lines depict merging events. In this tree, a few links
    between two dots are longer than others.  These are cases where
    clumps merge temporarily, but then re-emerge later as separate
    clumps (see the example shown in Figure~\ref{fig:jumper-demo}).
    We discuss these cases and how they are dealt with in
    Section~\ref{sect:jumpers}. }
  \label{fig:mergertree}
\end{figure}

The aim of a merger tree code is to link haloes from an earlier
snapshot to haloes in the consecutive snapshot, i.e. to find all the
descendants of the haloes in the earlier snapshot. If we do this
successfully each snapshot, then we can follow the formation history of
haloes throughout the simulation. In particular, this will enable us
to track the mass growth of haloes as well as the merging of different
sub-haloes during the course of the simulation.
To illustrate the idea, a merger tree of a main halo generated by
\acacia during a simulation is shown in Figure~\ref{fig:mergertree}.
In this particular case, we were able to track the formation history of the
main halo down to redshifts $z > 3$. By linking progenitors and descendants
throughout the simulation many branches of the tree are revealed.
Each branch represents a clump that eventually merged into the main halo
which we chose as the root of the tree at redshift zero.

Merger events occur when a clump, identified as such in a previous
snapshot, disappears from the list of clumps in the next snapshot. In
the case of a halo, it usually first becomes the sub-halo of another
halo and can be followed as such in many subsequent snapshots. After
some time, this sub-halo can merge into another sub-halo or dissolve
completely due to numerical over-merging. In both cases, the sub-halo
disappears completely from the clump catalogue.

A straightforward method to link progenitors with descendants in two
consecutive snapshots is to trace individual particles using their
unique particle ID. All merger tree codes use this simple technique
\citep[][]{behrooziGravitationallyConsistentHalo2013,
  springelSimulationsFormationEvolution2005a, jiangNbodyDarkMatter2014,
  knebeImpactBaryonicPhysics2010, tweedBuildingMergerTrees2009,
  elahiClimbingHaloMerger2019, jungEffectsLargescaleEnvironment2014,
  rodriguez-gomezMergerRateGalaxies2015} with the notable exception of
the code \codename{Jmerge} described and tested in
\cite{SUSSING_COMPARISON}.

Linking a progenitor to a descendant means checking how many particles
of the progenitor halo or sub-halo end up in the descendant halo or
sub-halo.  Naturally, these tracer particles may end up in multiple
clumps, giving multiple descendant candidates for a progenitor.  In
such cases, the most promising descendant candidate will be called the
\emph{main descendant}.  To find a main progenitor and a main
descendant, a merit function $\mathcal{M}$ has to be defined, which is
to be maximised or minimised, depending on its definition.  An
overview of the merit functions that are used in other merger tree
algorithms is given in Table~1 of \cite{SUSSING_COMPARISON}.  The
merit function used in our implementation is given in
Equation~\ref{eq:merit}.

Sometimes, unfortunately, linking progenitors to descendants is not as
straightforward as described so far. We now discuss two circumstances
where special care must be taken to define robust links between
different snapshots: fragmentation events and temporary merger events.


\begin{figure}
  \centering
  \includegraphics[width=.7\linewidth]{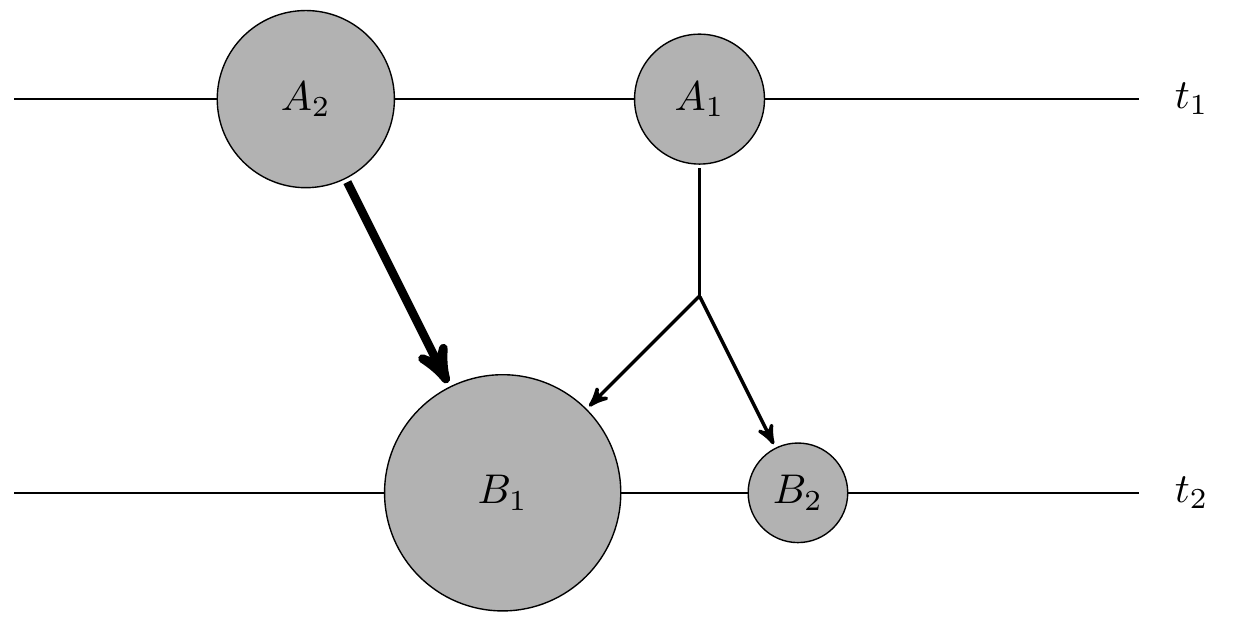}
  \caption{Illustration of a fragmentation event, where a
  	progenitor $A_1$ at time $t_1$ is partially merged together
  	with a second progenitor $A_2$ into a descendant $B_1$ at
  	time $t_2 > t_1$, while some fragmented part of $A_1$, $B_2$,
  	evaded the merging.
	}
  \label{fig:fracture}
\end{figure}

\section{Fragmentation Events}
\label{sect:frag}

In our current approach, each progenitor can have only one
descendant\footnote{Note that \cite{springelSimulatingCosmicStructure2021}
proposed another approach that allows explicitly fragmentation
events to be included in the formation history analysis, where they use
merger \emph{graphs} rather than merger trees.}. We therefore need
to pick only one descendant within a possibly large ranked list of
descendant candidates, that all contain particles coming from the
progenitor.

Normally, this choice is performed according to the ranking provided
by the merit function, where the main descendant is ranked number 1.
Problems arise for example when the progenitor $A_1$ is not the main
progenitor of its main descendant $B_1$, but also has fragmented into
another viable descendant candidate $B_2$.  This situation is
schematically shown in Figure~\ref{fig:fracture}.

Relying only on the merit function \eqref{eq:merit}, progenitor $A_1$
will seem to have merged with $A_2$, the direct progenitor of $B_1$,
in order to form $B_1$.  The other fragment, $B_2$, will be treated as
a newly formed clump and the entire formation history of $B_2$ would
be lost. In order to preserve this history, we choose to prioritize the
link from $A_1$ to $B_2$ over of merging progenitor $A_1$ into $B_1$.

It is simpler to deal with this case directly in the algorithm
than via the merit function. The resulting logic can be summarized as
follows: If $A_1$ is not the main progenitor of its main descendant
$B_2$, then we don't merge it into $B_2$ but we link it instead with
the first secondary descendants that considers $A_1$ as its
main progenitor.


\section{Temporary Merger Events}
\label{sect:jumpers}

\input{figures/ACACIA/jumper-demo/jumper-demo-images.tex}

When a sub-halo travels towards the core of its parent halo, it will
merge with the central clump and disappear from the sub-halo lists.  It
can however re-emerge at a later snapshot and will be added back
to the list as a newly born halo. Such a scenario is shown in
Figure~\ref{fig:jumper-demo}.  Indeed, when this occurs, the merger
tree code will deem the sub-halo to have merged into the main halo, and
will likely find no progenitor for the re-emerged sub-halo, thus
treating it as newly formed.

This is a problematic case because we lose track of the growth history
of the sub-halo, regardless of its size, and massive clumps may be
found to just appear out of nowhere in the simulation.  This is a well
known problem for configuration-space halo finders
\citep{onionsSubhaloesGoingNotts2012}, and phase-space halo finders
like \codename{Rockstar} \citep{behrooziRockstarPhaseSpaceTemporal2013}
have been developed precisely to alleviate this issue.  While they
typically perform better than configuration-space halo finders,
\cite{SUSSING_COMPARISON} found that phase-space halo finders aren't
infallible in recovering all such missing haloes, and strongly
recommend checking for links between progenitors and descendants in
non-consecutive snapshots as well.

As our merger tree code works on the fly, future snapshots will not be
available at the time of the merger tree analysis, so it will be
necessary to check for progenitors of a descendant across multiple
snapshots.  This can be achieved by keeping track of the most bound
particles of each clump when it is merged into some other clump.
These tracer particles are also used to track \emph{orphan galaxies}.
\footnote{In the context of SAM, orphan galaxies are galaxies born at
the center of dark matter haloes that merged later into bigger haloes
and eventually dissolved due to over-merging. As a consequence, these
galaxies don't have a parent halo or sub-halo anymore.}  For this
reason, we call these tracer particles ``orphan particles'', and
progenitor-descendant links over non-adjacent snapshots ``jumpers''.

These jumper links between different haloes widely separated in time
are less reliable than proper links between progenitors and
descendants from adjacent snapshots.  As we will discuss
in Section~\ref{chap:testing-nmb}, the quality of the merger trees
increases with the number of tracer particles used.  Using jumpers
corresponds to using only one tracer particle over a large time interval,
much larger than the one between two adjacent snapshots.

For this reason, priority is given to direct progenitor candidates in
adjacent snapshots.  Only if no direct progenitor candidates have been
found for some descendant, then progenitor candidates from
non-adjacent snapshots are searched for.  Because these progenitors
from non-adjacent snapshots are only tracked by one single particle,
we don't use the merit function to rank them.  Instead, we find the
orphan particle within the descendant clump which is the most tightly
bound.

In conclusion, although not ideal, using jumpers remains a necessity
to track these temporary merger events.  As a bonus, it allows us to
track orphan galaxies as will be discussed in
Chapter~\ref{chap:mock_catalogues}.

\chapter{Details Of Our New Algorithm \acacia}\label{chap:my_code}

\begin{figure}
	\includegraphics[width=\textwidth]{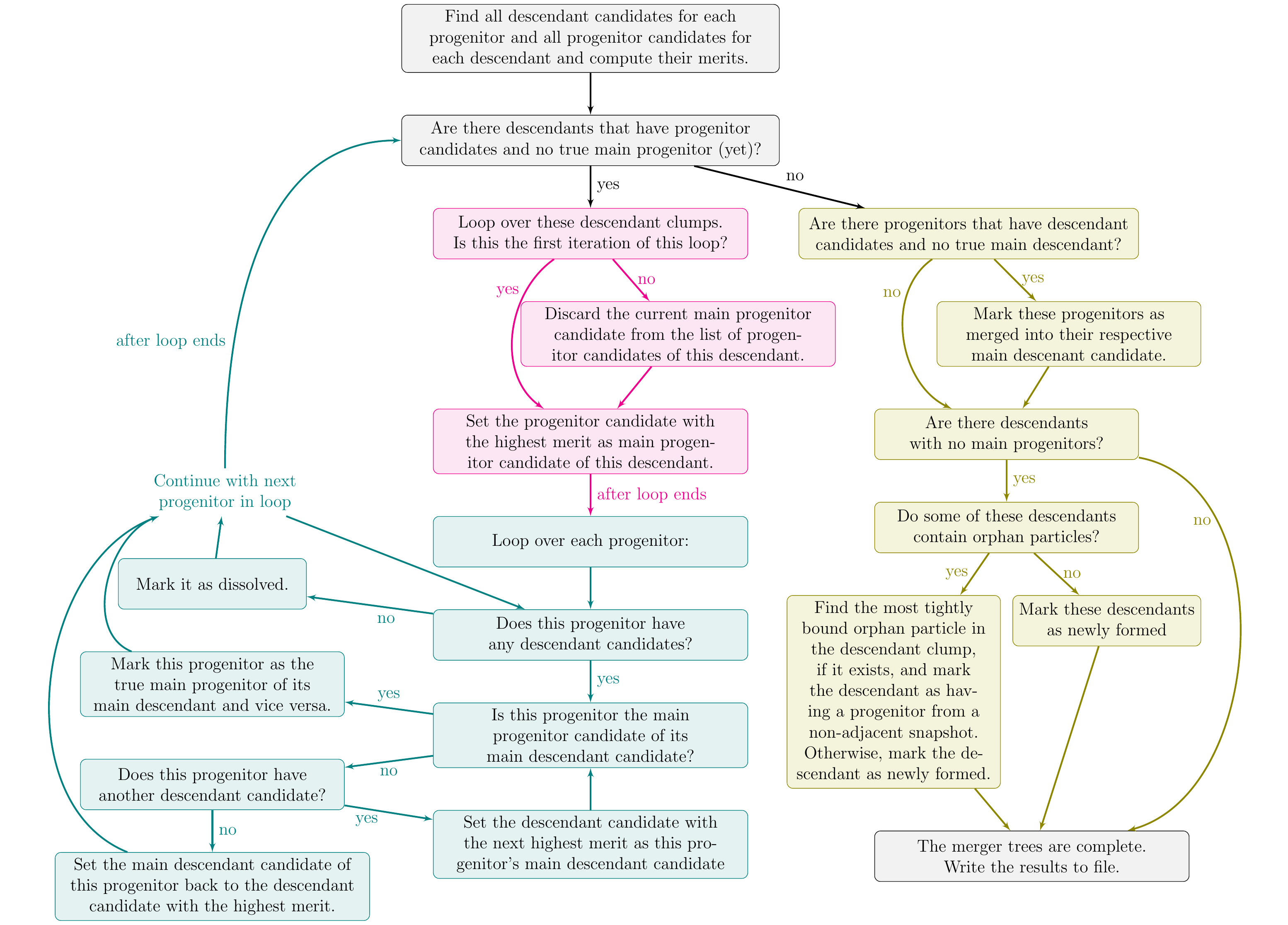}%
	\caption{\label{fig:flowchart}
    	Flowchart of the tree making algorithm. For more details, please refer to Section~\ref{chap:my_code} in the main text.
  }
\end{figure}

For clarity, the algorithm that we describe in what follows
is also shown in a flow diagram in Figure \ref{fig:flowchart}.
The first step of our algorithm is to identify plausible progenitor
candidates for descendant clumps, as well as descendant candidates for progenitor clumps. We achieve this by tracking tracer particles
across simulation snapshots. For any given snapshot, the tracer particles
of each clump are selected within the list of all particles belonging to
the clump, ranked from most bound to least bound. Indeed, the most bound particles are expected to remain well within the clump boundary between
two snapshots. The main parameter of our method is the maximum number of
these tracer particle used per progenitor clump, called $n_{\mathrm{mb}}$. The minimum number of
tracer particles is obviously equal to $n_{\mathrm{min}}$, the minimum mass threshold in units of
particle masses adopted by the \phew\ clump finder.

So for every clump in the current snapshot, the $n_{\mathrm{mb}}$
most bound tracer particles are found and written to file. In the following
output step, those files are read in and the tracer particles
are used to determine which clumps of the previous snapshot are the
progenitor candidates of the clumps of the current snapshot.
For each progenitor clump, we compile a list of descendant clump candidates by finding all
descendant clumps which contain tracer particles of said progenitor. Conversely, we also compile a
list of progenitor candidates for each descendant clump by finding all tracer particles amongst the
descendant clump's particles and by noting which progenitor clump they are tracing.

As explained earlier, in order to generate the merger trees the \textit{main} progenitor of each
descendant and the \textit{main} descendant of each progenitor need to be found. Commonly at this
point however, multiple progenitor candidates have been found for every descendant, as well as
multiple descendant candidates for each progenitor. Therefore, we somehow need to select the
``best'' candidate amongst those with the aim of generating reliable merger trees. To be able to
select the ``best'', we first quantify how ``good'' a candidate is by assigning a merit to each
descendant candidate of every progenitor, as well as every progenitor candidate of each descendant.
We define the merit as follows:

Let $\mathcal{M}_{\mathrm{pd}}(A,B_i)$ be the merit function to be
maximised for a list of descendant candidates $B_i$ of a progenitor
$A$. Let $n_{\mathrm{mb}}$ be the total number of particles of progenitor
$A$ that are being traced. Note that $n_{\mathrm{mb}}$ can be smaller than the
total number of particles in clump $A$.  A straightforward ansatz for
the merit function would be to based on the fraction of particle traced from the
progenitor to the descendant candidate:
\begin{equation}
\mathcal{M}_{\mathrm{pd}}(A,B_i) \propto \frac{n_{A \cap B_i}}{n_{\mathrm{mb}}}
\end{equation}
where $n_{A \cap B_i}$ is the number of tracer particles of $A$ found
in $B$.  Similarly, we define $\mathcal{M}_{\mathrm{dp}}(A_i,B)$ as the
merit function to be maximised for a list of progenitor candidates
$A_i$ of a descendant $B$. Another straightforward ansatz would be
based on the fraction of particle traced from the progenitor
candidates to the descendant:
\begin{equation}
\mathcal{M}_{\mathrm{dp}}(A_i,B) \propto \frac{n_{A_i \cap B}}{n_B}
\end{equation}
where $n_B$ is the total number of particles in the descendant $B$.
In these two merit functions, $n_{\mathrm{mb}}$ and $n_B$ are just
normalizing factors. They are independent of the properties of the
candidate and hence won't affect the selection process.  We can
therefore define a generic merit function as
\begin{equation}
\mathcal{M}(A,B) \propto n_{A \cap B}
\end{equation}

The \phew\ clump finder in \ramses\ identifies the main halo as the
clump with the highest density maximum.  During a major merger event,
the halo will have two clumps with similar masses and comparable
maximum densities.  It is then quite common that small variations in
the value of the density maxima will cause the identification of the
main halo to jump between these two clumps.  Indeed, the particle
unbinding algorithm will identify particles that are not bound to the
sub-halo and pass them on to the main halo, modifying the resulting
mass for the two clumps.  This effect is particularly strong if one
uses the \emph{strictly bound} definition for the particle
assignment. As a consequence, because the identification of the main
halo varies, strong mass oscillations are expected. To counter this
spurious effect, we modify the merit function to preferentially select candidates with similar masses:
\begin{equation}
\mathcal{M}(A,B) = \frac{n_{A \cap B}}{n_{\mathrm{max}} - n_{\mathrm{min}}} \label{eq:merit}
\end{equation}
where $n_{\mathrm{max}}=\max(n_A,n_B)$ and $n_{\mathrm{min}}=\min(n_A,n_B)$.  An
overview of other merit functions used in the literature is given in
Table~1 of \cite{SUSSING_COMPARISON}.

With the merit function defined, let's return to the tree making algorithm. We left off with the commonly encountered situation in which we have identified multiple progenitor candidates for descendants and vice versa, and are now
looking to find a matching progenitor-descendant pair, where the \textit{main} progenitor candidate of a descendant has precisely this descendant as its \textit{main} descendant candidate. This search is performed iteratively.
At every iteration, we first look at all progenitors that haven't found their respective match yet, and check all of their descendant candidates for a match. The loop over descendant candidates for a given progenitor is performed in order of decreasing merit of the descendant candidates, and the checks are  stopped either when a progenitor has found a match, or has run out of candidates. (In case a progenitor has no descendant candidates at all, we consider it as dissolved.)
Once all the progenitors have been checked, the second step of the iteration begins. We now look at all descendants that haven't found their respective match yet. For these descendants, we discard the current, unsuccessful main progenitor candidate in favour of the progenitor candidate with the next highest merit, and the first step of the iteration begins anew: All progenitors without a matching main descendant again loop over all their descendant candidates in search for a match. This two-step iteration is repeated until either all descendants have found their match, or have run out of progenitor candidates.

After the iteration finishes, we may have both progenitors as well as descendants which aren't part of a matching pair. We deal with them as follows:

\begin{enumerate}

\item Progenitors that have not found any available descendant will be
  considered to have merged into the descendant candidate with the highest merit. These progenitors are recorded in the merger tree as
  \emph{merged progenitors}.  In this case, only one tracer particle is
  kept for future use, the most strongly bound particle in the list of $n_{\mathrm{mb}}$ tracer
particles.  This single particle is referred to as the
  \emph{orphan particle} of the merged progenitor.  It is used to
  check whether the merger event was a final merger or only a
  temporary merger. It is also used to track orphan galaxies.

\item Descendants that have not found any available progenitor will be
  checked against non-consecutive past snapshots. The particles of the
  descendant are compared to the orphan particles in the list of past
  merged progenitors\footnote{There is an option to remove past merged
  progenitors from the list if they have merged into their main
  descendant too many snapshots ago.  By default, however, the
  algorithm will store them all until the very end of the
  simulation.}.  The most strongly bound orphan particle will be used
  to restore the broken link with its main progenitor.  Finally,
  remaining descendants without a progenitor are considered as being
  newly formed.

\end{enumerate}

Finally, we note that the choice to first check all progenitor candidates of descendants and only
merge progenitors into descendants later is how we deal with fragmentation
events (see Section~\ref{sect:frag}) in an attempt to preserve the formation
history of clumps. Effectively, this procedure assigns more weight to a
descendant having progenitor candidates \textit{at all} over the merging of a progenitor into its main descendant candidate, as the merit function would suggest.

%% file: figures/ACACIA/jumper-demo/jumper-demo-images.tex
\begin{figure*}
    {
        \setlength\tabcolsep{0em}
    	\centering
    	\begin{tabular}{p{4.3cm}p{4.3cm}p{4.3cm}}
            \centering
    		%
    		\fbox{\includegraphics[width = 4.3cm]{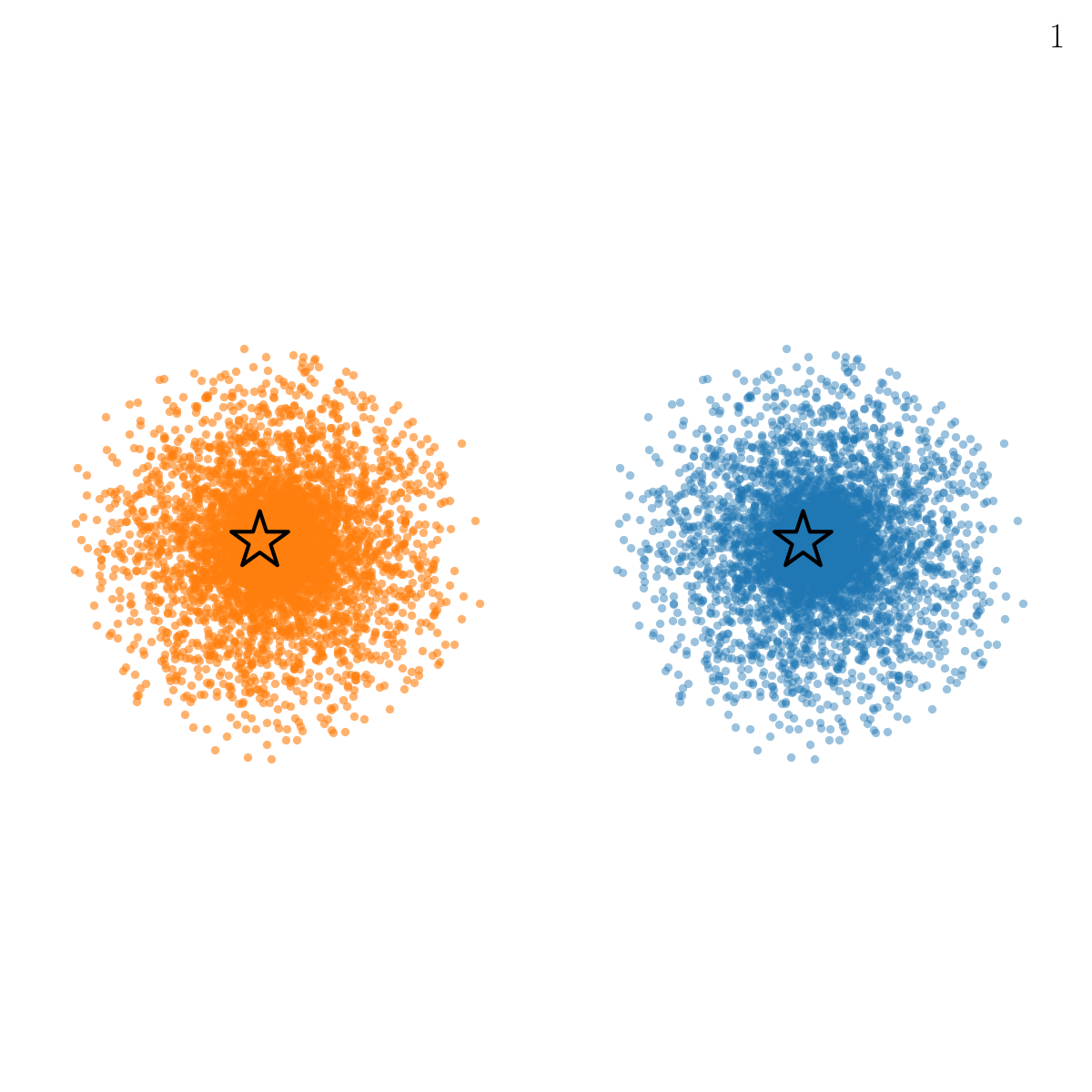}}	&
    		\fbox{\includegraphics[width = 4.3cm]{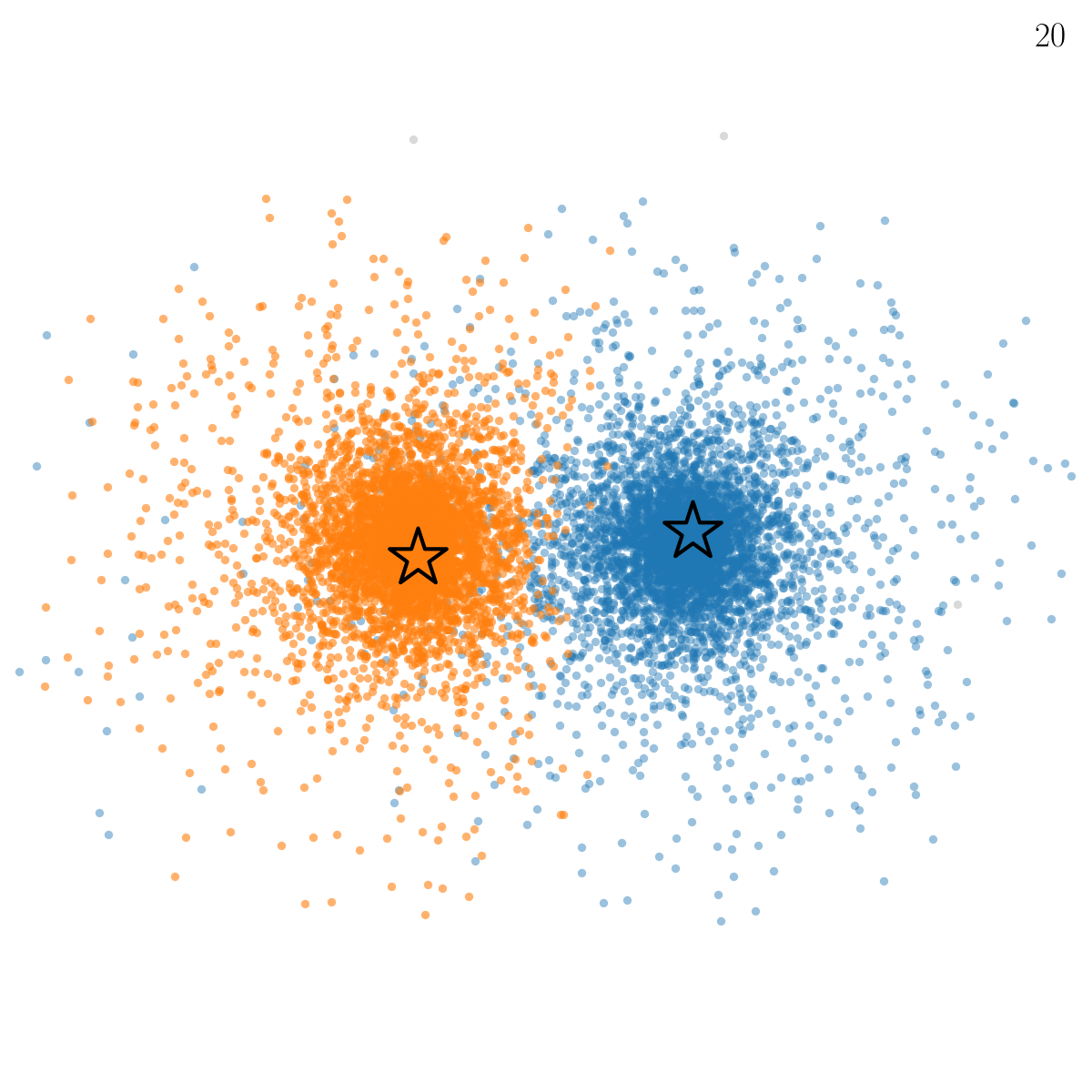}}	&
    		\fbox{\includegraphics[width = 4.3cm]{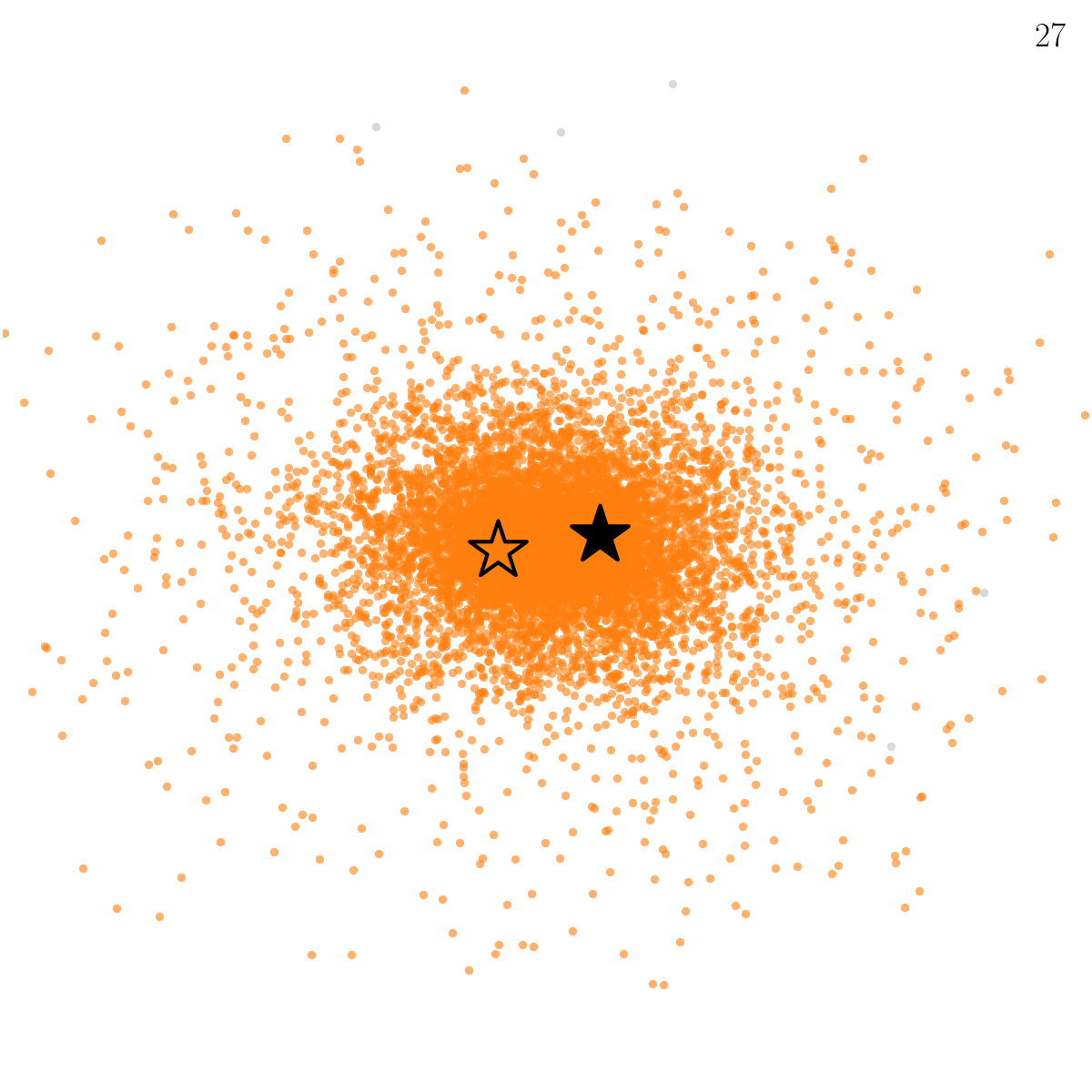}}  \\
    		\fbox{\includegraphics[width = 4.3cm]{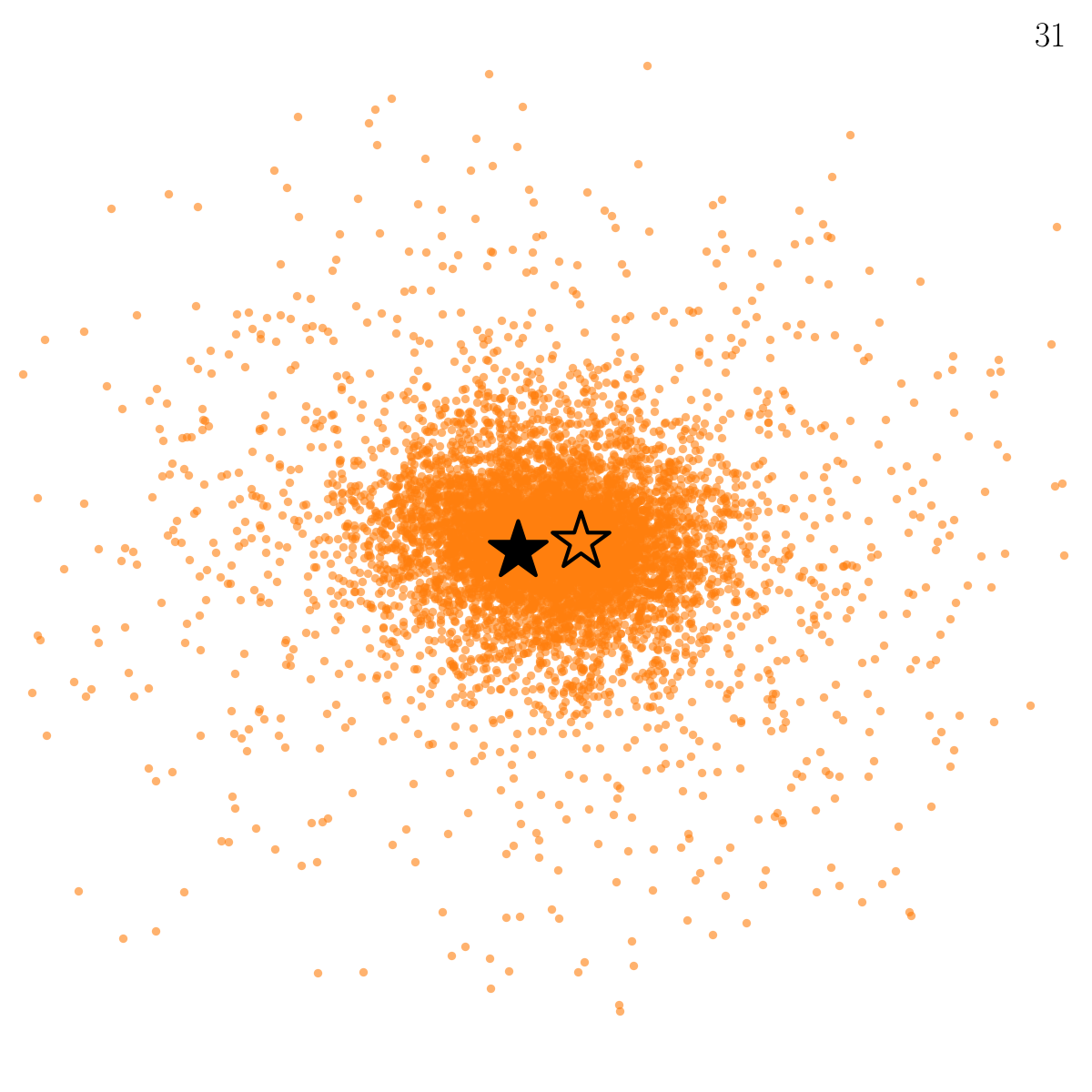}}	&
    		\fbox{\includegraphics[width = 4.3cm]{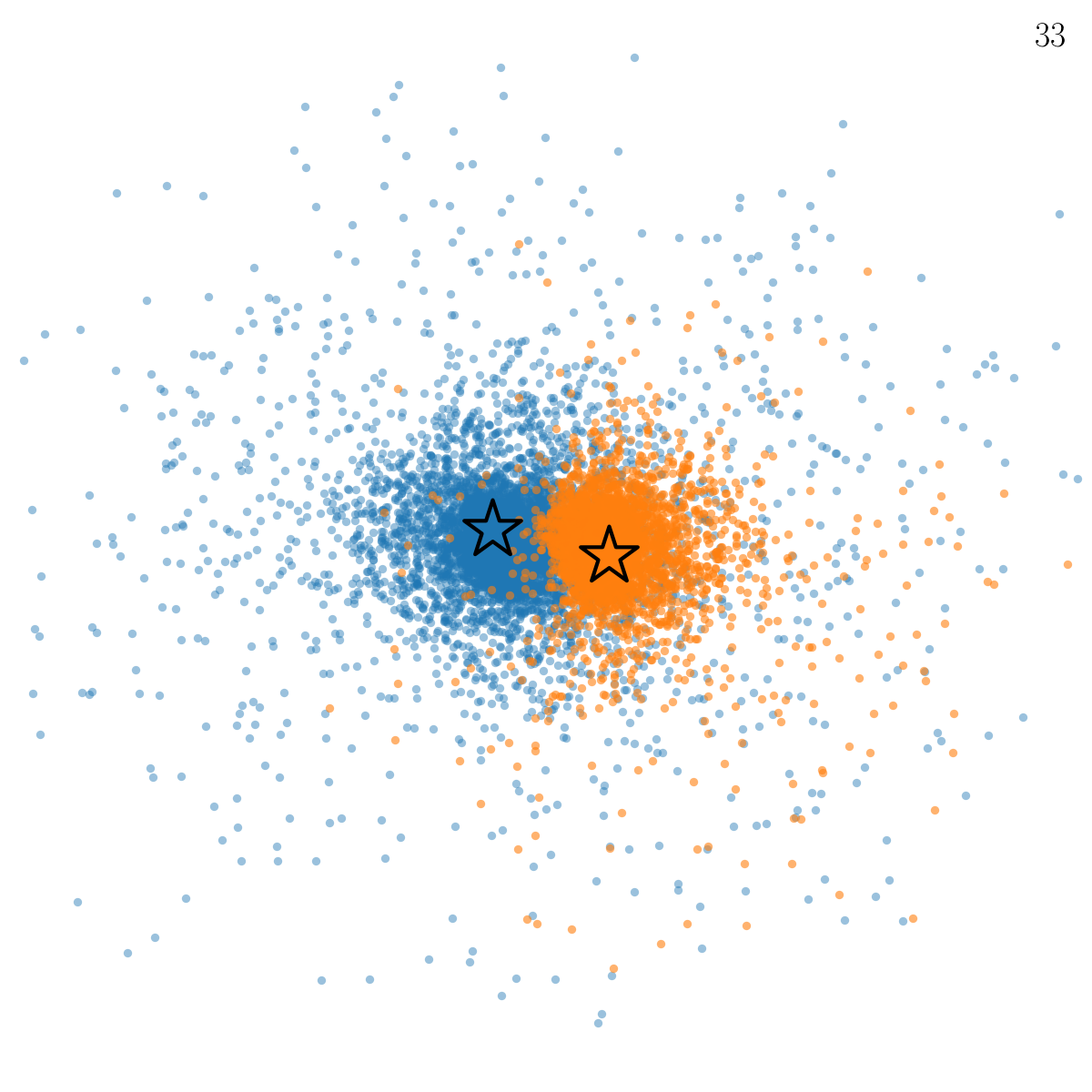}}	&
            \fbox{\includegraphics[width = 4.3cm]{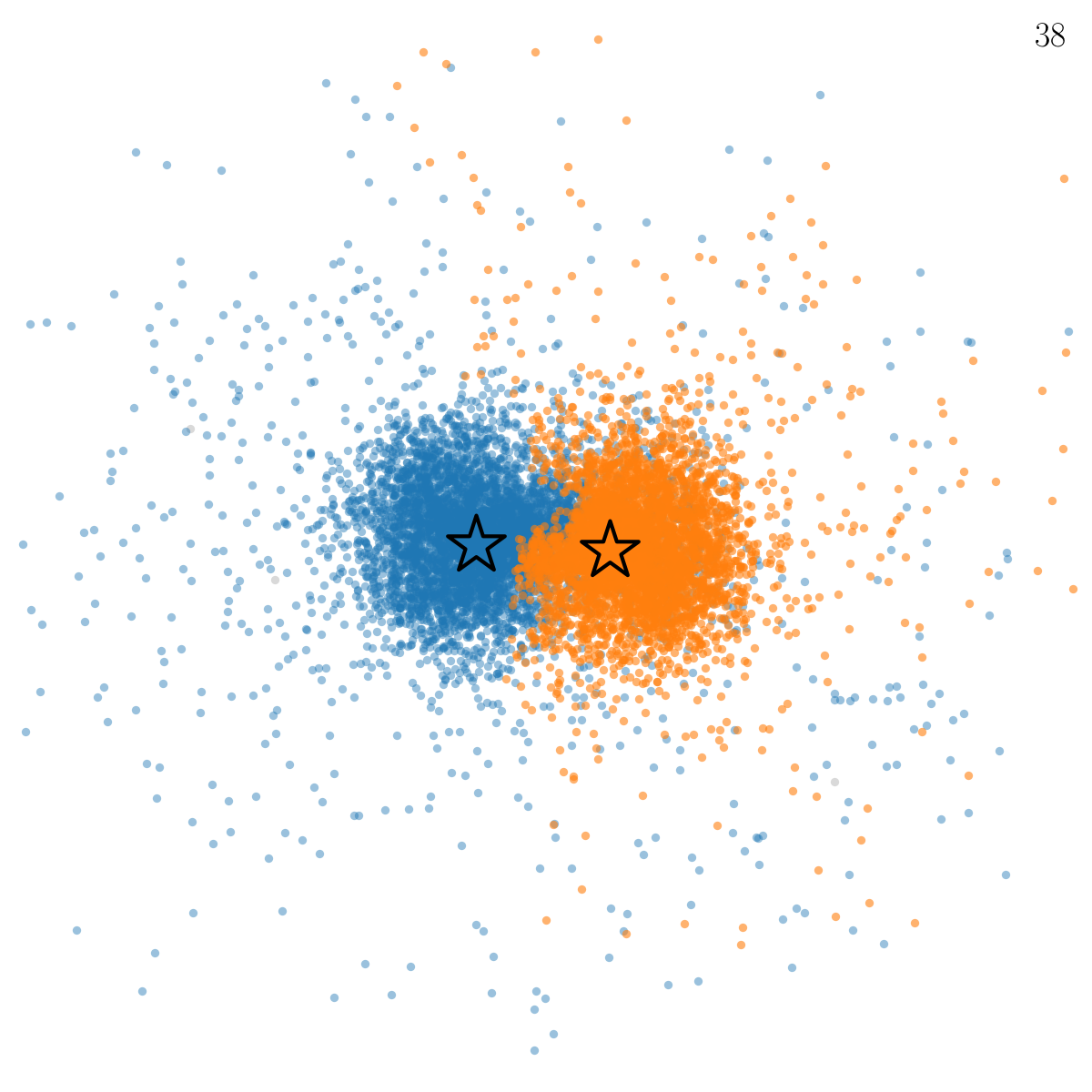}}	\\
    	\end{tabular}
    }
	\caption{\label{fig:jumper-demo}
        Illustration of how haloes can seemingly merge into another one and re-appear a few snapshots later.
        The blue and orange particles are two initially distinct haloes that pass through each other.
        The galaxies assigned to them are marked by a star with the same colour as the particles.
        Fully black stars mark orphan galaxies, which have lost their unique host (sub)halo.
        The number in the upper right corner of each plot is the snapshot number that is depicted.
        In snapshots 27-32, the halo-finding algorithm doesn't identify both haloes as distinct objects.
        However by tracing the blue halo's orphan galaxy it was possible to link the halo in snapshot 33 all the way back to snapshot 26. \hspace{\textwidth}
        The initial conditions were created using \texttt{DICE} \citep{DICE}.
        }
\end{figure*}

%% file: main/ACACIA/ACA3-mergertree_testing.tex
\chapter{Testing and Optimizing the Algorithm}\label{chap:tests}

The current implementation of our merger tree algorithm needs several
free parameters to be chosen by the user. We present in this section
multiple tests that reveal the recommended values for these
parameters.  We use for this a single reference cosmological
simulation and analyze the merger trees we obtained for different set
of parameters.  A similar methodology was used in the Sussing Merger
Trees Comparison Project \citep{SUSSING_COMPARISON,
  SUSSING_CONVERGENCE, SUSSING_HALOFINDER,leeSussingMergerTrees2014}.
This is why we adopt in this section several tests from this seminal
work, as they are quite efficient at testing the strengths and the
weaknesses of merger tree codes. They also allow for a direct
comparison of our new implementation with many other state-of-the-art
merger tree codes in the community.

\section{Test Suite}\label{chap:testing_methods}

We now list the different diagnostics we use to characterize the quality of
our merger tree algorithm.

\begin{enumerate}
  \setlength\itemsep{1em}
\item \emph{Length of the Main Branch of a Halo}

  The length of the main branch of a halo is simply defined as the
  number of snapshots in which a halo and all its progenitors are
  detected.  A halo at $z=0$ without any progenitors will be
  considered as newly formed and thus will have a main branch of
  length $1$.  If a halo appears to merge temporarily into another and
  re-emerges at a later snapshot, the missing snapshots will be
  counted towards the length of the main branch as if they weren't
  missing. Traditionally, finding long main branches is considered
  as a good thing for a merger tree code.

\item \emph{Number of Branches of a Halo}

  Another popular quantity is the number of branches of the tree
  leading to the formation of a halo at $z=0$.  The main branch is
  included in this count, thus the minimal number of branches is
  $1$. If a different choice of parameters leads to a reduction of the
  number of branches, it usually corresponds to an increase of the
  average length of the main branches and a smaller number of merger
  events. For example, intuitively if we compare the merger trees where
  we use only one tracer particle per clump to the trees that were built
  using several hundreds tracer particles per clump, we would expect to
  be able to detect more fragmentation events with the increased number of
  tracer particles. If the fragmentation remained undetected, we would
  instead have found a newly formed clump (the fragment) alongside a
  merging event. Both the ``newly formed'' fragment as well as its
  progenitor, which is now merged into a descendant, will have shorter
  main branches. Conversely, the descendant will have an increased
  number of branches compared to the scenario where the fragmentation
  was detected.
  Finally, in the hierarchical picture of structure formation, one
  would expect more massive clumps to have longer main branches and a
  higher number of branches.

\item \emph{Logarithmic Mass Growth of a Halo}

  The logarithmic mass growth rate of a halo is computed using the following
  finite difference approximation:
  \begin{equation}
    \frac{\de \log M}{\de \log t} \simeq
    \frac{(t_{k+1}+t_{k})(M_{k+1} -M_{k})}{(t_{k+1} - t_k)(M_{k+1} + M_{k})} \equiv \alpha_M(k, k+1)
  \end{equation}
  where $k$ and $k+1$ are two consecutive snapshots, with the
  corresponding halo mass $M_k$ and $M_{k+1}$ and times $t_k$ and
  $t_{k+1}$. A convenient approach was proposed by \cite{SUSSING_CONVERGENCE}
  to reduce the range of values to the interval $(-1, 1)$ using the new variable
  \begin{equation}
    \beta_M = \frac{2}{\pi}\arctan(\alpha_M) \label{eq:massgrowth}
  \end{equation}
  Note that we expect the mass of dark matter haloes to increase
  systematically with time.  We also expect in some cases the mass to
  remain constant or even to decrease slightly.  We nevertheless
  expect the distribution of $\beta_M$ to be skewed towards $\beta_M >
  0$.  $\beta_M \rightarrow \pm 1$ imply $\alpha_M \rightarrow \pm
  \infty$, indicating suspiciously extreme cases of mass growth or
  mass loss.

\item \emph{Mass Growth Fluctuation of a Halo}

  Mass growth fluctuations are defined similarly as
  \begin{equation}
    \xi_M = \frac{\beta_M(k, k+1) - \beta_M(k-1, k)}{2} \label{eq:massfluct}
  \end{equation}
  where $k-1$, $k$, $k+1$ are three consecutive snapshots.  A smooth
  mass accretion history generally leads to $\xi_M \simeq 0$.  Strong
  deviations from zero could indicate an erratic behaviour, indicating
  extreme mass loss followed by extreme mass growth and vice versa.
  Within the standard model of structure formation, this behaviour is
  expected only during major merger events. Otherwise, it might
  indicate either a misidentification by the merger tree code or a
  misdetection by the halo finder.


\end{enumerate}

Ideally, we should have tested \acacia on the dataset used in
\citet{SUSSING_COMPARISON} and \citet{SUSSING_HALOFINDER} (S13 and A14
from here on, respectively).  This would
have enabled a direct comparison of the performance to other merger
tree codes.  However, \acacia was designed to work on the fly
with the \ramses\ code.  Using it as a post-processing tool would
defeat its purpose and as a matter of fact handling other halo
catalogues  has proven technically impossible. \acacia is
tightly coupled to the \phew\ halo finder, and relies heavily on
already existing internal structures and tools, in particular the
explicit communications which are necessary for parallelism on
distributed memory architectures, as well as the structures and
their hierarchies as they are defined by \phew. Attempting to use
other halo catalogues would require us to re-write a significant
portion of the \phew\ halo finder. If we instead used only particle
data, which is possible, we would still find a different halo catalogue
compared to other structure finding codes, and we would still not be
able to do an exact comparison. Furthermore, we also want to demonstrate that the
\phew\ halo finder can be used within the \ramses\ code to produce
reliable merger trees.  For these various reasons, we have decided to
perform the same tests but using our own dataset generated on the fly by
\ramses.

Despite this limitation, we have performed a direct comparison to
other halo finders and merger tree codes using the exact same merger
tree parameters as in A14. The results are
given in Chapter~\ref{chap:performance_comparison}.  Our results are
comparable to e.g. the \codename{MergerTree}, \codename{TreeMaker} and
\codename{VELOCIraptor} tree builders with \codename{AHF}, \codename{Subfind},
or \codename{Rockstar} halo finders as presented in A14, demonstrating that
\acacia performs similarly to other state-of-the-art tools.

In this section, we would like to explore different parameters and see how they affect the quality
of the merger tree.  Our tests are performed on a single DMO simulation with $256^3 \approx
1.7\times 10^7$ particles of identical mass $m_p = 1.6\times 10^9\msol$. To enable a comparison
with
A14, we adapted the same cosmology and snapshot output times as them at a comparable, but slightly
lower resolution. The cosmological parameters used are taken from the WMAP-7
\citep{komatsuSevenYearWilkinsonMicrowave2011}, while the snapshot times are identical to the ones
used for Millenium Simulation \citep{springelSimulationsFormationEvolution2005a}, starting at
redshift 50 and being roughly uniformly spaced in log $a$ in 61 steps. At redshift zero, the
simulation was then continued for 3 further snapshots to ensure that the merging events at $z = 0$
are actual mergers and not temporary mergers that will re-emerge later.

This choice of spacings between the snapshots was relatively arbitrarily in the sense that we did
not take into account any further underlying physical considerations that would be important in e.g.
semi-analytical models. For different snapshot spacings, we recommend to follow the suggestions
found by \citet{SUSSING_CONVERGENCE}:
\begin{itemize}
\item  Sequences of snapshots with very rapidly changing time intervals
between them should be avoided as they can lead to very poor trees.
\item  Increasing the number of outputs from which the tree is generated
results in shorter trees. This is because, due to limitations in the input halo
catalogue, tree-builders may face difficulties caused by the fluctuating center
and size of the input haloes, and the frequency of detected temporary merging
events increases with the number of snapshots, resulting in haloes missing from
the catalogue. For merger trees built from
an order of 100 or more snapshots, they recommend using an algorithm capable of
dealing with these problems, which \acacia is able to do, although at the
moment this patching of missing haloes in the catalogue isn't based on a physical
timescale, but on a user defined number of snapshots.
\item  To facilitate this patching at the end of the simulation, snapshots should
be generated beyond the desired endpoint. This would entail typically running past
$z = 0$, as we did with our test suite.
\end{itemize}

For the clump finder, we have adopted an outer density
threshold of 80$\bar{\rho}$ and a saddle surface density threshold of
200$\bar{\rho}$, where $\bar{\rho} = \Omega_m \frac{3 H^2}{8 \pi G}$ is
the average density.  The minimal mass for clumps is set to 10
particles. Note that for the histogram of the logarithmic mass growth
and mass growth fluctuations, we adopt a threshold for the clump mass
of 200 particles for sake of visibility. Choosing a smaller mass
threshold would indeed give too much weight to small mass, poorly
resolved haloes in our statistical analysis.

\section{Varying the Clump Mass Definition}\label{chap:varying_clump_mass_definition}

In our current implementation, there are two important parameters that
can have a strong effect on the halo catalogue (beside the mass and
the density thresholds mentioned earlier) and the corresponding merger
tree.

The first one is the exact definition adopted for the mass of the
sub-haloes in the merit function.  For main haloes, there is no
ambiguity as the mass is defined as the sum of the masses of all
particles contained within the boundary of the halo (set by the outer
density isosurface). This is not the case for sub-haloes, because of the
unbinding process described in Chapter~\ref{chap:phew}. Indeed,
unbound particles are removed from their original sub-halo and passed
to the parent sub-halo in the hierarchy.  Clump masses are therefore
defined as the sum of the mass of all bound particles.  In the merit
function evaluation, we however consider two different cases to
compute the mass: 1- the mass is equal to the sum of the masses of
only the bound particles, like for sub-haloes or 2- the mass is equal
to the sum of the masses of all particles within their boundaries (set
by the saddle surface with neighbouring clumps), like for main haloes.
In the former case, the mass used in the merit function is identical
to the clump mass. It is referred to as the \exc\ case.  In the latter
case, the mass in the merit function is different that the sub-halo
mass definition but identical to the main halo mass definition.  We
refer to this case as \inc.

The second important definition is the exact boundedness criterion
adopted for the unbinding process.  As discussed in
Chapter~\ref{chap:phew}, we explored two different cases: When
particles are allowed to leave the outer boundary of their host clump
(and possibly come back later) or when particles are not allowed to
cross the saddle surface during their orbital evolution. In the first
case, we only require the binding energy to be negative, while in the
second case, the binding energy has to be smaller than the
gravitational potential of the nearest saddle point.  We call the
first case \nosad\ and the second case \sad.

\begin{figure}
  \centering
  \includegraphics[width=\textwidth,keepaspectratio]
	{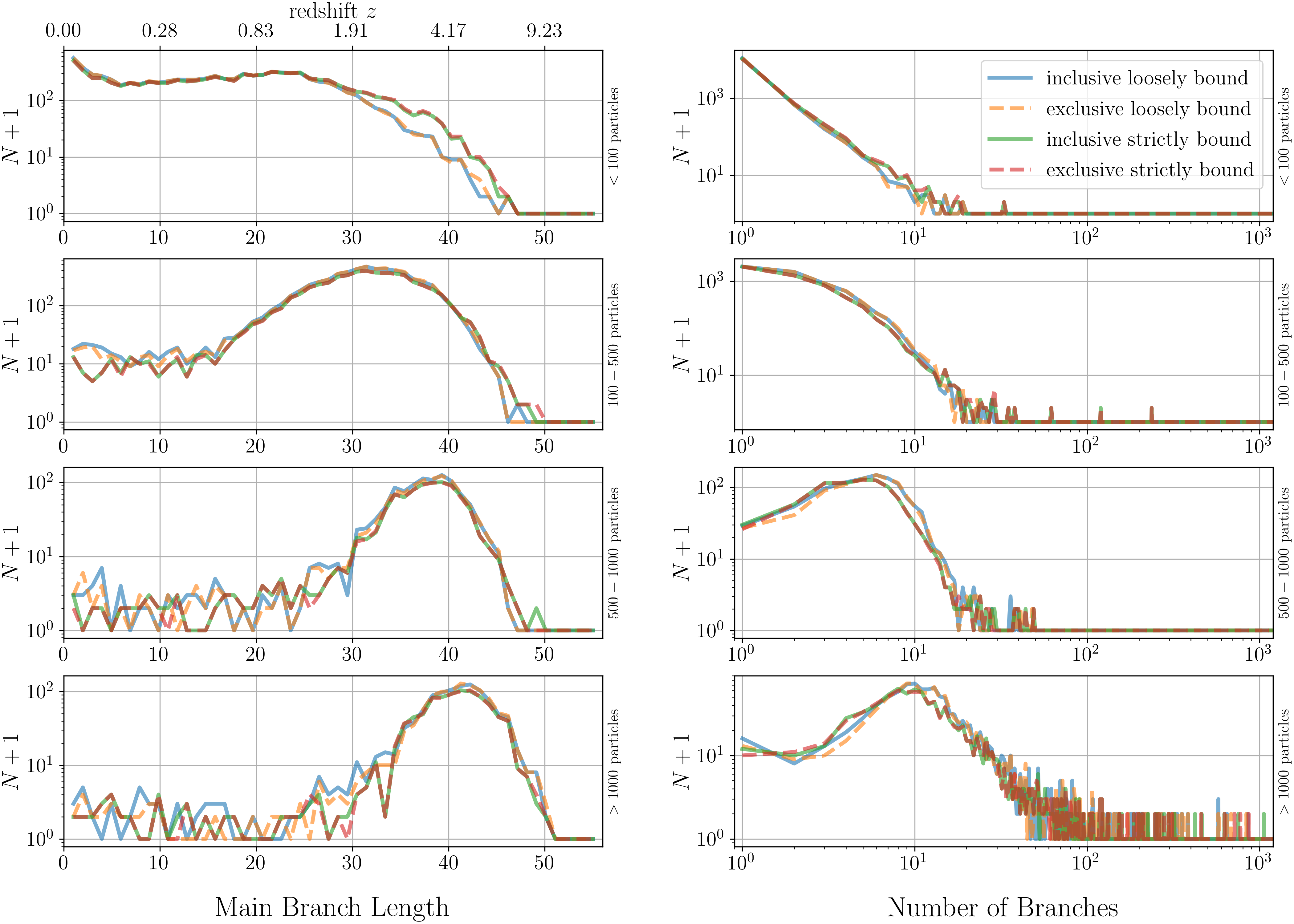}
  \caption{Histogram of the length of main branch (left) and of the
    number of branches (right) for all clumps (halo and sub-halo)
    detected at $z=0$. Each row corresponds to a different range of
    clump masses (expressed in particle numbers): less then 100 (top),
    100-500, 500-1000 and more than 1000 (bottom). We compare these
    histograms for four different cases: whether unbound particles are
    included (\inc) or excluded (\exc) in the evaluation of the merit
    function, and whether bound particles are \nosad\ or \sad.
  }%
  \label{fig:saddle_nosaddle_mbl_nbranch}
\end{figure}

\begin{figure}
  \centering
  \includegraphics[width=.6\linewidth, keepaspectratio]
	{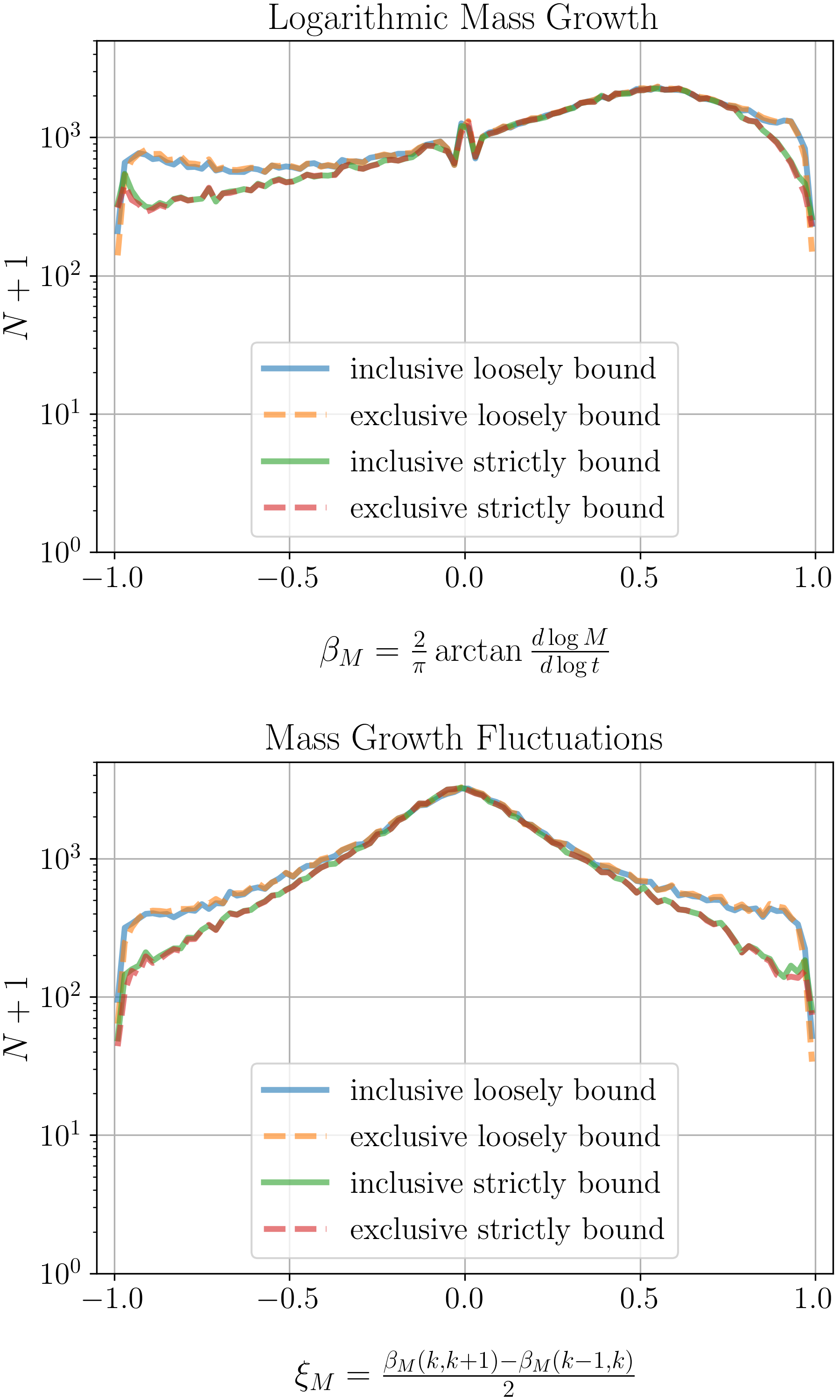}%
  \caption{Histogram of the logarithmic mass growth (top) and the mass
    growth fluctuation (bottom) for all clumps (halo and sub-halo)
    detected in two (top) or three (bottom) consecutive snapshots of
    the simulation and with more than 200 particles. We compare these
    histograms for four different cases: whether unbound particles are
    included (\inc) or excluded (\exc) in the evaluation of the merit
    function, and whether bound particles are \nosad\ or \sad.
  }%
  \label{fig:saddle_nosaddle_masses}
\end{figure}

\input{tables/ACACIA/saddle_nosaddle_global_stats}

We now test our algorithm with these four different options for the clump
masses, using the simulation presented in the previous section. Note
that we used here $n_{\mathrm{mb}}=200$ tracer particles to identify links
in the merger tree. We will study the impact of this other important
parameter in the next section.

We show in Figure~\ref{fig:saddle_nosaddle_mbl_nbranch} the histogram
of the length of the main branch and the histogram of the number of
branches for each clump (halo and sub-halo) at $z=0$ and for each of
our four different mass definitions. In all four cases, we see that more
massive clumps tend to have longer main branches and a higher number
of branches.  This is also visible from the average length of the main
branch and the average number of branches in different bins of halo
masses given in Table~\ref{tab:saddle_nosaddle}. This is a well-known
property of cosmological simulations in the hierarchical scenario of
structure formation.

Whether clump masses are defined in an \exc\ manner (like for
sub-haloes) or in an \inc\ manner (like for main haloes) in the merit
function has negligible effect on these two statistics. This means
that this a priori large difference in the mass definition of the
merit function has no effect on the linking process of the merger
tree.  The distinction between \sad\ and \nosad\ particles does not
change much for larger clumps but does change the length of the main
branch (and to a lesser extent the number of branches) for small mass
clumps (less than 100 particles). Our first idea was that
\sad\ particles might be better at identifying robust links between
snapshots. It turned out that the main effect of changing the mass
definition from \nosad\ to \sad\ is to reduce the mass of the clump and
to promote them systematically from a larger mass bin to a smaller
mass bin. We see indeed in
Figure~\ref{fig:saddle_nosaddle_mbl_nbranch} and
Table~\ref{tab:saddle_nosaddle} that the number of clumps is reduced
in the large mass bins and increased in the smallest mass bin,
explaining that this change in the mass definition merely transfers clumps
between different bins and affects the statistics accordingly.

A qualitative comparison of the length of the main branches of the most
massive haloes that we obtain in Figure~\ref{fig:saddle_nosaddle_mbl_nbranch}
to Figure~3 of A14 shows that our results are in good agreement with what
the other codes find: The distribution peaks around the length of 45, it
is about 20 snapshots wide, and there are only few cases with main branch
lengths below 30. This is in good agreement with what e.g. the
\codename{MergerTree}, \codename{TreeMaker}, and \codename{VELOCIraptor} tree builders
find in combination with the \codename{Rockstar} or \codename{Subfind} halo
finders. We note that in Figure~3 of A14, both the peak of the
distribution and the maximal value of the main branch lengths they found are
at slightly higher values than ours. We attribute this to the slightly lower
resolution of our simulations: The first identifiable clumps we find are at
snapshot 10, leading to a maximal main branch length of 51, compared to
$\sim 53$ that A14 find. Compared to Figure~3 of S13, the distributions we
find for the lower mass clumps are also in very good agreement. Our high
clump mass distribution however is much narrower around the peak value of
$\sim 45$. This difference is due to the different halo finders employed,
as is demonstrated in Figure~3 of A14. The \texttt{AHF} halo finder, which
was used in S13, displays the same differences in the distribution of
main branch lengths for nearly all tree codes.

We also note in Figure~\ref{fig:saddle_nosaddle_mbl_nbranch} that a few
large clumps (with mass larger than 500 particles) at $z=0$ have a
main branch length of unity.  These large clumps don't have any
progenitor and thus essentially appeared out of nowhere. As explained
in S13, this effect is present in many state-of-the-art merger tree
codes and is due to fragmentation events at the periphery of large
haloes leading to a misidentification of a few rare progenitor-descendant
links. We have however identified a second culprit, which is the way that
\phew\ establishes substructure hierarchies and the subsequent particle
unbinding. The hierarchy is determined by the density of the density peak
of each clump: A clump with a lower peak density will be considered lower
in the hierarchy of substructure. So in situations where two adjacent
clumps have similarly high density peaks, their order in the hierarchy
might swap. The unbinding algorithm then strips the particles from the
sub-haloes that have the lowest level in the hierarchy and passes it on to
the next level, amplifying the particle loss which these sub-haloes
experience. This loss of particles is essential here because it prevents
the algorithm to establish links between progenitors and descendants.
About half of the main branches that we tracked back in time were cut
short for this reason: The leaf of the main branch was a sub-halo with
much fewer particles ($\sim 100$) whose progenitor the algorithm was
not able to identify and who in subsequent snapshots was found to be
the main halo, gaining a lot of mass in a very short time. So this issue
arises due to the halo finder, not due to the tree builder.

The histogram of the logarithmic mass growth shown in
Figure~\ref{fig:saddle_nosaddle_masses} is indeed skewed towards
$\beta_M > 0$, demonstrating that clump masses are on average growing.
Based on the shoulder of the histogram around $\beta_M \sim 0.6$ our
results are comparable to those of the \codename{HBThalo} and \codename{Subfind}
halo finders in Column~A of Figure~8 of A14 for all tree codes. We do
find more extreme events with $\beta_M \rightarrow \pm 1$, which are
due to the smaller mass haloes and sub-haloes that we used compared
to A14. Indeed, when we apply the appropriate mass thresholds in
Appendix~\ref{chap:performance_comparison}, these extreme events are
significantly reduced. The small wiggle around $\beta_M = 0$ is due to
the discrete particle masses and the linear binning of the histogram.
Adopting a merit function based on the \inc\ or \exc\ mass definition
has here also no effect on the mass growth and mass growth fluctuation
statistics.  At first sight, using the \sad\ instead of the
\nosad\ definition would have led to more robust links and a smoother
mass growth. In Figure~\ref{fig:saddle_nosaddle_masses}, we do see in
the latter case more extreme mass growth around $\beta_M \rightarrow
\pm 1$ and mass growth fluctuations around $\xi_M \rightarrow \pm
1$. We verified that the increase in the number of these extreme
events for the \nosad\ case is in fact due to a larger number of small
sub-haloes that satisfy the adopted mass threshold of 200 particles.
This just means that mass growth statistics is more robust for large,
well resolved haloes, while smaller clumps, closer to the resolution
limit (between 10 and 100 particles; see discussion below) are less
reliable.

In conclusion, whether to use \inc\ or \exc\ mass definitions in the
merit function has no effect on the final merger trees, while using a
\sad\ definition for the clump mass is preferable. We expected the
\sad\ clumps to allow more stable tracking, but the only effect we
noticed was that it systematically promotes
sub-haloes to lower masses and so naturally selects better resolved,
higher mass clumps from the halo catalogue.

\section{Varying the Number of Tracer Particles} \label{chap:testing-nmb}

\input{tables/ACACIA/ntracers_global_stats}
\input{tables/ACACIA/ntracers_pruning}
\input{tables/ACACIA/ntracers_jumpers}

\begin{figure}
  \centering
  \includegraphics[width=\textwidth, keepaspectratio]
	{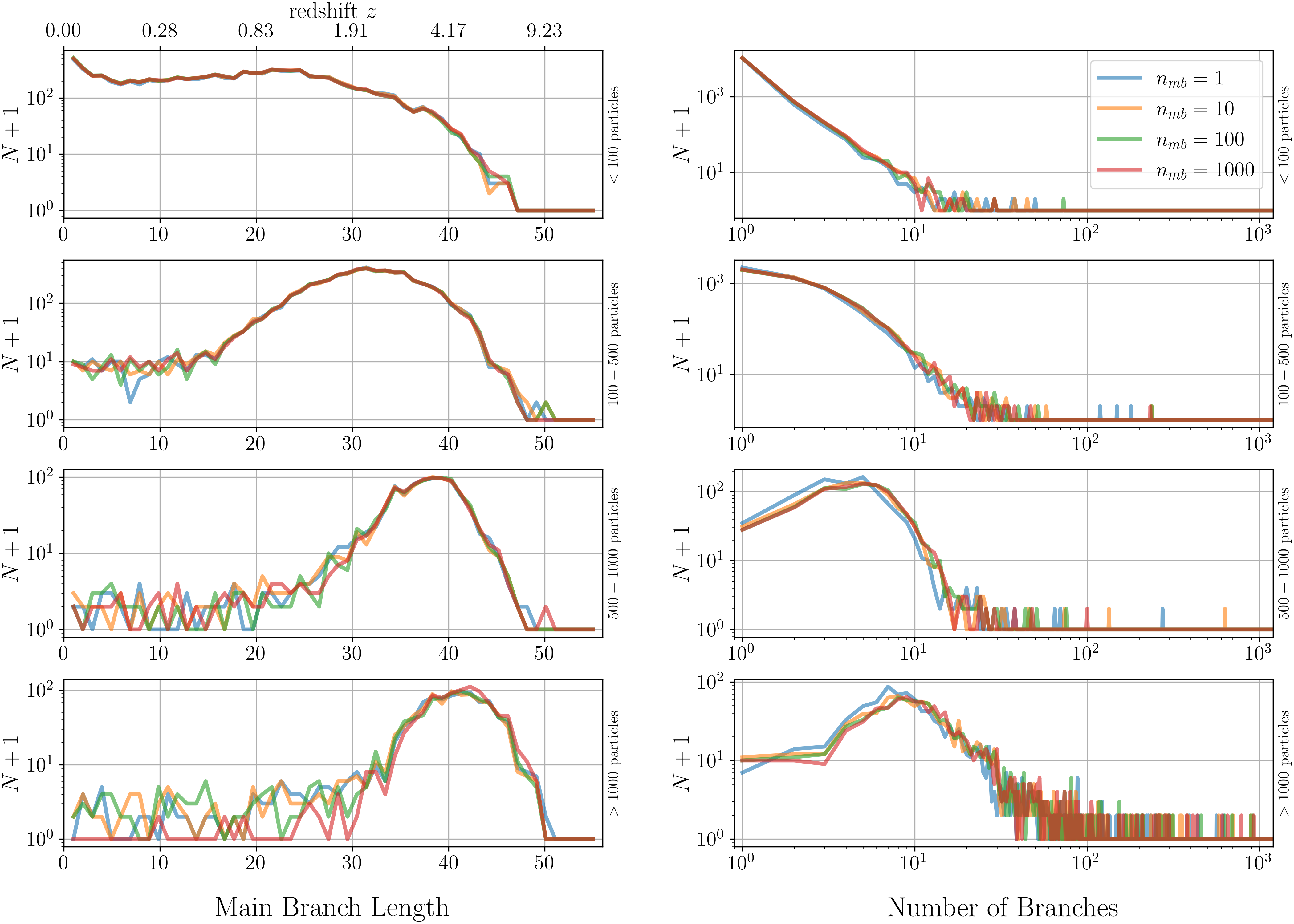}%
  \caption{ Histogram of the length of the main branch (left) and
    histogram of the number of branches (right) for all clumps (haloes
    and sub-haloes) detected at $z=0$ for different numbers of tracer
    particles $n_{\mathrm{mb}}$ indicated in the legend.  Each row
    corresponds to a different range of clump masses (expressed in
    particle numbers): less then 100 (top), 100-500, 500-1000 and more
    than 1000 (bottom).  In all cases, we used the \exc\ and
    \sad\ clump mass definitions.
  }%
  \label{fig:ntracers_mbl_nbranch}
\end{figure}

\begin{figure}
  \centering
  \includegraphics[width=.6\linewidth, keepaspectratio]
	{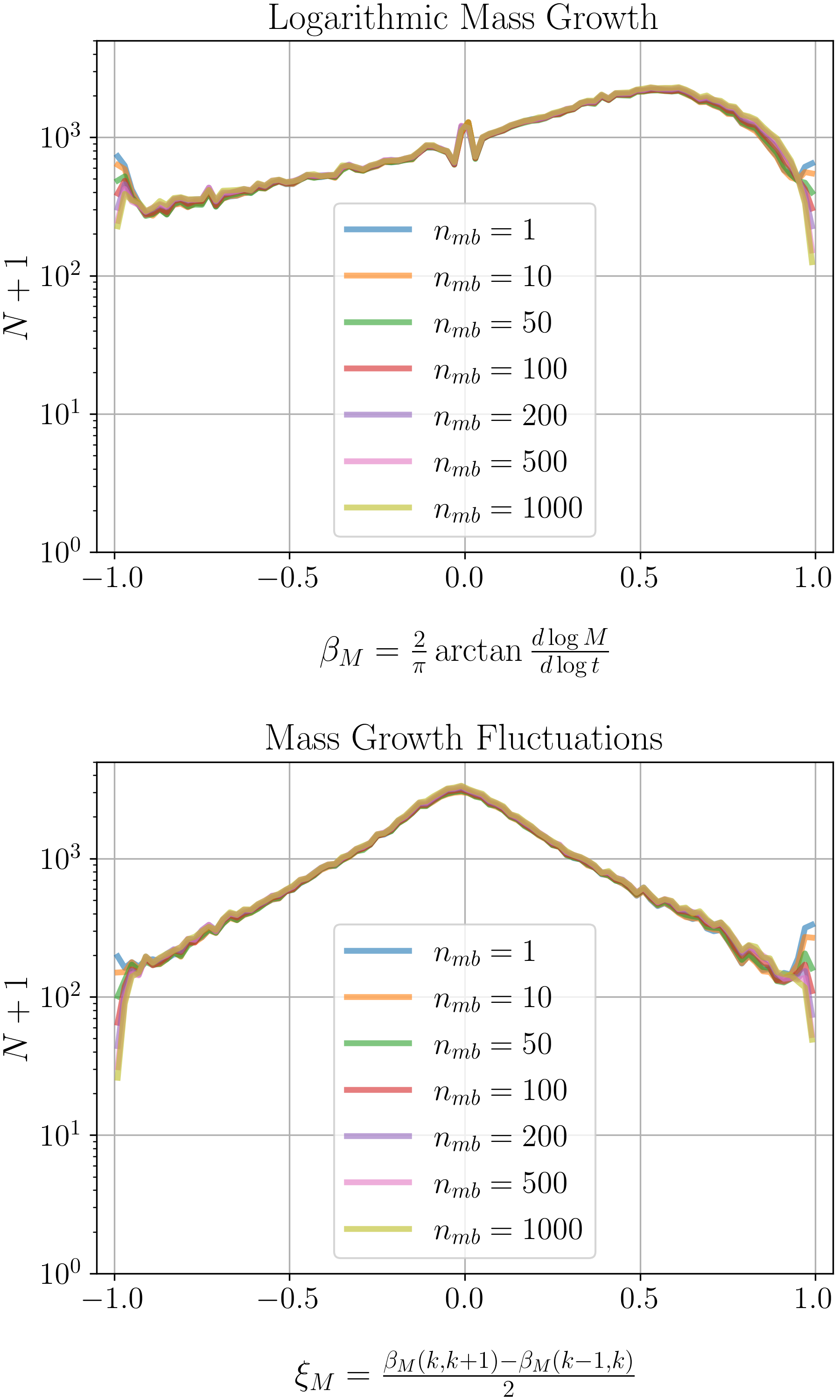}%
  \caption{ Histogram of the logarithmic mass growth (top) and
    histogram of the mass growth fluctuation (bottom) for all clumps
    (halo and sub-halo) detected in two (top) or three (bottom)
    consecutive snapshots of the simulation and with more than 200
    particles.  We compare these histograms for four different numbers of
    tracer particles $n_{\mathrm{mb}}$ as indicated in the legend.  In all
    cases, we used the \exc\ and \sad\ clump mass definitions.
  }%
  \label{fig:ntracers_masses}
\end{figure}

\begin{figure}
  \centering
  \includegraphics[width=.7\linewidth, keepaspectratio]
	{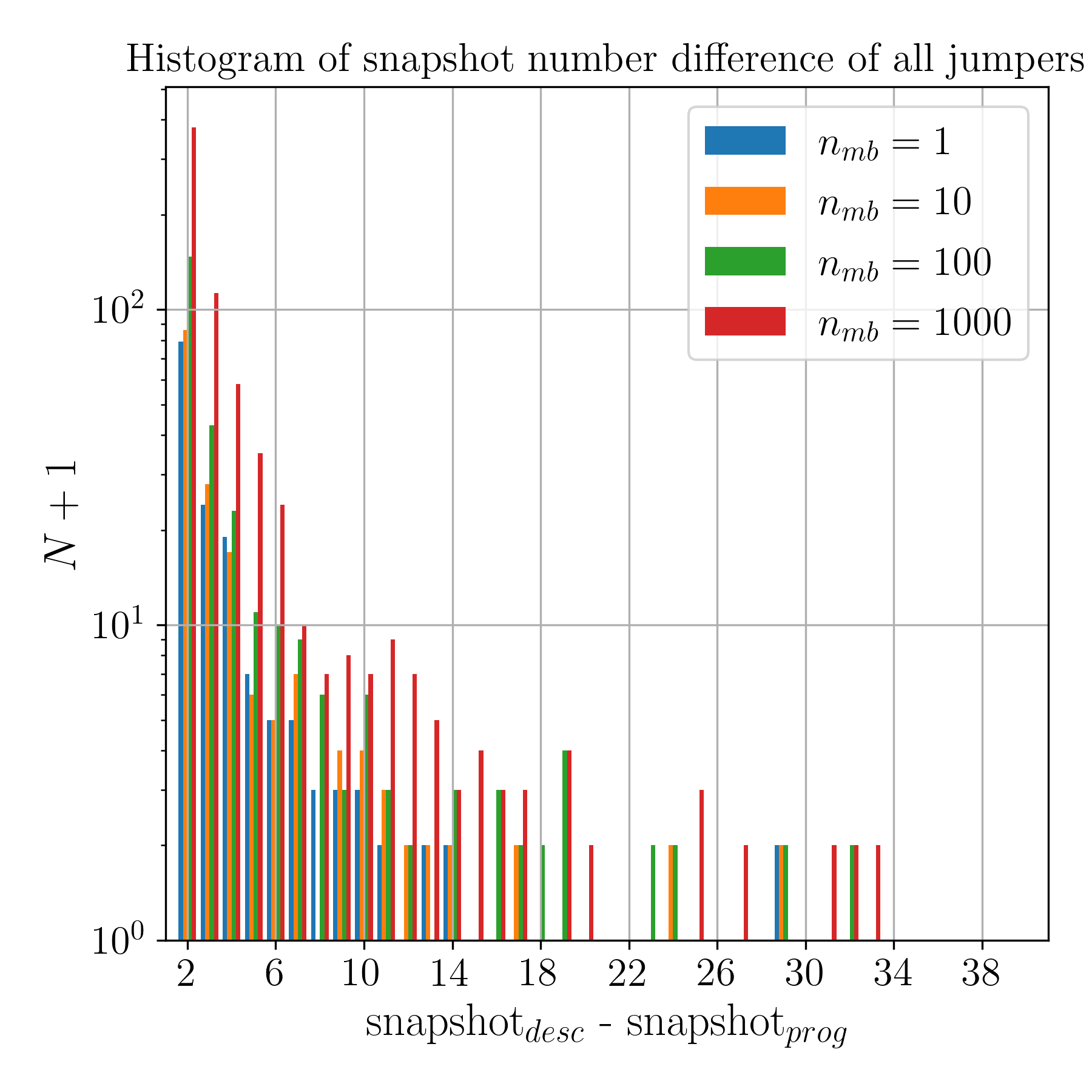}%
  \caption{ Histogram of the difference in snapshot numbers for
    jumpers, i.e.  progenitor-descendant links which are made across
    non-adjacent snapshots.  Only pairs where both the progenitor's
    and the descendant's masses exceed 200 particle masses were
    included in this plot.  We compare these histograms for four
    different numbers of tracer particles $n_{\mathrm{mb}}$ as indicated in
    the legend.  In all cases, we used the \exc\ and \sad\ clump mass
    definitions.
  }%
  \label{fig:jumper-distances}
\end{figure}

\begin{figure}
  \centering
  \includegraphics[width=.7\linewidth, keepaspectratio]
	{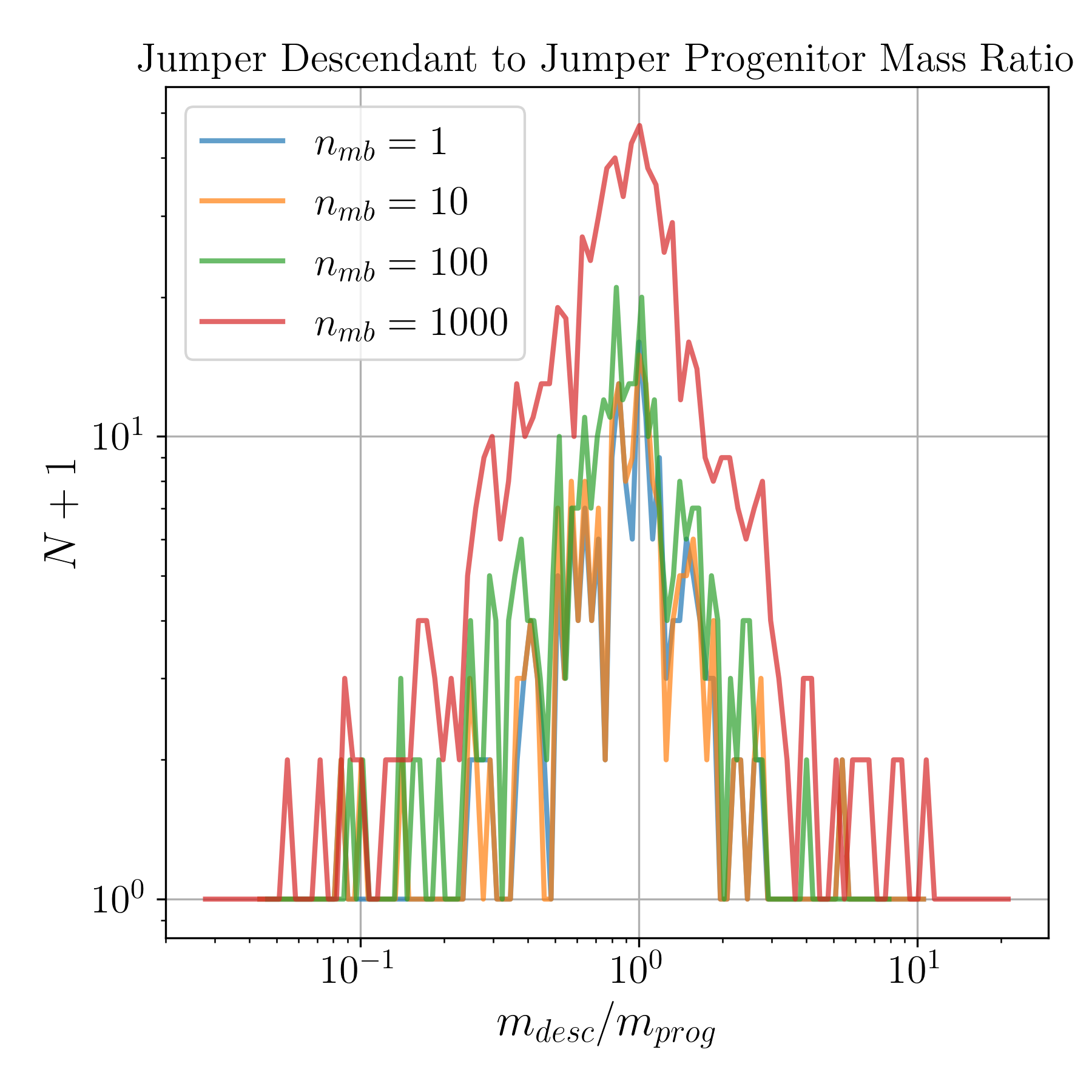}%
  \caption{ Histogram of the ratio of descendant mass to progenitor
    mass for jumpers, i.e.  progenitor-descendant links which are made
    across non-adjacent snapshots.  Only pairs where both the
    progenitor's and the descendant's masses exceed 200 particle
    masses were included in this plot.  We compare these histograms
    for four different numbers of tracer particles $n_{\mathrm{mb}}$ as
    indicated in the legend.  In all cases, we used the \exc\ and
    \sad\ clump mass definitions.
  }%
  \label{fig:jumper-mass-ratio}
\end{figure}

In this section, we study the effect of varying the number of tracer
particles on our various diagnostics of tree quality.
Even though we performed the simulations with
up to 1000 tracer particles, the mass threshold for clumps was always
kept constant at 10 particles. The number of tracer particles per clump
is an upper limit, not a lower limit. For clumps that contain less than
$n_{\mathrm{mb}}$, this means that they will be traced by every single
particle they consist of. In effect, we expect that this assigns greater
weight to clumps which are more massive than $n_{\mathrm{mb}}$ to be identified
as the main progenitor, and should hopefully decrease extreme
mass growths and mass fluctuations. To illustrate, consider for example the
merging of two clumps with unequal masses, where
all of the tracer particles of both clumps are found inside the resulting
merged descendant. Raising the number of tracer particles above the less
massive clump's particle number in this scenario means that the number of
its tracer particles inside the descendant will remain constant, while
the number of tracer particles stemming from the more massive clump will
increase, and thus raising its merit to be the main progenitor. This is
also the desired outcome.
Should however both clumps have masses above $n_{\mathrm{mb}}$ particle masses,
then our inclusion of the clump masses in the merit function should
nevertheless find the more massive clump to be the main progenitor if it
had a mass closer to the resulting descendants mass. So we expect that
increasing the number of tracer particles should enhance this effect,
and hence lead to at least as smooth mass growths and fluctuations.

We show in
Table~\ref{tab:ntracers} the average number of branches and the
average length of the main branch for all our detected clumps at $z=0$
organised in different mass bins. We see that the effect of the number
of tracer particles used is quite mild. Even with as few as one tracer
particle do we manage to recover the correct average main branch
length. This is also true for small mass haloes, although with a
slightly reduced accuracy. This also validates our orphan particle
technique to track temporary merger events.

On a closer look, the average number of branch is however more
affected by the number of tracer particles. If we look at the two highest
mass bins for clumps in the two bottom rows of Table~\ref{tab:ntracers},
we can see that the average number of branches converges towards
the values of 1000 tracer particles used, and is only slightly lower in
the cases when 200 or 500 tracer particles were used. Comparing these
converged results to the ones with only one tracer particle,
we loose $\sim 30\%$ of the average number of branches, which are
links associated to merger events.  This
can be easily explained by the fact that too few tracer particles
cannot be distributed across enough descendant candidates to identify
potential links. It appears that using 100 tracer particles is enough
to recover most of the otherwise broken links. These
conclusions remain the same after looking at the histogram of the
number of branches and the histogram of the main branch length for the
same clumps in Figure~\ref{fig:ntracers_mbl_nbranch}. Here again, we
see the peak of the histogram of the number of branches being shifted
to the right when increasing the number of tracer particles. We also
see that using 100 tracer particles seems enough to almost recover the
correct distribution.

We now examine the effect of the number of tracers on the mass growth
(and on the mass growth fluctuations) of all our detected clumps
within the entire redshift range. We see in
Figure~\ref{fig:ntracers_masses} that the effect is very weak, except
for the extreme cases $\beta_M \simeq \pm 1$ and $\xi_M \simeq \pm 1$,
corresponding to spurious links in the merger tree. These extreme mass
growth cases correspond to broken links due to the
small number of tracer particles. Here again, using more than 100
tracer particles seem to get rid of most of these spurious cases.
Note that we include in these histograms only clumps with more than
200 particles.

We now study in details another spurious effect of our merger tree
algorithm, shared by many other merger tree code in the literature,
namely dead tree branches.  A dead branch arises when no descendant
could have been identified after a certain redshift, even after
looking for all subsequent snapshots using the corresponding orphan
particle. Such an event is called a ``Last Identifiable Descendant In
Tree'' or LIDIT.  When such a case occurs, it is customary to prune
the corresponding tree from the tree catalogue. When not enough tracer
particles are used, we expect such spurious dead links to
appear. Table~\ref{tab:ntracers-pruning} shows the statistics of these
LIDITs (or tree pruning events), which confirms that the number of
LIDITs decreases strongly when using more and more tracer
particles. We also show in the same table the typical and maximum mass
of the LIDITs. Interestingly, the maximum mass also strongly decreases
when using more tracer particles. When enough tracer particles are
used, we see that LIDITs are typically less massive than 200
particles. We believe they correspond to poorly resolved clumps that
are subject to all sorts of spurious numerical effects. We found that
taking all LIDITs into account, over 80\% of them were main haloes.
LIDITs containing more than 50 particles however were over 95\% sub-haloes.
This suggests that the number of very low mass LIDITs is dominated
by poorly resolved small clumps in low density environments, since
in overdense environments clumps wouldn't have been identified as main
haloes, but as sub-haloes instead. Conversely, with increased resolution
of the clumps, the overwhelming majority of LIDITs are sub-haloes, and as
such in overdense regions. We conclude that a
conservative resolution limit of 200 particles per clump removes all the
LIDITs from our catalogue, as long as one uses more than 200 tracer
particles.

We finally study a specific aspect of our merger tree algorithm,
namely the possibility to follow the temporary mergers of clumps that
travel through another clumps to emerges later as a distinct object.
We show in Table~\ref{tab:jumpers} the number of these temporary
mergers that we call ``jumpers'' as they represent links across
non-adjacent snapshots that we are able to ``repair'' using orphan
particles. On average, the number of jumpers increases with the
number of tracer particles. This is a similar behavior than for the
number of branches: More tracer particles allows more merger events to
be detected, and every merger event results in a new orphan particle.
More orphan particles allow more non-adjacent descendant candidates
to be found. We here also recommend to use 200 tracer particles as a
compromise between speed and proper detection of jumpers in the
simulation.

We show in Figure~\ref{fig:jumper-distances} the histogram of the
distance in time between the two non-adjacent snapshot of all jumpers
in our merger tree. We see that most jumpers have a distance of only 2
snapshots. They corresponds to clumps traversing another clump and
re-emerging a snapshot later as a distinct halo. We also see in
this histogram that the number of jumpers increases with the number of
tracer particles. But overall, the statistics of the distance between
jumpers is relatively robust, with only very rare cases with a
distance larger than 10 snapshots. Figure~\ref{fig:jumper-mass-ratio}
shows the histogram of the mass ratio between the jumper progenitor
and the jumper descendant.  As expected, it peaks at one, which means
that the mass of the clump that re-emerges in a later snapshot is
close to the mass of the clump that disappeared in an earlier
snapshot. Note that this is not due to the merit function, because for
jumpers we only use a single orphan particle to repair the link. This
is a clear sign that the same clump is identified before and after the
temporary merger.  We also see that the distribution is slightly
skewed toward mass ratio smaller than 1, with values always bounded
between 0.3 and 3. Only a few very rare cases show more extreme mass
ratios. This means that clumps hosting our orphan particles either
preserve their mass (over 2 snapshots) or loose mass (on average),
usually when the time between non-adjacent snapshots increases.

In conclusion, we found that $n_{\mathrm{mb}} \simeq 200$ is a safe choice
to obtain robust results for our merger tree algorithm, in light of
the diagnostics we have used in this section.  We also recommend
adopting a conservative mass threshold of 200 particles per clumps to
get rid of a few rare spurious dead branches that would need to be
pruned from the halo catalogue anyway.

%% file: tables/ACACIA/saddle_nosaddle_global_stats.tex
\begin{table*}
	\caption{
        Average data for all clumps at $z=0$ depending on whether to consider particles which might wander off into another clump as bound (\nosad) or not (\sad).
        The results shown are for the \exc\ mass definition, which show no significant difference to when the \inc\ mass definition is used.
        \label{tab:saddle_nosaddle}
    }
        
%

	{\small 
		\begin{tabular}[c]{l | p{2.8cm} | p{2.8cm} |}
													&	 \sad\  &   \nosad\ \\ 
			
			\hline
			total clumps	 													&	 17115	& 	18247 	\\			
			median number of particles in a clump 	&	 77			& 	85 		\\			
			\hline
			average main branch length & & \\
			clumps with < 100 particles			&	14.7	& 	13.0 	\\			
			clumps with 100-500 particles		&	31.4	& 	31.0 	\\			
			clumps with 500-1000 particles	&	37.5	& 	37.3 	\\			
			clumps with > 1000 particles		&	40.7	& 	40.9 	\\			
			\hline
			average number of branches & & \\
			clumps with < 100 particles			&	1.2		& 	1.1 	\\			
			clumps with 100-500 particles		&	2.8		& 	2.8 	\\			
			clumps with 500-1000 particles	&	6.2		& 	6.7 	\\			
			clumps with > 1000 particles		&	25.4	& 	26.1 	\\				
			\hline	
		\end{tabular}
	}
\end{table*}

%% file: tables/ACACIA/ntracers_global_stats.tex
\begin{table*}
  \caption{Average length of main branch and average number of
    branches for clumps in different mass bins at $z=0$ and for
    varying numbers of clump tracer particles $n_{\mathrm{mb}}$.}
  \label{tab:ntracers}

  {\small
    \begin{tabular}[c]{l | p{0.9cm} | p{0.9cm} | p{0.9cm} | p{0.9cm} | p{0.9cm} | p{0.9cm} | p{0.9cm} |}
      $n_{\mathrm{mb}}=$				&	1 		& 	10 		& 	50 		& 	100 	& 200 	&
500 		& 1000 \\
      \hline
      Average main branch length & & & & & & &\\
      clumps with < 100 particles		&	24.2	& 	24.3	& 	23.6	&	23.4 	& 23.2 	& 22.9 	& 22.7  \\
      clumps with 100-500 particles		&	50.4	& 	50.1	& 	49.5	&	49.1 	& 48.8 	& 48.8 	& 48.8  \\
      clumps with 500-1000 particles		&	55.2	& 	54.9	& 	53.3	&	54.1 	& 54.7 	& 54.3 	& 54.2  \\
      clumps with > 1000 particles		&	56.7	& 	54.9 	& 	52.3	&	52.9 	& 54.0 	& 55.8 	& 56.4  \\
      \hline
      Average number of branches & & & & & & &\\
      clumps with < 100 particles		&	1.2	& 	1.3	& 	1.3	&	1.3 	& 1.3  & 1.4 	& 1.4  \\
      clumps with 100-500 particles		&	2.7	& 	3.0	& 	3.3	&	3.3 	& 3.4  & 3.6 	& 3.6  \\
      clumps with 500-1000 particles		&	6.6	& 	7.2	& 	8.1	&	8.2 	& 8.7  & 8.9 	& 9.1  \\
      clumps with > 1000 particles		&	20.4& 	25.2& 	27.3&	28.6 	& 29.5 & 30.4 	& 31.4 \\
      \hline
    \end{tabular}
  }


\end{table*}

%% file: tables/ACACIA/ntracers_pruning.tex
\begin{table*}

  \caption{Number of dead trees pruned from the merger tree catalogue
    for varying numbers of tracer particles $n_{\mathrm{mb}}$ throughout
    all snapshots.  ``LIDIT'' is an abbreviation for ``last
    identifiable descendant in tree''.  For a LIDIT, no descendant
    could have been identified throughout the simulation and
    consequently the corresponding tree is considered dead and pruned
    from the merger tree catalogue. LIDITS are obviously a spurious
    feature of the merger tree algorithm.
    \label{tab:ntracers-pruning}
  }

  {\small
    \begin{tabular}[c]{l | p{0.8cm} | p{0.8cm} | p{0.8cm} | p{0.8cm} | p{0.8cm} | p{0.8cm} | p{0.8cm} |}
      $n_{\mathrm{mb}}=$				&	1 		& 	10 	& 	50 	& 100 	& 200 	& 500 	& 1000
      \\
      \hline
      dead trees pruned from tree catalogue	&	23617	&	15438	&	14467	& 14433 & 14432 & 14432 & 14433
      \\
      highest particle number of a LIDIT		&	6418	&	674		&	182		&	182 	& 182 	& 182 	& 182
      \\
      median particle number of a LIDIT			&	19		&	20		&	20		&	20 		& 20 		& 20 		& 20
      \\
      LIDITs with >100 particles pruned 		&	493		&	61		&	32		&	28 		& 26 		& 26 		& 26
      \\
      \hline
    \end{tabular}
  }
\end{table*}

%% file: tables/ACACIA/ntracers_jumpers.tex
\begin{table*}
  \caption{Number of ``jumpers'' (progenitor-descendant links found
    across non-adjacent snapshots) during the entire simulation for
    varying number of tracer particles $n_{\mathrm{mb}}$.  }
  \label{tab:jumpers}
  {\small

\begin{tabular}[c]{l | p{0.9cm} | p{0.9cm} | p{0.9cm} | p{0.9cm} | p{0.9cm} | p{0.9cm} | p{0.9cm} |}
    $n_{\mathrm{mb}}=$                   & 1           & 10          & 50          & 100         &
    200         & 500         & 1000        \\
\hline
    Total Jumpers                   &    14738    &    15500    &    17654    &    18995    &
    20505    &    20305    &    20407    \\
\hline
    Jumper Progenitors & & & & & & & \\
    clumps with <  100 particles    &    13372    &    14064    &    15696    &    16448    &
    17233    &    16795    &    16779    \\
    clumps with  100- 500 particles &     1295    &     1353    &     1833    &     2383    &
    3024    &     3153    &     3214    \\
    clumps with  500-1000 particles &       52    &       60    &       87    &      121    &
    176    &      251    &      278    \\
    clumps with > 1000 particles    &       19    &       23    &       38    &       43    &
    72    &      106    &      136    \\
\hline
\end{tabular}

} 
\end{table*}

%% file: main/ACACIA/ACA4-mergertree_testing_other_codes.tex
\section{Tree Statistics Using \citet{SUSSING_HALOFINDER} Selection Criteria}\label{chap:performance_comparison}

\begin{table}
\centering
\caption{
	Comparison of simulation and evaluation parameters used in this work and of A14, where the parameters of the latter have been converted using $h = 0.704$.
	$m_m$ is the mass threshold for main haloes, $m_s$ is the mass threshold for sub-haloes.
	\label{tab:parameter-comparison}
}
	\begin{tabular}[c]{l l l}
																	&	This work		&	A14 \\
		\hline
		particle mass	[$10^9 \msol$]	&	$1.55$			& $1.32$						\\
		particles used								& $256^3$			& $270^3$ 					\\
		box size [Mpc/h]							& $62.5$			& $62.5$						\\
		snapshots until $z = 0$				& 62					& 62								\\
		$m_m$ [$10^{12} \msol$]				&	$1.35$			& $1.12$ - $1.37$		\\
		$m_s$ [$10^{11} \msol$]				&	$4.03$			& $4.26$ - $9.72$		\\
		\hline
	\end{tabular}
\end{table}

\begin{figure}
	\centering
	\includegraphics[width=.7\linewidth,
	keepaspectratio]
	{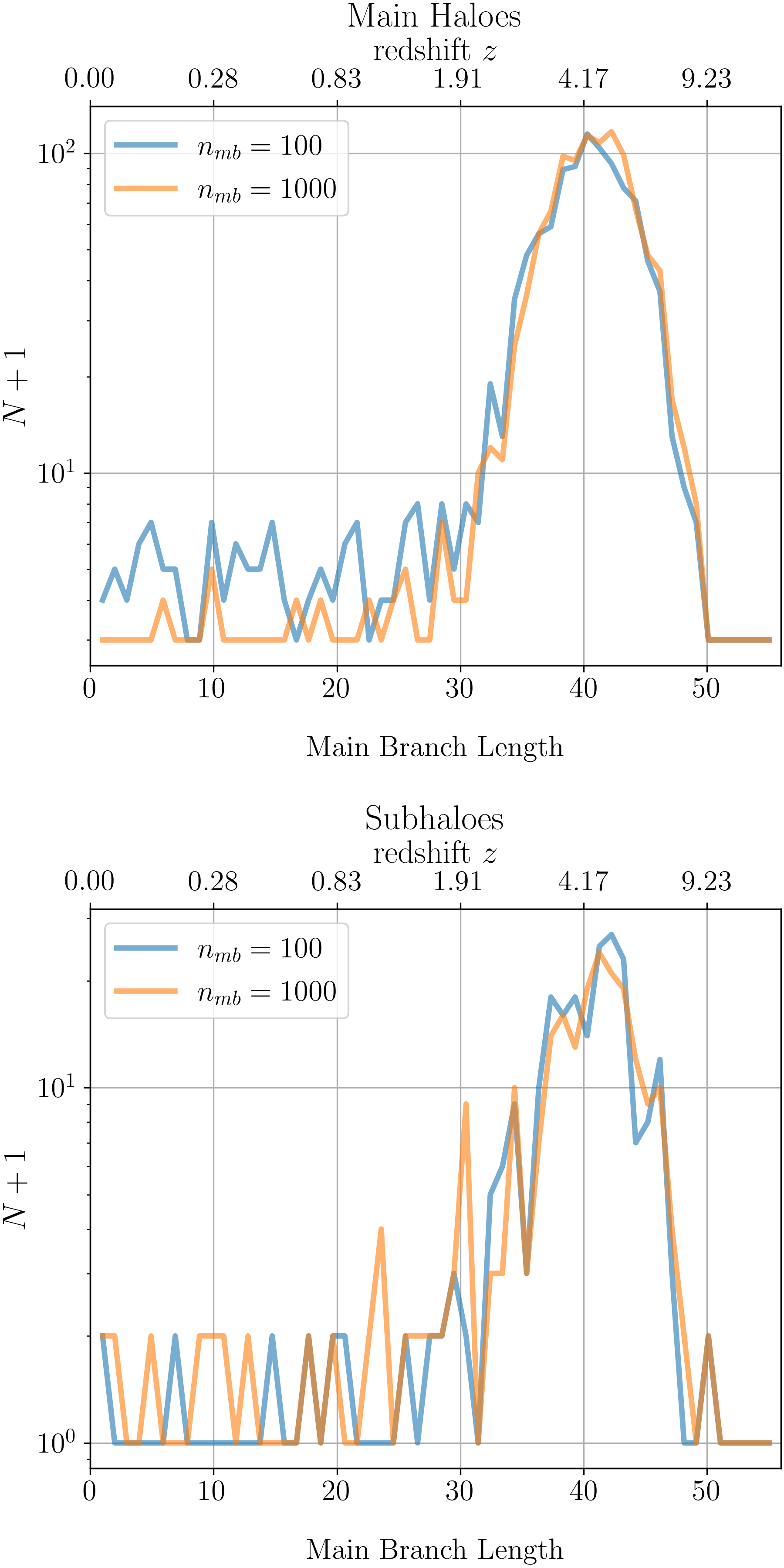}%
	\caption{
		Histograms of the length of the main branch.
		The top plot shows the length of the 1000 most massive haloes at $z = 0$, the bottom plot shows
		the length of the 200 most massive sub-haloes for $n_{mb} = 100$ and $1000$ tracer particles
		per clump.
	}%
	\label{fig:sussing-branch-lengths}
\end{figure}

\begin{figure}
	\centering
	\includegraphics[width=.7\linewidth, keepaspectratio]
	{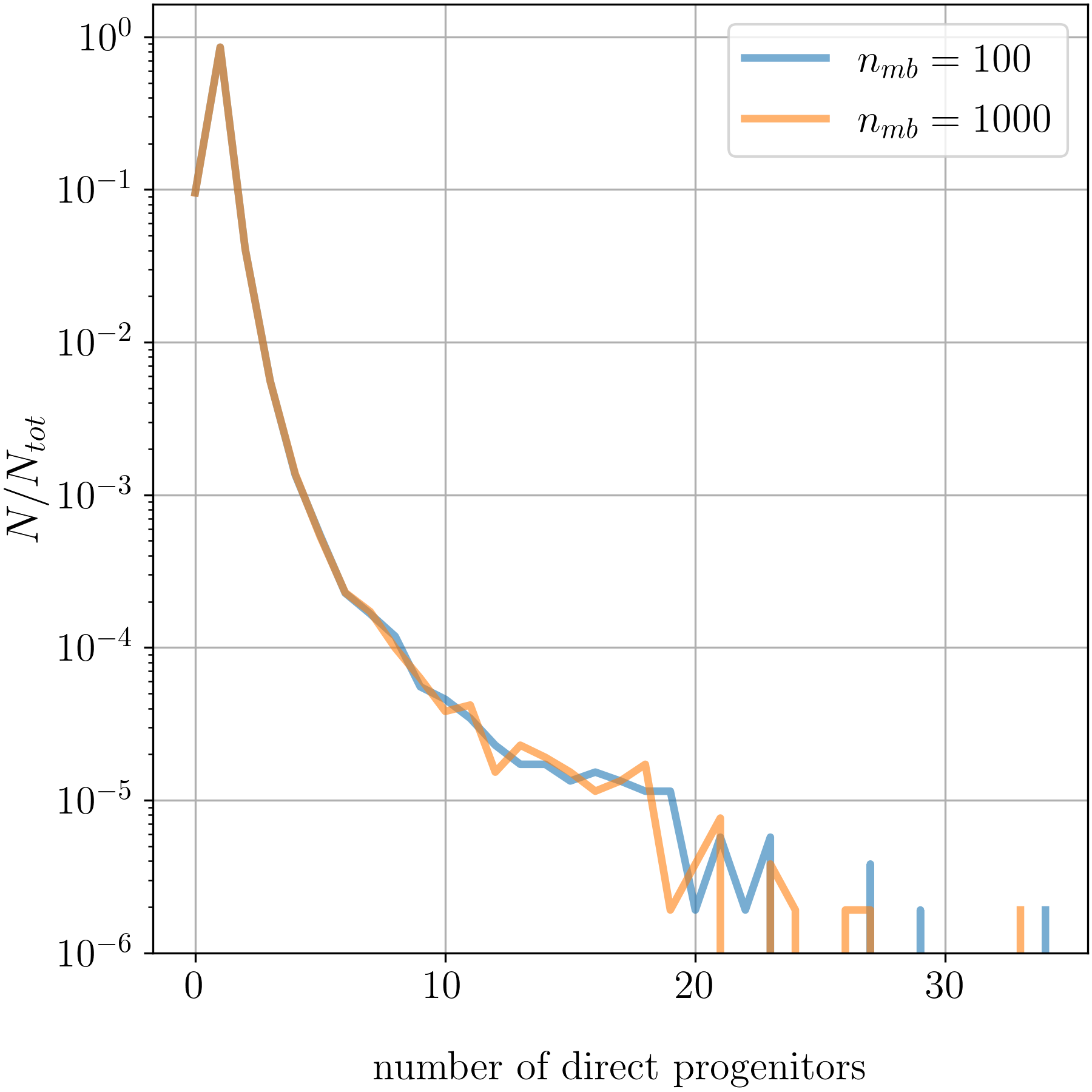}\\%
	\caption{
		Histogram of the number of direct progenitors for all clumps from $z = 0$ to $z = 2$ for $n_{mb} = 100$ and $1000$ tracer particles per clump.
		The histogram is normalized by the total number of events found.
	}%
	\label{fig:sussing-branching-ratio}
\end{figure}

\begin{figure}
	\centering
	\includegraphics[width=\textwidth, keepaspectratio]
	{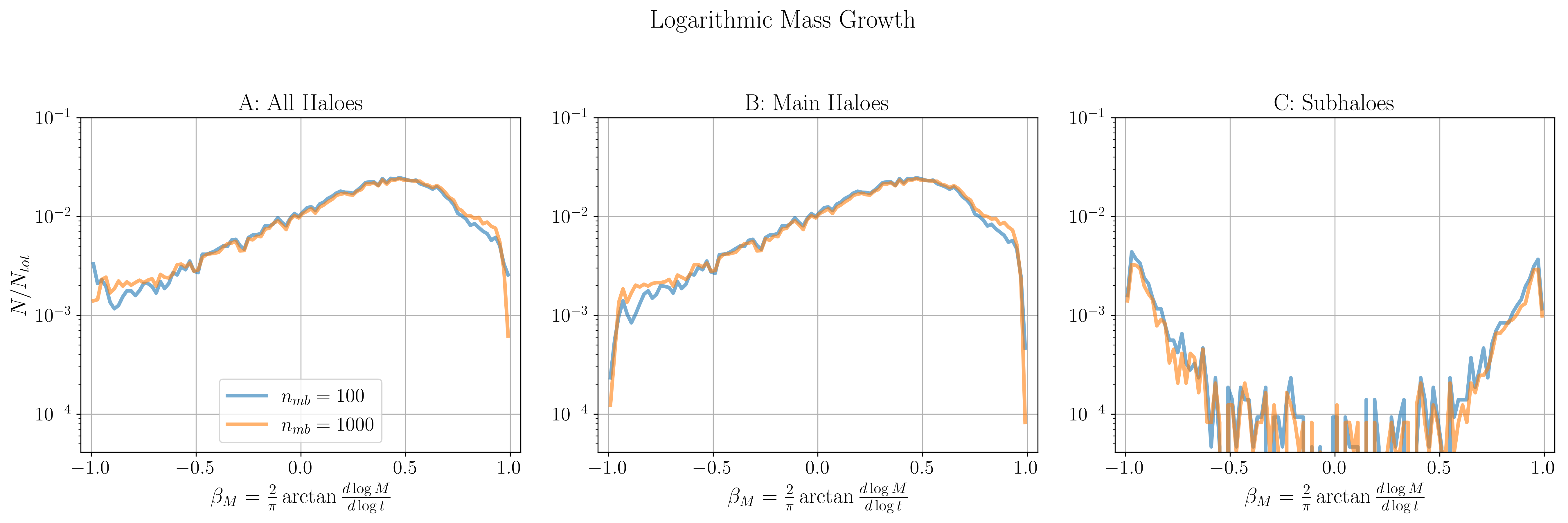}%
	\caption{
		Logarithmic mass growth for haloes and sub-haloes satisfying the mass thresholds.
		Group $A$ contains clumps that are either haloes or sub-haloes in consecutive snapshots $k$ and $k+1$ with masses $m \geq m_{m}$.
		Group $B$ contains clumps that are only haloes in two consecutive snapshots with mass above $m_{m}$, group $C$ contains only clumps that were sub-haloes in two consecutive snapshots with mass greater than $m_{s}$.
		The histogram is normalized by the total number of events found for group $A$.
	}%
	\label{fig:sussing-mass-growth}
\end{figure}

\begin{figure}
	\centering
	\includegraphics[width=\textwidth, keepaspectratio]
	{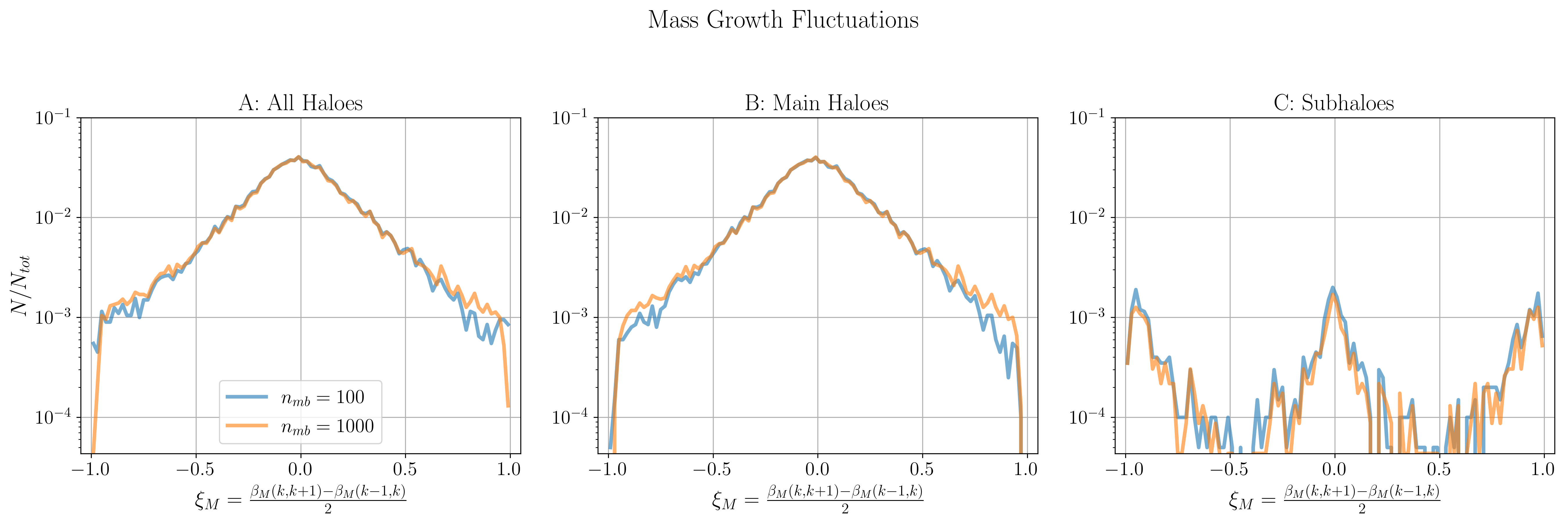}\\%
	\caption{
		Histogram of mass growth fluctuations for haloes and sub-haloes satisfying the mass thresholds.
		Group $A$ contains clumps that are either haloes or sub-haloes in three consecutive snapshots with masses $m \geq m_{m}$.
		Group $B$ contains clumps that are only haloes in three consecutive snapshots with mass above $m_{m}$, group $C$ contains only clumps that were sub-haloes in three consecutive snapshots with mass greater than $m_{s}$.
		The histogram is normalized by the total number of events found for group $A$.
	}%
	\label{fig:sussing-mass-fluct}
\end{figure}

In this Section, the merger tree statistics introduced in Section \ref{chap:tests} when following
the selection criteria that are used in \citet{SUSSING_HALOFINDER} (A14 from here on) are presented.
Ideally, \acacia should be tested on the same datasets and halo catalogues used in the
Comparison Project to enable a direct comparison to the performance of other merger tree codes.
However, since \acacia was designed to work on the fly, using it as a post-processing
utility would defeat its purpose.
Furthermore, \acacia is not necessarily compatible with other definitions of haloes and
sub-haloes. But most importantly, we also want to demonstrate that the halo finder \phew\ can be
used to produce reliable merger trees. So instead, the tests are performed on our own datasets and
halo catalogues, which are described in section \ref{chap:testing_methods}. A comparison of the used
parameters of our simulations and the ones used in A14 is given in Table
\ref{tab:parameter-comparison}. In the following, the results for $n_{mb} = 100$ and $n_{mb} = 1000$
are shown. Like before, when the influence of the number of tracer particles was investigated, the
\sad\ parameter and the \exc\ mass definition were used.

The difference to the results presented in the previous sections is that the mass thresholds are
set such that only the 1000 most massive main haloes and only the 200 most massive sub-haloes at $z = 0$ are included. This gives effective mass threshold $m_{m} = 1.35 \times 10^{12} \msol$ and
$m_{s} = 4.0 \times 10^{11}\msol$, which are on one hand comparable to the mass thresholds applied
in A14 (Table~\ref{tab:parameter-comparison}), but already show differences in the resulting halo
catalogue. \phew\ finds a mass threshold for main haloes that is close to the upper limit found in
A13, but a lower mass threshold for sub-haloes. This is consistent with the fact that the \sad\
parameter was used: Unbound particles are passed on to substructure that is higher up in the
hierarchy, and the unbinding is repeated until the top level, which are the main haloes, is reached.
The more strict unbinding criterion tends to assign more particles to the main haloes and remove
them from sub-haloes, which is reflected in the mass thresholds. Indeed, using the \nosad\ parameter
instead leads to $m_m = 1.27 \times 10^{12}\msol$ and $m_s = 1.92 \times 10^{12}\msol$.

The length of the main branches for haloes and sub-haloes individually are shown in Figure
\ref{fig:sussing-branch-lengths}. Compared to Figure 3 of A14, we note the following similarities
and differences:
\begin{enumerate}
	\item \acacia finds some main haloes with short (< 10) main branches.
		In A14, his only happens for the \codename{JMerge} and \codename{TreeMaker} tree builders
		regardless	of the halo finder employed, and the \codename{MergerTree}, \codename{SubLink},
		and \codename{VELOCIraptor} tree makers for \codename{AHF} and \codename{Subfind} halo finders.
	\item Like in nearly all cases in A14, the distribution of main branch lengths for
		main haloes peaks at high numbers and the bulk of the distribution is about 20 snapshots
		wide. The peak of the main branch length distribution of \acacia is at 40,
		while in most cases in A14, it's around 45 with the exception of the \codename{AHF}
		halo finder. This indicates that \acacia and \phew\ result in on average
		somewhat smaller main branch lengths than other codes. However, we also find the
		maximal main branch length of 50, while nearly all combinations of
		halo finders and tree makers in A14 find higher values. Both these differences can be
		explained by the slightly lower resolution that we used in our simulations, where we
		found no identifiable clumps before snapshot 11, which corresponds to a maximal main
		branch length of 51.
	\item The main branch lengths for sub-haloes also follow the trends noted above. Additionally,
		the distribution is flat for main branch lengths smaller than 30, which is not the case
		for \codename{Rockstar} and \codename{AHF} sub-haloes in A14 for nearly all tree builders. Instead,
		their main branch lengths peak again around unity.
\end{enumerate}
In summary, regarding the main branch lengths of main haloes, \phew\ and \acacia perform
comparably to \codename{Subfind} and \codename{Rockstar} halo finders and \codename{MergerTree},
\codename{TreeMaker}, and \codename{VELOCIraptor} tree builders. Concerning sub-haloes, the results are
most closely to \codename{Subfind} sub-haloes and the same tree builders as for the main haloes. This
is not surprising, because \codename{Subfind} employs a similar definition of substructure being
arbitrarily shaped self-bound structure that is truncated at the isodensity contour that is
defined by the density saddle point between the sub-halo and the main halo.

In Figure \ref{fig:sussing-branching-ratio} the number of direct progenitors for all clumps between
$z = 0$ and $z = 2$ are shown. Comparing to Figure 5 of A14, \acacia gives very comparable
results: $\sim 10^{-1}$ haloes have no direct progenitor, almost all have one, and the distribution
follows an exponential decay with the maximal number of direct progenitors lying around 20-25, save
for a very few outliers. Many tree makers and halo finders in A14 exhibit the same kind of
behaviour, particularly so for the \codename{AHF}, \codename{Subfind}, and \codename{Rockstar all} halo
finders in Figure 5 of A14.

For the logarithmic mass growth (Figure \ref{fig:sussing-mass-growth}) and the mass growth
fluctuations (Figure \ref{fig:sussing-mass-fluct}), the statistics are separated into three groups.
Group $A$ contains clumps that are either haloes or sub-haloes in \emph{consecutive} snapshots $k$
and $k+1$ with masses greater than the mass threshold $m_m$ in both snapshots. Group $B$ contains
clumps that are exclusively main haloes in two consecutive snapshots with mass above $m_{m}$, group
$C$ contains only clumps that were sub-haloes in two consecutive snapshots with mass greater than
$m_{s}$. We follow clumps of the $z = 0$ snapshot along the main branch only.

The logarithmic mass growth resulting from \acacia follows the general trend that the tree
makers in A14 exhibit too. The growth for groups $A$ and $B$ increases steadily and peaks around
$\beta_M \sim 0.5$, where the peak is $\sim2 \times 10^{-2}$. For $n_{mb} = 100$, the extreme mass
loss  with $\beta_M = -1$ increases for group $A$, which is an undesirable property, but is also
exhibited by \codename{Sublink} in A14. For $n_{mb} = 1000$, it drops to about $10^{-3}$ ($10^{-4}$
for group $B$), which is comparable behaviour to \codename{MergerTree}, \codename{Sublink}, and
\codename{VELOCIraptor}, particularly so in combination with \codename{AHF} and \codename{Subfind} halo
finders. Group $C$, containing only mass growths of clumps that have been sub-haloes in two
consecutive snapshots, shows a distribution peaking around extreme mass growths $\beta_M
\rightarrow \pm 1$ at $\sim 5 \times 10^{-3}$, which can again be seen in A14 for almost all tree
makers, albeit not for all halo finders. More noticeably, almost no sub-haloes are found with $-0.5
< \beta_M < 0.5$ with \phew\ and \acacia in Figure \ref{fig:sussing-mass-growth}.
This is due to the fact that once a halo is merged into another, it quickly loses its
outer mass due to the strict unbinding method used here. However, the distribution that
\acacia finds displays some differences with respect to the results in A14. Firstly,
we find almost no mass growth with $-0.5 < \beta_M < 0.5$, similar only to the results of
\codename{JMerge} in A14. This is partially due to the strict selection criteria used for this
analysis: We only include clumps that are classified as sub-haloes in two consecutive snapshots and
satisfy the mass threshold $m_s$ in both snapshots as well. In particular, this excludes all
non-adjacent ``jumper'' links that we find, since we don't modify the halo catalogue like e.g.
\codename{ConsistentTrees}. Furthermore, we employ the \sad\ unbinding criterium, which strips more
particles from sub-haloes and assigns them to main haloes, leaving the halo catalogue with fewer
sub-haloes that satisfy the mass threshold. When we instead use the \nosad\
criterium, we find that the distribution is on average around $3 \times 10^{-4}$ for $-0.5 <
\beta_M < 0.5$, albeit noisy, which is in good agreement with most halo finders and tree builders
in A14.
Secondly, the distribution \acacia finds looks remarkably symmetric w.r.t. $\beta_M = 0$.
While e.g. \codename{MergerTree} and \codename{TreeMaker} trees with \codename{AHF}, \codename{Subfind},
and \codename{Rockstar} haloes find peaks close to $\beta \pm 1$ of similar height, they are also
always have distributions skewed towards mass losses $\beta_M < 0$ and the peaks at $\beta_M = -1$
higher than the one at $\beta_M = 1$.

We found that the reason why our distribution looks so
symmetrical is due to the particle unbinding method and the way sub-halo hierarchies are
established in \phew, similarly to what we have found to be a reason for the short main branch
lengths in Section~\ref{chap:varying_clump_mass_definition}. The hierarchy is determined by the
density of the density peak of each clump:
A clump with a lower peak density will be considered lower in the hierarchy of substructure. So in
situations where two adjacent sub-haloes have similarly high density peaks, their order in the
hierarchy might change in between two snapshots due to small changes. The unbinding algorithm then
strips the particles from the sub-haloes that have the lowest level in the hierarchy and passes it
on to the next level, amplifying the mass loss which these sub-haloes experience. If in the next
snapshot the order in the hierarchy for these two clumps are inverted, the clump which experienced
a mass loss previously will now experience a strong mass growth and vice versa. In
Figure~\ref{fig:sussing-mass-growth} such an oscillation over two snapshots will simultaneously add
a strong mass growth and a strong mass loss twice in place of a net smoother mass loss, leading to
the symmetry of the distribution. We verified that about 10$\%$ of strong mass growth events with
$\beta_M > 0.75$ are also accompanied by the respective sub-haloes increasing their level in the
hierarchy. Similarly, about 10$\%$ of strong mass loss events with $\beta_M < -0.75$ are
accompanied by the respective sub-haloes decreasing their level in the hierarchy.

The mass growth fluctuations (Figure \ref{fig:sussing-mass-fluct}) of \acacia share the
general trend with the ones from Figure 8 in A14, in that they peak around $\xi_M = 0$ and decrease
outwards towards $\xi_M = \pm 1$. In A14, in all cases groups $A$ and $B$ peak just below
$10^{-1}$, while our results peak around $4 \times 10^{-2}$. However, similarly to the results of
e.g. \codename{Sublink}, \codename{TreeMaker}, and \codename{VELOCIraptor} with the \codename{AHF} or
\codename{Subfind} halo finders, the distribution around $\xi_M \sim \pm 0.5$ drops to $\sim 5 \times
10^{-3}$, and then continues dropping below $10^{-4} - 10^{-3}$ at $\xi_M \sim \pm 1$.
Group $B$ shows a steeper drop around the extreme values $\xi_M \sim \pm 1$ compared to group $A$,
dropping below $10^{-4}$ at these values, similarly to the behaviour of many tree makers and halo
finders in A14.
The sub-halo group $C$ of this work shows three main peaks, around $-1$, $0$, and $1$.
These peaks also appear in the A14 results.
However, the peaks at the extreme values in A14 are lower than the ones of this work, while the
peaks around $0$ is higher. The reason why these peaks are so pronounced in our results is the same as for why the mass growths in Figure~\ref{fig:sussing-mass-growth} is remarkably symmetric
compared to others: it's sub-haloes and their respective sub-sub-haloes switching their order in the substructure hierarchy repeatedly and the particle unbinding algorithm stripping particles from the lower level substructure and assigning it to the higher level substructure.
The missing values around $\xi_M \sim \pm 0.5$ that were also seen in the mass growth in Figure
\ref{fig:sussing-mass-growth} remain unsurprisingly, and are mitigated if the \nosad\ unbinding
criterion is applied instead.
Similar distributions are obtained by e.g. the \codename{AHF} and \codename{Rockstar} halo finders in
combination with the \codename{MergerTree}, \codename{JMerge}, \codename{Sublink}, and \codename{TreeMaker}
tree builders.

In summary, we find that \acacia and \codename{PHEW} produce merger tree statistics which are
similar to what multiple other state-of-the-art codes find as well. The results coincide most
commonly with those of the \codename{AHF}, \codename{Rockstar}, and \codename{Subfind} halo finders in
combination with the \codename{MergerTree}, \codename{TreeMaker}, \codename{Sublink}, and
\codename{VELOCIraptor} tree builders. One notable difference in our result however is that we
recover more extreme mass growths and losses as well as fluctuations for sub-haloes due to the way
the sub-halo hierarchy is established by \phew, specifically in cases where a sub-halo and its
subsub-halo switch their order in the hierarchy between snapshots.

%% file: main/ACACIA/ACA5-SHAM_testing.tex
\chapter{Application of the Merger Tree Algorithm: Creating a Mock Galaxy Catalogue}
\label{chap:mock_catalogues}

Now that we  know the optimal parameters to create  a merger tree with
\acacia, we  use it to  generate a mock galaxy  catalogue.  We
summarize here the main data products generated by our code.
\begin{enumerate}
\item For  every snapshot of the  N body simulation, we  have the full
  clump catalogue  generated by \phew. Every  clump is uniquely
  classified as a main halo or as a sub-halo.
\item Every dark  matter particle is given the clump  index it belongs
  to, or zero if it belongs to the smooth background.
\item For  each clump, we follow  and store the index  of the $n_{mb}$
  most strongly bound particles, the position of the density peak, the
  clump  bulk  velocities,  centre  of mass,  mass,  and  other  clump
  properties.
\item For each clump, we store the index of the direct progenitors (in
  particular its  main progenitor) and  its peak mass over  its entire
  past formation history.
\item We augment our clump database with orphan particles, storing for
  each of  them the index  of the last  known main progenitor  and its
  peak mass.
\end{enumerate}
To generate  the mock  galaxy catalogue,  we use  the well-established
technique of  Sub-Halo Abundance  Matching (SHAM). This  technique was
introduced  more than  ten  years  ago as  a  surprisingly simple  and
accurate  method  to  populate  a pure  dark  matter  simulation  with
galaxies  with the  correct  clustering statistics
\citep{valeNonparametricModelLinking2006, shankarNewRelationshipsGalaxy2006,
conroyModelingLuminositydependentGalaxy2006}.

Although  several   implementations  of  SHAM  exist   in  the  recent
literature \citep{guoHowGalaxiesPopulate2010,
wetzelWhatDeterminesSatellite2010,
mosterConstraintsRelationshipStellar2010, trujillo-gomezGalaxiesLCDMHalo2011,
nuzaClusteringGalaxiesSDSSIII2013,
zentnerGalaxyAssemblyBias2014,
chaves-monteroSubhaloAbundanceMatching2016} we use here  the variant
based on the peak clump mass as a proxy for the stellar mass
\citep{reddickConnectionGalaxiesDark2013},  using          the
Stellar-Mass-to-Halo-Mass (SMHM) relation of 	\cite{behrooziAVERAGESTARFORMATION2013}.

Using the  peak clump mass  is believed  to mimick the  actual stellar
mass growth of a galaxy, first as a central galaxy when the host clump
was a main halo, then as a satellite galaxy when the halo was accreted
and became  a sub-halo.  After  infall, although the clump  mass might
decrease quickly due to interactions within the parent main halo, this
model assumes  that the  stellar mass in  the galaxy  remains constant
\citep{nagaiRadialDistributionGalaxies2005}.

Note that in  the SHAM methodology, the merger tree  algorithm plays a
central role.
\begin{enumerate}
\item In order to compute the clump peak mass, we need the entire mass
  growth past history.
\item  In order  to follow  galaxies even  when the  parent clump  has
  dissolved due  to numerical  overmerging, we  need to  follow orphan
  particles and their peak clump mass.
\end{enumerate}

Our parent DMO simulation uses $512^3 \simeq 1.3\times 10^8$ particles
and a box size of 100 comoving  Mpc with a particle mass resolution of
$m_p \simeq  3.1 \times  10^8\msol$.  The cosmological  parameters are
taken from  the 2015 Planck Collaboration  results \citep{planckcollaborationPlanck2015Results2016},
with  Hubble constant  $H_0  = 67.74$  km s$^{-1}$Mpc$^{-1}$,  density
parameters  $\Omega_m  =  0.309$,  $\Omega_\Lambda  =  0.691$,  scalar
spectral index  $n_s = 0.967$,  and fluctuation amplitude  $\sigma_8 =
0.816$.  The initial conditions  were created using the \codename{MUSIC}
code \citep{hahnMultiscaleInitialConditions2011}.   As explained  before, the density  threshold for
clump finding was  chosen to be 80 times the  mean background density,
$\bar{\rho} =  \Omega_m \rho_c$, and  the saddle threshold  for haloes
was set to  200$\bar{\rho}$, where $\rho_c = \frac{3  H_0^2}{8 \pi G}$
is the cosmological critical density.

As a  first application of our  merger tree code, we  will now compute
the two point correlation functions and the average radial profiles of
galaxy  clusters in  our simulation.  We will  compare our  results to
observational data, and demonstrate that  this is only when we include
orphan  galaxies  that   our  results  are  in   good  agreement  with
observations.

\section{The Stellar Mass Correlation Function}\label{chap:correlation}

\begin{figure}
  \centering
  \includegraphics[width=.7\linewidth, keepaspectratio]
	{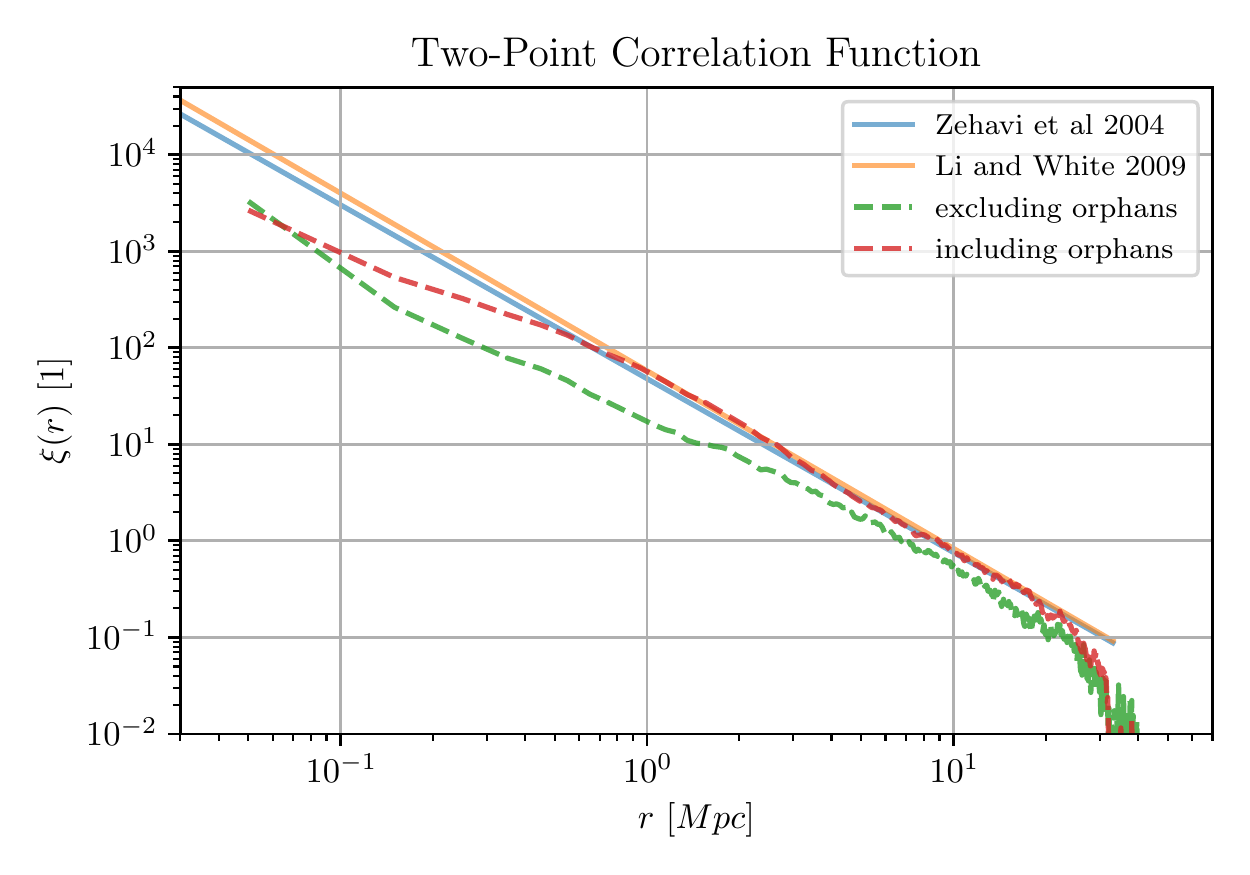}%
  \caption{The  predicted stellar  mass  2-point correlation  function
    (2PCF) $\xi(r)$ of our SHAM  model, including and excluding orphan
    galaxies, compared to  the power law fits of the  observed 2PCF in
    \citet{liDistributionStellarMass2009} and \citet{zehaviDeparturesPowerLaw2004}.
  }%
  \label{fig:correlations}
\end{figure}

The  stellar mass  two-point correlation  function (2PCF)  $\xi(r)$ is
computed via  inverse Fourier transform  of the power  spectrum $P(k)$
\citep[e.g.][]{moGalaxyFormationEvolution2010},
which itself  can  be obtained  from the  Fourier
transform    of   the    stellar   mass    density   contrast    field
$\delta(\mathbf{r})$:
\begin{align}
  \delta_\mathbf{k} = \frac{1}{V}\int e^{i\mathbf{kr}} \delta(\mathbf{r}) \de ^3 \mathbf{r}
\end{align}
with
\begin{align}
	\delta(\mathbf{r}) = \frac{\rho(\mathbf{r})}{\langle \rho(\mathbf{r}) \rangle } - 1
\end{align}
Where $\rho(\mathbf{r})$ is the galaxy  stellar mass density field and
$\langle \rho(\mathbf{r}) \rangle$ is  the corresponding mean density,
$V = L^3$ is the volume of our large box on which the density field is
assumed  periodic,  and  $\mathbf{k}=\frac{2\pi}{L}(i_x,  i_y,  i_z)$,
where  $i_x,  i_y,  i_z$  are  integers.   The  Fourier  transform  is
performed using  the FFTW library \citep{frigoDesignImplementationFFTW32005}.
The power spectrum $P(k)$ and the 2PCF $\xi(r)$ are given by
\begin{align}
  P(k)    &= V \langle |\delta_\mathbf{k}|^2 \rangle \\
  \xi(r)  &= \frac{1}{(2\pi)^3} \int e^{-i\mathbf{kr}} P(k) \de^3 \mathbf{k}
\end{align}
The simulation box is divided in  a uniform grid of $1024^3$ cells and
the  stellar mass  is  deposited  on the  grid  using a  cloud-in-cell
interpolation scheme.  The cloud-in-cell scheme consists  of assigning
each  galaxy a  cubic  volume  (``cloud'') the  size  of  a grid  cell
centered on the galaxy's position.  The galaxy stellar mass is assumed
to be uniformly distributed within the  cloud, and is deposited on the
uniform grid cells according to the  volume fraction of the cloud that
resides within each cell.

We only  include galaxies with  masses above $10^9 \msol$.   Using our
adopted SMHM relation,  these galaxies are hosted in  haloes with mass
larger than  $\sim 10^{11} \msol$,  or more than 300  particles.  This
threshold  ensures that  we  only  use well  resolved  clumps for  our
analysis, as discussed in the previous sections.

The predicted 2PCF $\xi(r)$ is shown in Figure \ref{fig:correlations},
and is  compared to  the observational  results of  \cite{liDistributionStellarMass2009} and
\cite{zehaviDeparturesPowerLaw2004}.   Note  that  the observed  galaxy  catalogue  is
presented as  complete down  to $10^8 \msol$,  one order  of magnitude
smaller than our  simulated catalogue.  To highlight  the influence of
orphan galaxies,  we have  computed the predicted  2PCF both  with and
without  orphan  galaxies.   Including   orphan  galaxies  produces  a
correlation function in much  better agreement with observations.  Our
theoretical 2PCF obtained  reproduces the observed power  law fit over
two orders  of magnitude  in scale  of $r\sim 0.3  - 25$~Mpc.   On the
smallest  scales, below  0.3~Mpc,  our predictions  are  likely to  be
affected by our limited mass resolution and the grid resolution which
is used for the Fourier Transform ($\sim 0.1$~Mpc per cell). We have
verified that using a lower mass threshold for the stellar masses of
galaxies makes no visible difference in the resulting correlation function.

The  role  of  the  orphan   galaxies  is  particularly  important  on
intermediate scales $\sim  0.2$ Mpc $<\ r \ <  2$ Mpc.  This behaviour
is explained by the fact that  orphan galaxies are located within host
haloes, thus contributing to the  correlations at small distances, the
so-called 1-halo term.  Our conclusion are in agreement  with those of
\cite{campbellGalaxyClusteringCrisis2018}, who  have found that  the inclusion of  orphan galaxies
for mass-based SHAM models improves  the clustering statistics of mock
galaxy catalogues, particularly so at small scales.

\section{Radial Profiles of Satellites in Large Clusters}\label{chap:radial-profiles}

\begin{figure}
  \centering
  \includegraphics[width=.7\linewidth, keepaspectratio]
	{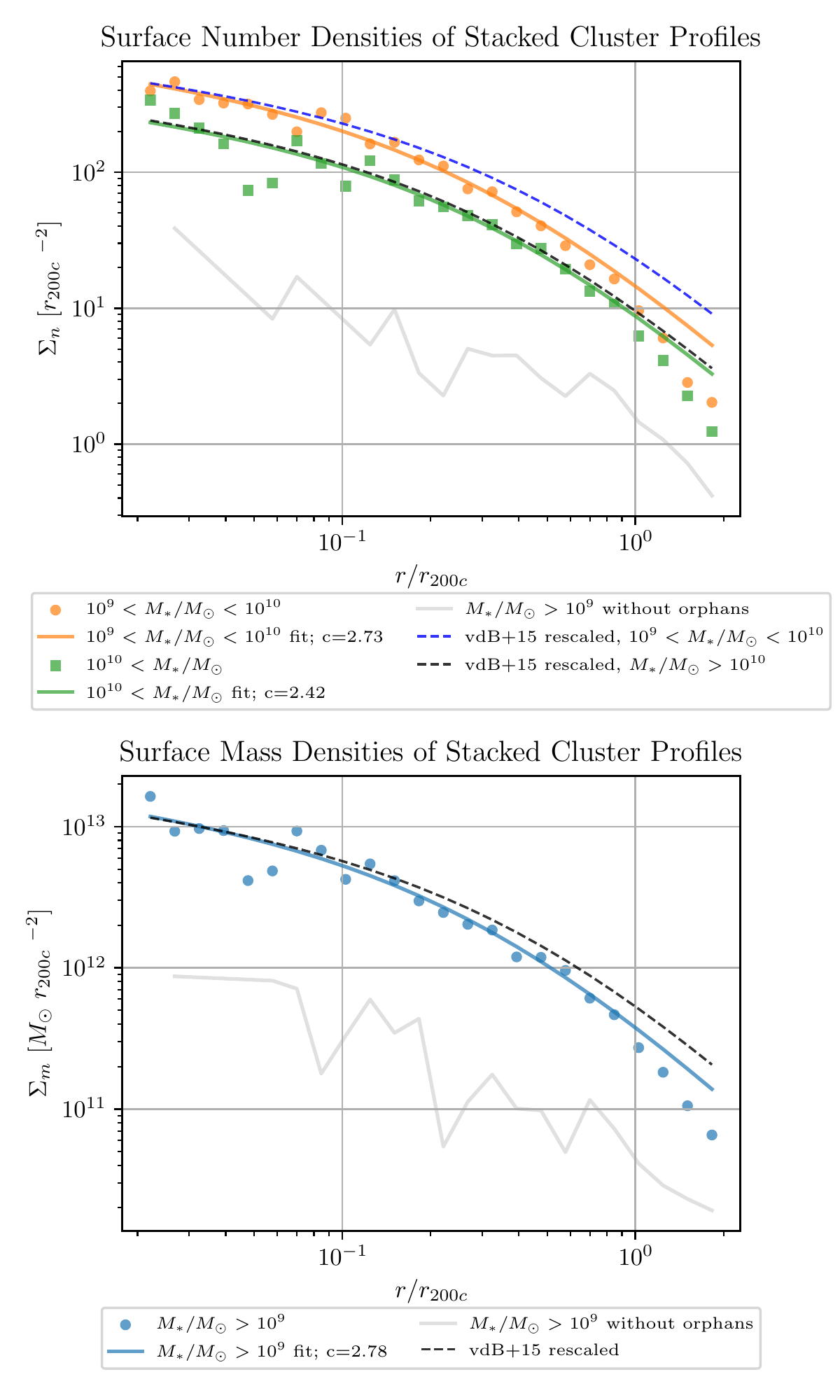}%
  \caption{Upper panel: the light grey  solid line shows the satellite
    number density profile averaged over  the 10 largest haloes in our
    simulation \emph{including only  satellite galaxies with a detected
    parent clump with masses above $10^9 \msol$}.   Satellites are
    then grouped  in two  mass bins
    indicated in  the legend. The colored  symbols show  the average
    number  density  profile  \emph{including  also  orphan  satellite
      galaxies}.  The colored solid  lines show the corresponding best
    fit projected NFW  profiles, while the  dashed lines  show the best
    fit projected NFW profiles of the observed sample in
    \citet{vanderburgEvidenceInsideoutGrowth2015}.  Note  that for the
    latter,  we have  renormalized the  surface density  to match  the
    lower median mass of the  simulated sample (see text for details).
    The lower  panel shows the  stellar mass surface  density averaged
    over our  simulated sample. Here  again the light grey  solid line
    shows the  mass density profile \emph{only for  satellite galaxies
      within a detected  clump} while the blue symbols  shows the same
    quantity \emph{also including  orphan satellite galaxies}. The blue
    solid  line corresponds  to our  best  fit projected NFW  profile while  the
    dashed line shows the best fit NFW profiles of the observed sample
    in \citet{vanderburgEvidenceInsideoutGrowth2015}.  Here again this
    last profile has  been renormalized to account  for the difference
    in  median mass  between the  simulated and  the observed  cluster
    samples.}
  \label{fig:radial-profiles}
\end{figure}

In this  section, we compute  the radial  profiles of number  and mass
densities  of satellite  galaxies in  the 10  largest clusters  in our
simulation. In  order to  compare to observations,  we will  follow as
closely      as     possible      the     method      described     in
\citet{vanderburgEvidenceInsideoutGrowth2015}, who analyzed 60 massive
clusters between 0.04<$z$<0.26 in  the Multi-Epoch Nearby Cluster Survey
and the Canadian Cluster Comparison Project.

We identified 10  haloes at $z=0$ with the highest  mass.  The mass is
defined here like  in \citet{vanderburgEvidenceInsideoutGrowth2015} as
$M_{200c}$, the  mass included within  radius $R_{200c}$ at  which the
average  enclosed  mass   density  of  the  halo  is   200  times  the
cosmological critical density $\rho_c$.   The $M_{200c}$ masses of our
10 selected  haloes range between  $1.4 \times 10^{14}\msol$  and $4.8
\times 10^{14}\msol$ with a median value of $1.6 \times 10^{14}\msol$.
Due to our limited box size, these  haloes are on the lower end of the
sample observed  in \citet{vanderburgEvidenceInsideoutGrowth2015}, who
reported masses ranging from $0.8 \times 10^{14}\msol$ and $1.6 \times
10^{15}\msol$ with a median value of $8.6 \times 10^{14}\msol$.

We compute  the projected number  density profiles and  projected mass
density  profiles as  follows: For  each  halo, we  first project  the
galaxies  (excluding the  central galaxy)  along each  coordinate axis
obtaining three  images.  We  then compute cylindrical  profiles using
radial shells equally space in log  radius in units of $r_{200c}$.  We
then average  the 30 profiles  (three projections for 10 clusters) to
obtain the final average radial surface density profile.  Each profile
is then fitted to a projected NFW profile \citep{navarroStructureColdDark1996b} using a
standard least square fitting procedure.   We obtain in particular the
concentration parameter that we can compare to the observed value.

The resulting profiles  are shown in Figure~\ref{fig:radial-profiles},
again including  and excluding  orphan galaxies. Orphan  galaxies play
here  also   a  crucial   role,  as   the  profiles   without  orphans
underestimate  the true  value by  an order  of magnitude.   Following
\citet{vanderburgEvidenceInsideoutGrowth2015},   we  adapt   the  same
galaxy  stellar   mass  $M_*$  thresholds   of  $10^9\msol  <   M_*  <
10^{10}\msol$ and $M_* > 10^{10}\msol$  for the surface number density
profile, and  a stellar mass threshold  of $M_* > 10^9  \msol$ for the
surface mass  density profile. It  is worth stressing that  this lower
mass threshold correspond exactly to  the mass resolution limit of our
mock galaxy catalogue.

As already noted above, the median  mass of the simulated and observed
catalogues  widely  differ.  In   order  to  facilitate  a  meaningful
comparison,  we  assume that  the  total  stellar mass  in  satellites
roughly  scales with  $M_{200c}$ in  the halo  mass range  of interest
here. We then adopt the following simple scaling relation
\begin{equation}
\Sigma \propto \frac{M_{200c}}{R_{200c}^2} \propto M_{200c}^{1/3}
\end{equation}
and    rescale    the    observed    average    profile    found    by
\citet{vanderburgEvidenceInsideoutGrowth2015}   using   this   scaling
relation  and  the  ratio  of  the  two  median  masses.  We  plot  in
Figure~\ref{fig:radial-profiles} the corresponding best fit projected NFW
profiles, showing excellent agreement with our simulated results.

Interestingly, the  observed surface density profiles  have a slightly
smaller concentration than  the simulated ones.  Although  we find for
the  number densities  of  satellite galaxies  above $10^{10}\msol$  a
concentration $c  = 2.4$,  in strikingly  good agreement  with $c=2.3$
found  by \citet{vanderburgEvidenceInsideoutGrowth2015}  for the  same
mass range, our results differ for the  mass range $10^9 \msol < M_* <
10^{10}        \msol$:       we        find       $c=2.7$        while
\citet{vanderburgEvidenceInsideoutGrowth2015} found  $c=1.8$. The same
mismatch is  found using the  mass density profile.  we  find $c=2.8$,
while \citet{vanderburgEvidenceInsideoutGrowth2015} found $c=2.0$.

We  believe  that this  mismatch  in  the concentration  parameter  is
consistent  with  the   difference  we  have  in   the  median  sample
mass. Indeed, the  theory predicts a larger  concentration for smaller
mass haloes, roughly in the amplitude observed here \citep{zhaoAccurateUniversalModels2009}.
We therefore  could in principle improve the agreement
between   our    simulation   results   and   the    observations   of
\citet{vanderburgEvidenceInsideoutGrowth2015}  by  also rescaling  the
radial direction according to the theoretical expectations. We believe
this is  beyond the  scope of this  paper to try  and fit  exactly the
data.

In   addition,    we   believe   there    is   much   more    to   the
story. \citet{vanderburgEvidenceInsideoutGrowth2015} found that larger
mass  satellite galaxies  have  a  significantly larger  concentration
parameter   than    the   low   mass   bin    ($c=2.3$   compared   to
$c=1.8$). Moreover, they have also  found a strong excess of satellite
galaxies compared to the best fit NFW  profile in the centre ($r < 0.1
R_{200c}$).   These observations  are  consistent with  the effect  of
dynamical friction  bringing the more  massive galaxies faster  to the
central regions of  the cluster. Dynamical friction is  expected to be
sufficiently efficient  for the  most massive  sub-haloes, with  a mass
larger than a few  percent of that of the host  halos
\citep[e.g.][]{binneyGalacticDynamicsSecond2008a,moGalaxyFormationEvolution2010}.
In our case, this  translates into
sub-haloes  more  massive  than  a  few  $10^{12}\msol$  and  satellite
galaxies  more  massive than  a  few  $10^{10}\msol$.  Our  simulation
clearly suffers from  numerical overmerging in this  mass range
\citep{vandenboschDisruptionDarkMatter2018}, as highlighted  by the
importance of including  orphan galaxies in
our  methodology.  Moreover,  our  pure DMO  parent simulation  cannot
follow precisely the many baryonic  effects that are needed to predict
accurately the  individual trajectory  of these higher  mass satellite
galaxies.

With these  caveats in mind,  we conclude  that using our  merger tree
code and a  state-of-the-art SHAM method, we can  model reasonably well
the cluster satellite galaxy number density and mass density profiles.

%% file: main/ACACIA/ACA6-conclusion.tex
\chapter{Conclusion}\label{chap:acacia-conclusion}

We presented \acacia, a new  algorithm to identify dark matter
halo  merger trees,  which is  designed  to work  \emph{on-the-fly}  on
systems  with  distributed  memory architectures,  together  with  the
adaptive mesh refinement code \ramses\ with its \emph{on-the-fly} clump
finder  \phew.  Clumps  of dark  matter are  tracked across  snapshots
through a user-defined  maximum number of most bound  particles of the
clump  $n_{\mathrm{mb}}$.   We found  that  using  $n_{\mathrm{mb}} \simeq  200$  tracer
particles is  a safe choice  to obtain  robust results for  our merger
tree  algorithm,  while   not  being  computationally  unrealistically
expensive.  We  also recommend adopting a  conservative mass threshold
of 200  particles per  clump to get  rid of a  few rare  spurious dead
branches that would need to be pruned from the halo catalogue anyway.

Additionally,  we examined  the  influence of  various definitions  of
substructure  properties on  the  resulting merger  trees. Whether  we
define substructures to contain their respective substructures' masses
or not  had negligible effect  on the merger trees.   However defining
particles  to  be  strictly  gravitationally  bound  to  their  parent
substructure  (by requiring  that  particles can't  leave the  spatial
extent of that  substructure) leads to better results,  with much less
extreme  mass growths  and extreme  mass growth  fluctuations of  dark
matter clumps.  We recommend to  use this strictly bound definition as
the  preferred  definition for  robust  merger  trees.  The  resulting
merger  trees  are  in   agreement  with  the  bottom-up  hierarchical
structure formation picture  for dark matter haloes.  The merger trees
of massive  haloes at $z=0$ have  more branches than their  lower mass
counterparts.  Their  formation history  can often  be traced  to very
high redshifts.

Once a progenitor  clump is merged into  a descendant, \acacia
keeps track of the progenitor's most strongly bound particle, called the
``orphan particle''.  It is possible for a temporarily merged sub-halo
to re-emerge from its host halo  at a later snapshot because it hasn't
actually dissolved  or merged completely,  but only because  it wasn't
detected  by the  clump finder  as a  separate density  peak.  Such  a
situation  is illustrated  in Figure  \ref{fig:jumper-demo}. In  these
cases,  orphan  particles  are  used   to  establish  a  link  between
progenitor  and  descendant   clumps  across  non-adjacent  snapshots.
By  default, \acacia will track orphans until the end of the
simulation, and orphans are only removed after they have indeed established
a link between a progenitor and descendant and thus have served their
purpose. Nonetheless, the current implementation
offers the option to remove orphan particles after a user defined number
of snapshots has passed. Keeping track of orphan particles indefinitely
might lead to misidentifications of progenitor-descendant pairs and
therefore to wrong formation histories. Our analysis shows however that
matches between progenitor-descendant pairs over an interval greater
than 10 snapshots are quite rare, so we expect this type of
misidentifications to be a negligible issue.

Compared to the test results in \citet{SUSSING_HALOFINDER}, our results
are comparable to e.g. the \codename{MergerTree}, \codename{TreeMaker} and
\codename{VELOCIraptor} tree builders with \codename{AHF}, \codename{Subfind},
or \codename{Rockstar} halo finders as presented in A14, demonstrating that
\acacia performs similarly to other state-of-the-art tools.
However, the performance is inferior to the one of the \codename{HBTtree}
algorithm, which together with the \codename{HBThalo} halo finder follows
structure from one timestep to the next and makes use of this information
when constructing both halo catalogues and trees. Furthermore we have encountered issues, e.g. main branch lengths of massive haloes being cut
short, due to failures in the \phew\ halo finder that we used. In those
cases, substructure changed their order in the hierarchy, and the subsequent
particle unbinding stripped particles from clumps in lower levels in the
hierarchy, preventing \acacia to establish any links between
progenitor and descendant clumps. The resolution of these issues with the
clump finder will be a high priority in future work, where we plan on
modifying the way \phew\ creates the hierarchies.
For example, we could define the peak hierarchy not based on the peak
density, but rather on the peak mass \citep[e.g. similarly to
\codename{AdaptaHOP},][]{aubertOriginImplicationsDark2004}. Additionally,
with the structure information from previous snapshots available now through
\acacia, further improvements can be made by taking this information
into account when constructing the sub-halo hierarchies, in a similar spirit
as \codename{HBThalo} does.

Orphan particles also serve a second purpose besides tracing disappearing
clumps. If we also
want to produce a mock galaxy catalogue on-the-fly using a dark matter
only simulation, the orphan particles are also used to track orphan
\emph{galaxies}. Those are  galaxies that  don't have  an associated
dark  matter  clump  any  longer  because  of  numerical  overmerging.
If we interpret orphan particles as orphan galaxies, there could be
additional reasons to consider stopping tracking them. For example, the
effects of dynamical friction makes them fall towards the central
galaxies. Once the orphan galaxies have lost enough energy, they may
find themselves in close proximity to the central galaxies, even below
the resolution limit. In these cases, it makes little sense to keep
track of these orphans as individual galaxies. They should rather be
regarded as merged into the central galaxy, and for that reason removed
from the list of tracked orphans. Given that the model we employ
doesn't provide us with the galaxy radii, this approach requires some
form of galaxy-galaxy merging cross-sections to compute the probability
of a collision between galaxies that will result in a galaxy merger.
A different approach that other models use is to estimate the time for
orphan galaxies to merge into the parent structure. This estimate could
be e.g. the dynamical friction time (as is done in \citet{mosterGalacticStarFormation2013}), or
the fitting formula for the merger timescale of galaxies in cold dark
matter models by \citet{jiangFittingFormulaMerger2008}. These physically motivated
approaches to remove orphans will be the subject of future work, where
we also intend to make use of the tools and methods presented in
\citet{poultonObservingMergerTrees2018} in order to improve our resulting
merger trees and mock galaxy catalogues.

Finally, as a  proof of concept and using the  known formation history
of  dark matter  clumps from  the merger  trees and  a widely  adopted
stellar-mass-to-halo-mass  relation  \citep{behrooziAVERAGESTARFORMATION2013}, we  generate  a
mock  galaxy  catalogue  from  a dark  matter  only  simulation.   The
influence of the merger trees on  the quality of the galaxy catalogues
is twofold.   First, while the stellar-mass-to-halo-mass  relation can
be directly  applied to  central galaxies  associated to  main haloes,
using the  peak clump  mass for  sub-haloes is  a better  approach for
satellite galaxies.   The reason is  that tidal stripping  of galaxies
inside  a dark  matter halo  sets in  much later  than for  their host
sub-halo \citep{nagaiRadialDistributionGalaxies2005}.  Second, without properly keeping track of all
merging  events, no  orphan  galaxies  can be  traced,  nor can  their
stellar  mass  be   estimated  through  the  stellar-mass-to-halo-mass
relation  unless  it's  known  from   which  halo  the  orphan  galaxy
originated from, and what properties this halo had in the past.

To highlight  the impact of  the merger trees, we  compute observables
from our  mock galaxy catalogue,  both including and  excluding orphan
galaxies.  Specifically,   we  compute  the  stellar   mass  two-point
correlation  functions   and  radial  profiles  of   projected  number
densities and projected stellar mass densities in galaxy clusters. When orphan
galaxies are included in the analysis, we obtain correlation functions
and radial  profiles in  good agreement with  observations, validating
the different steps in our overall methodology.


%
%
%

%% file: appendix/app1-primitive-variable-riemann.tex
\chapter{Approximate Linearized Primitive Variable Riemann Solver for the Euler Equations}
\label{app:riemann-primitive-variables}

Consider the one dimensional Euler equations in the primitive variable formulation:

\begin{align}
    \deldt \W + A(\W) \deldx \W = 0 &&
\end{align}

with the primitive state vector $\W$ and Jacobi matrix $A(\W)$

\begin{align}
    \W = \begin{pmatrix}
          \rho \\ v \\ p
         \end{pmatrix}
&&
    A(\W) = \frac{\del \F(\W)}{\del \W} = 
            \begin{pmatrix}
              v & \rho & 0 \\
              0 & v & 1/\rho \\
              0 & \rho c_s^2 & v
             \end{pmatrix}
\end{align}

The corresponding Eigenvalues of $A(\W)$ are

\begin{align}
 \lambda_1 = v - c_s \ , && \lambda_2 = v \ , && \lambda_3 = v + c_s
\end{align}

A simple linearized solution of the Riemann problem for the Euler equations can be obtained by 
assuming that the initial left and right states, $\W_L$ and $\W_R$, as well as the state of the 
star region for $t > 0$, $\W_*$, are close to some \emph{constant} state $\overline{\W}$, and that 
the Jacobi matrix can hence be approximated as

\begin{align}
\overline{A} = A(\overline{\W}) = \CONST
\end{align}

Using an expression for a ``primitive flux'' $\F_{\W} \equiv \overline{A} \W$, for a constant 
$\overline{A}$ we can write the primitive variable form of the Euler equations as a conservation 
law:

\begin{align}
 &\deldt \W + \overline{A} \deldx W = 0 \\
= &\deldt \W + \deldx \left(\overline{A}  W \right)  \\
= &\deldt \W + \deldx \F_{\W}
\end{align}

In this form (and furthermore approximating emanating waves as jump discontinuities), we can make 
use of the Rankine-Hugeniot relations (eq.~\ref{eq:rankine-hugeniot}) to relate fluxes $\F_{\W, k}$ 
and states $\W_k$ across each wave with associated characteristic speed $\overline{\lambda_k}$:

\begin{align}
    & \F_{\W, k + 1} - \F_{\W, k} = \lambda_k (\W_{k+1} - \W_k) \\
    = & \overline{A} \W_{k+1} - \overline{A} \W_k = \overline{A} (\W_{k+1} - \W_k) 
\end{align}

Applying the Rankine-Hugeniot relations across wave 1, which separates the states $\W_L$ and 
$\W_{*L}$ through a discontinuity with characteristic velocity $\overline{\lambda_1} = \overline{v} 
- \overline{c}_s$ gives us:

\begin{align*}
    \overline{v} (\rho_* - \rho_L)  + \overline{\rho} (v_* - v_L) &= 
        (\overline{v} - \overline{c}_s) (\rho_{*L} - \rho_L) \\
    \overline{v} (v_* - v_L) + \frac{1}{\overline{\rho}} (p_* - p_L) &= 
        (\overline{v} - \overline{c}_s) (v_* - v_L) \\
    \overline{\rho} \ \overline{c}_s^2 (v_* - v_L) + \overline{v} (p_* - p_L) &=
        (\overline{v} - \overline{c}_s) (p_* - p_L)
\end{align*}

Using the relation $c_s^2 = p / \rho$, it's easy to show that the third relation is exactly equal 
to the second one, so we'll leave it out in what follows. The first two relations can be simplified 
to:

\begin{align}
    \overline{\rho} (v_* - v_L) + \overline{c}_s (\rho_{*L} - \rho_L) &= 0 \label{eq:app-PV1}\\
    \frac{1}{\overline{\rho}} (p_* - p_L) + \overline{c}_s (v_* - v_L) &= 0 \label{eq:app-PV2}
\end{align}

Similarly, from the application of the Rankine-Hugeniot relations across wave 3, which has the 
characteristic velocity $\overline{\lambda}_3 = \overline{v} + \overline{c}_s$ we obtain:

\begin{align}
    \overline{\rho} (v_R - v_*) - \overline{c}_s (\rho_{R} - \rho_{*R}) &= 0 \label{eq:app-PV3} \\
    \frac{1}{\overline{\rho}} (p_R - p_*) - \overline{c}_s (v_R - v_*) &= 0 \label{eq:app-PV4}
\end{align}

Combining these four equations, we obtain relations for the primitive states of the star region:

\begin{align}
    p_* &= \frac{1}{2}(p_L + p_R) + 
            \frac{1}{2} (v_L - v_R) \overline{\rho} \ \overline{c}_s \\
    v_* &= \frac{1}{2}(p_L - p_R) + 
            \frac{1}{2} (v_L + v_R) \frac{1}{\overline{\rho} \ \overline{c}_s }\\
    \rho_{*L} &= \rho_L + (v_L - v_*) \frac{\overline{\rho}}{ \overline{c}_s }\\
    \rho_{*R} &= \rho_R + (v_* - v_R) \frac{\overline{\rho}}{ \overline{c}_s }
\end{align}

which only depend on the constant values $\overline{\rho}$ and $\overline{c}_s$. A sensible choice 
is to take the arithmetic mean:

\begin{align}
    \overline{\rho} &= \frac{1}{2} (\rho_L + \rho_R) \\
    \overline{c}_s &= \frac{1}{2} (c_{s,L} + c_{s,R})
\end{align}

which finally gives us

\begin{align}
    p_* &= \frac{1}{2}(p_L + p_R) + 
            \frac{1}{8} (v_L - v_R) (\rho_L + \rho_R)(c_{s,L} + c_{s, R}) \\
    v_* &= \frac{1}{2}(p_L - p_R) + 
             \frac{2 (v_L + v_R)}{ (\rho_L + \rho_R)(c_{s,L} + c_{s, R}) }\\
    \rho_{*L} &= \rho_L + (v_L - v_*) \frac{\rho_L + \rho_R}{c_{s,L} + c_{s,R} }\\
    \rho_{*R} &= \rho_R + (v_* - v_R)  \frac{\rho_L + \rho_R}{c_{s,L} + c_{s,R} }
\end{align}

%% file: appendix/app2-roes-theorem.tex
\chapter{Roe's Theorems On The Accuracy Of A Method For Scalar Conservation Laws}
\label{app:roe}

In this Appendix, Roe's theorems on the accuracy of a numerical method for the linear advection
equation are derived. The original reference for these theorems,
\cite{roep.l.NumericalAlgorithmsLinear1981}, is unfortunately unpublished, so a more practical
reference would be \cite{billetAccuracyStabilityExplicit1997}, who have derived the theorems for
two and three dimensional problems as well. Here, we only look at the one dimensional case.

Consider the linear advection equation

\begin{align}
 \deldt \uc + a \deldx \uc = 0 \label{eq:app-roe-linear-advection}
\end{align}

for a constant wave speed $a$. Consider a general scheme to solve the linear advection equation

\begin{align}
 \uc_i^{n+1} = \sum_\alpha A_\alpha u_{i+\alpha} ^ n \label{eq:app-roe-general-scheme}
\end{align}

where $\uc_i^n = \uc (i \Delta x, n \Delta t)$, and $\{A_\alpha \}$ is a finite set of constant
nonzero coefficients. $\alpha$ denotes the indices of the stencil of the method. For example in
Godunov's method (eq.~\ref{eq:godunov-advection}), the update formula for $\uc_i^{n+1}$ depends on
the states $\uc_{i-1}$ and $\uc_{i}$, and hence for that example $\alpha \in \{-1, 0 \}$.

The first theorem is as follows: If $\uc_i^n$ is a polynomial of degree $p$ in $i$,
scheme~\ref{eq:app-roe-general-scheme} will give the exact solution
to~\ref{eq:app-roe-linear-advection} if and only if

\begin{align}
 \sum_\alpha A_\alpha \alpha^q = (- C_{CFL})^q \label{eq:app-roe-theorem1}
\end{align}

for all integers $0 \leq q \leq p$, where $C_{CFL} = \frac{a \Delta t}{\Delta x}$.

To prove this theorem, we take some initial $\uc_i^{n} = \uc(i \Delta x, n \Delta t)$ to be a
polynomial of degree $p$:

\begin{align}
    \uc_i^{n} = \sum_{q = 0}^p \beta_q (i \Delta x)^q
\end{align}

where $\beta_q$ are some constant coefficients.

Performing one time step of the scheme~\ref{eq:app-roe-general-scheme} gives

\begin{align}
 \uc_i^{n+1}
    = \sum_\alpha A_\alpha \uc_{i+\alpha}^n
    =  \sum_\alpha A_\alpha \sum_{q = 0}^p \beta_q ((i + \alpha) \Delta x)^q
\end{align}

whereas the exact solution is given by

\begin{align}
\uc_{i,exact}^{n+1}
    = \uc_{i}^{n} (i \Delta x - a \Delta t, n \Delta t)
    = \sum_{q = 0}^p \beta_q (i \Delta x - a \Delta t)^q
\end{align}

Therefore the scheme gives the exact solution if and only if

\begin{align}
    &\uc_i^{n+1} = \uc_{i, exact}^{n+1} \\
    &\sum_\alpha A_\alpha \uc_{i+\alpha}^n =
    \sum_\alpha A_\alpha \sum_{q = 0}^p \beta_q ((i + \alpha) \Delta x)^q
    = \sum_{q = 0}^p \beta_q (i \Delta x - a \Delta t)^q
\end{align}

This equality must hold for each exponent $q$ individually, so we can look at each
summand with index $q$ individually:

\begin{align}
\sum_\alpha A_\alpha \beta_q ((i + \alpha) \Delta x)^q = \beta_q (i \Delta x - a \Delta t)^q
\end{align}

Dividing the equation by $\beta_q$ and $\Delta x^q$ gives

\begin{align}
\sum_\alpha A_\alpha (i + \alpha)^q = \left(i - \frac{a \Delta t}{\Delta x}\right)^q \ .
\end{align}

Since this must hold for all cell indices $i$, we can without loss of generality select $i = 0$ and
write

\begin{align}
\sum_\alpha A_\alpha \alpha^q = \left( - \frac{a \Delta t}{\Delta x}\right)^q = (- C_{CFL})^q
\end{align}

This relation must hold for each exponent $0 \leq q \leq p$ individually, and hence the first
theorem is proven.

It remains to demonstrate the second theorem, which states that a scheme of
form~\ref{eq:app-roe-general-scheme} that satisfies condition~\ref{eq:app-roe-theorem1} is $p$-th
order accurate.

A scheme is called $p$-th order accurate if the leading term of the point-wise error is proportional
to $\Delta x^p$. The point-wise error $P_i^{n+1}$ is defined as the difference between the exact
solution and the numerical solution divided by the time step:

\begin{align}
 P_i^{n+1} = \frac{\uc_{i,exact}^{n+1} - \uc_i^{n+1}}{\Delta t}
\end{align}

Without loss of generality, we once again consider the case for cell index $i = 0$. The point-wise
error is $P_0^{n+1}$ is then given by

\begin{align}
P_0^{n+1}
&= \frac{1}{\Delta t} \left[
    \uc_(i=0, (n+1) \Delta t) - \uc_{i=0}^{n+1}
    \right] \\
&= \frac{1}{\Delta t} \left[
    \uc(-a\Delta t, n \Delta t) - \sum_\alpha A_\alpha \uc_\alpha^n
    \right]
\end{align}

Now we Taylor-expand both the terms $\uc(-a \Delta t, n \Delta t)$ and $\uc_\alpha^n$ around $x =
0$ to obtain

\begin{align}
P_0^{n+1}
&= \frac{1}{\Delta t} \left[
    \sum_{q = 0}^\infty \frac{(-a \Delta t)^q}{q!} \frac{\del^q \uc^n}{\del x^q} -
    \sum_\alpha A_\alpha
        \left( \sum_{q=0}^\infty \frac{(\alpha \Delta x)^q}{q!} \frac{\del^q \uc^n}{\del x^q}
\right)
    \right] \\
&= \frac{1}{\Delta t} \sum_{q = 0}^\infty  \frac{1}{q!} \left[
    (-a \Delta t)^q - \Delta x^q \sum_\alpha A_\alpha \alpha^q
    \right] \frac{\del^q \uc^n}{\del x^q}  \\
&= \frac{1}{\Delta t} \sum_{q = 0}^\infty  \frac{\Delta x^q}{q!} \left[
    \left(\frac{-a \Delta t}{\Delta x} \right)^q - \sum_\alpha A_\alpha \alpha^q
    \right] \frac{\del^q \uc^n}{\del x^q} \\
&= \frac{1}{\Delta t} \sum_{q = 0}^\infty  \frac{\Delta x^q}{q!} \left[
    \left(- C_{CFL} \right)^q - \sum_\alpha A_\alpha \alpha^q
    \right] \frac{\del^q \uc^n}{\del x^q}
\end{align}

Using the property~\ref{eq:app-roe-theorem1}, it is obvious that the term in parentheses vanishes
for all $q \leq p$, and hence the leading term of the error is given by $q = p +1$:

\begin{align}
\frac{1}{\Delta t} \frac{\Delta x^{p + 1}}{(p + 1)!} \left[
    \left(- C_{CFL} \right)^{p + 1} - \sum_\alpha A_\alpha \alpha^{p + 1}
    \right] \frac{\del^{p + 1} \uc^n}{\del x^{p + 1}} \\
=
\frac{a \Delta x^p}{C_{CFL} (p + 1)!} \left[
    \left(- C_{CFL} \right)^{p + 1} - \sum_\alpha A_\alpha \alpha^{p + 1}
    \right] \frac{\del^{p + 1} \uc^n}{\del x^{p + 1}}
\end{align}

The leading term is proportional to $\Delta x^p$, and therefore a method that satisfies
condition~\ref{eq:app-roe-theorem1} will be $p$-th order accurate.

%% file: appendix/app3-psi_gradients_full_expressions.tex
\chapter{Full Expressions For Partitions of Unity And Their Gradients}
\label{app:psi-gradients-full}

In this appendix, the full expressions required to compute the partitions of unity, normalizations,
and derivatives of the partitions of unity are given. In what follows, we assume that the kernels
$W$ are defined as

\begin{align}
	W_i(\x) = W(\x - \x_i, h(\x)) = \frac{1}{h(\x)^\nu} w\left(\frac{| \x - \x_i |}{h(\x)} \right)
\end{align}

To compute the normalizations $\omega(\x_i)$ (eq.~\ref{eq:omega}) for a specific particle
position $\x_i$, we need to sum over all neighboring particles $j$:

\begin{align}
    \omega(\x_i) 	&= \sum_j W(\x_i - \x_j, h(\x_i))
\end{align}

with $h_i = h(\x_i)$.

For the \cite{ivanovaCommonEnvelopeEvolution2013} expression of the effective surfaces \Aij
(eq.~\ref{eq:IvanovaAij}), we need analytical gradients in Cartesian coordinates of $\psi_i(\x_j)$.

From eq.~\ref{eq:psi} we have that
\begin{align*}
	\psi_j(\x_i) &= \frac{ W(\x_i - \x_j, h(\x_i))}{\omega(\x_i)}\\
\end{align*}

Let $r_{ij} \equiv |\x_i - \x_j|$ and $q_{ij} \equiv \frac{r_{ij}}{h_i}$. Then the gradients are

\begin{align}
	\deldx \psi_j(\x_i) 	&= \deldx \frac{ W(\x_i - \x_j, h(\x_i))}{\omega(\x_i)}
							=  \deldx \frac{ W(r_{ij}, h_i)}{\omega(\x_i)}		\nonumber \\
							&= \frac{
									\DELDX{W}(r_{ij}, h_i) \  \omega(\x_i) -
									W(r_{ij}, h_i) \DELDX{\omega}(\x_i)
									}{\omega(\x_i)^2}		\nonumber  \\
							&= \frac{1}{\omega(\x_i)} \DELDX{W}(r_{ij}, h_i) -
							\frac{1}{\omega(\x_i)^2} W(r_{ij}, h_i) \DELDX{}\sum_k W(r_{ik}, h_i)		\nonumber \\
							&= \frac{1}{\omega(\x_i)} \DELDX{W}(r_{ij}, h_i) -
									\frac{1}{\omega(\x_i)^2} W(r_{ij}, h_i) \sum_k \DELDX{W}(r_{ik}, h_i)	\label{grad_psi}
\end{align}

The gradient of the kernel $W$ is given by

\begin{align}
	\deldx W_j (\x_i) 	&= \deldx \left( \frac{1}{h_i^\nu} w ( q_{ij} )	\right)	\nonumber \\
						&= \frac{1}{h_i^\nu} \frac{\del w(q_{ij})}{\del q_{ij}} \frac{\del q_{ij}(r_{ij})}{\del r_{ij}} \frac{\del r_{ij}}{\del x} 	\label{grad_kernel}
\end{align}

We now use
\begin{align}
	\frac{\del q_{ij}(r_{ij})}{\del r_{ij}} 	&= \frac{\del}{\del r_{ij}} \frac{r_{ij}}{h_i} = \frac{1}{h_i}		\label{dqdr} \\
	\DELDX{r_{ij}}		&= \deldx \sqrt{(\x_i - \x_j)^2}
							= \frac{1}{2} \frac{1}{\sqrt{(\x_i - \x_j)^2}} \cdot 2 (\x_i - \x_j) \nonumber \\
						&= \frac{\x_i - \x_j}{r_{ij}} 	\label{drdx}
\end{align}

To be perfectly clear, we should in fact write
\begin{align*}
	r_j(\x) & \equiv | \x - \x_j | \\
\end{align*}

which again leads to

\begin{align*}
	\DELDX{r_{ij}}		&= \DELDX{r_j(\x_i)} = \DELDX{r_j(\x)} \bigg{|}_{\x = \x_i} \\
						&= \deldx \sqrt{(\x - \x_j)^2} \big{|}_{\x = \x_i}
							= \frac{1}{2} \frac{1}{\sqrt{(\x - \x_j)^2}} \cdot 2 (\x - \x_j)  \big{|}_{\x = \x_i} \\
						&= \frac{\x - \x_j}{r_{j}(\x)} \big{|}_{\x = \x_i}
							= \frac{\x_i - \x_j}{r_{ij}}
\end{align*}

Inserting expressions \ref{dqdr} and \ref{drdx} in \ref{grad_kernel}, we obtain
\begin{equation}
	\deldx W_j (\x_i)  = \frac{1}{h_i^{\nu + 1}} \frac{\del w(q_{ij})}{\del q_{ij}}  \frac{\x_i - \x_j}{r_{ij}} 	\label{grad_kernel_final}
\end{equation}

Finally, inserting \ref{grad_kernel_final} in \ref{grad_psi} we get
\begin{align}
	\deldx \psi_j (\x_i) &=
		\frac{1}{\omega(\x_i)} \frac{1}{h_i^{\nu + 1}} \deldr{w}(r_{ij}, h_i)  \frac{\x_i -
\x_j}{r_{ij}} 	 - \nonumber\\
		& \frac{1}{\omega(\x_i)^2} W(r_{ij}, h_i) \sum_k \frac{1}{h_i^{\nu + 1}} \deldr{w}(r_{ik},
h_i)  \frac{\x_i - \x_k}{r_{ik}}
\end{align}

The definition of $r_{ij}$ requires a bit more discussion.
Since the kernels are required to be spherically symmetric, we might as well have defined

\begin{align}
r'_{ij} = |\x_j - \x_i |
\end{align}

which would leave the evaluation of the kernels invariant, but the gradients would have the opposite
direction:

\begin{align}
\DELDX{r'_{ij}}
=   \frac{\x_j - \x_i}{r'_{ij}}
= - \frac{\x_i - \x_j}{r_{ij}}
= - \DELDX{r_{ij}}
\end{align}

To show why the first choice is the correct one, consider a one-dimensional case where we choose
two particles $i$ and $j$ such that $x_j > x_i$ and $q_{ij} = | x_j - x_i | / h_i < H / h_i$, where
$H$ is the compact support radius of the kernel of choice.
Because we're considering a one-dimensional case with $x_j > x_i$, we can now perform a simple
translation such that particle $i$ is at the origin, i.e. $x'_i = 0$ and $x'_j = x_j - x_i = | x_j -
x_i | = r_{ij}$.
In this scenario, the gradient in Cartesian coordinates and in spherical coordinates must be the same:

\begin{align}
	\frac{\del}{\del x'} W (|x_i' - x'|, h_i)  \big{|}_{x' = x_j' } &= \frac{\del}{\del r_{ij}} W (r_{ij}, h_i) 		\nonumber\\
	\Rightarrow \quad \frac{1}{h_i^\nu} \frac{\del w(q'_{ij})}{\del q'_{ij}} \frac{\del q'_{ij}(r'_{ij})}{\del r'_{ij}} \frac{\del r'_{i}(x')}{\del x'}	  \big{|}_{x' = x_j' }  &=
		\frac{1}{h_i^\nu} \frac{\del w(q_{ij})}{\del q_{ij}} \frac{\del q_{ij}(r_{ij})}{\del r_{ij}}				\label{spher_cart}
\end{align}

We have the trivial case where
\begin{align*}
	r'_{ij} &= |x_i' - x_j'| = | x_i - x_j | = r_{ij} \\
	q'_{ij} &= r'_{ij}/h_i = q_{ij} \\
	\Rightarrow \quad \frac{\del w(q'_{ij})}{\del q'_{ij}} &= \frac{\del w(q_{ij})}{\del q_{ij}}, \quad \frac{\del q'_{ij}(r'_{ij})}{\del r'_{ij}} = \frac{\del q_{ij}(r_{ij})}{\del r_{ij}}
\end{align*}

giving us the condition from \ref{spher_cart}:

\begin{equation}
	\frac{\del r'_{i}(x')}{\del x'}	  \big{|}_{x' = x_j' } = 1 = \frac{\del r_{i}(x)}{\del x}
\end{equation}

this is satisfied for
\begin{align*}
	r_j(\x) &= | \x - \x_j |
\end{align*}

but not for

\begin{align*}
	r_j(\x) &= | \x_j - \x |
\end{align*}

%% file: appendix/app4-photon-number-to-luminosity.tex
\chapter{Converting Photon Number Emission Rates to Photon Energy Emission 
Rates}\label{app:number-to-luminosity}

Some radiative transfer codes use photon \emph{number} injection rates $\dot{N}_{\gamma}$ for their 
emission rates rather than the \emph{energy} injection rates $\dot{E}_\gamma$, or equivalently 
luminosities $L$. To convert between these two quantities, we need to assume that the emission 
follows some spectrum $J(\nu)$.

In the case of a single photon group, the conversion is quite simple: We first need to compute the 
average photon energy $\overline{E}_\gamma$:

\begin{align}
	\overline{E}_\gamma = \frac{\int J(\nu) \ \de \nu }{\int J(\nu) / (h \nu) \ \de \nu}
\end{align}

then the emitted luminosity (energy per unit time) is

\begin{align}
	L = \overline{E}_\gamma \ \dot{N}_{\gamma}
\end{align}

Note that in many cases, the given emission photon number rate is the number rate of \emph{ionizing} 
photons. This means that we need to start the integrals at the lowest ionizing frequency 
$\nu_{\text{ion, min}}$ in order to have the correct translation to the luminosity of the 
\emph{ionizing} energy:

\begin{align}
	\overline{E}_\gamma = \frac{\int\limits_{\nu_{\text{ion, min}}}^\infty J(\nu) \ \de \nu 
}{\int\limits_{\nu_{\text{ion, min}}}^\infty J(\nu) / (h \nu) \ \de \nu}
\end{align}

In the case of several photon frequency groups being used, the conversion requires a little 
adaptation in order to preserve the correct number of photons emitted. For each photon group $i$, 
the average photon energy is given by

\begin{align}
	\overline{E}_i = \frac{\int\limits_{\nu_{i \text{, min}}}^{\nu_{i \text{, max}}} J(\nu) \ \de 
\nu }{\int\limits_{\nu_{i \text{, min}}}^{\nu_{i \text{, max}}} J(\nu) / (h \nu) \ \de \nu}
\end{align}

Secondly, we need to compute the fraction $f_i$ of ionizing photons in each bin, which is given by

\begin{align}
	f_i = \frac{\int\limits_{\nu_{i \text{, min}}}^{\nu_{i \text{, max}}} J(\nu) / (h \nu)  \ \de 
\nu }{\int\limits_{\nu_{min}}^{\infty} J(\nu) / (h \nu) \ \de \nu}
\end{align}

Then the number of emitted photons in each bin is given by

\begin{align}
\dot{N}_i = f_i\ \dot{N}_\gamma
\end{align}

And the luminosities are given by

\begin{align}
	L_i &= \overline{E}_i \ \dot{N}_i \\
			&= \frac{\int\limits_{\nu_{i \text{, min}}}^{\nu_{i \text{, max}}} J(\nu) \ \de \nu 
}{\int\limits_{\nu_{min}}^{\infty} J(\nu) / (h \nu) \ \de \nu} \ \dot{N}_\gamma
\end{align}

%% file: appendix/app5-zero-flux-nonzero-energy.tex
\chapter{A Time Step For A Particle With Nonzero Radiation Energy Density And Zero Radiation Flux}
\label{app:zero-flux-nonzero-energy}

We could encounter cases where we have nonzero radiation energy, but zero radiation flux, for
example through diffusion, or when exception handling unphysical scenarios, or by intentionally only
injecting energy density.) In these cases, recall that (index $i$ below is for photon
frequency group, not particle index!)

\begin{align}
	\DELDT{\Fbf_i} + c^2 \ \nabla \cdot \mathds{P}_i &=
		- \sum\limits_{j}^{\mathrm{HI, HeI, HeII}} n_j \sigma_{i j} c \Fbf_i \\
	\mathds{P}_i &=
		\mathds{D}_i E_i \\
	\mathds{D}_i &=
		\frac{1- \chi_i}{2} \mathds{I} + \frac{3 \chi_i - 1}{2} \mathbf{n}_i \otimes \mathbf{n}_i\\
	\mathbf{n}_i &=
		\frac{\Fbf_i}{|\Fbf_i|} \\
	\chi_i &=
		\frac{3 + 4 f_i ^2}{5 + 2 \sqrt{4 - 3 f_i^2}} \\
	f_i &=
		\frac{|\Fbf_i|}{c E_i}
\end{align}

For $\Fbf_i = 0$, we get
\begin{align}
	f_i &= 0 \\
	\chi_i &= \frac{3}{5 + 2 \sqrt{4}} = \frac{1}{3} \\
	\mathds{D}_i &= \frac{1- \frac{1}{3}}{2} \mathds{I} = \frac{1}{3} \mathds{I} \\
	\mathds{P}_i &= \mathds{D}_i E_i =  \frac{1}{3} E_i \mathds{I}
\end{align}

which is the solution of the optically thick limit, where the radiation pressure tensor is
isotropic.
Now let's look at what happens when a particle $k$ has some nonzero energy $E_0$, but zero flux,
and interacts with particle $l$, which has both zero energy density and zero flux. The initial
state is then:

\begin{align}
    \mathcal{U}_k(t = 0) &= \left( \begin{matrix}
                    E_0 \\
                    \mathbf{0}
                  \end{matrix} \right)
    \quad & \quad \quad
    \F_k(t = 0)  &= \left( \begin{matrix}
                   \mathbf{0} \\
                   c^2 \mathds{P}
                  \end{matrix} \right)
                = \left( \begin{matrix}
                   \mathbf{0} \\
                   \frac{c^2}{3} E_0 \mathds{I}
                  \end{matrix} \right)\\
    \mathcal{U}_l(t = 0) &= \left( \begin{matrix}
                    0 \\
                    \mathbf{0}
                  \end{matrix} \right)
    \quad & \quad \quad
    \F_l(t = 0)  &= \left( \begin{matrix}
                   \mathbf{0} \\
                   0
                  \end{matrix} \right)
\end{align}

To simplify matters, let's assume both particles have equal volumes, $V_k = V_l = V$, and let's
omit the gradient extrapolation to the interface position that makes the method second order
accurate. Let's also assume that the particles are aligned parallel to a coordinate axis,  thus
making the projection along the normal vector to their interaction surface trivial.

Then the intercell flux given by the GLF Riemann solver (eq. \ref{eq:riemann-GLF}) is

\begin{align}
	\F_{1/2}(\mathcal{U}_L, \mathcal{U}_R) &=
		\frac{\F_{L} + \F_{R}}{2} -
		\frac{c}{2} \left(\mathcal{U}_R - \mathcal{U}_L \right) \\
	&=	\left(\begin{matrix}
        \frac{0 - 0}{2} - \frac{c}{2} (0 - E_0)\\
        \frac{c^2 / 3 \ E_0 - 0}{2} - \frac{c}{2} (0 -0)
	  	\end{matrix} \right)
	=	\left(\begin{matrix}
        \frac{c}{2} E_0 \\
        \frac{c^2}{6} E_0
	  	\end{matrix} \right)
\end{align}

Let $\beta \equiv \frac{\Delta t \mathcal{A}_{kl} c}{2 V}$. Using the update formula given in eq.
\ref{eq:transport-update-explicit}, the states at $t = \Delta t$ will be

\begin{align}
	\mathcal{U}_k(t = \Delta t) &= \mathcal{U}_k (t = 0) - \frac{\Delta t}{V} \F_{1/2}
\mathcal{A}_{kl} \\
	&=	\left(\begin{matrix}
        E_0 -  \frac{\Delta t \mathcal{A}_{kl}}{V} \frac{c}{2} E_0 \\
        \mathbf{0} - \frac{\Delta t \mathcal{A}_{kl}}{V} \frac{c^2}{6} E_0
	  	\end{matrix} \right)
    = \left( \begin{matrix}
             E_0 (1 - \beta) \\
             -\frac{\beta}{3} c E_0
             \end{matrix} \right) \\
	\mathcal{U}_l(t = \Delta t) &= \mathcal{U}_l (t = 0) - \frac{\Delta t}{V} (-F_{1/2})
\mathcal{A}_{kl} \label{eq:zero-flux-update-l}\\
	&=	\left(\begin{matrix}
        0 +  \frac{\Delta t \mathcal{A}_{kl}}{V} \frac{c}{2} E_0 \\
        \mathbf{0} + \frac{\Delta t \mathcal{A}_{kl}}{V} \frac{c^2}{6} E_0
	  	\end{matrix} \right)
    = \left( \begin{matrix}
             \beta E_0 \\
             \frac{\beta}{3} c E_0
             \end{matrix} \right)
\end{align}

where the minus sign for the inter-cell flux $(-\F_{1/2})$ in eq. \ref{eq:zero-flux-update-l} stems
from the orientation of the effective surface $\mathcal{A}_{kl}$.

The takeaway here is that after a single time step, both particles $k$ and $l$ will have nonzero
energy \emph{and} nonzero photon flux $\Fbf$. For the photon fluxes of the particle $l$, we see that
$|\Fbf_l| = \frac{1}{3} c E_l$ with $E_l = \beta E_0$, meaning that the photon flux corresponds to
the optically thick, diffusion limit. For particle $k$ we have $|\Fbf_k| = \frac{\beta}{1 - \beta} c
E_k$ with $E_k = (1 - \beta) E_0$. Comparing the particle volume $V$ and effective surface
$\mathcal{A}_{kl}$ to a regular cell of edge length $L$, then $V = L^3$ and $\mathcal{A}_{kl} =
L^2$. Simultaneously this gives us a comparative CFL condition: $L \geq c \Delta t$. This simple
comparison limits the possible values of $\beta$ to $0 \leq \beta \leq \frac{1}{2}$, given that all
terms of $\beta$ must be positive. The limits translate to $0 \leq |\Fbf_k| \leq c E_k$ for the
photon flux, meaning that particle $k$ can end up in any state between no net flux, and the free
streaming, optically thin limit.